\documentclass{report}
\title{Algorithmic Information Theoretic Issues in Quantum Mechanics}
\author{Gavriel Segre - PHD thesis}
\date{29-1-2004 12:53}
\usepackage{amsmath,amssymb,graphicx}
\numberwithin{equation}{section}
\newtheorem{definition}{DEFINITION}[section]
\newtheorem{theorem}{Theorem}[section]
\newtheorem{lemma}{Lemma}[section]
\newtheorem{corollary}{Corollary}[section]
\newtheorem{conjecture}{Conjecture}[section]
\newtheorem{remark}{Remark}[section]
\newtheorem{axiom}{AXIOM}[section]
\newtheorem{constraint}{CONSTRAINT}[section]
\newtheorem{diagram}{DIAGRAM}[section]
\newenvironment{hypothesis}{HP: \begin{center}} {\end{center}}
\newenvironment{thesis}{TH: \begin{center}} {\end{center}}
\newenvironment{proof}{\begin{center}PROOF: \end{center}} {$ \blacksquare $}
\newtheorem{example}{Example}[section]
\begin{document}
\maketitle
\tableofcontents
\part{Preliminaries}
\section{Warning}
The project of this work, going  beyond the possibility of
realization (at least mine) during doctoral studies, has not been
completed and has to be considered as \textbf{open}.

To underline the intellectual path I tried to pursue, I have
conserved the title and mentioning to the unrealized sections
instead of eliminating them, thinking that they add anyway some
cbit of classical information.
\section{Acknowledgments}
First of all I would like to thank my PHD-tutor, prof. Rimini, for
having  believed I was worth to be allowed to follow my own way.

\smallskip

Then I would like to thank doct. Benatti for having read,
analyzed and often constructively criticized many parts of this
work.

In particular I am grateful to him for clarifying me the key
point why some diagonalization-proofs don't generalize
noncommutatively.

\smallskip

I then want  to thank strongly prof. Jona-Lasinio for many remarks
and teachings and, in particular, for having suggested me the idea
of considering sequences of free coin tosses in view of
Voiculescu's Central Limit Theorem.

\smallskip

Then I want to thank prof. Rasetti who introduced me to Word
Problems, as well as to the Turing Barrier Problem of Quantum
Mechanics and all the related issues.

\smallskip

I am very grateful to  prof. Van Lambalgen for having given me a
copy of his wonderful dissertation.

\smallskip

Then I would like to thank doct. Winter for having suggested me to
partecipate to the Second Bielefeld Workshop on Quantum
Information and Computation, giving  me the motivations to pursue
walking on my way.

\smallskip

I would like to thank prof. Calude and prof. Svozil for many
precious teachings.

\smallskip

I would like to thank prof. Peres for useful indications.

\smallskip

I would like to thank prof. Landi and dr. Achieri for clarifying
me some key point about Noncommutative Geometry

\smallskip

Last but not least, I would like to thank prof. Odifreddi for
useful suggestions.

\newpage
\section{Notation}

 \bigskip
\begin{tabular}{|c|c|}
  $ \forall $ & for all (universal quantificator) \\
  $ \exists $ & exist (existential quantificator) \\
  $ \exists \; ! $ & exist and is unique \\
  $ x \; = \; y $ & x is equal to y \\
  $ x \; := \; y $ & x is defined as y \\
  $ 2^{S} $ & power-set of the set S \\
  $ S^{0} $ & interior of the topological space S \\
  $ \bar{S} $ & closure of the topological space S  \\
  $ {\mathcal{D}} ( A_{1} \, , \,  A_{2} ) $ & description methods
  of $ A_{2} $ through  $ A_{1} $ \\
  $ {\mathcal{R}} $ & universe of description \\
  $ \Sigma $ & binary alphabet $ \{ 0 , 1 \} $  \\
  $ \Sigma_{n} $ & n-letters' alphabet \\
  $ \Sigma^{\star} $ & strings on the alphabet $ \Sigma $  \\
  $ \Sigma^{\infty} $ &  sequences on the alphabet $ \Sigma $ \\
  $ \vec{x} $ & string \\
  $ \lambda $ & empty string \\
  $ | \vec{x} | $ &  length of the string $ \vec{x} $ \\
  $ string(n) $ & $ n^{th} $ string in quasilexicographic order \\
  $ |n | $ & length of the $ n^{th} $ string in quasilexicographic order \\
  $ \bar{x} $ & sequence \\
  $ <_{p} $ & prefix order relation \\
  $ \cdot $ & concatenation operator \\
  $ x_{n} $ & $ n^{th} $ digit of the string $\vec{x} $ or of the sequence $ \bar{x} $ \\
  $ \vec{x}(n) $ & prefix of length n of the string $ \vec{x} $ or of the sequence $ \bar{x} $  \\
  $ \vec{x}(n,m) $ & substring of the sequence $ \bar{x}
  $ obtained taking the digits from the $n^{th}$ to the $m^{th}$ \\
  $ \vec{x}^{n} $  &  string made of n repetitions
of the string  $ \vec{x} $  \\
  $ \vec{x} ^{\infty}  $ & sequence made of infinite repetitions
of the string $ \vec{x} $ \\
 $ S \, \Sigma^{ \infty } $ & sequences having the
strings of S as prefixes \\
$ \vec{x} \Sigma^{\infty} $ & sequences having
the string  $ \vec{x} $ as prefix \\
  $  N_{i}( \vec{y}) $ & number of occurence of the letter i in the string $
  \vec{y} $ \\
  $ N_{i}^{n} ( \bar{x} ) $ & number of successive letters i ending in position n of the sequence $ \bar{x} $ \\
  $ {\mathcal{I}} ( a , n , \vec{b} ) $ & string obtained inserting the letter a at the $ n^{th} $ place of
the string $ \vec{b} $ \\
 $ L_{D,P} $ & average code-word length w.r.t. the code D and the  distribution P \\
 $ H(P) $ & Shannon's entropy of the distribution P \\
 $ K( \vec{x} ) $ & simple algorithmic entropy of the string $
 \vec{x} $ \\
 $ I( \vec{x} ) $ & prefix algorithmic entropy of the string $
 \vec{x} $ \\
 $ \Sigma(n) $ & busy-beaver function \\
 $ P_{U} ( \vec{x} ) $ & universal probability of the string $
 \vec{x}$ w.r.t. the universal Chaitin computer U \\
 $ \Omega_{U}  $ & halting probability of the universal Chaitin computer U
 \\ \hline
 \end{tabular}
 \newpage
\begin{tabular}{|c|c|}
 $ CHAITIN-m-RANDOM ( \Sigma^{\star}) $ & Chaitin-m-random strings \\
$ CHAITIN-RANDOM ( \Sigma^{\star}) $ & Chaitin-random strings \\
 $ {\mathcal{N}} ( \bar{x} ) $ & numeric representation of the  sequence $ \bar{x} $ \\
 $ CHAITIN-RANDOM(\Sigma^{\infty} )$ & Martin L\"{o}f- Solovay -Chaitin-random sequences \\
 $ BRUDNO-RANDOM(\Sigma^{\infty} )$ & Brudno-random sequences \\
 $ q-PSEUDORANDOM ( \Sigma^{\star} , V ) $ & pseudorandom strings of level q  w.r.t. V
 \\
$ MARTINL\ddot{O}F-q-RANDOM( \Sigma^{\star} ) $ & Martin
L\"{o}f  pseudorandom strings of level q \\
$ \mu-RANDOM( \Sigma^{\infty} \, , \, \delta ) $ & Martin L\"{o}f
$\mu$-random sequences w.r.t.  $ \delta $ \\
$ \mu-RANDOM( \Sigma^{\infty}) $ & Martin L\"{o}f $\mu$-random
sequences \\
$ {\mathcal{P}} (M) $ & unary predicates over the set M \\
$ {\mathcal{P}}_{TYPICAL} (CPS) $ & typical
properties of  CPS \\
$ {\mathcal{L}}_{RANDOMNESS} (CPS) $ & laws of
randomness  of  CPS \\
$ P-CONF-RANDOM( \Sigma^{\infty} ) $ & conformistically-random
sequences w.r.t.  P \\
$ CONF-RANDOM( \Sigma^{\infty} ) $ &  conformistically-random sequences \\
$ EXT[S] $ & subsequence extraction function w.r.t. S \\
$ CHURCH-RANDOM(\Sigma^{\infty} )$ & Church-random sequences \\
 $ A \, \bigvee \, B $ & coarsest refinement of the partitions A
 and B \\
 $ h_{CDS} $ & Kolmogorov-Sinai entropy of CDS \\
  $ {\mathbb{N}} $ & natural numbers  \\
  $ {\mathbb{Z}} $ & integers numbers \\
  $ {\mathbb{A}}_{n} $ & algebraic numbers of order n \\
  $  {\mathbb{A}} $ & algebraic numbers  \\
  $ {\mathbb{Q}} $ & rational numbers \\
  $ {\mathbb{R}} $ &  real numbers \\
  $ {\mathbb{C}} $ & complex numbers \\
  $ {\mathbb{C}}P^{n} $ & complex projective space of order n \\
  $ G_{k,n} ({\mathbb{C}}) $ & complex Grassmann manifold of order
  (k,n) \\
  $ \aleph_{n} $ & $ (n+1)^{th} $ infinite cardinal \\
  $ \Re(z) $ & real part of the complex number z \\
  $ \Im(z) $ & imaginary part of the complex number z \\
  $ f_{1} \, \stackrel{ + }{\leq} \, f_{2} $ & $ f_{1} $ is
  additively less or equal to $ f_{2} $ \\
  $ f_{1} \, \stackrel{ + }{=} \, f_{2} $ & $ f_{1} $ is additively equal to $ f_{2} $ \\
  $ f_{1} \, \stackrel{ \times }{\leq} \, f_{2} $ & $ f_{1} $ is
  multiplicatively less or equal to $ f_{2} $ \\
  $ f_{1} \, \stackrel { \times }{=} \, f_{2} $ & $ f_{1} $ is
  multiplicatively equal to $ f_{2} $ \\
  $ a \, | \, b $ & a divides b \\
  $ a \, \nmid \, b $ & a does not divide b  \\
  gcd(a,b) & greatest common divisor of (a,b) \\
  lcm(a,b) & least common multiple of (a,b) \\
  $ \lfloor x \rfloor $ & floor of x: the greater integer less than or equal to x \\
  $ \lceil x \rceil $ & ceiling of x: the least integer greater than or equal to x  \\
  $ x \; mod \, n $ & remainder: $ x - n  \lfloor \frac{x}{n} \rfloor $ \\
  $ MAP(A,B) $ & maps from A to B \\
  $ f : A \mapsto B $ & map from A to B \\
  $ \stackrel{ \circ } {MAP}(A,B) $ & partial maps from A to B \\ \hline
\end{tabular}
\newpage
\begin{tabular}{|c|c|}
  $ f : A \stackrel{ \circ } { \mapsto } B $ & partial map from A to B \\
  $ C_{M} \, ( NC_{M} ) $ & mathematically-classical (mathematically-nonclassical) \\
  $ C_{\Phi} \, ( NC_{\Phi} ) $ & physically-classical (physically-nonclassical) \\
  $ REC $ & recursive \\
  $ \Delta_{0}^{0} $ & computable \\
  $ \varphi_{e}^{(n)} $ & n-ary partial recursive function with
  G\"{o}del number e \\
  $ {\mathcal{W}}_{e}^{(n)} $ & halting set of $  \varphi_{e}^{(n)} $
  \\
  $ {\mathcal{L}}({\mathcal{H}}) $ & lattice of all the closed linear subspaces of the Hilbert space $ {\mathcal{H}} $ \\
  $ {\mathcal{O}} ({\mathcal{H}} ) $ & linear operators on the Hilbert space $ {\mathcal{H}}
  $ \\
  $ {\mathcal{B}}({\mathcal{H}}) $ & bounded linear operators on the Hilbert space $ {\mathcal{H}} $ \\
  $ \| \cdot \|_{n} $ & $ n^{th} $ operatorial norm on $ {\mathcal{B}}({\mathcal{H}})
  $ \\
  $ | a | $ & modulus of the bounded operator a \\
  $ {\mathcal{C}}_{n} ({\mathcal{H}}) $ &  n-class bounded operators  on the Hilbert space $ {\mathcal{H}} $ \\
  $ {\mathcal{C}}_{1} ({\mathcal{H}}) $ &  trace-class bounded operators  on the Hilbert space $ {\mathcal{H}} $ \\
  $ {\mathcal{C}}_{2} ({\mathcal{H}}) $ &  Hilbert-Schmidt bounded operators  on the Hilbert space $ {\mathcal{H}} $ \\
  $ {\mathcal{C}} ({\mathcal{H}}) $ &  noncommutative infinitesimals  on the Hilbert space $ {\mathcal{H}} $ \\
  $ {\mathcal{I}}_{\alpha} ({\mathcal{H}}) $ &  noncommutative infinitesimals  of order $ \alpha $ on the Hilbert space $ {\mathcal{H}} $ \\
  $  {\mathcal{D}}(M) $ & classical probability
  distributions over M \\
  $  {\mathcal{D}} ({\mathcal{H}}) $ & density
  operators over the Hilbert space $ {\mathcal{H}} $ \\
  $ D_{T} ( \vec{p}^{(A)} , \vec{p}^{(B)} ) $ & classical trace
  distance among  $  \vec{p}^{(A)} $ and $  \vec{p}^{(B)} $ \\
  $ D_{T} ( \rho^{(A)} , \rho^{(B)} ) $ & quantum  trace
  distance among  $  \rho^{(A)} $ and $  \rho^{(B)} $ \\
   $ F ( \vec{p}^{(A)} , \vec{p}^{(B)} ) $ & classical fidelity among  $  \vec{p}^{(A)} $ and $  \vec{p}^{(B)} $ \\
   $ F ( \rho^{(A)} , \rho^{(B)} ) $ & quantum fidelity among  $  \rho^{(A)} $ and $  \rho^{(B)} $ \\
   $ D_{A} ( \vec{p}^{(A)} , \vec{p}^{(B)} ) $ & classical angle
  distance among  $  \vec{p}^{(A)} $ and $  \vec{p}^{(B)} $ \\
  $ D_{A} ( \rho^{(A)} , \rho^{(B)} ) $ & quantum  angle
  distance among  $  \rho^{(A)} $ and $  \rho^{(B)} $ \\
  $ A_{+} $ & positive part of the $ \star $-algebra A \\
  $ A_{p.s.d} $ & part with discrete spectrum of the $ C^{\star} $-algebra
  A \\
  $ A_{sa} $ & self-adjoint part of the $ \star $-algebra A \\
  $ PUR( \rho \, , \, {\mathcal{H}} ) $ & purifications
  of the density operator $ \rho $ w.r.t. the Hilbert space $
  {\mathcal{H}} $ \\
  $ {\mathcal{U}}(A) $ & unitary group of the $ \star $-algebra A \\
  $ {\mathcal{P}}(A) $ & projections of the $ \star $-algebra A \\
  $ QL(A) $ & quantum logic of the $W^{\star} $-algebra A \\
  $ S(A) $ & states on the $ W^{\star} $-algebra A \\
  $ S(A)_{n} $ & normal states on the $ W^{\star} $-algebra A \\
  $ \rho_{\omega} $ & density operator of the normal state $ \omega $ \\
  $ \omega_{\mu} $ & state of the classical probability measure $ \mu $ \\
  $ \Xi(A) $ &  pure states on the $ W^{\star} $-algebra A \\
  $ POINTS(A) $ & points on the $ W^{\star} $-algebra A \\
  $ Sp(a) $ & spectrum of a \\
  $ S_{\tau} ( A ) $ &  $ \tau $ - invariant states on the $ W^{\star} $-algebra
  A \\
  $ \Xi_{\tau} ( A ) $ &  $ \tau $ - invariant pure states on the $ W^{\star} $-algebra
  A \\
  $ \Delta_{\omega}(a) $ & dispersion of the state $ \omega $ on
  the element a of the $ W^{\star} $-algebra A \\
  $ ( \mathcal{H}_{\omega} \, , \, \pi_{\omega} \, , \, | \Omega_{\omega} > ) $ &
  GNS-representation of the $C^{\star}$-algebra A w.r.t. the state
  $ \omega $ \\
  $ AUT(A) $ & automorphisms of the $ W^{\star} $-algebra A \\
  $ INN(A) $ & inner automorphisms of the $ W^{\star} $-algebra A \\
  $ OUT(A) $ & outer automorphisms of the $ W^{\star} $-algebra A \\  \hline
 \end{tabular}
\newpage
\begin{tabular}{|c|c|}
   $ GR-AUT( G \, , \, A) $ & automorphisms' groups of A representing
  G \\
  $ GR-INN( G \, , \, A) $ & inner automorphisms' groups of A representing
  G \\
  $ GR-OUT( G \, , \, A) $ & outer automorphisms' groups of A representing
  G \\
  $ CPU(A,B) $ & channels from the $ W^{\star} $-algebra A to
  the $ W^{\star} $-algebra B \\
  $ CPU(A) $ & channels on the $ W^{\star} $-algebra A
  \\
  $ \alpha_{\star} $ & dual of the channel $ \alpha $ \\
  $ OPU(A) $ & operational partition of unity over the $ W^{\star} $-algebra A \\
  $ \{ \alpha_{i}( {\mathcal{V}}) \} $ & channels' set of the
  operational partition of unity $ {\mathcal{V}} $ \\
  $ R ( {\mathcal{V}} ) $ & reduction channel  of the
  operational partition of unity $ {\mathcal{V}} $ \\
  $ A' $ & commutant of the Von Neumann algebra A \\
  $ {\mathcal{Z}}(A) $ & centre of the Von Neumann algebra A \\
  R & hyperfinite $ II_{1} $-type factor \\
  $ R_{\lambda} $ & hyperfinite $ III_{\lambda} $-type factor $ ( 0 \, < \, \lambda \, < \,  1 ) $ \\
  $ cardinality_{NC} (A) $ & noncommutative cardinality of the
  noncommutative set A \\
  $ \Sigma_{NC} $ & noncommutative binary alphabet \\
  $ \Sigma_{NC}^{\star} $ & noncommutative space of qubits' strings  \\
  $ \Sigma_{NC}^{\infty} $ &  noncommutative space of qubits'
  sequences \\
  $ M_{n} (a) $ & $ n^{th} $ moment of the algebraic random
  variable a \\
  $ E (a) $ &  expectation value of the algebraic random
  variable a \\
  $ Var(a) $ & variance of the algebraic random
  variable a \\
  $ Z_{a} $ & characteristic function of the algebraic random
  variable a \\
  $ \mu_{a} $ & classical probability measure of the self-adjoint
  algebraic random variable a \\
   $ v_{a} $ & result of the measurement of the self-adjoint
  algebraic random variable a \\
  $ ZQ_{a}(t) $ & characteristic function of the noncommutative
  random variable a \\
  $ ZC_{Ap}(t) $ & characteristic function of the classical
  approximation Ap \\
  $ END(  \, A \, , \, \omega \, ) $ & endomorphisms of
  the algebraic probability space $ (  \, A \, , \, \omega \, ) $ \\
  $ AUT(  \, A \, , \, \omega \, ) $ & automorphism of
  the algebraic probability space $ (  \, A \, , \, \omega \, ) $ \\
  $ tr_{\omega} $ & Dixmier trace \\
  $ O(n) $ & $ n^{th} $ orthogonal group \\
  $ Spin(n) $ & $ n^{th} $ spin group \\
  $ \nabla $ & Levi-Civita connection of the (pseudo)riemannian manifold $
  ( M \, , \, g ) $ \\
  $ \triangle_{g} $ & Laplace-Beltrami operator on the (pseudo)riemannian manifold $
  ( M \, , \, g ) $ \\
  $ d \mu (g) $ & metric measure of the (pseudo)riemannian manifold $
  ( M \, , \, g ) $ \\
  $ Is[ ( M \, , \, g )] $ & isometries' group of the (pseudo)riemannian manifold $
  ( M \, , \, g ) $ \\
  $ \Gamma ( M , E ) $ & sections of the fibre bundle $ E \, \stackrel{\pi}{\rightarrow} \, M
  $ \\
  $ O(M) \, \stackrel{\pi}{\rightarrow} \, M $ & orthonormal frame bundle
  of the riemannian manifold $ ( M \, , \, g ) $ \\
  $ S(M) \, \stackrel{\pi}{\rightarrow} \, M $ & spin bundle on the spin manifold  $ ( M \, , \, g ) $ \\
  $ C(M) \, \stackrel{\pi}{\rightarrow} \, M $ & Clifford bundle on the spin manifold  $ ( M \, , \, g ) $ \\
  $ \int_{NC} $ & noncommutative integral on the spectral triple $
  ( A \, , \, {\mathcal{H}} \, , \, D ) $ \\
  $ d_{NC} $ & noncommutative differential on the spectral triple $
  ( A \, , \, {\mathcal{H}} \, , \, D ) $ \\
  $ J[ p(z) ] $ & Julia set of the polynomial p(z) on the complex
  field \\
  $ d \Lambda_{D} $ & Hausdorff measure on a set with Hausdorff
  dimension D \\
  $ {\mathcal{M}} $ & Mandelbrot's set \\
  $ d ( \omega_{1} , \omega_{2} ) $ & noncommutative geodesic distance between two states \\
  $ ( \, D \omega_{1} \, : \,  D \omega_{1} \, )_{t} $ &
  noncommutative Radon-Nikodym derivative between two states \\
  $ \sigma^{\omega}_{t} $ & modular group of the state $ \omega $
  \\
  $ <  \chi \,  | \, {\mathcal{R}}  > $ & presentation with generating system $ \chi $ and defining relators $ {\mathcal{R}} $ \\ \hline
\end{tabular}
\newpage
\begin{tabular}{|c|c|}
   $ F_{n} $ & free group of rank n \\
  $ l^{2}(G) $ & Hilbert space of the discrete group G \\
  $ {\mathcal{L}} (G) $ & left group Von Neumann algebra of the
  discrete group G \\
  $ {\mathcal{R}} (G) $ & right group Von Neumann algebra of the
  discrete group G \\
  $ C_{n} $ & $ n^{th} $ Catalan number \\
  $ CSM ( M , N ) $ & classical statistical model w.r.t M and N \\
  $ g_{CSM ( M , N )} $ & Fisher-Rao riemannian metric w.r.t. CSM ( M , N ) \\
  $ \hat{\Pi}_{random} $ & Coleman-Lesniewski's operator \\
  $ I_{Q} ( | \psi > ) $ & Svozil's quantum algorithmic
  information of $ | \psi > $ w.r.t. Q \\
  $ {\mathcal{P}}_{C} (A) $ & commutative predicates over
  the  $ W^{\star}$-algebra A \\
   $ {\mathcal{P}}_{NC} (A) $ & noncommutative predicates over
  the  $ W^{\star}$-algebra A \\
  $ {\mathcal{P}}_{C}^{TYPICAL} (APS) $ & typical commutative properties of APS  \\
   $ {\mathcal{P}}_{NC}^{TYPICAL} (APS) $ & typical noncommutative properties of APS  \\
  $ KOLMOGOROV_{C}(APS) $ &  Kolmogorov
  commutatively-random elements of APS  \\
  $ KOLMOGOROV_{NC}(APS) $ & Kolmogorov
  noncommutatively-random elements of  APS  \\
  $ {\mathcal{L}}_{RANDOMNESS}^{C} (APS) $ & commutative laws of
  randomness of APS \\
  $ {\mathcal{L}}_{RANDOMNESS}^{C} (APS) $ & noncommutative laws of
  randomness of APS \\
  $ RANDOM( \Sigma_{NC}^{\infty} ) $ &  random sequences of qubits \\
  $ [ G \, , \, G  ] $ & commutator subgroup of the group G \\
  $ G_{1} \, \star \, G_{2} $ & free product of the groups $ G_{1}
  $ and $ G_{2} $ \\
  $ A_{1} \, \star \, A_{2} $ & free product of the algebraic spaces $ A_{1}
  $ and $ A_{2} $ \\
  $ ( A_{1} , \omega_{1} ) \, \star \, ( A_{2} , \omega_{2} )   $ & free product of $ ( A_{1} , \omega_{1} ) $ and $  ( A_{2} , \omega_{2} ) $ \\
  ENSEMBLE[ CPS \, , \, n ] & ensemble of random matrices of order
  n w.r.t. CPS \\
  $ \mu_{emp} (a) $ & empirical eigenvalue distribution of the
  random matrix a \\
 $ \mu_{mean} (a) $ & mean eigenvalue distribution of the
  random matrix a \\
  $ S_{Araki} ( \omega_{1} \, , \, \omega_{2} ) $ & Araki's
  relative entropy of $  \omega_{1} $ w.r.t. $
  \omega_{2} $ \\
  $ S( \omega ) $ & entropy of the state $ \omega $ \\
  $ H_{\omega}(A) $ & entropy of the sub-$W^{\star}$-algebra A w.r.t. the state $ \omega $  \\
  $ H_{\omega}(\alpha) $ & entropy of the the channel $ \alpha $  w.r.t. the state $ \omega $ \\
  $ I( \omega \, ; \, \alpha ) $ &  mutual entropy of the state $ \omega $ and the channel $ \alpha $ \\
  $ DEC( \omega ) $ & decompositions of the state $ \omega $ \\
  $ DEC_{EXT} ( \omega ) $ & extremal decompositions of the state $ \omega $ \\
  $ DEC_{\bot} ( \omega ) $ & orthogonal decompositions of the state $ \omega $ \\
  $ DEC_{Schatten} ( \omega ) $ & Schatten's decompositions of the normal state $ \omega $ \\
  $ I_{acc} ( {\mathcal{E}} ) $ & classical accessible information information of the decomposition $ {\mathcal{E}} $ \\
  $ I_{Holevo} ( {\mathcal{E}} ) $ & Holevo's information of the decomposition $ {\mathcal{E}} $ \\
  n-GOE & gaussian orthogonal ensemble of order n \\
  n-GUE & gaussian unitary ensemble of order n \\
  $ gauss(D , \vec{m} , \hat{C} ; \vec{x} ) $ & D-dimensional gaussian measure of mean $ \vec{m} $ and covariance $ \hat{C} $ \\
  $ gauss_{STANDARD} $ & standard gaussian measure \\
  $ sc(m ,r ; x ) $ & semi-circle measure of mean m and variance $ \frac{r^{2}}{4} $ \\
  $ sc_{STANDARD} $ & standard semi-circle measure \\ \hline
\end{tabular}
\newpage
\begin{tabular}{|c|c|}
   $ S_{Bennett} (P) $ & Bennett's entropy of the distribution P \\
   $ S_{Zurek} (\rho) $ & Zurek's entropy of the density matrix $ \rho $ \\
   $ S_{therm} ( \omega ) $ & thermodynamical entropy of the state $ \omega $ \\
   $ S_{\text{double approach}} ( \omega ) $ & double approach entropy of the state  $ \omega $ \\
   $ I_{skew} ( \rho , a ) $ & skew information of the density matrix $ \rho $
   w.r.t. the operator a \\
   $ REC_{O} $ & recursivity in the oracle O \\
   $ f \, \leq_{T} \, g $ & f is Turing reducible to g \\
   $ f \, \sim_{T} \, g $ & f is Turing equivalent to g \\
   $ ( {\mathcal{D}}_{T} \, , \,  \leq_{T} ) $ & Turing degrees \\
   $ I^{-}(S) $ & chronological past of the space-time's region's S \\
   $ I^{+}(S) $ & chronological future of the space-time's region's S \\
   $ J^{-}(S) $ & causal past of the space-time's region's S \\
   $ J^{+}(S) $ & causal future of the space-time's region's S \\
   $ D^{-}(S) $ & past domain of dependence of the space-time's region's S \\
   $ D^{+}(S) $ & future domain of dependence of the space-time's region's S \\
   $ D(S) $ & domain of dependence of the space-time's region's S \\
   $ \approx_{Dirac} $ & weak equality \\
   $ [ a \, , \, b ]_{EFF} $ & effective commutator of a and b \\
   $ I^{(EFF)}_{skew} ( \rho \, , \, a ) $ & effective
   skew-information of $ \rho $ w.r.t. a \\
   $ r-\lim_{n \rightarrow \infty} $ & recursive limit for $ n
   \, \rightarrow \, \infty $ \\
   $ REC_{Nielsen} (A) $ & Nielsen-computable part of the $ W^{\star}$-algebra A \\
   $ COMP-ST(B) $ & computability structures on the Banach space B \\
   $ REC_{Pour \; El} (B , {\mathcal{S}}) $ & Pour-El computable
   vectors of B w.r.t. $  {\mathcal{S}} $ \\
   $ REC_{Pour \; El}-{\mathcal{O}}( {\mathcal{H}}) $ &
   effectively determined linear operators on $ {\mathcal{H}} $ w.r.t. $  {\mathcal{S}} $ \\
   $ Bloch ( \vec{r} ) $ & one qubit density operator w.r.t. the Bloch-sphere's vector $
   \vec{r}$ \\
    L(G) &  language generated by the Chomsky's grammar G \\
    FIN  & finite languages \\
    REG  & regular languages \\
    LIN  & linear languages \\
    CF   & context-free languages \\
    CS   & context-sensitive languages \\
    RE   & recursively enumerable languages \\
    IGUS & information gathering and using system \\
  PRG & pseudorandom number generator \\
  r.e. & recursively enumerable \\
  w.r.t. & with respect to \\
  l.h.s. & left-hand side \\
  r.h.s. & right-hand side \\
  iff & if and only if \\
  i.e. & id est \\
  e.g. & exempli gratia \\ \hline
\end{tabular}
\newpage
\section{Introduction} \label{sec:Introduction}
The new exciting research field of Quantum Computation has opened
a cross-fertilization area among Theoretical Physics and
Theoretical Computer Science that, beside the intrinsic
technological difficulties in the  physical implementation
essentially owed to a not-sufficient technological ability in
contrasting decoherence, is expected to be a strategic point for
developments in both fields \cite{Preskill-98},
\cite{Nielsen-Chuang-00}, \cite{Barndorff-Nielsen-Gill-Jupp-01},
\cite{Gruska-01}.

From the other side Quantum Computation Theory, concerning the
algorithmic evolution of quantum information, may be seen as a
sub-discipline of Quantum Information Theory,  a research field
that, in spite of its recent exciting developments, is a very
older object of investigation in Mathematical Physics
\cite{Ohya-Petz-93}, \cite{Ingarden-Kossakowski-Ohya-97}.

From a foundational perspective the first natural question is:
\begin{center}
\textbf{What is quantum information?}
\end{center}

\smallskip

Such an innocent question is, surprisingly, still open.

\smallskip

The more reasonable way of proceeding to answer this question
could consist in following the same footsteps Classical
Information Theory undertook to become a well-established
mathematical theory \cite{Khinchin-57}, \cite{Billingsley-65},
\cite{Ihara-93},\cite{Kakihara-99} of common engeneering
application \cite{Cover-Thomas-91}.

Though the invention of Information Theory must be tributed with
no doubt to  Claude E. Shannon's 1948 paper \textit{"The
Mathematical Theory of Communication"} \cite{Shannon-Weaver-71}
the mathematical foundation of the concept of \textbf{classical
information} was given, among the fifthies and the sixties, by the
great mathematician Andrei Nicolaevich Kolmogorov
\cite{Shiryayev-94}, \cite{AMS-LMS-00}.

\medskip

He observed that there exist three conceptually different ways of
approaching the problem of defining the notion of \textbf{amount
of information}:
\begin{enumerate}
  \item the \textbf{combinatorial approach}
  \item the \textbf{probabilistic approach}
  \item the \textbf{algorithmic approach}
\end{enumerate}

\medskip

The \textbf{combinatorial approach} furnishes a definition of the
information content of an object that is \textbf{contextual}, i.e.
depends from the particular context (collection of objects) in
which such an object is considered, but is \textbf{weight
independent}, i.e. it doesn't depend on the specification of a
way of weighting the contribution of different elements of such
context.

So, given an object x belonging to a set X of N elements (the
context) the \textbf{combinatorial approach}, invented by R.
Hartley in 1928, defines the amount of information of x simply as:
\begin{equation}
  I_{combinatorial} (x) \; := \; \log_{2} N
\end{equation}
Let us suppose, for example that x is a n-letter word in an
alphabet of s letters containing $ m_{i} $ occurences of the $
i^{th} $ letter ($ m_{1} + \cdots + m_{s} \, = \, n $).

Since there are:
\begin{equation}
  C( m_{1}, \cdots , m_{s}) \; := \; \frac{n !}{ m_{1} ! \cdots m_{s} ! }
\end{equation}
words of this kind, one has that:
\begin{equation}
  I_{combinatorial} (x) \; = \; \log_{2} C( m_{1}, \cdots , m_{s})
\end{equation}
As $ n, m_{1} , \cdots  m_{s} $ tend to infinity, Stirling's
asymptotic formula implies that:
\begin{equation} \label{eq:asymptotic combinatorial information}
   I_{combinatorial} (x) \; \sim \; \sum_{i=1}^{s} \frac{ m_{i} }{n}
   \log_{2} \frac{ m_{i} }{n}
\end{equation}

\medskip
The \textbf{probabilistic approach} furnishes a definition of the
information content of an object that is both \textbf{contextual}
and \textbf{weight dependent}.

So given an object x belonging to a set X of N elements (the
context) such the the $ i^{th} $ element is considered with
weight (probability) $ p_{i} $ , the \textbf{probabilistic
approach}, that invented by Shannon  in 1948 and often considered
as Classical Information Theory tout court, defines the amount of
information of x simply as:
\begin{equation} \label{eq:probabilistic information}
  I_{probabilistic} (x) \; := \; - \sum_{i=1}^{N} p_{i} \log_{2} p_{i}
\end{equation}
It must be noted, at this point, that the asymptotic formula
eq.\ref{eq:asymptotic combinatorial information} can be obtained,
in the probabilistic approach by eq.\ref{eq:probabilistic
information} applying the Law of Large Numbers.

Anyway, at this point, Kolmogorov underlines the importance that
such a result can be obtained getting rid of the \textbf{weight
dependence}, observing that it is precisely what he guaranteed for
other two important notions he introduced time before, namely the
\textbf{$\epsilon$-entropy} $H_{\epsilon}(K) $  and the
\textbf{$\epsilon$-capacity}  of compact classes of functions
describing, respectively, the amount of information necessary for
distinguishing some individual function in the class of functions
K and the amount of information that can be coded by elements of
K under the condition that elements of K no closer than $
\epsilon $ to each other can be reliably distinguished.

In the same way Kolmogorov stressed the importance of getting-rid
of the \textbf{context dependence}, i.e. to find an
\textbf{intrinsic} notion of the amount of information of an
object.

This led him to introduce the \textbf{algorithmic approach} that
is, indeed, both \textbf{weight independent} and \textbf{context
independent}:

in the \textbf{algorithmic approach} the amount of information of
an object x with respect to a given computer C is defined as the
length of the shortest program for C computing (i.e.
algorithmically-describing ) x:
\begin{equation} \label{eq:naife algorithmic information}
  I_{algorithmic} ( C ; x) \; := \;
  \begin{cases}
    \min \{ length(p) , C(p) = x \} & \text{if x is computable by the computer C}, \\
    + \infty & \text{otherwise}.
  \end{cases}
\end{equation}
The independence of this notion from the particular computer C is
then established by Kolmogorov through the proof of the so called
\textbf{Invariance Theorem} certifying the existence of
\textbf{optimal computers}, i.e. of computers that, up to an
object-independent constant, give  algorithmic-descriptions always
shorter of those given by any other computer.

\smallskip

Kolmogorov, the father of the usual, standard, measure-theoretic
axiomatization, stressed  from the beginning the conceptual
importance of such an \textbf{intrinsic} definition of the
informational-amount of an object for the same Foundation of
Probability Theory, i.e. for the explanation why Probability
Theory applies to reality.

The key point is that the \textbf{intrinsic nature} of the
algorithmic definition of information allows to address the issue
of giving an \textbf{intrinsic} characterization of randomness:

an algorithmically-random object x is, informally speaking, an
\textbf{algorithmically-incompressible} object, i.e. an object
whose more concise algorithmic-description is its same
assignation.

So the grown up theory concerning the algorithmic approach to
information, Algorithmic Information Theory from here and beyond,
appeared from the beginning as the corner-stone for an
alternative Algorithmic Foundation of Classical Probability
Theory \cite{Chaitin-87}, \cite{Van-Lambalgen-87},
\cite{Calude-94}, \cite{Li-Vitanyi-97}.

\smallskip

Later, especially by the work of Cristian Calude, Gregory Chaitin
and the Auckland's Center of Discrete Mathematics and Theoretical
Computer Science, Algorithmic Information Theory revealed soon an
even more fundamental rule in the Foundations of Mathematics,
furnishing an extraordinarily clear information-theoretic
explanation of the mathematical phenomenon of Incompleteness
\cite{Odifreddi-89} (Chaitin's First Undecidability Theorem
states that a formal system  can't decide statements involving an
algorithmic-information's amount higher than its own
algorithmic-informational content for more than a fixed constant,
implying the recursive undecidability of algorithmic randomness),
defining the notion of Halting Probability  codifying in optimal
way all the undecidabilities of Mathematics (the knowledge of the
first n cbits in the binary expansion of Chaitin's $ \Omega $
number would allow to decide all the n-cbit mathematical
statements), stating precise bounds on its determination
(Chaitin's Second Undecidability Theorem  states  that a formal
system  can't decide more than a finite number of digits in the
binary expansion of $ \Omega $, such  a result having been
recentely streghtened by R.M. Solovay through the proof that, by a
proper choice of the fixed Chaitin Universal Computer and
considering as formal system the Zermelo-Fraenkel axiomatic system
endowed with the Axiom of Choice, this finite number of digits
reduces even to zero) and, last but not least, showing that
Randomness is a pervasive phaenomenon in Pure Mathematics through
a paradigmatical example, i.e. shelding new light on Jones and
Matjasevic's proof that Hilbert's Tenth Problem (i.e. the problem
of finding an algorithm deciding whether an arbitrary Diophantine
equation has integer solutions) is undecidable by the proof of
the existence of a one integer parameter, let's call it k,
exponential diophantine equation such that to decide if it has a
finite number of integer solutions is equivalent to decide the
$k^{th}$ cbit of the Halting Probability \cite{Bennett-88}
\cite{Chaitin-87}, \cite{Chaitin-90}, \cite{Chaitin-98},
\cite{Chaitin-99}, \cite{Chaitin-01}, \cite{Solovay-00}.

\bigskip

Furthermore Algorithmic Information Theory appeared soon to play a
key rule in the Theory of Chaotic Dynamical Systems:

in 1958 Kolmogorov introduced (only for K-systems, the
generalization for arbitrary dynamical systems having being
furnished later by Ya. Sinai) a notion that would have played a
key rule for the solution of the  problem of giving a metric
classification of dynamical systems (that is the problem of
finding a complete set of invariants that imply a metric
isomorphism between dynamical systems): the metric entropy of a
dynamical system, characterizing the maximal asymptotic rate of
information obtained through a coarse-grained observation of
dynamics;

a dynamical system is called chaotic if it has strictly positive
metric entropy (or Kolmogorov-Sinai entropy as such notion is
more often called).

For the link existing between probabilistic information and
probabilistically expected algorithmic information it appeared,
then, intuitive that the characterization of chaoticity in terms
of the probabilistic approach  to information should have a
counterpart in terms of the algorithmic approach:

a first formalization of such a link was established by A.A.
Brudno \cite{Brudno-78}, \cite{Brudno-83},
\cite{Alekseev-Yakobson-1981} by the proof of a theorem (usually
called Brudno's Theorem) stating that the Kolmogorov-Sinai
entropy of a dynamical system is equal to the asympotic rate of
\textbf{simple algorithmic information} of almost all its
trajectories.

The importance of such a link was later stressed with particular
emphasis by Joseph Ford  who advocated strongly what he called an
Algorithmic Approach to Chaos Theory \cite{Ford-92}.

It must be said, anyway, that since the version of classical
algorithmic information giving rise to the correct
characterization of classical algorithmic randomness is not
\textbf{simple algorithmic entropy} but \textbf{prefix algorithmic
entropy}, the Algorithmic Approach to Chaos Theory is equivalent
to the usual one only in a weak sense as we will extensively
discuss.

\smallskip

The most important reason why  Algorithmic Information Theory is
of physical relevance lies, anyway, in Thermodynamics.

Many generations of physicists has been educated that the correct
exorcism of Maxwell's demon \cite{Maxwell-71} was the Leon's
Brilloiun one \cite{Brillouin-90}: the \textbf{acquisition of
information} on the velocity of the molecules by the demon is
responsible of the fact that the Second Law of Thermodynamics is
not violated.

The recent developments of the Thermodynamics of Computation
\cite{Feynman-96}, \cite{Bennett-90b}, has shown, anyway, that
Brillouin's exorcism doesn't work: by Landauer's Principle
\cite{Landauer-90a} such an information's acquisition process may
be realized in a thermodynamically reversible way.

The correct exorcism was, instead proposed by Charles Bennett
\cite{Bennett-90a}: it is the \textbf{erasure of information} by
the demon that cannot e accomplished in a thermodynamically
reversible way and is responsible of the preservation of the
Second Law.

This has, anyway, dramatic conseguence concerning the same
Foundations of Statistical Mechanics:

introducing the issue concerning the compatibility between the
time-reversibility of motions'-equation and the phenomenological
time-irreversibility of thermodynamics with Ludwig Boltzmann's
own words:
\begin{center}
  \textit{"If therefore we conceive of the world as an enormously large
mechanical system composed of an enormously large number of
atoms, which starts from a completelly ordered initial state, and
even at present is still in a substantially ordered state, then
we obtain consequences which actually agree with the observed
facts; although this conception involves, from a purely
theoretical - I might say philosophical - standpoint, certain new
aspects with contradicts general thermodynamics based on a purely
phenomenological viewpoint. General thermodynamics proceeds from
the fact that, as far as we can tell from our experience up to
now, all natural processes are irreversible. Hence according to
the principles of phenomenology, the general thermodynamics of
the second law is formulated in such a way that the unconditional
irreversibility of all natural process is asserted as a so-called
axiom, just as general physics based on a purely phenomenological
standpoint asserts the unconditional divisibility of matter
without limits as an axiom". From the section 89 of
\cite{Boltzmann-95}}
\end{center}
let us observe that most of the answers it has received, such as
the one authoritatively supported by Giovanni Gallavotti seeing
in Lanford's Theorem a mathematical formalization and confirmation
of Boltzmann's point of view that no inconsistency exists owing
to the not observability of the time-scale on which reversibility
manifests \cite{Gallavotti-99}, agree in a thing: the link
between the thermodynamical entropy and the state of a dynamical
system is given by the \textbf{probabilistic information} of that
state.

As it has been strongly supported by Wojciech Zurek
\cite{Zurek-89}, \cite{Zurek-90a}, \cite{Zurek-90b},
\cite{Zurek-99}, instead, Bennett's exorcism ultimatively implies
that in presence of a particular kind of information gathering and
using systems (IGUS), also the \textbf{algorithmic information}
of the state contribute to the thermodynamical entropy and has,
conseguentially, to be taken into account.

\bigskip

As far as Quantum Information Theory is concerned, the whole
Kolmogorovian analysis concerning the three possible approaches
to Information Theory may be rephrased with no variation.

Anyway, nowadays, while the probabilistic approach has received a
massive attention, resulting in a theory developed almost as much
as the classical Shannon's theory, the situation is radically
different for the combinatorial as well as for the algorithmic
approach where all is available is not so much more than a
plethora of attempts.

The algorithmic approach to Quantum Information Theory, for its
context-independence as well as for its weight-independence, is
of particular importance for the mathematical foundations of such
a subject.

\medskip

The organization of this thesis is the following:
\begin{itemize}
  \item In part\ref{part:Equivalent characterizations of classical
algorithmic randomness} we review the various equivalent
characterizations of classical-algorithmic randomness
  \item In part\ref{part:The road for  quantum algorithmic
  randomness} the issue of formalizing the notion of quantum
  algorithmic randomness in the framework of Quantum Algorithmic
  Information Theory is analyzed
  \item In part\ref{part:Classical Algorithmic
Information Theory of the results of quantum measurements} the
complementary issue of analyzing the classical algorithmic
information status of quantum-measurements' results is discussed
\end{itemize}
\part{Equivalent characterizations of classical algorithmic
randomness} \label{part:Equivalent characterizations of classical
algorithmic randomness}
\newpage
\chapter{Classical algorithmic randomness as classical algorithmic
incompressibility} \label{chap:Classical algorithmic randomness
as classical algorithmic incompressibility}
\section{The distinction between mathematical-classicality and physical-classicality} \label{sec:The distinction between mathematical-classicality and physical-classicality}
The attribute \textbf{classicality} is used by two different
scientific communities with different meanings:
\begin{itemize}
  \item it is usually used by Theoretical Physicists to express
  that some physical system obeys the laws of Classical Mechanics;
  this is, for example the acception of the adjective \textbf{classical}
  intended in the title of the first two volumes \textit{"Classical Dynamical
  Systems"} and \textit{"Classical Field Theory"} \cite{Thirring-97} of Walter
  Thirring's  monography \textit{"A Course in Mathematical
  Physics"}
  \item it is usually used by logico-mathematicians to express
  the part of a theory concerning only mathematical objects with
  cardinality less or equal to $ \aleph_{0} $; this is, for example, the acception of the adjective \textbf{classical}
  intended in the title \textit{"Classical Recursion Theory"} of Piergiorgio Odifreddi's
  monography \cite{Odifreddi-89}, \cite{Odifreddi-99a}
\end{itemize}
Unfortunately such a double acception of the term
\textbf{classical} have generated many confusions in the
literature belonging to the intersection of the two disciplines.

At a foundational level the generated confusion may be seen as a
confusion between the \textbf{subject} and the \textbf{object} of
a computational process, i.e. between the \textbf{attributes of
the computational device} and the \textbf{attributes of the
computed mathematical objects}.

Hence some property (classicality/quantisticality i.e.
commutativity/noncommutativity) is used in two undistinguished
(and often interchanged) acceptions according to it refers:
\begin{itemize}
  \item to the \textbf{subject of the computation}, i.e. to the
  computational device
  \item to the \textbf{object of the computation}, i.e. to the computed mathematical
  objects
\end{itemize}

\smallskip

An elegant way of avoiding this kind of mistakes is to pursue the
following prescription:

any issue of Computability Theory must analyze separetely each
cell of the following:

\medskip

\begin{diagram}\label{di:diagram of computation}
\end{diagram}
DIAGRAM OF COMPUTATION:

\begin{tabular}{|c|c|c|c|}
  $ \frac{OBJECT}{SUBJECT} $  & $C_{M}$ & $NC_{M}$  \\
  $C_{\Phi}$  &  $\cdot_{11}$ &  $\cdot_{12}$       \\
  $NC_{\Phi}$ &  $\cdot_{21}$ &  $\cdot_{22}$       \\ \hline
\end{tabular}

\medskip

with:
\begin{description}
  \item[$C_{M}$ :] MATHEMATICALLY CLASSICAL
  \item[$NC_{M}$:] MATHEMATICALLY NONCLASSICAL
  \item[$C_{\Phi}$:] PHYSICALLY CLASSICAL
  \item[$NC_{\Phi}$:] PHYSICALLY NONCLASSICAL
\end{description}

\smallskip

Let us consider, first of all, the following issue:

\textmd{$ 1^{th}$ ISSUE: WHAT IS COMPUTABLE ?}

\begin{itemize}
  \item $cell_{11} \; : \; C_{M} \, \cap \, C_{\Phi} $

  There is complete agreement in the scientific community that,
  as to the computation by \textbf{physically classical
  computers} of the following set of functions:
\begin{definition}
\end{definition}
MATHEMATICALLY CLASSICAL FUNCTIONS:

    (partial) functions on sets  $ S \, : \, card(S) \, \leq \,
  \aleph_{0}$

\textbf{Church-Turing's Thesis} holds leading to the
identification of the computable (partial) functions with the
(partial) recursive functions \cite{Odifreddi-89},
\cite{Odifreddi-96} that we will now define.

Introducing a notation we will adopt from here and beyond, we
will denote the computability attribute relative to the cell $
cell_{i j} $ of the diagram\ref{di:diagram of computation} by the
symbol $cell_{i j} - \Delta_{0}^{0} $.

For example the above statement may be rephrased saying that the
set $ C_{M} - C_{\Phi} - \Delta_{0}^{0} -\stackrel{ \circ }
{MAP}(S,S) $ is the set of all the partial recursive functions
over S.

Let, clarify, first of all, what we mean by a partial function:

a \textbf{total} (i.e.  ordinary) $ f \, : \, A \,
\rightarrow \, B $ is a rule associating to every element x of the set A an element f(x) of the set B:
\begin{equation*}
  x \in A \; \stackrel{f}{\rightarrow} \; f(x) \in B
\end{equation*}
We will indicate the set of all the total functions from a set A to a set B by MAP(A,B).

A \textbf{partial function} $ f \, : \, A
\stackrel{\circ}{\rightarrow}  \, B $ is a rule associating to
each element x of a certain subset $ HALTING(f) \subseteq A $ of
A, said the \textbf{halting set of f}, an element f(x) of the set B:
\begin{equation*}
   x \in HALTING(f) \; \stackrel{f}{\rightarrow} \; f(x) \in B
\end{equation*}
We will say that:
\begin{definition}
\end{definition}
f HALTS ON $ x\ \in A  \; ( f(x) \downarrow ) $:
\begin{equation}
  x \; \in \; HALTING(f)
\end{equation}
\begin{definition}
\end{definition}
f DOESN'T HALT ON $ x\ \in A  \; ( f(x) \uparrow ) $:
\begin{equation}
  x \; \notin \; HALTING(f)
\end{equation}
We will indicate the set of all the partial functions from a set A to a set B by $ \stackrel{\circ}{MAP}(A,B) $.
Given two partial functions $ f_{1} , f_{2} \, \in \, \stackrel{\circ}{MAP}(A,B) $:
\begin{definition}
\end{definition}
$ f_{1} $ IS EQUAL TO $ f_{2} \; ( f_{1} \, = \, f_{2} ) $:
\begin{equation}
  ( HALTING( f_{1} ) \, = \, HALTING( f_{2} ) )  \; and \;  (f_{1}
  (x) \, = \, f_{2}(x) \; \;  \forall x \in HALTING( f_{1} ))
\end{equation}

The language of \textbf{partial functions} is of common use in
Mathematical-Logic; we adopt it, anyway, also in unusual
environments: from Classical (i.e. commutative) Measure Theory (in which measures' halting sets
will be suitable $ \sigma $-algebras) to Operator Theory on Hilbert spaces (in which  unbounded operators' halting sets will be dense subspaces)

Denoted by $ {\mathbb{N}}^{\star} \; := \; \bigcup_{n \in
{\mathbb{N}}} {\mathbb{N}}^{n}  $ the set of all the n-ples of
natural numbers:
\begin{definition} \label{def:partial recursive functions on numbers}
\end{definition}
CLASS OF PARTIAL RECURSIVE FUNCTIONS (ON NUMBERS)
$(REC-\stackrel{\circ} {MAP} ( {\mathbb{N}}^{n} \, , \,
{\mathbb{N}} ))$

the smallest class of partial functions:
\begin{enumerate}
  \item containing the initial functions:
\begin{align}
  {\mathcal{O}} & (x) \; := \; 0 \\
  {\mathcal{S}} & (x) \; := \; x+1 \\
  {\mathcal{I}}_{i}^{n} & (x_{1}, \cdots , x_{n}) \; := \; x_{i}
  \; \; i=1 , \cdots , n \, , \, n \in {\mathbb{N}}
\end{align}
  \item closed under \textbf{composition}, i.e. the schema that given $
  \gamma_{1} , \cdots , \gamma_{m} , \psi $ produces:
\begin{equation}
  \varphi ( \vec{x} ) \; := \;  \psi (  \gamma_{1} ( \vec{x} ) ,
  \cdots ,  \gamma_{m} ( \vec{x} ) )
\end{equation}
  \item closed under \textbf{primitive recursion}, i.e. the schema
  that given $ \psi \, , \, \gamma $ produces:
\begin{align}
  \varphi( \vec{x} , 0 ) \; & := \; \psi ( \vec{x} )  \\
  \varphi( \vec{x} , y+1 ) \; & := \; \gamma ( \vec{x} , y , \phi(
  \vec{x}, y ))
\end{align}
  \item closed under \textbf{unrestricted $ \mu $-recursion}, i.e. the
  schema  that given $ \psi $ produces:
\begin{equation}
  \varphi( \vec{x} ) \; := \; \min \{ y \, : \, ( \psi( \vec{x} , z ) \downarrow \; \forall z \leq y  ) \: and
  \:  ( \psi( \vec{x} , y ) \, = \, 0  ) \}
\end{equation}
  where $ \varphi( \vec{x} ) \, := \, \uparrow $ if there is no
  such function
\end{enumerate}

\smallskip

The more fundamental properties of partial recursive functions may be collected in the
following:
\begin{theorem}  \label{th:Goedel's numbering of partial recursive functions}
\end{theorem}
G\"{O}DEL'S NUMBERING OF PARTIAL RECURSIVE FUNCTIONS:

It is possible to enumerate all partial recursive functions:
\begin{equation*}
  \varphi_{e}^{(n)} \: : \: {\mathbb{N}}^{n} \,
  \stackrel{\circ}{\rightarrow} \, {\mathbb{N}}
\end{equation*}
(where the natural number \textbf{e} is called the
\textbf{G\"{o}del's number} of the $ e^{th}$ n-ary partial
recursive function) in such a way that the following conditions
are satisfied:
\begin{itemize}
  \item \textbf{Universality:} there is a partial recursive
  function of two variables $ \varphi_{z}^{(2)}(e,x) $ such that:
\begin{equation}
   \varphi_{z}^{(2)}(e,x) \; = \;  \varphi_{e}^{(1)}(e,x) \; \;
   \forall x \in {\mathbb{N}}
\end{equation}
  \item \textbf{Uniform Composition:} there is a (total) recursive
  function of two variables \emph{comp} such that:
\begin{equation}
   \varphi_{comp(x,y)}^{(1)}(z) \; = \;  \varphi_{x}^{(1)}(
   \varphi_{y}^{(1)} (z)) \; \; \forall x,y,z \in {\mathbb{N}}
\end{equation}
  \item \textbf{Fixed Point:} For every $ m \in {\mathbb{N}}_{+} $
  and every recursive function f there effectively exists an x
  (called the fixed point of f) such that:
\begin{equation}
     \varphi_{x}^{(m)} \; = \; \varphi_{f(x)}^{(m)}
\end{equation}
\end{itemize}

\smallskip
It is useful to introduce the following notation concerning the domains of partial recursive functions:
\begin{definition}  \label{def:domains of partial recursive functions}
\end{definition}
\begin{equation}
  {\mathcal{W}}_{e}^{n} \; := \; HALTING(  \varphi_{e}^{(n)}) \; \;
  e,n \in {\mathbb{N}}
\end{equation}
Given an n-ary relation $ R( x_{1} , \cdots , x_{n} ) $ on
$ {\mathbb{N}} $:
\begin{definition} \label{def:r.e. relations}
\end{definition}
R IS RECURSIVELY ENUMERABLE (R.E.):
\begin{equation}
  \exists e \in {\mathbb{N}}  \; : \; R \, = \, {\mathcal{W}}_{e}^{n}
\end{equation}

We will identify, from here and beyond, \textbf{sets} and
\textbf{unary relations} by posing:
\begin{equation}
   {\mathcal{W}}_{e} \; := \;   {\mathcal{W}}_{e}^{1}
\end{equation}
Given a set $ S \subset {\mathbb{N}}^{\star} $:
\begin{definition} \label{def:recursive set}
\end{definition}
S IS RECURSIVE:

the characteristic function $ \chi_{S} $:
\begin{equation}\label{eq:characteristic function of a set}
  \chi_{S} (x) \; := \;
  \begin{cases}
    1 & \text{if $ x \in S $}, \\
    0 & \text{otherwise}.
  \end{cases}
\end{equation}
is a total recursive function

\smallskip
Clearly one has that:
\begin{theorem} \label{th:recursivity is stronger than recursive enumerability}
\end{theorem}
RECURSIVITY IS STRONGER THAN RECURSIVE ENUMERABILITY:
\begin{align*}
  recursivity  &  \; \Rightarrow \; \text{ recursive enumerability} \\
  \text{ recursive enumerability} & \; \nRightarrow \;  recursivity
\end{align*}

\smallskip

\begin{remark} \label{Godel numbering and self-reference}
\end{remark}
G\"{O}DEL NUMBERING AND SELF-REFERENCE

G\"{o}del's numbering, introduced by Kurt G\"{o}del in his his
famous 1931's paper \emph{"On Formally Undecidable Propositions of the
 Principia Mathematica and Related Systems"} \cite{Davis-65}, is a
 deep concept since it creates that link between \textbf{language}
 and \textbf{meta-language} giving rise to self-reference and all
 the consequences it generates through Cantor's Diagonalization.

 Since recursivity is equivalent to representability in an arbitary consistent formal system extending Tarski-Montowski-Robinson Arithmetics
  G\"{o}del's numbering may be equivalentely seen as a way of
  enumerating all the logical propositions concerning natural
  numbers.

  So one has a hierarchy of levels:
\begin{enumerate}
  \item the \textbf{objects} of investigation, i.e. natural
  numbers
  \item the \textbf{language} by which properties of the objects
  are described, i.e. the logical propositions concerning the
  objects
  \item the \textbf{meta-language} by which properties of the
  \textbf{meta-objects}, i.e the logical propositions of the language (by which properties of the objects are described) are
  described; we will denote by \textbf{meta-proposition} a
  proposition of the \textbf{meta-language}
\end{enumerate}
Owing to G\"{o}del numbering, a \textbf{number} plays a double rule:
\begin{itemize}
  \item as an \textbf{object}
  \item as the G\"{o}del number identifying a proposition of the
  language, i.e. as a \textbf{meta-object}
\end{itemize}
This can be used to pass from \textbf{meta-language} to the
\textbf{language}, simply associating to the
\textbf{meta-proposition} $ \varphi_{e} (x) $  concerning the
\textbf{meta-object} x, i.e. the proposition of the language
with  G\"{o}del number x, the \textbf{proposition}  $ \varphi_{e}
(x) $ concerning the \textbf{object}, i.e. the number, x and,
viceversa, to pass from \textbf{language} to \textbf{meta-language}, associating to the
\textbf{proposition}  $ \varphi_{e}(x) $ concerning the \textbf{object}, i.e. the number, x the
\textbf{meta-proposition} $ \varphi_{e} (x) $   concerning the \textbf{meta-object} x, i.e. the proposition of the language
with  G\"{o}del number x.

But then self-reference immediately appears since $  \varphi_{e}(e) $ happens to speak about itself.

  \item $cell_{21} \; : \; C_{M} \, \cap \, NC_{\Phi} $

There is no universally accepted  answer in the scientific
community to the question if a \textbf{physically nonclassical
computer} can violate Church-Turing's Thesis, i.e. can compute
non-recursive \textbf{mathematically classical functions}.

In particular, as far as  the computation by \textbf{physically
quantistical computers} of \textbf{mathematically classical
functions} is concerned, the common opinion among the researchers
in Quantum Computation \cite{Feynman-99}, \cite{Deutsch-85},
\cite{Jozsa-98} is that \textbf{Nonrelativistic Quantum
Mechanics} and \textbf{Special-relativistic Quantum Mechanics
(Local Quantum Field Theories)} don't violate Church-Turing's
Thesis.

It must be cited, anyway, that the opposite thesis has been
asserted by various authors (cfr.
\cite{Castagnoli-Rasetti-Vincenti-92}, \cite{Mitchison-Jozsa-99},
\cite{Calude-Dinneen-Svozil-00} and the paragraphs 4.12 and 4.23
of \cite{Calude-Paun-01})

Furthermore it must be observed that in the Masanao Ozawa's final
formalization of Quantum Turing Machines \cite{Ozawa-98a} (
saying according, to us, the last word on the consistence's
problem of Deutsch's Halting Protocol) the satisfaction of the
Church-Turing's Thesis is posed by hand restricting the range of
the local transition function to recursive complex numbers.

We will, anyway, extensively return on this point in  section\ref{sec:The problem of characterizing mathematically the notion of a quantum algorithm}

Finally, when \textbf{Generally-relativistic Quantum Mechanics}
(both in the form of \textbf{Quantum Gravity} and in the form of
some suggested \textbf{gravitationally-modificated Quantum
Mechanics}) is considered, the whole story touches the strongly
debated ideas of Roger Penrose about a non-computable alteration
of the quantum unitary dynamics induced by gravity
\cite{Penrose-89}, \cite{Penrose-96}, \cite{Anandan-98},
\cite{Penrose-00}.

  \item $cell_{12} \; : \; NC_{M} \, \cap \, C_{\Phi} $

  As soon as one goes out from the boundaries of $ C_{M}$-Classical
  Recursion Theory the almost miracolous equivalence of  all the
  different approaches (recursivity, finitely definability, Herbrand-G\"{o}del Computability, representability in consistent formal system extending Tarski-Montowski-Robinson Arithmetics, $ \lambda$-definability in Church's $ \lambda $- Calculus, flowchart computability, computability by Classical Turing Machines, by cellular automata \cite{Odifreddi-89}, by Shepherdson-Sturgis register machines \cite{Cutland-80}, LISP computability \cite{Mc-Carthy-60}, $ \cdots $) that in such a  theory manifests the strong
  experimental verification of Church's Thesis, dramatically
  disappears.

  Just as to Computability Theory by \textbf{physically classical
  computers} of  (partial) functions on sets  $ S \, : \, cardinality(S) \, = \,
  \aleph_{1}$ while  many different inequivalent candidate theories have
  been proposed:
\begin{enumerate}
  \item the well extablished and almost always accepted theory named Computable Analysis, generated by the studies of Grzegorczyck -
  Lacombe \cite{Pour-El-Richards-89}
  \item the theory developed by the so called Markov School in the framework of
  Constructive Mathematics \cite{Odifreddi-89}
  \item the Blum - Shub - Smale 's
  Theory \cite{Smale-92}, \cite{Blum-Cucker-Shub-Smale-98}
\end{enumerate}
The relative popularity of the issue about the concurrence of
such candidate theories is owed to Penrose's question if
Mandelbrot's set is recursive \cite{Penrose-89}. We will partially
analyze it in section\ref{sec:Brudno algorithmic entropy versus
the Uspensky abstract approach}

\item $cell_{22} \; : \; NC_{M} \, \cap \, NC_{\Phi} $

  It's important to realize that, contrary to what is often claimed, Church-Turing's Thesis doesn't imply that the answer to the $ 1^{th} ISSUE $ contained in the cells
  $cell_{12}$ and $cell_{22}$ must be equal.

  For example Church-Turing's Thesis is not incompatible with an hypothetical situation in which
  Mandelbrot's set would be $ C_{\Phi} $ - incomputable but $ NC_{\Phi} $ -
  computable.

  Though some undecidability theorems  and conjectures still exist (cfr. e.g. Lloyd's
  arguments concerning uncomputable diagonalizations in Quantum
  Computation \cite{Loyd-89} as well as his general consideration about the physical limits of Computation \cite{Lloyd-01}, or Geroch and Hartle's speculations
  concerning the eventuality that the recursive undecidability of
  the Homeomorphism-problem for four-manifolds
  \cite{Collins-Zieschang-98} may lead to the
  recursive undecidability of quantizing Gravity) no general
  mathematically formalization has been realized yet.

  Particular importance has, according to us, Karl Svozil's
  suggestion that in Quantum Algorirhmic Information Theory there
  should exist undecidability theorems analogues to the classical
  Chaitin's ones (cfr. the problem17 of \cite{Calude-96}).
\end{itemize}
\newpage
\section{Uspensky's abstract definition of algorithmic information} \label{sec:Uspensky's abstract definition of algorithmic information}
The last contribution Andrei Nikolaevich
Kolmogorov left us before dying was his forum report
\textit{Algorithms and Randomness}, made with and exposed by his
student Vladimir Uspensky, at the First World Congress of the
Bernoulli Society (September 8-14, 1986) \cite{AMS-LMS-00}.

Later Unspensky formalized the Kolmogorovian approach to
Algorithmic Information Theory in a very general and elegant way
we will start from \cite{Uspensky-92}, \cite{Uspensky-Semenov-93}.

\begin{definition} \label{def:aggregate}
\end{definition}
AGGREGATE:

a couple $ ( X \, , \, R ) $ such that:
\begin{itemize}
  \item X is a set
  \item R, called a \textbf{concordance relation}, is a computable
  binary relation on X
\end{itemize}

\medskip

\begin{remark} \label{rem:context dependence of the computability constraint}
\end{remark}
CONTEXT-DEPENDENCE OF THE COMPUTABILITY CONSTRAINT

Let us observe that, in the definition def.\ref{def:aggregate} we
have imposed a \textbf{computability constraint} without
specifying its precise mathematical meaning.

This has been done in order of guaranteeing the maximal
generality: in the different contextes corresponding to the
different cells $ cell_{ij} $ of the diagram\ref{di:diagram of
computation} such a constraint is formalized by the proper $
cell_{ij} - \Delta_{0}^{0}$ condition

\bigskip

Given two aggregates $ A_{1} \, := \,  ( X_{1} \, , \, R_{1} ) $
and $ A_{2} \, := \, ( X_{2} \, , \, R_{2} ) $ it is natural to
ask under which conditions we can think to elements of $  A_{2} $
as descriptions of elements of $  A_{1} $ with respect to a proper
description mode;

the answer is given by the following definition:
\begin{definition}
\end{definition}
MODE OF DESCRIPTION (OF $  A_{2} $-ELEMENTS THROUGH  $  A_{1}
$-ELEMENTS):

a relation R between elements of $  X_{1} $ and elements of  $
X_{2} $ such that:
\begin{equation}\label{eq:mode of description}
  R_{1}( x_{1} , y_{1}) \, and \, R_{2}( x_{2} , y_{2}) \, and \,
  R( x_{1} , x_{2}) \; \Rightarrow \; R( y_{1} , y_{2}) \; \;
  \forall x_{1} , x_{2}  \in X_{1} ,  \forall y_{1} , y_{2}  \in X_{2}
\end{equation}

\smallskip

We will denote the set of all the mode of description of $  A_{2}
$-elements through  $  A_{1} $-elements by $ {\mathcal{D}}( A_{1}
,  A_{2} ) $.

\smallskip

Given a mode of description R among the aggregates $ A_{1} $ and $
A_{2} $:
\begin{definition}
\end{definition}
$ x_{2} \in X_{2}$ IS A DESCRIPTION OF $ x_{1} \in X_{1} $
THROUGH THE MODE R:
\begin{equation}
  R( x_{1} , x_{2} )
\end{equation}

\smallskip

All the ingredients introduced up to this point are of pure
set-theoretic nature (with some constructibility constraint).

The introduction of a notion characterizing the amount of
not-redundant, i.e. algorithmically incompressible, information of
an object $ x_{1} \in X_{1} $ with respect to the description mode
R requires the introduction of some point measure quantifying the
extension of the descriptions.

Let us, then, define, the following notion:
\begin{definition}
\end{definition}
METRIC AGGREGATE:

a couple $ ( A \, , \mu ) $ such that:
\begin{itemize}
  \item $ A \, := \, ( X \, , R ) $ is an aggregate
  \item $ \mu $ is a point measure on A
\end{itemize}

\smallskip

Given a metric aggregate $ A_{1} \, := \,  ( X_{1} \, , \, R_{1}
\, , \, \mu ) $, an aggregate  $ A_{2} \, := \, ( X_{2} \, , \,
R_{2} \, ) $ and a mode of description R among the aggregates $
A_{1} $ and $ A_{2} $ we can finally introduce the following
basic notion:
\begin{definition} \label{def:algorithmic information}
\end{definition}
ALGORITHMIC INFORMATION OF $ x_{2} \in X_{2} $ W.R.T. THE
DESCRIPTION MODE R:
\begin{equation}
  I_{R} ( x_{2} ) \; := \;
  \begin{cases}
    \min \{ \mu ( x_{1} ) \, : \, R( x_{1} , x_{2} ) \} & \text{if $ \exists \, x_{1} \in X_{1} \, : R( x_{1} , x_{2} )$}, \\
    + \infty & \text{otherwise}.
  \end{cases}
\end{equation}

\smallskip

Clearly the definition def.\ref{def:algorithmic information}
depends on the particular chosen description mode R.

It is clear, anyway, that the whole consistence of Algorithmic
Information Theory lies on the possibility of getting-rid of such
a dependence.

The formalization of this issue is given by the following notions:
\begin{definition}
\end{definition}
UNIVERSE OF DESCRIPTION OF $ A_{2} $ THROUGH  $ A_{1} $:

a set $ {\mathcal{R}} $ of description modes of the aggregate $
A_{2} $ through the metric aggregate $ A_{1} $:
\begin{equation}
  {\mathcal{R}} \;  \subseteq \; {\mathcal{D}}( A_{1} ,  A_{2} )
\end{equation}

\smallskip

The intuitive idea we are going to formalize is that Algorithmic
Information Theory is meaningful provided the involved universes
of descriptions admit optimal mode of descriptions, i.e. mode of
descriptions that are always more concise of all the others, up to
an object-independent additive constant.

This requires the introduction of two ordering relation we will
use extensively in the whole dissertation.

Given two real-valued partial function  $ f_{1} : A \stackrel
{\circ}{\rightarrow} {\mathbb{R}} \, , \, f_{2} : A \stackrel
{\circ}{\rightarrow} {\mathbb{R}}  $ we will say that:
\begin{definition} \label{def:addittively less or equal}
\end{definition}
$ f_{1} $ IS ADDITIVELY LESS OR EQUAL TO $ f_{2} $ ( $ f_{1} \,
\stackrel{ + }{\leq} \, f_{2} $ )
\begin{equation}
  \exists c \in {\mathbb{R}}_{+} \; : \; f_{1} (x) \, \leq \, f_{2}
  (x) + c \; \; \forall x \in HALTING(f_{1}) \bigcap HALTING(f_{2})
\end{equation}
\begin{definition} \label{def:addittively equal}
\end{definition}
$ f_{1} $ IS ADDITIVELY EQUAL TO $ f_{2} $ ( $ f_{1} \,
\stackrel{ = }{\leq} \, f_{2} $ )
\begin{equation}
  f_{1} \, \stackrel{ + }{\leq} \, f_{2} \; and \;  f_{2} \, \stackrel{ + }{\leq} \, f_{1}
\end{equation}
\begin{definition} \label{def:multiplicatively less or equal}
\end{definition}
$ f_{1} $ IS MULTIPLICATIVELY LESS OR EQUAL TO $ f_{2} $ ( $ f_{1}
\, \stackrel { \times }{\leq} \, f_{2} $ )
\begin{equation}
  \exists c \in {\mathbb{R}}_{+} \; : \; f_{1} (x) \, \leq \, f_{2}
  (x) \times c \; \; \forall x \in HALTING(f_{1}) \bigcap HALTING(f_{2})
\end{equation}
\begin{definition} \label{def:multiplicatively equal}
\end{definition}
$ f_{1} $ IS MULTIPLICATIVELY EQUAL TO $ f_{2} $ ( $ f_{1} \,
\stackrel { \times }{=} \, f_{2} $ )
\begin{equation}
  f_{1} \, \stackrel{ \times }{ \leq } \, f_{2} \; and \;  f_{2} \, \stackrel{ \times }{\leq} \, f_{1}
\end{equation}

\medskip

Let us now consider an aggregate $ A_{1} $, a metric aggregate $
A_{2} $, a universe of description $ {\mathcal{R}} $ of $ A_{1} $
through $ A_{2} $ and a particular mode of description belonging
to such a universe $ U \in {\mathcal{R}} $.

We will say that:
\begin{definition} \label{def:optimal mode of description}
\end{definition}
U IS OPTIMAL W.R.T. $ {\mathcal{R}} $:
\begin{equation}
  U \; \stackrel{ + }{\leq} \; f \; \; \forall f \in {\mathcal{R}}
\end{equation}
We can then introduce the following notion:
\begin{definition}
\end{definition}
ALGORITHMIC INFORMATION THEORY IS MEANINGFUL W.R.T. $
{\mathcal{R}} $:
\begin{equation}
  \exists \; U \in {\mathcal{R}} \; optimal
\end{equation}
Adhering to Uspensky's terminology let us introduce the following
notion:
\begin{definition}
\end{definition}
ALGORITHMIC ENTROPY:

a function I equal to the algorithmic information w.r.t. a
description mode that is optimal w.r.t to some universe of
description modes.

\medskip

In order to discuss the first  fundamental examples, let us
introduce some basic notions.

Given a set $ \Sigma $:
\begin{definition} \label{def: classical strings on an alphabet}
\end{definition}
SET OF THE STRINGS ON $ \Sigma $ :
\begin{equation}
\Sigma^{\star} \; \equiv \;  \{ \lambda \} \; \bigcup \; \cup_{ k
\in {\mathbb{N}}} \Sigma^{k}
\end{equation}
\begin{definition} \label{def: classical sequences on an alphabet}
\end{definition}
SET OF THE SEQUENCES ON $ \Sigma $:
\begin{equation}
\Sigma^{\infty} \; \equiv \; \{ \lambda \} \; \bigcup \; \{
\bar{x} : {\mathbb{N}}_{+} \, \rightarrow \, \Sigma \}
\end{equation}
where $ \lambda $ denotes the \textit{empty string}.

Given $ \vec{x} \in \Sigma^{\star} $ let us denote by $
\vec{x}^{n} \in \Sigma^{\star} $ the string made of n repetitions
of $ \vec{x} $ and by $ \vec{x} ^{\infty} \in \Sigma^{\infty} $ the
sequence made of infinite repetitions of $ \vec{x} $.

It is important to remark that \cite{Calude-94}:
\begin{theorem} \label{th:cardinalities of strings and sequences}
\end{theorem}
ON THE CARDINALITIES OF STRINGS AND SEQUENCES OVER A FINITE
ALPHABET

\begin{hypothesis}
\end{hypothesis}
\begin{equation*}
  cardinality ( \Sigma ) \; \in \; {\mathbb{N}}
\end{equation*}
\begin{thesis}
\end{thesis}
\begin{align*}
  cardinality(\Sigma^{\star}) \; & = \; \aleph_{0}   \\
  cardinality(\Sigma^{\infty}) \; & = \;  \aleph_{1}
\end{align*}
We will assume from here and beyond that $ \Sigma \; := \; \{ 0
,1 \}$.

The total-ordering $ 0 \; < \; 1 $ induces the following:
\begin{definition}
\end{definition}
QUASI-LEXICOGRAPHIC ORDERING ON $\Sigma^{\star}$
\begin{multline}
\lambda \, < \, 0  \, < \, 1  \, < \, 00  \, < \, 01  \, < \, \\
10 \, < \, 11  \, < \, 000  \, < \, 001  \, < \, \cdots 111  \, <
\, \cdots
\end{multline}
We can then introduce the following bijection:
\begin{definition}
\end{definition}
QUASI-LEXICOGRAPHIC MAP:
\begin{equation}
\begin{split}
 string : {\mathbb{N}} & \rightarrow \Sigma^{\star}  \\
 string(n) \, & \, = \text{the $ n^{th} $ string in quasi-lexicographic ordering}
\end{split}
\end{equation}

\smallskip

Let us now introduce the following ordering relation on $
\Sigma^{\star} $
\begin{definition}
\end{definition}
PREFIX-ORDER RELATION   $ <_{p} $ ON $ \Sigma^{\star} $:
\begin{equation}
\vec{x} <_{p} \vec{y} \; := \;  \exists \vec{z} \in \Sigma^{\star}
\, : \; \vec{y} = \vec{x} \cdot \vec{z}
\end{equation}
Give a set $ S \, \subset \, \Sigma^{\star} $ we will say that:
\begin{definition}
\end{definition}
S IS PREFIX-FREE:
\begin{equation}
  ( \vec{x} <_{p} \vec{y} \Rightarrow \vec{x} = \vec{y} ) \,
\forall \vec{x},\vec{y} \in S
\end{equation}

\medskip

\begin{example} \label{ex:simple algorithmic entropy}
\end{example}
SIMPLE ALGORITHMIC ENTROPY

Let us consider the case in which $ A_{1} \; = \; A_{2} \; = \; (
{\mathbb{N}} \, , \, = \, , \, | \cdot | ) $ with:
\begin{equation}
  | n | \; := \; | string^{- 1} (n) | \; = \; \llcorner \log_{2} (
  n + 1 ) \lrcorner
\end{equation}

Kolmogorov started considering as universe of mode of
descriptions the whole $ {\mathcal{D}}( A_{1} , A_{2} ) $.

But he immediately realized that:

\begin{theorem} \label{th:simple algorithmic information theory w.r.t. all the partial functions is not meaningful}
\end{theorem}
ALGORITHMIC INFORMATION THEORY W.R.T. $ {\mathcal{D}}( A_{1} ,
A_{2} ) $ IS NOT MEANINGFUL

\begin{proof}
Following \cite{Li-Vitanyi-97} let us us suppose by abdurdum that
there exist a function $ U \in \bigcap {\mathcal{D}}( A_{1} ,
A_{2} )$ such that:
\begin{equation}
  U \; \stackrel{ + }{\leq} \; f \; \; \forall f \in
  {\mathcal{D}} ( A_{1} \, , \,  A_{2} )
\end{equation}
Let us then consider an infinite sequence $ X \, := \, \{ x_{n}
\in {\mathbb{N}} \}_{n \in {\mathbb{N}}} $ such that:
\begin{equation}
  i < j \; \Rightarrow \;  x_{i} < x_{j} \; \, \forall i,j \in {\mathbb{N}}
\end{equation}
Considered a subsequence $ Y \, := \,  \{ y_{n} \in {\mathbb{N}}
\}_{n \in {\mathbb{N}}} $ of the sequence X such that:
\begin{equation}
  \log y_{n} \; < \; \frac{ \log x_{n}  }{2} \; \; \forall n \in {\mathbb{N}}
\end{equation}
let us introduce the function $ f \in {\mathcal{D}}( A_{1} ,
A_{2} ) $ conciding with U everywhere but for the points of the
sequence X where it is defined as:
\begin{equation}
  f( x_{n} ) \; := \; U (  y_{n} ) \; \; \forall n \in {\mathbb{N}}
\end{equation}
We have clearly that:
\begin{equation}
  cardinality ( \{ n \in {\mathbb{N}} \, : I_{f} ( n ) \, \leq \, \frac{ I_{U} ( n
  )}{2} \} ) \; = \; \aleph_{0}
\end{equation}
that contradict the absurdum hypothesis
\end{proof}

\medskip

Theorem\ref{th:simple algorithmic information theory w.r.t. all
the partial functions is not meaningful} is the first of a set of
theorems we will meet in this dissertation showing that certain
quantities of Algorithmic Information Theory are meaningful only
by effectivizing some notion.

Indeed, requiring that to describe objects must be an effective
property, one is led by Church-Turing's thesis, for reasons that
will be clarified in the next section, to restrict the universe of
modes of descriptions to the set  $ C_{M}-C_{\Phi} -\Delta_{0}^{0}
[ {\mathcal{D}}( A_{1} , A_{2} )] $ of the \textbf{partial
recursive ones}.

Kolmogorov realized that in this way the problem was overcome
proving the following \cite{Calude-94}:
\begin{theorem} \label{th:invariance theorem for simple algorithmic entropy}
\end{theorem}
INVARIANCE THEOREM FOR SIMPLE ALGORITHMIC ENTROPY:

Algorithmic Information Theory w.r.t. $ C_{M}-C_{\Phi}-
\Delta_{0}^{0} - [ {\mathcal{D}}( A_{1} , A_{2} )] $  is
meaningful

\smallskip

As we have preannounced, the resulting algorithmic entropy,
w.r.t. an optimal description mode that we will call from here and
beyond a \textbf{simple universal computer}, is called the
\textbf{simple algorithmic entropy} and is denoted by K.

\medskip

\begin{example} \label{ex:monotone algorithmic entropy}
\end{example}
MONOTONE ALGORITHMIC ENTROPY

Let us consider the case in which $ A_{1} \; = \; A_{2} \; = \; (
\Sigma^{\star} \, , \, <_{p} \, , \, | \cdot | ) $.

Exactly as in the example\ref{ex:simple algorithmic entropy} it
may be proved that:
\begin{theorem} \label{th:monotone algorithmic information theory w.r.t. all the partial functions is not meaningful}
\end{theorem}
ALGORITHMIC INFORMATION THEORY W.R.T. $  {\mathcal{D}}( A_{1} ,
A_{2} ) $ IS NOT MEANINGFUL

\smallskip

but:
\begin{theorem}
\end{theorem}
INVARIANCE THEOREM FOR MONOTONE ALGORITHMIC ENTROPY:

Algorithmic Information Theory w.r.t. $ C_{M}-C_{\Phi}-
\Delta_{0}^{0} - [ {\mathcal{D}}( A_{1} , A_{2} )] $  is
meaningful

\smallskip

As we have preannounced, the resulting algorithmic entropy is
called the \textbf{monotone algorithmic entropy}

\medskip

\begin{remark}
\end{remark}
FROM BINARY STRINGS TO NATURAL NUMBERS AND VICEVERSA

For pure simplicity reasons we have defined simple algorithmic
entropy for natural numbers and monotone algorithmic information
for binary strings.

By the quasi-lexicographic bijection the corrispondent notions,
namely simple algorithmic entropy of strings and prefix
algorithmic entropy of natural numbers are immediately obtained.

For the same reason from here and beyond everything stated for $
\Sigma^{\star} $ may be immediately translated in terms of $
{\mathbb{N}} $ and viceversa.

\medskip

Up to now we have considered the case in which the two metric
aggregates coincide.

Anyway one can clearly introduce also the following mixed notions:
\begin{example} \label{ex:decision algorithmic entropy}
\end{example}
DECISION ALGORITHMIC ENTROPY:

Let us assume $ A_{1} \; = ( {\mathbb{N}} \, , \, = \, , \, |
\cdot | ) $ while $ A_{2} \; = \; ( \Sigma^{\star} \, , \, <_{p}
\, , \, | \cdot | )$.

Exactly as in the example\ref{ex:simple algorithmic entropy} it
may be proved that:
\begin{theorem} \label{th:decision algorithmic information theory w.r.t. all the partial functions is not meaningful}
\end{theorem}
ALGORITHMIC INFORMATION THEORY W.R.T. $  {\mathcal{D}}( A_{1} ,
A_{2} ) $ IS NOT MEANINGFUL

\smallskip

but:
\begin{theorem}
\end{theorem}
INVARIANCE THEOREM FOR DECISION ALGORITHMIC ENTROPY:

Algorithmic Information Theory w.r.t. $ C_{M}-C_{\Phi}-
\Delta_{0}^{0} - [ {\mathcal{D}}( A_{1} , A_{2} )] $ is meaningful

\smallskip

As we have preannounced, the resulting algorithmic entropy is
called the \textbf{monotone algorithmic entropy}

\medskip

\begin{example} \label{ex:prefix algorithmic entropy}
\end{example}
PREFIX ALGORITHMIC ENTROPY:

Let us assume  $ A_{1} \; = \; ( \Sigma^{\star} \, , \, <_{p} \, ,
\, | \cdot | )$ while $ A_{2} \; = ( {\mathbb{N}} \, , \, = \, ,
\, | \cdot | ) $.

Exactly as in the example\ref{ex:simple algorithmic entropy} it
may be proved that:
\begin{theorem} \label{th:prefix algorithmic information theory w.r.t. all the partial functions is not meaningful}
\end{theorem}
ALGORITHMIC INFORMATION THEORY W.R.T. $  {\mathcal{D}}( A_{1} ,
A_{2} ) $ IS NOT MEANINGFUL

\smallskip

but:
\begin{theorem} \label{th:invariance theorem for prefix algorithmic entropy}
\end{theorem}
INVARIANCE THEOREM FOR PREFIX ALGORITHMIC ENTROPY:

Algorithmic Information Theory w.r.t. $ C_{M}-C_{\Phi}-
\Delta_{0}^{0} - [ {\mathcal{D}}( A_{1} , A_{2} )] $ is meaningful

\smallskip

As we have preannounced, the resulting algorithmic entropy, w.r.t.
an optimal description mode that we will call from here and
beyond a \textbf{Chaitin universal computer}, is called the
\textbf{prefix algorithmic entropy} and is denoted by I.

\medskip

While the \textbf{decision entropy} and \textbf{monotone entropy}
are of scarce utility, \textbf{simple entropy} and \textbf{prefix
entropy} are of fundamental importance
\newpage
\section{Why prefix algorithmic entropy is better than simple algorithmic entropy} \label{sec:Why prefix entropy is better than simple entropy}
Let us now compare simple algorithmic entropy and prefix
algorithmic entropy.

Though more intuitive, simple algorithmic entropy has a list of
inconveniences that, after decades of debates among different
attempts, led the scientific community to realize that the
correct way of formulating Classical Algorithmic Information
Theory involves prefix algorithmic entropy:
\begin{enumerate}
  \item \textbf{classical probabilistic information}, namely \textbf{Shannon entropy}, satisfies the \textbf{subadditivity property}:
\begin{equation} \label{eq:subadditivity of probabilistic information}
  I_{probabilistic} ( X , Y ) \; \leq \; I_{probabilistic} (X) +  I_{probabilistic} (Y)
\end{equation}
with the equality holding iff the classical random variables X and
Y are independent, where the joint probabilistic information $ I_{probabilistic} ( X , Y ) $ will be defined in  section\ref{sec:From the communicational-compression of the Quantum Coding Theorems to the algorithmic-compression in Quantum Computation}.

As we will see therein the subadditivity property remain preserved in the
noncommutative generalization, i.e. eq.\ref{eq:subadditivity of
probabilistic information} holds also in Quantum Probability
Theory, where $ I_{probabilistic} $ is the \textbf{quantum
probabilistic information}, namely \textbf{Von Neumann entropy},
( X , Y ) denotes a state over a tensor product Von Neumann
algebra $ A_{1} \bigotimes A_{2} $ having X and Y as marginal
states, the equality holding iff ( X , Y ) is not entangled.

The intuitive meaning of eq.\ref{eq:subadditivity of probabilistic
information} (the information of a compound system is less or
equal to the information of its parts) lead to think that such a
condition should hold also for \textbf{classical algorithmic
information}.

As to \textbf{simple algorithmic entropy}, anyway, the
subadditivity condition is violated by a disturbing logarithmic
addendum causing that:
\begin{theorem} \label{th:not subadditivity of simple algorithmic entropy}
\end{theorem}
NOT SUBADDITIVITY OF SIMPLE ALGORITHMIC ENTROPY
\begin{equation}
  K ( ( \vec{x} , \vec{y} )) \; \not \stackrel{ + }{\leq} \; K(
  \vec{x}) \, + \, K(\vec{y})
\end{equation}
The subadditivity property is instead satisfied by \textbf{prefix
algorithmic entropy}:
\begin{theorem} \label{th:subadditivity of prefix algorithmic entropy}
\end{theorem}
SUBADDITIVITY OF PREFIX ALGORITHMIC ENTROPY
\begin{equation}
  I ( ( \vec{x} , \vec{y} )) \; \stackrel{ + }{\leq} \; I(
  \vec{x}) \, + \, I(\vec{y})
\end{equation}
  \item intuitive reasoning suggests that $ C_{M} - C_{ \Phi } $-algorithmic
  information should be \textbf{monotonic over prefixes}.

Anyway one has that:
\begin{theorem} \label{th:not monotonicity over prefixes of simple algorithmic entropy}
\end{theorem}
NOT MONOTONICITY OVER PREFIXES OF SIMPLE ALGORITHMIC INFORMATION
\begin{equation}
  \vec{x} \, <_{p} \,  \vec{y} \; \nRightarrow \; K ( \vec{x} ) \,
   \stackrel{ + }{\leq} \, K ( \vec{y} )
\end{equation}
while:
\begin{theorem} \label{th:monotonicity over prefixes of
prefix algorithmic entropy}
\end{theorem}
MONOTONICITY OVER PREFIXES OF PREFIX ALGORITHMIC INFORMATION
\begin{equation}
  \vec{x} \, <_{p} \,  \vec{y} \; \Rightarrow \; I( \vec{x} ) \,
   \stackrel{ + }{\leq} \, I ( \vec{y} )
\end{equation}

 \item since the \textbf{probabilistic approach} and the \textbf{algorithmic
 approach} to $ C_{M} - C_{ \Phi } $ - Information Theory are
 different ways of formalizing the same object of investigation, it
 is natural to suppose that \textbf{$ C_{M} - C_{ \Phi } $-probabilistic
 information} and \textbf{$ C_{M} - C_{ \Phi } $-algorithmic
 information} are strictly connected notions.

 While the link is very clear in term of \textbf{prefix algorithmic
 entropy}, anyway, it is much obscure in terms of \textbf{simple algorithmic
 entropy}.

 To show this it is necessary to introduce some notion of  $ C_{M} $ - Coding
 Theory:
\begin{definition} \label{def:mathematically classical code}
\end{definition}
$ C_{M} $ - CODE:

a partial function $ D : \Sigma^{\star} \stackrel{ \circ } {
\mapsto } \Sigma^{\star} $ of decoding associating to each word $
\vec{x} $ belonging to the set HALTING(D) of \textbf{code words}
its \textbf{source word} $ D(\vec{x})$.

\medskip

Given a $ C_{M} $ - code D and a source word $ \vec{x} \in
\Sigma^{\star} $:
\begin{definition}
\end{definition}
SET OF THE D - CODE WORDS OF $\vec{x}$:

the (eventually empty) set $ D^{-1} (\vec{x})$.

\medskip

Let us observe that the definition\ref{def:mathematically
classical code} doesn't require nor the surjectivity of a code
(i.e. that each source word is codificable) neither the
injectivity of a code (i.e. that each source word has only one
code word).

\medskip

Let us now introduce a particular fundamental kind of code:
\begin{definition}
\end{definition}
PREFIX-CODE:

a code $ D : \Sigma^{\star} \stackrel{ \circ } { \mapsto }
\Sigma^{\star} $ such that HALTING(D) is prefix-free

\medskip

The more fundamental property of prefix-codes is given by the
following:
\begin{theorem} \label{th:Kraft inequality}
\end{theorem}
KRAFT'S INEQUALITY

\begin{hypothesis}
\end{hypothesis}
\begin{align*}
  I \text{  index set} & :  cardinality(I) \leq \aleph_{0} \\
  \{ l_{i} \in  & {\mathbb{N}} \}_{i \in I}
\end{align*}
\begin{thesis}
\end{thesis}
\begin{equation}
  \exists \,  D : \Sigma^{\star} \stackrel{ \circ } { \mapsto }
\Sigma^{\star} \text{ prefix code}  :   \{ | \vec{x} | , \vec{x}
\in HALTING(D) \} \, = \,  \{ l_{i} \in  {\mathbb{N}} \}_{i \in
I} \; \Leftrightarrow \; \sum_{ i \in I} 2^{- l_{i}} \, \leq \, 1
\end{equation}

We will appreciate the importance of Kraft Inequality as soon as
we will introduce the \textbf{universal algorithmic probability}
and the \textbf{Halting Probability}.

\medskip

Let us now start the probabilistic analysis of  $ C_{M} $ - Coding
Theory.

Let us suppose that the generic source-word $ \vec{x} $ occur
with probability $ P ( \vec{x} ) $.

Given an injective prefix-code $ D : \Sigma^{\star} \stackrel{ \circ }
{ \mapsto } \Sigma^{\star} $ we can then introduce the:
\begin{definition}
\end{definition}
AVERAGE CODE WORD LENGTH OF THE CODE D W.R.T. THE SOURCE CODE
DISTRIBUTION P:
\begin{equation}
  L_{D,P} \; := \; \sum_{\vec{x} \in HALTING(D)} P ( \vec{x} )
  | D (\vec{x}) |
\end{equation}

It is clear that, in a communicational situation, the objective
of a transmitter is to minimize the average code word length.

Clearly a coding strategy will be the more clever the more it will
assign short code words to highly probable source words and
viceversa, in order to minimize the average code word length.
\begin{definition}
\end{definition}
MINIMAL AVERAGE CODE WORD LENGTH ALLOWED BY THE DISTRIBUTION P:
\begin{equation}
  L \; := \; \min \{ L_{D,P} \, , \, D \, prefix-code \}
\end{equation}
\begin{definition}
\end{definition}
OPTIMAL PREFIX-CODE W.R.T. THE SOURCE CODE DISTRIBUTION P:

a prefix-code D such that:
\begin{equation}
  L_{D,P} \; = \; L
\end{equation}

\smallskip

The probabilistic approach to $ C_{M}  - C_{\Phi} $ Information
Theory is based on the following notion:
\begin{definition} \label{def:Shannon entropy of a distribution}
\end{definition}
SHANNON ENTROPY OF THE DISTRIBUTION P:
\begin{equation}
  H(P) \; := \; - \sum_{\vec{x} \in \Sigma^{\star}} P ( \vec{x} )
  \log_{2} P ( \vec{x} )
\end{equation}
The corner stone of $ C_{M}  - C_{\Phi} $ Probabilistic Information Theory is
the following:
\begin{theorem} \label{th:mathematically classical mathematically physical noiseless coding theorem}
\end{theorem}
$ C_{M}  - C_{\Phi} $ NOISELESS CODING THEOREM
\begin{equation}
  H(P) \; \leq \; L  \; \leq \; H(P)+1
\end{equation}

\smallskip
Let us now observe that \textbf{prefix algorithmic entropy} may be
used to define a particular code:

by definition we have that:
\begin{equation}
  I( \vec{x} ) \, = \, | \vec{x}^{\star} |
\end{equation}
where $ \vec{x}^{\star} $ is the shortest input for the fixed
universal Chaitin computer giving $ \vec{x} $ as output (or the
first one in quasi-lexicographic order if there are many).

The map $ D_{I} : \Sigma^{\star} \stackrel{ \circ } { \mapsto }
\Sigma^{\star} $ defined by:
\begin{equation}
  D_{I}( \vec{x} ) \; := \; \vec{x}^{\star}
\end{equation}
is by construction a prefix-code.

\smallskip

Since the code $ D_{I} $ is of pure algorithmic nature, it would
be very reasonable to think that it may me optimal only for some
ad hoc probability distribution, i.e. that for a generic
probability distribution P the average code word length of $
D_{I} $ w.r.t. P:
\begin{equation}
  L_{ D_{I} , P } \; = \; \sum_{\vec{x} \in HALTING( D_{I} )} P(
  \vec{x}) I( \vec{x})
\end{equation}
won't achieve the optimal bound of H(P) stated by
Theorem\ref{th:mathematically classical mathematically physical
noiseless coding theorem}

\smallskip

But  here the deep link between the
\textbf{probabilistic-approach} and the
\textbf{algorithmic-approach} makes the miracle: under mild
assumptions about the distribution P the code $ D_{I} $ is
optimal as is stated by the following:
\begin{theorem} \label{th:link between mathematically classical mathematically physical probabilistic information and mathematically classical mathematically physical algorithmic information}
\end{theorem}
LINK BETWEEN $C_{M}-C_{\Phi}$ PROBABILISTIC INFORMATION AND
$C_{M}-C_{\Phi}$ ALGORITHMIC INFORMATION

\begin{hypothesis}
\end{hypothesis}
P  $ C_{M}-C_{\Phi} - \Delta_{0}^{0} $  probability distribution
over $ \Sigma^{\star} $

\begin{thesis}
\end{thesis}
\begin{equation}
  \exists c_{P} \in {\mathbb{R}}_{+} \; : \; 0 \, \leq  \,
  L_{D_{I},P} - H(P) \, \leq  \, c_{P}
\end{equation}

\smallskip
Such a result of a substantial equivalence between
\textbf{Shannon entropy} and \textbf{average algorithmic prefix
entropy} has a strongly weaker counterpart in terms of
\textbf{algorithmic simple entropy}.

Indeed the two algorithmic entropies are linked by the following:
\begin{theorem} \label{th:first Solovay theorem}
\end{theorem}
FIRST SOLOVAY'S THEOREM:
\begin{align}
  I( \vec{x} ) \; & \; = \; K( \vec{x} ) \, + \, K( string^{- 1} ( K (\vec{x} )) \, +  \, O ( K ( string^{- 1}  K( string^{- 1} ( K (\vec{x} )))))   \\
  K( \vec{x} ) \; & \; = \; I( \vec{x} ) \, - \, K( string^{- 1} ( K (\vec{x} )) \, -  \, O ( I ( string^{- 1}  I( string^{- 1} ( I (\vec{x} )))))
\end{align}
that substituted in the Theorem\ref{th:link between
mathematically classical mathematically physical probabilistic
information and mathematically classical mathematically physical
algorithmic information} gives:
\begin{equation}
  - c_{P} \; \leq \; L_{D_{K},P} - H(P) \; \leq \; \sum_{\vec{x}}
  P ( \vec{x} ) K( string ^{-1} (C ( string ^{-1} ( \vec{x}))))
\end{equation}
which is bounded only if $ \sum_{\vec{x}}  P ( \vec{x} ) K( string
^{-1} (C ( string ^{-1} ( \vec{x})))) $
  converges.
 \item called U the fixed Chaitin universal computer let us
 introduce the main actors of some of the most fascinating developes of $ C_{M} - C_{\Phi} $ Algorithmic Information Theory:
\begin{definition} \label{def:universal algorithmic probability}
\end{definition}
UNIVERSAL ALGORITHMIC PROBABILITY OF $ \vec{x} \in \Sigma^{\star}
$:
\begin{equation}
  P_{U} (\vec{x}) \; := \; \sum_{\vec{y} \in \Sigma^{\star} : U( \vec{y} ) = \vec{x}
  } 2^{- |\vec{y} | }
\end{equation}
\begin{definition} \label{def:halting probability}
\end{definition}
HALTING PROBABILITY:
\begin{equation}
  \Omega_{U} \; := \; \sum_{ \vec{x} \in \Sigma^{\star}}  P_{U} (\vec{x})
\end{equation}
These notion has a very intuitive meaning:
\begin{itemize}
  \item $ P_{U} (\vec{x}) $ is
the probability that the computer U gives as output the string $
\vec{x} $ under an uniformely random distributed input.
  \item $ \Omega_{U} $ gives the probability that the computer U halts
  under an uniformely random distributed input.
\end{itemize}
Such a probabilistic meaning,anyway, lies on the fact that U is a
\textbf{Chaitin computer} so that its halting set is prefix-free
and hence Theorem\ref{th:Kraft inequality} implies that:
\begin{equation} \label{eq:universal algorithmic probability takes values in the unitary interval}
  0 \; \leq \; P_{U} (\vec{x})  \; \leq \; 1 \; \; \forall \vec{x}
  \in \Sigma^{\star}
\end{equation}
and that:
\begin{equation} \label{eq:halting probability takes values in the unitary interval}
  0 \; \leq \; \Omega_{U}   \; \leq \; 1
\end{equation}
If we considered \textbf{simple algorithmic information} instead
of \textbf{prefix algorithmic information} and hence we adopted a
\textbf{non Chaitin computer}, anyway, the halting set of U
wouldn't be  prefix-free anymore, so that Theorem\ref{th:Kraft
inequality} wouldn't imply eq.\ref{eq:universal algorithmic
probability takes values in the unitary interval} and
eq.\ref{eq:halting probability takes values in the unitary
interval}.

\item Unlike \textbf{prefix algorithmic information}, \textbf{simple algorithmic information} is affected by
oscillations that exclude the possibility of using it to define the
notion of algorithmic randomness for sequences in an enough robust
way as we will show in the next section
\end{enumerate}
\newpage
\section{Chaitin random strings and sequences of cbits}
Let us suppose to make 100 independent tosses of a  coin.

If we obtained head all times we would certainly claim the the
used coin is not fair.

But let us observe that, assuming that the coin is fair,  the
string of 100 heads have the same exact probability, i.e. $ 2^{-
100} $, of any other binary string of 100 cbits.

So, which foundation could we give at our claim that the coin is
not fair?

The first to analyze this problem was Laplace that
dedicated to this issue the Fifth Principle among the
\textit{"General Principles of the calculus of probabilities"}
making the content of the third chapter of his pioneering work
\cite{Laplace-51}; it is worth to report his own words:
\begin{center}
  \textit{"Sixth Principle: Each of the causes to which an observed event may be attributed is indicated with just as much likelihood as there is probability that the event will take place supposing the event to be constant.
  The probability of the existence of any one of these causes is then a fraction whose numerator is the probability of the event resulting from this cause and whose denominator is the sum of the similar probabilities relative to all the causes; if these various causes considerated \`{a} priori, are unequally probable, it is necessary in place of the probability of the event resulting from each cause, to employ the product of this probability by the possibility of the cause itself.
  This is the fundamental principle of this branch of the analysis of chances which consists in passing from events to causes.}
  \textit{This principle gives the reason why we attribute regular events to a particular cause. Some philosophers have thought that these events are less possible than others and that at the play of heads and tails, for example, the combiantion in which heads occurs twenty successive times is less easy in its nature than those where head and tails are mixed in irregular manner.
   But this opinion suppose that past events have an influence on the possibility of future events which is not at all admissible. The regular combinations occur more rarely only because they are less numerous. $ \cdots $}
  \textit{Thus at the play of head and tail the occurence of heads a hundred successive times appears to us extraordinary because of the almost infinite number of combinations which may occur in a hundred throws; and if divide the combinations in two regular series containing an order easy to comprehend, and into irregular series, the latter are incomparably more numerous"}
\end{center}
Laplace catchs the following basic points:
\begin{itemize}
  \item what the string made of one hundred heads have of
  particular is to possess some kind of regularity
  \item this string has the same probability $ 2^{- 100} $ of every other string
  \item the fact that if this string of results occurs we can claim the coin was unfair is founded by the observation that the
  fraction of the set of 100 cbit strings made by strings having some kind of
  regularity, i.e. the  probability that a string of this kind occurs, is enormously low and , conseguentially, the
  probability that a string of this kind occurs is extraordinarily
  low.
\end{itemize}
The only thing Laplace wasn't able to explain, as anyone else for
little less than two centuries, was the exact meaning of the
locution \emph{\textbf{"to possess some regularity"}}.

In this dissertation we will see how Classical Algorithmic
Information Theory gives many equilavent mathematical
characterization of this \textbf{absence of regularity} or, as we
say it nowadays , of this \textbf{algorithmic randomness}.

Among these characterization the more important one is with no
doubt that as \textbf{algorithmic incompressibility}.

As an \textbf{algorithmically incompressible} object we mean, informally speaking , an object whose more concise algorithmic description is substantially its own assignation.

So one could be to tempted to say that the string $ \vec{x} \in
\Sigma^{\star} $ is algorithmically random iff:
\begin{equation} \label{eq:naife simple algorithmic random string}
  K( \vec{x} ) \; = \; | \vec{x} |
\end{equation}
or iff:
\begin{equation} \label{eq:naife prefix algorithmic random string}
   I( \vec{x} ) \; = \; | \vec{x} |
\end{equation}
The meaningness of these definitions, anyway, appear evident as
soon as one keeps into account , as to eq.\ref{eq:naife simple algorithmic random string},  the issue of the additive constant
involved in the passage from a \textbf{universal computer} to another \textbf{universal computer} and, as to  eq.\ref{eq:naife prefix algorithmic random string}, the issue of the additive constant involved in the passage from a \textbf{Chaitin universal computer} to another \textbf{Chaitin universal computer}.

The notion of random string originally introduced by Kolmogorov in 1965 was the following: given a constant $ c \in {\mathbb{R}}_{+} $:
\begin{definition} \label{def:Kolmogorov c-random string}
\end{definition}
$ \vec{x} \in \Sigma^{\star} $ IS c-KOLMOGOROV-RANDOM:
\begin{equation}
  K( \vec{x} ) \; \geq \; | \vec{x} | \, - \, c
\end{equation}

\smallskip

Before of analyzing the analogous notion involving \textbf{prefix
algorithmic information} instead of \textbf{simple algorithmic information} let us introduce the following preliminary notion:
\begin{definition} \label{def:busy beaver function}
\end{definition}
BUSY BEAVER FUNCTION:

the function $  \Sigma : {\mathbb{N}} \, \rightarrow \,
{\mathbb{N}} $:
\begin{equation}
   \Sigma(n) \; := \; \max_{ \vec{x} \in \Sigma^{n}} I( \vec{x} )
\end{equation}

It obeys the following \cite{Calude-94}:
\begin{theorem}
\end{theorem}
\begin{equation}
   \Sigma(n) \; \stackrel{ + }{=} \; n \, + \, I ( string(n))
\end{equation}

\smallskip

Chaitin's idea was that of defining the random strings of length n to be the strings with maximal prefix entropy among the strings of length n. So, given a natural number m:
\begin{definition} \label{def:Chaitin m-random string}
\end{definition}
 $ \vec{x} \in \Sigma^{\star} $ IS CHAITIN m-RANDOM:
\begin{equation}
  I( \vec{x} ) \; \geq \;  \Sigma( | \vec{x} | ) \, - \, m
\end{equation}
We will denote the set of al the Chaitin m-random binary strings
by $CHAITIN-m-RANDOM( \Sigma^{\star}$.

a 0-Chatin random string is often called simply a \textbf{Chaitin random}. Following this terminology we will pone:
\begin{equation}
  CHAITIN-RANDOM( \Sigma^{\star}) \; := \;  CHAITIN-0-RANDOM(
  \Sigma^{\star})
\end{equation}

\smallskip

\begin{remark} \label{rem:impossibility of a sharp distinction between regularity and randomness for strings}
\end{remark}
IMPOSSIBILITY OF A SHARP DISTINCTION BETWEEN REGULARITY AND RANDOMNESS FOR STRINGS

It is essential to observe that the introduction of additive
constants in both definition\ref{def:Kolmogorov c-random string}
and definition\ref{def:Chaitin m-random string} solves the
problem of the inconsistence of, respectively,
definition\ref{eq:naife simple algorithmic random string} and
definition\ref{eq:naife prefix algorithmic random string} only in
a partial way: indeed definition\ref{def:Kolmogorov c-random
string} and definition\ref{def:Chaitin m-random string} continue
to depend upon, respectively, the fixed \textbf{universal
computer} and the fixed  \textbf{universal Chaitin computer}.

The improvement is that under these ansatzs one doesn't lose algorithmic randomness of strings but passes simply from algorithmic randomness relative
to a certain constant to algorithmic randomness relative to a different constant.

But in this way one has to look at the transition from  regular to random strings as a continuous, asymptotic one:

indeed the effective connotation of randomness given by the
specification that a certain string $ \vec{x}\in \Sigma^{\star} $
is m-Chaitin random is the more significative the more high is
the difference $ | \vec{x} | - m $.

A sharp distinction is possible only for sequences.

In chapter\ref{chap:Classical algorithmic randomness as stability of the relative frequences under proper classical algorithmic place selection rules} we will give a clear, intuitive explanation of this fact in terms of Classical Gambling Theory.

Unfortunately, as we will  show in part\ref{part:The road for  quantum algorithmic randomness}, this point hasn't received the necessary consideration in most the attempts of defining quantum algorithmic randomness, that , erroneously to our opinion, concentrate the analysis to strings
of qubits considering this case as simpler and only later, in a derivate mode, pass to analyze sequences of qubit.

We anticipate here that our point of view is opposite: since exactly as in the classical case a sharp distinction between regularity and randomness is possible only for sequences of qubits, the analysis of quantum algorithmic randomness has to start from sequences of qubit.

\bigskip

Let us now observe that there exist many reasons to prefer Chaitin-randomness to Kolmogorov-randomness:
\begin{enumerate}
  \item the adoption of Chaitin randomness allows to give a clear
  quantitative foundation to the observation Laplace himself realized almost two centuries ago, namely that the strings not having any kind of regularity, the patternless ones, are the overwhelming majority:
\begin{theorem}
\end{theorem}
\begin{equation}
  \exists c \in {\mathbb{R}}_{+} \; : \; cardinality ( CHAITIN-RANDOM(
  \Sigma^{n})) \, > \, 2^{n - c} \; \; \forall n \in {\mathbb{N}}
\end{equation}

 \item Robert Solovay has proved that Chaitin randomness is
 stronger than Kolmogorov randomness

 \item  Chaitin randomness may be easily extended to binary
 sequences defining, informally speaking, an algorithmic random
 sequence as one whose prefixes are all Chaitin algorithmic
 random.

 As we will now show, the \textbf{phenomenon of the oscillations of simple algorithmic
 entropy} avoid this possibility for Kolmogorov randomness.
\end{enumerate}

\medskip
Let us introduce , first of all, a useful notation.

Given a sequence $ \bar{x} \in \Sigma^{\infty} $ let us denote by
$ x_{n} $ its $ n^{th} $ digit, by $ \vec{x}(n) $ its prefix of
length n and by $  \vec{x}(n , m) $ ($ n \leq m$) the substring
of $ \bar{x} $ obtained taking its digits from the $ n^{th} $ to
the $ m^{th} $, namely:
\begin{equation}
    \vec{x}(n , m) \; := \; x_{n} \cdots  x_{m} \, \in \, \Sigma^{m-n}
\end{equation}

Let us then observe that by identifying the generic string $
\vec{x} \in \Sigma^{\infty} $ with the sequence $  \vec{x}
0^{\infty} \in  $ we can look at $ \Sigma^{\star} $ as a proper subset of $ \Sigma^{\infty} $.

Let us then introduce the following useful map:
\begin{definition} \label{def:numeric representation}
\end{definition}
NUMERIC REPRESENTATION:

$ {\mathcal{N}} : \Sigma^{\infty}  \, \mapsto \, [ 0 ,1) $:
\begin{equation}
  {\mathcal{N}} ( \bar{x} ) \; := \; \sum_{n = 1}^{\infty} \frac{x_{n}}{2^{n}}
\end{equation}
whose restriction $ {\mathcal{N}} |_{ \Sigma^{\infty} -
\Sigma^{\star}} $ is a bijection and allows, conseguentially, to
identify $ \Sigma^{\infty} $ with the set $ [ 0 ,1 ) $.

Let us then introduce the probability measure:
\begin{definition}
\end{definition}
UNBIASED PROBABILITY MEASURE ON $ \Sigma^{\infty} $:

$ P_{unbiased} \, : \, 2^{ \Sigma^{\infty}} \; \stackrel{\circ
}{\rightarrow} \; [0,1]$ :
\begin{align}
  HALTING(P_{unbiased}) \; & = \; {\mathcal{F}}_{cylinder} \\
  P_{unbiased} ( \Gamma_{\vec{x}} ) \; & \equiv \;  \frac{1}{2^{| \vec{x} |}} \; \;
  \forall \, \vec{x} \, \in \, \Sigma^{\star}
\end{align}
where:
\begin{definition}
\end{definition}
CYLINDER SET W.R.T. $ \vec{x} \, =  ( x_{1} , \ldots , x_{n} ) \,
\in \, \Sigma^{\star} $:
\begin{equation} \label{eq:cylinder set}
\Gamma_{\vec{x}} \; \equiv \; \{ \bar{y} = ( y_{1} , y_{2} ,
\ldots ) \in \Sigma^{\infty} \; : \; y_{1} = x_{1} , \ldots ,
y_{n} = x_{n} \}
\end{equation}
\begin{definition}
\end{definition}
CYLINDER - $ \sigma $ - ALGEBRA ON $ \Sigma^{\infty} $:
\begin{equation}
  {\mathcal{F}}_{cylinder} \; \equiv \;  \sigma- \text{algebra generated by}  \{ \Gamma_{\vec{x}} \, : \, \vec{x} \in \Sigma^{\star}
  \}
\end{equation}

\smallskip

In the numeric representation of $ \Sigma^{\infty} $ as the real
interval $ [ 0 ,1) $, $P_{unbiased} $ is, clearly, nothing but
Lebesgue measure \cite{Lebesgue-73}.

Denoted by $ N_{i}^{n} ( \bar{x} ) $ the number of successive  $ i
\in \Sigma $ ending in position n of the sequence $ \bar{x} $ the
First Borel-Cantelli's Lemma implies that (cfr. the fifth section
of the fourth chapter of \cite{Billingsley-95}):
\begin{theorem} \label{th:on the islands of regularity of almost all sequences}
\end{theorem}
\begin{align}
  P_{unbiased} & ( \{ \bar{x} \in \Sigma^{\infty} \, : \, \limsup_{ n \rightarrow \infty} N_{0}^{n} ( \bar{x} ) \, = \, 1 \} ) \; = \; 1  \\
  P_{unbiased} & ( \{ \bar{x} \in \Sigma^{\infty} \, : \, \limsup_{ n \rightarrow \infty} N_{1}^{n} ( \bar{x} ) \, = \, 1 \} ) \; = \; 1
\end{align}

\smallskip

Theorem\ref{th:on the islands of regularity of almost all
sequences} tells us that for  $ P_{unbiased} $-almost all
sequences $ \bar{x} \in \Sigma^{\infty} $ there exist infinitely many n for which:
\begin{equation}
  \vec{x}(n) \; \stackrel{ + }{=} \; \vec{x} ( 1 , n - \log_{2} n) 0^{n}
\end{equation}
i.e.:
\begin{equation} \label{eq:oscillations of simple algorithmic entropy}
  K( \vec{x}(n) ) \stackrel{ + }{=} \; n - \log_{2} n
\end{equation}
This suggest that if we adopted the following definition of algorithmic randomness for sequences:
\begin{definition} \label{def:Kolmogorov random sequence}
\end{definition}
$ \bar{x} \in \Sigma^{\infty} $  IS KOLMOGOROV RANDOM:
\begin{equation}
  \exists c \in {\mathbb{R}}_{+} \; : \; K ( \vec{x}(n) ) \, > \,
  n - c \; \; \forall n \in  {\mathbb{N}}
\end{equation}
there wouldn't exist Kolmogorov random sequences.

That this is indeed the case may be rigorously proved observing
that  the existence of infinitely many n such that
eq.\ref{eq:oscillations of simple algorithmic entropy} holds may
be proved to hold for all sequences (not only with $ P_{unbiased}$- probability one).

\smallskip

This doesn't happen if we use prefix algorithmic entropy.
\begin{definition} \label{def:Chaitin random sequence}
\end{definition}
$ \bar{x} \in \Sigma^{\infty} $  IS CHAITIN RANDOM:
\begin{equation}
  \exists c \in {\mathbb{R}}_{+} \; : \; I ( \vec{x}(n) ) \, > \,
  n - c \; \; \forall n \in  {\mathbb{N}}
\end{equation}
We will denote the set of all the Chaitin random sequences by $
CHAITIN -RANDOM(\Sigma^{\infty})$.

By the numeric representation's map the notion of Chaitin
randomness may be immediately extended to reals numbers in the
following way:

\begin{definition} \label{def:Chaitin random real number}
\end{definition}
 $ x \in [ 0 \, , \, 1 ) $ IS CHAITIN RANDOM:
\begin{equation}
   ({\mathcal{N}} | _{\Sigma^{\infty} \, -  \, \Sigma^{\star}})^{- 1} (\Omega)
\end{equation}
We will denote the set of all the random real numbers by $ CHAITIN
-RANDOM( [ 0 , 1 )) $

As we will prove later, \textit{"almost all"} the numbers in the
unitary interval are Chaitin-random.

In particular one has the following:
\begin{theorem} \label{th:Chaitin randomness of the halting probability}
\end{theorem}
CHAITIN RANDOMNESS OF THE HALTING PROBABILITY:
\begin{equation}
  \Omega \; \in \;  CHAITIN-RANDOM( [ 0 , 1 ))
\end{equation}

Supposing now to let the fixed Chaitin universal computer to
vary, Theodore A. Slaman has recentely proved the followin
remarkable  \cite{Kucera-Slaman-01}:

\begin{theorem} \label{th:Slaman's theorem}
\end{theorem}
SLAMAN'S THEOREM:
\begin{equation}
  \{ \, \Omega_{U} \, , \, : \, U \text{ Chaitin's universal computer }
  \, \} \; = \; CHAITIN-RANDOM( [ 0 \, , \, 1 ) ) \, \bigcap \,
  REC-EN( {\mathbb{R}})
\end{equation}

\newpage
\section{Brudno random sequences of cbits} \label{sec:Brudno random sequences of cbits}
As to the definition of algorithmically random binary sequences
we have seen in the previous section that the phenomenon of
oscillations of simple algorithmic entropy causes that, denoted
by $KOLMOGOROV-RANDOM(\Sigma^{\infty})$ the set of all the
Kolmogorov-random binary sequence, one has that:
\begin{theorem} \label{th:not existence o Kolmogorov random sequences}
\end{theorem}

NOT EXISTENCE OF KOLMOGOROV RANDOM SEQUENCES:
\begin{equation}
  KOLMOGOROV-RANDOM( \Sigma^{\infty} ) \; = \; \emptyset
\end{equation}

\smallskip

One could , at this point, argue that the existence of
infinite islands of regularity in a generic sequence resulting in
the logarithmic deficit of simple algorithmic entropy showed by
an infinite number of its prefixes is a false problem since what
is really relevant is the  rate of plain algorithmic entropy for
digit of the prefixes, i.e. the ratio $ \frac{K( \vec{x}(n)) }{n} $ for whose asymptotic
behaviour the logarithmic deficits are irrilevant since obviuosly:
\begin{equation}
  \lim_{n \rightarrow \infty} \frac{ \log_{2}(n) }{n} \; = \; 0
\end{equation}
This way of reasoning led A.A. Brudno to introduce the following notions:
\begin{definition} \label{def:Brudno algorithmic entropy of a sequence}
\end{definition}
BRUDNO ALGORITHMIC ENTROPY OF $ \bar{x} \in \Sigma^{\infty} $:
\begin{equation}
  B( \bar{x} ) \; := \; \lim_{n \rightarrow \infty} \frac{K( \vec{x}(n)) }{n}
\end{equation}

\smallskip
At this point one could think that considering the asympotic rate
of prefix entropy instead of simple entropy would result in a
different definition of the algorithmic entropy of a sequence.

That this is not the case is the claim of the following:
\begin{theorem} \label{th:plain-prefix insensitivity of Brudno algorithmic entropy}
\end{theorem}
\begin{equation}
   B( \bar{x} ) \; = \; \lim_{n \rightarrow \infty} \frac{I( \vec{x}(n))}{n}
\end{equation}
\begin{proof}
It immediately follows by the fact that \cite{Staiger-99}:
\begin{equation}
   I( \vec{x}(n) ) \, - \, K( \vec{x}(n) ) \; = \; o(n)
\end{equation}
\end{proof}

\begin{definition} \label{def:Brudno random sequence}
\end{definition}
$ \bar{x} \in \Sigma^{\infty} $ IS BRUDNO RANDOM:
\begin{equation}
  B( \bar{x} ) \; > \; 0
\end{equation}
We will denote the set of all the Brudno random binary sequences
by $ BRUDNO( \Sigma^{\infty})  $.

\smallskip

Anyway one is here faced to a problem almost always misunderstood that is the main source of a  sort of incomunicability between the scientific community of mathematical physicists
studying Dynamical Systems Theory and the scientific community of the logico-mathematicians and Theoretical-Computer scientists studying Algorithmic Information Theory:
\begin{theorem} \label{th:Brudno randomness is weaker than Chaitin randomness}
\end{theorem}
BRUDNO RANDOMNESS IS WEAKER THAN CHAITIN RANDOMNESS:
\begin{equation}
  BRUDNO-RANDOM( \Sigma^{\infty}) \; \supset \;  CHAITIN -RANDOM(\Sigma^{\infty})
\end{equation}

\smallskip

Such a theorem was proven by Brudno himself in the last section
of \cite{Brudno-83} by the explicit presentation of a
Brudno-random sequence doesn't passing a Martin-L\"{o}f test.

We postpone the presentation of such a proof to
section\ref{sec:Equivalence between passage of a Martin Lof
universal statistical test and Chaitin randomness} where the
involved properties of universal sequential Martin-L\"{o}f test are introduced.

\newpage
\section{Brudno algorithmic entropy versus the Uspensky abstract approach}  \label{sec:Brudno algorithmic entropy versus the Uspensky abstract approach}
In this section we will  show that definition\ref{def:Brudno algorithmic entropy of a sequence}
is not compatible with Uspensky's abstract approach of defining algorithmic information discussed in section\ref{sec:Uspensky's abstract definition of algorithmic information}.

Uspensky's abstract approach would, indeed, require the specification of:
\begin{enumerate}
  \item a \textbf{concordance relation} $ {\mathcal{R}} $
  \item a \textbf{point measure}  $ \mu $
\end{enumerate}
on $ \Sigma^{\infty} $ such to constitute a  \textbf{metric aggregate} $ ( \Sigma^{\infty} \, , \, {\mathcal{R}} \, , \, \mu )$.

Both these points are highly not-trivial.

Remembering remark\ref{rem:context dependence of the
computability constraint}  we have to keep attention on the
meaning of the computability constraint. Indeed, by
theorem\ref{th:cardinalities of strings and sequences}, we see
that the definition of the concordance relation $ {\mathcal{R}}
$  exit from the boundaries of $ C_{M}$-Classical Recursion Theory.

As we have anticipated in section\ref{sec:The distinction between mathematical-classicality and physical-classicality}, just as to Computability Theory by \textbf{physically classical
  computers} of  (partial) functions on sets  $ S \, : \, card(S) \, = \,
  \aleph_{1}$ many different inequivalent candidate theories have
  been proposed:
\begin{enumerate}
  \item the Orthodox Theory generated by the studies of Grzegorczyck -
  Lacombe \cite{Pour-El-Richards-89}
  \item the theory developed by the so called Markov School in the framework of
  Constructive Mathematics \cite{Odifreddi-89}
  \item the Blum - Shub - Smale 's
  Theory \cite{Smale-92}, \cite{Blum-Cucker-Shub-Smale-98}
\end{enumerate}

The basic notion of the Othodox Theory, namely the definition of
a  \textbf{recursive real number}, seems rather robust:

starting from the  $  C_{\Phi} $-Computability of the whole $
{\mathbb{Q}} $ the strategy of defining a \textbf{recursive real
number} consists in effectivizing a method for constructing $
{\mathbb{R}} $ from  $ {\mathbb{Z}} $; as shown by R.M. Robinson
whichever of these methods one effective:
\begin{enumerate}
  \item the construction of $ {\mathbb{R}} $ from  $ {\mathbb{Z}} $
  through Cauchy sequences
  \item the construction of $ {\mathbb{R}} $ from  $ {\mathbb{Z}}$
  through nested intervals
  \item  the construction of $ {\mathbb{R}} $ from  $
  {\mathbb{Z}}$ through the Dedekind Cut
  \item the construction of $ {\mathbb{R}} $ from  $ {\mathbb{Z}}$
  through the expansion to base b, where b is an integer $ > 1 $.
\end{enumerate}
one results in the the same set $ REC ({\mathbb{R}}) $
\cite{Pour-El-Richards-89}, \cite{Pour-El-99}.

Let us review the first of these strategies:

given a sequence $ \{ r_{n}  \}_{ n \in {\mathbb{N}}} $ of rational numbers:
\begin{definition}
\end{definition}
$ \{ r_{n}  \} $ IS RECURSIVE:
\begin{equation}
  \exists b , c , s  \in  C_{M} - C_{\Phi} - \Delta_{0}^{0}-map({\mathbb{N}},{\mathbb{N}}) \; : \;
  r_{n} \, = \, (- 1) ^{s(n)} \frac{ b(n)  }{ c(n) }
\end{equation}
\begin{definition}
\end{definition}
$ \{ r_{n}  \} $ CONVERGES EFFECTIVELY  TO $ x \in {\mathbb{R}} $:
\begin{equation}
  \exists e \in  C_{M} - C_{\Phi} - \Delta_{0}^{0}-map({\mathbb{N}},{\mathbb{N}})
  \; : \;  m \geq e(n) \, \Rightarrow \, | r_{m} - x | < \frac{1}{2^{n}}
\end{equation}

\smallskip

Given a real number $ x \in {\mathbb{R}} $:
\begin{definition} \label{def:r.e. real numbers}
\end{definition}
RECURSIVELY ENUMERABLE REAL NUMBERS:
\begin{multline}
  REC-EN(  {\mathbb{R}} )  \; := \; \{ x \in {\mathbb{R}} )  \, : \\
   \text{ there is a increasing, computable sequence of rationals
  which converges to x}  \}
\end{multline}

\begin{definition} \label{def:recursive real number}
\end{definition}
RECURSIVE  REAL NUMBERS:
\begin{multline}
  REC(  {\mathbb{R}} )   \; := \; \{ x \in {\mathbb{R}} ) \, : \\
   \text{ there is a computable sequence of rationals
  which converges effectively to x } \}
\end{multline}

\smallskip

Given a complex number  $ z \in {\mathbb{C}} $:
\begin{definition} \label{def:recursive complex number}
\end{definition}
 z IS RECURSIVE:
\begin{equation}
   \Re(z) \in  REC( {\mathbb{R}} )  \; and \; \Im(z) \in  REC( {\mathbb{R}} )
\end{equation}
\smallskip

We will denote the set of all the recursive complex numbers by $
REC ( {\mathbb{R}} ) $.

It may be proved that:
\begin{theorem}
\end{theorem}
BASIC PROPERTIES OF $ REC( {\mathbb{R}} ) $
\begin{enumerate}
  \item  $ REC ( {\mathbb{R}} ) $ is a \textbf{closed field}
  \item
\begin{equation}
    cardinality( REC ( {\mathbb{R}} ) ) \; = \; \aleph_{0}
\end{equation}
\end{enumerate}
Obviously this immediately implies that:
\begin{corollary} \label{cor:enumerability of the set of all the recursive complex numbers}
\end{corollary}
\begin{equation}
  cardinality( REC ( {\mathbb{C}} )) \; = \; \aleph_{0}
\end{equation}

\medskip

To understand in which sense Corollary\ref{cor:enumerability of the set of all the recursive complex numbers}
may be seen highly unsatisfactory and even suggest the necessity of an Eterodox Theory, let us start from the analysis Roger Penrose dedicates to the issue:
\begin{center}
 \textbf{Is Mandelbrot's set recursive?}
\end{center}
in the seventh section of the fourth chapter of his book
\cite{Penrose-89} and its reformulation by Lenore Blum, Felipe
Cucker, Michael Shub and Steve Smale in the first two chapters of
their book \cite{Blum-Cucker-Shub-Smale-98} significatively
reporting a picture of Mandelbrot's set on its front cover.

Given a generic complex number $ c \in {\mathbb{C}} $ let us
introduce the polynomial $ p_{c} (z) \; := \; z^{2} + c $ and let
us denote by $ p_{c}^{(n)} (z) $ its $ n^{th} $ iterate.
\begin{definition} \label{def:Mandelbrot set}
\end{definition}
MANDELBROT'S SET:
\begin{equation}
  {\mathcal{M}} \; := \; {\mathbb{C}} \, - \, \{ c \in {\mathbb{C}} \, : \, \lim_{n \rightarrow \infty } p_{c}^{(n)} (0) = \infty  \}
\end{equation}
A key property of Mandelbrot's set is stated by the following
\cite{Falconer-90}:
\begin{theorem}
\end{theorem}
$ {\mathcal{M}} $ is the halting set of the following algorithm:
\begin{center}
Label[start] Input c

$ x^{2} + c \; \rightarrow \; x $

If $ | x | \, \leq \, 2 $ then Goto start

Output 1

Halt

\end{center}

To answer Penrose's query one needs an algorithm that, given the
input $ c \in {\mathbb{C}} $, will decide in a finite number of steps whether or not $ c \in {\mathcal{M}} $.

Penrose appeals to the Orthodox Theory, but immediately refutes it:
\begin{center}
    \textit{"One implication of this is that even with such a simple set as the unit disc $ \cdots $ there would be no algorithm for deciding for sure $ \cdots $ whether the computable number $ x^{2} + y^{2} $ is actually equal to 1 or not, this being the criterion for deciding whether or not the computable complex number $ x + i y $ lies on the unit circle $ \cdots $ Clearly that is not what we want"}
\end{center}

Then he tries to follow other approaches but at the end he concludes that:
\begin{center}
  \textit{"One is left with the strong feeling that the correct viewpoint has not yet been arrived at"}
\end{center}

Blum, Cucker, Shub and Smale settle Penrose's question in the
framework of their foundation of Computability Theory over a
generic ring \cite{Smale-92} as a generalization of the Goldstine
- Von Neumann axiomatization of Flowchart Theory (cfr. par.1.5
of  \cite{Odifreddi-89}).
Introduced the following notion:
\begin{definition}
\end{definition}
$ S \; \subset \; {\mathbb{C}} $  IS A SEMI-ALGEBRAIC SET:

it is a Boolean combination of sets defined by polynomial equalities and disequalities

\smallskip

Then they prove that a necessary condition for the decidability
of a set $ S \; \subset \; {\mathbb{C}} $ w.r.t. Computation
Theory over the ring $ {\mathbb{R}} $  is that it is the
countable union of semi-algebraic sets over $ {\mathbb{R}} $.

That this is not the case for the Mandelbrot's set follows by
Shishikura's Theorem stating that the boundary of $ {\mathcal{M}}
$ has Hausdorff dimension two, resulting in the following:
\begin{theorem} \label{th:Blum Cucker Shub Smale incomputability of Mandelbrot set}
\end{theorem}

$ {\mathcal{M}} $ is not Blum, Cucker, Shub and Smale computable

\medskip

Following Uspensky abstract approach of defining algorithmic
information, one would infer from Theorem\ref{th:Blum Cucker Shub
Smale incomputability of Mandelbrot set} that  $ {\mathcal{M}} $
has infinite algorithmic information, constrasting which what is
claimed by another book reporting a picture of (part of) the
Mandelbrot's set on its front cover, namely
\cite{Cover-Thomas-91}, and stating that its information content
is essentially zero.

That the applicability of the Uspensky abstract approach to this
particular case might be problematic, anyway, results by the
obsevation that it doesn't seem to exist some natural
\textbf{point measure} to use in the the specification of the
\textbf{metric aggregate} $ ( \Sigma^{\infty} \, , \,
{\mathcal{R}} \, , \, \mu )$.

This throws a shadow on the same foundation of $ NC_{M} -
C_{\Phi} $ - Algorithmic Information that is, as has been
recentely remarked by Chaitin in the first paragraph of the fourth
chapter of \cite{Chaitin-01}, mostly an unexplored field.

Chaitin discusses this subject in the usual concretelly LISP
programming attitude he followed in his other two Springer books
\cite{Chaitin-98}, \cite{Chaitin-99}, adding to his version of
LISP a new primitive function \textbf{display} that allow to get
the partial outputs of a non-halting computation, and hence
considering the algorithmic information of (so produced) infinite
sets of S-expressions: in this way he concretely shows that the
infinite version of $ C_{\Phi} $ - Algorithmic Information Theory
differs for the finite version in many respects, an example being
the violation of Theorem\ref{th:subadditivity of prefix
algorithmic entropy}.

\newpage
\section{The algorithmic approach to Classical Chaos Theory and Brudno's Theorem}
Let us review the basic notions of Classical Ergodic Theory:

given a classical probability space $ ( X \, , \, \mu ) $:
\begin{definition}  \label{def:endomorphism of a classical probability space}
\end{definition}
ENDOMORPHISM OF $ ( X \, , \, \mu ) $:

$T \, : \, HALTING(\mu) \rightarrow HALTING(\mu) $ surjective :
\begin{equation}
  \mu ( A ) \; = \;   \mu ( T^{-1} A ) \; \; \forall A \in HALTING(\mu)
\end{equation}

\smallskip

\begin{definition} \label{def:automorphism of a classical probability space}
\end{definition}
AUTOMORPHISM OF $ ( X \, , \, \mu ) $:

$T \, : \, HALTING(\mu) \rightarrow HALTING(\mu) $ injective endomorphism of  $ ( X \, , \, \mu ) $

\smallskip

\begin{definition} \label{def:classical dynamical system}
\end{definition}
CLASSICAL DYNAMICAL SYSTEM:

a therne $ ( X \, , \, \mu \, , \, T ) $ such that:
\begin{itemize}
  \item  $ ( X \, , \, \mu ) $ is a classical probability space
  \item  $T \, : \, HALTING(\mu) \rightarrow HALTING(\mu) $ is an
  endomorphism of  $ ( X \, , \, \mu ) $
\end{itemize}

\medskip

Given a classical dynamical system  $ ( X \, , \, \mu \, , \, T ) $:
\begin{definition}  \label{def:reversible classical dynamical system}
\end{definition}
$ ( X \, , \, \mu \, , \, T ) $ IS REVERSIBLE:

T is an automorphism

\smallskip

\begin{definition} \label{def:ergodic classical dynamical system}
\end{definition}
$ ( X \, , \, \mu \, , \, T ) $ IS ERGODIC:
\begin{equation}
lim_{n \rightarrow \infty} \frac{1}{n} \sum_{k=0}^{n-1} \, \mu ( A
\cap T^{k}(B)) \; = \; \mu(A) \,  \mu(B) \; \forall \, A,B
\in HALTING( \mu )
\end{equation}
\begin{definition} \label{def:mixing classical dynamical system}
\end{definition}
$ ( X \, , \, \mu \, , \, T ) $ IS MIXING:
\begin{equation}
  lim_{n \rightarrow \infty} \, \mu ( A \cap T^{n}(B)) \; = \;
    \mu(A) \,  \mu(B) \; \forall \, A,B \in  HALTING( \mu )
\end{equation}

\medskip

\begin{example}
\end{example}
CLASSICAL SHIFTS

\begin{definition} \label{def:def:classical shift}
\end{definition}
CLASSICAL SHIFT OVER $ \Sigma $:

a classical dynamical system  $ ( \, \Sigma^{\infty} \,  , \, \sigma \, , \, \mu ) $ such that:
\begin{equation}
\begin{split}
  \sigma & : \Sigma^{\infty} \rightarrow \Sigma^{\infty} \\
   ( \sigma & \bar{x} ) _{n} \; := \; x_{n+1}
\end{split}
\end{equation}
and:
\begin{equation}
  HALTING( \mu ) \; = \; {\mathcal{F}}_{cylinder}
\end{equation}
\begin{remark} \label{rem:classical shift is synonimous of discrete-time stationary classical stochastic process}
\end{remark}
CLASSICAL SHIFT IS SYNONIMOUS OF DISCRETE-TIME STATIONARY CLASSICAL  STOCHASTIC PROCESS

The notion of a classical shift  is nothing but a way of inglobing the Theory of Classical
Stationary Stochastic Processes as a sub-discipline of Classical Ergodic Theory.

As we will see in section\ref{sec:From the communicational-compression of the Quantum Coding Theorems to the algorithmic-compression in Quantum Computation} an analogous inglobation is possible in a quantum case.

That the notion of a classical shift over $ \Sigma $ is indeed equivalent to the notion of a classical stationary stochastic process over $ \Sigma $
follows immediately by the following two facts:
\begin{enumerate}
  \item every classical probability measure $ \mu $ on $ \Sigma^{\infty} $
  such that $ HALTING( \mu ) \; = \; {\mathcal{F}}_{cylinder} $
  individuates the classical stationary stochastic process over $ \Sigma
  $ with \textbf{occurence probability of strings}:
\begin{equation} \label{eq:occurence probability of strings}
  p_{k} ( i_{1}  ,  \cdots ,  i_{k}  )  \; \equiv \; \mu ( \Gamma_{( i_{1} , \cdots , i_{k}
  )}) \; \;  i_{1}  ,  \cdots \ i_{k} \, \in \Sigma  ,  k \in { \mathbb{N}}
\end{equation}
satisfying the conditions:
\begin{equation} \label{eq:first constraint on the occurence probability of strings}
   p_{k} ( i_{1} \, , \, \cdots \,  i_{k} \, )  \; \geq \; 0
\end{equation}
\begin{equation} \label{eq:second constraint on the occurence probability of strings}
  \sum_{i \in \Sigma} p_{k+1} ( i_{1} \, , \, \cdots \,  i_{k} \, i ) \; = \; p_{k} ( i_{1}  \, , \, \cdots \,  i_{k} \, )
\end{equation}
\begin{equation} \label{eq:third constraint on the occurence probability of strings}
     \sum_{i \in \Sigma } p_{1} ( \, i \, ) \; = \; 1
\end{equation}
 \item the collection of functions:
 \begin{equation*}
   p_{k} ( i_{1} \, , \, \cdots \,  i_{k} \, )  \; \; i_{1} \, , \, \cdots \,  i_{k} \, \in \Sigma  \, , \, k \in {\mathbb{N}}
 \end{equation*}
expressing the \textbf{occurence probability of strings} of a classical stationary stochastic process (and, hence, satisfying the constraints eq.\ref{eq:first constraint on the occurence probability of
 strings}, eq.\ref{eq:second constraint on the occurence probability of
 strings}, eq.\ref{eq:third constraint on the occurence probability of
 strings}) individuates the $ \sigma$-invariant classical probability measure $ \mu $ on $ \Sigma^{\infty} $ such that $ HALTING( \mu ) \; = \; {\mathcal{F}}_{cylinder} $
 and:
 \begin{equation}
   p_{k} ( i_{1}  ,  \cdots ,  i_{k}  )  \; = \; \mu ( \Gamma_{( i_{1} , \cdots , i_{k}
   )}) \; \;  i_{1}  ,  \cdots \ i_{k} \, \in \Sigma  ,  k \in { \mathbb{N}}
 \end{equation}
\end{enumerate}

\smallskip

Let us introduce some useful notion:
\begin{definition} \label{def:stochastic vector}
\end{definition}
STOCHASTIC VECTOR OVER $ \Sigma $:

$ \vec{P} \; = \;
\begin{pmatrix}
  p_{0} \\
  p_{1}
\end{pmatrix}$
such that:
\begin{align}
  p_{i} & \geq 0 \; \; i=0,1 \\
  \sum_{i \in \Sigma } & p_{i} \; = \; 1
\end{align}
i.e. a column vector specifying a probability distribution over $ \Sigma $.

\smallskip

\begin{definition} \label{def:stochastic matrix}
\end{definition}
STOCHASTIC MATRIX OVER $ \Sigma  $:

 $ 2 \, \times \, 2 $ matrix:

$ \hat{P} =
\begin{pmatrix}
  p_{0,0} & p_{0 , 1 } \\
  p_{1,0} & p_{1,1}
\end{pmatrix} $
such that:
\begin{align}
  p_{i,j} & \geq 0 \; \; i,j=0,1 \\
  \sum_{j \in \Sigma} & p_{i,j} \; = \; 1 \; \; i=0,1
\end{align}

\smallskip

We can now introduce some basic classical shift:
\begin{definition} \label{def:classical bernoulli shift}
\end{definition}
CLASSICAL BERNOULLI SHIFT OVER $ \Sigma $ W.R.T. THE STOCHASTIC VECTOR $ \vec{P} \; = \; \begin{pmatrix}
p_{0} \\
p_{1}
\end{pmatrix}$:

the classical shift over $ \Sigma $ with  measure $ \mu $ :
\begin{equation}
p_{k} ( \, i_{1} \, , \, \cdots \, i_{k} \, ) \; = \;
\prod_{l=1}^{k} p ( \, i_{l} \, ) \; \; \forall k \in {\mathbb{N}
}
\end{equation}

\smallskip

Given a stochastic vector $ \vec{e} \; := \; \begin{pmatrix}
  e_{0} \\
  e_{1}
\end{pmatrix}$ and a stochastic matrix $ \hat{P} \; := \begin{pmatrix}
  p_{00} & p_{01} \\
  p_{10} & p_{11}
\end{pmatrix}$ such that:
\begin{equation}
  ( \vec{P} )^{T} \hat{P} \; = \; \hat{P}
\end{equation}

\begin{definition} \label{def:classical markov shift}
\end{definition}
CLASSICAL MARKOV SHIFT OVER $ \Sigma $ W.R.T. THE STOCHASTIC VECTOR $ \vec{e} $ AND THE STOCHASTIC MATRIX $ \hat{P} $

the classical shift over $ \Sigma $ with  measure $ \mu $ :
\begin{equation}
  p_{k} ( i_{1} \, , \, \cdots \,  i_{k} ) \; = \; e_{i_{1}} \, p_{i_{1} ,
  i_{2}} \, \cdots \, p_{i_{k-1} , i_{k}}
\end{equation}

\smallskip

\begin{remark}  \label{rem:classical shift is synonimous of stationary  classical information source}
\end{remark}
CLASSICAL SHIFT IS SYNONIMOUS OF STATIONARY CLASSICAL INFORMATION
SOURCE

We have seen in remark\ref{rem:classical shift is synonimous of discrete-time stationary classical stochastic process}  that the
notion of classical shift over $ \Sigma $ is equivalent to the notion of a stationary classical stochastic process
over $ \Sigma $.

But a classical stochastic process $ \{ x_{n} \} $
may be equivalentely seen as a classical information source,
considering the random variable $ x_{n} $ as the letter trasmitted at time n by a sender (Alice) to a receiver (Bob)
through a proper communicational channel.

The resulting equivalence between the notion of a classical shift and the notion of a stationary classical information
source persists at the quantum level as we will see in the section\ref{sec:The algorithmic approach to Quantum Chaos Theory: quantum algorithmic information versus quantum dynamical entropies}

\medskip

Given a classical probability space $ ( X \, , \mu ) $:
\begin{definition} \label{def:partition of a classical probability space}
\end{definition}
FINITE MEASURABLE PARTITION OF $ ( X \, , \, \mu ) $:
\begin{equation}
\begin{split}
  A \, &  = \; \{ \, A_{1} \, , \, \cdots \, A_{n} \} \; n \in
{\mathbb{N}} \, : \\
  A_{i} & \in  HALTING(\mu) \; \; i \, = \, 1 \, , \, \cdots \, n \\
  A_{i} & \, \cap \, A_{j} \, = \, \emptyset \; \; \forall \, i \, \neq \,
  j \\
  \mu  &  ( \, X  \,- \, \cup_{i=1}^{n} A_{i} \, ) \; = \; 0
\end{split}
\end{equation}

\smallskip

We will denote the set of all the finite measurable partitions of $ ( X \, , \, \mu ) $ by $ \mathcal{P} ( \, X \, , \, \mu \, ) $.
\begin{definition}
\end{definition}
$ A \in  \mathcal{P} ( X , \mu ) $ IS FINER  THAN  $ B \in \mathcal{P} ( X , \mu ) $:

every atom of B is the union of atoms by A

\smallskip

\begin{definition}
\end{definition}
COARSEST REFINEMENT OF $ A = \{ A_{i} \}_{i=1}^{n} $ AND  $ B = \{ B_{j} \}_{j=1}^{m} \in {\mathcal{P}}( \; X \, , \mu \; ) $:
\begin{equation}
  \begin{split}
      A \, & \vee \, B \; \in {\mathcal{P}}( X  , \mu )  \\
      A \, & \vee \, B \; \equiv \; \{ \, A_{i} \, \cap \, B_{j} \, \;  i =1 , \cdots , n \; j = 1 , \cdots , m  \}
  \end{split}
\end{equation}

Clearly $ \mathcal{P} ( X , \mu ) $ is closed both under coarsest refinements and under endomorphisms of $ ( X , \mu ) $.

\medskip

\begin{remark}
\end{remark}
COARSE-GRAINED MEASUREMENTS ON A CLASSICAL PROBABILITY SPACE: THE
OPERATIONAL MEANING OF A CLASSICAL PARTITION

Beside its abstract, mathematical formalization, the definition
\ref{def:partition of a classical probability space} has a precise
operational meaning.

Given the classical probability space $  ( X , \mu ) $ let us
suppose to make an experiment on the probabilistic universe it
describes using an instrument whose resolutive power is limited in
that it is not able to distinguigh events belonging to the same
atom of a partition $ A = \{ A_{i} \}_{i=1}^{n} \in \mathcal{P} (
X , \mu ) $.

Conseguentially the outcome of such an experiment will be a number
\begin{equation}
  r \in \{ 1 , \cdots , n \}
\end{equation}
specifying the observed atom $ A_{r} $ in our coarse-grained
observation of $ ( X, \mu ) $.

\smallskip

We will call such an experiment an \textbf{operational observation
of $ ( X , \mu ) $ through the partition A}.

\smallskip

Considered another partition $ B = \{ B_{j} \}_{i=1}^{n} \in
\mathcal{P} ( X , \mu ) $ we have obviously that the operational
observation of $ ( X , \mu ) $ through the partition $  A \vee B
$ is the conjuction of the two experiments consisting in the
operational observations of $ ( X , \mu ) $ through the
partitions, respectively, A and B.

Conseguentially we may consistentely call an \textbf{operational
observation of $ ( X , \mu ) $ through the partition A} more
simply an \textbf{A experiment}.

\medskip

\begin{remark}
\end{remark}
THE DOUBLE MEANING OF THE CLASSICAL PROBABILISTIC SHANNON ENTROPY
OF AN EXPERIMENT

The experimental outcome of an operational observation of $ ( X ,
\mu ) $ through the partition $ A = \{ A_{i} \}_{i=1}^{n} \in
\mathcal{P} ( X , \mu ) $ is a classical random variable having
as distribution the stochastic vector $ (
\begin{pmatrix}
  \mu(A _{1} ) \\
  \vdots       \\
  \mu(A _{n})
\end{pmatrix} $ whose Shannon entropy we will call the entropy of the partition
A, according to the following:
\begin{definition}
\end{definition}
ENTROPY OF $ A = \{ A_{i} \}_{i=1}^{n} \in \mathcal{P} ( X , \mu
) $:
\begin{equation}
  H(A) \equiv H ( \begin{pmatrix}
    \mu ( A _{1} ) \\
      \vdots       \\
    \mu ( A _{n} ) \
  \end{pmatrix} )
\end{equation}
with the right hand side expressed in terms of the
definition\ref{def:Shannon entropy of a distribution} we
introduced in section\ref{sec:Why prefix entropy is better than
simple entropy}.

It is fundamental, at this point, to observe that, given an
experiment, one has to distinguish between two conceptually
different concepts:
\begin{enumerate}
  \item  the \textbf{uncertainty of the experiment}, i.e. the amount of
  uncertainty on the outcome of the experiment before of realizing it
  \item  the \textbf{information of the experiment}, i.e. amount
  of information gained by the outcome of the experiment
\end{enumerate}
The fact that in Classical Probabilistic Information Theory both
these concepts are quantified by the Shannon entropy of the
experiment is a conseguence of the following (cfr. pag. 62 of
\cite{Billingsley-65}):
\begin{theorem} \label{th:the soul of Classical Information Theory}
\end{theorem}
THE SOUL OF CLASSICAL INFORMATION THEORY
\begin{equation}
  \text{information gained} \; = \; \text{uncertainty removed}
\end{equation}

\smallskip

Theorem\ref{th:the soul of Classical Information Theory} applies,
in particular, as to the partition-experiments we are discussing.

\medskip

Let us now consider a classical dynamical system $  CDS \, := \, (
X \, , \, \mu \, , \, T ) $.

The T-invariance of $ \mu $ implies the the partitions $ A = \{
A_{i} \}_{i=1}^{n} \in \mathcal{P} ( X , \mu ) $ and $ T^{-1}A $
have equal probabilistic structure. Conseguentially the \textbf{A
experiment} and the \textbf{ $T^{-1}A $-experiment} are repliques
, \textbf{not necessarily independent}, of the same experiment ,
made at succesive times.

In the same way the \textbf{$ \vee_{k=0}^{n-1} \, T^{-k} A
$-experiment} is the compound experiment consisting in n
replications $ A \, , \, T^{-1} A \, , \, , \cdots , \,
T^{-(n-1)}A  $ of the experiment corresponding to $ A \in
{\mathcal{P}}(X , \mu) $.

The amount of classical information per replication we obtain in
this compound experiment is clearly:
\begin{equation*}
  \frac{1}{n} \, H(\vee_{k=0}^{n-1} \, T^{-k} A )
\end{equation*}
It may be proved (cfr. e.g. the second paragraph of the third
chapter of \cite{Kornfeld-Sinai-00}) that when n grows this amount
of classical information acquired for replication converges, so
that the following quantity:
\begin{equation}
  h( A , T ) \; \equiv \; lim_{n \rightarrow \infty} \,  \frac{1}{n} \, H(\vee_{k=0}^{n-1} \, T^{-k} A )
\end{equation}
does exist.

In different words, we can say that $ h( A , T ) $ gives the
asymptotic rate of production of classical information for
replication of the A-experiment.

\begin{definition} \label{def:Kolmogorov-Sinai entropy}
\end{definition}
\begin{equation}
  h_{CDS} \; \equiv \; sup_{A \in {\mathcal{P}}(X , \mu)} \, h( A , T )
\end{equation}
By definition we have clearly that:
\begin{equation}
  h_{CDS} \; \geq \; 0
\end{equation}
\begin{definition} \label{def:classical chaoticity}
\end{definition}
CDS IS CHAOTIC:
\begin{equation}
  h_{CDS} \; > \; 0
\end{equation}

\smallskip

\begin{remark}
\end{remark}
INFORMATION-THEORETIC NATURE OF THE CONCEPT OF CLASSICAL CHAOS

Definition\ref{def:classical chaoticity} shows explicitly that
the concept of classical-chaos is an information-theoretic one:

a classical dynamical system is chaotic if there exist at least
one experiment on the system that, no matter how many times we
insist on repeating it, continue to give us classical information.

That such a meaning of classical chaoticity is equivalent to the
more popular one as the sensible (i.e. exponential) dependence of
dynamics from the initial conditions (the so called
\textbf{butterfly effect} for which the little perturbation of
the atmospheric flow produced here by  a butterfly's flight may
produce an hurricane in Alaska) is a conseguence of Pesin's
Theorem stating (under mild assumptions) the equality of the
Kolmogorov-Sinai entropy and the sum of the positive Lyapunov
exponents.

This inter-relation may be caught observing that:
\begin{itemize}
  \item if the system is chaotic we know that there exist an
  experiment whose repetition definitely continues to give information: such an information may be seen as the information on the initial condition
  that is necessary to furnish more and more with time if one want to keep
  the error on the prediction of the phase-point below a certain
  bound
  \item if the system is not chaotic the repetition of every
  experiment is useful only a finite number of times, after which
  every suppletive repetition doesn't furnish further information
\end{itemize}

Let us now consider the issue of symbolically translating the
coarse-gained dynamics:

the traditional way of proceeding is that described in the second
section of \cite{Alekseev-Yakobson-1981}:

given a positive integer $ n \in {\mathbb{N}} $ let us introduce
the:
\begin{definition}
\end{definition}
n-LETTERS ALPHABET:
\begin{equation}
  \Sigma_{n} \; := \; \{ 0 , \cdots , n - 1 \}
\end{equation}
Clearly:
\begin{equation}
  \Sigma_{2} \; = \; \Sigma
\end{equation}
Considered a partition $ A \, = \, \{ A_{i} \}_{i = 1}^{n} \in \,
{\mathcal{P}}(X , \mu) $
\begin{definition} \label{def:symbolic translator w.r.t. a partition}
\end{definition}
SYMBOLIC TRANSLATOR OF CDS W.R.T. A:

$ \psi_{A} \, : \, X  \rightarrow \Sigma_{n} $:
\begin{equation}
  \psi_{A} ( x  ) \; \equiv \; j  \, : \, x \in A_{j}
\end{equation}

In this way one associate to each point of X the letter, in the
alphabet having as many letters as the number of atoms of the
considered partition, labelling the atom to which the point
belongs.

Concatenating the letters corresponding to the phase-point at
different times one can then codify  $ k \in {\mathbb{N}}$ steps
of the dynamics:
\begin{definition} \label{def:n-point symbolic translator w.r.t. a partition}
\end{definition}
k-POINT SYMBOLIC TRANSLATOR OF CDS W.R.T. A:

$ \psi_{A}^{(k)} \, : \, X \, \rightarrow \Sigma_{n}^{k} $:
\begin{equation}
   \psi_{A}^{(k)} ( x ) \; \equiv \; \cdot_{j = 1}^{n}  \psi ( T^{j} x )
\end{equation}
and whole orbits:
\begin{definition} \label{def:orbit symbolic translator w.r.t. a partition}
\end{definition}
$ \psi_{A}^{(\infty)} \, : \, X \,  \rightarrow \,
\Sigma_{n}^{\infty} $:
\begin{equation}
   \psi_{A}^{( \infty)} ( x ) \; \equiv \; \cdot_{j = 1}^{\infty}  \psi ( T^{j} x )
\end{equation}

\medskip

The bug of this strategy of symbolic translation is the dependence
of the used alphabet from the cardinality of the partition:
\begin{multline}
  cardinality(A) \, \neq \, cardinality(B) \; \Rightarrow \\
  Range[ \psi_{A} ( x) ] \, \neq \, Range[ \psi_{B} ( x ) ] \; \;
  \forall x \in A \, , \, \forall A , B \in  {\mathcal{P}}( \, X \,
, \, \mu )
\end{multline}
As already done in \cite{Segre-02} I therefore adopt a different
strategy of symbolic-coding using only the binary alphabet $
\Sigma $ based on the following:
\begin{definition} \label{def:universal symbolic translator}
\end{definition}
UNIVERSAL SYMBOLIC TRANSLATOR OF CDS:

$ \Psi \, : \, X \, \times \,  {\mathcal{P}}( \, X \, , \, \mu )
\rightarrow {\Sigma^{\star} } $:
\begin{equation}
  \Psi ( x \, , \,  \{ A_{i} \}_{i = 1}^{n} ) \; \equiv \; string(
  j) \; \;  , \,  x \, \in \, A_{j}
\end{equation}
that can be again used to codify $ k \in {\mathbb{N}}$ steps of
the dynamics:
\begin{definition} \label{def:k-point universal symbolic translator}
\end{definition}
k-POINT UNIVERSAL SYMBOLIC TRANSLATOR OF CDS:

$ \Psi^{(k)} \, : \, X \, \times \,  {\mathcal{P}}( \, X \, , \,
\mu ) \rightarrow \Sigma^{\star} $:
\begin{equation}
   \Psi^{(k)} ( x \, , \,   \{ A_{i} \}_{i = 1}^{n} ) \; \equiv \;
   \cdot_{j = 1}^{k}  \psi ( T^{j} x \, , \,   \{ A_{i} \}_{i = 1}^{n} )
\end{equation}
and whole orbits:
\begin{definition} \label{def:orbit universal symbolic translator}
\end{definition}
ORBIT UNIVERSAL SYMBOLIC TRANSLATOR OF CDS:

$ \Psi^{(\infty)} \, : \, X \, \times \,  {\mathcal{P}}( \, X \,
, \, \mu ) \rightarrow \Sigma^{\infty} $:
\begin{equation}
   \Psi^{(\infty)} ( x \, , \,   \{ A_{i} \}_{i = 1}^{n} ) \; \equiv \;
   \cdot_{j = 1}^{\infty}  \psi ( T^{j} x \, , \,   \{ A_{i} \}_{i = 1}^{n} )
\end{equation}

Considered the binary sequences  obtained translating symbolically
the orbit generated by $ x in X $ through partitions   one is
naturally led to introduce the following notion:
\begin{definition} \label{def:Brudno algorithmic entropy of a point}
\end{definition}
BRUDNO ALGORITHMIC ENTROPY OF (THE ORBIT STARTING FROM) x:
\begin{equation}
  B_{CDS} (x) \; \equiv \; sup_{A \in {\mathcal{P}}(X , \mu)} B( \Psi^{(\infty)} ( x, A ) )
\end{equation}
linked to Kolmogorov-Sinai entropy by the celebrated:
\medskip
\begin{theorem} \label{th:Brudno theorem}
\end{theorem}
BRUDNO'S THEOREM
\begin{equation}
   h_{CDS} \; = \; B_{CDS} (x) \; \;  \forall - \mu - a.e. \, x
   \in X
\end{equation}
for whose proof and meaning I demand again to

\medskip

Let us now consider the \textbf{algorithmic approach to Classical
Chaos Theory} strongly supported by Joseph Ford, whose objective
is the characterization of the concept of chaoticity of a
classical dynamical system as the algorithmic-randomness of its
symbolically-translated trajectories.

To require such a condition for all the trajectories would be too
restrictive since it is reasonable to allow a chaotic dynamical
system to have a numerable number of periodic orbits.

Let us introduce then following two notions:
\begin{definition} \label{def:strong algorithmic chaoticity}
\end{definition}
CDS IS STRONGLY ALGORITHMICALLY-CHAOTIC:
\begin{equation}
  \forall-\mu-a.e. x \in X \, , \, \exists A \in {\mathcal{P}}(X ,
  \mu) \, : \, \Psi^{(\infty)} ( x \, , \,  A) \in CHAITIN-RANDOM(
  \Sigma^{\infty})
\end{equation}
\begin{definition} \label{def:weak algorithmic chaoticity}
\end{definition}
CDS IS WEAK ALGORITHMICALLY-CHAOTIC:
\begin{equation}
  \forall-\mu-a.e. x \in X \, , \, \exists A \in {\mathcal{P}}(X ,
  \mu) \, : \, \Psi^{(\infty)} ( x \, , \,  A) \in BRUDNO-RANDOM(
  \Sigma^{\infty})
\end{equation}

The difference between definition\ref{def:strong algorithmic
chaoticity} and definition\ref{def:weak algorithmic chaoticity}
follows by Theorem\ref{th:Brudno randomness is weaker than Chaitin
randomness}.

Clearly Theorem\ref{th:Brudno theorem} implies the following:
\begin{corollary}
\end{corollary}
\begin{align*}
  CHAOTICITY \; \; & = \; \; \text{WEAK ALGORITHMIC CHAOTICITY} \\
  CHAOTICITY \; \; & < \; \; \text{STRONG ALGORITHMIC CHAOTICITY}
\end{align*}
that shows that the algorithmic approach to Classical Chaos Theory
is equivalent to the usual one only in weak sense.
\newpage
\chapter{Classical algorithmic randomness as passage of all the classical algorithmic statistical
tests} \label{chap:Classical algorithmic randomness as passage of
all the classical algorithmic statistical tests}
\section{Pseudorandom generators}
We have seen in chapter\ref{chap:Classical algorithmic randomness
as classical algorithmic incompressibility} how the notion of
classical algorithmic randomness as classical algorithmic
incompressibility may be properly formalized.

The deepest way of introducing the characterization of classical
algorithmic randomness as passage of a certain battery of
statistical tests is to analyze the issue of \textbf{random
number generation}.

Such an expression is an oxymoron: the same fact that there exist
an algorithm by which we generate an object on a classical
deterministic computer means that such an object is
algorithmically-compressible trough the adopted algorithm and,
hence, is not algorithmically-random.

This observation was condensed by John Von Neumann in his famous
sentence:
\begin{center}
 \textit{"Anyone who considers arithmetic methods of producing random digits is, of course, in a state of sin"}
\end{center}
What a pseudorandom number generator (a PRG from here and beyond)
outputs are \textbf{pseudorandom numbers}, i.e. numbers who mimic
true algorithmic randomness up to a certain degree of accuracy,
i.e pass a suffiencetely extended battery of randomness tests.

Before embarking in abstract mathematical definitions about what
does it mean it is rather useful to make a previous breaf
historical analysis of the concrete problem of pseudo-random
number generation in Theoretical Computer Science.

Von Neumann himself introduced an arithmetic
pseudorandom-generation method known as the \textbf{middle square
method}:

supponing to want to generate m random integers of 10 digits
starting from a certain 10' digits integer \textbf{seed}  , such
method is defined through the following algorithm:
\begin{enumerate}
  \item set $ i \leftarrow 0 $
  \item set $ n_{i} \, \leftarrow \, seed $
  \item  \label{it:begin loop} square $ n_{i} $ to get an intermediate number M with 20
  or less digits
  \item set $ i \leftarrow i+1 $
  \item set $ n_{i} \, \leftarrow \, $ the middle ten digits of M
  \item if $ i \, < \, m $ then goto step\ref{it:begin loop}, else
  halt
\end{enumerate}

\medskip

A serious problem with the middle-square method is that the orbit
generated by many seeds is periodic with a very little period
\cite{Yan-00}.

\medskip

In 1949 D.H. Lehmer proposed to use the \textbf{Theory of
congruences}, i.e. the theory of the  residue classes  $ {
\mathbb{Z}}_{n} $ modulo n (a ring on integers being a field if
and only if n is a prime number) to generate pseudorandom numbers.

Fixed the following numbers:
\begin{itemize}
  \item n : the \textbf{modulus} , $ n \, > \, 0 $
  \item $x_{0}$ : the \textbf{seed}, $ 0 \, \leq \,  x_{0} \, \leq n
  $
  \item a : the \textbf{multiplier} $ 0 \, \leq \,  a \, \leq n
  $
  \item b : the \textbf{increment} $ 0 \, \leq \,  b \, \leq n
  $
\end{itemize}
the \textbf{linear congruential generator} generates the sequence
of pseudorandom numbers defined recursively by:
\begin{equation}
  x_{j} \; := \; a x_{j-1} + b \; (mod \, n) \, j >0
\end{equation}
for $ 1 \, \leq \, j   \, \leq \, l $ where $ l \in { \mathbb{N}}
$ is the least value such that $ x_{l+1} \, \equiv \, x_{j} \;
(mod \, n) $ i.e. is the \textbf{period} of the PRG.

Since the period is less or equal to the modulus $ l \, \leq \, n
$ to have a PRG of sufficient quality it is necessary to use high
enough moduli.

For fixed n one would, then like to optimize the situation
choosing the other parameters so that the period is equal to the
modulus.

A necessary and sufficient condition for this to happen is given
by the following \cite{Knuth-98}:
\begin{theorem}
\end{theorem}
THEOREM OF GREENBERGER, HULL, DOBELL
\begin{equation}
  l \, = \, n  \; \Leftrightarrow \; ( gcd(b,n) = 1 \, and \, a
  \equiv 1 (mod \, p) \forall primes \, p | n , \, and \, a \equiv 1
  (mod \, 4) \, if \, 4 | n )
\end{equation}

\medskip

The \textbf{linear congruential generator} had (and  continues to
have) an immense application:

for example RANDU, the random number generator common on IBM
mainframes computers in the sixties, was based on a linear
congruential generator with parameters $ a = 65539 \, , \, b = 0
\, , n = 2^{31} $.

However it was soon found that RANDU gave rise to insidious
correlations: if successive triples of the numbers that RANDU
generated were used as a set of coordinates in a three-dimensional
space, the generated distribution of point seems random from most
viewpoints but there exist a special orientation from wich one
can see that they lie in a set of planes
\cite{Williams-Clearwater-98}.

\medskip

The linear congruential method is still of great popularity: for
example it is often used the \textbf{minimal standard 32-bit
generator} obtained by the choice $  a = 16807 \, , \, b = 0 \, ,
\, n = 2^{31} - 1 $ adopted by many programming languages such as
TURBO PASCAL even if its lack of randomness may be easily
visualized \cite{Denker-Woyczysnki-98}.

\medskip

To avoid the bugs of linear congruential generators, a new kind
of generators, the \textbf{shift-register-generator}, was then
introduced.

In these generators each successive number depends upon many
preceding values.

The basic operation may still be  the \textbf{modular addition}
or other functions such as the \textbf{exlusive or}.

For example a commonly used algorithm is the following:
\begin{equation}
  x_{j} \; := \; x_{j-p} \, XOR \, x_{j-q}
\end{equation}

The best choice of the pair of integers p and q happens when p
and q are \textbf{Mersenne primes}  such that
\cite{Landau-Binder-00} :
\begin{equation}
  p^{2} + q^{2} + 1 \text{ is prime}
\end{equation}
A common choice of this kind is $ p = 250 \, , \, q = 103 $.

\medskip

Anyway it would be an error to think that this method (or their
evolutions such as the  \textbf{subtract with carry generators} or
 the \textbf{subtract with carry Weyl generators}) is, in
absolute, better than the linear congruential method: there exist
Monte-carlo simulations of the bidimensional Ising model for
which the minimal standard 32-bit generator is not inficiated by
systematic errors of younger algorithms producing deviations from
Onsager's known solution.

\medskip

Let us conclude this hystorical review mentioning Wolfram's
suggestion of using some TYPE 3 elementary cellular automata as
pseudo-random generators \cite{Wolfram-94} that he himself
adopted for the integer-random number generator of Mathematica
(Rule[30] while for real numbers it is used  a Marsaglia-Zaman
subtract with borrow \cite{Wolfram-96}).

\bigskip

This panoramic on  concrete pseudorandom number generators,
hasn't, anyway, touched the basis question: what is a
\textbf{pseudorandom generator}?

The answer that we have already anticipated, i.e. an algorithm
producing binary strings passing an enough number of
\textbf{randomness statistical tests}, may appear satisfying since
anyone, eventually from his undergraduate $ \chi^{2} $ - test
experiences, has an intuitive idea of what a randomness
statistical test is: a well-defined procedure that allows to catch
some kind of regularity of the statistical data.

\medskip

For all practical purposes the definition of the notion of
pseudo-random generator may indeed be accomplished concretely
specifying a set of randomness statistical tests it has to pass:
(e.g. the Kolmogorov-Smirnov test + the Frequency test + the
Serial test + the Gap test + the Partition test + the Coupon
collector's test + the Permutation test + the Run test + the
Maximum-of t-test + the Collision test + the Birthday spacing
test + Serial correlation test \cite{Knuth-98}).

The higher is the number of elements of this list of randomness
statistical tests, the higher is the quality of the generator.

\medskip

From a conceptual point of view, anyway, such a definition is
unsatisfactory since:

\begin{itemize}
  \item it ultimately doesn't say what a \textbf{statistical randomness
  test} is
  \item it doesn't clarify the structure of the set of the \textbf{statistical randomness
  tests} and, conseguentially, the meaning of considering  a
  subset of it.
\end{itemize}

A satisfactory answer to both these points may be given
introducing Per Martin -  L\"{o}f's Theory of Randomness
Statistical tests.

\medskip

The idea behind Martin -  L\"{o}f's Theory is the following:

in statistical practice we are given an element x of some sample
space (associated with some distribution that we will assume from
here and beyond to be the unbiased one) and we want to test the
hypothesis: \textbf{x is a typical outcome}.

Being \textbf{typical} means \textit{"belonging to every
reasonable majority"}.

An element will be \textbf{random} just in the case it lies in the
intersections of all such majorities.

A level of a  statistical test is a set of strings which are found
to be relatively not-random (by the test).

Each level is a subset of the previous level, containing less and
less strings, considered more and more not-random.

the number of strings decreases exponentially fast at each level.

So a test contains at level 0 all possible strings, at level 2 at
most $ \frac{1}{2} $ of the strings, at level three only $
\frac{1}{4} $ of all strings and so on; accordingly at level m
the test contains at most $ 2^{n-m} $ strings of length n.

\smallskip

\begin{definition} \label{def:uniform-test for strings}
\end{definition}

A recursively enumerable (r.e.) set $ V \, \subset \,
\Sigma^{\star} \, \times \, {\mathbb{N}}_{+} $ is a
\textbf{Martin - L\"{o}f test} if:
\begin{enumerate}
  \item
\begin{equation}
  V_{m+1} \; \subset \; V_{m} \; \; \forall m \geq 1
\end{equation}
where:
\begin{equation} \label{eq:critical region for the uniform-test}
  V_{m} \; := \; \{ \vec{x} \in \Sigma^{\star} \, : \, ( \vec{x} ,
  m ) \in V \}
\end{equation}
is called the \textbf{critical region of the test at level $
\frac{1}{2^{m} }$}
  \item
\begin{equation}
  cardinality( \Sigma^{n} \cap V_{m} ) \; < \; 2^{n-m} \; \; \forall n
  \geq m \geq 1
\end{equation}
\end{enumerate}

\medskip

Given a Martin - L\"{o}f test V and an integer number q:
\begin{definition}
\end{definition}
SET OF THE q-PSEUDORANDOM STRINGS FOR V:
\begin{equation}
  q-PSEUDORANDOM ( \Sigma^{\star} ; V ) \; := \; \{ \vec{x} \in
  \Sigma^{\star} \, : \,  \vec{x}  \notin V_{q} \, and \, q < | \vec{x}
  | \}
\end{equation}

\medskip

Given a Martin - L\"{o}f test U:
\begin{definition}
\end{definition}
U IS UNIVERSAL:

for every Martin - L\"{o}f test V there exist a constant c
(depending upon U and V) such that:
\begin{equation}
  V_{m+c} \; \subset \; U_{m} \; \; m = 1, 2, \cdots
\end{equation}

\medskip

A universal Martin - L\"{o}f statistical test is a Martin -
L\"{o}f statistical test that is as strong as any other
Martin-Martin - L\"{o}f test.

So it is reasonable to fix once and for all a universal  Martin -
L\"{o}f test U and, given an integer number q, define:
\begin{definition}
\end{definition}
SET OF THE MARTIN L\"{O}F - q - RANDOM STRINGS:
\begin{equation}
    MARTIN L\ddot{O}F - q - RANDOM( \Sigma^{\star} ) \; := \;  q-PSEUDORANDOM ( \Sigma^{\star} ; U )
\end{equation}

\medskip

We can then introduce the following basic notion:
\begin{definition} \label{def:Martin Lof PRG}
\end{definition}
MARTIN L\"{O}F PRG OF QUALITY $ q \in {\mathbb{N}}$:

a  PRG whose outputs belongs to $ MARTIN L\ddot{O}F - q - RANDOM(
\Sigma^{\star} ) $

\medskip

It must be remarked, any way, that this is not the only possible
way one can follow in order of formalizing the concept of a PRG.

For example, following Oded Goldreich \cite{Goldreich-01}, one
can found the Theory of Pseudorandom Generation on Structural
Complexity Theory \cite{Odifreddi-99a} defining a PRG as an
\textbf{efficient} algorithm that stretches short random strings
into longer strings that are computationally indistinguishable
from long random strings, in the sense that the difference can't
be certified in an  \textbf{efficient} way.

Let us shortly show how this can be formalized.

\smallskip

Adhering to Goldreich's terminology we will speak of
\textbf{Classical Turing machines} instead of \textbf{partial
recursive functions} remembering that by the
\textbf{Church-Turing's Thesis} that is exactly the same.

Given a Turing machine M:
\begin{definition} \label{def:polynomial time Turing machine}
\end{definition}
M IS POLYNOMIAL-TIME:

there exist a polynomial p such that for every $ \vec{x} \in
\Sigma^{\star} $, when invoked on input x, M halts after at most
$ p ( | \vec{x} | ) $ steps

\medskip

We will will consider the following particular kind of sequence of
probability distributions:

\begin{definition}
\end{definition}
PROBABILITY ENSEMBLE OVER $ \Sigma^{\star} $:

a sequence $ \{ P_{n} \}_{n \in {\mathbb{N}}} $ of probability
distributions over $ \Sigma^{\star} $ with the property that there
exist a polynomial p such that:
\begin{equation}
   P_{n} ( \vec{x} ) \, > \, 0 \; \Rightarrow \; | \vec{x} | \, =
   \, p(n)
\end{equation}

\smallskip

Given two probability ensembles $ \{ P_{n} \}_{n \in
{\mathbb{N}}} $ and $ \{ Q_{n} \}_{n \in {\mathbb{N}}} $:
\begin{definition} \label{def:computational indistinguishability}
\end{definition}
$ \{ P_{n} \}_{n \in {\mathbb{N}}} $ AND $ \{ Q_{n} \}_{n \in
{\mathbb{N}}} $ ARE COMPUTATIONAL INDISTINGUISHABLE:

for every probabilistic polynomial-time Turing machine M, for
every positive polynomial $ p ( \cdot ) $ and for all sufficentely
large n:
\begin{equation}
  | Pr[ M ( 1^{n} , P_{n} ) = 1 ] \, - \, Pr[ M ( 1^{n} , Q_{n} ) = 1
  ] | \; < \; \frac{1}{p(n)}
\end{equation}

\begin{definition} \label{def:Goldreich PRG}
\end{definition}
GOLDREICH PRG:

a polynomial-time Turing machine G such that there exist a
monotonically increasing function $ l : {\mathbb{N}} \rightarrow
{\mathbb{N}} $ such that the following two probability ensembles,
denoted by $ \{ G_{n} \}_{n \in {\mathbb{N}}} $ and $ \{ R_{n}
\}_{n \in {\mathbb{N}}} $, are computationally indistinguishable:
\begin{itemize}
  \item $ G_{n} $ is defined as  the output of G on a uniformely-selected n-bits string
  \item $ R_{n} $ is defined as  the uniform probability distribution on $ \Sigma^{l(n)} $
\end{itemize}

\smallskip

The link between definition\ref{def:Goldreich PRG} and Structural
Complexity Theory passes through \textbf{one-way functions}, i.e.
functions easy to compute but hard to invert:
\begin{definition} \label{def:one-way function}
\end{definition}
ONE-WAY FUNCTION:

a function $ f : \Sigma^{\star} \, \rightarrow  \, \Sigma^{\star}
$:
\begin{itemize}
  \item easy to compute: f is computable in polynomial time
  \item hard to invert: for every probabilistic polynomial-time Turing machine M, for
every positive polynomial $ p ( \cdot ) $ and for all sufficentely
large n and $ \vec{x} $ uniformely distributed over $ \Sigma^{n} $
\begin{equation}
  Pr[ M ( 1^{n} , f( \vec{x} ) ) \in f^{- 1} ( f ( \vec{x} ) ) = 1 ] \; < \; \frac{1}{p(n)}
\end{equation}
\end{itemize}

And here appears the following:
\begin{conjecture} \label{con:fundamental conjecture of Structural Complexity Theory}
\end{conjecture}
FUNDAMENTAL CONJECTURE OF STRUCTURAL COMPLEXITY THEORY:
\begin{equation}
   P \; \neq \; NP
\end{equation}
governing the following:
\begin{theorem}
\end{theorem}
\begin{equation}
  \exists G \, \text{ Goldreich PRG } \; \Leftrightarrow \; \exists \, f \, \text{ one-way function } \; \Rightarrow \;
  \text{ Conjecture\ref{con:fundamental conjecture of Structural Complexity Theory} holds }
\end{equation}

\medskip

The inter-relation between definition\ref{def:Martin Lof PRG} and
definition\ref{def:Goldreich PRG} has not been analyzed yet.

Ultimately it touches the issue of the link existing between
Structural Complexity Theory and Algorithmic Information Theory
that, though having been the subject of intensive investigation
since the pioneristic Levin's analysis of the inter-relation among
\textbf{perebor} (a russian term literally meaning \emph{"brute
force"} and adopted from the late fifthies by the sovietic
operations research community to denote the necessity of an
exhaustive search of all the alternatives in certain search
problems) and Kolmogororov's ideas (cfr. the sixth paragraph of
the second part of \cite{Shasha-Lazerre-98}, the $ 7^{th} $
chapter of ), is still uncertain (cfr. \cite{Longpre-92}, the $
7^{th} $ chapter of \cite{Li-Vitanyi-97} and the $ 10^{th} $
chapter of \cite{Balcazar-Diaz-Gabarro-90} )
\newpage
\section{Equivalence between passage of a Martin L\"{o}f universal sequential statistical test and Chaitin randomness} \label{sec:Equivalence between passage of a Martin Lof universal statistical test and Chaitin randomness}
In the last paragraph we have introduced the notion of a
Martin-L\"{o}f  test on strings only for the uniform
distribution, the only one necessary in order to define a PRG.

\smallskip

Since, anyway, we will front, in the next sections, also
not-uniform distributions it may be appropriate to give a more
general definition.

\smallskip

Given a recursive probability distribution P on $ \Sigma^{\star}
$:

\begin{definition} \label{def:P-test for strings}
\end{definition}
MARTIN L\"{O}F TEST OF P-RANDOMNESS (P-TEST) :

a function $ \delta : \Sigma^{\star} \, \rightarrow \,
{\mathbb{N}} $:
\begin{enumerate}
  \item the set $ V \; := \; \{ ( m , \vec{x} ) \, : \, \delta (
  \vec{x}) > m \} $ is recursively enumerable
  \item
\begin{equation}
  \sum_{\vec{x} \in \Sigma^{n}} \{ P( \vec{x} | | \vec{x} | = n
  \, : \, \delta ( \vec{x} ) \, \geq \, m \} \; \leq \; \frac{1}{ 2^{m}} \; \; \forall n
\end{equation}
\end{enumerate}

\medskip

Defined the \textbf{critical region of the test at level $
\frac{1}{2^{m} }$}, for any integer $ m \geq 1 $, as:
\begin{equation}
  V_{m} \; := \; \{ \vec{x} \in \Sigma^{\star} \, : \, \delta ( \vec{x} ) \geq m \}
\end{equation}
it is immediate to see that for $ P = P_{unbiased} $ the
definition\ref{def:P-test for strings} reduces to the
definition\ref{def:uniform-test for strings}.

\bigskip

We would like, now, to extend this definition from $
\Sigma^{\star} $ to $ \Sigma^{\infty} $.

Since an effective test can't be performed on an infinite
sequence it is necessary to introduce an effective process of
sequential approximations.

\medskip

So, given a recursive probability measure $ \mu $ on $
\Sigma^{\infty} $:
\begin{definition} \label{def:sequential mu-test for sequences}
\end{definition}
SEQUENTIAL MARTIN L\"{O}F TEST OF $ \mu $-RANDOMNESS (SEQUENTIAL $
\mu $-TEST) :

a function $ \delta : \Sigma^{\infty} \, \rightarrow \,
{\mathbb{N}} \bigcup \{ \infty \} $:
\begin{enumerate}
  \item
\begin{equation}
  \delta ( \bar{x} ) \; = \; \sup_{n \in {\mathbb{N}}} \{ \gamma ( \vec{x}(n) ) \}
\end{equation}
where $ \vec{x}(n) \in \Sigma^{n} $ denotes the prefix of length n
of the sequence $ \bar{x} $ while $ \gamma \, : \, {
\Sigma^{\star} } \, \rightarrow \, {\mathbb{N}} $ is a
\textbf{total enumerable function} (i.e. $ V \; := \; \{ ( m ,
\vec{y} ) : \gamma ( \vec{y} ) \geq m \} $ is a
\textbf{recursively enumerable set}
  \item
\begin{equation}
  \mu ( \{  \bar{x} \in \Sigma^{\infty} \, : \, \delta (  \bar{x} )
  \geq m \} \; \leq \; \frac{1}{2^{m}} \; \; \forall m \geq 0
\end{equation}
\end{enumerate}

\medskip

Given a \textbf{sequential $\mu$-test} $ \delta $ we have that a
sequence $ \bar{x} \in \Sigma^{\infty} $ passes the test if $
\delta ( \bar{x} ) \, < \, \infty $ while it doesn't passes the
test if $ \delta ( \bar{x} ) \, = \, \infty $.

\smallskip

The set of the sequences passing the test  $ \delta $  are those
that it declares random:
\begin{definition} \label{def:mu-random sequences w.r.t. to a mu-test}
\end{definition}
SET OF THE $ \mu $-RANDOM SEQUENCES W.R.T. $ \delta $:
\begin{equation}
   \mu - RANDOM ( \Sigma^{\infty} ; \delta ) \; := \; \{ \bar{x} \in
   \Sigma^{\infty} \, : \, \delta ( \bar{x} ) \, < \, \infty \}
\end{equation}

\medskip

Clearly the definition\ref{def:mu-random sequences w.r.t. to a
mu-test} depends on the particular sequential $ \mu $-test
considered.

\medskip

This relativization can be, anyway, eliminated by the usual
strategy of Algorithmic Information Theory:
\begin{definition} \label{def:universal sequential mu-test for sequences}
\end{definition}
UNIVERSAL SEQUENTIAL MARTIN L\"{O}F TEST OF $ \mu $-RANDOMNESS
(UNIVERSAL SEQUENTIAL $ \mu $-TEST) :

a sequential $ \mu $ -test f such that for every other sequential
$ \mu$ -test $ \delta $ there exist a constant $ c \geq 0 $ such
that:
\begin{equation}
  f ( \bar{x} ) \; \geq \; \delta ( \bar{x} )  - c
\end{equation}

\medskip

A  universal sequential $ \mu $-test is a sequential $ \mu $-test
that is as strong as any other sequential $ \mu $-test.

So it is reasonable to fix once and for all a universal
sequential $ \mu $-test $ \delta_{0} ( \cdot | \mu ) $ and define:
\begin{definition} \label{def:mu-random sequences}
\end{definition}
SET OF THE $ \mu $-RANDOM SEQUENCES:
\begin{equation}
  \mu - RANDOM ( \Sigma^{\infty} ) \; := \; \mu - RANDOM ( \Sigma^{\infty} ; \delta_{0} ( \cdot | \mu ) )
\end{equation}

\medskip

To be a $ \mu $-random sequence is the $ \mu $-rule in $
\Sigma^{\infty} $ since:
\begin{theorem} \label{th:foundation of the applicability of probability theory to reality}
\end{theorem}
FOUNDATION OF THE APPLICABILITY OF PROBABILITY THEORY TO REALITY
\begin{equation}
  \mu [  \mu - RANDOM ( \Sigma^{\infty} ) ] \; = \; 1
\end{equation}

\medskip

The most important case from which, in a certain proper sense,
all the others cases may be derived is when the measure $ \mu $
is the unbiased Lebesgue measure $ P_{unbiased} $ on $
\Sigma^{\infty} $.

\begin{definition}  \label{def:Martin-Lof random sequences}
\end{definition}
MARTIN-L\"{O}F RANDOM SEQUENCES:
\begin{equation}
  MARTIN-L\ddot{O}F - RANDOM ( \Sigma^{\infty} ) \; := \; P_{unbiased} - RANDOM( \Sigma^{\infty} )
\end{equation}

\smallskip

We want now to present one of the more fundamental results of
Algorithmic Information Theory: the Chaitin-Schnorr's Theorem.

This requires, anyway, the introduction of some technicalities.

Given a sequence $ \bar{x} \in \Sigma^{\infty} $ and a set of
strings $ S \subset \Sigma^{\star} $ let us denote by $ S \,
\Sigma^{ \infty } $ the set of all the sequences having the
strings of S as prefixes, i.e.:
\begin{equation}
  S \, \Sigma^{ \infty } \; := \; \{ \bar{x} \in \Sigma^{\infty}
  \, : \, \vec{x}(n) \in S \text{ for some natural } n \geq 1 \, \}
\end{equation}
lightening the notation for singletons by poning:
\begin{equation}
 \vec{x} \Sigma^{\infty} \; := \{ \vec{x} \} \Sigma^{\infty} \; \;
 \vec{x} \in \Sigma^{\infty}
\end{equation}

We need the following:
\begin{lemma} \label{lem:classical algorithmic randomness in terms of  coverings}
\end{lemma}
\begin{multline}
  \bar{x} \in MARTIN-L\ddot{O}F - RANDOM ( \Sigma^{\infty} ) \;
  \Leftrightarrow \\
   \forall Covering \in \Sigma^{\star} \times
  {\mathbb{N}} \text{ r.e. } : ( P_{unbiased} ( Covering_{n} \Sigma^{\infty} ) <
  \frac{1}{2^{n}} \, \forall n \geq 1 ) \; \\
  \exists m \in {\mathbb{N}}  \, :  \, \bar{x} \notin Covering_{m} \Sigma^{\infty}
\end{multline}
where, as with the same notation of eq.\ref{eq:critical region
for the uniform-test} that we will understand from here and
beyond:
\begin{equation}
  Covering_{n} \; := \; \{ \vec{x} \in \Sigma^{\star} \, : \, (
  \vec{x}, n ) \in Covering \}
\end{equation}

\smallskip

Indeed Lemma\ref{lem:classical algorithmic randomness in terms of
coverings} is the starting point of a path leading to Solovay' s
way of  characterizing classical algorithmic randomness.

The first step it to observe that one can always effectively pass
from an arbitrary covering to a prefix-free one, as is stated by
the following:
\begin{lemma} \label{lem:effective passage from arbitrary to prefix-free covering}
\end{lemma}
For every r.e. set $ B \subset \Sigma^{\star} \times
{\mathbb{N}}_{+} $, we can effectively find a r.e. set $  C
\subset \Sigma^{\star} \times {\mathbb{N}}_{+} $ such that:
\begin{align}
  C_{n}  & \; \text{ is prefix-free} \; \; \forall n \in {\mathbb{N}}_{+}   \\
  B_{n} \Sigma^{\infty} & \; = \; C_{n} \Sigma^{\infty} \; \; \forall n \in {\mathbb{N}}_{+}
\end{align}

\medskip

 Lemma\ref{lem:classical algorithmic randomness in terms of
coverings} and Lemma\ref{lem:effective passage from arbitrary to
prefix-free covering} immediately imply the following:
\begin{lemma} \label{lem:classical algorithmic randomness in terms of prefix-free coverings}
\end{lemma}
\begin{multline}
  \bar{x} \in MARTIN-L\ddot{O}F - RANDOM ( \Sigma^{\infty} ) \;
  \Leftrightarrow \\
   \forall Covering \in \Sigma^{\star} \times
  {\mathbb{N}} \text{ r.e. } : ( Covering_{n} \text{ is prefix-free} \, \forall n ) \, and \,( P_{unbiased} ( Covering_{n} \Sigma^{\infty} ) <
  2^{n} \, \forall n \geq 1 ) \; \\
  \exists m \in {\mathbb{N}}  \, :  \, \bar{x} \notin Covering_{m} \Sigma^{\infty}
\end{multline}
that allows to show the equivalence between the passage of a
universal sequential Martin-L\"{o}f test and Solovay randomness
defined as:
\begin{definition} \label{def:Solovay randomness}
\end{definition}
$ \bar{x} \in \Sigma^{\infty} $ IS SOLOVAY-RANDOM ( $ \bar{x} \in
SOLOVAY-RANDOM( \Sigma^{\infty} ) ) $:
\begin{multline}
  \forall X \subset \Sigma^{\star} \times {\mathbb{N}}_{+}  \,
  r.e. \, : \sum_{n=1}^{\infty} P_{unbiased} ( X_{n} \Sigma^{\infty} )  \, < \, \infty  \\
  \exists N \in {\mathbb{N}} : \bar{x} \notin X_{n}
  \Sigma^{\infty} \, \, \forall n > N
\end{multline}
as stated by the following:
\begin{theorem} \label{th:equivalence of Martin-Lof randomness and Solovay randomness}
\end{theorem}
\begin{equation}
  MARTIN-L\ddot{O}F - RANDOM ( \Sigma^{\infty} ) \; = \; SOLOVAY-RANDOM( \Sigma^{\infty} )
\end{equation}
\begin{proof}
\begin{itemize}
  \item $ SOLOVAY-RANDOM( \Sigma^{\infty} ) \; \subseteq \; MARTIN-L \ddot{O}F-RANDOM( \Sigma^{\infty} ) $

  Clearly to prove the thesis is equivalent to show that:
\begin{equation}
  \bar{x} \notin MARTIN-L \ddot{O}F-RANDOM( \Sigma^{\infty} ) \;
  \Rightarrow \; \bar{x} \notin SOLOVAY-RANDOM( \Sigma^{\infty} )
\end{equation}
Let us then assume that $ \bar{x} \notin MARTIN-L\ddot{O}F -
RANDOM ( \Sigma^{\infty} ) $.

  By Lemma\ref{lem:classical algorithmic randomness in terms of prefix-free
  coverings} it follows that there exist a r.e. set $ X \subset
  \Sigma^{\star} \times {\mathbb{N}}_{+} $ such that:
\begin{align}
  X_{n} & \; \text{ is prefix-free } \; \; \forall n \in {\mathbb{N}}_{+}  \\
 P_{unbiased} ( X_{n} \Sigma^{\infty} ) & \; < \;  \frac{1}{2^{n}} \\
 \bar{x} & \; \notin \; \bigcap_{n=1}^{\infty} X_{n} \Sigma^{\infty}
\end{align}
But then:
\begin{equation}
  \sum_{n=1}^{\infty} P_{unbiased} ( X_{n} \Sigma^{\infty} ) \;
  \leq \; \sum_{n=1}^{\infty} \frac{1}{ 2^{n} } \; = \; 1  \; < \;
  \infty
\end{equation}
and conseguentially $ \bar{x} \, \notin
SOLOVAY-RANDOM(\Sigma^{\infty} ) $
  \item $MARTIN-L \ddot{O}F-RANDOM( \Sigma^{\infty} )  \; \subseteq \; SOLOVAY-RANDOM( \Sigma^{\infty} ) $

Clearly to prove the thesis is equivalent to show that:
\begin{equation}
  \bar{x} \notin SOLOVAY-RANDOM( \Sigma^{\infty} )  \;
  \Rightarrow \; \bar{x} \notin MARTIN-L \ddot{O}F-RANDOM( \Sigma^{\infty} )
\end{equation}
Let us then assume that $ \bar{x} \notin MARTIN-L\ddot{O}F -
RANDOM ( \Sigma^{\infty} ) $.

Conseguentially there exist a r.e. set $ X \subset
  \Sigma^{\star} \times {\mathbb{N}}_{+} $ such that:
\begin{align}
  X_{n} & \; \text{ is prefix-free } \; \; \forall n \in {\mathbb{N}}_{+}  \\
 P_{unbiased} ( X_{n} \Sigma^{\infty} ) & \; < \;  \frac{1}{2^{n}} \\
 cardinality ( & \{ n \in {\mathbb{N}}_{+} \, : \, \bar{x} \, \in \, X_{n}
 \Sigma^{\infty} \} ) \; = \; \aleph_{0}
\end{align}
Given an arbitrary positive real number $ c \in {\mathbb{R}}_{+}
$ let us introduce the set:
\begin{equation}
  B \; := \; \{ ( \vec{y} , n ) \in \Sigma^{\star} \times
  {\mathbb{N}} \, : \, cardinality( \{ n \in {\mathbb{N}}_{+} : \vec{y}
  \in X_{n} \Sigma^{\star} \} ) \: > \: 2^{n+c} \}
\end{equation}
By construction:
\begin{equation}
  P_{unbiased} ( B_{n} \Sigma^{\infty} ) \; < \; 2^{- n} \; \;
  \forall n \in {\mathbb{N}}_{+}
\end{equation}
Furthermore $ \bar{x} \in \bigcap_{n=1}^{\infty} B_{n}
\Sigma^{\infty} $, i.e. for every natural $ n \geq 1 $ there
exist a natural $ m \geq 1 $ such that:
\begin{equation}
  cardinality( \{ n \in {\mathbb{N}}_{+} : \vec{x}(m) \in X_{n}
  \Sigma^{\star} \} \; > \; 2^{n+c}
\end{equation}
Just take $ m \, = \, \max \{ i_{1} , i_{2} , \cdots , i_{t} \}$,
where $ t \; > \; 2^{n+c} $ and:
\begin{equation}
  \bar{x} \; \in \; \bigcap_{j=1}^{t} X_{i_{t}} \Sigma^{\infty}
\end{equation}
\end{itemize}
\end{proof}

\medskip
An other ingredient required for proving Chaitin-Schnorr's Theorem
is (a slighty streghtened form of the) Chaitin-Levin's Theorem
expressing the deep link existing between \textbf{prefix
algorithmic entropy} and the \textbf{universal algorithmic
probability} introduced by definition\ref{def:universal
algorithmic probability}, namely the following:
\begin{theorem} \label{th:strengthened Chaitin-Levin theorem}
\end{theorem}
\begin{equation}
  \exists c \in {\mathbb{R}}_{+} \; : \; 0 \: \leq \: I ( \vec{x}) \,
  + \, \log_{2} P_{U} ( \vec{x}) \: \leq \: c \; \; \forall
  \vec{x} \in \Sigma^{\star}
\end{equation}
\begin{corollary} \label{cor:Chaitin-Levin theorem}
\end{corollary}
CHAITIN-LEVIN'S THEOREM
\begin{equation}
  I( \vec{x}) \; \stackrel{ + }{=} \; - \log_{2} P_{U} ( \vec{x} )
\end{equation}

\medskip

The last ingredient required for proving Chaitin-Schnorr's Theorem
is the following generalization of  Theorem\ref{th:Kraft
inequality} to arbitrary r.e. sets.
\begin{theorem} \label{th:Kraft-Chatin theorem}
\end{theorem}
KRAFT-CHAITIN'S THEOREM

\begin{hypothesis}
\end{hypothesis}
\begin{equation*}
  \phi \in C_{M}-C_{\Phi}-\Delta_{0}^{0}-\stackrel{ \circ } {MAP} ( {\mathbb{N}} ,
  {\mathbb{N}} ) \; : \; HALTING(\phi) \text{ is an initial segment of
  } {\mathbb{N}}_{+}
\end{equation*}
\begin{thesis}
\end{thesis}
The following statements are equivalent:
\begin{enumerate}
  \item We can effectively construct a function $ \theta \in
  C_{M}-C_{\Phi}-\Delta_{0}^{0}- \stackrel{ \circ } {MAP} ( {\mathbb{N}}_{+} ,
  \Sigma^{\star}) $ such that:
\begin{align}
   HALTING(\theta)  &  \; = \; HALTING(\phi) \\
   | \theta ( n ) |  & \; = \; \phi (n) \; \; \forall n \in
   HALTING( \phi) \\
   Range(\theta) & \; \text{ is prefix-free}
\end{align}
  \item
\begin{equation}
  \sum_{n \in HALTING(\phi)} 2^{- \phi(n)} \; \leq \; 1
\end{equation}
\end{enumerate}

\medskip

Let us finally afford our objective:
\begin{theorem} \label{th:Chaitin-Schnorr theorem}
\end{theorem}
CHAITIN-SCHNORR'S THEOREM
\begin{equation}
  MARTIN-L \ddot{O}F-RANDOM( \Sigma^{\infty} ) \; = \; CHAITIN-RANDOM( \Sigma^{\infty} )
\end{equation}
\begin{proof}
\begin{itemize}
  \item $ MARTIN-L \ddot{O}F-RANDOM( \Sigma^{\infty} ) \; \subseteq \; CHAITIN-RANDOM( \Sigma^{\infty}
  ) $
Clearly to prove the thesis is equivalent to show that:
\begin{equation}
  \bar{x} \notin CHAITIN-RANDOM( \Sigma^{\infty} ) \; \Rightarrow \;
   \bar{x} \notin  MARTIN-L \ddot{O}F-RANDOM( \Sigma^{\infty} )
\end{equation}
Let us assume that for every $ m  > 0 $ there exists an $ n_{m} $
such that $ I( \vec{x} ( n_{m} ) ) \, < \, n_{m} $. By
theorem\ref{th:strengthened Chaitin-Levin theorem}  we know we can
choose a natural number  $ c > 0 $ such that:
\begin{equation}
  \exists c \in {\mathbb{R}}_{+} \; : \; 0 \: \leq \: I ( \vec{x}) \,
  + \, \log_{2} P_{U} ( \vec{x}) \: \leq \: c \; \; \forall
  \vec{x} \in \Sigma^{\star}
\end{equation}
Let us introduce the set:
\begin{equation}
  Covering \; := \; \{ ( \vec{y} ,t ) \in \Sigma^{\star} \times
  {\mathbb{N}}_{+} \: : \: I ( \vec{y} ) \, < \, | \vec{y} | - t -
  c - 1 \}
\end{equation}
Clearly the set Covering is r.e. and:
\begin{multline}
  P_{unbiased} ( Covering_{t} \Sigma^{\infty} ) \; \leq \; \sum_{\vec{y} \in
  Covering_{t}} 2^{- |\vec{y}|} \; = \\
  = \;  \sum_{ \{ \vec{y} \in \Sigma^{\star} \, : \, I ( \vec{y} ) \, < \, | \vec{y} | - t -
  c - 1 \}} 2^{- |\vec{y}|} \; \leq  \\
\leq \; \sum_{ \{ \vec{y} \in \Sigma^{\star} \, : \, I ( \vec{y} )
\, < \, | \vec{y} | - t - c - 1 \}} 2^{ - I ( \vec{y} ) -  t - c
- 1 }
\end{multline}
so that:
\begin{multline}
  P_{unbiased} ( Covering_{t} \Sigma^{\infty} ) \; \leq \; \sum_{ \vec{y} \in \Sigma^{\star}
  } 2^{ - I ( \vec{y} ) -  t - c
- 1 } \; = \\
= \; 2^{t-c-1} \sum_{\vec{y} \in \Sigma^{\star}} 2^{ - I (
\vec{y} )} \; \leq \; 2^{- t - 1} \sum_{\vec{y} \in
\Sigma^{\star}} P_{U} ( | \vec{y} | ) \; = \\
= \; 2^{t - 1} \; < \; 2^{ - t}
\end{multline}
We prove now that $ \bar{x} \in \bigcap_{t = 1}^{\infty}
Covering_{t} \Sigma^{\infty} $. Indeed, given $ t > 0 $, construct
$ m_{t} \, := \, n_{t + c + 1} $ and use the hypothesis:
\begin{equation}
  I ( \vec{x} ( m_{t} ) ) \; = \; I ( \vec{x} ( n_{t + c + 1} ))
  \; < \; n_{t + c + 1} \, - \, ( t + c + 1 ) \; = \; m_{t} - t -
  c - 1
\end{equation}
i.e. $ \vec{x}( m_{t} ) \,  \in \, Covering_{t} $.

By Lemma\ref{lem:classical algorithmic randomness in terms of
coverings} $  \bar{x} \notin  MARTIN-L \ddot{O}F-RANDOM(
\Sigma^{\infty} ) $
  \item $  CHAITIN-RANDOM( \Sigma^{\infty}
  ) \; \subseteq \;  MARTIN-L \ddot{O}F-RANDOM( \Sigma^{\infty} ) $

To prove the thesis is equivalent to show that:
\begin{equation}
  \bar{x} \notin  MARTIN-L \ddot{O}F-RANDOM( \Sigma^{\infty} ) \; \Rightarrow \;
   \bar{x} \notin CHAITIN-RANDOM( \Sigma^{\infty} )
\end{equation}

Let us assume that $ \bar{x} \notin  MARTIN-L \ddot{O}F-RANDOM(
\Sigma^{\infty} ) $. By Lemma\ref{lem:classical algorithmic
randomness in terms of coverings} there exist a r.e. set $
Covering \, \subset \, \Sigma^{\star} \times {\mathbb{N}} $ such
that:
\begin{align}
  P_{unbiased} & ( Covering_{n} \Sigma^{\infty} ) \;   < 2^{ - n} \; \; \forall n \in {\mathbb{N}}_{+} \\
   \bar{x} & \in \bigcap_{n=1}^{\infty} Covering_{n} \Sigma^{\infty}
\end{align}
Furthermore, by Lemma\ref{lem:effective passage from arbitrary to
prefix-free covering}, we may assume that $ Covering_{n} $ is
prefix-free for all $ n \, geq \, 1 $. Then:
\begin{multline}
  \sum_{n=2}^{\infty} \sum_{\vec{y} \in Covering_{n^{2}}} 2^{ - ( | \vec{y} | - n
  )} \; = \; \sum_{n=2}^{\infty} 2^{n} \sum_{\vec{y} \in
  Covering_{n^{2}}} 2^{ - | \vec{y} | }  \: = \\
  = \; \sum_{n=2}^{\infty} 2^{n} P_{unbiased} ( Covering_{n^{2}}
  \Sigma^{\infty} ) \; \leq \; \sum_{n=2}^{\infty} 2^{ n - n^{2} }
  \; \leq \; 1
\end{multline}
By theorem\ref{th:Kraft-Chatin theorem} we get a Chaitin computer
C satisfying the following requirement:
\begin{equation}
  \forall n \geq 2 ,  \forall \vec{y} \in Covering_{n^{2}} \;
  \exists \vec{u} \in \Sigma^{ | \vec{y} | -n } \, : \, C( \vec{u}
  , \lambda ) \, = \, \vec{y}
\end{equation}
By the Invariance Theorem for Prefix Algorithmic Entropy, namely
theorem\ref{th:invariance theorem for prefix algorithmic
entropy}, there exists a positive constant c such that:
\begin{equation} \label{eq:application of the invariance theorem}
  I( \vec{y} ) \; \leq \; | \vec{y} | - n + c \; \; \forall n \geq
  2 , \forall  \vec{y} \in Covering_{n^{2}}
\end{equation}
Next we prove that for all natural $ n \geq 1 $ there exist
infinitely many m such that $ \vec{x}(m) \, \in \,
Covering_{n^{2}} $.

By hypothesis:
\begin{equation}
  \vec{x} \; \in \; \bigcap_{k=1}^{\infty} Covering_{k} \Sigma^{\infty}
\end{equation}
so for every n we can find a natural $ m_{ n^{2} } $ with $
\vec{x} ( m_{n^{2}}) \, \in \, Covering_{n^{2}} $.

We have to prove that we can choose these numbers $ m_{n^{2}} $ as
large as we wish.

Assume, for the sake of a contradiction, that $ m_{n^{2}} \; \geq
\; N $, for all n and  some fixed N. This means the existence of
a string $ \vec{y} $ of length less than N such that $ \vec{y} \,
in \, Covering_{n^{2}} $ , for all $ n \,  \geq \, 1 $.
Accordingly, for every $ n \geq 1 $ one has:
\begin{equation}
  \vec{y} \Sigma^{\infty} \; \subset \; Covering_{n^{2}} \Sigma^{\infty}
\end{equation}
and:
\begin{equation}
  2^{ - n^{2}} \; > \; P_{unbiased} ( Covering_{ n^{2}})
  \Sigma^{\infty} \; \geq \; P_{unbiased} ( \vec{y}
  \Sigma^{\infty} ) \; = \, 2^{ - | \vec{y} | } \; \geq \;2^{ - N }
\end{equation}
that is a contradiction.

In conclusion, given $ d > 0 $ we pick $ i > d + c $ and $ m \geq
2 $ in order to get $ \vec{x}(m) \in Covering_{n^{2}} $: by
eq.\ref{eq:application of the invariance theorem}:
\begin{equation}
  I ( \vec{x} (m) ) \; \leq \; m - n + c \; < \; m - d
\end{equation}
\end{itemize}
\end{proof}

\medskip

Summing up, theorem\ref{th:equivalence of Martin-Lof randomness
and Solovay randomness} and theorem\ref{th:Chaitin-Schnorr
theorem} show that Martin-L\"{o}f randomness, Solovay randomness
and Chaitin randomness are equivalent notions, characterizing
what is nowadays \emph{almost} universally considered as the the
correct characterization of the concept of
\textbf{$C_{\Phi}$-algorithmic randomness}.

The above \emph{almost} is owed to to a problem we mentioned at
the end of section\ref{sec:Brudno random sequences of cbits},
almost always misunderstood and that is the main source of a sort
of incomunicability between the scientific community of
mathematical physicists studying Dynamical Systems Theory and the
scientific community of the logico-mathematicians and
Theoretical-Computer scientists studying Algorithmic Information
Theory: Theorem\ref{th:Brudno randomness is weaker than Chaitin
randomness} stating that:
\begin{equation}
  BRUDNO-RANDOM( \Sigma^{\infty}) \; \subset \;  CHAITIN -RANDOM(\Sigma^{\infty})
\end{equation}
whose proof, as promized, we report here:
\begin{proof}
Given a universal computer C, let us introduce another computer $
C' $ defined in the following way:
\begin{equation}
  C'( \vec{x} ) \; := \;
  \begin{cases}
    \cdot_{i=1}^{ | \vec{y}| } {\mathcal{I}} ( y_{i} , 2^{i}, C ( \vec{z} )) & \text{if $ \exists \vec{y}, \vec{z} \in \Sigma^{\star} : \vec{x} = \cdot_{i=1}^{ | \vec{y}| } y_{i}^{2}  \cdot 01 \cdot \vec{z} $  }, \\
    C ( \vec{x} ) & \text{otherwise}.
  \end{cases}
\end{equation}
where, generally, $ {\mathcal{I}} ( a , n , \vec{b} ) $ denotes
the string obtained inserting the letter a at the $ n^{th} $ place
of the string $ \vec{b} $, i.e.:
\begin{equation}\label{eq:insertion operator}
  {\mathcal{I}} ( a , n , \vec{b} ) \; := \; b_{1} \ldots b_{n -
  1} \cdot a \cdot b_{n + 1} \ldots b_{| \vec{b} | } \; \; a \in
  \Sigma , \vec{b} \in \Sigma^{\star} , n \in {\mathbb{N}}_{+} : n
  \leq | \vec{b} |
\end{equation}
Clearly $ C' $ is a universal computer too.

Given $ \bar{u} \in CHAITIN -RANDOM(\Sigma^{\infty}) $ let us
consider the sequence $ \bar{u}' $ defined in the following way:
\begin{equation}
  \bar{u}'_{i} \; := \;
  \begin{cases}
    0 & \text{if $ i = 2^{k} , k \in { \mathbb{N}}$ }, \\
   \bar{u}_{i} & \text{otherwise}.
  \end{cases}
\end{equation}
Since:
\begin{equation}
  K_{C'} ( \vec{u}_{n} ) \; \leq \; K_{C'} ( \vec{u}'_{n} ) + 2 +
  \log_{2} n
\end{equation}
It follows that:
\begin{equation}
  \bar{u}' \; \in \; BRUDNO-RANDOM(\Sigma^{\infty})
\end{equation}
Let us now consider the Martin-L\"{o}f sequential test $ V \,
\subset \, \Sigma^{\star} \times {\mathbb{N}}_{+} $ whose $
n^{th}$ section is given by:
\begin{equation}
  V_{n} \; := \; \{ \vec{x} \in \Sigma^{\star} \, : \, x_{2^{k}} =
  0 \, \, k = 0 , \cdots , i - 1 \}
\end{equation}
Since by construction one has that:
\begin{equation}
  \bar{u}' \notin \bigcap_{n=1}^{\infty} V_{n}
\end{equation}
it follows that:
\begin{equation}
  \bar{u}' \; \notin \; CHAITIN-RANDOM(\Sigma^{\infty})
\end{equation}
\end{proof}
\newpage
\chapter{Classical algorithmic randomness as satisfaction of all the classical algorithmic typical
properties} \label{chap:Classical algorithmic randomness as
satisfaction of all the classical algorithmic typical properties}
\section{Typical properties of a classical probability space} \label{sec:Typical properties of a classical probability space}
We have seen in chapter\ref{chap:Classical algorithmic randomness
as passage of all the classical algorithmic statistical tests}
that classical algorithmic randomness may be characterized as the
passage of all the classical algorithmic statistical tests, i.e.
of all the effectively implementable testes designed to catch some
kind of regularity:

in this chapter we will show in which sense the absense of any
kind of regularity may be interpreted as a condition of maximal
conformism, i.e. as the ownership to all the overwhelming
majorities.

Let us consider a collectivity S made of $ N \in {\mathbb{N}} $
people.

Given a  property $ p( \cdot ) $ will say that it is a
\textbf{majoritary} property of S if the people in S having such a
property are more than those not having it, i.e. iff:
\begin{equation}
  cardinality( \{ x \in S : p(x) \; holds \} ) \; > \; \frac{N}{2}
\end{equation}
We will say that $ p( \cdot ) $ is a \textbf{typical property} of
S if the people in S having such a property are very more than
those not having it, i.e. iff:
\begin{equation}
  cardinality( \{ x \in S : p(x) \; holds \} ) \; \gg \; \frac{N}{2}
\end{equation}
Of course this last notion is only an informal one, owing to the
informal nature of the ordering relation \textbf{very greater
than}.

\smallskip

Let us now consider the case in which the collectivity S is
infinite but countable, i.e. $ cardinality(S) \, = \, \aleph_{0}
$; in this case the same notion of a majoritary property loses
its meaning \footnote{Let us observe, by the way, that this
inficiates the meaning of the generalization to infinite
collectivities of the celebrated Nobel-prize for-Economics-winning
Kenneth Arrow's theorem on the impossibility of democracy stating
, in technical terms, that under the assumption that the decisive
sets form an ultrafilter on the set of voters they form a
principal filter too (and so there exist a dictator, i.e. a voter
whose vote alone determines the result of any election). Such a
generalization may be obtained in the same way the impossibility
of applying the theorem stating the principality of any
ultafilter on finite sets was overcome by Kurt Go\"{e}del in his
mathematical formalization of  Anselm of Aosta's ontological
proof simply by adding the assumption that being God is a
positive property: appealing to the theorem stating that an
ultrafilter on a (finite or infinite set) containing the
intersection of all its elements is principal \cite{Manin-98}
\cite{Odifreddi-00b}}.

In the case of an infinite and uncountable community, i.e. $
cardinality(S) \, \geq \, \aleph_{1} $, if S admits an unbaised
probability measure $ P_{unbiased} $  the notion of a
\textbf{typical property} of S may be rigorously defined as a
property holding $ P_{unbiased} $ almost everywhere in S.

So the characterization of classical algorithmic randomness as
\textbf{absolute conformism}, i.e. as the ownership of all the
typical properties, would seem to be precisely formalized.

Such a formalization, anyway, results in an empty notion:
asbolute conformism is impossible.

The solution to such a bug consists in requiring only the
ownerhip of all the \textbf{effectively-refutable} typical
properties.

The fact that, once again, Classical Measure Theory appears not
to be self-consistent as to the characterization of classical
algorithmic randomness has a great foundational relevance.

\medskip

Given a  classical  probability space $ CPS \; := \; ( \, M \, ,
\, \mu  \, ) $:
\begin{definition} \label{def:classical null set}
\end{definition}
$ S \; \subset  \; M $  IS A NULL SET OF CPS:
\begin{equation}
  \forall \epsilon > 0 \; \exists F_{ \epsilon } \in HALTING(\mu)  \; : \; S
  \subset F_{ \epsilon } \; and \; \mu( F_{ \epsilon } ) < \epsilon
\end{equation}

Let us introduce the following notions:
\begin{definition}
\end{definition}
UNARY PREDICATES ON M:
\begin{equation}
  {\mathcal{P}} ( M ) \; \equiv \; \{ p( x ) \, : \, \text{ predicate about } x \in M  \}
\end{equation}
\begin{definition} \label{def:typical properties of a classical probability space}
\end{definition}
TYPICAL PROPERTIES OF CPS:
\begin{equation}
 {\mathcal{P}}  (  CPS  )_{TYPICAL} \; \equiv \; \{ \, p ( x )  \in  {\mathcal{P}} ( M ) \, :
   \{ x \in M \, : \, p ( x ) \text{ doesn't hold } \}
   \; \text{is a null set} \}
\end{equation}
\begin{example}
\end{example}
TYPICAL PROPERTIES OF A DISCRETE CLASSICAL PROBABILITY SPACE

If CPS is discrete-finite $ ( \, M \; = \; \{ a_{1} \, , \ldots \,
a_{n} \} \, ) $  or  discrete-infinite  $ ( \, M \; = \; \{ a_{n}
\}_{n \in \mathbb{N}} \, ) $ it is natural to assume that $ \mu (
\{ a_{i} \} ) > 0  \; \forall i $  since an element whose
singleton has zero probability can be simply thrown away from the
beginning.

It follows, than, that CPS has no null sets and, conseguentially,
typical properties are simply the holding properties.

\begin{example} \label{ex:some typical property of the unbiased space of cbit's sequences}
\end{example}
SOME TYPICAL PROPERTY OF THE UNBIASED SPACE OF CBITS' SEQUENCES:

Among the typical properties of the \textbf{unbaised space of
binary sequences} $ ( \Sigma^{\infty} \, , \, P_{unbaised} ) $
there are the following:
\begin{itemize}
  \item \textbf{Borel normality of order $ m \in {\mathbb{N}}$}:
\begin{equation}
  p_{m-Borel} ( \bar{x} ) \; := \; <<  \, \lim_{n \rightarrow \infty} \frac{ N_{i}( \vec{x}(n))}{ \lfloor \frac{n}{m} \rfloor
  } \; = \; \frac{1}{2^{m}} \, >>
\end{equation}
where $  N_{i}( \vec{y}) \, i \in \Sigma $ denotes the number of
occurence of the letter $ i \in \Sigma $ in the string $ \vec{y}
\in \Sigma^{\star} $
  \item \textbf{infinite recurrence}
\begin{equation}
  p_{infinite \; recurrence} ( \bar{x} ) := << cardinality \{ n \in {\mathbb{N}} \; :
  \;
\frac{ N_{1}( \vec{x} (n))}{n} \; = \; \frac{1}{2} \} \; = \;
\aleph_{0} >>
\end{equation}
  \item \textbf{iterated-logarithm property}
\begin{equation}
   p_{iterated \; logarithm} ( \bar{x} ) := << \, \lim \sup_{n \rightarrow \infty} \frac{ \sum_{i=1}^{n} x_{i} \, - \, \frac{n}{2}}{ \sqrt{n \log \log
   n}} \; \leq \; \frac{1}{\sqrt{2}} \, >>
\end{equation}
  \item \textbf{transcendence}
\begin{equation}
  p_{trascendence} ( \bar{x} ) := << \, {\mathcal{N}} (
  \bar{x}) \, \notin  \, {\mathbb{A}} >>
\end{equation}
  \item \textbf{irrecursivity}
\begin{equation}
  p_{irrecursivity} ( \bar{x} ) := << \, {\mathcal{N}} (
  \bar{x}) \, \notin \, \Delta_{0}^{0} ( [ 0 ,1 ) ) \, >>
\end{equation}
  \item \textbf{irrationality}
\begin{equation}
  p_{irrationality} ( \bar{x} ) := << \, {\mathcal{N}} (
  \bar{x}) \, \notin \, {\mathbb{Q}} \, >>
\end{equation}
 \item \textbf{ownership of all substrings}
\begin{equation}
  p_{\text{ownership of all substrings}} ( \bar{x} ) := << \, \forall \vec{y}
  \in \Sigma^{\star} \, \exists n,m \in {\mathbb{N}}_{+} : \vec{x}
  (n,m) \, = \, \vec{y} \, >>
\end{equation}
  \item \textbf{difference from $ \bar{y} \in \Sigma^{\infty} $}
\begin{equation}
  p_{\text{difference from $ \bar{y}$}} ( \bar{x} ) := << \, \bar{x} \, \neq \,
  \bar{y} >>
\end{equation}
\end{itemize}
\newpage
\section{Impossibility of absolute conformism in Classical Probability Theory}
Kolmogorov's original idea about the characterization of the
\textbf{intrinsic randomness} of an \textbf{individual object}
was to consider it as more random as more it is conformistic, in
the sense of conforming itself to the collectivity belonging to
all the overwhelming majorities, i.e. possessing all the typical
properties \cite{Li-Vitanyi-97}, \cite{Calude-94} .

Such an attitude results in the following:
\begin{definition}
\end{definition}
SET OF THE KOLMOGOROV-RANDOM ELEMENTS OF $ ( \Sigma^{\infty} \, ,
\, P_{unbaised} ) $:
\begin{multline}
  KOLMOGOROV-RANDOM [ ( \Sigma^{\infty} \, ,
\, P_{unbaised} )  ] \; \equiv \\
   \{ \, x \, \in \, M \, : p ( x ) \; holds \; \; \forall p \in {\mathcal{P}}  [  ( \Sigma^{\infty} \, ,
\, P_{unbaised} )
  ]_{TYPICAL} \; \}
\end{multline}
But an immediate application of the Cantorian
diagolization-proof's technique \cite{Odifreddi-89} lead to the
following:
\begin{theorem} \label{th:not existence of Kolmogorov random sequences of cbits}
\end{theorem}
NOT EXISTENCE OF KOLMOGOROV RANDOM SEQUENCES OF CBITS
\begin{equation}
KOLMOGOROV-RANDOM [ ( \Sigma^{\infty} \, , \, P_{unbaised} )  ]
\; = \; \emptyset
\end{equation}
\begin{proof}
Let us consider again the family of unary predicates $
p_{\text{difference from $ \bar{y}$}} $ over $ \Sigma^{\infty} $
depending on the parameter $ \bar{y} \in \Sigma^{\infty} $
introduced in the example\ref{ex:some typical property of the
unbiased space of cbit's sequences}

We already saw that they are all typical properties of $ (
\Sigma^{\infty} \, , \, P_{unbaised} ) $
\begin{equation}
   p_{\text{difference from $ \bar{y}$}}  (\bar{x}) \; \in \; {\mathcal{P}} [  ( \Sigma^{\infty} \, ,
\, P_{unbaised} )]_{TYPICAL} \; \;
   \forall \bar{y} \in \Sigma^{\infty}
\end{equation}
Let us now observe that:
\begin{equation}
   p_{\text{difference from $ \bar{x}$}}  (\bar{x}) \; \text{doesn't hold} \; \; \forall \bar{x} \in \Sigma^{\infty}
\end{equation}
So $ p_{\text{difference from $ \bar{x}$}}  $ is a typical
property that is not satisfied by any element of $
\Sigma^{\infty} $, immediately implying the thesis
\end{proof}

\bigskip

The theorem\ref{th:not existence of Kolmogorov random sequences of
cbits} shows that we have to relax the condition that a random
sequence of cbits possesses \textbf{all the typical properties}
requiring only that it satisfies \textbf{a proper subclass of
typical properties}.

The right subclass was proposed by P. Martin L\"{o}f  who observed
that all the Classical Laws of Randomness, i.e. all the properties
of Classical Probability Theory that are known to hold with
probability one (such as the \textit{Law of Large Numbers}, the
\textit{Law of Iterated Logarithm} and so on ) are
\textbf{effectively-falsificable} in the sense that we can
effectively test whether they are violated (though we cannot
effectively certify that they are satisfied).

This leads, assuming the Church-Turing's Thesis
\cite{Odifreddi-89} and endowed $ \Sigma^{\infty} $ with the
\textbf{product topology} induced by the \textbf{discrete
topology} of $ \Sigma $, to introduce the following notions:
\begin{definition}
\end{definition}
$S \; \subset \; \Sigma^{\infty} $ IS ALGORITHMICALLY-OPEN:
\begin{equation}
  ( S \text{ is open } ) \; and \; ( S \, = \, X \Sigma^{\infty}
  \,
  {\mathbf{X \; recursively-enumerable}})
\end{equation}
\begin{definition}
\end{definition}
ALGORITHMIC SEQUENCE OF ALGORITHMICALLY-OPEN  SETS:

a sequence $ \{ S_{n} \}_{n \geq 1} $ of algorithmically open
sets $  S_{n} \; = \; X_{n} \Sigma^{\infty} $ : $ \exists X \;
\subset \; \Sigma^{\star} \times {\mathbb{N}} $
\textbf{recursively enumerable} with:
\begin{equation*}
   X_{n} \; = \; \{ \vec{x} \in \Sigma^{\star} \, : \, ( \vec{x} , n ) \in
   X \} \; \; \forall n \in {\mathbb{N}}_{+}
\end{equation*}

\smallskip

Given the classical probability space $ CPS \, := \, (
\Sigma^{\infty} , P ) $:
\begin{definition}
\end{definition}
$ S \; \subset \; \Sigma^{\infty} $  IS AN ALGORITHMICALLY-NULL
SUBSET OF CPS:

$ \exists  \{ G_{n} \}_{n \geq 1} $ algorithmic sequence of
algorithmically-open sets :
\begin{equation*}
  S \; \subset \; \cap_{n \geq 1} G_{n}
\end{equation*}
and:
\begin{equation*}
  alg - \lim_{n \rightarrow \infty} P ( G_{n} ) \; = \; 0
\end{equation*}
i.e. there exist and increasing, unbounded, \textbf{recursive}
function $ f \, : \, {\mathbb{N}} \rightarrow {\mathbb{N}} $ so
that $ P ( G_{n} ) \; < \; \frac{1}{2^{k}} $ whenever $ n \; \geq
\; f(k) $

\begin{definition}
\end{definition}
LAWS OF RANDOMNESS OF CPS:
\begin{multline}
 \mathcal{L}_{randomness} (CPS) \; \equiv \; \{ \, p ( \bar{x} )  \in  {\mathcal{P}} ( \Sigma^{\infty} ) \,
 : \\
   \{ \bar{x} \in \Sigma^{\infty} \, : \, p ( \bar{x} ) \text{ doesn't hold } \}
   \; \text{is an algorithmically null set of CPS} \}
\end{multline}
\begin{example} \label{ex:some law of randomness of the unbiased space of cbit's}
\end{example}
SOME LAW OF RANDOMNESS OF THE UNBIASED SPACE OF CBITS' SEQUENCES:
Let us consider again the typical properties of the classical
probability space $ ( \Sigma^{\infty} , P_{unbiased} ) $
introduced  in the example\ref{ex:some typical property of the
unbiased space of cbit's sequences}.

\textbf{Borel normality of order $ m \in {\mathbb{N}} $},
\textbf{infinite recurrence}, the \textbf{iterated-logarithm
property}, \textbf{transcendence} and \textbf{irrationality} are
all effectively-refutable and, hence, are all Laws of Randomness.

To refute that a sequence has the property of
\textbf{irrecursivity}, or of \textbf{ownership of all
substrings}, or of  \textbf{difference from $ \bar{y} \in
\Sigma^{\infty} $} would, instead, require the inspection the
analysis of an infinite number of of its digits.

Hence all such typical properties are not effectively-refutable
and, hence, are not laws of randomness.

\bigskip

We can now introduce the following:
\begin{definition}\label{def:P-conformistically randomness}
\end{definition}
P-CONFORMISTICALLY RANDOM ELEMENTS OF CPS:
\begin{equation}
  P-CONF-RANDOM(\Sigma^{\infty}) \; := \; \{ \bar{x} \in \Sigma^{\infty} \: p ( \bar{x} ) \text{ holds } \; \forall p \in  {\mathcal{L}_{randomness}} (CPS) \}
\end{equation}
As usual the case of the unbaised measure deserves an ad hoc
definition:
\begin{definition}\label{def:conformistically random sequences}
\end{definition}
CONFORMISTICALLY-RANDOM SEQUENCES:
\begin{multline}
  CONF-RANDOM(\Sigma^{\infty}) \; := \\
   P_{unbaised}-CONF-RANDOM(\Sigma^{\infty})
\end{multline}

\smallskip

\begin{remark} \label{rem:why the diagonalization proof doesn't apply to P-conformistically randomness}
\end{remark}
WHY THE DIAGONALIZATION PROOF OF THEOREM\ref{th:not existence of
Kolmogorov random sequences of cbits} DOESN'T APPLY TO
P-CONFORMISTICALLY RANDOMNESS

Let us observe that the diagonalization proof of
theorem\ref{th:not existence of Kolmogorov random sequences of
cbits} is based on the one-parameter family of typical properties
$ p_{\text{difference from $ \bar{y}$}} \, , \,\bar{y} \in
\Sigma^{\infty} $ . Since noone of these is a law of randomness
the argument falls down.
\newpage
\section{Equivalence between Martin L\"{o}f-conformistical randomness and Chaitin randomness}
Summing up, we have seen that Per Martin L\"{o}f introduced two
approaches to the mathematical characterization of classical
algorithmic randomness:
\begin{itemize}
  \item the \textbf{statistical approach} discussed in chapter\ref{chap:Classical algorithmic randomness as passage of
all the classical algorithmic statistical tests} and resulting in
the definition of the set $ MARTIN
L\ddot{O}F-RANDOM(\Sigma^{\infty}) $ whose equality with the set
$ CHAITIN-RANDOM(\Sigma^{\infty}) $ is stated by
theorem\ref{th:Chaitin-Schnorr theorem}
  \item the \textbf{logical approach} discussed in the previous
  sections and resulting in the definition of the set $
  CONF-RANDOM(\Sigma^{\infty}) $
\end{itemize}
In this section we will prove Martin-L\"{o}f's Theorem showing the
the complete equivalence of these approaches.

This requires the introduction of some technical ingredient,
starting from the following:
\begin{lemma} \label{lem:cardinality inequality of a sequential Martin-Lof test}
\end{lemma}
For every sequential $ P_{unbaised}$-test  V and for every
natural $ m \geq 1 $:
\begin{equation}
  \sum_{\vec{x} \in \Sigma^{n} \bigcap V_{m}} 2^{- | \vec{x} |} \,
  < \, 2^{- m}
\end{equation}
\begin{proof}
It follows immediately from the cardinality inequality in the
definition of a sequential Martin-L\"{o}f test
\end{proof}

\smallskip

Then we need the following:
\begin{lemma} \label{lem:asymptotic condition for a sequential Martin-Lof test}
\end{lemma}
Let V be a sequential $ P_{unbaised}$-test. Then
\begin{equation}
  \lim_{m \rightarrow \infty} P_{unbaised} ( V_{m} \Sigma^{\infty}
  ) \; = \; 0 \; \; constructively
\end{equation}
\begin{proof}
Take V and define for every natural $ m \geq 1 $ the sets $ W_{m}
\, := \, V_{m} \Sigma^{\infty}$. It is seen that for each $ m \geq
1 $, $ W_{m} \, = \, \bigcup_{n=2}^{\infty} X_{n}$ , where:
\begin{equation}
  X_{n} \; := \; \bigcup_{\vec{x} \in \Sigma^{n} \bigcap V_{m}}
  \vec{x} \Sigma^{\infty}
\end{equation}
Furthermore, $ X_{n} \; \subset \; X_{n+1} $ and:
\begin{multline}
  P_{unbaised} ( X_{n} ) \; = \; \sum_{\vec{x} \in \Sigma^{n} \bigcap
  V_{m}} P_{unbaised} ( \vec{x} \Sigma^{\infty} ) \\
  = \; \sum_{\vec{x} \in \Sigma^{n} \bigcap
  V_{m}} 2^{- | \vec{x} | } \; = \; \frac{cardinality( \Sigma^{n} \bigcap
  V_{m})}{2^{n}} \\
  < 2^{- m}
\end{multline}
in view of lemma\ref{lem:asymptotic condition for a sequential
Martin-Lof test} and of the fact that the sets $ \{ \vec{x}
\Sigma^{\infty} : \vec{x}  \in \Sigma^{n} \bigcap V_{m} \} $ are
mutually disjoint. So:
\begin{equation}
  P_{unbiased} ( W_{m} ) \; = \; \lim_{n \rightarrow \infty} P_{unbiased} ( X_{n}
  ) \; \leq \; 2^{- m}
\end{equation}
Finally, put $ H( m ) \; := \; m +1 $ and notice that if $ m \;
\geq \; H(k) $, then $ P_{unbiased} ( W_{m} ) \; \leq \; 2^{- k} $
\end{proof}

The last ingredient required for proving Martin-L\"{o}f's Theorem
is the following:
\begin{lemma} \label{lem:algorithmic null set associated to a sequential Martin Lof test}
Let V be a sequential $ P_{unbaised}$-test. Then $
\bigcap_{m=1}^{\infty} ( V_{m} \Sigma^{\infty} ) $ is an
algorithmically-null subset of $ ( \Sigma^{\infty} \, , \,
P_{unbaised} ) $
\end{lemma}
\begin{proof}
Take V and define for every natural $ m \geq 1 $ the sets $ W_{m}
\, := \, V_{m} \Sigma^{\infty}$. Since V is r.e. it follows that
the sequence $ \{ V_{m} \}_{m \in {\mathbb{N}}_{+}} $ is an
algorithmic sequence of algorithmically opens sets.

By lemma\ref{lem:asymptotic condition for a sequential Martin-Lof
test} it follows the thesis
\end{proof}

\smallskip

We have at last all the ingredients required to prove the
following:
\begin{theorem} \label{th:Martin-Lof theorem}
\end{theorem}
MARTIN-L\"{O}F'S THEOREM:
\begin{equation}
  CONF-RANDOM(\Sigma^{\infty}) \; = \; CHAITIN-RANDOM(\Sigma^{\infty})
\end{equation}
\begin{proof}
Fix a universal sequential $ P_{unbaised}$-test U. Since:
\begin{equation}
  \Sigma^{\infty} \, -  \, CHAITIN-RANDOM( \Sigma^{\infty} ) \; =
  \; \bigcap_{m=1}^{\infty} U_{m} \Sigma^{\infty}
\end{equation}
we may apply lemma\ref{lem:algorithmic null set associated to a
sequential Martin Lof test} to conclude that $ \Sigma^{\infty} \,
-  \, CHAITIN-RANDOM( \Sigma^{\infty} ) $ is an
algorithmically-null set.

Next let $ S \, \subset \, \Sigma^{\infty} $ be an arbitrary
algorithmically-null set. We shall prove that:
\begin{equation}
  S \; \subset \; \Sigma^{\infty} \, -  \, CHAITIN-RANDOM( \Sigma^{\infty} )
\end{equation}
To this aim let us consider an algorithmic sequence of
algorithmically open sets $ ( G_{m} )_{m \geq 1} $ such that:
\begin{equation}
  S \; \subset \; \bigcap_{m=1}^{\infty} G_{m}
\end{equation}
and:
\begin{equation}
  P_{unbaised} ( G_{t} ) \; < \; 2^{-m} \; \; \forall t \, \geq \, H(m)
\end{equation}
where $ H : {\mathbb{N}} \, \mapsto \, {\mathbb{N}} $ is a fixed
increasing, unbounded recursive function.

Write:
\begin{equation}
  G_{m} \; := \; X_{m} \Sigma^{\infty} \; = \; ( X_{m}
  \Sigma^{\star} ) \Sigma^{\infty}
\end{equation}
for all $ m \, \geq \, 1 $, where $ X_{m} \; \subset \;
\Sigma^{\star} $ is an r.e. set. We have to construct a sequential
$ P_{unbiased} $-test V such that:
\begin{equation} \label{eq:sequential Martin-Lof test to construct}
  \bigcap_{m=1}^{\infty} V_{m} \Sigma^{\infty} \; = \; \bigcap_{m=1}^{\infty} G_{m}
\end{equation}
Put:
\begin{equation}
  V_{m} \; := \; \bigcap_{i=1}^{H(m)} X_{i} \Sigma^{\star} \; \; m
  \in {\mathbb{N}}_{+}
\end{equation}
Clearly the set V defined as:
\begin{equation}
  V \; := \; \{ ( \vec{x} , m ) \in \Sigma^{\star} \times
  {\mathbb{N}}_{+} \, : \, \vec{x} \in V_{m} \}
\end{equation}
is r.e., $ V_{m+1} \; \subset \; V_{m} $  and is such that if $
\vec{x} <_{p} \vec{y} $ and $ \vec{x} \in V_{m} $ then $ \vec{y}
\in V_{m} $.

Fixed $ n, m \in {\mathbb{N}}_{+} $:
\begin{multline}
  cardinality( \Sigma^{n} \bigcap V_{m} ) \; \leq \;  cardinality(
  X_{H(m)} \Sigma^{\star} \bigcap \Sigma^{n} \\
  = \; 2^{n} \, cardinality( X_{H(m)} \Sigma^{\star} \bigcap
  \Sigma^{n}) \, 2^{- n} \; = \; 2^{n} P_{unbaised} ((( X_{H(m)}
  \Sigma^{\star}) \, \bigcap \, \Sigma^{n} ) \Sigma^{\infty} ) \\
  \leq \; 2^{n} P_{unbaised} (( X_{H(m)}
  \Sigma^{\star} ) \Sigma^{\infty} ) \; \leq \; 2^{n - m}
\end{multline}
So V is a sequential $ P_{unbaised} $-test and, hence,
eq.\ref{eq:sequential Martin-Lof test to construct} holds by
virtue of the strict monotonicity of H.

According to the universality of U one can find a natural c such
that:
\begin{equation}
  V_{m + c} \; \subset \; U_{m} \; \; \forall m \in {\mathbb{N}}_{+}
\end{equation}
Then:
\begin{multline}
  S \; \subset \;  \bigcap_{m=1}^{\infty} V_{m} \Sigma^{\infty} \; \subset \;  \bigcap_{m=1}^{\infty} V_{m+c}
  \Sigma^{\infty} \\
  \subset \bigcap_{m=1}^{\infty} U_{m} \Sigma^{\infty} \; = \; \Sigma^{\infty} \, -  \, CHAITIN-RANDOM( \Sigma^{\infty} )
\end{multline}
\end{proof}

\smallskip

\begin{remark} \label{rem:on why Classical Probability Theory applies to reality}
\end{remark}
ON WHY CLASSICAL PROBABILITY THEORY APPLIES TO REALITY

We can now fully appreciate the conceptual relevance of
theorem\ref{th:foundation of the applicability of probability
theory to reality} (and the name we gave to it): it tells us that
extracted at random a sequence according to the probability
distribution $ \mu $ the occured sequence will satisfy all the $
\mu$ Laws of Randomness with certainty.

We indeed have to bless such a theorem: it is only for his
courtesy that it is possible to give certain mathematical
predictions concerning the statistical behaviour of
classically-non deterministic phenomena.

This is particularly rilevant in the case in which $ \mu $ is the
unbaised probability measure $ P_{unbiased} $:

it is only because making infinite independent tosses of a fair
coin we obtain with certainty a sequence without intrinsic
regularity that we can find a mathematical regularity in
classical-nondeterminism.

This clarifies why, as we will discuss in chapter\ref{chap:The
irreducibility of quantum probability both to classical
determinism and to classical nondeterminism} is very raesonable
to expect that an  analogous situation must happen also in
Quantum Probability Theory.
\chapter{Classical algorithmic randomness as stability of the relative frequences under proper classical
algorithmic place selection rules} \label{chap:Classical
algorithmic randomness as stability of the relative frequences
under proper classical algorithmic place selection rules}
\section{Von Mises' Frequentistic Foundation of Probability} \label{sec:Von Mises' Frequentistic Foundation of Probability}
The mirable features of the Kolmogorovian measure-theoretic
axiomatization of  Classical Probability Theory
\cite{Kolmogorov-56} has led to consider it as the last word
about Foundations of Classical Probability Theory, leading to the
general attitude of forgetting the other different
axiomatizations and, in particular, von Mises' Frequentistic one
\cite{von-Mises-81}.

Richard Von Mises' axiomatization of Classical Probability Theory
lies on the mathematical formalization of the following two
empirical laws:
\begin{enumerate}
  \item \textbf{Law of Stability of Statistic Relative
  Frequencies}
\begin{center}
\textit{"It is essential for the theory of probability that
experience has shown that in the game of dice, as in all other
mass phenomena which we have mentioned, the relative frequencies
of certain attributes become more and more stable as the number of
observations is increased"} (cfr. pag.12 of \cite{von-Mises-81})
\end{center}

  \item \textbf{Law of Excluded Gambling Strategies}
\begin{center}
\textit{"Everybody who has been to Monte Carlo, or who has read
descriptions of a gambling bank, know how many 'absolutely safe'
gambling systems, sometimes of an enormously complicated
character, have been invented and tried out by gamblers; and new
systems are still suggested every day. The authors of such
systems have all, sooner or later, had the sad experience of
finding out that no system is able to improve their chance of
winning in the long run,i.e. to affect the relative frequencies
with which different colours of numbers appear in a sequence
selected from the total sequence of the game. This experience
forms the experimental basis of our definition of probability"}
(cfr. pagg.25-26 of \cite{von-Mises-81})
\end{center}
\end{enumerate}
According to Von Mises Probability Theory concerns properties of
collectivities, i.e. of sequences of identical objects.

Considering each individual object as a letter of an alphabet $
\Sigma$, we can then say that Probability Theory concerns
elements of the set $ \Sigma^{\infty} $ of the sequences of
letters from $ \Sigma $ or, more properly, a certain subset  $
{\mathcal{C}}ollectives\; \subset \;  \Sigma^{\infty} $ whose
elements are called \textbf{collectives}.

\medskip

Let us then introduce the set $ {{\mathcal{A}}ttributes} ( \Sigma
) $ of the \textbf{attributes} of $ \mathcal{C} $'s elements
defined as the set of unary predicates about the generic $ C \;
\in  \; {\mathcal{C}}ollectives $.

The mathematical formalization of the \textbf{Law of Stability of
Statistic Relative Frequencies} results in the following:
\begin{axiom} \label{ax:axiom of convergence}
\end{axiom}
AXIOM OF CONVERGENCE

\begin{hypothesis}
\end{hypothesis}
\begin{equation*}
   C \; \in \; {\mathcal{C}}ollectives
\end{equation*}
\begin{equation*}
  A  \; \in \; {{\mathcal{A}}ttributes} ( \Sigma )
\end{equation*}
\begin{thesis}
\end{thesis}
\begin{equation*}
  \exists \; \; \lim_{n \rightarrow \infty} \frac{ N( A | \vec{C}(n) ) }{n}
\end{equation*}

where $N( A | \vec{C}(n) )$ denotes the number of elements of the
prefix $ \vec{C}(n) $ of C of length n  for which the attribute A
holds.

Given an \textbf{attribute} $ A \in  {\mathcal{A}}ttributes (
\Sigma ) $ of a \textbf{collective} $ C \in
{\mathcal{C}}ollectives $ the axiom\ref{ax:axiom of convergence}
make consistent the following definition:
\begin{definition}
\end{definition}
VON MISES' FREQUENTISTIC PROBABILITY OF A IN C:
\begin{equation}\label{eq:von Mises' definition of probability}
  P_{VM} ( A | C ) := lim_{n \rightarrow \infty} \frac{ N( A | \vec{C}(n) ) }{n}
\end{equation}
Let us then introduce the following basic definition:
\begin{definition} \label{def:gambling strategy}
\end{definition}
GAMBLING STRATEGY:

$ S : \Sigma^{\star} \stackrel {\circ}{\rightarrow} \{ 0 , 1 \} $

\smallskip

Given a gambling strategy S:
\begin{definition} \label{def:subsequence extraction function}
\end{definition}
SUBSEQUENCE EXTRACTION FUNCTION INDUCED BY S:

$ EXT[S] : \Sigma^{\infty} \; \rightarrow \; \Sigma^{\infty} $ :

\begin{equation}
   EXT[S] (  x_{1}  x_{2}  \cdots  ) \; := \; \text{ordered concatenation}
   ( \{ x_{n} \, :
    \,  S ( x_{1}  \cdots x_{n-1} ) = 1 , n \in {\mathbb{N}}_{+} \} )
\end{equation}
The name in the definition\ref{def:subsequence extraction
function} is justified by the fact that obviously:
\begin{equation}
  EXT[S] ( \bar{x} ) \leq_{s}  \bar{x} \; \; \forall \bar{x}
  \in \Sigma^{\infty}
\end{equation}
where $  \leq_{s} $ is the following:
\begin{definition}
\end{definition}
SUBSEQUENCE ORDERING RELATION ON $ \Sigma^{\infty} $
\begin{equation}
  \bar{x} \leq_{s} \bar{y} := \text{$ \bar{x} $ is a subsequence of $ \bar{y} $ }
\end{equation}
\begin{example} \label{ex:bet on the last result}
\end{example}
BET EACH TIME ON THE LAST RESULT

Considered the binary alphabet $ \Sigma \; := \; \{ 0 ,1 \} $, let
us analyze the following gambling strategy:
\begin{equation}
  S ( x_{1}  \cdots  x_{n} ) \; := \;
  \begin{cases}
    \uparrow  &   \text{if $ n = 0 $}, \\
     x_{n}  &   \text{otherwise}
  \end{cases} \; \;  x_{1}  \cdots  x_{n} \in \Sigma^{n} , n \in \mathbb{N}
\end{equation}
and the \textit{subsequence extraction function} $EXT[S]$ it
gives rise to.

Clearly we have that:

\begin{tabular}{|c|c|}
  $\vec{x}$ & $S( \vec{x} )$ \\ \hline
  $\lambda$ & $\uparrow$ \\
  0 & 0 \\
  1 & 1 \\
  00 & 0 \\
  01 & 1 \\
  10 & 0 \\
  11 & 1 \\
  000 & 0 \\
  001 & 1 \\
  010 & 0 \\
  011 & 1 \\
  100 & 0 \\
  101 & 1 \\
  110 & 0 \\
  111 & 1 \\
  0000 & 0 \\
  0001 & 1 \\
  0010 & 0 \\
  0011 & 1 \\
  0100 & 0 \\
  0101 & 1 \\
  0110 & 0 \\
  0111 & 1 \\
  1000 & 0 \\
  1001 & 1 \\
  1010 & 0 \\
  1011 & 1 \\
  1100 & 0 \\
  1101 & 1 \\
  1110 & 0 \\
  1111 & 1 \\ \hline
\end{tabular}

\smallskip

Furthermore we have, clearly, that:
\begin{align*}
  EXT[S] ( 0^{\infty} ) \; & = \; \lambda \\
  EXT[S] ( 0^{\infty}) \; & = \;  1^{\infty} \\
  EXT[S] ( 01^{\infty}\cdots ) \; & = \; 0^{\infty} \\
  EXT[S] ( 10^{\infty} ) \; & = \; 0^{\infty} \\
  EXT[S] ( \bar{x}_{Champernowne} ) \; & = \; 0101\cdots
\end{align*}
where $ \bar{x}_{Champernowne} $ is the Champernowne sequence
defined as the lexicografic ordered concatenation of the binary
strings:
\begin{equation*}
  \bar{x}_{Champernowne} \; = \; 0100011011000001010011100101110111 \cdots
\end{equation*}

\begin{example} \label{ex:bet on the less frequent letter}
\end{example}
BET ON THE LESS FREQUENT LETTER

Considered again the binary alphabet $ \Sigma \; := \; \{ 0 ,1 \}
$, let us analyze the following gambling strategy:
\begin{equation}
  S ( \vec{x} ) \; = \;
  \begin{cases}
    \uparrow & \text{if $ \vec{x} = \lambda $ or $ N_{0} ( \vec{x} ) = N_{1} ( \vec{x} )$} , \\
    1 & \text{if  $N_{0} ( \vec{x} ) > N_{1} ( \vec{x} )$} , \\
    0 & \text{otherwise}.
   \end{cases}
\end{equation}
where $ N_{0} ( \vec{x} ) , N_{1} ( \vec{x} ) $ denote the number
of, respectively, zeros and ones in the string $ \vec{x} $.

We have that:

\smallskip

\begin{tabular}{|c|c|}
  $\vec{x}$ & $S( \vec{x} )$ \\ \hline
  $\lambda$ & $\uparrow$ \\
  0 & 1 \\
  1 & 0 \\
  00 & 1   \\
  01 & $\uparrow$ \\
  10 & $\uparrow$ \\
  11 & 0 \\
  000 & 1 \\
  001 & 1  \\
  010 & 1 \\
  011 & 0 \\
  100 & 1 \\
  101 & 0 \\
  110 & 0 \\
  111 &  0 \\
  0000 & 1 \\
  0001 & 0 \\
  0010 & 1 \\
  0011 & $\uparrow$ \\
  0100 & 1 \\
  0101 & $\uparrow$ \\
  0110 & $\uparrow$ \\
  0111 & 1 \\
  1000 & 1 \\
  1001 & $\uparrow$ \\
  1010 & $\uparrow$ \\
  1011 & 0 \\
  1100 & $\uparrow$ \\
  1101 & 0 \\
  1110 & 0 \\
  1111 & 0 \\ \hline
\end{tabular}

\smallskip

As to the extraction function of S:
\begin{align*}
  EXT[S] ( 0^{\infty} ) \; & = \; 0^{\infty} \\
  EXT[S] ( 1^{\infty} ) \; & = \;  \lambda \\
  EXT[S] ( 01^{\infty} ) \; & = \; 1^{\infty} \\
  EXT[S] ( 10^{\infty} ) \; & = \; \lambda \\
  EXT[S] ( \bar{x}_{Champernowne} ) \; & = \; 10011011\cdots
\end{align*}

Denoted by $ {\mathcal{S}}trategies ( {\mathcal{C}}ollectives ) $
the \textbf{set of gambling strategies} concerning $
{\mathcal{C}}ollectives $, we can formalize the \textbf{Law of
Excluded Gambling Strategies} by the following:
\begin{axiom} \label{ax:axiom of randomness}
\end{axiom}
AXIOM OF RANDOMNESS

\begin{hypothesis}
\end{hypothesis}
\begin{equation*}
  S \, \in \,  {{\mathcal{S}}trategies}_{admissible} ( {\mathcal{C}}ollectives )
\end{equation*}
\begin{equation*}
  C \, \in \, {\mathcal{C}}ollectives
\end{equation*}
\begin{equation*}
  A  \, \in \, {{\mathcal{A}}ttributes} ( \Sigma )
\end{equation*}
\begin{thesis}
\end{thesis}
\begin{equation*}
  P_{VM}( \,A  \,| \, EXT[S] (C) \,  ) \; = \;  P_{VM} ( A | C )
\end{equation*}
where $ {{\mathcal{S}}trategies}_{admissible} (
{\mathcal{C}}ollectives ) \; \subseteq \; {\mathcal{S}}trategies (
{\mathcal{C}}ollectives )  $ is the \textbf{set of admissible
gambling strategies} whose mathematical characterization will be
investigated in the next sections.
\newpage
\section{Classical Gambling in the framework of Classical Statistical Decision
Theory} \label{sec:Classical Gambling in the framework of
Classical Statistical Decision Theory}

Classical Statistical Decision Theory \cite{French-Rios-Insua-00}
concerns the following situation:

a \textit{decision maker} have to make a  single action $ a \in
{\mathcal{A}}ctions $ from a space $ {\mathcal{A}}ctions $ of
possible actions.

Features that are unknown about the external world are modelled
by an unknown state of nature $ s \in {\mathcal{S}}tates $ in a
set $ {\mathcal{S}}tates $ of possible states of nature.

The consequence $ c ( a , s ) \in {\mathcal{C}}onsequences $ of
his choice depends both on the action chosen and on the unknwown
state of nature.

Before making his decision the decision maker may observe an
outcome $ X = x $ of an experiment, which depends on the unknown
state s. Specifically the observation X is drawn from a
distribution $ P_{X} ( \cdot | s ) $.

His objectives are encoded in a real valued \textit{utility
function} $ u ( a , s ) $.

Let us assume that the \textit{decision maker} knows the
\textit{action space} $ {\mathcal{A}}ctions $, \textit{state
space} $ {\mathcal{S}}tates $ and \textit{consequence space} $
{\mathcal{C}}onsequences $, along with the probability
distribution and the \textit{utility function}.

His problem is:

\textbf{observe $ X = x $ and then choose an action $ d(x) \in
{\mathcal{A}}ctions $, using the information that $ X = x $, to
maximize, in some sense, $ u ( d ( x ) , s ) $ }.

Every decision process may obviously  be seen as a gambling
situation: the \textit{action space} $ {\mathcal{A}}ctions$ may
be seen as the set of possible bets of the decision maker, that
we will call from here and beyond the \textit{gambler}, while the
\textit{utility function} gives the \textit{payoff}.

Let us consider, in particular, the following gambling situation:

in the city's \textit{Casino} at each turn $ n \in \mathbb{N} $
the croupier tosses a fair coin.

Before the $ n^{th} $ toss the gambler can choose among one of the
possbile choices:
\begin{itemize}
  \item to bet one fiche on \textit{head}
  \item to bet one fiche on \textit{tail}
  \item not to play at that turn
\end{itemize}
Leaving all the philosophy behind its original foundational
purpose we can, now, from inside the standard Kolomogorovian
measure-theoretic formalization of Classical Probability Theory,
appreciate the very intuitive meaning lying behind Von Mises'
axioms.

Let us indicate  by $ X_{n} $ the random variable on the binary
alphabet $ \Sigma := \{ 0 , 1 \} $ (where we will assume from
here and beyond, that $ head = 1 $ and $ tail = 0 $)
corresponding to the $ n^{th} $ coin toss and by $ x_{n} \in
\Sigma $ the result of the $ n^{th} $ coin toss.

Let us, furthermore, denote by $ \bar{x} \; := \; (  x_{1} ,
x_{2} , , \cdots )\in \Sigma^{\infty} $ the sequence of all the
results of the coin tosses and by $ \vec{x}(n) \in \Sigma^{n} $
its $ n^{th} $ prefix.

By hypothesis $\{ X_{n} \}_{n \in \mathbb{N}} $ is a Bernoulli($
\frac{1}{2}$)  discrete-time stochastic process over $ \Sigma $.

A gambling strategy $ S : \Sigma^{\star} \stackrel
{\circ}{\rightarrow} \{ 0 , 1 \} $ determines the gambler's
decision at the $ n^{th} $ turn in the following way:
\begin{itemize}
  \item if $ S( \vec{x}(n-1) ) \; = \; 1 $ he bets on \textit{head}
  \item if $ S( \vec{x}(n-1) ) \; = \; 0 $ he bets on \textit{tail}
  \item if $  S( \vec{x}(n-1) ) \; = \; \uparrow $ he doesn't bet at that turn
\end{itemize}

\medskip

\begin{example}
\end{example}

APPLYING TO THE CASINO THE GAMBLING STRATEGY OF
EXAMPLE\ref{ex:bet on the last result}

Let us suppose that the first 10 coin tosses give the  following
string of results: $ \vec{x}(n) \; = \; 1101001001 $

Our evening to Casino may be told by the following table:

\smallskip

\begin{tabular}{|c|c|c|c|}
TOSS &  RESULT OF THE TOSS & BET MADE ABOUT THAT TOSS &  PAYOFF  \\
\hline
  1  &          1          &         no bet           &        0         \\
  2  &          1          &          1               &       +1        \\
  3  &          0          &          1               &        0       \\
  4  &          1          &          0               &       -1        \\
  5  &          0          &          1               &       -2        \\
  6  &          0          &          0               &       -1        \\
  7  &          1          &          0               &       -2        \\
  8  &          0          &          1               &       -3        \\
  9  &          0          &          0               &       -2        \\
  10 &          1          &          0               &       -3        \\ \hline
\end{tabular}

\smallskip

As we see $ PAYOFF(10) \; = \; -3 $.

\begin{example}
\end{example}
APPLYING TO THE CASINO THE GAMBLING STRATEGY OF EXAMPLE\ref{ex:bet
on the less frequent letter}

\smallskip

Let us suppose again that the first 10 coin tosses give the
following string of results: $ \vec{x}(n) \; = \; 1101001001 $.

Our evening to Casino may be told by the following table:

\smallskip

\begin{tabular}{|c|c|c|c|}
TOSS &  RESULT OF THE TOSS & BET MADE ABOUT THAT TOSS & PAYOFF
\\ \hline
  1  &          1          &         no bet           &       0          \\
  2  &          1          &           0              &      -1          \\
  3  &          0          &           0              &       0          \\
  4  &          1          &           0              &      -1          \\
  5  &          0          &           0              &       0          \\
  6  &          0          &           0              &      +1          \\
  7  &          1          &         no bet           &      +1           \\
  8  &          0          &           0              &      +2          \\
  9  &          0          &         no bet           &      +2          \\
  10 &          1          &           1              &      +3          \\ \hline
\end{tabular}

\smallskip

As we see $ PAYOFF(10) \; = \; +3  $.

\bigskip

The probability distribution of the string $ \vec{x}(n) $ is the
uniform distribution on $ \Sigma^{n} $ :
\begin{equation}
  Prob [ \vec{x}(n) = \vec{y} ] \; =  P_{unbaised \, , \, n} (  \vec{y} ) \; := \; \frac{1}{2^{n}} \; \; \forall
  \vec{y} \in \Sigma^{n} , \forall n \in \mathbb{N}
\end{equation}

When $ n \rightarrow \infty $ such a distribution tends to the
unbiased probability measure $ P_{unbiased} $ on $
\Sigma^{\infty} $.

\medskip

Clearly the possible \textbf{attributes} of a letter on the binary
alphabet are:
\begin{itemize}
  \item $ a_{1} \; := \;   << \text{to be 1} >> $
  \item $ a_{0} \; := \;  << \text{to be 0} >> $
\end{itemize}
so that:
\begin{equation}
  {{\mathcal{A}}ttributes} ( \Sigma ) \; = \; \{ a_{1} , a_{0} \}
\end{equation}
Whichever $ {\mathcal{C}}ollectives\; \subset \;  \Sigma^{\infty}
$ is the axiom\ref{ax:axiom of convergence} is, from inside the
standard kolmogorovian measure-theoretic foundation, an immediate
corollary of the Law of Large Numbers.

As far as axiom\ref{ax:axiom of randomness} is concerned, anyway,
the situation is extraordinarily subtler.

Every \textbf{intrinsic regularity} of $ \vec{x}(n) $ could have
been encoded by the gambler in a proper \textbf{winning strategy
up to the $ n^{th} $ turn}.

The same definition of what a \textbf{winning strategy} is
requires some caution:

we can, indeed, give two possible definitions of such a concept:
\begin{definition}[AVERAGE-WINNING STRATEGY UP TO THE $ n^{th} $
TOSS] \label{def:average winning strategy}

a strategy so that the expectation value of the payoff after the
first n tosses \textbf{payoff(n) } is greater than zero
\end{definition}

The fact the a strategy is average-winning doesn't imply that the
payoff after the $ n^{th} $ toss will be strictly positive with
certainty: it happens if we are lucky.

Let us now introduce a weaker notion of a winning strategy:

\begin{definition}[LUCKY-WINNING STRATEGY UP TO THE $ n^{th} $
TOSS] \label{def:lucky winning strategy}

a strategy so that the the probability that the  payoff after the
first n tosses \textbf{payoff(n) } is greater than zero is itself
greater than zero
\end{definition}

For finite n every strategy is obviously lucky-winning.

Let us now consider the limit $ n \rightarrow \infty $.

By purely measure-theoretic considerations we may easily prove the
following:
\begin{theorem} \label{th:weak law of excluded gambling systems}
WEAK LAW OF EXCLUDED GAMBLING STRATEGIES

For $ n \rightarrow \infty $ the set of  the
\textit{average-winning strategies} tends to the null set
\end{theorem}
\begin{proof}
Given a gambling strategy $ S : \Sigma^{\star} \stackrel
{\circ}{\rightarrow} \{ 0 , 1 \} $ we have clearly that the
conditional expectation of the payoff at the $ n^{th} $ turn
conditioned to the payoff at the $ (n-1)^{th} $ turn is the sum of
two addenda:
\begin{itemize}
  \item the payoff at the $ (n-1)^{th} $ turn
  \item the expectation value of the gain at the $ (n-1)^{th} $
  turn
\end{itemize}
This second addendum is clearly equal to zero if the adopted
gambling strategy prescribes not to bet at the $ n^{th} $ turn.
Otherwise its is itself given by the sum of two addenda:
\begin{itemize}
  \item one related to the case in which heads turn up and given
  by the probability of this fact, obviously equal to $
  \frac{1}{2}$, taken with positive sign if we betted on head and
  taken with negative sign if we betted on tail
  \item one related to the case in which tails turn up  and given
  by the probability of this fact, obviously equal to $
  \frac{1}{2}$, taken with positive sign if we betted on tail and
  taken with negative sign if we betted on head
\end{itemize}
But these last two addenda obviously compensate each other, so
that the conditional expectation of the payoff at the $ n^{th} $
turn conditioned to the payoff at the $ (n-1)^{th} $ turn is
simply given by the  payoff at the $ (n-1)^{th} $ turn.

This reasoning can be expressed in formulae as:
\begin{multline} \label{eq:conditional payoff at step n}
  E[payoff(n) | payoff(n-1) ] \; = \; payoff(n-1) + \\
  If[ S ( \vec{x}_{n-1} ) = \uparrow , 0 ,  \frac{1}{2} \\
If[ S ( \vec{x}_{n-1} )= 1 , 1 , -1 ] + \frac{1}{2} \\
   If[ S ( \vec{x}_{n-1} ) = 0 , 1 , -1]] \; = \; payoff(n-1)  \; \; \forall n \in \mathbb{N}
\end{multline}
(where I have adopted Mc Carthy's LISP conditional notation
\cite{Mc-Carthy-60} popularized by Wolfram's \textit{Mathematica}
\cite{Wolfram-96}).

Furthermore:
\begin{equation} \label{global payoff at step n}
    E[payoff(n)] \; = \; \sum_{k = -n+1}^{n-1} P[payoff(n-1) = k] \,  E[payoff(n) | payoff(n-1)
    ] \; \; \forall n \in \mathbb{N}
\end{equation}

We will prove that $ \lim_{n \rightarrow \infty} E[ payoff(n) ]
\; = \; 0 $ by proving by induction on n that $ E[ payoff(n) ] \;
= \; 0 \; \; \forall n \in {\mathbb{N}} $.

That $ E[payoff(1)] \; = \; 0 $ follows immediately by the fact
that $ S ( \lambda ) \; = \; \uparrow \; \; \forall S $.

We have, conseguentially, simply to prove that $  E[payoff(n-1)]
\; = \; 0 \; \; \Rightarrow \; \; E[payoff(n)] \; = \; 0 \; \;
\forall S $.

This is, anyway, an obvious conseguence of the equations
eq.\ref{eq:conditional payoff at step n} and eq.\ref{global payoff
at step n}
\end{proof}

\bigskip

Theorem\ref{th:weak law of excluded gambling systems} is not,
anyway, a great assurance for Casino's owner:

in fact it doesn't exclude that the gambler, if  enough lucky, may
happen to get a positive payoff for $ n \rightarrow \infty $.

What will definitely assure him is the following:
\begin{theorem} \label{strong law of excluded gambling strategies}
STRONG LAW OF EXCLUDED GAMBLING STRATEGIES

For $ n \rightarrow \infty $ the set of  the \textit{lucky-winning
strategies} tends to the null set
\end{theorem}
And here comes the astonishing fact: Theorem\ref{strong law of
excluded gambling strategies} can't be proved with purely
measure-theoretic concepts.

Our approach will consist in taking von Mises' axiom\ref{ax:axiom
of randomness} as a definition of the set of subsequences   to
which such an axiom applies.

Let us then define the \textbf{set of collectives} $
{\mathcal{C}}ollectives\; \subset \; \Sigma^{\infty} $ as the set
of sequences having not enough intrinsic regularity to allow, if
they occur, a lucky-winning strategy. Clearly such a definition
depends on the class $ {{\mathcal{S}}trategies}_{admissible} (
{\mathcal{C}}ollectives )$ of admissible gambling strategies.

It would appear natural ,at first, to admit every gambling
strategy.

But such a choice would lead immediately to conclude that $
{\mathcal{C}}ollectives \; = \; \emptyset $ since given two
gambling strategies $ S_{0} $ and $ S_{1} $  so that:
\begin{equation}
  EXT[S_{i}] ( \bar{x} ) \; \text{ is made only of i } \; i=0,1 \;
  \forall \bar{x} \in \Sigma^{\infty}
\end{equation}
we would have clearly that:
\begin{equation}
   P_{VM}( a_{i}  \,| \, EXT[S_{1}] (\bar{x}) \, ) \; \neq P_{VM}( a_{i}  \,| \, EXT[S_{2}]
   (\bar{x}) \; \; \forall \bar{x} \in \Sigma^{\infty}
\end{equation}
The history of the attempts of characterizing in a proper way the
class of the admissible gambling strategies is  very long and
curious \cite{Van-Lambalgen-87}, \cite{Li-Vitanyi-97},
\cite{Gillies-00} and involved many people: Church, Copeland,
D\"{orge}, Feller, Kamke, Popper, Reichenbach, Tornier, Waismann
and Wald; I will report here only the conceptually more important
contributions:

in the thirties Abraham Wald showed that:
\begin{theorem} \label{th:Wald theorem}
\end{theorem}
WALD'S THEOREM
\begin{equation}
   ( cardinality (  {{\mathcal{S}}trategies}_{admissible} ( {\mathcal{C}}ollectives
  ) ) \, = \, \aleph_{0} ) \; \Rightarrow \;( {\mathcal{C}}ollectives
  \neq \emptyset )
\end{equation}

In the fourties,  basing on the observation that gambling
strategies must be effectively followed, Alonzo Church  proposed,
according to the Church-Turing's Thesis \cite{Odifreddi-89}, to
consider admissible a gambling strategy if and only if it is a
\textbf{partial recursive function}.

With such an assumption:
\begin{equation} \label{eq:Church admissible gambling strategies}
  {{\mathcal{S}}trategies}_{admissible} ( {\mathcal{C}}ollectives
  )   \; := C_{\Phi}-C_{M}-\Delta_{0}^{0}-\stackrel{ \circ } {MAP}( \Sigma^{\star},\Sigma^{\star}  )
\end{equation}
it can be proved that:
\begin{equation}
  P_{unbiased} (  {\mathcal{C}}ollectives
   )  \; = \; 1
\end{equation}
immediately implying Theorem\ref{strong law of excluded gambling
strategies}.

Let us the introduce the following:
\begin{definition} \label{Church random sequences}
\end{definition}
CHURCH RANDOM SEQUENCES:
\begin{equation}
  CHURCH-RANDOM( \Sigma^{\star} ) \; := \; {\mathcal{C}}ollectives
   \; \text{ with the assumption of eq.\ref{eq:Church admissible gambling
  strategies} }
\end{equation}

\smallskip

\begin{remark}
\end{remark}
MARTINGALES AND THE REASON WHY REAL CASINOS RESULT IN ACTIVE

It is important to observe that Theorem\ref{strong law of excluded
gambling strategies} was proved under the assumption that the
gambler bets always at a fixed odd.

Assuming a more general definition of a gambling strategy in which
the odd betted each time is adjustable in function of a recursive
function of the history up to that bet, it may be easily proved
that winning gambling strategies do exist.

An example is given by \textbf{martingales}:

let us suppose that the gambler plays in the following way:
\begin{itemize}
  \item he insists on  betting
always on \emph{head}, doubling the stake after a loss
  \item he stops to bet for ever after the first win
\end{itemize}
In analyzing such a gambling situation Daniel Bernoulli
introduced the so called \textbf{Saint Petersburg paradox}:

since the gambler bets 1 fiche that heads will turn up on the
first throw, 2 fiches that heads will turn up on the second throw
if it didn't turn up on the first, 4 fiches that heads will turn
up on the third throw if it didn't turn up in the first two
throws and so on, one could conclude the gabler's expected pay off
is infinite:
\begin{equation} \label{eq:Saint Petersburg paradox}
  \frac{1}{2} (1) \, + \, \frac{1}{4}(2) \, + \, \frac{1}{8} (4)
  \, + \cdots \; = \;  \frac{1}{2} \, + \,  \frac{1}{2} \, + \,
  \frac{1}{2} \, + \, \cdots \; = + \infty
\end{equation}
To see clearly where the mistake is, let us proceed by steps.

First of all let us observe  that, since \textbf{ownerhip of all
subsequences} is a Law of Randomness, we have in
particular that \textbf{ownerhip of the subsequence 1} is a Law of
Randomness too.

Hence heads will certainly turn up one day.

Conseguentially it is legitimate to express the expected payoff as
a sum on the first time heads turn up, as it was done in
eq.\ref{eq:Saint Petersburg paradox}:
\begin{equation}
  \lim_{ n \rightarrow \infty } E[payoff(n)] \; = \; \sum_{n=1}^{\infty} P_{unbaised \, n} ( 0^{n-1}
  1 ) gain(n)
\end{equation}
But the gain corresponding to the situation in which heads turn up
for the first time at the $ n^{th} $ throw must take into account
of all the fiches he lost in the previous $ n - 1 $ turns.

So:
\begin{equation}
  gain(n) \; = \; 2^{n-1} \, - \, \sum_{k=1}^{n-1} 2^{k} \; = \;
  2^{n} \, - \, ( 2^{n-1} - 1) \; = \; 1
\end{equation}
Hence:
\begin{equation}
    \lim_{ n \rightarrow \infty } E[payoff(n)] \; = \;
    \sum_{n=1}^{\infty} \frac{1}{2^{n}}  \; = \; 2
\end{equation}

But, if allowing to rule also the stakes, one can violate even
the Weak Law of Excluded Gambling Systems,  why don't Casinos go
all in ruin?

The reason is that a gambling strategy as the displayed
martingale requires an unbounded budget, i.e. that the gambler
cannot go broke.

\newpage
\section{The weakness of Church randomness with respect to Chaitin randomness}
The Law of Excluded  Classical Gambling System could be seen, at a
foundational level, as the corner stone for a mathematical
characterization of the concept of classical algorithmic
randomness.

With the intuitivelly compelling choice of eq.\ref{eq:Church
admissible gambling strategies} for the class of admissible
gammblig strategies, this results in the notion of Church
randomness introduce in the previous section.

It appears then natural to ask ourselves which inter-relation
exists between the resulting notion of Church randomness and the
notion of Martin-L\"{o}f Solovay Chaitin randomness we have
arrived to recognize as the correct notion of classical
algorithmic randomness.

We stressed in the remark\ref{rem:on why Classical Probability
Theory applies to reality} the importance of the fact that $
P_{unbaised} (  CHAITIN-RANDOM( \Sigma^{\infty} ) ) \; = \; 1 $.

So we can appreciate the fact that:
\begin{theorem} \label{th:the occured sequence of  infinite independent tosses of a fair coin is certainly Church random}
\end{theorem}
THE OCCURED SEQUENCE OF INFINITE INDEPENDENT TOSSES OF A FAIR
COIN IS CERTAINLY CHURCH-RANDOM
\begin{equation}
  P_{unbaised} (  CHURCH-RANDOM( \Sigma^{\infty} ) ) \; = \; 1
\end{equation}
\begin{proof}
Given a generic $ S \in \Delta_{0}^{0} - \stackrel{ \circ } {MAP}
(\Sigma^{\star} , \{0,1 \}) $  let us consider the unary predicate
\textbf{failure of the gambling-system S} $ p_{failure \;
gambling-system \; S} \, \in \, {\mathcal{P}} ( \Sigma^{\infty} )
$ defined as:
\begin{multline}
  p_{failure \; gambling-system \; S} ( \bar{x} ) := << \lim_{ n \rightarrow \infty} \frac{ N_{i} (EXT[S] ( \vec{x} ) (n) )}{n}  \; =  \\
  \lim_{ n \rightarrow \infty} \frac{ N_{i}( \vec{x} (n))}{n}  \; \; i \in \Sigma  >>
\end{multline}
The thesis follows immediately by the observation that:
\begin{equation}
  p_{failure \; gambling-system \; S} \; \in \; {\mathcal{P}}_{TYPICAL} (
  \Sigma^{\infty}) \; \; \forall S \in \Delta_{0}^{0} - \stackrel{ \circ } {MAP}
(\Sigma^{\star} , \{0,1 \})
\end{equation}
\end{proof}

\smallskip

Let us , now, observe that:
\begin{theorem} \label{Chaitin randomness implies Church randomness}
\end{theorem}
\begin{equation}
  CHURCH-RANDOM( \Sigma^{\infty} ) \; \supseteq \;   CHAITIN-RANDOM( \Sigma^{\infty} )
\end{equation}
\begin{proof}
The generic predicate $ p_{failure \; gambling-system \; S}$ is
effectively-refutable. Together with theorem\ref{th:the occured
sequence of  infinite independent tosses of a fair coin is
certainly Church random} this implies that:
\begin{equation}
  p_{failure \; gambling-system \; S} \; \in \; {\mathcal{L}}_{RANDOMNESS} [ ( \Sigma^{\infty} \, , \, P_{unbiased} ) ]      \; \; \forall S \in \Delta_{0}^{0}- \stackrel{ \circ } {MAP}
(\Sigma^{\star} , \{0,1 \})
\end{equation}
from which the the thesis follows immediately
\end{proof}

\smallskip

Church randomness is, anyway, weaker than
Martin-L\"{o}f-Solovay-Chaitin randomness as it was proved by J.
Ville in 1939. Demanding to the wonderful Michael Van Lambalgen's
dissertation thesis \cite{Van-Lambalgen-87}  (in particular to the
section2.6 for an hystorical analysis of the decline of von Mises
axiomatization of Classical Probability Theory after the Geneva
conference of 1937, and the collection of objection , both
philosophical and formal, it received by Frechet, to section3.1
for a deep analysis of the philosophical differences between
Church randomness and Martin-L\"{o}f-Solovay-Chaitin randomness
and to the fourth chapter for the more advanced available analysis
of the formal differences between such notions) for further information, let us simply report the statement of Ville's result:
\begin{theorem} \label{th:Ville theorem}
\end{theorem}
VILLE'S THEOREM:
\begin{align}
  \exists \,  & \,  \bar{x} \in CHURCH-RANDOM( \Sigma^{\infty} ) \; : \; p_{\text{infinite recurrence}} ( \bar{x} ) \text{ doesn't hold}  \\
  \exists \,  & \,  \bar{y} \in CHURCH-RANDOM( \Sigma^{\infty} ) \; : \; p_{\text{iterated logarithm}} ( \bar{y} ) \text{ doesn't hold}
\end{align}

\smallskip

Since the \textbf{infinite recurrence property} and the
\textbf{iterated logarithm property} are Laws of Randomness,
Ville Theorem immediately implies that:
\begin{corollary}
\end{corollary} \label{cor:Church randomness is weaker than Chaitin randomness}
\begin{equation}
  CHURCH-RANDOM( \Sigma^{\infty} ) \; \supset \;   CHAITIN-RANDOM( \Sigma^{\infty} )
\end{equation}
\part{The road for  quantum algorithmic randomness} \label{part:The road for  quantum algorithmic randomness}
\chapter{The irreducibility of quantum probability both to classical determinism and to classical
nondeterminism} \label{chap:The irreducibility of quantum
probability both to classical determinism and to classical
nondeterminism}
\section{Why to treat sequences of qubits one has
to give up the Hilbert-Space Axiomatization of Quantum
Mechanics}\label{sec:Why to treat sequences of qubits one has to
give up the Hilbert-Space Axiomatization of Quantum Mechanics}

The problem of giving a mathematical foundation, i.e. a rigorous
mathematical axiomatization, of Quantum Mechanics was first faced
by John Von Neumann through his 1932's masterpiece
\cite{Von-Neumann-83} in which he introduced Hilbert spaces,
codifying the rule they play in Quantum Mechanics.

So he introduced his, nowadays standard, Hilbert space
axiomatization of Quantum Mechanics, where:

\begin{definition} \label{def:Hilbert space axiomatization of Quantum Mechanics}
\end{definition}
HILBERT SPACE AXIOMATIZATION OF QUANTUM MECHANICS:

any axiomatization of Quantum Mechanics assuming the following
two axioms:
\begin{axiom} \label{ax:Hilbert space axiom on states}
\end{axiom}
HILBERT-SPACE'S AXIOM ON STATES:

The \textbf{pure states} of a \textbf{quantum mechanical systems}
are \textbf{rays} in an Hilbert space $ {\mathcal{H}} $

\medskip

\begin{axiom} \label{ax:Hilbert space axiom on observables}
\end{axiom}
HILBERT-SPACE'S AXIOM ON OBSERVABLES:

The \textbf{observables} of a \textbf{quantum mechanical systems}
are \textbf{self-adjoint operators} on $ {\mathcal{H}} $. The
expected value of the observable $ \hat{O} $ in the state $ | \psi
> $ is:
\begin{equation}
  E_{ | \psi > } ( \hat{O} ) \; = \; \frac{ < \psi | \hat{O} | \psi > }{ < \psi | \psi >   }
\end{equation}

\medskip

The success and influence of the book \cite{Von-Neumann-83} was so
great that the point of view therein exposed  became suddenly the
\emph{koin\'{e}} about the foundation of Quantum Mechanics,
taught in all undergraduate courses.

This had the curious effect of throwing a shadow on Von Neumann's
successive intellectual path that led him to doubt not only about
his 1932's Hilbert space axiomatization, but of the same fact
that Quantum Mechanics may be formalized through an an Hilbert
space formalization of some kind.

\medskip

To understand the corner-stone of Von Neumann's post-32 doubts
let us consider the \textbf{Separability Issue}.

Given the many subtilities involved it is useful to recall even
the more elementary notions:

\begin{definition}
\end{definition}
HILBERT SPACE:

a complete inner-product space

\smallskip

Given an Hilbert space $ {\mathcal{H}} $ we shall say that
\cite{Reed-Simon-80}:
\begin{definition}
\end{definition}
$ {\mathcal{H}} $ IS SEPARABLE: it has a finite or countable
orthonormal basis

\medskip

Given two Hilbert spaces $ {\mathcal{H}}_{1} $ and $
{\mathcal{H}}_{2} $ their tensor product, i.e. the Hilbert space $
{\mathcal{H}}_{1} \; \bigotimes \;
  {\mathcal{H}}_{2}$ is defined in the following way \cite{Reed-Simon-80}:
\begin{enumerate}
  \item to any couple $ ( \, | \phi_{1} >   \, , \, | \phi_{2} >
  \, ) $ with $ | \phi_{1} > \in {\mathcal{H}}_{1} $ and   $ | \phi_{2} > \in {\mathcal{H}}_{2}
  $ one can associate the conjugate bilinear form $ | \phi_{1} > \,
  \bigotimes \, | \phi_{2} > $ defined on $ {\mathcal{H}}_{1}
  \times {\mathcal{H}}_{2} $ as:
\begin{equation}
  ( \, | \phi_{1} > \,
  \bigotimes \, | \phi_{2} > ) ( | \psi_{1} > , | \psi_{2} > ) \;
  := \; < \phi_{1} | \psi_{1} > < \phi_{2} | \psi_{2} >  \; \; | \psi_{1} > \in {\mathcal{H}}_{1} \, ,
  \,  | \psi_{2} > \in {\mathcal{H}}_{2}
\end{equation}
   \item one considers the set $ {\mathcal{E}} $ of the finite
   linear combinations of such conjugate linear forms
   \item one defines on $ {\mathcal{E}} $ an inner product $ <
   \cdot | \cdot > $ by defining:
\begin{equation}
  <  \phi_{1} \bigotimes \phi_{2}  |  \phi_{3} \bigotimes \phi_{4}
  > \; := \; < \phi_{1} | \phi_{3} > < \phi_{2} | \phi_{4} >
\end{equation}
and extending by linearity to $ {\mathcal{E}} $
  \item one defines $ {\mathcal{H}}_{1}
  \times {\mathcal{H}}_{2} $ as the completion of $ {\mathcal{E}} $
  under such an inner product
\end{enumerate}

\medskip

Such a definition of the  tensor product of two Hilbert spaces
trivially generalizes to define the tensor product $
{\mathcal{H}}_{1}  \; \bigotimes \; \cdots \; \; \bigotimes \;
{\mathcal{H}}_{n} $ of a finite number  of Hilbert spaces.

In particular one can consider the case in which the Hilbert
spaces $ {\mathcal{H}}_{1} \, , \, \cdots \, , \,
  {\mathcal{H}}_{n} $ are equal:
\begin{equation}
  {\mathcal{H}}_{i} \; = \; {\mathcal{H}} \; \; i \, = \, 1 , \cdots
  , n
\end{equation}
in which the above construction results in the following:
\begin{definition}
\end{definition}
n-FOLD TENSOR PRODUCT OF THE HILBERT SPACE $ {\mathcal{H}} $:
\begin{equation}
  {\mathcal{H}}^{\bigotimes n } \; := \; \bigotimes_{k=1}^{n} {\mathcal{H}}
\end{equation}

\medskip

If, anyway, one tries to generalize such a procedure to
  define the $ \infty $-fold tensor product $ {\mathcal{H}}^{\bigotimes \infty } $
  of an Hilbert space $ {\mathcal{H}} $ one immediately sees that the business doesn't work \cite{Thirring-83}:

 the squared norm of a vector $ | \psi > \;  :=  \; | \psi_{1} > \, \bigotimes
 \cdots \, \bigotimes \, | \psi_{n} > \; \in \;  {\mathcal{H}}^{\bigotimes n } $
 is given by:
\begin{equation} \label{eq:squared norm in the n-fold tensor product}
  \| \psi \|^{2} \; = \; < \psi | \psi > \; = \; \prod_{k=1}^{n} \, < \psi_{k} | \psi_{k} >
\end{equation}
Now if $ n \, = \, \infty $ the productory can, in general,
diverge so one has to restrict only to those vectors for which the
 r.h.s. of eq.\ref{eq:squared norm in the n-fold tensor
product} converges.

Furthermore, on the remaining vectors, the productory can converge
to zero  even in those particular cases in which $ < \psi_{k} |
\psi_{k} > \,
> \, 0 \; \; \forall k \in {\mathbb{N}} $.

In order to take the quotient space with respect to the zero
vectors it is then necessary to form the equivalences classes not
only of vectors with some factor zero, but also containing the
vectors for which $ \; \prod_{k=1}^{n} \, < \psi_{k} | \psi_{k}
> $ converges to zero.

On such a quotient space the eq.\ref{eq:squared norm in the n-fold
tensor product} defines a separating norm that can be used to
complete it resulting in the required Hilbert space  $
{\mathcal{H}}^{\bigotimes \infty } $, with the linear structure
defined in the usual way.

This does not yet, however, suffice to define the scalar product
of different vector $ | \psi > $ and $  | \phi > $. Though only
vectors such that:
\begin{equation}
 < \psi_{k} | \psi_{k} > \, =  \, < \phi_{k} | \phi_{k} > \, = \; 1  \; \; \forall k \in {\mathbb{N}}
\end{equation}
need to be considered, there are still two possibilities, namely:
\begin{itemize}
  \item \textbf{CASE-I}:
\begin{equation}
  \prod_{k=1}^{\infty} | < \psi_{k} | \phi_{k} > | \; \rightarrow
  \; c \,  > \, 0
\end{equation}
  \item \textbf{CASE-II}:
\begin{equation}
  \prod_{k=1}^{\infty} | < \psi_{k} | \phi_{k} > | \; \rightarrow
  \; 0
\end{equation}
\end{itemize}
where $ \rightarrow $ means unconditional convergence.

In case-2 $ \prod_{k=1}^{\infty}  < \psi_{k} | \phi_{k} >  \;
\rightarrow  \; 0 $ as well, and the vectors may be considered
orthogonal.

In case-II, on the other hand, there is no guarantee that  $
\prod_{k=1}^{\infty}  < \psi_{k} | \phi_{k} > $ converges. If $ |
< \psi_{k} | \phi_{k} > | \: = \: \exp ( i \theta_{k} )  | <
\psi_{k} | \phi_{k} > | $, then their product is said to converge
if not only $ \prod_{k=1}^{\infty} | < \psi_{k} | \phi_{k} > | $
but also $ \sum_{k} | \phi_{k} | $ converges.

One now encounters the convention that vectors may be deemed
orthogonal whenever $ \sum_{k} | \phi_{k} | \: \rightarrow \:
\infty $ (we will indicate this situation as the
\textbf{case-I.B})

Let us then agree on the following definition of the inner
product:

\begin{definition} \label{def:non zero inner product in the infinite tensor product Hilbert space}
\end{definition}
$ < \psi | \phi >  \: =  \: c \, \neq \, 0$ (\textbf{case-I.A})
\begin{equation}
  \lim_{n \rightarrow \infty} < \psi_{n} | \phi_{n} >  \, = \, c
\end{equation}
\begin{definition}  \label{def:zero inner product in the infinite tensor product Hilbert space}
\end{definition}
$ < \psi | \phi >  \: =  \: 0 $ (\textbf{case-II} or
\textbf{case-I.B}))
\begin{equation}
  \lim_{n \rightarrow \infty} < \psi_{n} | \phi_{n} >  \, = \, 0
\end{equation}

\bigskip

Let us now observe that \textbf{separability} is a rather robust
property, i.e. a property that preserves under many operations.

Given an Hilbert space  ${\mathcal{H}}$:
\begin{theorem} \label{th:separability preservation under finite tensor product}
\end{theorem}
PRESERVATION OF SEPARABILITY UNDER FINITE-FOLD TENSOR PRODUCT
\begin{equation}
 {\mathcal{H}} \; \text{ is separable } \; \; \Rightarrow \; \;  ( {\mathcal{H}}^{ \bigotimes
 n} \; \text{ is separable }  \; \; \forall n \in {\mathbb{N}} )
\end{equation}

\medskip

Given a sequence of Hilbert spaces $ \{ {\mathcal{H}}_{n} \}_{ n
\in { \mathbb{N}}} $ we have furthermore the following:
\begin{theorem} \label{th:separability preservation under infinite direct sum}
\end{theorem}
PRESERVATION OF SEPARABILITY UNDER INFINITE DIRECT SUM:
\begin{equation}
 ( {\mathcal{H}}_{n} \; \text{ is separable }  \; \forall n \in
 {\mathbb{N}}) \; \; \Rightarrow \; \; \bigoplus_{ n
\in { \mathbb{N}}} \, {\mathcal{H}}_{n} \; \text{ is separable }
\end{equation}

These theorems are sufficient to guarantee the separability of
almost all the Hilbert spaces appearing in Theoretical Physics:

For the theorem\ref{th:separability preservation under finite
tensor product} this is certainly the case when, in
Nonrelativistic Quantum Mechanics, one considers a finite number
of particles of spin s: since the Hilbert space for one particle
is $ {\mathcal{H}} := L^{2} ( {\mathbb{R}}^{3} \, d \vec{x} ) \,
\bigotimes  \, {\mathbb{C}}^{2 s + 1} $, the n - particle Hilbert
space is $ S_{n} {\mathcal{H}}^{\bigotimes n} $ if s is integer
(i.e. if the particles are \textbf{bosons} ) and $ A_{n}
{\mathcal{H}}^{\bigotimes n} $ if s is half-integer (i.e. if the
particles are \textbf{fermions} ) where $ S_{n} $ and $ A_{n} $
are, respectively, the \textbf{n-simmetrization},
\textbf{n-antisimmetrization} operators.

The underlying Hilbert space continues to remain separable even
allowing an infinite number of particles as follows immediately
introducing the following Hilbert spaces:
\begin{definition}
\end{definition}
FOCK SPACE ASSOCIATED TO $ {\mathcal{H}} $:
\begin{equation}
 {\mathcal{H}} ^{ \bigotimes \star } \; :=  {\mathcal{F}} ( {\mathcal{H}} ) \; := \; \bigoplus_{ n
\in { \mathbb{N}}} {\mathcal{H}}^{\bigotimes n}
\end{equation}
\begin{definition}
\end{definition}
BOSONIC FOCK SPACE ASSOCIATED TO $ {\mathcal{H}} $:
\begin{equation}
  {\mathcal{F}}_{S} ( {\mathcal{H}} ) \; := \; \bigoplus_{ n
\in { \mathbb{N}}} S_{n} {\mathcal{H}}^{\bigotimes n}
\end{equation}
\begin{definition}
\end{definition}
FERMIONIC FOCK SPACE ASSOCIATED TO $ {\mathcal{H}} $:
\begin{equation}
  {\mathcal{F}}_{A} ( {\mathcal{H}} ) \; := \; \bigoplus_{ n
\in { \mathbb{N}}} A_{n} {\mathcal{H}}^{\bigotimes n}
\end{equation}
and observing that, for $ {\mathcal{H}} := L^{2} (
{\mathbb{R}}^{3} \, d \vec{x} ) \, \bigotimes  \, {\mathbb{C}}^{2
s + 1} $,  they are separable owing to
theorem\ref{th:separability preservation under finite tensor
product} and theorem\ref{th:separability preservation under
infinite direct sum}.

Let us now pass to Relativistic Quantum Mechanics, i.e. to
Quantum Field Theory:

for a free-field theory the separability of the proper Fock spaces
follows again immediately from theorem\ref{th:separability
preservation under finite tensor product} and
theorem\ref{th:separability preservation under infinite direct
sum}.

For interacting field theories the situation is more complicated
owing to the fact that a  general mathematically-rigorous
formalization of Quantum Field Theory, beside its exceptional
developments
\cite{Deligne-Etingof-Freed-Jeffrey-Kazhdan-Morgan-Morrison-Witten-99a},\cite{Deligne-Etingof-Freed-Jeffrey-Kazhdan-Morgan-Morrison-Witten-99b}
and all the work of the Constructivists \cite{Glimm-Jaffe-87},
\cite{Jaffe-00}, is unfortunately still lacking \cite{Witten-95}.

One could simply assert that Wightman Axioms constraint the
underlying Hilbert space to be separable \cite{Reed-Simon-75} but
such an answer would sound as a rather dogmatical one.

A more convincing argument consists in considering that in the
Lehmann-Symanzik-Zimmerman formalism  the involved Hilbert spaces
are only the asympotic In and Out Fock spaces \cite{Strocchi-93}.

\medskip
Unfortunately the robustness of \textbf{separability} is not
complete.

In fact:
\begin{theorem} \label{th:separability not preservation under infinite tensor product}
\end{theorem}
NOT PRESERVATION OF SEPARABILITY UNDER INFINITE-FOLD TENSOR
PRODUCT

$ {\mathcal{H}}^{ \bigotimes
 \infty} $ \textbf{is not separable even if} $ {\mathcal{H}} $ \textbf{is separable}

\begin{example} \label{ex:the Hilbert space of Quantum Information Theory}
\end{example}
THE HILBERT SPACES OF QUANTUM INFORMATION THEORY

How much classical information is contained in a  state:
\begin{equation}
  | \psi > \; := \; \alpha | + > \, + \, \beta | - >  \; \; \alpha
  , \beta \in {\mathbb{C}} \, : \, | \alpha |^{2} + | \beta |^{2}
  = 1
\end{equation}
of a $ spin \frac{1}{2} $ system ?

Since the bidimensional complex projective space has the
continuum power:
\begin{equation}
  cardinality( {\mathbb{C}} P^{2} ) \; = \;\aleph_{1}
\end{equation}
the specification of a point P on it requires the assignation of a
whole sequence $ \bar{x}_{P} \in \Sigma^{\infty} $.

In this way one is led to to argue that:
\begin{equation} \label{eq:conclusion that one qubit is equal to infinite cbits}
  information( | \psi > ) \; = \;  \infty \, bits
\end{equation}

But, given a $ spin \frac{1}{2} $ system prepared in the state  $
| \psi > $, let us now suppose to make a measurement of the
operator $ \hat{S}_{z} $. The information gained by the knowledge
of the experimental outcome is only of one bit.

So, from this reasoning, one is led to argue that:
\begin{equation} \label{eq:conclusion that one qubit is equal to one cbits}
  information( | \psi > ) \; = \; 1 \, bit
\end{equation}

Obviously eq.\ref{eq:conclusion that one qubit is equal to
infinite cbits} and eq.\ref{eq:conclusion that one qubit is equal
to one cbits} are incompatible.

This simple reasoning shows that the quantification of the
informational content of the state  $ | \psi > $ must be given in
terms of a measure's unity not commensurable with that of
\textbf{classical information}.

This is a a conceptually extremelly deep concept: there doesn't
exist a unique, mathematically charaterizable, notion of
information, resulting in a measure's unity , the \textbf{bit},
in terms of which one can analyze the informational content of
both classical and quantum physical systems:

\textbf{quantum information} is not commensurable with
\textbf{classical information}.

Hence one has to give up the universal notion of \textbf{bit},
replacing it with the following couple of notions:
\begin{itemize}
  \item the \textbf{cbit}, i.e. the measure's unity of \textbf{classical information}
  \item the \textbf{qubit}, i.e. the measure's unity of \textbf{quantum information}
\end{itemize}

The quantum-informational amount of the state $ | \psi > $ gives
the operational definition of the \textbf{qubit}.

A more formal definition will be given, anyway, in the
remark\ref{rem:the noncommutative combinatory information and the
definition of the qubit} in terms of the notion of
\textbf{combinatorial quantum information}.

As we will see in section\ref{sec:From the communicational-compression of the Quantum Coding Theorems to the algorithmic-compression in Quantum Computation} the incommensurability
of \textbf{classical information} and \textbf{quantum information} is deeply linked with the No-Cloning Theorem

\smallskip

\begin{definition}
\end{definition}
ONE QUBIT HILBERT SPACE:
\begin{equation}
  {\mathcal{H}}_{2} \; := \; {\mathbb{C}}^{2}
\end{equation}
Given an $ n \in {\mathbb{N}} $:
\begin{definition} \label{def:space of the quantum strings of n qubits}
\end{definition}
n QUBITS HILBERT SPACE:
\begin{equation}
  {\mathcal{H}}_{2}^{\bigotimes n} \; := \; {\mathbb{C}}^{2^{n}}
\end{equation}
\begin{definition} \label{def:space of the quantum strings of qubits}
\end{definition}
HILBERT SPACE OF QUBITS' STRINGS:
\begin{equation}
  {\mathcal{H}}_{2}^{\bigotimes \star} \; := \;  {\mathcal{F}} ( {\mathcal{H}}_{2})
\end{equation}

\medskip

On all these \textbf{separable} Hilbert spaces it is useful to
introduce orthonormal complete bases, said the
\textbf{computational basis} that embeds the strings of cbits in
the quantum domain:
\begin{definition} \label{def:computational basis}
\end{definition}
COMPUTATIONAL BASIS OF $ {\mathcal{H}}_{2} $:
\begin{equation}
    {\mathbb{E}}_{2} \; := \; \{ | 0 > , | 1 >  \} \; : \; | 0 > \, :=
    \,
    \, \begin{pmatrix}
      1 \\
      0 \
    \end{pmatrix} \; , \; | 1 > \, := \, \begin{pmatrix}
      0 \\
      1 \
    \end{pmatrix}
\end{equation}

\smallskip

\begin{remark} \label{rem:on the qubit operator}
\end{remark}
ON THE QUBIT OPERATOR

The adoption of the \textbf{computational language} requires some
caution.

As to definition\ref{def:computational basis}, it is only a
renaming of the usual language of spin 1/2 system:
\begin{align}
  | 0 > &  \; := \; | \uparrow_{z} > \\
  | 1 > &  \; := \; | \downarrow_{z} > \\
\end{align}
The correspondence clearly continues considering the projectors:
\begin{equation}
  | 0 > < 0 | \; = \; \begin{pmatrix}
    1 & 0 \
  \end{pmatrix} \: \begin{pmatrix}
    1 \\
    0 \
  \end{pmatrix} \; = \; \begin{pmatrix}
    1 & 0 \\
    0 & 0 \
  \end{pmatrix} \; = \; | \uparrow_{z} > < \uparrow_{z} |
\end{equation}
\begin{equation}
  | 1 > < 1 | \; = \; \begin{pmatrix}
    0 & 1 \
  \end{pmatrix} \: \begin{pmatrix}
    0 \\
    1 \
  \end{pmatrix} \; = \; \begin{pmatrix}
    0 & 0 \\
    0 & 1 \
  \end{pmatrix} \; = \; | \downarrow_{z} > < \downarrow_{z} |
\end{equation}
The problem arises if one tries to introduce a \textbf{qubit
operator} having the binary alphabet $ \Sigma := \{ 0 ,1 \} $ as
eigenvalues;

in fact,owing to the vanishing of the addendum concerning the
zero eigenvalue, one has obviously that:
\begin{equation} \label{eq:wrong definition of the qubit operator}
  \hat{q} \; := \; 0 \, | 0 > < 0 | \: + \:  1 \, | 1 > < 1 | \: =
  \: | 1 > < 1 | \: = \; | \downarrow_{z} > < \downarrow_{z} |
\end{equation}
So, in order of introducing a qubit operator, one has to avoid
the zero eigenvalue, e.g. assuming the  spectrum  of the
\textbf{qubit operator} to be the binary alphabet $ \{ +1 , - 1
\} $, with the convention that the eigenvalue $ + 1 $ corresponds
to zero and the eigenvalue $ - 1 $ corresponds to one.

With these conventions one has that:
\begin{equation} \label{eq:right definition of the qubit operator}
  \hat{q} \; := \; + 1 \, | 0 > < 0 | \: + \:  ( - 1  ) \, | 1 > < 1 | \: =
  \: \hat{\sigma}_{z} \: = \: \begin{pmatrix}
    1 & 0 \\
    0 & -1  \
  \end{pmatrix}
\end{equation}

\smallskip

Given any positive integer number $ n \geq 3 $:
\begin{definition}
\end{definition}
COMPUTATIONAL BASIS OF $ {\mathcal{H}}_{2}^{\bigotimes n} $:
\begin{equation}
  {\mathbb{E}}_{n} \; := \; \{ \:  | \vec{x} > \, , \,
  \vec{x} \in \Sigma^{n} \: \}
\end{equation}
\begin{definition}
\end{definition}
COMPUTATIONAL BASIS OF $ {\mathcal{H}}_{2}^{\bigotimes \star} $:
\begin{equation}
  {\mathbb{E}}_{\star} \; := \; \{ | \vec{x} > \, , \,
  \vec{x} \in \Sigma^{\star} \: \}
\end{equation}
The generic \textbf{string of qubits}, i.e. the generic vector of
$ {\mathbb{H}}_{2}^{\bigotimes \star} $ is then given by a  linear
combination of the form $ \sum_{ \vec{x} \in \Sigma^{\star}}
c_{\vec{x}} | \vec{x} > $.

And what about \textbf{sequences of qubits}?

We can indeed introduce the following notions:
\begin{definition} \label{def:space of the quantum sequences of qubits}
\end{definition}
HILBERT SPACE OF QUBITS' SEQUENCES:
\begin{equation}
  {\mathcal{H}}_{2}^{\bigotimes \infty} \; := \; \bigotimes_{n \in
  {\mathbb{N}}} {\mathcal{H}}_{2}
\end{equation}

By theorem\ref{th:separability not preservation under infinite
tensor product} $ {\mathcal{H}}_{2}^{\bigotimes \infty} $ is not
separable.

\bigskip

The \textbf{Separability Issue} consists in the following
question:

\smallskip

\begin{center}
\textbf{is it necessary to modify the axiom\ref{ax:Hilbert space
axiom on states} adding the constraint that the Hilbert space $
{\mathcal{H}} $ is separable?}
\end{center}

\smallskip
The thesis that the correct answer is affermative has been
asserted by authoritative voices; for example Walter Thirring
remembers that \cite{Thirring-01}:
\begin{center}
  \textit{"For finite tensor products the dimension of the spaces is
  multiplicative and for infinite tensor product is is uncountable
  even if the individual spaces have only dimension $ \, = \, 2$.
  This casts some doubt on whether there is a mathematically
  valid description of infinite quantum systems. Schr\"{o}dinger
  once told me that the corresponding non-separable Hilbert
  space did not make sense to him.
  To determine N components of his $ \psi $-function one needs N
  experiments and in a non-separable space one would need an
  uncountable number of measurements and this is nonsense.
  However such an opinion means that Schr\"{o}dinger did not get the main message of Von Neumann's celebrated
  paper on infinite tensor products ."}
\end{center}
Keeping aside for a moment Thirring's last remark, let us observe
that there exist also  many other arguments supporting a positive
answer to the \textbf{Separability Issue}: for example on an
not-separable Hilbert space the Gram-Schmidt orthogonalization
process can be adopted only appealing to the Axiom of Choice
\cite{Reed-Simon-75}.

\smallskip

But let us now analyze Thirring's last remark: is Thirring right
in claiming that Von Neumann's celebrated paper on infinite tensor
products lead to give a negative answer to the Separability Issue?

Our point of view, though  predictively completelly equivalent  to Thirring's
authoritative one, is philosophically different and lead, as to
the Separability Issue, to the opposite answer, as we will arrive
to discuss at the end of this section.

To  show why, anyway, it may be useful to follow the
reconstruction of Von Neumann's intellectual path on the
Foundations of Quantum Physics made by Miklos Redei in
\cite{Redei-98}, emerging from \label{Petz-Redei-95} and
condensated in \cite{Redei-01}:
\begin{center}
     Hilbert spaces  $ \rightarrow $ orthocomplemented modular lattices  $ \rightarrow
       W^{\star}$-algebras
\end{center}
that, as we will show, correspond conceptually to the path:
\begin{center}
     Quantum Mechanics  $ \rightarrow $ Quantum Logic  $ \rightarrow
     $ Quantum Probability
\end{center}

This requires , anyway the introduction of a whole abstract
algebraic machinery.

\begin{definition} \label{def:partially ordered set}
\end{definition}
PARTIALLY ORDERED SET (POSET)

a  couple $ ( {\mathcal{L}} \, , \, \preceq ) $ such that $
{\mathcal{L}} $ is a set, while $ \leq $ is a partial ordering on
$ {\mathcal{L}} $, i.e. a reflexive, transitive, antisimmetric
relation on $ {\mathcal{L}} $

\smallskip

Given a poset $ ( {\mathcal{L}} \, , \, \preceq ) $ and  two
element $ a , b \in {\mathcal{L}} $:
\begin{definition}
\end{definition}
\begin{equation}
  a \, \prec \, b \; := \; ( a \preceq b) \, and \, a \neq b
\end{equation}
\begin{definition}
\end{definition}
\begin{equation}
  a \, \succ \, b \; := \; ( a \succeq b) \, and \, a \neq b
\end{equation}

Given a set $ S \subseteq {\mathcal{L}} $:
\begin{definition} \label{def:upper bound}
\end{definition}
a IS AN UPPER BOUND OF S IN THE POSET $ ( {\mathcal{L}} \, , \,
\preceq ) $:
\begin{equation}
  b \; \preceq a \; \; \forall b \in S
\end{equation}
\begin{definition} \label{def:lower bound}
\end{definition}
a IS A LOWER BOUND OF S IN THE POSET $ ( {\mathcal{L}} \, , \,
\preceq) $:
\begin{equation}
  b \; \succeq a \; \; \forall b \in S
\end{equation}
\begin{definition} \label{def:least upper bound}
\end{definition}
a IS THE LEAST UPPER BOUND OF S IN THE POSET $ ( {\mathcal{L}} \,
, \, \preceq ) $:
\begin{equation}
  b \; \preceq a \; \; \forall b \text{ upper bound of S in } ( {\mathcal{L}} \, , \,
\leq )
\end{equation}
\begin{definition} \label{def:least lower bound}
\end{definition}
a IS THE LEAST LOWER BOUND OF S IN THE POSET $ ( {\mathcal{L}} \,
, \, \preceq ) $:
\begin{equation}
  b \; \succeq  \; a \; \; \forall b \text{ lower bound of S in } ( {\mathcal{L}} \, , \,
\leq )
\end{equation}
\begin{definition} \label{def:lattice}
\end{definition}
LATTICE:

a poset $ ( {\mathcal{L}} \, , \, \preceq ) $ such that:
\begin{align}
  \forall \,  & a , b  \in {\mathcal{L}} \; , \; \exists a \bigvee b \, := \, \text{ least upper bound of $ \{ a , b \} $ in } {\mathcal{L}}  \\
  \forall \,  & a , b  \in {\mathcal{L}} \; , \; \exists a \bigwedge b \, := \, \text{ greatest lower bound of $ \{ a , b \} $ in }
  {\mathcal{L}} \\
  \exists \, & \, 0_{{\mathcal{L}}}  \in {\mathcal{L}} \; : \; 0_{{\mathcal{L}}} \, \preceq a \; \; \forall a \in {\mathcal{L}} \\
  \exists \, & \, 1_{{\mathcal{L}}} \in {\mathcal{L}} \; : \; a \, \preceq 1_{{\mathcal{L}}} \; \; \forall a \in {\mathcal{L}}
\end{align}

\smallskip

Given a lattice $ ( {\mathcal{L}} \, , \, \preceq ) $:
\begin{definition} \label{def:atom in a lattice}
\end{definition}
$ a \in {\mathcal{L}} $ IS AN ATOM OF  $ ( {\mathcal{L}} \, , \,
\preceq ) $:
\begin{equation}
  b \, \preceq \, a \; \Rightarrow \; b = a \, or \, b = 0_{{\mathcal{L}}}
\end{equation}
\begin{definition} \label{def:atomic lattice}
\end{definition}
$ ( {\mathcal{L}} \, , \, \preceq ) $ IS ATOMIC:
\begin{equation}
  \forall b \in {\mathcal{L}} \; \exists b \in {\mathcal{L}} \;
  atom \; : \; a \, \preceq \, b
\end{equation}
\begin{definition} \label{def:logical dimension function on a lattice}
\end{definition}
LOGICAL DIMENSION FUNCTION ON $ ( {\mathcal{L}} \, , \, \preceq )
$:

a function $ d \, : \, {\mathcal{L}} \; \mapsto \; [ 0 , + \infty
] $ such that:
\begin{align}
  a  \, & \preceq \, b  \; \Rightarrow \; d(a) \, \leq \, d(b) \; \; \forall a , b \in {\mathcal{L}}  \\
  d(a) \, & \, + \, d(b) \; = \; d ( a \bigwedge b ) \,+ \, d ( a \bigvee b ) \; \; \forall a , b \in {\mathcal{L}}
\end{align}
\begin{definition} \label{def:distributive lattice}
\end{definition}
$ ( {\mathcal{L}} \, , \, \preceq ) $ IS DISTRIBUTIVE:
\begin{equation}
  a \bigvee ( b \bigwedge c ) \; = \; ( a \bigvee b ) \, \bigwedge
  \, ( a \bigvee c ) \; \; \forall a , b , c \in {\mathcal{L}}
\end{equation}
\begin{definition} \label{def:modular lattice}
\end{definition}
$ ( {\mathcal{L}} \, , \, \preceq ) $ IS MODULAR:
\begin{equation}
  a \preceq b \; \Rightarrow \; a \bigvee ( b \bigwedge c ) \; = \; ( a \bigvee b ) \, \bigwedge
  \, ( a \bigvee c ) \; \; \forall a , b , c \in {\mathcal{L}}
\end{equation}
\begin{definition} \label{def:orthocomplementation on a lattice}
\end{definition}
ORTHOCOMPLEMENTATION ON $ ( {\mathcal{L}} \, , \, \preceq ) $:

a map $ \bot : {\mathcal{L}} \mapsto {\mathcal{L}} $ such that:
\begin{align}
  ( a & ^  {\bot})^{ \bot } \; = \; a \; \; \forall a \in {\mathcal{L}}  \\
  a \,  & \preceq \, b  \; \Rightarrow \; b^{\bot} \, \preceq \,
  a^{\bot} \; \; \forall a , b \in {\mathcal{L}}  \\
  a \, & \bigwedge \, a^{\bot} \; = \; 0_{{\mathcal{L}}} \; \; \forall a \in {\mathcal{L}}  \\
  a \, & \bigvee \, a^{\bot} \; = \; 1_{{\mathcal{L}}} \; \; \forall a \in {\mathcal{L}}
\end{align}

\smallskip

\begin{definition} \label{def:orthocomplemented lattice}
\end{definition}
ORTHOCOMPLEMENTED LATTICE:

a therne $ ( ( {\mathcal{L}} \, , \, \preceq \, , \, \bot ) $
such that $ ( {\mathcal{L}} \, , \, \preceq ) $ is a lattice
while $ \bot $ is an orthocomplementation on $ ( {\mathcal{L}} \,
, \, \preceq ) $

\smallskip

Given an orthocomplemented lattice $ ( ( {\mathcal{L}} \, , \,
\preceq \, , \, \bot ) $ an two its elements $ a,b \in
{\mathcal{L}} $
\begin{definition} \label{def:orthogonality in an orthocomplemented lattice}
\end{definition}
a IS ORTHOGONAL TO B:
\begin{equation}
  a \, \bot \, b \; := \; a \, \preceq b^{\bot}
\end{equation}
Clearly, by the definition\ref{def:orthocomplementation on a
lattice}, one has that orthogonality is a simmetric relation:
\begin{equation}
  a \, \bot \, b \; \Leftrightarrow \; b \bot a \; \; \forall a,b \in
{\mathcal{L}}
\end{equation}
\begin{definition} \label{def:orthomodular lattice}
\end{definition}
$ ( ( {\mathcal{L}} \, , \, \preceq \, , \, \bot ) $ IS
ORTHOMODULAR:
\begin{equation}
  a \preceq b \; and \; a \bot c \; \Rightarrow \; a \bigvee ( b \bigwedge c ) \; = \; ( a \bigvee b ) \, \bigwedge
  \, ( a \bigvee c ) \; \; \forall a , b , c \in {\mathcal{L}}
\end{equation}
Orthomodularity is a weakening of modularity that is a weakening
of distributivity as is stated by the following:
\begin{theorem}
\end{theorem}
\begin{align}
  distributivity &  \; \Rightarrow \; modularity \; \Rightarrow  \; orthomodularity \\
  orthomodularity &  \; \nRightarrow \; modularity \; \nRightarrow
  \; distributivity
\end{align}

\smallskip

We will soon see the utility of the following:
\begin{theorem} \label{th:on the finite dimension}
\end{theorem}
THEOREM ON THE FINITE DIMENSION:

\begin{hypothesis}
\end{hypothesis}
\begin{equation*}
   ({\mathcal{L}} \, , \, \preceq )  \; \; lattice
\end{equation*}
\begin{equation*}
  \exists \, d \, \text{ logical dimension function } \; : \infty \notin Range(d)
\end{equation*}
\begin{thesis}
\end{thesis}
\begin{center}
  $ {\mathcal{L}} $ is modular
\end{center}

\smallskip

In a fundamental 1936's paper with G. Birkhoff
\cite{Birkhoff-Von-Neumann-95}, Von Neumann suggested the idea,
yet implicitely advanced in the fifth section of the third
chapter of \cite{Von-Neumann-83}, that the difference between
Quantum Mechanics and Classical Mechanics could be ascribed to
the fact the algebraic structure of the set of all the
propositions concerning a quantum system violates the laws of
Classical Logic, obeying a new kind of logic.

This was the seed of the 65 year old research-field of Quantum
Logic.

It must be remarked that the original Birkhoff and Von Neumann's
definition of a quantum logic was more restrictive than that
choram-populi later assumed in such a research field; to
distinguish the two notion we will speak, respectively, of weak
and strong quantum logics.

Anyway, exactly as the Theoretical Physicist's community
foxilized on the 1932's snapshot of Von Neumann's intellectual
path, the same happened to the Quantum Logicist's community that
foxilized on the 1936's snapshot (mostly altering it), so don't
catching all the reasons led Von Neumann to make the
phase-transition:
\begin{center}
   Quantum Logic  $ \rightarrow $ Quantum Probability
\end{center}
that, as we will briefly point in section\ref{sec:On the rule
Noncommutative Measure Theory and Noncommutative Geometry play in
Quantum Physics}, can be seen as the starting point of the of the
open intellectual challenge summarized by the path:

\begin{center}
   Quantum Mechanics as Nondistributive Logic  $ \rightarrow $
   Quantum Mechanics as Noncommutative Probability  $ \rightarrow $
   Quantum Mechanics as Noncommutative Geometry
\end{center}

\smallskip

The difference between Classical Logic and Quantum Logic is
enclosed in the different algebraic structure that the set of all
the propositions concerning a physical system obey as far as the
\textbf{conjunction} $ \bigwedge $, \textbf{disjunction} $
\bigvee $ and \textbf{negation} $ ' $ are concerned:
\begin{definition} \label{def:classical logic}
\end{definition}
CLASSICAL LOGIC:

a \textbf{distributive}, \textbf{orthocomplemented} lattice

\smallskip

\begin{definition} \label{def:strong quantum logic}
\end{definition}
STRONG QUANTUM LOGIC:

a \textbf{modular}, \textbf{orthocomplemented} lattice

\medskip

\begin{definition} \label{def:weak quantum logic}
\end{definition}
WEAK QUANTUM LOGIC:

an \textbf{orthomodular}, \textbf{orthocomplemented} lattice
\begin{example} \label{ex:the classical logic of power-sets}
\end{example}
THE CLASSICAL LOGIC OF POWER-SETS

Given an arbitrary set S let us introduce on its power-set $ 2^{S}
$ the partial-ordering relation:
\begin{equation}
    a \,\preceq \, b \; := \; a \, \subseteq \, b \; \; a,b \in 2^{S}
\end{equation}
It may be easily verified that $  ( 2^{S} , \preceq ) $ is a
lattice, with:
\begin{align}
  a  \, & \bigwedge \, b \; := \; a \, \bigcap \, b \; \; \forall a , b \in 2^{S}  \\
  a  \, & \bigvee \, b \; := \; a \, \bigcup \, b \; \; \forall a , b \in 2^{S}  \\
  0_{2^{S}} & \; = \; \emptyset \\
  1_{2^{S}} & \; = \; 2^{S}
\end{align}
Introduced on  $  ( 2^{S} , \preceq ) $ the orthocomplementation
map $ \bot \, : \, 2^{S} \mapsto 2^{S} $:
\begin{equation}
  a^{\bot} \; := \; S \, - \, a \; \; a \in 2^{S}
\end{equation}
$ ( 2^{S} \, , \, \preceq \, , \, \bot ) $ is a classical logic.

If $ cardinality(S) \, \in \,  {\mathbb{N}} $ the map  $ d :
2^{S} \mapsto {\mathbb{R}}_{+} $ defined as:
\begin{equation}
   d(S) \; := \; cardinality(S)
\end{equation}
is a logical dimension function.

This is not the case, anyway, if $ cardinality(S) \, \geq \,
\aleph_{0} $, even generalizing definition\ref{def:logical
dimension function on a lattice} in order of allowing infinite
cardinal values of a logical dimension function, as can be seen,
for example, observing that:
\begin{align}
  {\mathbb{Z}} \, & \, \preceq \, {\mathbb{Q}} \; but \; cardinality({\mathbb{Z}}) \, = \, cardinality({\mathbb{Q}})  \, = \, \aleph_{0} \\
  {\mathbb{R} } & - {\mathbb{Q}}  \, \preceq \, {\mathbb{R}} \; but \; cardinality({\mathbb{R} }  - {\mathbb{Q}}) \, = \, cardinality({\mathbb{R}})  \, = \, \aleph_{1}
\end{align}

\medskip

\begin{example} \label{ex:the strong quantum logic of the n-qubits Hilbert space}
\end{example}
THE  STRONG QUANTUM LOGIC OF THE n-QUBITS HILBERT SPACE

Given the n-qubits Hilbert space $ {\mathcal{H}}_{2}^{ \bigotimes
n} $ let us consider its \textbf{projective geometry}, i.e. the
set $ {\mathcal{L}} ( {\mathcal{H}}_{2}^{ \bigotimes n} ) $ of
all its linear subspaces:
\begin{equation}
   {\mathcal{L}} ( {\mathcal{H}}_{2}^{ \bigotimes n} ) \; := \;
   \bigcup_{k=0}^{2^{n}} G_{k , 2^{n} } ({\mathbb{C}})
\end{equation}
Introduced on $  {\mathcal{L}} ( {\mathcal{H}}_{2}^{ \bigotimes
n} ) $ the partial ordering relation:
\begin{equation}
  a \, \preceq \, b \; := \;  a \, \subseteq \, b \; \; a , b \in {\mathcal{L}} ( {\mathcal{H}}_{2}^{ \bigotimes n} )
\end{equation}
the poset $ ( {\mathcal{L}} \, , \, \preceq ) $ is an atomic
lattice, with:
\begin{align}
   a \, &  \bigwedge \, b \; := \;  a \, \bigcap \, b \; \; a , b \in {\mathcal{L}} ( {\mathcal{H}}_{2}^{ \bigotimes n} )  \\
   a \, &  \bigvee \, b \; := \;  a \, \bigoplus \, b \; \; a , b \in {\mathcal{L}} ( {\mathcal{H}}_{2}^{ \bigotimes n} )  \\
\end{align}
whose atoms are the one-dimensional subspaces, namely the elements
of $ G_{1 , 2^{n} } ({\mathbb{C}}) \, = \, {\mathbb{C}}P^{ 2^{n}
-1 }$, said the \textbf{points} of the \textbf{projective
geometry}, i.e. the \textbf{rays} of $ {\mathcal{H}}_{2}^{
\bigotimes n} $.

It may be easily verified that the following map:
\begin{equation}
  d(a) \; := \; dim(a) \; \; a \in {\mathcal{L}} ( {\mathcal{H}}_{2}^{ \bigotimes n} )
\end{equation}
is a logical dimension function. Since:
\begin{equation}
  \infty \; \notin \; Range(d) \; = \; \{ 0 , 1 , \cdots , 2^{n} \}
\end{equation}
it follows by theorem\ref{th:on the finite dimension} that $  (
{\mathcal{L}} \, , \, \preceq ) $  is modular.

So, introduced on  $ ( {\mathcal{L}} ({\mathcal{H}}_{2}^{
 \bigotimes n} ) \, , \, \preceq ) $ the orthocomplementation map:
\begin{equation}
  a^{\bot} \; := \; \{ | \psi_{1} > \in {\mathcal{H}}_{2}^{ \bigotimes
  n} \, : \, < \psi_{1} | \psi_{2} > = 0 \; \forall | \psi_{2} > \, \in \, a
  \} \; \; a \in {\mathcal{L}} ( {\mathcal{H}}_{2}^{ \bigotimes n} )
\end{equation}
$ ( ( {\mathcal{L}} ( {\mathcal{H}}_{2}^{ \bigotimes n} ) \, , \,
\preceq  \, , \, \bot ) $ is a  strong quantum logic.

\medskip

\begin{example} \label{ex:the weak quantum logic of the qubits' strings Hilbert space}
\end{example}
THE  WEAK QUANTUM LOGIC OF THE QUBITS' STRINGS HILBERT SPACE

Given the qubits' strings Hilbert space $
{\mathcal{H}}_{2}^{\bigotimes \star} $ let us consider the set $
{\mathcal{L}} ( {\mathcal{H}}_{2}^{\bigotimes \star}  ) $ of all
its  closed linear subspaces.

Introduced on $  {\mathcal{L}} ( {\mathcal{H}}_{2}^{ \bigotimes
n} ) $ the partial ordering relation of example\ref{ex:the strong
quantum logic of the n-qubits Hilbert space}:
\begin{equation}
  a \, \preceq \, b \; := \;  a \, \subseteq \, b \; \; a , b \in {\mathcal{L}} ( {\mathcal{H}}_{2}^{\bigotimes \star} )
\end{equation}
the poset $ ( {\mathcal{L}} \, , \, \preceq ) $ is again an
atomic lattice with atoms the \textbf{rays} of $
{\mathcal{H}}_{2}^{ \bigotimes \star} $.

As in example\ref{ex:the strong quantum logic of the n-qubits
Hilbert space} the following map:
\begin{equation}
  d(a) \; := \; dim(a) \; \; a \in {\mathcal{L}} ( {\mathcal{H}}_{2}^{\bigotimes \star} )
\end{equation}
is a logical dimension function. But since now:
\begin{equation}
  \infty \; = \; dim( {\mathcal{H}}_{2}^{\bigotimes \star} )  \; \in \; Range(d) \; = \; \{ 0 , 1 , \cdots , \infty \}
\end{equation}
we can't apply  theorem\ref{th:on the finite dimension} anymore.

Indeed it may be proved that the lattice $ ( {\mathcal{L}} (
{\mathcal{H}}_{2}^{\bigotimes \star} ) $ is not modular but only
orthomodular.

So, introduced on  $ ( {\mathcal{L}}
({\mathcal{H}}_{2}^{\bigotimes \star} ) \, , \, \preceq ) $ the
orthocomplementation map:
\begin{equation}
  a^{\bot} \; := \; \{ | \psi_{1} > \in {\mathcal{H}}_{2}^{ \bigotimes
  \star} \, : \, < \psi_{1} | \psi_{2} > = 0 \; \forall | \psi_{2} > \, \in \, a
  \} \; \; a \in {\mathcal{L}} ( {\mathcal{H}}_{2}^{ \bigotimes \star} )
\end{equation}
$ ( ( {\mathcal{L}} ( {\mathcal{H}}_{2}^{ \bigotimes \star} ) \, ,
\, \preceq  \, , \, \bot ) $ is a  not a strong quantum logic but
only a weak quantum logic.

\smallskip

The situation delineated by example\ref{ex:the strong quantum
logic of the n-qubits Hilbert space} and example\ref{ex:the weak
quantum logic of the qubits' strings Hilbert space} is a
particular case of Hilbert lattices' theory.

Given an arbitrary Hilbert space $ {\mathcal{H}} $:
\begin{definition} \label{def:Hilbert lattice}
\end{definition}
HILBERT LATTICE OF $ {\mathcal{H}} $:

the orthocomplemented lattice $ ( {\mathcal{L}} ( {\mathcal{H}})
\, , \, \preceq \, , \, \bot ) $ where, as usual, $ {\mathcal{L}}
( {\mathcal{H}}) $ is the set of all the closed linear subspaces
of $ {\mathcal{H}} $, and:
\begin{align}
  a \, & \preceq \, b \; := \;  a \, \subseteq \, b \; \; a , b \in {\mathcal{L}} ( {\mathcal{H}}) \\
  a^{\bot} & \; := \; \{ | \psi_{1} > \in {\mathcal{H}} \, : \, < \psi_{1} | \psi_{2} > = 0 \; \forall | \psi_{2} > \, \in \, a
  \} \; \; a \in {\mathcal{L}} ( {\mathcal{H}} ) \\
\end{align}
\begin{theorem} \label{th:Hilbert lattices as quantum logics}
\end{theorem}
\begin{align}
( & {\mathcal{L}} ( {\mathcal{H}}) \, , \, \preceq \, , \,
 \bot ) \text{ is a weak quantum logic} \\
 ( & {\mathcal{L}} ( {\mathcal{H}}) \, , \, \preceq \, , \,
 \bot ) \text{ is a strong quantum logic } \; \Leftrightarrow \;
 dim({\mathcal{H}}) < \infty
\end{align}

\smallskip

Theorem\ref{th:Hilbert lattices as quantum logics} can be applied
also to the qubit sequences' Hilbert space  $ {\mathcal{H}}_{2}^{
\bigotimes \star} $ to infer that $ ( {\mathcal{L}} (
{\mathcal{H}}_{2}^{ \bigotimes \star} ) \, , \, \preceq  \, , \,
\bot ) $ is a weak quantum logic.

But here comes the great conceptual shock, to prepare which, let
us observe, first of all, that the \textbf{Separability Issue} and
Thirring's claim that Schrodinger negative answer to it was owed
to a not-comprehension of Von Neumann's paper (written with J.
Murray) on infinite tensor product clashes with the Von Neumann's
1936-dated confession to Birkhoff (partially reprinted in the
paragraph7.1.2 of \cite{Redei-98}):
\begin{center}
   \textit{"I would like to make a confession which may seem immoral: I do not believe absolutely in Hilbert
   spaces any more. After all Hilbert space (as far as quantum mechanical things are concerned) was obtained by generalizing
   Euclidean space, footing on the principle of 'conserving the validity al all formal rules'. $ \cdots $ Now we begin to believe
   that it is not the vectors which matter, but the lattice of all linear (closed) subspaces.
   Because
\begin{enumerate}
  \item The vectors ought to represent the physical states, but
  they do it redundantly, up to a complex factor only,
  \item and besides, the states are merely a derived notion, the
  primitive (phenomenologically given) notions being the qualities
  which correspond to the linear closed subspaces
\end{enumerate}
   But if we wish to generalize the lattice of all linear closed subspaces from a Euclidean space
   to infinitely many dimensions, then one does not obtain Hilbert space, but the configuration which
   Murray and I called the 'case $ II_{1}$'.(The lattice of all linear closed subspace of Hilbert space is our '$I_{\infty}$' case)"}
\end{center}
What is Von Neumann speaking about?

To answer this question it is necessary to introduce some notion
concerning $W^{\star}$-algebras; demanding to the  immense
literature (e.g. \cite{Sunder-87}, \cite{Kadison-Ringrose-97a},
\cite{Kadison-Ringrose-97b}, \cite{Sakai-98}, \cite{Bing-Ren-92}
from the purely mathematical side, to \cite{Thirring-81},
\cite{Thirring-83}, \cite{Simon-93}, \cite{Bratteli-Robinson-87},
\cite{Bratteli-Robinson-97}, \cite{Connes-94}, \cite{Ruelle-99}
for physically motivated treatment or to the reviews
\cite{Petz-Redei-95}, \cite{Kadison-90} if you want to survive
the experience) for any further $ C_{\Phi}$-information we will
here briefly recall the basic facts:
\begin{definition} \label{def:algebra}
\end{definition}
ALGEBRA:

a couple  $( \, A \, , \, \circ  \, ) $ such that:
\begin{itemize}
  \item A is as linear space on the complex field $ {\mathbb{C}}$
  \item $ \circ \, A \, \times \, A \, \rightarrow \, A \; \; ( a , b
  ) \rightarrow a  b \, \equiv \,  a \circ b \, : $
\begin{align}
  ( a b ) c \; & = \; a ( b c ) \; \; \forall a,b,c \, \in \, A \\
   ( a + b ) + c \; & = \; a + ( b + c ) \; \; \forall a,b,c \, \in \,
   A \\
   \exists \, & I \in A \, : \, a I \, = \, I a \; \; \forall a \, \in \, A
\end{align}
\end{itemize}

\smallskip

\begin{definition} \label{def:involutive algebra}
\end{definition}
INVOLUTIVE ALGEBRA $ ( \star-ALGEBRA ) $:

a couple  $( \, A \, , \, \star  \, ) $ such that:

\begin{itemize}
  \item  A is an algebra
  \item $ \star \, : \, A \, \rightarrow \, A $ is an involution on A, i.e.:
\begin{align}
  (a^{\star})^{\star} \; & = \; a \; \; \forall \, a \, \in \, A  \\
  ( a + \lambda b )^{\star} \; & = \;  a^{\star} + \lambda^{\star}
  b^{\star} \; \; \forall \, a , b \in \, A \, \forall \lambda \in
  { \mathbb{C}} \\
  ( a b ) ^{\star} \; & = \; b^{\star} a ^{\star} \; \; \forall \,
  a , b \in A
\end{align}
\end{itemize}

\smallskip

Given a $ \star - algebra $ A:
\begin{definition} \label{def:unitary group of an involutive algebra}
\end{definition}
UNITARY GROUP OF A:
\begin{equation}
  {\mathcal{U}}(A) \; := \; \{ u \in A \, : \, u u^{\star} = u^{\star}
  u = I \}
\end{equation}
\begin{definition} \label{def:positive part of an involutive algebra}
\end{definition}
POSITIVE PART OF A:
\begin{equation}
  A_{+} \; := \; \{ a \in A \, : \, a = b b^{\star} \, b \in A \}
\end{equation}
\begin{definition} \label{def:self-adjoint part of an involutive algebra}
\end{definition}
SELF-ADJOINT PART OF A:
\begin{equation}
  A_{sa} \; := \; \{ a \in A \, : \, a^{\star} = a \}
\end{equation}
\begin{definition} \label{def:set of the projections of an involutive algebra}
\end{definition}
SET OF THE PROJECTIONS OF A:
\begin{equation}
  {\mathcal{P}}(A) \; := \; \{ a \in A \, \, : \, a = a^{\star} = a^{2} \}
\end{equation}

\smallskip

Given an algebra A:
\begin{definition} \label{def:norm on an algebra}
\end{definition}
NORM ON A:

a map $ \| \cdot \| \,  : \, A \rightarrow {\mathbb{R}}_{+} $
such that:
\begin{align}
  \|  a \| \;  \geq  0 \; & \; \forall a \in A  \\
  \|  a \| \;  = 0   \; \Leftrightarrow & \; a = 0 \; \; \forall a \in A  \\
  \|  \lambda a \| \;  = | \lambda | \| a \| \ \; & \; \forall a \in A \, \forall \lambda \in  {\mathbb{C}}  \\
  \| a + b \| \leq \| a \| + &  \| b \| \; \; \forall a , b \in A
\end{align}
\begin{definition} \label{def:normed algebra}
\end{definition}
NORMED ALGEBRA:

a couple ( A , $ \| \cdot \| \; $ ) such that:
\begin{itemize}
  \item A is an algebra
  \item $ \| \cdot \| $ is a norm on A :
\begin{equation}
  \| a b \| \; \leq \; \| a \| \, \| b \| \; \; \forall a , b \in A
\end{equation}
\end{itemize}
\begin{definition} \label{def:Banach algebra}
\end{definition}
BANACH ALGEBRA:

a normed algebra $ ( A \, , \,  \| \cdot \|  ) $ such that A is
complete w.r.t. $ \| \cdot \| $

\smallskip

\begin{definition} \label{def:Banach involutive algebra}
\end{definition}
BANACH INVOLUTIVE ALGEBRA  $ ( B^{\star}-ALGEBRA ) $:

a couple ( A , $\star$ ) such that:
\begin{itemize}
  \item A is a Banach algebra
  \item $\star$ is an involution on A such that:
\begin{equation}
  \| a^{\star} \| \; = \;  \| a \| \; \; \forall a \in A
\end{equation}
\end{itemize}

\smallskip

\begin{definition} \label{def:C-star algebra}
\end{definition}
$C^{\star}-ALGEBRA$:

a Banach $\star-algebra$ A such that:
\begin{equation}
  \| a^{\star} a  \| \; = \;  \| a \|^{2} \; \; \forall a \in A
\end{equation}

\smallskip

Given a $C^{\star}$-algebra A:
\begin{definition} \label{def:spectrum of an element of a C-star algebra}
\end{definition}
SPECTRUM OF $ a \in A $:
\begin{equation}
  Sp(a) \; := \; {\mathbb{C}} \, - \, \{ z \in  {\mathbb{C}} \, :
  \, \exists ( a - z ) ^{- 1} \in A \}
\end{equation}
\begin{definition} \label{def:part with discrete spectrum of a C-star algebra}
\end{definition}
PART WITH DISCRETE SPECTRUM OF A:
\begin{equation}
  A_{p.s.d} \; := \; \{ a \in A \: : \: cardinality( Sp(a)) \,  \leq \,
  \aleph_{0} \}
\end{equation}
\begin{definition} \label{def:linear functional on a C-star algebra}
\end{definition}
LINEAR FUNCTIONAL ON  A:
\begin{equation}
\begin{split}
  \varphi \, : \, A  & \rightarrow {\mathbb{C}} \\
   \varphi ( \lambda a + \mu b ) \; & = \; \lambda \varphi ( a ) + \mu \varphi ( b )
     \; \; \forall a,b \in  A \, , \, \forall \lambda , \mu  \in  {\mathbb{C}}
\end{split}
\end{equation}
\begin{definition} \label{def:dual of a C-star algebra}
\end{definition}
DUAL OF A:
\begin{equation}
  A^{\star} \; := \; \{ \varphi \, : \, \text{ linear
  functional on A} \}
\end{equation}
The dual $ A^{\star} $  of a $ C^{\star}$-algebra A is itself a
normed space w.r.t the following norm:
\begin{definition} \label{def:norm of a linear functional on a C-star algebra}
\end{definition}
NORM OF $ \varphi \in A^{\star} $
\begin{equation}
  \| \varphi \| \; := \; \varphi ( I )
\end{equation}
\begin{definition} \label{def:positive linear functionals on a C-star algebra}
\end{definition}
POSITIVE LINEAR FUNCTIONALS ON a:
\begin{equation}
  A^{\star}_{+} \; := \; \{ \varphi ( a^{\star} \, a ) \; \geq \; 0 \; \; \forall a \in
  A \}
\end{equation}
\begin{definition} \label{def:states on a C-star algebra}
\end{definition}
STATES ON A:
\begin{equation}
  S(A) \; = \; \{ \omega \in  A^{\star}_{+}  \, : \, \| \omega \| = 1 \, \}
\end{equation}
\begin{definition} \label{def:mixed state on a C-star algebra}
\end{definition}
THE STATE $\omega \in S(A)$ IS MIXED:
\begin{multline}
  \exists \, \lambda \in (0,1) \, , \, \exists \omega_{1} ,
  \omega_{2} \in S(A) : \\
    \omega_{1} \neq \omega_{2} \;  e \; \omega = \lambda
    \omega_{1} + ( 1 - \lambda )  \omega_{2}
\end{multline}
\begin{definition} \label{def:pure states on a C-star algebra}
\end{definition}
PURE STATES OF A:
\begin{equation}
  \Xi (A) \; \equiv \; \{ \omega \, \in \, S(A) \, : \, \omega \, \text{ is not mixed } \}
\end{equation}
A useful property we shall use in the sequel is the following:
\begin{theorem} \label{th:Cauchy-Schwarz inequality}
\end{theorem}
CAUCHY-SCHWARZ INEQUALITY:
\begin{equation}
  | \omega( b^{\star} a ) | ^{2} \; \leq \;   \omega( b^{\star} b )
  \, \omega( a^{\star} a ) \; \; \forall a,b \in A \, , \, \forall
  \omega \in S(A)
\end{equation}

\smallskip

Given a $ C^{\star}$-algebra A let us consider the set of all the
linear functionals over $ A^{\star}$, namely the dual of the dual
$ A^{\star \star} $.

Since an element of A $ a \in A $ may be identified with the
following linear function over $ A^{\star}$:
\begin{equation}
  a ( \varphi ) \; := \; \varphi ( a ) \; \; \varphi \in  A^{\star}
\end{equation}
it follows that:
\begin{equation}
  A \; \subseteq \; A^{\star \star}
\end{equation}
\begin{definition}
\end{definition}
$ W^{\star}$-TOPOLOGY ON $ A^{\star} $:

the coarsest topology on $ A^{\star} $ w.r.t. which all the
elements of A (seen as linear functionals over $ A^{\star} $) are
continuous

\smallskip

Given two $C^{\star}-algebras$ A and B :
\begin{definition} \label{def:involutive morphism between C-star algebras}
\end{definition}
INVOLUTIVE MORPHISM ($\star$-MORPHISM) FROM A TO B:

a map $ \tau : A \rightarrow B $ such that:
\begin{align}
  \tau ( \lambda a + \mu b ) \; = & \; \lambda \tau ( a ) + \mu \tau ( b ) \; \; \forall a , b \in A \, , \, \forall \lambda , \mu \in {\mathbb{C}} \\
  \tau ( a b ) \; = & \; \tau ( a ) \tau ( b ) \; \; \forall a , b \in
  A \\
  \tau ( a^{\star} ) \; = & \; \tau ( a )^{\star} \; \; \forall a  \in
  A
\end{align}
\begin{definition} \label{def:involutive isomorphism between C-star algebras}
\end{definition}
INVOLUTIVE ISOMORPHISM ($\star$-ISOMORPHISM) FROM A TO B:

an involutive morphism $ \tau : A \rightarrow B $ that is
bijective

\smallskip

Given a $C^{\star}$-algebra A and an Hilbert space ${\mathcal{H}}
$:
\begin{definition} \label{def:representation of a C-star algebra}
\end{definition}
REPRESENTATION OF A ON ${\mathcal{H}}$:

an involutive morphism $ \pi \, :  \, A \rightarrow B (
{\mathcal{H}} ) $ from A to $B ({\mathcal{H}})$

\smallskip

Given a representation  $ \pi $ of A on the Hilbert space
${\mathcal{H}}$:
\begin{definition} \label{reducibility of a representation}
\end{definition}
$ \pi $ IS REDUCIBLE:
\begin{equation}
  \exists \; V \subset {\mathcal{H}} \text{ linear subspace} \; : \; \pi (
  V ) \: \subseteq \: V
\end{equation}

Given two representations $ \pi_{1}$ and $  \pi_{2}$ of A on the
Hilbert spaces, respectively, ${\mathcal{H}}_{1}$ and
${\mathcal{H}}_{2}$:
\begin{definition} \label{def:equivalent representations of a C-star algebra}
\end{definition}
$ \pi_{1}$ AND $ \pi_{2}$ ARE EQUIVALENT:
\begin{multline}
  \exists \, U \, : \, {\mathcal{H}}_{1} \, \rightarrow  {\mathcal{H}}_{2} \, isomorphism \, : \\
   \pi_{2} ( a )  \; = \; U \pi_{1} ( a ) U^{-1} \; \; \forall a \in A
\end{multline}

\smallskip

Any state $ \omega \in S(A) $ over a $C^{\star}$-algebra A gives
rise to a particularly important representation of A we are going
to introduce.

Defined the following subset of A:
\begin{equation}
  {\mathcal{N}} \; := \; \{ a \in A \, : \, \omega ( a^{\star} a ) = 0 \}
\end{equation}
let us define on the quotient space $ \frac{A}{ {\mathcal{N}}} $
the inner product:
\begin{equation} \label{eq:GNS inner product}
  < a | b > \; := \; \omega (a^{\star} b ) \; \; [ a ] , [ b ] \in \frac{A}{ {\mathcal{N}}}
\end{equation}
Let us observe, furthermore, that the canonical embedding $ i \,
: \, A \, \mapsto \,  \frac{A}{ {\mathcal{N}}} $:
\begin{equation}
  i ( a  ) \; := \; [ a ] \; = \; \{ b \in A : b = a
  + c \, , \,  c \in {\mathcal{N}} \}
\end{equation}
is a continuous application from A (endowed with the
norm-topology) to $ \frac{A}{ {\mathcal{N}}} $ (endowed with the
norm topology induced by the inner-product of eq.\ref{eq:GNS inner
product}).

We can then introduce the following:
\begin{definition} \label{def:GNS representation}
\end{definition}
GELFAND-NAIMARK-SEGAL REPRESENTATION (GNS REPRESENTATION) OF A
W.R.T. $ \omega $

the representation $ \pi_{\omega} $ of A on the Hilbert space $
{\mathcal{H}}_{\omega} $:
\begin{itemize}
  \item $ {\mathcal{H}}_{\omega} $ is the completion of the inner
  product space $ ( \, \frac{A}{ {\mathcal{N}}} \, , \, < a | b
  > ) $
  \item $ \pi_{\omega} $ is the continuous extension of the
  application:
\begin{equation}
  [ \pi_{\omega} ( a ) ] | [ b ] >  \; := \;  a \, b \; \; [ b ]
  \in \frac{A}{ {\mathcal{N}}}
\end{equation}
\end{itemize}

One has that:
\begin{theorem} \label{th:basic properties of the GNS representation}
\end{theorem}
BASIC PROPERTIES OF THE GNS REPRESENTATION:
\begin{enumerate}
  \item the vector:
\begin{equation}
  | \Omega_{\omega} > \; := \; | [ I ] >
\end{equation}
is cyclic, i.e. $ \pi_{\omega} ( A)  | \Omega_{\omega} > $ is
dense in $ {\mathcal{H}}_{\omega} $
  \item any representation $ \pi $ of A admitting a ciclic vector
  $ | \Phi > $ is equivalent to the GNS representation $ \pi_{\varphi} $ w.r.t.
  the state:
\begin{equation}
   \varphi ( a ) \; := \; < \Phi | \pi ( a ) | \Phi >  \; \; a \in A
\end{equation}
  \item
\begin{equation}
  \pi_{\omega} \text{ is irreducible } \; \Leftrightarrow \; \omega
  \in \Xi (A)
\end{equation}
\end{enumerate}

\smallskip

Let us now present the following fundamental:
\begin{theorem} \label{th:Gelfand isomorphism at C-star algebraic level}
\end{theorem}
GELFAND'S ISOMORPHISM AT $ C^{\star}$-ALGEBRAIC LEVEL:

\begin{hypothesis}
\end{hypothesis}
A  abelian $ C^{\star}$-algebra

$ C( X ( A )) \;  C^{\star}$-algebra of the complex-valued
 continuous (w.r.t. the $ W^{\star}$-topology) on $ X ( A ) $

\begin{thesis}
\end{thesis}
A and $ C( X ( A )) $ are $ \star $-isomorphic

\medskip

Indeed also the converse property holds, and
theorem\ref{th:Gelfand isomorphism at C-star algebraic level} may
be considerably streghtened, resulting in the following:
\begin{theorem} \label{th:category isomorphism at the basis of Noncommutative Topology}
\end{theorem}
CATEGORY EQUIVALENCE AT THE BASIS OF NONCOMMUTATIVE TOPOLOGY

The \textbf{category} having as \textbf{objects} the \textbf{
Hausdorff compact topological spaces} and as \textbf{morphisms}
the \textbf{continuous maps} on such spaces is equivalent to the
category having as \textbf{objects} the \textbf{abelian
$C^{\star}$-algebras} and as \textbf{morphisms}  the
\textbf{involutive morphisms} of such spaces.

\bigskip

\begin{remark} \label{rem:the metaphore by which we can speak about noncommutative sets from within ZFC}
\end{remark}
THE METAPHORE BY WHICH WE CAN SPEAK ABOUT NONCOMMUTATIVE SETS
FROM WITHIN ZFC:

Theorem\ref{th:category isomorphism at the basis of
Noncommutative Topology} says, in particular, that an abelian $
C^{\star} $-algebra may be always seen as an algebra of function
over a suitable topological space X.

This suggest to introduce a metaphore, that we call the
\textbf{noncommutative metaphore} from here and beyond, according
to which one looks at a noncommutative algebra A as if it was an
algebras of functions over an hypothetic \textbf{noncommutative
set}:
\begin{equation} \label{eq:the impossible equation}
  A \; =_{METAPHORE} \; C ( X_{NC} )
\end{equation}
Of course this is only a metaphore, but it is the only way we can
speak about \textbf{noncommutative sets} from within the formal
system of \textbf{commutative set theory} (namely the formal
system ZFC of Zermelo  Fraenkel endowed with the Axiom of Choice)
giving foundations to Mathematics.

Since there is no possibility  inside ZFC  of formalizing
eq.\ref{eq:the impossible equation} and so to speak directly of $
X_{NC} $ , we will never mention it and we will directly refer to
A as a \textbf{noncommutative set}.

It should be even supreflous to remark that, of course, from the
same fact we can speak about it inside ZFC, A  is also a
\textbf{commutative set}.

So it is fundamental, when we speak about A, to specify if we are
looking at it as an ordinary \textbf{commutative set} or as a
\textbf{noncommutative set}, i.e. as a way of speaking about the
non formalizable-in-ZFC object $ X_{NC} $.

Such a double nature, of course, reflects itself at different
levels:
\begin{itemize}
  \item at a logical level if we look at A as a \textbf{noncommutative set}, one can formalize its propostion calculus through the quantum logic
  QL(A).

 If instead one look at A as a \textbf{commutative set}, one can,
 of course, apply to it the ordinary classical (i.e. distributive)
 set-theoretical predicative calculus

  \item we will soon introduce the notion of \textbf{noncommutative
  cardinality} of a noncommutative set;

  according to such a notion we will arrive to characterize the
  noncommutative binary alphabet $ \Sigma_{NC} \, = \, M_{2} ( {\mathbb{C}}) $ by the condition:
\begin{equation}
  cardinality_{NC} ( \Sigma_{NC} ) \; = \; 1
\end{equation}
Anyway, obviously, looking at $ M_{2} ( {\mathbb{C}}) $ simply as
a commutative set, we can consider its commutative cardinality:
\begin{equation}
  cardinality ( M_{2} ( {\mathbb{C}}) ) \; = \; \aleph_{1}^{4} \;
  = \; \aleph_{1}
\end{equation}
\end{itemize}

\medskip

Theorem\ref{th:category isomorphism at the basis of
Noncommutative Topology} introduces a \textbf{topological
structure} over \textbf{noncommutative sets}.

It is just the first of a collection of Category Equivalence
Theorems that allow to introduce an high hierarchy of more and
more refined structures on  \textbf{noncommutative sets},
resulting in the wonderful conceptual tower, namely
Noncommutative Geometry, built by that genius named Alain Connes
\cite{Jaffe-91}.

\smallskip

\begin{definition} \label{def:W-star algebra}
\end{definition}
$W^{\star}$-ALGEBRA (or ALGEBRAIC SPACE):

a $ C^{\star}-ALGEBRA $ A such that $ ( A^{\star} \, , \, \|
\cdot \| ) $ is a Banach space

\smallskip

\begin{definition} \label{def:commutative space}
\end{definition}
COMMUTATIVE SPACE:

a commutative algebraic space

\smallskip

\begin{definition} \label{def:noncommutative space}
\end{definition}
NONCOMMUTATIVE SPACE:

a noncommutative algebraic space

\smallskip

\begin{example} \label{ex:the W-star algebra of the bounded operators on an Hilbert space}
\end{example}
THE $ W^{\star}$-ALGEBRA OF THE BOUNDED OPERATORS ON AN HILBERT
SPACE

Given an arbitrary Hilbert space $ {\mathcal{H}} $ the space $
B({\mathcal{H}}) $ of all the bounded linear operators on $
{\mathcal{H}} $ is a $ W^{\star}$-algebra

\medskip
\begin{definition} \label{def:automorphisms of a W-star algebra}
\end{definition}
AUTOMORPHISMS OF A:
\begin{equation}
  AUT(A) \; := \; \{  \tau : A \rightarrow A  \text{ involutive isomorphism of A
  } \}
\end{equation}
\begin{definition} \label{def:inner automorphisms of a W-star algebra}
\end{definition}
INNER AUTOMORPHISMS OF A:
\begin{equation}
  INN(A) \; := \; \{ \tau \in AUT(A) \, :  \exists u \in {\mathcal{U}}(A) \, , \, \tau (a) = u a
  u^{\star} \, \, \forall a \in A \}
\end{equation}
\begin{definition} \label{def:outer automorphisms of a W-star algebra}
\end{definition}
OUTER AUTOMORPHISMS OF A:
\begin{equation}
  OUT(A) \; := \; \frac{AUT(A)}{INN(A)}
\end{equation}
Given two algebraic spaces A and B:
\begin{definition} \label{def:positive maps among two  W-star algebras}
\end{definition}
POSITIVE MAPS FROM A TO B:
\begin{equation}
 P ( A , B ) \; := \; \{ \, \tau \, : \, A \rightarrow B \, linear \, :
 \, \tau ( A_{+} ) \; \subseteq \; B_{+} \, \}
\end{equation}
\begin{definition} \label{def:completely positive maps among two W-star algebras}
\end{definition}
COMPLETELY POSITIVE MAPS FROM A TO B:
\begin{equation}
 CP ( A, B ) \; \equiv \; \{ \, \tau \in P (A,B) \, : \, \tau \otimes {\mathbb{I}}_{n} \in P ( A,B ) \; \; \forall n
  \in  {\mathbb{N}} \}
\end{equation}
\begin{definition} \label{def:CPU-maps among two W-star algebras}
\end{definition}
COMPLETELY POSITIVE UNITAL MAPS (CPU-MAPS OR CHANNELS) FROM A TO
B:
\begin{equation}
  CPU ( A,B ) \; \equiv \; \{ \, \tau \in CP (A,B) \, : \, \tau ( {\mathbb{I}} ) \,
 = \, {\mathbb{I}} \}
\end{equation}
In particular:
\begin{definition} \label{def:positive maps on a W-star algebra}
\end{definition}
POSITIVE MAPS ON A:
\begin{equation}
 P ( A ) \; := \; P( A , A)
\end{equation}
\begin{definition} \label{def:completely positive maps on a W-star algebra}
\end{definition}
COMPLETELY POSITIVE MAPS ON A:
\begin{equation}
 CP ( A ) \; \equiv \; CP( A , A)
\end{equation}
\begin{definition} \label{def:CPU-maps on a W-star algebra}
\end{definition}
COMPLETELY POSITIVE UNITAL MAPS (CPU-MAPS OR CHANNELS) ON A:
\begin{equation}
  CPU ( A ) \; := \; CPU( A , A)
\end{equation}
One has the following:
\begin{theorem} \label{th:on the relationship between positivity and complete-positivity}
\end{theorem}
ON THE RELATIONSHIP BETWEEN POSITIVITY AND COMPLETE-POSITIVITY:
\begin{enumerate}
  \item
\begin{equation}
  A \; commutative  \; \Rightarrow \; CP(A) \, = \, P(A)
\end{equation}
  \item
\begin{equation}
  A \; noncommutative  \; \Rightarrow \; CP(A) \, \subset \, P(A)
\end{equation}
\end{enumerate}

The analysis of channels is particularly simplified by the
following:
\begin{theorem} \label{th:Kraus-Stinespring's theorem}
\end{theorem}
KRAUS-STINESPRING'S THEOREM:

\begin{hypothesis}
\end{hypothesis}

\begin{center}
$ A \subseteq B( {\mathcal{H}}) $   Von Neumann algebra acting on
the Hilbert space $ {\mathcal{H}} $

$ \alpha \, : \, A \rightarrow A $ normal linear map

\end{center}

\begin{thesis}
\end{thesis}
\begin{multline}
   \alpha \in CPU(A) \; \; \Leftrightarrow  \; \;  \exists \, \{ V_{i} \}_{ i \in I} \in {\mathcal{B}}( {\mathcal{H}}) \} \;
   : \\
\alpha (a ) \;  = \; \sum_{ i \in I} V_{i}^{\star} a V_{i} \;
\forall a \in A \\
   \sum_{ i \in I} V_{i}^{\star}  V_{i} \; = \; {\mathbb{I}}
\end{multline}
where the convergence is in the weak topology.

\smallskip

\begin{remark} \label{rem:identificability of involutively-isomorphic W-star algebras}
\end{remark}
IDENTIFICABILITY OF $\star$-ISOMORPHIC $W^{\star}$-ALGEBRAS:

Since, from an algebraic viewpoint, $\star$-isomorphic $
W^{\star}$-algebras are identical, they can be identified.

So, for example, the $W^{\star}$-algebra $ B ({\mathcal{H}}_{n})
$ of all the bounded linear operators on an n-dimensional Hilbert
space $ {\mathcal{H}}_{n} $, with $ n \, \in \, {\mathbb{N}} $
may be identificated with the $W^{\star}$-algebra $ M_{n}
({\mathbb{C}}) $ of all the $ n \, \times \, n $ matrices with
complex entries.

So, in particular, the algebra $ B({\mathcal{H}}_{2}^{ \bigotimes
n}) $ of all the bounded linear operators on the n-qubits Hilbert
space $ {\mathcal{H}}_{2}^{ \bigotimes n} $ may be identified
with the  $W^{\star}$-algebra $ M_{2^{n}} ({\mathbb{C}}) $

\smallskip

Given a $ C^{\star}-algebra $ we will, obviously, say that:
\begin{definition} \label{def:abelian C-star algebra}
\end{definition}
A IS ABELIAN:
\begin{equation}
  [ a , b ] \; := \; a b \, - \, b a \; = \; 0 \; \; \forall a,b
  \in A
\end{equation}
Given an abelian  $ C^{\star}-algebra $ A:
\begin{definition} \label{def:characters of an abelian C-star algebra}
\end{definition}
\begin{equation}
  X(A) \; := \; \{ \pi \, : \, \text{ representation of A on }
  {\mathbb{C}} \}
\end{equation}

\smallskip

Given a $ C^{\star}-algebra $ A and a linear subspace $ B
\subseteq A $:
\begin{definition} \label{sub-C-star algebra of a C-star algebra}
\end{definition}
B IS A SUB-$C^{\star}$-ALGEBRA OF A:

B is a $ C^{\star}-algebra $ w.r.t. to the restriction to B of the
$ C^{\star}-algebraic  $ structure of A

\smallskip

Given an Hilbert space $ \mathcal{H} $ ed and a  $ sub-
C^{\star}-algebra \; A \, \subseteq \, B({\mathcal{H}}) $ of $
B({\mathcal{H}}) $:
\begin{definition} \label{def:commutant}
\end{definition}
COMMUTANT OF A:
\begin{equation}
  A' \; = \; \{ a \in A \, : \, [ a , b ] \, = \, 0 \; \; \forall
  b \in B({\mathcal{H}}) \}
\end{equation}
\begin{definition} \label{def:centre}
\end{definition}
CENTRE OF A:
\begin{equation}
  {\mathcal{Z}}(A) \; := \; A \cap A'
\end{equation}
\begin{definition} \label{def:Von Neumann algebra}
\end{definition}
A IS A VON NEUMANN ALGEBRA:
\begin{equation}
 A'' \; := \; (A')' \; = \; A
\end{equation}

\smallskip

The two notions of a $ W^{\star}$-algebra and of a Von Neumann
algebra introduced, respectively, in definition\ref{def:W-star
algebra} and definition\ref{def:Von Neumann algebra} would seem
to have anything in common: the first is a purely abstract
algebraic notion while the latter is a concrete notion concerning
operators on an Hilbert space.

So it may appear rather shocking that these notions are indeed
equivalent (remember remark\ref{rem:identificability of
involutively-isomorphic W-star algebras}) , as is stated by the
following:
\begin{theorem} \label{th:Sakai theorem}
\end{theorem}
SAKAI'S THEOREM:

\begin{hypothesis}
\end{hypothesis}
\begin{equation*}
  A \; C^{\star}-algebra
\end{equation*}
\begin{thesis}
\end{thesis}
\begin{center}
  A is $ \star$-isomorphic to a Von Neumann algebra $
  \Leftrightarrow $ A is a $ W^{\star}-algebra $
\end{center}

Up to now all these operator-algebraic stuff would seem no to have
anything in common with Quantum Logic.

Anyway it may be proved that:
\begin{theorem} \label{th:the quantum logic of a Von Neumann algebra}
\end{theorem}
ON THE QUANTUM LOGIC OF A VON NEUMANN ALGEBRA:

\begin{hypothesis}
\end{hypothesis}
\begin{align*}
   {\mathcal{H}} \, & \, \text{ Hilbert space }
  A \, &  \, \subseteq \, B({\mathcal{H}}) \text{ Von Neumann algebra} \\
\end{align*}
\begin{thesis}
\end{thesis}
\begin{align*}
  QL (A) \; := \;  ( {\mathcal{P}}(A)  \, , \, \preceq \, , \, \bot ) \text{ is a quantum logic } \\
  {\mathcal{P}}(A)'' & \; = \;A
\end{align*}
where $ \preceq $ and $ \bot $ are, respectively, the
partial-ordering relation and the orthocomplementation inherited
from $ {\mathcal{L}}( {\mathcal{H}}) $.

Theorem\ref{th:the quantum logic of a Von Neumann algebra} tells
us that, substantially, a Von Neumann algebra is generated by the
quantum logic it gives rise to.

Indeed, as we will now show, the underlying quantum logics
substantially govern the classification of Von Neumann algebras.

Given a Von Neumann algebra $ A \subseteq B ({\mathcal{H}}) $:
\begin{definition} \label{def:factor}
\end{definition}
A IS A FACTOR:
\begin{equation}
  {\mathcal{Z}}(A) \; = \; \{ \lambda I \, , \, \lambda \in {\mathbb{C}} \}
\end{equation}

Factors are, substantially, the building blocks of Von Neumann
algebras: any Von Neumann algebra A may, actually, be expressed
as a direct integral of factors:
\begin{multline*} \label{eq:factor decomposition of a Von Neumann algebra}
  A \; = \; \int_{{\mathcal{Z}}(A)}^{\otimes} A_{\lambda} \, d \nu ( \lambda
  )  \\
    {\mathcal{Z}}(A_{\lambda}) \; = \; \{ {\mathbb{C}} \, {\mathbb{I}} \} \; \; \forall \lambda \in  {\mathcal{Z}}(A)
\end{multline*}
Hence the analysis of a Von Neumann algebra may be reduced to the
analysis of its building blocks.

Given, now, a generic Von Neumann algebra $ A \subseteq
B(\mathbb{H}) $  and two its projections $ a , b \in
{\mathcal{P}}(A) $:
\begin{definition} \label{def:equivalence of projections in a Von Neumann algebra}
\end{definition}
a AND b ARE EQUIVALENT IN A:
\begin{multline}
  a \, \sim_{A} \, b \; := \; \exists o \in A \, : \, ( o | \psi > = 0
  \, \forall | \psi > \in Range(a)^{\bot} ) \\
  and \; ( \| o | \psi
  > \| = \| | \psi > \| \, \,  \forall | \psi > \in Range(a) \}
\end{multline}

\begin{remark} \label{rem:intuitive meaning of the equivalence of projections in a Von Neumann algebra}
\end{remark}
INTUITIVE MEANING OF $ \sim_{A} $: EQUALITY OF THE DIMENSION
RELATIVE TO A

Definition\ref{def:equivalence of projections in a Von Neumann
algebra} may appear rather counter-intuitive. Its meaning is that
the existence of a partial (since it acts so only on Range(a)
being identically null on its complement) isometry between
Range(a) and Range(b) that belongs to A may be interpreted,
informally speaking, as the fact that the \textbf{dimension
relative to A} of the subspace a projects to is equal to the
\textbf{dimension relative to A} of the subspace b projects to

\smallskip

The equivalence relation $ \sim_{A} $ over $ {\mathcal{P}}(A) $
may be used to introduce a new partial ordering on projections
(different from that of the quantum logic of A)
\begin{equation}\label{eq:total ordering of projections in a Von Neumann
algebra}
   a \, \trianglelefteq \, b \; := \; \exists c \in  {\mathcal{P}}(A)
     \: : \: a \, \sim_{A} \,  c \, \leq \, b
\end{equation}

\begin{remark} \label{rem:intuitive meaning of the total ordering of projections in a Von Neumann algebra}
\end{remark}
INTUITIVE MEANING OF $ \trianglelefteq $: ORDERING ACCORDING TO
THE RELATIVE DIMENSION

The intuitive meaning of the partial ordering $ \trianglelefteq $
is induced by that of the equivalence relation $ \sim_{A} $.

So the condition $ a \, \trianglelefteq \, b $ means,intuitively,
that the \textbf{dimension relative to A} of a is less or equal
to the \textbf{dimension relative to A} of b.

But here comes the following:
\begin{theorem} \label{th:total ordering of a factor's projections}
\end{theorem}
TOTAL ORDERING W.R.T. $ \trianglelefteq $ OF A FACTOR'S
EQUIVALENCE CLASSES OF PROJECTIONS
\begin{equation}
 {\mathcal{Z}}(A) = {\mathbb{C}} I \; \Rightarrow \; (  a \, \trianglelefteq \, b \; or \; b \, \trianglelefteq \, a \;
  \; \forall a , b \in  \frac{{\mathcal{P}}(A)}{\sim_{A}} )
\end{equation}
whose importance is owed to an immediate conseguence of its:
\begin{corollary} \label{cor:order type is an invariant for factors}
\end{corollary}
ORDER TYPE OF $ \frac{{\mathcal{P}}(A)}{\sim_{A}}  $ IS AN
INVARIANT FOR FACTORS
\begin{equation}
  A, B \, \star-isomorphic \, factors \; \Rightarrow \;
  Type-order(  \frac{{\mathcal{P}}(A)}{\sim_{A}} ) ) \, = \, Type-order(  \frac{{\mathcal{P}}(B)}{\sim_{B}} ) )
\end{equation}

\smallskip

To determine the order type of a $
\frac{{\mathcal{P}}(A)}{\sim_{A}}  $  of a factor A a key concept
is the finiteness of projections:
\begin{definition} \label{def:finite projection on a lattice}
\end{definition}
$ a \in {\mathcal{P}}(A) $ IS FINITE:
\begin{equation}
  a \, \sim_{A} \, b \, \trianglelefteq \, a  \; \Rightarrow \: a
  \, = \, b \; \; \forall b \in {\mathcal{P}}(A)
\end{equation}
We can, at last, formalize the intuitive statements concerning the
\textbf{relative dimension} of remark\ref{rem:intuitive meaning of
the equivalence of projections in a Von Neumann algebra} and
remark\ref{rem:intuitive meaning of the total ordering of
projections in a Von Neumann algebra}introducing the following
fundamental notion:
\begin{definition} \label{def:relative dimension w.r.t. a factor}
\end{definition}
RELATIVE DIMENSION W.R.T. A:

a map $ d \, : \, {\mathcal{P}}(A) \, \mapsto \, [ 0 , + \infty ]
$ such that:
\begin{align}
  d(a) \, &  \, = \, 0 \; \Leftrightarrow \; a \, = \, 0 \; \; \forall a \in {\mathcal{P}}(A)  \\
  a \, &  \bot \, b \; \Rightarrow d ( a + b ) \, = \, d(a) + d(b) \; \; \forall a \in {\mathcal{P}}(A)  \\
  d(a) \, & \, < d(b) \; \Leftrightarrow \; a \, \trianglelefteq
  \, b \; \; \forall a,b \in {\mathcal{P}}(A)  \\
   d(a) \, & \, + \infty \; \Leftrightarrow \; a \text{ is finite
   } \; \; \forall a \in {\mathcal{P}}(A)  \\
    d(a) \, & \, = \, d(b) \; \Leftrightarrow \; a \, \sim_{A}
  \, b \; \; \forall a,b \in {\mathcal{P}}(A)  \\
  d(a) \, & \, + \, d(b) \; = \; d ( a \bigwedge b ) \, + \,  d ( a \bigvee b
  ) \; \; \forall a,b \in {\mathcal{P}}(A)
\end{align}

We will denote, from here and beyond, the relative dimension
w.r.t. a factor A by $ d_{A} $.

\smallskip

The astonishing fact is that:
\begin{theorem} \label{th:unicity of the relative dimension w.r.t. a factor}
\end{theorem}
UNICITY OF THE RELATIVE DIMENSION W.R.T. A FACTOR:
\begin{equation}
  d_{1} \, , \,  d_{2} \text{ relative dimensions w.r.t. A
  } \; \Rightarrow \; \exists c \, \in  \, {\mathbb{R}}_{+} \, : \, (
  d_{1}(a) \, = \, c \,  d_{2}(a) \; \forall a \in
  {\mathcal{P}}(A) )
\end{equation}

The importance of theorem\ref{th:unicity of the relative
dimension w.r.t. a factor} lies in that it implies that the order
type of $ \frac{{\mathcal{P}}(A)}{\sim_{A}}  $ can be read off the
order type of $  d_{A} $'s range.

Murray and Von Neumann determined the possible ranges of $  d_{A}
$ (suitably normalized) resulting in the following classification:
\begin{definition} \label{def:discrete finite factor}
\end{definition}
A IS OF TYPE FINITE, DISCRETE $(  Type (A) \, = \, I_{n}   )$
\begin{equation*}
  Range( d_{A} ) \; = \; \{ 0 , 1 , \ldots , n
  \} \; \; n \in {\mathbb{N}}_{+}
\end{equation*}
\begin{definition} \label{def:discrete infinite factor}
\end{definition}
A IS OF TYPE INFINITE, DISCRETE ($  Type (A) \, = \,  I_{\infty}
$):
\begin{equation}
  Range( d_{A} ) \; = \; \{ 0 , 1 , \ldots , + \infty \}
\end{equation}
\begin{definition} \label{def:continuous finite factor}
\end{definition}
A IS OF TYPE FINITE, CONTINUOUS ($  Type (A) \, = \, II_{1} $):
\begin{equation}
  Range( d_{A} ) \; = \;  [ 0 , 1 ]
\end{equation}
\begin{definition} \label{def:continuous infinite factor}
\end{definition}
A IS OF TYPE INFINITE, CONTINUOUS ($  Type (A) \, = \,
II_{\infty}   $):
\begin{equation}
  Range( d_{A} ) \; = \;  [ 0 , + \infty ]
\end{equation}
\begin{definition} \label{def:purely infinite factor}
\end{definition}
A IS OF TYPE PURELY INFINITE ($  Type (A) \, = \, III  $)
\begin{equation}
  Range( d_{A} ) \; = \;  \{ 0 , + \infty \}
\end{equation}

\smallskip

\begin{remark}
\end{remark}
THE ORDER TYPE OF $ \frac{{\mathcal{P}}(A)}{\sim_{A}}  $ DOESN'T
ALLOW A COMPLETE CLASSIFICATION OF FACTORS

It is important, at this point, to remark that the the above
classification of factors based on the order type of the
algebraic dimension function's range is not complete.

The complete classification, furnished almost fourty years later
by the great Alain Connes, will be briefly introduced in
section\ref{sec:On the rule Noncommutative Measure Theory and
Noncommutative Geometry play in Quantum Physics}

\smallskip

It is, now, conceptually important to observe that the relative
dimension with respect to factors, underlying their Murray-Von
Neumann classification, is a Quantum-Logic's notion:
\begin{theorem} \label{th:relative dimension w.r.t. a factor is a lattice dimension function}
\end{theorem}
RELATIVE DIMENSION W.R.T. A FACTOR IS A MATTER OF QUANTUM LOGIC

$  d_{A} $ is a lattice dimension function  over the weak quantum
logic QL(A)

\begin{corollary} \label{cor:the quantum logic of a finite factor is strong}
\end{corollary}
THE QUANTUM LOGIC OF A FINITE FACTOR IS STRONG:

\begin{hypothesis}
\end{hypothesis}
\begin{equation*}
  A  \; factor \; : \; Type(A) \, \in \, \{  I_{n}  \}_{n \in {\mathbb{N}}}
  \, \bigcup \{ II_{1} \}
\end{equation*}
\begin{thesis}
\end{thesis}
\begin{center}
  QL(A) is a strong quantum logic
\end{center}
\begin{proof}
Since $ d_{A} $ assumes  only finite values  this happens,
obviously, also to its restriction to $ {\mathcal{P}}(A) $ that,
by theorem\ref{th:relative dimension w.r.t. a factor is a lattice
dimension function}, is an algebraic dimension function over
QL(A).

By theorem\ref{th:on the finite dimension} the thesis immediately
follows
\end{proof}

We have, at last, almost all the ingredients required to
understand the confession of Von Neumann to Birkhoff.

The last ingredients required are those deriving from the
following:
\begin{theorem} \label{th:characterization of discrete factors}
\end{theorem}
CHARACTERIZATION OF DISCRETE FACTORS:
\begin{equation}
  Type(A) \: \in  \: \{  I_{n}  \}_{n \in \{0 , 1 , \cdots , \infty \}} \; \Leftrightarrow \;
  \exists {\mathcal{H}} \text{ Hilbert space } : \, A \, = B (
  \mathcal{H})
\end{equation}
\begin{proof}
  A is $ \star$-isomorphic to a $ B (
  {\mathcal{H}}) $, for a proper Hilbert space $ {\mathcal{H}} $,
  iff the quantum logic of A is an Hilbert Lattice.

But a logic dimension function of an Hilbert lattice may be
easily defined as the  dimensionality, as linear subspaces, of
its elements and, conseguentially, can assume only integer values.

By theorem\ref{th:unicity of the relative dimension w.r.t. a
factor}and theorem\ref{th:relative dimension w.r.t. a factor is a
lattice dimension function} it immediately follows the thesis
\end{proof}

\begin{corollary} \label{cor:discreteness of a factor is equivalent to the atomicity of its quantum logic}
\end{corollary}
ATOMICITY OF A FACTOR'S QUANTUM LOGIC
\begin{equation}
  QL(A) \text{ is atomic } \; \Leftrightarrow \; Type(A) \: \in  \: \{  I_{n}  \}_{n \in \{0 , 1 , \cdots , \infty \}}
\end{equation}
\begin{proof}
By theorem\ref{th:characterization of discrete factors} we have
that QL(A) is an Hilbert lattice iff A is discrete.

The thesis immediately follows
\end{proof}

\smallskip

Given a $ W^{\star}$-algebra A:
\begin{definition} \label{def:trace on a W-star algebra}
\end{definition}
TRACE ON A:

A linear map $ \tau \, : \, A_{+} \, \mapsto \, [ 0 , + \infty ]
$ such that:
\begin{equation}
  \tau  \: \circ \: \alpha \; = \;  \tau  \; \; \forall \alpha \in INN(A)
\end{equation}
Given  a trace $ \tau $ on A:
\begin{definition} \label{def:finite trace on a W-star algebra}
\end{definition}
$ \tau $ IS FINITE:
\begin{equation}
  \infty \; \notin \; Range( \tau )
\end{equation}

A very important property of finite factors is stated by the
following:
\begin{theorem} \label{th:extensibility of the relative dimension w.r.t. a finite factor to a finite trace}
\end{theorem}
EXTENSIBILITY OF THE RELATIVE DIMENSION W.R.T. A FINITE FACTOR TO
A FINITE TRACE

\begin{hypothesis}
\end{hypothesis}
\begin{equation*}
  A  \; factor \; : \; Type(A) \: \in \: \{  I_{n}  \}_{n \in {\mathbb{N}}}
  \, \bigcup \{ II_{1} \}
\end{equation*}
\begin{thesis}
\end{thesis}
\begin{equation}
  \exists ! \tau_{A} \text{ finite trace on A } \; : \; \tau_{A} |
  _{{\mathcal{P}}(A)} \, = \, d_{A}
\end{equation}
\begin{theorem} \label{th:extensibility of the finite state to a state}
\end{theorem}
\begin{hypothesis}
\end{hypothesis}
\begin{equation*}
  A  \; factor \; : \; Type(A) \: \in \: \{  I_{n}  \}_{n \in {\mathbb{N}}}
  \, \bigcup \{ II_{1} \}
\end{equation*}
\begin{thesis}
\end{thesis}
\begin{equation}
  \exists ! \omega_{A} \in S(A) \; : \; \omega_{A}|_{A_{+}} \, =
  \,  \tau_{A}
\end{equation}

\smallskip

\begin{example} \label{ex:the finite trace on the n-qubits' W-star algebra}
\end{example}
THE FINITE TRACE ON THE n-QUBITS' $ W^{\star} $-ALGEBRA

We know that:
\begin{equation}
  B( \mathcal{H}_{2}^{ \bigotimes n}) \; = \;
  \bigotimes_{k=1}^{n} M_{2} ( {\mathbb{C}} ) \; = \;  M_{2^{n}} ( {\mathbb{C}} )
\end{equation}
The finite trace on the $ n  \times n $ matrix algebra $ M_{n} (
{\mathbb{C}} ) $ is simply the normalized matricial trace:
\begin{equation}
  \tau_{n} \; := \; \frac{1}{n} Tr_{n} \; = \;
  \bigotimes_{k=1}^{n} \tau_{2}
\end{equation}
so that the finite trace on $  B( \mathcal{H}_{2}^{ \bigotimes n})
$ is $ \tau_{2^{n}} $.

\smallskip

Let us now consider an arbitrary $ W^{\star}$-algebra A.

We want to characterize the condition stating that A is $
C_{\Phi}-NC_{M}$-computable, i.e. is approximable with arbitrary
precision by finite-dimensional matrix algebras.

Mathematically formalized such a constraint results in the
following:
\begin{definition} \label{def:hyperfinite W-algebra}
\end{definition}
A IS HYPERFINITE:

there exists an increasing sequence $ \{ A_{n} \}_{n \in
{\mathbb{N}}} $ of $ W^{\star} $-algebras of A such that:
\begin{equation}
  A \; = \; ( \bigcup_{n \in {\mathbb{N}}} A_{n} )''
\end{equation}

\begin{example} \label{ex:the hyperfinite finite continuous factor}
\end{example}
THE HYPERFINITE $ II_{1} $ FACTOR R

Given the one dimensional lattice $ {\mathbb{Z}} $ let us attach
to the $ n^{th} $ lattice site the 1-qubit $ W^{\star}$-algebra:
\begin{equation}
  A_{n} \; := \: M_{2} ({\mathbb{C}}) \; \; n \in {\mathbb{Z}}
\end{equation}
Given an arbitary set of sites $ \Lambda  \, \subseteq  \,
{\mathbb{Z}} $ let us define:
\begin{equation}
  A_{\Lambda} \; := \; \bigotimes_{n \in \Lambda} A_{n}
\end{equation}
Clearly we have that:
\begin{equation} \label{eq:increasing nature of the local algebras}
   \Lambda_{1} \, \subseteq  \, \Lambda_{2} \; \Rightarrow \; A_{\Lambda_{1}} \, \subseteq \, A_{\Lambda_{2}}
\end{equation}
\begin{equation}\label{eq:commutativity of the local algebras of disjoint regions}
  \Lambda_{1} \,  \bigcap \, \Lambda_{2} \: = \: \emptyset \; \Rightarrow
   \; [ A_{\Lambda_{1}} , A_{\Lambda_{2}} ] \, = \, 0
\end{equation}
Let us now consider the state $ \tau_{\Lambda} \, \in \, S (
A_{\Lambda} )$ defined as:
\begin{equation}
  \tau_{\Lambda} \; := \;  \bigotimes_{n \in \Lambda} \tau_{2}
\end{equation}
Clearly we have, in particular, that the state:
\begin{equation}
  \tau_{ \{ 0 , 1 , \cdots , n \}} \; = \; \tau_{2^{n}}
\end{equation}
is nothing but the finite tracial state on the n-qubit $
W^{\star}$-algebra $  B( \mathcal{H}_{2}^{ \bigotimes n}) $ we saw
in the example\ref{ex:the finite trace on the n-qubits' W-star
algebra}

We are at last ready to introduce one of the main actors of this
dissertation, namely the $ W^{\star}$-algebra:
\begin{equation} \label{eq:the hyperfinite finite continuous factor}
  R \; := \;  \pi_{ \tau_{{\mathbb{Z}}}} ( A_{{\mathbb{Z}}} ) ''
\end{equation}
By eq.\ref{eq:commutativity of the local algebras of disjoint
regions} and the linearity of the GNS-representations we infer
that R is itself a factor.

Eq.\ref{eq:increasing nature of the local algebras} implies that
R is hyperfinite.

We already know that the n-qubit $W^{\star}$-algebra is a type $
I_{2^{n}}$ factor, i.e. that, taking into account the
normalization coefficient $ \frac{1}{2^{n}} $ of $ \tau_{2^{n}}
$(not considered, for convenience, in definition\ref{def:discrete
finite factor}), we have that:
\begin{equation}
  Range ( d_{B ( \mathcal{H}_{2}^{ \bigotimes n})}) \; = \; \{ 0 \,
  , \,  \frac{1}{2^{n}} \, , \, \frac{2}{2^{n}} \, , \, \cdots \,
  1 \}
\end{equation}
Since in the limit $ n \rightarrow \infty $ the dyadic rationals
fill the interval $ [ 0 \, , \, 1 ] $ it follows that:
\begin{equation}
  Range( \tau_{\mathbb{Z}} |_{ {\mathcal{P}} ( A_{\mathbb{Z}})})
  \; = \; [ 0 \, , \, 1 ]
\end{equation}
implying that R is a $ II_{1}$-factor.

Since, as we will explain more clearly in section\ref{sec:On the
rule Noncommutative Measure Theory and Noncommutative Geometry
play in Quantum Physics}, the hyperfinite $ II_{1}$-factor is
unique (obviously, remembering remark\ref{rem:identificability of
involutively-isomorphic W-star algebras}, up to $ \star
$-isomorphism) it is precisely R.

\medskip

We can, at last, face Von Neumann's confession to Birkhoff,
clarifying why the right the right noncommutative space of qubits'
sequences is not $ B ( {\mathcal{H}}_{2}^{\bigotimes \infty} ) $
but R.

\smallskip

This requires, first of all, to understand  that the equivalence
relation of definition \ref{def:equivalence of projections in a
Von Neumann algebra} is nothing but the noncommutative analogue
of the Theory of Cardinal Numbers, i.e. the the theory of the
\textbf{noncommutative cardinal numbers} describing the
infinity's degree of \textbf{noncommutative sets}.

In the words of Von Neumann:
\begin{center}
   \textit{"$ \cdots $ the whole algorithm of Cantor theory is such that the most of it
   goes over in this case. One can prove various theorems on the additivity of equivalence and the
   transitivity of equivalence, which one would normally expect, so that one can introduce a theory of alephs here, just as in set
   theory. $ \cdots $ I may call this dimension since for all matrices of the ordinary space, is nothing else but dimension"
   (Unpublished, cited in \cite{Redei-98})}

   \textit{"One can prove most of the Cantoreal properties of finite and
   infinite, and, finally, one can prove that given a Hilbert space and a ring in it , a simple ring in it,
   either all linear sets except the null sets are infinite (in which case this concept of alephs gives you
   nothing new), or else the dimensions, the equivalence classes, behave exactly like numbers
   and there are two qualitatively different cases. The dimensions either behave like integers, or else
   they behave like all real numbers. There are two subcases, namely there is either a finite top or
   there is not"
   (Unpublished, cited in \cite{Redei-98})}
\end{center}

Given a factor A, let us uniformize the commutative and
noncommutative terminology introducing the following notation:
\begin{definition}
\end{definition}
A HAS NONCOMMUTATIVE CARDINALITY EQUAL TO $ n \in {\mathbb{N}}$:
\begin{equation}
  cardinality_{NC}(A) \, = \, n \; := \; Type(A) \, = \, I_{n}
\end{equation}
\begin{definition}
\end{definition}
A HAS NONCOMMUTATIVE CARDINALITY EQUAL TO $ \aleph_{0} $:
\begin{equation}
  cardinality_{NC}(A) \, = \,\aleph_{0}   \; := \; Type(A) \, = \, I_{\infty}
\end{equation}
\begin{definition}
\end{definition}
A HAS NONCOMMUTATIVE CARDINALITY EQUAL TO $ \aleph_{1} $:
\begin{equation}
  cardinality_{NC}(A) \, = \,\aleph_{1}   \; := \; Type(A) \, \in \,
  \{ II_{1} , II_{\infty} \}
\end{equation}
A HAS NONCOMMUTATIVE CARDINALITY EQUAL TO $ \aleph_{2} $:
\begin{equation}
  cardinality_{NC}(A) \, = \,\aleph_{2}   \; := \; Type(A) \, = \,
  III
\end{equation}

The definition of noncommutative cardinality may then be extended to arbitary $ W^{\star}$-algebras rqeuiring its additiviy w.r.t. the factor decomposition.

Let us then observe  that in the commutative case (see
theorem\ref{th:cardinalities of strings and sequences}) the
passage from strings to sequences implies an increasing of one
\textbf{commutative cardinal number}; this implies that:
\begin{enumerate}
  \item it doesn't exist a bijection $ b : \Sigma^{\star}
  \mapsto \Sigma^{\infty} $, i.e.:
\begin{equation}
  cardinality( \Sigma^{\star} ) \; \neq \; cardinality( \Sigma^{\infty})
\end{equation}
  \item it does exist an injection $ i : \Sigma^{\star}
  \mapsto \Sigma^{\infty} $, i.e.:
\begin{equation}
   cardinality( \Sigma^{\star} ) \; \leq \; cardinality( \Sigma^{\infty})
\end{equation}
  \item the degree of infinity of $ \Sigma^{\infty} $ is that
  immediately successive to the the degree of infinity of $
  \Sigma^{\star} $ \footnote{Such a constraint that there does not exist intermediate degrees of infinity
  requires the assumption of the following:
\begin{axiom} \label{ax:continuum hypothesis}
\end{axiom}
CONTINUUM HYPOTHESIS:
\begin{equation}
  2^{\aleph_{0}} \; = \; \aleph_{1}
\end{equation}
that is well known to be \textbf{consistent} but
\textbf{independent} from the formal system ZFC giving foundation
to Mathematics \cite{Odifreddi-89}}, i.e.:
\begin{equation}
    \nexists \; S \: : \;  cardinality( \Sigma^{\star} ) \: <
    \:  cardinality(S)  \: < \:   cardinality( \Sigma^{\infty} )
\end{equation}
\end{enumerate}

Denoted by $ \Sigma_{NC}^{\star} $  the \textbf{$
W^{\star}$-algebra of qubits's strings} and by $
\Sigma_{NC}^{\infty} $ the \textbf{$ W^{\star}$-algebra of qubits'
sequences}, theorem\ref{th:category isomorphism at the basis of
Noncommutative Topology} requires that the same conditions hold
for noncommutative cardinality:
\begin{enumerate}
  \item
\begin{equation}
  cardinality_{NC}(  \Sigma_{NC}^{\star} ) \; \neq \; cardinality_{NC}( \Sigma_{NC}^{\infty})
\end{equation}
  \item
\begin{equation}
  cardinality_{NC}(  \Sigma_{NC}^{\star} ) \; \leq \; cardinality_{NC}( \Sigma_{NC}^{\infty})
\end{equation}
  \item
\begin{equation}
  \nexists \; A   \, factor \;  \: : \;  cardinality_{NC}( \Sigma_{NC}^{\star} ) \: <
    \:  cardinality_{NC}(A)  \: < \:   cardinality_{NC}( \Sigma_{NC}^{\infty} )
\end{equation}
\end{enumerate}
Since:
\begin{equation}
  \Sigma_{NC}^{\star} \; = \; B ( \mathcal{H}_{2}^{\bigotimes
  \star} )
\end{equation}
we have, conseguentially, that:
\begin{equation}
   cardinality_{NC}( \Sigma_{NC}^{\infty}) \, = \,  cardinality_{NC}(
   \Sigma_{NC}^{\star}) + 1 \, = \, \aleph_{1}
\end{equation}
Since we already know that $ cardinality_{NC}
(B(\mathcal{H}_{2}^{\bigotimes \infty} )    ) \; = \; \aleph_{0}
$ it follows that:
\begin{equation}
  \Sigma_{NC}^{\infty} \; = \; R \; \neq \; B(\mathcal{H}_{2}^{\bigotimes \infty} )
\end{equation}

\smallskip

\begin{remark} \label{rem:difference between the raising of commutative cardinality ad the raising of noncommutative cardinality}
\end{remark}
DIFFERENCE BETWEEN THE RAISING OF COMMUTATIVE CARDINALITY AND THE
RAISING OF NONCOMMUTATIVE CARDINALITY

The fact that the passage from a separable to a non-separable
Hilbert space involves a kind of passage from the
\textbf{discrete} to the \textbf{continuum} may be highly
misleading:

introduced the:
\begin{definition} \label{def:compuational rigged basis of the Hilbert space of qubits' sequences}
\end{definition}
COMPUTATIONAL RIGGED-BASIS OF $ {\mathcal{H}}_{2}^{\bigotimes
\infty} $:
\begin{align}
  {\mathbb{E}}_{\infty} & \; := \; \{ | \bar{x} > \, , \,
  \bar{x} \in \Sigma^{\infty} \: \}  \\
  < \bar{x} & | \bar{y} > \; = \; \delta ( \bar{x} -  \bar{y}
  ) \; \;  \bar{x} , \bar{y} \in \Sigma^{\infty} \\
\int_{ \Sigma^{\infty} } & d P_{unbiased}  | \bar{x} > < \bar{x} |
\; = \; \hat{{\mathbb{I}}}
\end{align}
theorem\ref{th:cardinalities of strings and sequences} may be
restated as:
\begin{align} \label{al:cardinalities of the computational bases}
  cardinality( & {\mathbb{E}}_{\star}) \; = \; \aleph_{0} \\
  cardinality( & {\mathbb{E}}_{\infty}) \; = \; \aleph_{1}
\end{align}
So eqs.\ref{al:cardinalities of the computational bases} are
simply the reformulation in an Hilbert space setting on the
constraint imponing the raising of one cardinal number in passing
from the \textbf{commutative cardinality} of the
\textbf{commutative space} of cbits' stings to the
\textbf{commutative space} of cbits' sequences.

It is not the constraint  imponing the raising of one cardinal
number in passing from the \textbf{noncommutative cardinality} of
the \textbf{noncommutative space} of cbits' strings to the
\textbf{noncommutative space} of cbits' sequences, namely the
following:
\begin{theorem} \label{th:on the noncommutative cardinality of strings and sequences of qubits}
\end{theorem}
ON THE NONCOMMUTATIVE CARDINALITIES OF STRINGS AND SEQUENCES OF
QUBITS:
\begin{align*}
  cardinality_{NC}(\Sigma_{NC}^{\star}) \; & = \; \aleph_{0}   \\
  cardinality_{NC}(\Sigma_{NC}^{\infty}) \; & = \;  \aleph_{1}
\end{align*}

\smallskip

\begin{remark} \label{rem:the phenomenon of continuous dimension from a logical point of view}
\end{remark}
THE PHENOMENON OF CONTINUOUS DIMENSION FROM A LOGICAL POINT OF
VIEW

The phenomenon of continuous dimension involved in the passage
from noncommutative cardinality $ \aleph_{0} $ to noncommutative
cardinality $ \aleph_{1} $ has an intuitive logic meaning: the
lost of \textbf{atomicity} of the underlying quantum logic states
the disapperance of the \textbf{atomic propositions} as stated by
corollary\ref{cor:discreteness of a factor is equivalent to the
atomicity of its quantum logic}.

We want here to give a more intuitive picture of what does it
means.

In the quantum logic of $ QL(B ( \mathcal{H}_{2}^{\bigotimes
\star} )) $ the propositions of the form:
\begin{multline}
  p_{\int_{\Sigma^{\infty}} c (\bar{x})| \bar{x} >  } \; := \; \int_{\Sigma^{\infty}} c (\bar{x})^{\star} < \bar{x} | \; \int_{\Sigma^{\infty}} c (\bar{x})  | \bar{x} >
\end{multline}
are \textbf{atomic proposition}, i.e. correspond to the
elementary statements from which all the others are generated
through the logical connectives.

This is not the case in $ QL ( \Sigma_{NC}^{\infty} ) $; one
could think that the rule of elementary propositions is therein
played by  projections of the form:
\begin{equation}
   p_{k \,  , \, \vec{n}} \; := \; \bigotimes_{i \in {\mathbb{Z}}} a_{i \,  , \, \vec{n}}
\end{equation}
\begin{equation}
  p_{i \,  , \, \vec{n}} \; :=  \;
  \begin{cases}
    \frac{1}{2} ( I_{2} \, + \, \vec{\sigma} \cdot \, \vec{n} ) & \text{if $ i = k $}, \\
     I_{2} & \text{otherwise}.
  \end{cases}
\end{equation}
(where $ I_{2} $ denotes the bidimensional identity matrix).

that, interpreting $ \Sigma_{NC}^{\infty} $ as the $
W^{\star}$-algebra of a quantum spin-1/2 chain at infinite
temperature, correspond to the statement $ << $ the spin in the $
k^{th} $ lattice site points in the direction $ \vec{n} \: >> $.

Anyway $ p_{k \, , \, \vec{n}} $  is not atomic as can be
inferred by the fact that the  $ \aleph_{1}$ noncommutative
cardinality of $ \Sigma_{NC}^{\infty} $ implies the existence of
a projection $ p^{1}_{k \,  , \, \vec{n}} \, \in \, {\mathcal{P}}(
\Sigma_{NC}^{\infty}) $ such that:
\begin{equation}
  d_{\Sigma_{NC}^{\infty}} ( p^{1}_{k \,  , \, \vec{n}} ) \; = \; \frac{d_{\Sigma_{NC}^{\infty}} ( p_{k \,  , \, \vec{n}} )}{2}
\end{equation}
and so:
\begin{equation}
  p^{1}_{k \,  , \, \vec{n}} \; \trianglelefteq \; p_{k \, , \, \vec{n}}
\end{equation}
But then there exist a projection $ p^{2}_{k \,  , \, \vec{n}} \,
\in \, {\mathcal{P}}( \Sigma_{NC}^{\infty}) $ such that:
\begin{equation}
  p^{2}_{k \,  , \, \vec{n}} \, \trianglelefteq \, p_{k \, , \,
  \vec{n}} \; and \; p^{2}_{k \,  , \, \vec{n}} \,
  \sim_{\Sigma_{NC}^{\infty}} \, p^{1}_{k \,  , \, \vec{n}}
\end{equation}
and hence:
\begin{equation}
 d_{\Sigma_{NC}^{\infty}} ( p^{2}_{k \,  , \, \vec{n}} ) \; = \;  d_{\Sigma_{NC}^{\infty}} ( p^{1}_{k \,  , \, \vec{n}} )  \; = \; \frac{d_{\Sigma_{NC}^{\infty}} ( p_{k \,  , \, \vec{n}} )}{2}
\end{equation}
The projection $ p^{2}_{k \,  , \, \vec{n}} $ is thus a non-zero
projection strictly smaller than $ p_{k \, , \, \vec{n}} $, so
that, conseguentially, $ p_{k \, , \, \vec{n}} $ is not an atom.

\medskip

Let us finally, as promized, discuss Walter Thirring's reasoning
that lead him to give a negative answer to the
\textbf{Separability Issue} despite of Shr\"{o}dinger opposite
position \cite{Thirring-01}:
\begin{center}
   \textit{"However such an opinion means that Shr\"{o}dinger did not get the main message of
   Von Neumann's celebrated paper on infinite tensor products. There he shows
   that the corresponding operator algebras are highly reducibly represented in this vast non-separable
   space and there are many (inequivalent) subrepresentations which act on a separable subspace"}
\end{center}

What Thirring is speaking about is the analysis he explicitly
reports in the section1.4 of \cite{Thirring-83}:

the cases \textbf{case-I.A} and \textbf{case-I} discussed in view
of definition\ref{def:non zero inner product in the infinite
tensor product Hilbert space} and definition\ref{def:zero inner
product in the infinite tensor product Hilbert space} give rise
to the following two equivalence relations inside $
\mathcal{H}_{2}^{\bigotimes \infty}$:

\begin{definition} \label{def:strong equivalence relation on the qubits' sequences Hilbert space}
\end{definition}
$ | \psi > , | \phi > \; \in \; \mathcal{H}_{2}^{\bigotimes
\infty} $ ARE STRONGLY-EQUIVALENT:
\begin{equation}
   | \psi > \, \sim_{S} \, | \phi > \; := \; \prod'_{n} < \psi_{n}
   | \phi_{n} > \, \rightarrow \, c \neq 0
\end{equation}
\begin{definition} \label{def:weak equivalence relation on the qubits' sequences Hilbert space}
\end{definition}
$ | \psi > , | \phi > \; \in \; \mathcal{H}_{2}^{\bigotimes
\infty} $ ARE WEAKLY-EQUIVALENT:
\begin{equation}
   | \psi > \, \sim_{W} \, | \phi > \; := \; \prod'_{n} < \psi_{n}
   | \phi_{n} > \, \rightarrow \, c > 0
\end{equation}
where the symbol $ \prod' $ means that any finite number of
factors 0 are to be left out.

Both the quotient spaces $ \frac{\mathcal{H}_{2}^{\bigotimes
\infty}}{\sim_{S}} $ and $ \frac{\mathcal{H}_{2}^{\bigotimes
\infty}}{\sim_{W}} $ are linear spaces.

Furthermore one has that:
\begin{equation}
    cardinality (  \frac{\mathcal{H}_{2}^{\bigotimes
\infty}}{\sim_{W}} ) \; > \; \aleph_{0}
\end{equation}
\begin{equation}
   [ | \psi > ]_{W} \, \neq  \,  [ | \phi > ]_{W} \; \Rightarrow \;
   < \psi | \phi > \, = \, 0 \; \; \forall | \psi > , | \phi > \in
   {\mathcal{H}}_{2}^{\bigotimes \infty}
\end{equation}
\begin{equation}
   [  [ | \psi > ]_{W} ]_{S} \, \neq  \, [  [ | \phi > ]_{W} ]_{S}
   \; \Rightarrow \;  < \psi | \phi > \, = \, 0 \; \; \forall | \psi > , | \phi > \in
   {\mathcal{H}}_{2}^{\bigotimes \infty} :  [ | \psi > ]_{W} \, =  \,  [ | \phi > ]_{W}
\end{equation}

Adopting the notation used in example\ref{ex:the hyperfinite
finite continuous factor}, the key point is that
\begin{theorem} \label{th:on the high reducibility of the representations of the chain spin algebra over the qubits' sequences' Hilbert space}
\end{theorem}
ON THE HIGH REDUCIBILITY OF THE REPRESENTATIONS OF $
A_{{\mathbb{Z}}} $ ON $ {\mathcal{H}}_{2}^{\bigotimes \infty} $

\begin{hypothesis}
\end{hypothesis}
$ \pi $ representation of $ A_{{\mathbb{Z}}} $ over $
 {\mathcal{H}}_{2}^{\bigotimes \infty} $

\begin{thesis}
\end{thesis}
\begin{equation}
  [ | \psi > ] _{S}  \text{ is separable }  \;
  \; \forall | \psi > \in  {\mathcal{H}}_{2}^{\bigotimes \infty}
\end{equation}
\begin{equation}
   \pi ( [ | \psi > ] _{S} )  \; \subseteq \;  [ | \psi > ] _{S} \;
  \; \forall | \psi > \in {\mathcal{H}}_{2}^{\bigotimes \infty}
\end{equation}
\begin{equation}
  \text{ the sub-representations arising from different weak equivalence classes are inequivalent}
\end{equation}

According to Thirring  theorem\ref{th:on the high reducibility of
the representations of the chain spin algebra over the qubits'
sequences' Hilbert space} implies that one has to answer
negatively to the Separability issue.

According to him, one can consistentely assume that qubits
sequences are described by rays of the not-separable Hilbert
space $ {\mathcal{H}}_{2}^{\bigotimes \infty}$.

Simply, as always happens in the limit of infinite degrees of
freedom, different representations of the observables' algebra may
become inequivalent so that one has to select the representation
suitable to the physical situation he is studying.

Since such a representation lives on a separable subspace of $
{\mathcal{H}}_{2}^{\bigotimes \infty}$ everything seems ok.

This way of recovering an Hilbert space axiomatization of Quantum
Mechanics is, anyway, apparent:

one start a priori with the observable's algebra and returns to
the usual Hilbert space formalism only a posteriori, after a
suitable representation is used.

Such an attitude to the algebraic approach (precisely codified in
the section1.8 of \cite{Strocchi-85}), though FAPP completelly
equivalent to the one we support, is philosophically disappealing
since it founds on axioms based on  posteriori derived quantities.

According to theorem\ref{th:Sakai theorem} one can, in a
completelly equivalent way, forget concrete algebras of operators
on Hilbert spaces, forget Hilbert spaces themselves, and speak
only about $ W^{\star}$-algebras.

Futhermore there is no reason to represent these $
W^{\star}$-algebras, returning in this way to an Hilbert space
setting.

Such a view point corresponds, substantially, to give a positive
answer to the Separability Issue, to infer from that that an
Hilbert space axiomatization in terms of the not-separable
Hilbert space $ {\mathcal{H}}_{2}^{\bigotimes \infty} $ is not
acceptable and, conseguentially, to conclude that, as to qubits'
sequences, one has to give up the idea to remain inside the
boundaries of an Hilbert space axiomatization.

\medskip

We would like to end this section clarifying more  properly the
concept of qubit, through the following:
\begin{remark} \label{rem:the noncommutative combinatory information and the definition of the qubit}
\end{remark}
THE NONCOMMUTATIVE COMBINATORY INFORMATION AND THE DEFINITION OF
THE QUBIT

It should be clear, at this point, the the \textbf{qubit} may be
defined, in a conceptually more satisfying way, in the following
way:

Given a \textbf{noncommutative set} A let us introduce the
following notion:
\begin{definition} \label{def:noncommutative combinatory information}
\end{definition}
NONCOMMUTATIVE COMBINATORY INFORMATION:
\begin{equation}
  I_{NC \; combinatory} (A) \; := \; \log_{2} cardinality_{NC} (A)
\end{equation}
By theorem\ref{th:on the noncommutative cardinality of strings and
sequences of qubits} we have, via axiom\ref{ax:continuum
hypothesis}, that:
\begin{equation}
  I_{NC \; combinatory}  ( \Sigma_{NC}^{\star} ) \; = \; \log_{2} ( \aleph_{0} )
\end{equation}
\begin{equation}
   I_{NC \; combinatory}  ( \Sigma_{NC}^{\infty} ) \; = \; \log_{2} ( \aleph_{1}
   ) \; = \; \aleph_{0}
\end{equation}
Then we can at last define precisely the qubit as:
\begin{definition} \label{def:qubit}
\end{definition}
QUBIT:
\begin{equation}
  \text{ 1 qubit } \; := \; I_{NC \; combinatory} (\Sigma_{NC})
\end{equation}
where:
\begin{equation}
 \Sigma_{NC} \; := \; {\mathcal{B}} ( {\mathcal{H}}_{2} )
\end{equation}
is the \textbf{noncommutative binary alphabet}
\newpage
\section{On the rule Noncommutative Measure Theory and Noncommutative
Geometry play in Quantum Physics} \label{sec:On the rule
Noncommutative Measure Theory and Noncommutative Geometry play in
Quantum Physics}

In the previous section, precisely in the  remark\ref{rem:the
metaphore by which we can speak about noncommutative sets from
within ZFC} we have introduced the Noncommutative Metaphore
allowing to speak about noncommutative sets from inside the ZFC
axiomatization of commutative set theory.

By theorem\ref{th:category isomorphism at the basis of
Noncommutative Topology} we have then given foundations to
Noncommutative Topology, i.e. the analysis of topological
structure on noncommutative sets.

The next floor in the construction of the noncommutative tower is
the introduction of Noncommutative Measure Theory
\cite{Streater-95},\cite{Streater-00a}.

\begin{definition} \label{def:algebraic probability space}
\end{definition}
ALGEBRAIC PROBABILITY SPACE:   $ ( \, A \, , \, \omega \, ) $,
where:
\begin{itemize}
  \item A is a  Von Neumann algebra
  \item $ \omega $ is a state on A
\end{itemize}
The  notion of  \textbf{algebraic probability space}  is a
noncommutative generalization of the notion of \textbf{classical
probability space}  as is implied by the following:
\begin{theorem} \label{th:Gelfand isomorphism at W-star algebraic level}
\end{theorem}
GELFAND'S ISOMORPHISM AT $ W^{\star}$-ALGEBRAIC LEVEL:
\begin{enumerate}
  \item a generic \textbf{classical probability space}   $ ( \; X \, , \, \mu \; ) $
 may be equivalententely seen as the algebraic probability space $ ( \,  L^{\infty} ( X , \mu ) \, , \, \omega_{\mu} \,)
  $, with:
\begin{equation}
\begin{split}
  \omega_{\mu} ( A ) & \in  S ( A ) \\
  \omega_{\mu} ( a ) & \; := \; \int_{X} a ( x ) d \mu (x)
\end{split}
\end{equation}
  \item  given a generic \textbf{abelian algebraic probability space}
 $ ( \, A \, , \, \omega \, ) $ there exist a \textbf{classical probability space}
  $ ( \; X \, , \, \mu \; ) $ and a
 $ \star $ - isomorphism $ {\mathcal{I}}_{GELFAND} \, : A \rightarrow  L^{\infty} ( X , \mu \,) $, namely the \textbf{Gelfand isomorphism}, under which the state $ \omega  \in S(A) $
 corresponds to the state $ \omega_{\mu} \in L^{\infty} ( X , \mu )$.
\end{enumerate}

\medskip

\begin{definition} \label{def:quantum probability space}
\end{definition}
NONCOMMUTATIVE PROBABILITY SPACE

a non-abelian algebraic probability space.

\smallskip

Given a finite factor A:
\begin{definition} \label{def:unbaised algebraic probability space}
\end{definition}
UNBIASED ALGEBRAIC PROBABILITY SPACE ON A:

the algebraic probability space $ ( A \, , \, \tau_{unbiased} )
$, where:
\begin{equation}
  \tau_{unbiased} \; := \; \tau_{A}
\end{equation}

\medskip

Let us now clarify an important point.

Given a Von Neumann algebra $ A \; \subseteq \; B({\mathcal{H}})
$:
\begin{definition} \label{def:normal states on a Von Nuemann algebra}
\end{definition}
NORMAL STATES ON A:
\begin{multline}
  S(A)_{n}  \; := \; \{ \omega \in  S(A) \; : \; \sup_{\alpha}
  \omega( a_{\alpha} ) \; = \; \omega ( \sup_{\alpha}  a_{\alpha} ) \\
   \forall \text{ bounded increasing net } \{ a_{\alpha} \} \text{ in A } \}
\end{multline}
The importance of normal states is owed to a key feature of them,
to formalize which let us first introduce the following sequence
of norms on $ B ({\mathcal{H}}) $:
\begin{definition} \label{def:sequence of operatorial norms}
\end{definition}
$n^{th}$ OPERATORIAL NORM ON $ B ({\mathcal{H}}) $
\begin{equation}
  \| a \|_{n} \; := \; (Tr | a |^{n}) ^{\frac{1}{n}}
\end{equation}
where:
\begin{equation} \label{eq:modulus of a bounded operator}
  | a | \; := \; \sqrt{a ^{\star} a }
\end{equation}
is the \textbf{modulus} of the operator a, whose name is owed to
its rule in the polar decomposition $ a  \; = \; U | a | $  by
which any bounded operator on $ {\mathcal{H}} $ can be expressed
as a product of a partial isometry U times the modulus of a, as
in the usual polar decomposition $ z \; = \; e^{ i arg(z) } | z |
$ of complex numbers.

The definition\ref{def:sequence of operatorial norms} contains the
usual norm $ \| \cdot \| $ of the $ W^{\star} $-algebraic
structure of  $ B ({\mathcal{H}}) $ as the $ n = \infty $ case:
\begin{equation}
 \| a \|_{\infty} \; = \; \| a \| \; \; a \in B ({\mathcal{H}})
\end{equation}
Introduced the subspace of the finite-rank bounded operators on $
{\mathcal{H}} $:
\begin{equation}
  {\mathcal{E}}( {\mathcal{H}} ) \; := \; \{ a \in B
  ({\mathcal{H}}) \, : \, dim ( Range(a) ) < + \infty \}
\end{equation}
let us introduce the following sequence of operators' classes:
\begin{definition}  \label{def:classes of bounded operator}
\end{definition}
n-CLASS BOUNDED OPERATORS ON  $ {\mathcal{H}} $:
\begin{equation}
   {\mathcal{C}}_{n}( {\mathcal{H}} ) \; := \; completion( {\mathcal{E}}( {\mathcal{H}}
   ) , \| \cdot \|_{n} )
\end{equation}
In particular, the operators in $ {\mathcal{C}}_{1}(
{\mathcal{H}} ) $ are called \textbf{trace class}, those in $
{\mathcal{C}}_{2}( {\mathcal{H}} ) $ \textbf{Hilbert-Schmidt},
while the operators in $  {\mathcal{C}}( {\mathcal{H}} ) \; := \;
{\mathcal{C}}_{\infty}( {\mathcal{H}} ) $ are called
\textbf{compact} or \textbf{infinitesimal}, this last name being
owed to the rule they play, as we will see, in Noncommutative
Differential Calculus.

Let us then introduce the following notion:
\begin{definition}  \label{density operators}
\end{definition}
DENSITY OPERATORS ON $ {\mathcal{H}} $:
\begin{equation}
  {\mathcal{D}}( {\mathcal{H}} ) \; := \; \{ \rho \in {\mathcal{C}}_{1}( {\mathcal{H}} ) \bigcap  ({\mathcal{B}}( {\mathcal{H}}))_{+} \, : \, Tr \rho =1  \}
\end{equation}

The key feature of the normal states over a Von Neumann algebra $
A \; \subseteq \; B({\mathcal{H}}) $ may then be stated as
follows:
\begin{theorem}  \label{th:on the density operators of normal states}
\end{theorem}
ON THE DENSITY OPERATORS OF NORMAL STATES:
\begin{equation}
  \omega \in S(A)_{n} \; \Leftrightarrow \; ( \exists \rho_{\omega}
  \in  {\mathcal{D}}( {\mathcal{H}} ) \; : \; \omega( a ) \, = \,
  Tr (  \rho_{\omega} a ) \, \; \; \forall a \in A )
\end{equation}
and is important essentially owing to the following:
\begin{theorem} \label{th:normality of the states of noncommutative spaces with finite noncommutative cardinality}
\end{theorem}
NORMALITY OF THE STATES OF NONCOMMUTATIVE SPACES WITH FINITE
NONCOMMUTATIVE CARDINALITY
\begin{equation}
  cardinality_{NC} (A) \, \in \, {\mathbb{N}} \; \Rightarrow \;
  S(A) _{n} \, = \, S(A)
\end{equation}

\smallskip

\begin{example} \label{ex:n qubits' noncommutative probability spaces}
\end{example}
n QUBITS' NONCOMMUTATIVE PROBABILITY SPACES

Given an n qubit probability space $ ( B
({\mathcal{H}}_{2}^{\bigotimes n} ) \, , \, \omega ) $
theorem\ref{th:normality of the states of noncommutative spaces
with finite noncommutative cardinality} implies that everything
can be rephrased in terms of the more popular couple $ (
{\mathcal{H}}_{2}^{\bigotimes n} \, , \, \rho_{\omega} ) $.

So, in this case when can make any noncommutative-probabilistic
analysis avoiding all the algebraic machinery,e.g. according the
lines developed in \cite{Parthasarathy-92}.

This applies, in particular, for the n qubit unbaised
noncommutative probability space $ ( B
({\mathcal{H}}_{2}^{\bigotimes n} ) \, , \, \tau_{unbiased}) $.

Given an  \textbf{algebraic random variable}  $ a \in A $ over
the \textbf{algebraic probability space}  $ ( A , \omega )$:
\begin{definition}
\end{definition}
$ n^{th} $ MOMENT OF a:
\begin{equation}
  M_{n} (a) \; := \; \omega ( a ^{n} )
\end{equation}
Of particular rilevance are the following:
\begin{definition}
\end{definition}
EXPECTATION VALUE OF a:
\begin{equation}
  E(a) \; := \; M_{1} (a)
\end{equation}
\begin{definition}
\end{definition}
VARIANCE OF a:
\begin{equation}
    Var(a) \; := \; \sqrt{ E(a ^{2}) - ( E(a) )^{2} }
\end{equation}

The information  contained in the moments'- sequence of a may be
usefully incorporated in the following:
\begin{definition}
\end{definition}
CHARACTERISTIC FUNCTION OF a:

$ ZQ_{a} \; : \; Convergence-circle(a) \; \rightarrow \;
{\mathbb{C}} $
\begin{equation}
   ZQ_{a}(t) \; := \; \sum_{n=0}^{\infty} M_{n} (a) \frac{ t^{n}}{ n !}
\end{equation}
where $ \; Convergence-circle(a) \; := \{ z \in {\mathbb{C}} \, :
\, | z | \leq R_{Convergence-circle(a)} \} $ is the circle in the
complex plane with centre the origin inside which the sum
converges.

When $ R_{Convergence-circle(a)} \; > \; 0 $ we have that:
\begin{equation}
  M_{n} (a) \; = \;  \frac{d^{n} ZQ_{a}(t) }{ d t^{n}} ( t=0 )
\end{equation}

\medskip

Given two \textbf{algebraic random variables} a e b on the
\textbf{algebraic probability space} $ ( A , \omega )$:
\begin{definition}
\end{definition}
a and b are INDEPENDENT:
\begin{equation}
  E( a^{n} b^{m} ) \; = \;  E( a^{n} ) E( b^{m} ) \; \; \forall n , m
  \in {\mathbb{N}}
\end{equation}

\medskip

Given two sets $ Q_{1} $ and $ Q_{2} $ of algebraic random
variables on \textbf{algebraic probability space} $ ( A , \omega
)$:
\begin{definition}
\end{definition}
$ Q_{1} $ and $ Q_{2} $  ARE INDEPENDENT:

$ q_{1} $ e $ q_{2} $ are independent $ \forall  q_{1} \in Q_{1}
\, , \, \forall q_{2} \in Q_{2} $

\bigskip

We have the following:
\begin{theorem} \label{th:dependence of noncommutating random variables}
\end{theorem}
DEPENDENCE OF NONCOMMUTATING RANDOM VARIABLES:
\begin{equation}
  a \,  and \,  b \; \; independent \; \; \Rightarrow  \; \; [ a , b
  ] \, = \, 0
\end{equation}

\bigskip
Given a self-adjoint algebraic random variable $ a \in A_{sa} $:
\begin{definition}
\end{definition}
CLASSICAL PROBABILITY MEASURE OF a:

the classical probability measure $ \mu_{a} $ induced by $ \omega
$ on the \textbf{spectrum} Sp(a) of a

\begin{definition}
\end{definition}
RESULT OF A MEASUREMENT OF a:

The classical random variable $ v_{a} $ on the \textbf{spectrum}
Sp(a) of a  having $ \mu_{a} $ as classical probability
distribution.

\medskip

Given two noncommutative probability spaces $ ( A_{1} \, , \,
\omega_{1} ) $ and $ ( A_{2} \, , \, \omega_{2} ) $ let us
introduce the following notion:
\begin{definition}
\end{definition}
TENSORIAL PRODUCT  OF $ ( A_{1} \, , \, \omega_{1} ) $ AND $ (
A_{2} \, , \, \omega_{2} ) $:
\begin{equation}
  ( A_{1} \, , \, \omega_{1} ) \, \bigotimes \, ( A_{2} \, , \, \omega_{2}
  ) \; := \; ( A_{1} \bigotimes A_{2} \, , \, \omega_{1} \cdot
  \omega_{2} )
\end{equation}
where:
\begin{equation}
  \omega_{1} \cdot \omega_{2} ( a_{1} \bigotimes  a_{2} ) \; := \;
   \omega_{1} ( a_{1} ) \omega_{2} ( a_{2} ) \; \; \forall a_{1}
   \in A_{1} \, , \, \forall a_{2}  \in A_{2}
\end{equation}

\medskip

Clearly we have the following:

\begin{theorem} \label{th:automatic independence on tensorial products}
\end{theorem}

AUTOMATIC INDEPENDENCE ON TENSORIAL PRODUCTS:

$ a_{1} \bigotimes {\mathbb{I}} $ and $ {\mathbb{I}} \bigotimes
a_{2} $ are  \textbf{independent algebraic random variables} on $
( A_{1} \, , \, \omega_{1} ) \, \bigotimes \, ( A_{2} \, , \,
\omega_{2} ) \; \; \forall a_{1} \in A_{1} \, , \, \forall a_{2}
\in A_{2} $

Given an \textbf{algebraic probability space} $ ( A \, , \, \omega
) $:
\begin{definition}
\end{definition}
$ ( A \, , \, \omega ) $ IS FACTORIZABLE:
\begin{equation}
  \exists A_{1} , A_{2} \text{ $W^{\star}$-algebras } \, , \,  \omega_{1} \in S( A_{1} ) \, , \,  \omega_{2} \in S( A_{2}
  ) \; : \; ( A \, , \, \omega ) \, = \, ( A_{1} \, , \, \omega_{1} ) \, \bigotimes \, ( A_{2} \, , \, \omega_{2})
\end{equation}
\begin{definition}
\end{definition}
$ ( A \, , \, \omega ) $ IS ENTANGLED:

it is not factorizable ed A is not an $ I_{2} $ - factor.

\bigskip

Let us consider, now, the following problem:

\textbf{given a noncommutative probability space is it possible to
approximate it through a classical probability space up to a given
perturbative order ?}.

\bigskip

To assign an  \textbf{algebraic random variable} a on the
\textbf{algebraic probability space} $ ( A \, , \omega \, ) $ is
equivalent to assign the moments'- sequence $ \{ M_{n} (a) \}_{n
\in {\mathbb{N}}} $;

\smallskip

Given a \textbf{noncommutative probability space}  $ ( A \, ,
\omega \, ) $ and a collection  $ Q \; := \; \{ q_{1} \, , \,
\cdots q_{n} \} $ of \textbf{noncommutative random variables} on
it, one could think of trying to  approximate the quantum random
variables contained in Q by classical random variables reproducing
correctly the moments up to a certain order.

This may be formalized in the following way:
\begin{definition} \label{def:approssimazione classica ad un dato ordine}
\end{definition}
CLASSICAL APPROXIMATION OF Q UP TO THE $ n^{th} $ ORDER:

a bijection $ Ap : Q \mapsto C $, where C  is a collection of
classical random variables on a suitable classical probability
space $ ( M \, ,  \, P ) $ such that:
\begin{equation}
E( q_{1}^{i_{1}} \, \cdots  \, q_{n}^{i_{n}})    \; = \; \int_{M}
d P Ap(q_{1})^{i_{1}}   \, \cdots Ap(q_{n})^{i_{n}}  \; \; i_{1}
, \cdots  i_{n} \in {\mathbb{N}} \; : \, \sum_{k=1}^{n} i_{k} \;
\leq \; n
\end{equation}

Given a classical approximation  $ Ap : Q \mapsto C $ of Q up to
the  $n^{th}$ order, let us introduce the following notion:
\begin{definition}
\end{definition}
CHARACTERISTIC FUNCTION ASSOCIATED TO Ap:

$ ZC_{Ap} \, : \, {\mathbb{C}}^{n}  \; \rightarrow \;
{\mathbb{C}} $
\begin{equation}
  ZC_{Ap} ( t_{1} , \cdots , t_{n} ) \; := \; \int_{M}
d P  e^{  \sum_{i=1}^{n} t_{i} Ap(q_{i})}
\end{equation}
We will have, then, clearly that:
\begin{equation}
  E( q_{1}^{i_{1}} \, \cdots  \, q_{n}^{i_{n}}) \; = \;  \frac{d^{\sum_{k=1}^{n} i_{k}} ZC_{Ap} ( t_{1} , \cdots , t_{n} ) } {d t_{1}^{i_{1}} \, \cdots  \, d
  t_{n}^{i_{n}}} ( t_{1} = 0, \cdots , t_{n} = 0  )  \; \; i_{1}
, \cdots  i_{n} \in {\mathbb{N}} \: : \: \sum_{k=1}^{n} i_{k} \;
\leq \; n
\end{equation}
It appears , then, clear that a classical approximation up to the
$ n^{th} $ order of Q involves precisely the consideration of a
series-expansion of the associated characteristic function up to
the  $ n^{th} $ order.

\medskip

Let us introduce, now, the following fundamental notion:
\begin{definition}
\end{definition}
Q IS  IRRIDUCIBLE TO CLASSICAL PROBABILITY UP TO THE $ n^{th} $
ORDER :

it doesn't exist a classical approximation of Q up to the $ n^{th}
$ order.

\bigskip

Demanding to \cite{Ohya-Petz-93}, \cite{Benatti-93} for any
suppletive notion, let us now briefly recall here the basic
notions concerning  the theory of noncommutative dynamical
systems.

Given an \textbf{algebraic probability space } $ ( \, A \, , \,
\omega \, )$:
\begin{definition} \label{endomorphism of an algebraic probability space}
\end{definition}
ENDOMORPHISMS OF $ ( \, A \, , \, \omega \, ) $:
\begin{equation}
  END(  \, A \, , \, \omega \, ) \; := \; \{  \tau \, : \, A \rightarrow A \; \text{ surjective $ \star $ -
  morphism of A} \; : \; \omega \, \in \, S_{\tau} ( A ) \}
\end{equation}

where $ S_{\tau} ( A ) $ is the set of the $ \tau $ - invariant
states on A.
\begin{definition} \label{automorphism of an algebraic probability space}
\end{definition}
AUTOMORPHISMS OF $ ( \, A \, , \, \omega \, ) $:
\begin{equation}
  AUT( \, A \, , \, \omega \, ) \; := \; \{  \tau \, : \, A \rightarrow A  \text{ bijective endomorphism  of } ( \, A \, , \, \omega \,
  ) \}
\end{equation}
\begin{definition} \label{algebraic dynamical system}
\end{definition}
ALGEBRAIC DYNAMICAL SYSTEM :

$ ( \, A \, , \, \omega \, , \, \tau ) $ such that:
\begin{itemize}
  \item  $ ( \, A \, , \, \omega \, ) $  is an algebraic
  probability space
  \item $\tau$  is an endomorphism of $ ( \, A \, , \, \omega \, ) $
\end{itemize}

\begin{remark}
\end{remark}
ON THE PASSAGE FROM THE HEISENBERG-PICTURE TO THE SCHR\"{O}DINGER-PICTURE OF DYNAMICS:

We have implicitely assumed Heisenberg's picture of
dynamics  (in which states are fixed while observable evolve with
time).

The passage to the Schr\"{o}dinger picture (in which
observables are fixed while states evolve with time) is, anyway, straightforward:

given a $ \star $-morphism C from a $ W^{\star}$-algebra B to a $
W^{\star}$-algebra A:
\begin{definition} \label{def:dual of an involutive morphism}
\end{definition}
DUAL OF C:

the map $ C_{\star} \, : S(A) \, \rightarrow \,  S(B) $:
\begin{equation}
  (  C_{\star} \alpha ) (b) \; := \; \alpha ( C b ) \; \; \forall
  \alpha \in S(A) \, , \, \forall b \in B
\end{equation}

\medskip

Given an algebraic dynamical system  $ ( \, A \, , \, \omega \, ,
\, \tau ) $:
\begin{definition} \label{reversible algebraic dynamical system}
\end{definition}
$ ( \, A \, , \, \omega \, , \, \tau ) $ IS REVERSIBILE

$\tau$ is  an automorphism of $ ( \, A \, , \, \omega \, ) $.

\medskip

\begin{definition} \label{def:quantum dynamical system}
\end{definition}
$ ( \, A \, , \, \omega \, , \, \tau ) $ IS NONCOMMUTATIVE

$ ( \, A \, , \, \omega \; ) $ is noncommutative

\smallskip

The notion of  \textbf{(reversibile) algebraic dynamical system}
is a noncommutative generalization of the notion of
\textbf{(reversibile) classical dynamical system}.

In fact:

\begin{enumerate}
  \item $( \, X  \, , \, {\mathcal{F}} \,  , \, \mu  \, , \,  T \, ) $ (reversibile) classical dynamical system $\Rightarrow \; ( \,  L^{\infty} ( X , \mu
  )  ,  \omega_{\mu} , \tau_{T}) $ (reversibile) algebraic dynamical system

where:
\begin{equation} \label{eq:endomorfismo associato a una mappa classica}
  \begin{split}
  \tau_{T} \; \; &  \text{automorphism of $L^{\infty} ( X , \mu
  )$} \\
  \tau_{T} & ( a ) \; \equiv \;  a \cdot T^{-1}
\end{split}
\end{equation}
  \item  given a  (reversibile)  algebraic dynamical system
 $ ( \, A , \omega , \tau \, )$ with  A   abelian $W^{\star}$-algebra ,
then equation eq.\ref{eq:endomorfismo associato a una mappa
classica} univoquely individualizes an endomorphism (automorphism)
T of the assoociated classical probability space.
\end{enumerate}

\medskip

Such a result may be enunciated in the abstract language of
Categories' Theory as the following:
\begin{theorem} \label{th:category isomorphism at the basis of Noncommutative Probability}
\end{theorem}
CATEGORY EQUIVALENCE AT THE BASIS OF NONCOMMUTATIVE PROBABILITY

The \textbf{category} having as \textbf{objects} the
\textbf{classical probability spaces} and as \textbf{morphisms}
the  \textbf{endomorphisms (automorphisms)} of such spaces is
equivalent to the \textbf{category} having as objects the
\textbf{abelian algebraic probability  spaces} and as\textbf{
morphisms}  the \textbf{endomorphisms (automorphisms)} of such
spaces.

\medskip
Let us now analyze the symmetries of algebraic dynamical systems:
on discussing derivations on a $ C^{\star}$-algebra A  we have
already met strongly-continuous one-parameter subgroups of AUT(A).

Given, in general, a Lie group G:
\begin{definition} \label{def:set of the automorhisms' groups of a C-star algebra representing a Lie group}
\end{definition}
SET OF THE AUTOMORHISMS' GROUPS OF A REPRESENTING G:
\begin{equation}
  GR-AUT( G \, , \, A) \; := \, \{ \{ \alpha_{g} \}_{g \in G} \,
  \text{ strongly-continuous subgroup of AUT(A) } \}
\end{equation}
\begin{definition} \label{def:set of the inner automorhisms' groups of a C-star algebra representing a Lie group}
\end{definition}
SET OF THE INNER AUTOMORHISMS' GROUPS OF A REPRESENTING G:
\begin{equation}
  GR-INN( G \, , \, A) \; := \, \{ \{ \alpha_{g} \}_{g \in G} \, \in \, G-AUT( G \, , \, A)  \; : \alpha_{g} \in INN(A) \, \forall g \in G \}
\end{equation}
\begin{definition} \label{def:set of the outer automorhisms' groups of a C-star algebra representing a Lie group}
\end{definition}
SET OF THE OUTER AUTOMORHISMS' GROUPS OF A REPRESENTING G:
\begin{equation}
  GR-OUT( G \, , \, A) \; := \, \frac{GR-AUT( G \, , \, A)}{GR-INN( G \, , \, A)}
\end{equation}
Mackey's notion of \textbf{system of imprimitivity} as well as
its unsharp generalization, i.e the notion of \textbf{generalized
system of imprimitivity} often also called \textbf{system of
covariance} (cfr. e.g the $ 3^{th}$ chapter of
\cite{Prugovecki-92} and the section2.3 of \cite{Holevo-99}), may
be generalized to the Quantum Probability's framework in the
following way:
\begin{definition} \label{def:covariance system on a W-star algebra w.r.t. a Lie Group}
\end{definition}
COVARIANCE SYSTEM ON A W.R.T. G:

a couple $ ( E \, , \,  \{ \alpha_{g} \}_{g \in G} ) $ such that:
\begin{itemize}
  \item $  E \, \in \, \stackrel{\circ}{MAP} ( X , A_{+} ) $  is
  a POVM on A
  \item
\begin{equation}
  \{ \alpha_{g} \}_{g \in G} \; \in \; GR-AUT( G \, , \, A)
\end{equation}
  \item
\begin{equation}
  \alpha_{g} \, E(  B ) \; = \; E ( g \, B ) \; \; \forall B \in
  HALTING(E) \, , \, \forall g \in G
\end{equation}
\end{itemize}

\smallskip

\begin{remark} \label{rem:quantum physics versus noncommutative sets}
\end{remark}
QUANTUM PHYSICS VERSUS NONCOMMUTATIVE SETS

He have seen in section\ref{sec:Why to treat sequences of qubits
one has to give up the Hilbert-Space Axiomatization of Quantum
Mechanics} how Von Neumann's investigations on the foundations of
Quantum Mechanics led him to implicitely introduce Noncommutative
Set Theory.

This was the starting point of a foundational school aimed at
explicating the transition from Classical Physics to Quantum
Physics in terms of the ansatz:
\begin{equation*}
  \text{commutative spaces} \; \rightarrow \; \text{noncommutative spaces}
\end{equation*}
The school looking at the modification of Probability Theory
involved in such an ansatz as the root of the quantum peculiarity
is called Quantum Probability.

Such a position is exemplified by the following words of Raymond
F. Streater \cite{Streater-00a}:
\begin{center}
   \textit{"It took some time before it was understood that quantum theory is a
   generalization of probability, rather than a modification of the laws of mechanics.
   This was not helped by the term quantum \emph{mechanics}; more, the Copenhagen interpretation
   is given in terms of probability, meaning as understood at the time. Bohr has said that
   the interpretation of microscopic measurements must be done in terms of classical terms, because the
   measuring instruments are large, and are therefore describe by classical laws. It is
   true, that the springs and cogs making up a measuring instrument themselves obey classical laws;
   but this does not mean that the \emph{information} held on the instrument, in the numbers indicated by the
   dials, obey classical statistics. If the instruments faithfully
   measures an atomic variable, then the numbers indicated by the
   dials should be analyzed by quantum probability, however large
   the instruments is"}
\end{center}

Such a viewpoint pervaded Richard Feynman's
thought\cite{Feynman-Hibbs-65}:
\begin{center}
   \textit{"But far more fundamental was the discovery that in nature the laws of combining
   probabilities were not those of the classical probability
   theory of Laplace. The quantum-mechanical laws of the physical
   world approach very closely the laws of Laplace as the size of
   the objects involved in the experiment increases. Therefore the
   laws of probabilities which are conventionally applied are
   quite satisfactory in analyzing the behaviour of the roulette
   wheel but no the behaviour of a single electron or a photon of
   light"}
\end{center}
being at the heart of his path-integral formalism
based on the observation that the \textbf{Law of Composed
Probabilities}:
\begin{equation} \label{eq:law of composed probabilities}
  P ( x_{1} | x_{3} ) \; = \; \sum_{x_{2} \in {\mathcal{E}}  }  P ( x_{1} | x_{2} )  P ( x_{2} | x_{3} )
\end{equation}
(with $ {\mathcal{E}} $ denoting the space of events) doesn't
hold in Quantum Probability where it is replaced by the
\textbf{Law of Composed Probability-amplitudes}:
\begin{equation} \label{eq:law of composed probabilities-amplitudes}
   AP ( x_{1} | x_{3} ) \; = \; \sum_{x_{2}\in {\mathcal{E}} }  AP ( x_{1} | x_{2} )  AP ( x_{2} | x_{3} )
\end{equation}
that, owing to the link between \textbf{probabilities} and
\textbf{probabilities-amplitudes}:
\begin{equation}
  P ( x_{1} | x_{3} ) \; = \; | AP ( ( x_{1} | x_{3} ) |^{2}
\end{equation}
implies that:
\begin{multline}
   P ( x_{1} | x_{3} ) \; = \; | \sum_{x_{2}\in {\mathcal{E}} }  AP ( x_{1} | x_{2} )  AP ( x_{2} | x_{3} ) |^{2} \;
   = \\
    ( \sum_{x_{2}\in {\mathcal{E}} }  AP ( x_{1} | x_{2} )  AP ( x_{2} | x_{3} )
) ^{\star} \, ( \sum_{x_{2}\in {\mathcal{E}} }  AP ( x_{1} |
x_{2} )  AP ( x_{2} | x_{3} ) ) = \\
\sum_{x_{2} \in {\mathcal{E}}  }  P ( x_{1} | x_{2} )  P ( x_{2}
| x_{3} ) \; + CT(x_{1} | x_{3})
\end{multline}
where the non-null \textbf{correction-term}:
\begin{equation}
  CT(x_{1} | x_{3}) \; \neq \; 0
\end{equation}
is the basis of the interference between different paths
contributing to a path-integral.

It was always such a viewpoint that led Feynman \cite{Feynman-99}
to geniously perceive that the difference between Quantum
Probability from Classical Probability implies the irreducibility
of Quantum Computational Complexity Theory to Classical
Computational Complexity Theory , catching the essence of Quantum
Computation, as we will discuss more completely  in
section\ref{sec:Irreducibility of Quantum Computational Complexity
Theory to Classical Computational Complexity Theory}.

\bigskip

To see why not only the Measure Theory, but also the geometry of
noncommutative sets play a fundamental role for Quantum Physics,
let us analyze the metric aspects of Quantum Information Theory.

At this regard our point of view is rather different from the
usual one, being based on the following:
\begin{remark} \label{rem:it is wrong to apply commutative geometry to noncommutative sets}
\end{remark}
IT IS WRONG TO APPLY COMMUTATIVE GEOMETRY TO NONCOMMUTATIVE SETS

The metric aspect of Classical Information Theory is required in
order of formalizing the concept of \textbf{distance} among
\textbf{classical probability distributions}.

This may be done in terms of some conceptually appealing
\textbf{metric} one can introduce on the \textbf{space of
classical probability distributions}

Clearly the same situation appears in Quantum Information Theory,
where one needs to quantify the  \textbf{distance} among
\textbf{quantum probability distributions}.

This has led to introduce suitable \textbf{metrics} on the space
of \textbf{quantum probability distributions} generalizing the
classical ones in a nice way.

In \textbf{Information Geometry}, furthermore, one goes further
introducing a suitable \textbf{riemannian metric} on the
\textbf{space of classical probability distributions} such to
formulate many Classical Statistics' issues in a purely
\textbf{riemannian-geometric} context.

So it has appeared natural to mimic such an attitude in Quantum
Information Theory, giving rise to the discipline of
\textbf{Quantum Information Geometry}, in which one introduces a
suitable \textbf{riemannian metric} on the \textbf{space of
quantum  probability distributions} properly generalizing the
classical one, again recasting many Quantum Statistics's issues in
a \textbf{riemannian-geometric} context.

According to us, anyaway, these approaches are unsatisfactory, in
that they ultimatively apply the usual \textbf{Commutative
Geometry} to \textbf{noncommutative spaces}:

since the space of \textbf{space of quantum probability
distributions} is  a  \textbf{noncommutative space} its metric
properties should be formalized in terms of \textbf{noncommutative
metrics}.

The same reasoning applies to \textbf{Quantum Information
Geometry}: since the \textbf{space of quantum  probability
distributions} is a \textbf{noncommutative space} one should
introduce on it a \textbf{noncommutative riemannian-geometric
structure} rather than a \textbf{commutative riemannian-geometric
structure}.

In the sequel we will see how this leads to an application of
Alain Connes's Noncommutative Geometry in all its power and
beauty.

\medskip

Given the set of n elements $ M \; := \; \{ 1  , \cdots , n \} $
let us denote by $ \mathcal{D} ( M ) $ the set of the probability
distributions over M (endowed with the Borel-$\sigma$-algebra
derived from the discrete topology).

Since:
\begin{equation}
   {\mathcal{D}} (M) = \{ \vec{p} = ( p_{1} , \cdots , p_{n} ) \in
   {\mathbb{R}}^{n} \, : \, \sum_{i=1}^{n} p_{i} = 1 \, , \, p_{i}
   \geq 0 \; i = 1 , \cdots , n \}
\end{equation}

we have that $ {\mathcal{D}} (M) $ is an $ (n-1)$-simplex of $
{\mathbb{R}}^{n} $.

A  first reasonable distance over $  \mathcal{D} ( M ) $ it is
natural to take in consideration is the following:
\begin{definition} \label{def:naife classical trace distance}
\end{definition}
CLASSICAL TRACE DISTANCE ON $ \mathcal{D} ( M ) $:
\begin{equation}
  D_{T} ( \vec{p}^{(A)} ,  \vec{p}^{(B)} ) \; := \; \frac{1}{2} \sum_{i \in M} (  | p^{(A)}_{i} - p^{(B)}_{i} | )
\end{equation}
The intuitive meaning of the definition\ref{def:naife classical
trace distance} is clarified by the
following\cite{Nielsen-Chuang-00}:
\begin{theorem} \label{th:physical meaning of the naife classical
trace distance}
\end{theorem}
CLASSICAL TRACE DISTANCE AS DISTANCE OF THE CLASSICAL PROBABILITY
OF ANTIPODAL EVENTS:
\begin{equation}
  D_{T}  ( \vec{p}^{(A)} ,  \vec{p}^{(B)} ) \; = \; \max_{e \in 2^{M}} | p^{A} (e)
  \, - \, p^{B} (e) |
\end{equation}

The natural quantum corrispective of definition\ref{def:naife
classical trace distance} could seem the following:

given an  n-dimensional Hilbert space $ {\mathcal{H}} $:
\begin{definition} \label{def:naife quantum trace distance}
\end{definition}
QUANTUM TRACE DISTANCE ON $ \mathcal{D} ( { \mathcal{H}} ) $:
\begin{equation}
  D_{T}  ( \rho^{(A)} , \rho^{(B)} ) \; := \; \frac{1}{2} \| \rho^{(A)} - \rho^{(B)} \|_{1}
\end{equation}

It is remarkable that also in the quantum case an analogous of
theorem\ref{th:physical meaning of the naife classical trace
distance} holds \cite{Nielsen-Chuang-00}:
\begin{theorem} \label{th:physical meaning of the naife quantum
trace distance}
\end{theorem}
QUANTUM TRACE DISTANCE AS DISTANCE OF THE QUANTUM PROBABILITY OF
ANTIPODAL EVENTS:
\begin{equation}
  D_{T} ( \rho^{(A)} , \rho^{(B)} )  \; = \; \max_{P \in B({\mathbb{H}})_{+} }
  Tr P ( \rho^{(A)} - \rho^{(B)} )
\end{equation}

Theorem\ref{th:physical meaning of the naife quantum trace
distance} has an immediate operational  interpretation to
appreciate which we have to enter the highly insidious lands of
quantum measurements.

Given a $ W^{\star}$-algebra A:
\begin{definition} \label{def:obsevational channel on A}
\end{definition}
OBSERVATIONAL CHANNEL ON A:
\begin{equation*}
  \alpha \in CPU(C,A) \; : \; C \text{ commutative space}
\end{equation*}
\begin{definition} \label{def:POVM on a W-star algebra}
\end{definition}
POSITIVE OPERATOR VALUED MEASURE (POVM) ON A :

a partial map $ E \, \in \, \stackrel{\circ}{MAP} ( X , A_{+} ) $
such that:
\begin{itemize}
  \item $ HALTING( E ) $ is a $ \sigma $-algebra over a set X
  \item
\begin{multline}
  \sum_{i} \, E ( F_{i} ) \;  = \; E ( \cup_{i}  F_{i} ) \\
  \forall \{ F_{i} \in HALTING( E ) \} \, : \, F_{i} \cap  F_{j} = \emptyset \, , \, \forall i \neq j
\end{multline}
  \item
\begin{equation}
  E(X) \; = \; {\mathbb{I}}
\end{equation}
\end{itemize}
\begin{definition} \label{def:PVM on a W-star algebra}
\end{definition}
PROJECTION VALUED MEASURE (PVM) ON A:

a POVM E on A such that:
\begin{equation}
  E(F) \; \in \; {\mathcal{P}}(A) \; \; \forall F  \in HALTING( E )
\end{equation}
\begin{theorem} \label{th:Naimark's theorem}
\end{theorem}
NAIMARK'S THEOREM:

\begin{hypothesis}
\end{hypothesis}
\begin{equation*}
  {\mathcal{H}} \text{ Hilbert space}
\end{equation*}
\begin{equation*}
   E   \text{ POVM on } {\mathcal{B}} ( {\mathcal{H}} )
\end{equation*}
\begin{thesis}
\end{thesis}

There exist an Hilbert space  $ {\mathcal{K}} \, \supset \, {\mathcal{H}} $  and a PVM  $ \tilde{E} $ on $ {\mathcal{B}} ( {\mathcal{H}} ) $ such that:
\begin{equation*}
  \tilde{E} (F) \; = \; P_{{\mathcal{H}}} \, E(F) \, P_{{\mathcal{H}}} \; \; \forall F \in HALTING( E )
\end{equation*}
where $ P_{{\mathcal{H}}} $  is the projector from $ {\mathcal{K}} $ to $ {\mathcal{H}} $.

\smallskip

\begin{definition} \label{def:operational partition of unity on a W-star algebra}
\end{definition}
OPERATIONAL PARTITIONS OF UNITY ON A:
\begin{multline}
  OPU(A) \; := \; \{ {\mathcal{V} } \, := \, ( V_{1} \, , \, \cdots \,
V_{n} ) \, ( n \in {\mathbb{N}}) \, : \\
 V_{i}  \; \in  \; A_{+} \; \; i=1, \cdots , n \; and \\
 \sum_{i=1}^{n}  V_{i}^{\star} V_{i} \; = \; I \}
\end{multline}

Given a operational partition of unity $ {\mathcal{V} } \, := \,
( V_{1} \, , \, \cdots \, V_{n} ) \, \in \, OPU(A) $:
\begin{definition} \label{def:channels' set of an operational partition of unity}
\end{definition}
CHANNELS' SET OF $ {\mathcal{V} } $:

the set $ \{ \alpha_{1}( {\mathcal{V} }) \, , \, \cdots \, ,
\alpha_{n}( {\mathcal{V} }) \} $, where $ \alpha_{i}( {\mathcal{V}
})$  is the channel (owing to theorem\ref{th:Kraus-Stinespring's
theorem}) of A :
\begin{equation}
  \alpha_{i}( {\mathcal{V}}) (a) \; := \; V_{i}^{\star} a V_{i} \;
  \; a \, \in A
\end{equation}
\begin{definition} \label{def:reduction channel of an operational partition of unity}
\end{definition}
REDUCTION CHANNEL OF $ {\mathcal{V} } $:

the channel (owing to theorem\ref{th:Kraus-Stinespring's theorem})
$ R({\mathcal{V}}) \in CPU(A) $:
\begin{equation}
  R({\mathcal{V}}) \; := \; \sum_{i=1}^{n}  \alpha_{i}( {\mathcal{V}})
\end{equation}

The interrelation among these notions is the following:
\begin{itemize}
  \item a POVM $ E \, : \, X \, \stackrel{\circ}{\rightarrow} \, A_{+} $ whose
  halting set is the Borel-$\sigma$-algebra of a topology on X may
  be seen as an observational channel  $ E \, : \, C(X) \rightarrow
  A $
  \item an observational channel  $ \alpha \in CPU( C,A) $
  such that $ cardinality_{NC} ( C ) \, \in \, {\mathbb{N}} $
  indivuates an operational partition of unity

\end{itemize}

We have spent much efforts, in section\ref{sec:Why to treat
sequences of qubits one has to give up the Hilbert-Space
Axiomatization of Quantum Mechanics}, to discuss a statement by
Walter Thirring in which he claimed that Sch\"{o}dinger's
positive answer to the separability issue was owed to the fact he
didn't understand Von Neumann's paper on infinite tensor products.

But we are perfectly aware that if there is someone knowing
exceptionally all the mirabilities of the noncommutative approach
is precisely Walter Thirring, who has given in \cite{Thirring-81}
and \cite{Thirring-83} one of its most authoritative presentations.

We have implicitely adopted  such an approach in
section\ref{sec:On the rule Noncommutative Measure Theory and
Noncommutative Geometry play in Quantum Physics}, though not
exlicitely presenting its underlying axiomatization.

We will do this here, partly moving away from the basic-assumption-2.2.32 of \cite{Thirring-81}.

\begin{definition} \label{def:noncommutative axiomatization of Quantum Mechanics}
\end{definition}
NONCOMMUTATIVE AXIOMATIZATION OF QUANTUM MECHANICS:

any axiomatization of Quantum Mechanics assuming the following
two axioms:
\begin{axiom} \label{ax:noncommutative axiom on observables}
\end{axiom}
NONCOMMUTATIVE AXIOM ON OBSERVABLES:

The \textbf{observables} of a \textbf{quantum mechanical systems}
are POVM's over a noncommutative space A, called its
\textbf{observables' algebra}

\begin{axiom} \label{ax:noncommutative axiom on states}
\end{axiom}
NONCOMMUTATIVE AXIOM ON STATES:

The \textbf{states} of a \textbf{quantum mechanical systems} are
states over its \textbf{observables' algebra}

\medskip

We will assume, from here and beyond, a Noncommutative Axiomatization of Quantum Mechanics endowed with the other following axioms:

\begin{axiom} \label{ax:noncommutative axiom on closed dynamics}
\end{axiom}
NONCOMMUTATIVE AXIOM ON DYNAMICS OF A CLOSED SYSTEM:

The dynamical evolution of a \textbf{closed} \textbf{quantum
mechanical system} S  is given by a \textbf{strongly-continuous group of inner automorphisms}
of its \textbf{observables' algebra}

\smallskip

\begin{axiom} \label{ax:noncommutative axiom on measurement}
\end{axiom}
NONCOMMUTATIVE AXIOM ON MEASUREMENT:

If on a  \textbf{quantum mechanical system} S, prepared in the
state $ \omega \in S(A) $, it is performed the measurement
mathematically described by the operational partition of unity $
{\mathcal{V}} \, := \, \{ V_{1} \, , \, \cdots \, , V_{n} \} $
then:
\begin{itemize}
  \item during the measurement S is \textbf{open}
  \item by definition one says that the \textbf{$ i^{th}$-experimental outcome occurs} if on a suitable  \textbf{classical display} one  reads the number  $ i \in \{ 1 , \cdots , n \} $
  \item if the \textbf{$ i^{th}$-experimental outcome occurs} then
  then S's observables' algebra evolves according to the the
  \textbf{channel} $ \alpha_{i}( {\mathcal{V}}) $
  \item the \textbf{$ i^{th}$-experimental outcome occurs} with
  probability:
\begin{equation}
  p_{i} \; := \; \omega( \alpha_{i}(I) )
\end{equation}
\end{itemize}

\medskip

Axiom\ref{ax:noncommutative axiom on closed dynamics} and axiom\ref{ax:noncommutative axiom on measurement} are consistent owing to the following:

\begin{theorem} \label{th:dynamics of an open quantum mechanical system}
\end{theorem}
DYNAMICS OF AN OPEN QUANTUM MECHANICAL SYSTEM:

The dynamical evolution of an \textbf{open} \textbf{quantum
mechanical system} S is given by a \textbf{one-parameter family of channels} of its
\textbf{observables' algebra}

\smallskip

\begin{remark} \label{rem:noncommutative axiomatizations and unbounded operators}
\end{remark}
NONCOMMUTATIVE AXIOMATIZATIONS AND UNBOUNDED OPERATORS:

A first natural reaction to definition\ref{def:noncommutative
axiomatization of Quantum Mechanics} is to ask what about
unbounded operators:

if the \textbf{observables' algebra} of a quantum system is
assumed to be a noncommutative space $ A \, \subseteq \,
{\mathcal{B}} ( {\mathcal{H}}) $  isn't one arbitarily throwing
away all the self-adjoints elements of $ {\mathcal{O}} (
{\mathcal{H}} ) \, - \, {\mathcal{B}} ( {\mathcal{H}} ) $?

The answer to such an objection touches the original argument that
led Irving Segal in 1947 to introduce the algebraic approach (cfr.
the introduction of \cite{Haag-96}): given a self-adjoint
unbounded operator T on an Hilbert space $ {\mathcal{H}} $ the
Spectral Theorem allows us to define the operator $ f(T) $ for
every Lebesgue-integrable function $ f \in {\mathcal{B}} (
{\mathbb{R}}) $, the set of linears operators obtained varying f
being called the \textbf{abelian $ W^{\star}$-algebra generated by
T} (cfr. the section7.2 of \cite{Reed-Simon-80}.

Since one can look at the passage from  T to f(T) simply as a
relabeling of the possible experimental outcomes, the physically
relevant notion is that of the \textbf{abelian $
W^{\star}$-algebra generated by T}, of which one can always
choose a bounded element such as  $ e^{T} $.

\smallskip

\begin{remark} \label{rem:on superselection rules}
\end{remark}
ON SUPERSELECTION RULES

Let us observe that axiom\ref{ax:noncommutative axiom on observables} says that an observable of a quantum mechanical system is a POVM on its observables' algebra, but it
doesn't say that any POVM on such an observable's algebra is an observable of the system.

Similarly, axiom\ref{ax:noncommutative axiom on states} says that a physical state of a quantum mechanical system is a state on its observables' algebra, but it doesn't say
that any state on such an observable's algebra is a physical state of the system.

Finally, axiom\ref{ax:noncommutative axiom on measurement} tells us what happens when on a quantum mechanical system it is performed
the measurement mathematically described by a partitional operation of unity, but it doesn't say that any  operational partition of unity
describes a possible measurement.

If on a quantum mechanical systems with observables' algebra A there exist a self-adjoint operator $ a \in A_{sa} $ such that the  PVM associated to it by the Spectral Theorem cannot be physical observable, one
says that the system has Superselection Rules.

\begin{remark} \label{rem:if God plays dices is a kinematical issue and not a dynamical one}
\end{remark}
IF GOD PLAYS DICES IS A KINEMATICAL ISSUE AND NOT A DYNAMICAL ONE

The way  axiom\ref{ax:noncommutative axiom on measurement} is
formalized may be partially misleading owing to the fact that it
would seem to introduce a \textbf{dynamical indeterminism} in the
theory.

This is not the case: we spoke about of collection $ \{ \alpha
_{i} \} $ of possible dynamical evolutions, each with a classical
probability $ p_{i} $ of occurring, only for simplicity, but it
is exactly the same as saying the that the observable's algebra
evolves according to the \textbf{reduction channel} $ R(
{\mathcal{V}} ) $ of $ {\mathcal{V}} $

\medskip

We can now appreciate  the physical meaning of
Theorem\ref{th:physical meaning of the naife quantum trace
distance}: it tells us that $ D_{T} ( \rho_{1} , \rho_{2} ) $ is
the maximal distance of the classical probabilities of a
measurement outcome between the case in which the state before the
measurement is $ \rho_{1} $ and the case in which the state
before the measurement is $ \rho_{2} $.

\medskip

\textbf{Trace distance} is not, anyway, the only reasonable
distance over $ \mathcal{D} ( M ) $ one can introduce.

An other example is the following:
\begin{definition} \label{def:naife classical angle distance}
\end{definition}
CLASSICAL ANGLE DISTANCE ON $ \mathcal{D} ( M ) $:
\begin{equation}
  D_{A} ( \vec{p}^{(A)} , \vec{p}^{(B)} ) \; := \; \arccos F ( \vec{p}^{(A)} , \vec{p}^{(B)} )
\end{equation}
where:
\begin{definition} \label{def:naife classical fidelity}
\end{definition}
CLASSICAL FIDELITY ON $ \mathcal{D} ( M ) $:
\begin{equation}
   F ( \vec{p}^{(A)} , \vec{p}^{(B)} ) \; := \;  \sum_{i \in M} \sqrt{ p^{(A)}_{i} \, p^{(B)}_{i}   }
\end{equation}

The name \textbf{angle distance} is justified by the following
considerations:

the vectors $  \vec{\xi}^{(A)} = \begin{pmatrix}
  \sqrt{p^{A}_{1}} \\
  \vdots \\
  \sqrt{p^{A}_{n}}
\end{pmatrix} , \vec{\xi}^{(B)}  = \begin{pmatrix}
  \sqrt{p^{B}_{1}} \\
  \vdots \\
 \sqrt{ p^{B}_{n}}
\end{pmatrix} $
belong to the n-sphere of unitary radius $ S^{(n)} $.

So $ D ( p_{1} , p_{2} ) $ is precisely the angle between $
\vec{\xi}_{1} $ and $ \vec{\xi}_{2} $, i.e. the geodesic distance
between them on the riemannian manifold $ ( S^{(n)} , g_{S^{(n)}}
) $, $ g_{S^{(n)}} := i^{\star} \delta $ being the  metric
induced on $ S^{(n)} $ by its inclusion's embedding $ i : S^{(n)}
\rightarrow {\mathbb{R}}^{n}$ in the euclidean space $ (
{\mathbb{R}}^{n} , \delta )$ \cite{Nakahara-95}.

The quantum corrispective of definition\ref{def:naife classical
angle distance} used by the Quantum Computation's community is the
following:
\begin{definition} \label{def:naife quantum angle distance}
\end{definition}
QUANTUM ANGLE DISTANCE ON $ \mathcal{D} ( { \mathcal{H}} ) $:
\begin{equation}
   D_{A} ( \rho^{(A)} , \rho^{(B)} ) \; := \;  \arccos F ( \rho^{(A)} , \rho^{(B)} )
\end{equation}
where:
\begin{definition} \label{def:naife quantum fidelity}
\end{definition}
QUANTUM FIDELITY ON $ \mathcal{D} ( { \mathcal{H}} ) $:
\begin{equation}
   F ( \rho^{(A)} , \rho^{(B)} ) \; := \; Tr \sqrt{ \sqrt{\rho^{(A)}} \rho^{(B)} \sqrt{\rho^{(A)}}}
\end{equation}

A geometric interpretation of the \textbf{quantum angle distance}
analogous to the classical one is furnished by the following:
\begin{theorem} \label{th:first Uhlmann's theorem}
\end{theorem}
FIRST UHLMANN'S THEOREM:

\begin{hypothesis}
\end{hypothesis}
\begin{equation*}
   \rho^{(A)} \, , \, \, \rho^{(B)} \; \in \; \mathcal{D}({\mathcal{H}})
 \end{equation*}
\begin{thesis}
\end{thesis}
\begin{equation*}
  F ( \rho^{(A)} \, , \, \rho^{(B)} ) \; = \; \max_{ | \psi_{A} > \in PUR( \rho^{(A)} ,
  {\mathcal{H}} ) \; , \; | \psi_{B} >  \in PUR( \rho^{(B)} ,
  {\mathcal{H}} )} \, | < \psi_{A} | \psi_{B} > |
\end{equation*}

where, given two generical Hilbert spaces $ \mathcal{H}_{A} $ and
$ \mathcal{H}_{B} $ and a density matrix $ \rho \in {\mathcal{D}}(
{\mathcal{H}}_{A})$ :
\begin{definition}
\end{definition}
PURIFICATIONS OF $ \rho $ WITH RESPECT TO $  \mathcal{H}_{B} $:
\begin{equation}
  PUR( \rho \, , \, {\mathcal{H}_{B}}) \; := \; \{ | \psi > \in
  {\mathcal{H}_{A}} \bigotimes {\mathcal{H}_{B}} \, : \,
  Tr_{{\mathcal{H}_{B}}} | \psi > \; = \;  \rho \}
\end{equation}
So the cosin of the angle distance between two density matrices is
equal to the maximum inner product between purifications of such
density matrices.

\smallskip

The \textbf{quantum angle distance} is not, anyway, the only
possible quantum corrispective of definition\ref{def:naife
classical angle distance} that, as many other notions, had been
extensively studied in the Mathematical-Physics' literature, many
years before the Quantum Computation's community rediscovered it.

Indeed the geometric interpretation of the \textbf{classical angle
distance} between two distributions as the geodesic distance on
the unit sphere among their square-root densities, may be seen as
the first taste of \textbf{Classical Information Geometry},
namely the approach to Classical Probability Theory studying the
set of all the probability measures on a given sample space from a
differential-geometric viewpoint \cite{Cencov-82},
\cite{Amari-85}.

Indeed, the more relevant application of Information Geometry
concerns \textbf{Statistical Estimation}:

given a  submanifold  $ N \subset {\mathbb{R}}^{m} $.

\begin{definition} \label{def:classical statistical model}
\end{definition}
CLASSICAL STATISTICAL MODEL WITH SAMPLE SPACE M AND PARAMETER
SPACE N:
\begin{multline}
  CSM ( M , N ) \; := \; \{ \vec{\xi}(\theta) \, = \, \begin{pmatrix}
    \xi_{1} (\theta) \\
    \vdots \\
    \xi_{n} (\theta)
  \end{pmatrix} \: : \\
   \xi_{i} (\theta) = \sqrt{p_{i} (\theta)}
  \, , \, \vec{p}( \vec{\theta} ) :=  \begin{pmatrix}
    p_{1} (\theta) \\
    \vdots \\
    p_{n} (\theta)
  \end{pmatrix} \in \mathcal{D} ( M ) \, \forall \theta := \begin{pmatrix}
    \theta_{1} \\
    \vdots \\
    \theta_{m}
  \end{pmatrix} \in N \}
\end{multline}

So a \textbf{classical statistical model} $ CSM ( M , N ) $ with
\textbf{sample space} M and \textbf{parameter space} N is a
collection of square-roots probability distributions over M
parametrized through points of N.

By construction $ CSM ( M , N ) $ is a submanifold of $ S^{n} $.

As we will know show, definition\ref{def:naife classical angle
distance} naturally induces a riemannian metric on $ CSM ( M , N
) $, called the \textbf{Fisher-Rao riemannian metric}, playing a
key rule, through the Cramer-Rao Inequality, in the theory of the
statistical inference of $ \vec{\theta} $ (or, more generally, a
suitable function of it), from a finite set of statistical data.

Given a  function $ f \in  C^{\infty} ( N , {\mathbb{R}}) $ and a
parametrized family of classical random variables $
X(\vec{\theta} ) $ over M with distribution $ p(x | \vec{\theta}
)$:
\begin{definition} \label{def:unbiased estimator}
\end{definition}
$ X(\vec{\theta} ) $ IS AN UNBAISED ESTIMATOR OF THE FUNCTION f:
\begin{equation}
  E[ X(\vec{\theta})] \; = \; f(\vec{\theta}) \; \; \forall
  \vec{\theta}\in N
\end{equation}

The meaning of definition\ref{def:unbiased estimator} lies in
that from sampling of X we get some information about the
function $ f(\vec{\theta}) $ we want to estimate.

Clearly, the smaller is the variance of $ X(\vec{\theta} ) $ the
higher is the classical information we gain by the estimation
process.

The Cramer-Rao inequality states the existence of an upper bound
about such information.

What is surprising in that, is the simple geometric nature
underlying such a bound.

Let us introduce the following:
\begin{definition} \label{Fisher riemannian metric}
\end{definition}
FISHER-RAO RIEMANNIAN METRIC:

the riemannian metric over CSM ( M , N ):
\begin{equation}
  g_{ CSM ( M , N )} \; := \; g_{N} |_{CSM ( M , N )} \; = \;
  i^{\star} ( \delta )
\end{equation}
where $  \delta $ is again the euclidean metric on $
{\mathbb{R}}^{n} $ while $ i : CSM ( M , N ) \rightarrow
{\mathbb{R}}^{n} $ is the inclusion-embedding of  $ CSM ( M , N )
$ in $ {\mathbb{R}}^{n} $.

Then one has that:
\begin{theorem} \label{th:Cramer Rao inequality}
\end{theorem}
CRAMER RAO INEQUALITY:
\begin{equation}
  Var[ X(\vec{\theta} ) ] \; \geq \; g_{ CSM ( M , N )}^{i j}
  \partial_{i} f  \partial_{j} f
\end{equation}
where we have expressed the Fisher-Rao riemannian metric through
the global coordinates $ \{ \theta^{i} \} $ over N:
\begin{equation}
  g_{ CSM ( M , N )} \; = \;   ( g_{ CSM ( M , N )})_{i j} d
  \theta^{i} \bigotimes d \theta^{j}
\end{equation}
and where $ \partial_{i} f \; := \frac{\partial f}{ \partial
\theta^{i}} $.

\smallskip

We can now appreciate how the issue of finding a quantum
generalization of  definition\ref{def:naife classical angle
distance} fits in the more ambitious process of constructing a
\textbf{Quantum Information Geometry} playing in \textbf{Quantum
Estimation Theory} the same rule \textbf{Classical Information
Geometry} plays in \textbf{Classical Estimation Theory}.

Such a project has been pursued extensively by many authors
\cite{Hasegawa-Petz-97}, \cite{Petz-Sudar-99}
\cite{Brody-Hughston-98}, \cite{Streater-00b}, conceptually in the
framework of Helstrom's Quantum Statistical Decision Theory for a
modern presentation of which we demand to the section2.2 of
\cite{Holevo-99}.

All the proposed approaches, anyway, or reconduct the quantum
case to the classical one (e.g. the approach by Brody and
Hughston based on an application of the Fisher riemannian metric
to the horizontal lift  of paths in the Stiefel bundle underlying
the Aharonov-Anandan geometric phase \cite{Bohm-93}) or introduce
suitable riemannian geometric structures on the space of the
quantum states (e.g. the riemannian geometric structure
underlying Hasegawa's $ \alpha $-divergence, or Petz's monotone
riemannian metrics \cite{Lesniewski-Ruskai-98} on which we will return in section\ref{sec:The problem of characterizing mathematically the notion of a quantum algorithm} on discussing the rule
of the Wigner-Araki-Dyson skew information for superselection-rules).

But then the considerations of remark\ref{rem:it is wrong to apply
commutative geometry to noncommutative sets} apply.

To sketch an idea of how \textbf{Quantum Information Geometry}
should be constructed in terms of noncommutative riemannian
spaces, we have to go further in the construction of the
noncommutative tower.

The next floor after Noncommutative Topology and Noncommutative
Measure Theory is Noncommutative Differential Calculus
\cite{Connes-92}, \cite{Connes-94}, \cite{Connes-98}.

Commutative Differential Calculus started with Leibniz's
\textbf{infinitesimals} (or equivalentely  Newton's fluxions).

The difficulty of furnishing a rigorous mathematical formalization
of the notion of \textbf{infinitesimal} inside Commutative
Analysis, led the fathers of Commutative Calculus to replace them
by well-defined objects, i.e integrals, recasting the foundations
of Commutative Analysis inside the boundaries of Commutative
Measure Theory.

In fact, though usually used as a linguistic shortcut by
physicists, statements like the following:

\begin{center}
  " Let us call $ dp(x) $ the  probability that a particle is found in the interval $ [ x ,
  x + dx ]" $
\end{center}
has no rigorous meaning, as can be seen observing that the natural
condition that the \textbf{commutative infinitesimal} dp(x)
should satisfy, namely:
\begin{equation} \label{eq:impossible constraint on a commutative infinitesimal}
  dp(x) \; < \; \epsilon \; \; \forall \epsilon > 0
\end{equation}
obviously implies that:
\begin{equation} \label{eq:catastrophic conseguence of the impossible constraint on a commutative infinitesimal}
  dp(x) \; = \; 0
\end{equation}
The well-defined quantities are only the integrals:
\begin{equation}
  \int dp(x) f
\end{equation}
of suitable functions.

So the fate of the poor infinitesimal dp(x) was very sick until
Robinson's Nonstandard Analysis gave it a rigorous mathematical
status as a nonstandard-real, though at the price of high
logico-mathematical sophistications.

So it is very curious that, as we will now show the issue of
defining an \textbf{infinitesimal} in a \textbf{noncommutative
space} is highly simpler.

Given an Hilbert space $ {\mathcal{H}} $ let us analyze if there
is some natural way of defining an infinitesimal element of the
noncommutative space $ {\mathcal{B}} ( \mathcal{H} ) $. Exactly
as in the commutative case, then natural condition that one would
require in order of considering an element $ a \in  {\mathcal{B}}
( {\mathcal{H}} ) $ as an infinitesimal is that:
\begin{equation} \label{eq:impossible constraint on a noncommutative infinitesimal}
  \| a \|  \; < \; \epsilon \; \; \forall \epsilon > 0
\end{equation}
But in the same way eq.\ref{eq:impossible constraint on a
commutative infinitesimal} implies eq.\ref{eq:catastrophic
conseguence of the impossible constraint on a commutative
infinitesimal} one has that eq.\ref{eq:impossible constraint on a
noncommutative infinitesimal} implies that:
\begin{equation} \label{eq:catastrophic conseguence of the impossible constraint on a noncommutative infinitesimal}
  a \; = \; 0
\end{equation}

Contrary to the commutative case, anyway, the condition of
eq.\ref{eq:impossible constraint on a noncommutative
infinitesimal} may be slightly  modified in order of becoming
meaningful, substituting the condition of eq.\ref{eq:impossible
constraint on a noncommutative infinitesimal} by the condition:
\begin{equation} \label{eq:constraint on a noncommutative infinitesimal}
  \forall \epsilon > 0 \; , \; \exists \text{ a subspace }  E_{\epsilon} \subset
  {\mathcal{H}} \: : \: dim(E) < \infty \; and \; \| a |_{ E^{\bot}
  } \| \, < \,  \epsilon
\end{equation}
Since an operator a satisfies the condition of
eq.\ref{eq:constraint on a noncommutative infinitesimal} iff it
is compact, it follows that the set of the infinitesimals
elements of $ {\mathcal{B}} ( {\mathcal{H}} ) $ are exactly the
the set  $ {\mathcal{C}} ( {\mathcal{H}} ) $ of the compact
operators on $ {\mathcal{H}} $:
\begin{definition}  \label{def:noncommutative infinitesimal}
\end{definition}
$ a \in  {\mathcal{B}} ( {\mathcal{H}} ) $ IS INFINITESIMAL:
\begin{equation}
  a \; \in \;  {\mathcal{C}} ( {\mathcal{H}} )
\end{equation}
Given a  noncommutative infinitesimal $ a \in  {\mathcal{C}} (
{\mathcal{H}} ) $:
\begin{definition} \label{def:characteristic values of a noncommutative infinitesimal}
\end{definition}
$ n^{th} $ CHARACTERISTIC VALUE OF a:
\begin{equation}
  \mu_{n} (a) \; := \; \inf \{  \| a \|_{ E^{\bot} }  \, , \, dim(E)
  \leq n \}
\end{equation}
We can then classify the order of noncommutative infinitesimals
in the following way:

given an  $ \alpha \in {\mathbb{R}}_{+} $:
\begin{definition} \label{def:noncommutative infinitesimal of given order}
\end{definition}
INFINITESIMALS OF ORDER $ \alpha  $:
\begin{equation}
  {\mathcal{I}}_{\alpha} ({\mathcal{H}}) \; := \;  \{ a \in {\mathcal{C}} (
{\mathcal{H}} ) \, : \,   \mu_{n} (a) \; = \; O ( \frac{1}{
n^{\alpha}}) \; for \; n \rightarrow \infty  \}
\end{equation}

\smallskip

The next step in the construction of Noncommutative Calculus is
the definition of noncommutative integration.

Given a trace-class operator $ a \in  {\mathcal{C}}_{1} (
{\mathcal{H}} ) $ one would be tempted to define its
noncommutative integral over $ {\mathcal{B}} ( {\mathcal{H}} ) $
simply as its trace:
\begin{equation}
  Tr(a) \; := \; \sum_{n} < \psi_{n} | a | \psi_{n} >
\end{equation}
that is independent from the choice of the orthonormal basis $ \{
| \psi_{n} > \} $ of $  {\mathcal{H}} $.

Since the characteristic values $ \mu_{0} (a) \, \geq \, \mu_{1}
(a) \, \geq \, \cdots \,  \mu_{n} (a) \rightarrow 0 $ of an
operator $ a \in  {\mathcal{B}} ( {\mathcal{H}} ) $ are nothing
but the eigenvalues of $ | a | $ one has that:
\begin{equation}
  Tr(a) \; = \; \sum_{n=0}^{\infty} \mu_{n} (a) \; \; \forall a
  \in ({\mathcal{B}} ( {\mathcal{H}}) )_{+}
\end{equation}

But let us now observe that a reasonable notion of noncommutative
integration has to satisfy the following constraints:
\begin{enumerate}
  \item \textbf{the integral of infinitesimals of order one  must converge}
\begin{equation} \label{eq:the integral of infinitesimals of order one  must converge}
  \int_{NC} a \; < \; + \infty \; \; \forall a \in
  {\mathcal{I}}_{1}  ( {\mathcal{H}})
\end{equation}
  \item \textbf{the integral of infinitesimals of order greater than one must vanish}
\begin{equation} \label{eq:the integral of infinitesimals of order greater than one must converge}
   \int_{NC} a \; < \; 0 \; \; \forall a \in
   {\mathcal{I}}_{\alpha}  ( {\mathcal{H}}) , \alpha > 1
\end{equation}
\end{enumerate}
Since $ {\mathcal{I}}_{1}  ( {\mathcal{H}})  \, \nsubseteq \,
{\mathcal{C}}_{1} ( {\mathcal{H}} ) $ the simple trace Tr(a)
doesn't satisfy the constraint of eq.\ref{eq:the integral of
infinitesimals of order one  must converge}. Furthermore it
doesn't satisfy also eq.\ref{eq:the integral of infinitesimals of
order greater than one must converge}.

Hence it doesn't work.

A good notion of noncommutative-integral was, instead, obtained
by J. Dixmier starting from the observation that:
\begin{equation}
  \sum_{n=0}^{\infty} \mu_{n-1} (a) \; \stackrel{ \times }{\leq}
  \; \log n
\end{equation}
So he introduced a quantity, now called the Dixmier trace, that,
informally speaking, extracts the coefficient of the logarithmic
divergence.

Though in all the more important cases it is given simply by:
\begin{equation} \label{eq:naife Dixmier trace}
  \lim_{n \rightarrow \infty} \frac{1}{\log n } \sum_{n=0}^{n-1} \mu_{n} (a)
\end{equation}
in the general case such an expression presents two problems: its
linearity and its convergence.

Given an infinitesimal $ a \in  {\mathcal{C}}( {\mathcal{H}} ) $
let us consider the argument of the limit at the r.h.s. of
eq.\ref{eq:naife Dixmier trace}, namely:
\begin{equation}
  \gamma_{n} (a) \; := \;  \frac{1}{\log n } \sum_{n=0}^{n-1} \mu_{n} (a)
\end{equation}
Since it obeys the relation \cite{Connes-94}:
\begin{equation} \label{eq:asymptotic additivity a Dixmier sequence}
  \gamma_{n} (a_{1} + a_{2}) \; \leq \;  \gamma_{n} (a_{1}) + \gamma_{n}
  (a_{2}) \; \leq \; \gamma_{2 n} (a_{1} + a_{2}) ( 1 +
  \frac{ \log 2}{ \log n } )
\end{equation}
we see that linearity would follow from convergence.

Unfortunately, though always bounded, the sequence $ \{
\gamma_{n} \} $ doesn't always converge.

Considered the Banach space of bounded sequences $ l^{\infty} (
{\mathbb{N}} ) $ let us introduce the space of all the linear
forms $ \lim_{\omega} $ on it such that:
\begin{enumerate}
  \item
\begin{equation}
   \gamma_{n} \, \leq \, 0 \; \Rightarrow \; \lim_{\omega}  \gamma_{n} \, \leq \, 0
\end{equation}
  \item
\begin{equation}
  \exists \lim_{ n \rightarrow + \infty} \gamma_{n} \;
  \Rightarrow \; \lim_{\omega}  \gamma_{n} \, = \, \lim_{ n \rightarrow + \infty} \gamma_{n}
\end{equation}
  \item
\begin{equation}
  \lim_{\omega} \{ (\gamma_{n} ) ^{n} \} \; = \; \lim_{\omega}  \gamma_{n}
\end{equation}
  \item
\begin{equation}
   \lim_{\omega}  \gamma_{2 n} \; = \; \lim_{\omega}  \gamma_{2 n}
\end{equation}
\end{enumerate}
To each of such linear forms $ \lim_{\omega} $ (they are
infinite) it is associated a \textbf{Dixmier trace}, according to
the following:
\begin{definition} \label{def:Dixmier trace}
\end{definition}
DIXMIER TRACE OF  $ a \, \in \,  ({\mathcal{B}} ( {\mathcal{H}}
))_{+} \, \bigcap \,  {\mathcal{I}}_{1} ( {\mathcal{H}} )  $:
\begin{equation}
  tr_{\omega} (a) \; := \; \lim_{\omega} \gamma_{n} \; = \;
  \lim_{\omega} \frac{1}{\log n } \sum_{n=0}^{n-1} \mu_{n} (a)
\end{equation}
By eq.\ref{eq:asymptotic additivity a Dixmier sequence} it
follows that a Dixmier trace is additive on positive
infinitesimals of order one, so that, owing to
eq.\ref{eq:asymptotic additivity a Dixmier sequence}, it can be
extended by linearity to the whole $  {\mathcal{I}}_{1} (
{\mathcal{H}} ) $.

That a Dixmier trace is indeed a trace,i.e. that:
\begin{equation}
  tr_{\omega} ( \alpha(a) ) \; = \;  tr_{\omega} (a) \; \;
  \forall a \in   {\mathcal{I}}_{1} (
{\mathcal{H}} ) , \forall \alpha \in  INN ( {\mathcal{B}}(
{\mathcal{H}} ))
\end{equation}
follows immediately by the unitary invariance of the
characteristic values of an infinitesimal a owed to the fact that
they are nothing but the eigenvalues of $ | a | $.

Since any linear form $ lim_{\omega} $ assumes only finite
values, a Dixmier trace satisfies by construction the first
constraint, namely eq.\ref{eq:the integral of infinitesimals of
order one  must converge}, we required for a reasonable notion of
noncommutative integration.

Furthermore, since the space of all infinitesimal of order higher
than one is a two-sided ideal whose elements satisfy the
condition:
\begin{equation}
  \lim_{n \rightarrow \infty} \mu_{n} (a) \; = \; 0
\end{equation}
and so:
\begin{equation}
  \lim_{n \rightarrow \infty} \gamma_{n}
\end{equation}
it follows that a Dixmier trace satisfies also the second
constraint, namely eq.\ref{eq:the integral of infinitesimals of
order greater than one must converge}, we ask to a noncommutative
integral.

\smallskip

After \textbf{noncommutative integration}  let us pass to
\textbf{noncommutative differentiation}.

Let us observe, at this purpose, that a key rule of commutative
differentiation is given by Leibniz's rule for the differential of
products:
\begin{equation}\label{eq:Leibniz rule}
  d ( f_{1} f_{2} ) \; = \; d ( f_{1} ) f_{2} \, + \, f_{1}  d ( f_{2} )
\end{equation}
So it appears natural to attempt to characterize the notion of
noncommutative differentiation imponing that eq.\ref{eq:Leibniz
rule} holds in the noncommutative case too.  The resulting notion, introduced by Kaplansky in 1953,
is that of a derivation \cite{Sakai-91}, namely the following:
\begin{definition} \label{def:derivation on a C-star algebra}
\end{definition}
DERIVATION ON A $C^{\star}$-ALGEBRA A:

a linear operator $ \delta \, : D( \delta ) \rightarrow A $ from
a $ \star $-subalgebra $ D( \delta ) $ to A such that:
\begin{equation}
  \delta (a b ) \; := \;  \delta (a ) b   \, + \,  a \delta ( b )
  \; \; \forall a , b \in D( \delta )
\end{equation}

Given a derivation $ \delta $ on a  $C^{\star}$-algebra A:
\begin{definition}   \label{def:involutive derivation on a C-star algebra}
\end{definition}
$ \delta $ IS AN INVOLUTIVE DERIVATION ( $ \star $- DERIVATION ):
\begin{equation}
  \delta ( a^{\star} ) \; = \; \delta (a)^{\star} \; \; \forall a
  \in D( \delta)
\end{equation}

\smallskip

Let us now observe that we are used, by Functional Analysis, to the
fact that the assignment of a one-parameter strongly continuous
group (or semigroup) of operators is equivalent to the assignment
of its generator:
\begin{itemize}
  \item by the Stone's Theorem \cite{Reed-Simon-80} we know that the
  assignment of a strongly continuous one-parameter group U(t) of unitary
  operators on an Hilbert space $ {\mathcal{H}} $ is equivalent to
  the assignment of its generator, namely the (unique)
  self-adjoint operator A such that:
\begin{equation}
  U(t) \; = \; e^{i t A}  \; \; \forall t \in {\mathbb{R}}
\end{equation}
  \item by the Hille-Yosida's Theorem \cite{Reed-Simon-75} we know that the
  assignment of a strongly continuous one-parameter semigroup C(t) of contractions on an Hilbert space $ {\mathcal{H}} $ is equivalent to
  the assignment of its generator, namely the (unique)
  self-adjoint operator A such that:
\begin{equation}
  C(t) \; = \; e^{-  t A}  \; \; \forall t \in {\mathbb{R}}_{+}
\end{equation}
\end{itemize}
So we are not surprised that a similar situation occurs also for
strongly continuous one-parameter subgroups of the automorphisms'
group AUT(A) of a generic $ C^{\star} $ algebra.

Demanding to the paragraph3.4 of \cite{Sakai-91} for details it
is sufficient here to recall that exactly as Edward Nelson's notion of
analytic vectors allows to  construct directly the exponential  $
e^{A} $ of a self-adjoint operator A as a power series, the same
happens in our operator-algebraic setting allowing to define, as a
series-power, the exponential $ e^{\delta} $ of an involutive
derivation $ \delta $ on a $ C^{\star}$-algebra.

Then one has the following:
\begin{theorem} \label{th:on the generators of strongly continuous one-parameter groups of automorphisms}
\end{theorem}
ON THE GENERATORS OF STRONGLY CONTINUOUS ONE-PARAMETER GROUPS OF
AUTOMORPHISMS

\begin{hypothesis}
\end{hypothesis}
\begin{equation*}
   A \; \;  C^{\star}-algebra
\end{equation*}
\begin{equation*}
  (\alpha_{t} )_{t \in {\mathbb{R}}} \text{strongly continuous one-parameter subgroup of AUT(A)}
\end{equation*}
\begin{thesis}
\end{thesis}
\begin{equation*}
  \exists ! \: \delta \text{ $\star$-derivation on A } \; : \;  (
  \alpha_{t} \: = \: e^{t \delta} \; \; \forall t \in {\mathbb{R}}
  )
\end{equation*}

\smallskip

Let us then consider the particular case of one-parameter groups
of inner automorphisms.

Theorem\ref{th:on the generators of strongly continuous
one-parameter groups of automorphisms} immediately implies the
following:
\begin{corollary} \label{th:on the generators of strongly continuous one-parameter
groups of inner automorphisms}
\end{corollary}
\begin{hypothesis}
\end{hypothesis}
\begin{equation*}
   A \; \;  C^{\star}-algebra
\end{equation*}
\begin{equation*}
  (\alpha_{t} )_{t \in {\mathbb{R}}} \text{ strongly continuous one-parameter subgroup of INN(A)}
\end{equation*}
\begin{equation*}
  \delta \text{ generator of the group } (\alpha_{t} )_{t \in {\mathbb{R}}}
\end{equation*}
\begin{thesis}
\end{thesis}
\begin{equation*}
  \exists ! \;  D \in A_{sa} \; : \; \delta( a ) \: = \: i \, [ D , a ]
  \; \forall a \in A
\end{equation*}

\smallskip

Let us now introduce the following notion:
\begin{definition} \label{def:spectral triple}
\end{definition}
SPECTRAL TRIPLE:

a therne $ ( A \, , \, {\mathcal{H}}  \, , \, D ) $ such that:
\begin{itemize}
  \item $ {\mathcal{H}} $ is an Hilbert space
  \item  $ A \subseteq {\mathcal{B}}( {\mathcal{H}} ) $ is a $
  \star $ - subalgebra of $ {\mathcal{B}}( {\mathcal{H}} ) $
  \item D is a self-adjoint operator on $ {\mathcal{H}} $ such
  that:
\begin{align*}
  [ & D , a ]  \; \in \; {\mathcal{B}}( {\mathcal{H}} ) \; \; \forall a \in A  \\
  ( & D - \lambda )^{- 1} \; \in \;  {\mathcal{C}}( {\mathcal{H}}
  )  \; \; \forall \lambda \in {\mathbb{C}} - {\mathbb{R}}
\end{align*}
\end{itemize}
Given a spectral triple $ ( A \, , \, {\mathcal{H}}  \, , \, D )
$:
\begin{definition} \label{def:even spectral triple}
\end{definition}
$ ( A \, , \, {\mathcal{H}}  \, , \, D ) $ IS EVEN:

there is a $ {\mathbb{Z}}_{2} $ grading on $ {\mathcal{H}}$,i.e.
an operator $ \Gamma $ on $ {\mathcal{H}} $ such that:
\begin{align}
  \Gamma^{\star} & \; = \; \Gamma  \\
  \Gamma^{2} & \; = \; 1 \\
  \{ \Gamma & , D \} \; := \;  \Gamma  D \, - \, D \Gamma \; =  \; 0 \\
   [ \Gamma & , a ] \; = \; 0 \; \; \forall a \in A
\end{align}
\begin{definition} \label{def:odd spectral triple}
\end{definition}
$ ( A \, , \, {\mathcal{H}}  \, , \, D ) $ IS ODD:

it is not even

\smallskip

Given an $ n > 0 $:
\begin{definition} \label{def:dimension of a spectral triple}
\end{definition}
$ ( A \, , \, {\mathcal{H}}  \, , \, D ) $  HAS DIMENSION $ n \;
\; ( \,  dim([ \, ( A \, , \, {\mathcal{H}}  \, , \, D ) \, ]) \;
= \; n  ) \,  $:
\begin{equation}
   | D |^{- 1} \in  {\mathcal{I}}_{ \frac{1}{n} } (
{\mathcal{H}} )
\end{equation}

\smallskip

We can at last formalize all the previously machinery about
noncommutative differentiation and integration in the following
way:

given an n-dimensional spectral triple $ ( A \, , \,
{\mathcal{H}}  \, , \, D ) $
\begin{definition} \label{def:noncommutative integration in a spectral triple}
\end{definition}
NONCOMMUTATIVE INTEGRAL OF $ a \in A$:
\begin{equation}
  \int_{NC}  a \; := \frac{1}{V} tr_{\omega} a | D |^{- n}
\end{equation}
where V is a normalization factor such that:
\begin{equation}
  \int_{NC}  I \; := \frac{1}{V} tr_{\omega} | D |^{- n} \; = \; 1
\end{equation}
By the previously discussed properties of the Dixmier trace one
has that:
\begin{theorem} \label{th:basic properties of the noncommutative integral in a spectral triple}
\end{theorem}
BASIC PROPERTIES OF THE NONCOMMUTATIVE INTEGRAL IN A SPECTRAL
TRIPLE:
\begin{enumerate}
  \item
\begin{equation}
  \int_{NC} a b \; = \; \int_{NC} b a \; \; \forall a , b \in A
\end{equation}
  \item
\begin{equation}
  \int_{NC} a^{\star} a \; \geq  \; 0  \; \; \forall a  \in A
\end{equation}
  \item
\begin{equation}
    \int_{NC} a \; = \; 0 \; \; \forall a \in
    {\mathcal{I}}_{\alpha} ( {\mathcal{H}} ) \, , \, \alpha > 1
\end{equation}
\end{enumerate}
Then:
\begin{definition} \label{def:noncommutative differential in a spectral triple}
\end{definition}
NONCOMMUTATIVE DIFFERENTIAL OF $ a \in A$:
\begin{equation}
  d_{NC} a \; := \; [ D , a ]
\end{equation}

\begin{example}
\end{example}
QUANTIZED CALCULUS ON THE CIRCLE AND MANDELBROT'S SET

Let us consider the spectral triple $ ( A \, , \, {\mathcal{H}}
\, , \, D ) $, where:

\begin{itemize}
  \item
\begin{equation}
  {\mathcal{H}} \; := \; L^{2} ( S^{(1)} , d \vec{x}_{Lebesgue} )
\end{equation}
  \item
\begin{equation}
  A \; := \; L^{\infty} ( S^{(1)} , d \vec{x}_{Lebesgue} )
\end{equation}
where a function $ f \in A $ is seen as a multiplication operator:
\begin{equation}
  ( f \psi ) (t) \; := \; f(t) \psi (t) \; \; f \in A , \psi \in {\mathcal{H}}
\end{equation}
  \item D is the linear operator on $ {\mathcal{H}} $ defined by:
\begin{equation}
  D e_{n} \; := \; sign(n) e_{n} \: , \: e_{n} ( \theta ) \; := e^{ i n \theta
  }  \; \forall \theta \in S^{(1)}
\end{equation}
\end{itemize}

Let us now consider again the quadratic maps on the complex plane
$ p_{c} (z) \, := \, z^{2} + c $ we considered in
section\ref{sec:Brudno algorithmic entropy versus the Uspensky
abstract approach}.

Demanding to $ 14^{th} $ chapter of \cite{Falconer-90} for
details it will be sufficient  to our purposes to remind that the
Julia set of the application $ p_{c} (z) $ may be simply expressed
as:
\begin{equation}
  J [ p_{c} (z) ] \; = \; \partial \, \{ z \in {\mathbb{C}} : \sup_{n \in
  {\mathbb{N}}} | p_{c}^{(n)} (z)| < \infty \}
\end{equation}
and that it may be proved that there exist an homeomorphism $ Z :
S^{(1)} \, \rightarrow J [ p_{c} (z) ] $.

Denoted by D the Hausdorff dimension of $ J [ p_{c} (z) ] $ Alain
Connes was able to prove that:
\begin{enumerate}
  \item $ | d_{NC} Z | $ is an infinitesimal of order $
  \frac{1}{D} $
  \item
\begin{equation}\label{eq:integration of continuous functions on Julia's
set}
  \exists \lambda > 0 \; : \; ( \int_{J [ p_{c} (z) ]} f d \Lambda_{D} ) \, = \;
  \lambda \int_{NC} f(Z) | D |^{ - 1}  | d_{NC} Z |^{D} \; \; \forall f \in C( J [ p_{c} (z)
  ] )
\end{equation}
where $ d \Lambda_{D} $ is the Hausdorff measure on $ J [ p_{c}
(z) ] $.
\end{enumerate}
The eq.\ref{eq:integration of continuous functions on Julia's
set} tells us that the integral w.r.t. the Hausdorff measure of
continuous functions over the Julia set $ J [ p_{c} (z) ] $ may
be computed as a noncommutative integral in the spectral triple $
( A \, , \, {\mathcal{H}} \, , \, D ) $.

Since the Mandelbrot's set $ {\mathcal{M}} $ we introduced by
definition\ref{def:Mandelbrot set} is linked to the family of
Julia sets $ J [ p_{c} (z) ] $ by the condition:
\begin{equation}
  {\mathcal{M}} \; = \; \{ c \in {\mathbb{C}} \,  : \,  J [ p_{c} (z) ]
  \text{ is connected } \}
\end{equation}
eq.\ref{eq:integration of continuous functions on Julia's set}
could be useful to investigate some of the still unknown
properties of $  {\mathcal{M}} $

\medskip

So, up to this point, we have seen how the notion of a spectral
triple implements Noncommutative Calculus.

We shall now see that it, indeed, makes much more: it implements
Noncommutative Riemannian Geometry.

Demanding to \cite{Nakahara-95}, \cite{Gilkey-95},
\cite{Landi-97}, \cite{Esposito-98},
\cite{Gracia-Bondia-Varilly-Figueroa-01} for details, let us
recall that a \textbf{spin structure} on an n-dimensional
riemannian manifold $ ( M , g ) $ is a lifting of its orthonormal
frame bundle $ O(M) \stackrel{\pi}{\rightarrow} M $ to a bundle $
S(M) \stackrel{\pi}{\rightarrow} M $, said a \textbf{spin bundle}
over M, in which the structure group $ O(n) $ is replaced by its
universal covering group (that is by definition the spin group
SPIN(n)) through the substitution of the transition functions $
t_{ij} $ by new transition functions $ \tilde{t}_{ij} $ such that:
\begin{equation}
  \phi ( \tilde{t}_{ij} ) \; = \; t_{ij}
\end{equation}
where $ \phi : SPIN(n) \mapsto SO(n) $ is the double covering.

\begin{definition} \label{def:spin manifold}
\end{definition}
SPIN MANIFOLD:

a riemannian manifold $ ( M , g ) $ which admit  a spin structure

A well known theorem of riemannian geometry states that $ ( M ,g
) $ is a spin manifold iff its first two Stiefel-Whitney classes $
w_{1}(M) $ and $ w_{2}(M) $ vanish (the vanishing of $ w_{1}(M) $
being equivalent to the orientability of M)

In this case it may, of course, admit different spin structures,
corresponding to different choices of the transition functions $
\tilde{t}_{ij} $.

\smallskip

Given an n-dimensional spin manifold $ ( M , g ) $ let us consider
a section $ \{ e_{a} , a = 1, \cdots , n \} $ of its orthonormal
frame bundle and let us relate it to the natural basis $ \{
\partial_{\mu} \} $ by the  n-beins, with components $
e_{a}^{\mu} $, so that the components $ \{ g^{\mu \nu} \} $ of the
metric g and the components  $ \{ \eta^{a b} \} $ of the flat
metric over M may be related by the equations:
\begin{align}
  g^{\mu \nu}  & \; = \; e_{a}^{\mu} e_{b}^{\nu} \eta^{a b} \\
  \eta^{a b}  & \; = \; e_{a}^{\mu} e_{b}^{\nu} g_{\mu \nu}
\end{align}
We will assume, from here and beyond, that the curve indices $ \{
\mu \} $ are raised and lowered by the curved metric g while the
flat indices $ \{ a \} $ are raised and lowered by the flat metric
$ \eta $.

Denoted by $ \nabla $ the Levi-Civita connection of $ ( M , g ) $
(i.e. the unique torsion free affine connection on M that is
compatible with g) let us introduce its connection coefficients $
\omega_{ \mu a }^{b} $ defined by the condition
\begin{equation}
 \omega_{ \mu a }^{b} e^{b} \; := \; \nabla_{\mu} e_{a} \; \; \mu , \nu = 1 , \cdots , n
\end{equation}
Let us now introduce the Clifford bundle $ C(M)
\stackrel{\pi}{\rightarrow} $ over M, whose fiber at $ x \in M $
is the complexified Clifford algebra $ Cliff_{{\mathbb{C}}} $, and
the space $ \Gamma ( M , C(M) ) $ of sections on it.

Called $ {\mathcal{H}} \, := \, L^{2} ( M , S ) $ the Hilbert
space of the irreducible spin bundle over M, with the inner
product given by:
\begin{equation}
  < \psi_{1} |  \psi_{2} > \; := \; \int_{M} d \mu (g)
  \bar{\psi}_{1}( x ) \psi_{2} (x)
\end{equation}
let us now introduce that the map $ \gamma \, : \, \Gamma ( M ,
C(M) ) \; \rightarrow \; {\mathcal{B}} ( {\mathcal{H}} ) $
defined by the condition:
\begin{equation} \label{eq:curved Dirac matrices}
  \gamma^{\mu} (x) \; := \;\gamma ( dx^{\mu} ) \; :=  \; \gamma^{a} e_{a}^{\mu}
\end{equation}
and extended as an algebra map requiring its linearity under
linear combinations with coefficients taking values in the algebra
$ A \; := \; {\mathcal{F}} (M) $ of complex valued smooth
functions over M.

$ \Gamma ( M , C(M) ) $ is as $ \star $-algebra and $ \gamma $ is
an involutive morphism.

By the definition of a Clifford algebra and by eq.\ref{eq:curved
Dirac matrices} the curved Dirac matrices $ \{ \gamma^{\mu}(x) \}
$ and the flat Dirac matrices $  \{ \gamma^{a} \} $ obey the
relations:
\begin{align}\label{eq:Clifford structure on Dirac matrices}
  \gamma^{\mu}(x) &  \gamma^{\mu}(x) \, + \,  \gamma^{\nu}(x)  \gamma^{\mu}(x) \; = \; -2 g( dx^{\mu} ,  dx^{\nu} ) \; = \; - 2 g^{\mu \nu} \; \; \mu , \nu = 1 , \cdots , n  \\
   \gamma^{a} & \gamma^{b} \, + \, \gamma^{b} \gamma^{a} \; = \; -
   2 \eta^{a b} \; \; a , b = 1 , \cdots , n
\end{align}
The lift of the Levi-Civita connection to the bundle of spinors
is then:
\begin{equation}
  \nabla^{S}_{\mu} \; = \; \partial_{\mu} \, + \, \omega^{S}_{\mu}
  \; = \; \partial_{\mu} + \frac{1}{4} \omega_{\mu a b} \gamma^{a}
  \gamma^{b}
\end{equation}
We can then introduce the following:
\begin{definition} \label{def:Dirac operator}
\end{definition}
DIRAC OPERATOR:

the linear operator D on the Hilbert space $ {\mathcal{H}} $
given by:
\begin{equation}
  D \; := \; \gamma \circ \nabla^{S}
\end{equation}

The properties of the Dirac operator justify the following:
\begin{definition} \label{def:canonical spectral triple of a spin manifold}
\end{definition}
CANONICAL SPECTRAL TRIPLE OF $ ( M , g )$:

the n-dimensional spectral triple $ ( A , {\mathcal{H}} , D ) $,
where (we recall that):
\begin{enumerate}
  \item $ A \; := \; {\mathcal{F}} (M) $ is the algebra of all
  complex valued smooth functions on M
  \item $ {\mathcal{H}} $ is the Hilbert space of square integral
  sections (w.r.t. the the metric measure $ d \mu (g) $ ) of the
  irreducible spinor bundle over M
  \item D is the Dirac operator of  $ ( M , g ) $
\end{enumerate}

If n is even the canonical spectral triple of $ ( M , g ) $ is
even, the $ {\mathbb{Z}}_{2} $-grading being given by:
\begin{equation}
  \Gamma \; := \; \gamma^{n+1} \; := \; i^{\frac{n}{2}} \gamma^{1}
  \cdots \gamma^{n}
\end{equation}

\smallskip

And now comes the first astonishing fact:
\begin{theorem} \label{th:the canonical spectral triple of a spin manifold encodes all its riemannian structure}
\end{theorem}
THE CANONICAL SPECTRAL TRIPLE OF A SPIN MANIFOLD ENCODES ALL ITS
RIEMANNIAN STRUCTURE

\begin{enumerate}
  \item  the geodesic distance d( p , q ) of two points of  $ ( M
  , g ) $ is given by:
\begin{equation}\label{def:geodesic distance in algebraic terms}
  d(p,q) \; = \; \sup_{f \in A} \{ | f(p) - f(q) | \, : \, \| [ D , f ] \| \leq 1
  \} \; \; \forall p ,q \in M
\end{equation}
  \item the integration of a function $ a \in A $ w.r.t. the metric
  measure of $( M , g) $ is  substantially given by its noncommutative integral:
  \begin{equation}
  \int_{M} d \mu ( g ) a \;  = \; c(n) \int_{NC} a    \; \; \forall a \in A
\end{equation}
where:
\begin{equation}
  c(n) \; := \; 2^{ n - \lfloor \frac{n}{2} \rfloor - 1} \, \pi^{\frac{n}{2}}
  \, \Gamma ( \frac{n}{2} )
\end{equation}
\end{enumerate}

By a suitable definition of an equivalence relation among spectral
triples, also the converse holds, i.e., given a commutative
spectral triple  $ ( \, A \, , \, {\mathcal{H}} \, , \, D ) $
there exist a closed finite-dimensional riemannian spin manifold
$ ( M , g ) $ whose canonical spectral triple is equivalent to $ (
\, A \, , \, {\mathcal{H}} \, , \, D ) $.

Furthermore:
\begin{itemize}
  \item given a diffeomorphism $ \phi \in Diff(M) $ of a closed finite-dimensional riemannian spin manifold $ ( M , g )
  $ we may associate to it the automorphism $ \alpha_{\phi} $ of the  corrispondent
  involutive algebra $ A \, := \, {\mathcal{F}}(M) $, defined as:
\begin{equation} \label{eq:automorphism associated to a diffeomorphism}
  \alpha_{\phi} (f) (x) \; := \; f( \phi^{- 1}(x)) \; \; \forall f \in {\mathcal{F}}(M) , \forall x \in M
\end{equation}
  \item given an automorphism $ \alpha \in AUT(A) $ of a
  commutative spectral triple $ ( \, A \, , \, {\mathcal{H}} \, , \, D )
  $ there exist a diffeomorphism $ \phi \in Diff(M) $ of the
  associated closed finite-dimensional riemannian spin manifold $ ( M , g
  ) $  such that $ \alpha \; = \; \alpha_{\phi} $
\end{itemize}

\smallskip

All these results may  then be enunciated in the abstract
language of Categories' Theory as the following:
\begin{conjecture} \label{con:the category isomorphism at the basis of Noncommutative Geometry}
\end{conjecture}
THE CATEGORY EQUIVALENCE AT THE BASIS OF NONCOMMUTATIVE GEOMETRY

The \textbf{category} having as \textbf{objects} the \textbf{closed finite-dimensional riemannian
spin manifolds} and as \textbf{morphisms} the \textbf{diffeomorphisms of such manifolds} is equivalent to the \textbf{category} having as \textbf{objects} the
\textbf{abelian spectral triples} and as \textbf{morphisms}  the \textbf{automorphisms of the involved involutive algebras}.

\smallskip

We can now, at last, see how Noncommutative Geometry allows to
afford Quantum Information Geometry.

Given a spectral triple $ ( A , {\mathcal{H}} , D ) $  over the $
W^{\star}-algebra $ A by eq.\ref{def:geodesic distance in
algebraic terms} and conjecture\ref{con:the category isomorphism
at the basis of Noncommutative Geometry} it results natural to
define the following noncommutative generalization of the geodesic
distance among probability distributions:
\begin{definition} \label{def:noncommutative geodesic distance among states}
\end{definition}
NONCOMMUTATIVE GEODESIC DISTANCE AMONG $ \omega_{1} \in S(A) $
AND $ \omega_{2} \in S(A) $ :
\begin{equation}
  d ( \omega_{1} \, , \, \omega_{2} ) \; := \; \sup_{ a \in A } \{ | \omega_{1} (a) -  \omega_{2} (a) | \, : \, \| [ D , a ] \| \leq 1  \}
\end{equation}

So, given a Von Neumann algebra $ A \, \subseteq \, {\mathcal{B}}
({\mathcal{H}}) $, our genuinely noncommutative approach to Quantum
Information Geometry is completely specified if we succeed in
individuating a \emph{"natural"} \textbf{noncommutative Dirac
operator} to use in order of defining the correct spectral triple
$ ( A \,  , \,  {\mathcal{H}}  \,  , \, D ) $ to use for
computing the distance among noncommutative probability measures by
eq.\ref{def:noncommutative geodesic distance among states}.

Let us consider, at this purpose, the commutative case:

given a spin manifold $ ( M , g) $ the correct operator to use in
order to obtain the correct expression for the geodesic distance, namely the Dirac operator D of (M ,g ), is that obtained
minimizing the action :
\begin{equation} \label{def:commutative spectral action}
  S ( x ,  \Lambda ) \; := \;
  Tr_{ L^{2}(M,S) } ( \chi ( \frac{ x^{2}}{ \Lambda^{2}} ) )
\end{equation}
where $ \Lambda $ is a cut-off with the dimensions of the inverse
of a length, $ \chi $ is a proper cut-off function throwing
away the contribution of the $ x ^{2} $'s eigenvalues greater than
$ \Lambda ^{2} $ that we will assume to be Heaviside's step
function and x denotes the unknown operatorial quantity.

The action of eq.\ref{def:commutative spectral action} is defined
on the set $ OP[  {\mathcal{F}}(M) \, , \, L^{2}(M,S) ] $ of all
the self-adjoint operators x  on $ L^{2}(M,S) $ such that $ (
{\mathcal{F}}(M) \, , \, L^{2}(M,S) \, , \, x ) $ is a spectral
triple.

The meaning of the cut-off $  \Lambda $ for the variational
problem of eq.\ref{def:commutative spectral action} is the
following:
\begin{enumerate}
  \item one impose  the variational condition:
\begin{equation}
  \frac{  \delta S ( x , , \Lambda )}{ \delta x }  \; = \; 0
\end{equation}
obtaining an equation of the form:
\begin{equation} \label{eq:cut-offed equation for the operator}
  F_{1} [ x , \Lambda ] \; = \; 0
\end{equation}
where $  F_{1} $ is a certain functional of the unknown quantity
x and the cut-off $ \Lambda $
  \item one takes the limit $ \Lambda \rightarrow \infty $ in eq.\ref{eq:cut-offed equation for the operator}
obtaining a new equation of the form:
\begin{equation} \label{eq:equation for the operator}
  F_{2} [ x ] \; = \; 0
\end{equation}
where $  F_{2} $ is another functional of the only unknown
operator x
\end{enumerate}
By conjecture\ref{con:the category isomorphism at the basis of
Noncommutative Geometry} is appears then natural to generalize
noncommutatively such a variational procedure, i.e. to choose the
operator x by which to compute, via eq.\ref{def:noncommutative
geodesic distance among states}, the noncommutative geodesic
distance between two states as the solution of the variational
problem for the following:
\begin{definition} \label{def:noncommutative spectral action}
\end{definition}
NONCOMMUTATIVE SPECTRAL ACTION FOR $ ( A , {\mathcal{H}}) $:

the map $ S \, : \,  OP[  {\mathcal{H}} \, , \, A ] \: \rightarrow
\: {\mathbb{R}} $ given by:
\begin{equation}
  S [ x \, , \, \Lambda ] \; := \;
  Tr_{{\mathcal{H}}} ( \chi ( \frac{ x^{2}}{ \Lambda^{2}} ) )
\end{equation}
where:
\begin{equation}
  OP[  {\mathcal{H}} \, , \, A ] \; := \; \{ x \, : \, (  {\mathcal{H}} \, , \,
  A \, , \, x ) \text{ is a spectral triple } \}
\end{equation}

\smallskip

\begin{example}
\end{example}
DISTANCE OF A NONCOMMUTATIVE PROBABILITY DISTRIBUTION ON THE
NONCOMMUTATIVE SPACE OF QUBITS' SEQUENCES BY THE UNBIASED ONE.

Let us apply our noncommutative-geometric approach to Quantum
Information Geometry to answer the following question:

\smallskip

how much a given noncommutative probability distribution $ \omega
\in S( \Sigma_{NC}^{\infty}) $ differs from the unbaised one $
\tau_{unbaised} $?

\smallskip
According to our strategy such a distance is given by:
\begin{equation}
  d ( \omega \, , \, \tau_{unbiased} ) \; := \; \sup_{ \bar{x}_{NC} \in \Sigma_{NC}^{\infty}  } \{ | \omega (\bar{x}_{NC}) -  \tau_{unbiased}( \bar{x}_{NC} ) | \, : \, \| [ D , \bar{x}_{NC} ] \| \leq 1  \}
\end{equation}
where D is the element of $ OP[  {\mathcal{H}}_{
\tau_{{\mathbb{Z}}}} \, , \, \Sigma_{NC}^{\infty} ] $ minimizing
the spectral action of definition\ref{def:noncommutative spectral
action}.

\medskip

Leaving, anyway, aside Quantum Information Geometry, let us
investigate a  more direct strategy of Quantum Statistical
Inference, namely Quantum Bayesianism.

The issue of formalizing Quantum Bayesian Theory has been
recentely analyzed by various authors \cite{Ozawa-97},
\cite{Schack-Brun-Caves-00}, \cite{Caves-Fuchs-Schack-01}.

Unfortunately no one of them adopts the general language of
Quantum Probability Theory, on which the mathematical foundations
of the whole matter, from a less empirical point of view, lies.

An exeption to this attitude is the authorithative  exposition of   Quantum Statistical Decision Theory
made by Alexander S. Holevo in the section2.2 of \cite{Holevo-99}, from which, as already happened as to Quantum Information Geometry,
we will implicitly move away.

Though well-knowing that, as we will see, Bayes' Formula is
nothing but a matter of conditional expectations, Holevo keeps
away from  its natural noncommutative generalization, namely
definition\ref{def:Bayes formula}.

Indeed, according to him:
\begin{center}
\textit{"Conditional expectations play a less important part in quantum than in classical probability, since in general
the conditional expectation into a given subalgebra $ {\mathcal{B}} $ with respect to a given state S exists only if
 $ {\mathcal{B}} $ and S are related in a very special way which in a sense reduces the situation to the classical one;
 for more see n. 1.3 in Chapter 3." (extracted from section1.3.2 of \cite{Holevo-99})}
\end{center}

Holevo's argument, clarified in section3.1.3, is based on Takesaki's theorem, namely our theorem\ref{th:Takesaki's theorem}; while the
conseguences we will infer from this theorem we be essentially:
\begin{enumerate}
  \item the impossibility of quantum-bayesian-subjectivism (as first remarked by Miklos Redei: cfr. the seection 8.2 of \cite{Redei-98})
  \item the existence of the constraint\ref{def:constraint for the feasibility of the bayesian statistical inference}
\end{enumerate}
Holevo's claim that Takesaki's Theorem implies a reduction to the classical case is wrong, being based on the
observation that, in our terminology, the constraint\ref{def:constraint for the feasibility of the bayesian statistical
  inference}, is certainly satisfied if the modular operator of S belongs to the commutant of $  {\mathcal{B}} $; such a condition, though sufficient, is indeed
far from being necessary.

\smallskip

Let us so start to analyze one of the deepest conseguences of
\textbf{Quantum Probability Theory}: the \textbf{Bayesian
Statistical Inference Theory}:

\bigskip

Let us consider a statistician having access only to the
information concerning the all algebraic random variables
belonging to a $W^{\star}$-sub-algebra $ A_{accessible} $ of a
quantum probability space $ ( A \, , \, \omega ) $.

This means that he doesn't know the state $ \omega \in S(A) $ but
only its restriction to the algebra $ A_{accessible} $ he can
test, i.e.:
\begin{equation}
  \omega_{accessible} \; := \; \omega \, |_{ A_{accessible} } \;
  \in S ( A_{accessible} )
\end{equation}

\medskip

Let us suppose that, at the beginning, he hasn't used even this
partial available information:

according to the Bayesian Theory the best estimation of the true
state $ \omega $ that he can make in this situation is to assume
as estimation the uniform algebraic probability distribution:
\begin{equation}
  \omega_{A \; PRIORI} \; := \; \tau_{unbiased}
\end{equation}

\medskip

Here it does arises the first problem, common both to the
classical case and to the quantum case:  the \textbf{canonical
trace} $ \tau_{unbiased} $ exists if and only if the  Von Neumann
algebra  A is finite.

Supposed, anyway, that this is the case let us consider the
\textbf{statistical-inference's problem }:

\textbf{which is the optimal way by which the statistician can
improve his estimation of the true state $ \omega $ using the
information  that is available to him, i.e. using $
\omega_{accessible} $ ?}

The answer of the Bayesian Theory is inclosed in the following:
\begin{definition} \label{def:Bayes formula}
\end{definition}
BAYES FORMULA:
\begin{equation}
   \omega_{A \; PRIORI} ( \cdot ) \, = \, \tau_{unbiased} ( \cdot ) \; \rightarrow \omega_{A \;
   POSTERIORI} ( \cdot ) \; := \;  \omega_{accessible} ( E_{unbiased} \cdot )
\end{equation}
where $ E_{unbiased} \, : \, A \mapsto A_{accessible} $ is the
\textbf{conditional expectation w.r.t. $ A_{accessible} \; \;
\tau_{unbiased}$-invariant} whose definition and properties we are
going to introduce.

Given a $ W^{\star}$-algebra A:
\begin{definition} \label{def:conditional expectation on a Von Neumann algebra}
\end{definition}
CONDITIONAL EXPECTATION ON A W.R.T. $ A_{accessible} $:

a linear map $ E \, : \,  A \; \rightarrow A_{accessible}  $ such
that:
\begin{enumerate}
  \item
\begin{equation}
  E(a) \; \geq \; 0 \; \; \forall a \in A_{+}
\end{equation}
  \item
\begin{equation}
  E(a) \; = \;  a \; \; \forall a \in A_{accessible}
\end{equation}
  \item
\begin{equation}
  E ( a b ) \; = \;  E(a) b \; \; \forall a \in A , \forall b \in A_{accessible}
\end{equation}
\end{enumerate}
Let us suppose that the Von Neumann algebra A  and its subalgebra
$ A_{accessible} $ act on the Hilbert space $ {\mathcal{H}} $,
i.e. $ A_{accessible} \; \subset \; A \; \subseteq \;
{\mathcal{B}} ({\mathcal{H}}) $. We will say that:
\begin{definition} \label{def:injective Von Neumann}
\end{definition}
A IS INJECTIVE:
\begin{equation}
  \exists \, E : {\mathcal{B}} ({\mathcal{H}}) \rightarrow A \text{ conditional expectation on
  } {\mathcal{B}} ({\mathcal{H}}) \; w.r.t. \; A
\end{equation}
In an epoch making result of 1976 Alain Connes proved that:
\begin{theorem} \label{th:equivalence of injectivity and hyperfiniteness}
\end{theorem}
\begin{equation}
  injectivity \; \Leftrightarrow \; hyperfiniteness
\end{equation}

Given a conditional expectation on A w.r.t. to $ A_{accessible}
$  and a state $ \omega  \in S(A) $:
\begin{definition} \label{def:state preserving conditional expectation on a Von Neumann algebra}
\end{definition}
E IS $ \omega $-PRESERVING:
\begin{equation}
  \omega \circ E \; = \; E
\end{equation}

The issue about the existence of state-preserving conditional
expectations is rather subtle involving the Tomita-Takesaki Modular
Theory \cite{Araki-97}.

Given a Von Neumann  algebra A  acting on a separable Hilbert
space $ {\mathcal{H}} $ and a  vector $ | \psi > \in
{\mathcal{H}} $:
\begin{definition} \label{def:cyclic vector for a Von Neumann algebra}
\end{definition}
$ | \psi > $ IS CYCLIC FOR A:
\begin{equation}
  A \, | \psi >  \; \text{ is dense in } {\mathcal{H}}
\end{equation}
\begin{definition} \label{def:separating vector for a Von Neumann algebra}
\end{definition}
$ | \psi > $ IS SEPARATING FOR A:
\begin{equation}
  ( a | \psi > \: = \: 0 \; and \; a \in A )  \; \Rightarrow \; a \,
  = \, 0
\end{equation}
Supposing the vector $ | \psi > $ to be cyclic  and separating
for A let us consider the linear operator  $ S_{| \psi >} $ on $ {\mathcal{H}} $
defined by the condition:
\begin{equation}
  S_{| \psi > } a | \psi > \; := a^{\star} | \psi > \; \; a \in A
\end{equation}
The operator $ S_{| \psi >} $  has a closure $ \bar{S}_{| \psi >
} $ that can be used to introduce the following:
\begin{definition} \label{def:modular operator}
\end{definition}
MODULAR OPERATOR W.R.T. A AND $ | \psi > $:
\begin{equation}
  \Delta_{ | \psi >} \; := \;  S_{| \psi > } \, \bar{S}_{| \psi >
}
\end{equation}
Let us then introduce the following:
\begin{definition} \label{def:conjugate modular operator}
\end{definition}
MODULAR CONJUGATION W.R.T. A AND $ | \psi > $:

the operator $ J_{ | \psi >} $ occurring in the polar
decomposition:
\begin{equation}
  S_{| \psi > } \; = \; J_{| \psi > } \, \Delta_{ | \psi >}^{\frac{1}{2}}
\end{equation}

\smallskip

The corner stone of the Modular Theory is the following:
\begin{theorem} \label{th:Tomita-Takesaki's theorem}
\end{theorem}
TOMITA-TAKESAKI'S THEOREM
\begin{enumerate}
  \item
\begin{equation}
  \Delta_{ | \psi >}^{it} \,  A \,  \Delta_{ | \psi >}^{- it} \; =
  \; A \; \; \forall t \in {\mathbb{R}}
\end{equation}
  \item
\begin{equation}
  J \, A \, J \; = \; A'
\end{equation}
\end{enumerate}
that, in particular, justifies the following:
\begin{definition} \label{def:group of modular automorphisms w.r.t. a pure state}
\end{definition}
GROUP OF MODULAR AUTOMORPHISMS OF A W.R.T. $ | \psi > $:

the one-parameter subgroup of AUT(A):
\begin{equation}
  \sigma^{| \psi >}_{t}(a) \; := \;  \Delta_{ | \psi >}^{it} \,  a \,  \Delta_{ | \psi >}^{- it}
\end{equation}
The group of modular automorphisms $  \sigma^{| \psi >}_{t} $
depends on the cyclic and separating vector $ | \psi > \in
{\mathcal{H}} $, i.e. from the normal state $ \omega_{| \psi > }
\in S(A)  $ with associated density operator $  \rho_{ | \psi > <
 \psi |}$.

If one looks at outer automorphisms, anyway, such a dependence
disappears:
\begin{theorem} \label{th:independence from the state of the group of outer modular automorphisms}
\end{theorem}
INDEPENDENCE FROM THE STATE OF THE GROUP OF OUTER MODULAR
AUTOMORPHIMS:

\begin{hypothesis}
\end{hypothesis}
\begin{equation*}
  | \psi_{1} > \, , \, | \psi_{2} >   \in {\mathcal{H}} \text{ cyclic and separating for A }
\end{equation*}
\begin{thesis}
\end{thesis}
\begin{equation}
  [  \sigma^{| \psi_{1} >}_{t} ]_{OUT(A)} \; = \;  [  \sigma^{| \psi_{2} >}_{t} ]_{OUT(A)}
\end{equation}
The proof of theorem\ref{th:independence from the state of the
group of outer modular automorphisms} allowed Connes to introduce
the following two $\star$-isomorphisms invariants of Von Neumann
algebras:
\begin{definition} \label{def:first Connes' invariant of a Von Neumann algebra}
\end{definition}
FIRST CONNES' INVARIANT OF A:
\begin{equation}
  Inv_{Connes}^{(1)}(A) \; := \; \bigcap_{ | \psi > } Spectrum( \Delta_{ | \psi >} )
\end{equation}
\begin{definition} \label{def:second Connes' invariant of a Von Neumann algebra}
\end{definition}
SECOND CONNES' INVARIANT OF A:
\begin{equation}
  Inv_{Connes}^{(2)}(A) \; := \; \{   t \in {\mathbb{R}}  \: : \:  \sigma^{| \psi >}_{t} \in INN(A)  \}
\end{equation}
by which he classified type-III factors:
\begin{definition}
\end{definition}
$  Type(A) \; := \; III_{0} $:
\begin{equation}
  cardinality_{NC}(A) = \aleph_{2} \: and \: Inv_{Connes}^{(1)}(A) = \{ 0 , 1 \}
\end{equation}
\begin{definition}
\end{definition}
$  Type(A) \; := \; III_{1} $:
\begin{equation}
  cardinality_{NC}(A) = \aleph_{2} \: and \: Inv_{Connes}^{(1)}(A) = {\mathbb{R}}_{+}
\end{equation}
\begin{definition}
\end{definition}
$  Type(A) \; := \; III_{\lambda}  \; (  0 < \lambda < 1 )$:
\begin{equation}
  cardinality_{NC}(A) = \aleph_{2} \: and \: Inv_{Connes}^{(1)}(A) = \{ \lambda^{n} \, , \,  n \, \in {\mathbb{Z}} \}
\end{equation}

Furthermore Connes proved that:
\begin{theorem}  \label{th:unicity of injective factors}
\end{theorem}
SINGLENESS OF INJECTIVE FACTORS OF EACH TYPE EXCEPT $ III_{0} $:

there exist a unique injective factor of each type $ I _{n} \, ,
\, n \in {\mathbb{N}} $, $ I_{\infty} $, $ II_{1} $, $
II_{\infty} $, $ III_{\lambda} \, , \, \lambda \in ( 0 , 1 ] $

\begin{example} \label{ex:Powers factors}
\end{example}
THE HYPERFINITE $ III_{\lambda} \; ( \, 0 \, < \, \lambda \, < \, 1
\, ) $ FACTOR

\smallskip

To get an intuitive insight into the structure of the
noncommutative space of qubits' sequences $ \Sigma_{NC}^{\infty}
= R $ there is nothing better than analyzing its differences with
a class of purely infinite factors, usually called the Powers
factors, defined in a way very similar to that we followed in
example\ref{ex:the hyperfinite finite continuous factor} to
define R.

It is useful, at this purpose to introduce a suitable, compact
notation concerning infinite tensor products of an algebraic
probability space $ ( A \, , \, \omega ) $:
\begin{definition}
\end{definition}
\begin{equation}
  \bigotimes_{n=1}^{\infty} ( A \, , \, \omega ) \; :=
  \pi_{\bigotimes_{n=1}^{\infty} \omega } (  \bigotimes_{n=1}^{\infty} A )
  ''
\end{equation}
With this notation our space of qubits' sequences may be compactly
expressed as:
\begin{equation}\label{eq:compact expression of the space of qubits' sequences}
  \Sigma_{NC}^{\infty} \; = \; R \; = \;  \bigotimes_{n=1}^{\infty} ( M_{2} ({\mathbb{C}}) \, , \tau_{unbiased} )
\end{equation}
where:
\begin{equation}
  \tau_{unbiased} ( \cdot) \; = \; Tr [ \begin{pmatrix}
    \frac{1}{2} & 0 \\
    0 & \frac{1}{2} \
  \end{pmatrix} \; \cdot ]
\end{equation}
A physical realization of the one-qubit unbiased quantum
probability space $  ( M_{2} ({\mathbb{C}}) \, , \tau_{unbiased}
) $ is given by a $ spin 1/2 $ system in thermal equilibrium at
temperature $ T \, = \; + \infty $, as can be seen observing that
the canonical-ensemble's state:
\begin{equation}
  \omega_{CAN} ( H , \beta )( \cdot ) \; := \; Tr[ \frac{e^{- \beta H} }{Tre^{- \beta
  H}} \; \cdot ]
\end{equation}
collapses to $ \tau_{unbiased} $ when $ k T \, = \, \beta^{ - 1}  \,
\rightarrow \, \infty $ for any self-adjoint, bounded from  below
hamiltonian operator H.

Let us now introduce the following:
\begin{definition}  \label{def:Powers factors}
\end{definition}
POWERS FACTORS:
\begin{equation}
  R_{ \lambda } \; := \; \bigotimes_{n=1}^{\infty} ( M_{2} ({\mathbb{C}}) \, , \omega_{\lambda}
  ) \; \; \lambda \in ( 0 , 1 )
\end{equation}
where:
\begin{equation}
  \omega_{\lambda} ( \cdot ) \; := \; Tr[ \begin{pmatrix}
    \frac{1}{1 +  \lambda} & 0 \\
    0 & \frac{\lambda}{1 +  \lambda}
  \end{pmatrix}  \, \cdot ]
\end{equation}
It may be proved that:
\begin{equation}
  Type( R_{ \lambda }) \; = \; \lambda \; \; \forall \lambda \in (
  0 , 1 )
\end{equation}
By the same considerations made in the example\ref{ex:the
hyperfinite finite continuous factor} we may infer that   each $
R_{ \lambda } \, , \, \lambda \in ( 0 , 1 )  $ is hyperfinite and,
hence, by theorem\ref{th:equivalence of injectivity and
hyperfiniteness} and theorem\ref{th:unicity of injective
factors}, it is the only hyperfinite, type $ III_{\lambda}$
factor.

Clearly, for $ \lambda \in [ 0 , 1 ) $, the  state $
\omega_{\lambda} $ is not unbiased: for $ \lambda = 0 $ it is
simply the pure state of density matrix $ | 0 > < 0 | $. Then,
when $ \lambda $ monotonically increases from 0 to 1, it becomes a
mixture of $ | 0 > < 0 | $ and $ | 1 > < 1 | $ with the bias
bestowing a privilege on $ | 0 > < 0 | $ decreasing so that it
vanishes in the limit $ \lim_{\lambda \rightarrow 1} R_{\lambda}
\, = \, R $.

\smallskip

Let us now observe that definition\ref{def:group of modular
automorphisms w.r.t. a pure state} defines the group of modular
automorphisms of the Von Neumann algebra $ A \subseteq
{\mathcal{B}} ( {\mathcal{H}} ) $ w.r.t. a pure, normal state $
\omega \, \in \, \Xi(A) \bigcap S(A)_{n} $. In order of
generalizing it to non-pure normal states we have to introduce a
generalization of the modular operator, the \textbf{spatial
derivative operator} and the associated \textbf{noncommutative
Radon-Nikodym derivative} \cite{Ohya-Petz-93}, \cite{Connes-94}.

Given an arbitrary state $  \psi  \, \in \, S(A) $ let us
introduce the following:
\begin{definition}
\end{definition}
LINEAL OF  $  \psi   $:
\begin{equation}
  D( {\mathcal{H}} ,   \psi  ) \; := \; \{ | \xi > \in
  {\mathcal{H}} \, : \, \| a | \xi > \| \, \leq \, C_{| \xi >}
  \psi ( a a^{\star} )  \; \; \forall a \in A \}
\end{equation}
Considered the GNS-triplet  $ ( {\mathcal{H}}_{\psi} \, , \,
\pi_{\psi} \, , \, | \Psi >_{\psi} $ corresponding to the state $
psi $ and taken any $ | \xi > \in D( {\mathcal{H}} , \psi ) $,
let us introduce the following operators:
\begin{definition}
\end{definition}
$ R^{\psi}( | \xi > )  \, : \,  {\mathcal{H}}_{\psi} \rightarrow
{\mathcal{H}} $:
\begin{equation}
  R^{\psi}( | \xi > )  \, \pi_{\psi}(a) \, | \Psi >_{\psi} \; :=
  \; a  | \xi > \; \; a \in A
\end{equation}
\begin{definition}
\end{definition}
\begin{equation}
  \Theta^{\psi} ( ( | \xi > ) ) \; := \; R^{\psi}( | \xi > ) ( R^{\psi}( | \xi >
  ))^{\star}
\end{equation}
Fixed a $ \varphi ' \, \in \, A ' $:
\begin{definition} \label{def:spatial derivative operator}
\end{definition}
SPATIAL DERIVATIVE OPERATOR W.R.T. $  \varphi ' $ AND $ \psi $:

the positive self-adjoint operator $ \Delta (  \varphi ', \psi )
$ associated by the Form Representation Theorem to the closure of
the quadratic form q:
\begin{equation}
  q ( | \xi > + | \eta > ) \; := \; \varphi ' ( \Theta^{\psi} ( ( | \xi > )
  ))
\end{equation}
such that:
\begin{enumerate}
  \item
\begin{equation}
  \| \Delta ( \varphi ' \, , \,  | \xi > ) ^{\frac{1}{2}} \, | \zeta
  > \| ^{2} \; = \; q( | \zeta
  > \| ) \; \; \forall | \zeta
  >  \in Dom(q)
\end{equation}
  \item
\begin{equation}
  Dom(q) \text{ is the core of }  \Delta ( \varphi ' \, , \,  | \xi > ) ^{\frac{1}{2}}
\end{equation}
\end{enumerate}
Given now two states $ \omega_{1} \, , \, \omega_{2} \, \in \,
S(A) $:
\begin{definition} \label{def:noncommutative Radon-Nikodym derivative}
\end{definition}
NONCOMMUTATIVE RADON NIKODYM DERIVATIVE OF $ \omega_{1} $ W.R.T. $
\omega_{2} $:
\begin{equation}
  ( D \omega_{1} \, : \,  D \omega_{2} )_{t} \; := \; \Delta (
  \omega_{1} / \varphi ' )^{it} \,  \Delta (
  \omega_{2} / \varphi ' )^{- it} \; \; t \in {\mathbb{R}}
\end{equation}
where the name remarks the independence from the state $  \varphi
' $.

One can now generalize definition\ref{def:group of modular
automorphisms w.r.t. a pure state} in the following way:

given a state $ \omega \in S(A)_{norm} $:
\begin{definition} \label{def:group of modular automorphisms w.r.t. a normal state}
\end{definition}
GROUP OF MODULAR AUTOMORPHISMS OF A W.R.T. $ \omega $:

the one-parameter subgroup of AUT(A):
\begin{equation}
  \sigma_{t}^{\omega} (a) \; := \; \Delta( \omega , \varphi ' ) ^{i t
  } a \Delta( \omega , \varphi ' ) ^{- i t
  } \; \; a \in A  \, , \,   t \in  {\mathbb{R}}
\end{equation}

We have seen how theorem\ref{th:equivalence of injectivity and
hyperfiniteness} poses some  constraint on the existence of
conditional expectations.

As to state-preserving conditional expectation, we have at last
all the required ingredients to state Takesaki's theorem ruling
the whole business:
\begin{theorem} \label{th:Takesaki's theorem}
\end{theorem}
TAKESAKI'S THEOREM

\bigskip

\begin{hypothesis}
\end{hypothesis}

\bigskip

\begin{equation*}
  ( A \, , \, \omega ) \; \; \text{ algebraic probability space}
\end{equation*}
\begin{equation*}
   A_{accessible}  \; \;  \text{ $W^{\star} $-subalgebra of A}
\end{equation*}

\bigskip

\begin{thesis}
\end{thesis}

\bigskip

\begin{enumerate}
  \item a \textbf{conditional expectation }$ E_{\omega} \, : \,
  A \rightarrow A_{accessible} $ \textbf{ w.r.t. $
A_{accessible} $} \; $ \omega $ - \textbf{invariant} exists
\textbf{if and only if  $ A_{accessible} $ is invariant under the
modular group of $ \omega $, namely}:
\begin{equation}
  \sigma^{\omega}_{t} ( a ) \; \in \; A_{accessible} \; \;  \forall a \in A_{accessible} \, \forall t \in {\mathbb{R}}
\end{equation}

  \item if it exists , the \textbf{conditional expectation }$ E_{\omega} \, : \,
  A_{accessible}
\rightarrow A $ \textbf{ w.r.t.  $ A_{accessible} $} \; $ \omega
$ - \textbf{invariant} is unique
\end{enumerate}

We have at last all the necessary technical machinery to analyze
how and when the Bayesian Strategy can be applied to our problem
of Statistical Inference.

It is important, first of all, to underline that the state on the
complete algebra A involved in the \textbf{Bayes formula} is not
the state $ \omega $, that the statistician doesn't know, but the
\textbf{a priori estimation of it} $ \omega_{A \; PRIORI} $.

Conseguentially, for the theorem\ref{th:Takesaki's theorem}, the
involved conditional expectation  $ E_{\omega_{A \; PRIORI}} \, :
\, A_{accessible} \rightarrow A $ exist (and in this case is
unique) under the following:
\begin{definition} \label{def:constraint for the feasibility of the bayesian statistical inference}
\end{definition}
NECESSARY AND SUFFICIENT CONDITION FOR THE FEASIBILITY OF THE
BAYESIAN STATISTICAL INFERENCE:
\begin{equation}
  \sigma^{\omega_{A \; PRIORI}}_{t} ( a ) \; \in \; A_{accessible} \; \;  \forall a \in A_{accessible} \, \forall t \in {\mathbb{R}}
\end{equation}

\medskip

In the classical case such a condition is always satisfied,
guaranteeing that bayesian statistical inference on finite
classical probability spaces is always feasible.

\medskip

This doesn't happen, instead, in the quantum case with the
following fundamental consequence lucidly discovered by Miklos
Redei (cfr. the cap.8 of \cite{Redei-98}) and confuting the point
of view exposed in \cite{Caves-Fuchs-Schack-01}.

\begin{theorem} \label{th:impossibility of a subjectivistic Bayesian foundation of Quantum Probability Theory}
\end{theorem}
IMPOSSIBILITY OF A SUBJECTIVISTIC BAYESIAN FOUNDATION OF QUANTUM
PROBABILITY THEORY

 as far as \textbf{Foundations of Probability Theory} is
concerned Quantum Bayesian Theory can't be used to give a
\textbf{subjectivist foundation} of Quantum Probability Theory as
it happens in the classical case \cite{Bernardo-Smith-00}.

\begin{example} \label{ex:Bayesian statistical inference for an EPR-pair}
\end{example}
BAYESIAN STATISTICAL INFERENCE FOR AN EINSTEIN-PODOLSKI-ROSEN PAIR

Let us consider the Einstein-Podolsky-Rosen's setting (in its
reformulation in terms of spin 1/2  given by David B\"{o}hm
\cite{Bohm-79}, \cite{Bell-93}):

Each among Alice and Bob receive from a proper source one of the
two spin 1/2 particles on an \textbf{EPR pair}.

The quantum probability space of the system is $ ( \, A := M_{4}
({\mathbb{C}}) \, , \, \omega ) $, where:
\begin{equation}
  \rho_{\omega}\; : = \; | \psi > < \psi |
\end{equation}
\begin{equation}
 | \psi > \; := \; \begin{pmatrix}
   0 \\
   \frac{1}{\sqrt{2}} \\
    - \frac{1}{\sqrt{2}} \\
   0 \
 \end{pmatrix}
\end{equation}
\begin{equation}
  \rho_{\omega} \; = \; \begin{pmatrix}
    0 & 0 & 0 & 0 \\
    0 & \frac{1}{2} & - \frac{1}{2} & 0 \\
    0 & - \frac{1}{2} & \frac{1}{2} & 0 \\
     0 & 0 & 0 & 0 \
  \end{pmatrix}
\end{equation}
Let us now consider Alice.

She has at her disposal only the information relative to the
subalgebra $ A _{accessible} \; := \; M_{2} ({\mathbb{C}} )$ given
by the state:
\begin{equation}
  \omega_{accessible} \; = \; \omega | _{A_{accessible} } \; =
  \; \tau_{2}
\end{equation}

The $ \tau_{unbaised} $ - conditional expectation $ E_{unbaised}
\, : \, A \, \mapsto A _{accessible} $ is, simply, the orthogonal
projection on$ A _{accessible} $ with respect to the following:
\begin{definition}
\end{definition}
HILBERT-SCHIMDT SCALAR PRODUCT ON $ M_{n} ({\mathbb{C}}) $:
\begin{equation}
  < a_{1} | a_{2} > \; := \; \tau_{n} ( a_{1}^{\star}  a_{2} )
\end{equation}

\medskip

Considered the basis $ {\mathbb{E}}_{2} \; := \; \{ e_{1} :=
\sigma_{1} \, , \, e_{2} := \sigma_{2} \, , \,  e_{3} :=
\sigma_{3} \, , \, e_{4} := {\mathbb{I}} \} $ of $ M_{2}
({\mathbb{C}}) $ and the basis  $ {\mathbb{E}}_{4} \; := \; \{
e_{i,j} := e_{i} \bigotimes e_{j} \, , \, i,j = 1 , \cdots , 4 \}
$  of $ M_{4} ({\mathbb{C}}) $, we have clearly that:
\begin{equation}
  E_{unbaised} ( \sum_{i=1}^{4} \sum_{j=1}^{4} c_{i,j} e_{i,j} )
  \; = \; \sum_{i=1}^{4} c_{i,4} e_{i}
\end{equation}
Conseguentially:
\begin{equation}
 \omega_{A \, POSTERIORI}  ( \sum_{i=1}^{4} \sum_{j=1}^{4} c_{i,j} e_{i,j}
 ) \; = \; \omega_{accessible} (  E_{unbaised} ( \sum_{i=1}^{4} \sum_{j=1}^{4} c_{i,j} e_{i,j}
 ))
\end{equation}
namely:
\begin{equation}
   \omega_{A \, POSTERIORI} (  \sum_{i=1}^{4} \sum_{j=1}^{4} c_{i,j} e_{i,j}
 ) \; = \; \omega_{accessible} (  \sum_{i=1}^{4} c_{i,4} e_{i}
 )
\end{equation}
and so:
\begin{multline}
 \omega_{A \, POSTERIORI}  ( \sum_{i=1}^{4} \sum_{j=1}^{4} c_{i,j} e_{i,j}
 ) \; = \; \tau_{2} (  \sum_{i=1}^{4} c_{i,4} e_{i}) \; = \\
 \sum_{i=1}^{4} c_{i,4} \tau_{2} ( e_{i} ) \; = \; c_{i,4}
\end{multline}

\smallskip

\begin{remark} \label{rem:noncommutative axiomatizations of Quantum Mechanics and Relativity Theory}
\end{remark}
NONCOMMUTATIVE AXIOMATIZATIONS OF QUANTUM MECHANICS AND RELATIVITY
THEORY:

Looking at definition\ref{def:noncommutative axiomatization of
Quantum Mechanics} one could ask what about Relativity Theory:

are we in the framework of  Nonrelativistic Quantum Mechanics, of
Special-relativistic Quantum Mechanics or of General Relativistic
Quantum Mechanics?

The answer is that it has been formulated in order of holding in
any case, adding suitable further axioms:
\begin{itemize}
  \item assuming conjecture\ref{con:the category isomorphism at the basis of Noncommutative
  Geometry} it appears natural \cite{Connes-98} to suppose that General Relativistic Quantum
  Mechanics is based on a quantum spacetime described by a
  spectral triple $ ( A_{\hbar} \, , \, {\mathcal{H}}_{\hbar} \, , \, D_{\hbar} ) $, where
  the observables' algebra of quantum space-time $ A_{\hbar} \, \subseteq
  \, {\mathcal{B}} ( {\mathcal{H}}_{\hbar}  ) $ is a Von Neumann algebra
  acting on the Hilbert space $ {\mathcal{H}}_{\hbar} $.

  Let us observe that, in the classical limit $ \hbar \,
  \rightarrow \, 0 $,  $ A_{\hbar} $ becomes commutative, so that
  by Conjecture\ref{con:the category isomorphism at the basis of Noncommutative
  Geometry}, the spectral triple $ ( A_{\hbar} \, , \, {\mathcal{H}}_{\hbar} \, , \, D_{\hbar} )
  $ tends to a riemannian manifold $ ( M \, , \, g_{Riemannian} )
  $

  The lorentzian manifold constituing the classical space-time  $ ( M \, , \, g_{Lorentzian} ) $  is then
  recovered by a suitable non-euclidean generalization of Wick's
  rotation \cite{Deligne-Etingof-Freed-Jeffrey-Kazhdan-Morgan-Morrison-Witten-99a}, based on the analytic continuation
  to the complex plane, and a suitable rotation,  of a \textbf{global time
  function} (i.e. of a function $ t \in \Omega^{0}(M) $  such
  that $ \nabla_{a} t $ is a past-directed time-like vector field,
  whose existence is assured by the assumption of axiom\ref{ax:axiom of strong cosmic
censorship} since \textbf{global-hyperbolicity} implies
\textbf{stable-causality}) having the property that its level's
surfaces $ \Sigma_{t} $ are Cauchy surfaces leading to the
foliation $ M \; = \; \bigcup_{t}  \Sigma_{t}  $ of M (cfr. the $
8^{th} $ chapter of \cite{Wald-84}).

Denoted by $ n^{a} $ the unit normal vector field to the
hypersurface $ \Sigma_{t} $ and called $ h_{a b } $ the
riemannian metric induced on it by  $ g_{a b } $ one can choose a
vector field $ t^{a} $ on M such that $ t^{a} \nabla_{a} t \, =
\, 1 $ such the the \textbf{lapse function}:
\begin{equation} \label{eq:lapse function}
 N \; := \; - t^{a} n_{a} \; = \; ( n^{a} \, \nabla_{a} t )^{ - 1 }
\end{equation}
and the \textbf{shift vector}:
\begin{equation}\label{eq:shift vector}
  N_{a} \; := \; h_{a b} t^{b}
\end{equation}
are those for coherent flows of classical test particles adapted
to the chosen \textbf{foliation}, i.e. (cfr. the sections 5.4 and
11.1 of \cite{Prugovecki-92}):
\begin{align}
  N &  \; = \; 1 \\
  N_{a} & \; = \; 0
\end{align}
so that $ g _{a b } $ may be expressed in terms of the
corresponding \textbf{syncronous (Gaussian normal) coordinates} $
( x^{0} \, = \, t \, , \, x^{1} \, , \, x^{2} \, , \,  x^{3} ) $
as:
\begin{equation}
   g_{Lorentzian} \; = \; d t \bigotimes  d t \, - \, h_{i j} d x^{i}
   \bigotimes d x^{j}
\end{equation}
Prolonging the coordinate t to the complex plane and evaluating it
on the imaginary axis one results in the required riemannian
manifold $ ( M \, , \, g_{riemannian} ) $.

As the Wick's rotation's operation is always named as the passage
from the \textbf{minkowskian} to the \textbf{euclidean} we will
refer to the introduced not-flat generalization as to the passage
from the \textbf{lorentzian} to the \textbf{riemannian}.

\smallskip

Since the Universe is closed by definition,it follows by
axiom\ref{ax:noncommutative axiom on closed dynamics} that it is
described by a strongly-continuous one-parameter group of inner
automorphisms.

Conjecture\ref{con:the category isomorphism at the basis of
Noncommutative Geometry} suggests that OUT(A) plays the rule of
the \textbf{quantum diffeomorphims' group} of the quantum
spacetime A, while inner fluctuations of the quantum spacetime,
i.e. elements of
  INN(A) corresponds to gauge transformations, as it is
  supported by the INN(A)-invariance of the minimally-coupled
  version of definition\ref{def:noncommutative spectral action}.

  Since the dynamics is made only by gauge transformations,
  one may conclude that such a picture respects Rovelli's
  suggestion of forgetting time \cite{Rovelli-88}, i.e. though not following Canonical Quantum
  Gravity but the:
\begin{center}
   \textit{" $ \cdots $ gnostic subculture of workers in quantum gravity who feel that that the structure of space and time may undergo
    radical changes at scales of the Planck length"; from \cite{Isham-93}   }
\end{center}
it may be catalogued in the category \emph{"Tempus Nihil Est"}
of  Chris Isham's classification of different approaches to the
Problem of Time in Quantum Gravity \cite{Isham-93}.
  \item an approximation to the complete quantum theory of fields
  coupled with gravity is that in which one considers quantum
  fields on a fixed, classical space-time $ ( M \, , \, g_{a b } )
  $ we will suppose to be globally-hyperbolic.

  A quantum field theory on $ ( M \, , \, g_{a b } )
  $ may be defined in terms of the so called
  Weyl algebra A of $ ( M , g_{a b} ) $  and the Hadamard's states on it (for whose definition we demand to the $ 4^{th} $ chapter of \cite{Wald-94}), i.e. by the   collection $ \{ A_{O} \} $ of
  $ C^{\star}$-sub-algebras of A, one for every open subset $ O \,
  \subseteq \, M $, with $ A_{O}$ representing  the local
  observables localized on O, satisfying  suitable natural conditions:
\begin{equation}
  O_{1}  \, \subseteq  \, O_{2} \; \Rightarrow \; A_{O_{1}} \, \subseteq \, A_{O_{2}}
\end{equation}
\begin{equation}
   O_{1} , O_{2} \text{ causally disconnected } \; \Rightarrow \;
  [ A_{O_{1}} \, , \,  A_{O_{2}} ] \, = \, 0
\end{equation}
\begin{multline}
   \exists   \, \{ \alpha_{g} \} \in GR-INN[Is[( M \, , \, g_{a b } )] \, , \, A] \,
   : \\
\alpha_{g} ( A_{O} ) \, = \, A_{g \, O}
   \forall g \in Is[( M \, , \, g_{a b } )] \, , \, \forall O \subseteq M \; open
\end{multline}
where $ Is[( M \, , \, g_{a b } )] $ is the isometries'-group of
$ ( M \, , \, g_{a b } )$.

In the particular case in which $ ( M = {\mathbb{R}}^{4} \, , \,
\eta = \eta_{\mu \nu } d x^{\mu} \bigotimes d x^{\nu} ) $:
\begin{equation}
  \eta_{\mu \nu} \; := \; \begin{pmatrix}
    -1 & 0 & 0 & 0 \\
    0 & 1 & 0 & 0 \\
    0 & 0 & 1 & 0 \\
    0 & 0 & 0 & 1 \
  \end{pmatrix}
\end{equation}
is the Minkowski space-time, the above conditions reduce to the
Haag-Kastler axioms \cite{Haag-96}.

   \item Non-relativistic Quantum Mechanics can then be recovered
   taking the  Inonu-Wigner's contraction $ c \,
   \rightarrow \,  + \infty $ of the isometries'-group of the
   minkowskian space-time, namely of the Poincar\'{e} group, thus
   obtaining the Galilei group (cfr. e.g. the $ 3^{th} $ chapter of
   \cite{De-Azcarrega-Izquierdo-95})
\end{itemize}

\smallskip

\begin{remark}
\end{remark}
QUBITS ENTERING AND EXITING BLACK-HOLES

Since, as a matter of principle, the theory of quantum-black holes
is nothing but a matter of bayesian statistical inference w.r.t. a
sub-$W^{\star}$-algebra $ A_{accessible} $ of the Weyl's algebra
of a space-time with an event-horizon, it is a matter of Quantum
Information Theory.

Our knowledges in this field are infimous, but there is a thing
that have often aroused our's curiosity:

the specialists of these matters (apart from some very timid
allusion in \cite{Bekenstein-01}), speak always of
\textbf{classical information} attached to certain geometrical
entities, \textbf{classical information} finishing lost inside or
exiting from the event-horizon though they are treating quantum
systems.

It seems to us that, with this regard, the high energy physics'
community have not catched the great conceptual revolution of
modern Quantum Information Theory:
\begin{center}
 \textbf{the irreducibility of quantum information to classical information, of the qubit to the
cbit}
\end{center}
When we will at last hear about qubits entering and exiting
black-holes ?

\newpage
\section{The problem of hidden points of a noncommutative space} \label{sec:The problem of hidden points of a noncommutative space}
Let us now analyze the concept of a point in a noncommutative
space.

Given an abelian $ W^{\star}$-algebra A  we know by
theorem\ref{th:Gelfand isomorphism at C-star algebraic level}
that it may be seen as (i.e.  it is $ \star$-isomorphic to) the $
C^{\star}$-algebra $ C(X(A)) $ of all the continuous (w.r.t the $
w^{\star}-topology$) functions on the set X(A) of its
\textbf{characters} that may thus be  seen as the \textbf{points}
of the \textbf{commutative space} A.

We want to characterize this concept more precisely.

Given a generic \textbf{algebraic space} A:
\begin{definition} \label{def:points of an algebraic space}
\end{definition}
POINTS OF A:
\begin{equation}
  POINTS(A) \; := \; \{ \omega \in S(A) \: : \: Var(a) \; = \; 0 \text{ in } (A , \omega ) \; \; \forall a \in A  \}
\end{equation}
The previous considerations may then be formalized as:
\begin{theorem} \label{th:on the points of a commutative space}
\end{theorem}
ON THE POINTS OF A COMMUTATIVE SPACE:
\begin{equation}
  A  \; commutative \; \Rightarrow \; POINTS(A) \; = \; \Xi(A)
  \; = \; X(A)
\end{equation}

\begin{remark} \label{rem:points of a commutative space as Dirac-delta measures}
\end{remark}
POINTS OF A COMMUTATIVE SPACE AS DIRAC-DELTA MEASURES

Given the commutative $ C^{\star}$-algebra C(X) of continuous
functions over the compact, Hausdorff topological space X the
points of C(X) are nothing but the Dirac delta measures over X,
i.e. the states of the form:
\begin{equation}
  \delta_{x} [ f(y) ] \; := \; f(x) \; \; f \in C(X) \, , \, x \in X
\end{equation}
Considered a probability measure $ \mu $ on X and represented the
classical probability space $ ( X \, , \, \mu ) $ as the
commutative probability space $ ( L^{\infty} ( X , \mu ) \, , \,
\omega_{\mu} ) $ one has that the projections  of $ L^{\infty} (
X , \mu ) $ are nothing but the characteristic functions of $
\mu$-measurable subsets of X forming a classical logic.

Obviously the values of $  L^{\infty} ( X , \mu ) $'s  points on
the projections is given by:
\begin{equation}
   \delta_{x} [ \chi_{A} ] \; = \;
  \begin{cases}
    1 & \text{if $ x \in A$}, \\
    0 & \text{otherwise}.
  \end{cases}
\end{equation}

\smallskip

Given an algebraic probability space $ ( A , \omega ) $:
\begin{definition} \label{def:determinism}
\end{definition}
$ ( A , \omega ) $ IS DETERMINISTIC:
\begin{equation*}
  \omega \; \in \; POINTS(A)
\end{equation*}
\begin{definition} \label{classical-nondeterminism}
\end{definition}
$ ( A , \omega ) $ IS CLASSICALLY-NONDETERMINISTIC:

$ ( A , \omega ) $ is nondeterministic and  A is commutative

\smallskip

\begin{definition} \label{quantistically-nondeterminism}
\end{definition}
$ ( A , \omega ) $ IS QUANTISTICALLY-NONDETERMINISTIC:

$ ( A , \omega ) $ is nondeterministic and  A is noncommutative

\smallskip

where, obviously, the form of
definition\ref{classical-nondeterminism} and
definition\ref{quantistically-nondeterminism} is owed to
theorem\ref{th:category isomorphism at the basis of
Noncommutative Probability}.

\medskip

The existence of points on noncommutative spaces is inficiated by
the following:
\begin{theorem} \label{th:indetermination's theorem}
\end{theorem}
INDETERMINATION'S THEOREM:
\begin{equation}
  | E( \frac{ [ a , b ] }{2 i } ) | \; \leq \; \sqrt{Var(a)}
  \sqrt{Var(b)} \; \; \forall a , b \in A
\end{equation}
\begin{proof}
Introduced the quantity:
\begin{equation}
  O(a,b) \; := \; \frac{a-E(a)}{\sqrt{Var(a)}} \, + \, i
  \frac{b-E(b)}{\sqrt{Var(b)}} \; \; a , b \in A
\end{equation}
we have clearly that:
\begin{equation}
  O(a , b) \, O(a , b)^{\star} \; \in \; A_{+} \; \; \forall a , b
  \in A
\end{equation}
from which the thesis immediately follows
\end{proof}

Theorem\ref{th:indetermination's theorem} implies that:
\begin{corollary} \label{cor:corollary of the indeterminations' theorem}
\end{corollary}
\begin{equation}
  ( A \, , \, \omega ) \; deterministic \;  \Rightarrow \; ( E( [
  a \, , \, b ] ) \; = \; 0 \; \; \forall a , b \in A )
\end{equation}
from which it follows that:
\begin{theorem} \label{th:first Von Neumann's theorem}
\end{theorem}
FIRST VON NEUMANN'S THEOREM:

\begin{hypothesis}
\end{hypothesis}
\begin{equation*}
  A   \text{ noncommutative space }
\end{equation*}
\begin{equation*}
  cardinality_{NC}(A) \, = \, \aleph_{0}
\end{equation*}
\begin{thesis}
\end{thesis}
\begin{equation*}
   POINTS(A) \; = \; \emptyset
\end{equation*}
\begin{proof}
By hypothesis A is ( $ \star$-isomorphic to) the algebra $
{\mathcal{B}} ( {\mathcal{H}}) $ of all bounded operator on a
infinite-dimensional Hilbert-space $ {\mathcal{H}}$.

Let us assume, for simplicity, that $ {\mathcal{H}}$ is separable.

Fixed a complete orthonormal basis $ {\mathbb{E}} \, := \, \{ | n
> \} $ of $ {\mathcal{H}}$ let us consider the sequence  $ \{
P_{n} \} $ of projectors defined by:
\begin{equation}
 P_{n} \; := \; \sum_{k=1}^{n} | k > < k |
\end{equation}
We have clearly that:
\begin{equation}
  P_{0} \; \preceq \; P_{1} \;  \preceq \; P_{2} \;  \preceq \;
  \cdots \;  \preceq \; I
\end{equation}
Furthermore there exist hermitian operators $ \{ a_{n} \} $ and $
\{ b_{n} \} $ such that:
\begin{equation}
  P_{n} \; = \; [ a_{n} \, , \, b_{n} ] \; \; \forall n \in {\mathbb{N}}
\end{equation}
Let us then suppose ad absurdum that there exist a
dispersion-free state $ \omega \in POINTS(A) $.

By the corollary\ref{cor:corollary of the indeterminations'
theorem} one has that:
\begin{equation}
  \omega( P_{n} ) \; = \; \omega( [ a_{n} \, , \, b_{n} ] ) \; =
  \; 0 \; \; \forall n \in {\mathbb{N}}
\end{equation}
from which it follows that $ \omega \, = \, 0 $.
\end{proof}

As we will now show theorem\ref{th:first Von Neumann's theorem}
can be generalized to higher noncommutative cardinality.

\smallskip

Given an algebra A and a subalgebra $ B \; \subset \; A $:
\begin{definition}
\end{definition}
B IS A LEFT IDEAL OF A:
\begin{equation}
  a \in A \; , \; b \in B \; \Rightarrow \; a  b \, \in \, B
\end{equation}
\begin{definition}
\end{definition}
B IS A RIGHT IDEAL OF A:
\begin{equation}
  a \in A \; , \; b \in B \; \Rightarrow \;   b a \, \in \, B
\end{equation}
\begin{definition}
\end{definition}
B IS A TWO-SIDED IDEAL OF A:

B is both a left ideal and a right ideal of A

\medskip
Given a $ C^{\star}$-algebra A:
\begin{definition} \label{simple C-star algebra}
\end{definition}
A IS SIMPLE:
\begin{equation*}
  \nexists B \subset A \: : \: \text{not-trivial two-sided ideal}
\end{equation*}
Given an algebraic probability space:
\begin{definition} \label{simple algebraic probability space}
\end{definition}
$ ( A \, , \, \omega ) $ IS SIMPLE:

A is simple

\smallskip

\begin{example} \label{ex:example of ideals}
\end{example}
The space $ {\mathcal{C}}_{1} ({\mathcal{H}}) $ of
\textbf{trace-class operators}, the space $ {\mathcal{C}}
({\mathcal{H}}) $ of \textbf{infinitesimals operators}, the space
$ {\mathcal{I}}_{\alpha} ({\mathcal{H}}) $ of
\textbf{order-$\alpha$ infinitesimals operators} are all
two-sided ideals of the $W^{\star}$-algebra $ {\mathcal{B}}
({\mathcal{H}}) $ of all bounded operators on an Hilbert space $
{\mathcal{H}} $.

The rule of ideals for hidden variables issues is owed to the
following:
\begin{lemma} \label{lem:on the not-trivial ideals}
\end{lemma}
ON THE NOT-TRIVIAL IDEALS

\begin{hypothesis}
\end{hypothesis}
\begin{equation*}
  A \;  not-trivial \; C^{\star}-algebra \; : \;
\end{equation*}
\begin{thesis}
\end{thesis}
\begin{multline}
  POINTS(A) \, \neq \, \emptyset \; \Leftrightarrow \\
\exists \, J  \, \text{not-trivial two-sided ideal in A} \, :
\frac{A}{J} \text{ is abelian}
\end{multline}
\begin{proof}
\begin{enumerate}
  \item Given $ \phi \, \in \, POINTS(A) $ we will prove that:
\begin{equation}
  J \; := \;  \{ a \in A \, : \, \phi(a) \, = \, 0 \}
\end{equation}
is a not-trivial two-sided ideal such that $ \frac{A}{J} $ is
commutative.

Obviously J is a linear subspace of A and is a subset of $ A_{sa}
$.

Furthermore, the hypothesis of not-triviality of A implies that
also A is not trivial, since the existence of an $ a \in A \, : \,
a \, \neq \, I $ implies that also $ a - \phi(a) I \; \neq \; I $
so that $  a - \phi(a) I $ is a not-trivial element of J.

Let us now show that J is an ideal:

a generic element $ a \in A $ may be expressed as linear
combination of self-adjoint elements:
\begin{equation}
  a \; = \; \sum_{n} c_{n} x_{n} \;  c_{n} \in {\mathbb{C}} , x_{n}
  \in A_{sa} \; \forall n
\end{equation}
Given $ a , b \in J $ we have, by theorem\ref{th:Cauchy-Schwarz
inequality} and the fact that $ \phi $ is a point, that:
\begin{multline}
  | \phi ( x_{n} b ) | \; \leq \;   \phi ( x_{n} x_{n}^{\star} )
  \phi ( b b^{\star} ) \\
  = \; \phi( x_{n}^{2} ) \phi ( b^{2} ) \; = \; 0 \; \; \forall n
\end{multline}
so that:
\begin{equation}
   \phi ( a b ) \; = \; 0
\end{equation}
and hence $ a b \, \in \, J $, so that it is proved that J is a
left-ideal.

The proof that J is also a right-ideal is specular.

To prove, finally, that $ \frac{A}{J} $ is abelian let us observe
that the map $ h : \frac{A}{J} \rightarrow {\mathcal{C}} $
defined by:
\begin{equation}
  h( [ a ]_{ \frac{A}{J} } ) \; := \; \phi ( a )
\end{equation}
is a $ \star $ -isomorphism from $ \frac{A}{J} $ to $
{\mathcal{C}} $; in fact if $ \phi( a )  \, = \, \phi( b) $ then
$ \phi( a - b) \, = \, 0 $ so that $ [ a ]_{\frac{A}{J}} \, = \,
[ b ]_{\frac{A}{J}} $ and, consequentially, h is invertible;
furthermore:
\begin{equation}
  h ( [ a ]_{\frac{A}{J}} , [ b ]_{\frac{A}{J}} ) \; = \; h ( [ a b
  ]_{\frac{A}{J}}) \; = \; \phi( a b ) \; = \; \phi(a) \phi(b) \;
  \; \forall a, b, \in A
\end{equation}
and hence h preserves the product
  \item let suppose that there exist a two-sided ideal J such that
  $ {\frac{A}{J}} $ is abelian. Then, by theorem\ref{th:on the points of a commutative
  space}, it follows that $ POINTS(A) \, \neq \, \emptyset $.
\end{enumerate}
\end{proof}

immediately implying the following:
\begin{corollary} \label{cor:indeterminism of simple algebraic probability spaces}
\end{corollary}
INDETERMINISM OF SIMPLE ALGEBRAIC PROBABILITY SPACES

\begin{equation*}
  ( A \, , \, \omega ) \text{ simple } \; \Rightarrow \; ( A \, , \, \omega ) \text{ nondeterministic }
\end{equation*}

Corollary\ref{cor:indeterminism of simple algebraic probability
spaces}, anyway, is only the tip of an iceberg, as is stated by
the following generalization of theorem\ref{th:first Von Neumann's
theorem}
\begin{theorem} \label{th:indeterminism of noncommutative probability spaces}
\end{theorem}
INDETERMINISM OF NONCOMMUTATIVE PROBABILITY SPACES:

\begin{hypothesis}
\end{hypothesis}
\begin{equation*}
  ( A \, , \, \omega ) \; \; \text{algebraic probability space}
\end{equation*}
\begin{thesis}
\end{thesis}
\begin{equation*}
  ( A \, , \, \omega ) \; noncommutative \; \Rightarrow \; ( A \, , \, \omega ) \;
  nondeterministic
\end{equation*}
\begin{proof}
Proceeding  exactly as  in the proof of theorem\ref{th:first Von
Neumann's theorem} let us consider a sequence $ \{ p_{n} \}_{n \in
{\mathbb{N}}} $ such that:
\begin{align}
  p_{n} & \; \in  \; {\mathcal{P}}(A) \; \; \forall n \in
{\mathbb{N}} \\
   p_{i} & \; \preceq \; p_{j} \; \; \forall i \, <  \, j
\end{align}
There will exist hermitian operators $ \{ a_{n} \} $ and $ \{
b_{n} \} $ such that:
\begin{equation}
  p_{n} \; = \; [ a_{n} \, , \, b_{n} ] \; \; \forall n \in {\mathbb{N}}
\end{equation}
Let us then suppose ad absurdum that there exist a
dispersion-free state $ \omega \in POINTS(A) $.

By the corollary\ref{cor:corollary of the indeterminations'
theorem} one has that:
\begin{equation}
  \omega( p_{n} ) \; = \; \omega( [ a_{n} \, , \, b_{n} ] ) \; =
  \; 0 \; \; \forall n \in {\mathbb{N}}
\end{equation}
from which it follows that $ \omega \, = \, 0 $.
\end{proof}

\smallskip

\begin{remark} \label{rem:usnharp localization of a noncommutative space}
\end{remark}
UNSHARP LOCALIZATION ON A NONCOMMUTATIVE SPACE:

In the commutative case a point may be characterized through a
monotonically descreasing sequence of projections.

Given, for example, the unbaised probability space of cbits'
sequences $ ( \Sigma^{\infty} \, , \, P_{unbiased} ) $ let us use
the numeric representation map of definition\ref{def:numeric
representation} to visualize it as the the classical probability
space $ ( [ 0 , 1 ) \, , \, \mu_{Lebesgue} ) $.

A point $ x \in ( 0 , 1) $ is completelly specified by a nested
sequence of measurable $ ( 0 , 1) $'s subsets $ \{ A_{n} \} $
whose intersection is the singleton containing x :
\begin{align}
  A_{i} &  \supset A_{j} \; \; \forall i < j  \\
  \{ x \} & \; = \; \bigcap _{n \in {\mathbb{N}}} A_{n}
\end{align}
For example one can take:
\begin{equation}
  A_{n} \; := \; ( x - \frac{1}{2^{n}} \, , \, x + \frac{1}{2^{n}})
\end{equation}
Let us now look at the classical probability space $  ( [ 0 , 1 )
\, , \, \mu_{Lebesgue} ) $ as at the commutative probability space
$ ( A := L^{\infty} ( [ 0 , 1 ) \, , \, \mu_{Lebesgue} ) \, , \,
\omega_{\mu_{Lebesgue}})$.

As we saw in the remark\ref{rem:points of a commutative space as
Dirac-delta measures} the projections of $ {\mathcal{P}}(A) $ are
nothing but the characteristic functions of measurable $ [ 0 , 1
) $'s subsets constituting  a classical logic.

The sequence  $ \{ A_{n} \} $ corresponds to the sequence of
projections $ \{ \chi_{A_{n}} \} $ satisfying the condition:
\begin{equation}
   \chi_{A_{i}} \, > \, \chi_{A_{j}} \; \; \forall i < j
\end{equation}
The point $ x \in A $, i.e the point $ \delta_{x} \in POINTS(A)
$, is then characterized by the condition:
\begin{equation}
  \delta_{x} ( \chi_{A_{n}} ) \; = \; 1 \; \; \forall n \in {\mathbb{N}}
\end{equation}

Given, now a noncommutative space X, one could think that, though
$ POINTS(X) \,= \, \emptyset $, the characterization of the
concept of an X's point can be recovered generalizing the above
procedure, i.e finding a monotonically increasing sequence of
projections $ \{ p_{n} \} $ over X:
\begin{align}
   p_{n} &  \, \in \, {\mathcal{P}}(X)  \; \; \forall n \in {\mathbb{N}} \\
   p_{i} &  \, > \, p_{j} \; \; \forall i < j
\end{align}
and a state $ \omega \in \Xi(X) $ such that:
\begin{equation}
  \omega( p_{n} ) \; = \; 1 \; \; \forall n \in {\mathbb{N}}
\end{equation}
One could, in fact, think that in such a situation it is
possible, after all, to look at  the sequence $ \{ p_{n} \} $ as
a sequence of propositions stating the localization  in a
monotonically-increasing way, so that a state $ \omega $ giving
value one to all these propositions, i.e. in quantum-logic
language , stating the truth of all these propositions, can
assume a geometrical meaning as an unsharped-localized region of
the noncommutative space X.

\smallskip

\begin{remark}
\end{remark}
ON HIDDEN POINTS OF A NONCOMMUTATIVE SPACE

Given a noncommutative space $ A_{accessible} $ one could
 think of completing it, i.e. of considering a larger
noncommutative space A of which $  A_{accessible} $  is a
sub-$W^{\star}$-algebra, such that $ POINTS(A) \; \neq \;
\emptyset $.

In such a situation  the indeterminism of any noncommutative
probability space $ ( A , \omega ) $ on A could then be simply
attributed to the not accessibility of the algebraic random
variables belonging to $ A \, - \, A_{accessible} $.

That this in not the case is  stated by the following obvious
corollary of theorem\ref{th:indeterminism of noncommutative
probability spaces}:
\begin{corollary} \label{cor:not existence of hidden points of a noncommutative space}
\end{corollary}
NOT EXISTENCE OF HIDDEN POINTS OF A NONCOMMUTATIVE SPACE:
\begin{equation}
  POINTS(A) \; = \; \emptyset \; \; \forall A \supset  A_{accessible}
\end{equation}
\begin{proof}
Since $ A_{accessible} $ is noncommutative, A is noncommutative
too.

The thesis follows immediately by theorem\ref{th:indeterminism of
noncommutative probability spaces}
\end{proof}

Though not leading to a sharp localization, one could anyway
think that completions can anyway improve the unsharp
localization.

If this is possible or not depends sensibly from the definition of
completion one assume, as we will now show.

Given a state $ \beta \in S(B) $:
\begin{definition}
\end{definition}
CLASSICAL PROBABILITY MEASURES WITH BARYCENTER $ \beta $:

the set $ M_{\beta}[ S(B)] $:
\begin{multline}
  M_{\beta}[ S(B)] \; := \; \{ \mu \in M[S(B)] \, : \\
   \beta (b) \, = \, \int_{S(B)} \omega(b) \, d \mu(\omega) \; \; \forall b \in B  \}
\end{multline}

Given a channel $ \beta \in CPU(B,A) $:
\begin{definition} \label{def:completion-channel}
\end{definition}
C IS A COMPLETION-CHANNEL:

$ \forall \alpha \in S(A) \; , \; \exists \mu \in M_{C^{\star}
\alpha}[ S(B)] $ such that:
\begin{itemize}
  \item \textbf{in the completion the uncertainty descreases},
  i.e.:
\begin{multline}
   \sqrt{Var_{\alpha} ( C b )} \; > \;  \sqrt{Var_{\omega} ( b )}
   \; \; \forall \omega \in supp( \mu ) \, , \\
    \forall b \in B \, : \, ( C b \, \in A_{sa} \: and \: \sqrt{Var_{\alpha} ( C b )}
   \, > \, 0 )
\end{multline}
  \item \textbf{in the completion the certainty remains certain},
  i.e.:
  \begin{multline}
   \sqrt{Var_{\alpha} ( C b )} \; = \;  \sqrt{Var_{\omega} ( b )} \; =
   \; 0 \; \; \forall \omega \in supp( \mu ) \, , \\
    \forall b \in B \, : \, ( C b \, \in A_{sa} \: and \: \sqrt{Var_{\alpha} ( C b )}
   \, = \, 0 )
\end{multline}
\end{itemize}
\begin{definition} \label{def:deterministic-completion-channel}
\end{definition}
C IS A DETERMINISTIC-COMPLETION-CHANNEL:

\begin{itemize}
  \item C is a completion-channel
  \item
\begin{multline}
 \sqrt{Var_{\omega} ( b )} \; = 0 \; \; \forall \omega \in supp( \mu ) \, , \\
    \forall b \in B \, : \, ( C b \, \in A_{sa} \: and \: \sqrt{Var_{\alpha} ( C b )}
   \, > \, 0 )
\end{multline}
\end{itemize}
Von Neumann himself was the first to investigate the possibility
of deterministic completion channels, though only of the
following particular kind:
\begin{definition} \label{def:Von Neumann's completion}
\end{definition}
VON NEUMANN'S COMPLETION:

a deterministic-completion-channel of the form $ {\mathbb{I}} \in
CPU(A) $ such that:
\begin{equation}
  cardinality_{NC} ( A )  \; \leq \; \aleph_{0}
\end{equation}

formulating the first no-go theorem on hidden variables:
\begin{theorem} \label{th:Von Neumann's no-go theorem}
\end{theorem}
VON NEUMANN'S NO-GO THEOREM:
\begin{equation}
  \{ C \; : \; \text{ Von Neumann's completion } \} \; = \; \emptyset
\end{equation}
\begin{proof}
It immediately follows by theorem\ref{th:first Von Neumann's
theorem}
\end{proof}

The generalization involved in the passage from
theorem\ref{th:first Von Neumann's theorem} to
theorem\ref{th:indeterminism of noncommutative probability
spaces} induces the following generalization of
theorem\ref{th:first Von Neumann's theorem}:
\begin{theorem} \label{th:first algebraic no-go theorem}
\end{theorem}
FIRST ALGEBRAIC NO-GO THEOREM:

\begin{hypothesis}
\end{hypothesis}
\begin{equation*}
  A \; \; \text{ algebraic space}
\end{equation*}
\begin{thesis}
\end{thesis}
\begin{equation*}
  {\mathbb{I}}_{A} \text{ is a deterministic-channel completion }
  \; \Leftrightarrow \; \text{A is commutative}
\end{equation*}
\begin{proof}
\begin{itemize}
  \item A commutative $ \; \Rightarrow \; {\mathbb{I}}_{A} \text{ is a deterministic-channel completion
  } $

  If A is commutative we know by theorem\ref{th:Gelfand isomorphism at C-star algebraic
  level} that it may be seen (i.e. it is $ \star$-isomorphic to) the space $ C(X(A)) $
  of the continuous functions over the characters, i.e. the points,
  of A.

  By the Riesz-Markov theorem (cfr. the section 4.4 of
  \cite{Reed-Simon-80}) for every state $ \phi \in S(C(X)) $ there
  exist a measure $ \mu_{\phi} $ on X such that:
\begin{equation}
  \phi (f) \; = \; \int_{X} d \mu_{\phi} f \; \; \forall f \in C(X)
\end{equation}
By theorem\ref{th:on the points of a commutative space} if
follows that:
\begin{equation}
  supp(\mu) \; \subseteq \; POINTS(A)
\end{equation}
for which the thesis follows

  \item  $ {\mathbb{I}}_{A} \text{ is a deterministic-channel completion
  }  \; \Rightarrow \; $ A commutative

Let us assume that $ {\mathbb{I}}_{A} $ is a deterministic-channel
completion.

Given $ x , y \, \in \, A \; : \; x \, > \, y \, \leq \, 0 $ one
has, by the  definition\ref{def:completion-channel}, that for
every state $ \phi \in S(A) $ there exists a measure $ \mu \in M[
S(A)] $ such that:
\begin{align}
  \phi( x^{2} ) & \; = \; \int_{S(A)} d \mu ( \omega ) \omega ( x^{2}) \; = \; \int_{S(A)} d \mu ( \omega ) \omega ( x )^{2} \\
  \phi( y^{2} ) & \; = \; \int_{S(A)} d \mu ( \omega ) \omega (
  y^{2}) \; = \; \int_{S(A)} d \mu ( \omega ) \omega ( y )^{2}
\end{align}
Since $ \omega ( x ) \; \leq \; \omega ( y ) $, it follows that:
\begin{equation}
   \phi( x^{2} ) \, - \,  \phi( y^{2} ) \; = \; \int_{S(A)} d \mu ( \omega
   ) ( \omega ( x )^{2}  \, - \, \omega ( y )^{2} ) \; \leq \; 0
\end{equation}
that implies that $ x^{2} \, \geq \, y^{2} $.

The thesis follows immediately from the property:
\begin{equation}
  ( x \, \geq \, y \: \rightarrow \: x^{2} \, \geq \, y^{2} \, \,
  \forall x , y \in A ) \; \Rightarrow \; A \; commutative
\end{equation}
\end{itemize}
\end{proof}

\smallskip

Let us now return to the Noncommutative Bayesian Statistical
Inference Theory we have outlined in section\ref{sec:On the rule
Noncommutative Measure Theory and Noncommutative Geometry play in
Quantum Physics}:

one could think that the process of statistical inference
corresponds to an improvement in the localization on a
noncommutative space.

This is not ,anyway, the case, as it is stated by the following
\cite{Redei-98}:
\begin{theorem} \label{th:bayesian statistical inference doesn't noncommutatively-localize}
\end{theorem}
SECOND ALGEBRAIC NO-GO THEOREM (BAYESIAN STATISTICAL INFERENCE
DOESN'T NONCOMMUTATIVELY-LOCALIZE):

\begin{hypothesis}
\end{hypothesis}
\begin{equation*}
  A \text{ noncommutative space}
\end{equation*}
\begin{equation*}
  A_{accessible} \; \subset \; A \; \; \text{sub-$W^{\star}$-algebra
  of A satisfying the condition of definition\ref{def:constraint for the feasibility of the bayesian statistical
  inference}}
\end{equation*}
\begin{thesis}
\end{thesis}
\begin{equation*}
  E_{unbaised} \, : \, A \: \rightarrow \: A_{accessible} \text{ is not a channel-completion}
\end{equation*}

\smallskip

Let us conclude this section by an analysis of John Bell's
contribution to the Hidden Variables' Issue.

This involves the dicussion of an (apparentely) different kind of
completion, concerning the degree oif irreducibility of
noncommutative probabilities to the commutative ones.

An immediate conseguence of theorem\ref{th:category isomorphism
at the basis of Noncommutative Probability} is the following:

\begin{theorem} \label{th:irreducibility of noncommutative probability to commutative probability to any order}
\end{theorem}
IRREDUCIBILITY OF NONCOMMUTATIVE PROBABILITY TO COMMUTATIVE
PROBABILITY TO ANY ORDER:

\begin{hypothesis}
\end{hypothesis}
\begin{equation*}
  ( A \, , \, \omega ) \text{ noncommutative probability space }
\end{equation*}
\begin{thesis}
\end{thesis}
\begin{equation*}
  \exists m \in {\mathbb{N}} \; : \; \text{A is irreducible to Classical Probability Theory up to $ m^{th}$ order}
\end{equation*}

\begin{remark} \label{rem:impossibility of the Osterwalder-Schrader program}
\end{remark}
IMPOSSIBILITY OF THE OSTERWALDER-SCHRADER'S PROGRAM

A consequence of theorem\ref{th:irreducibility of noncommutative
probability to commutative probability to any order}  is the
impossibility of founding Quantum Field Theory on the
Osterwalder-Schrader axiomatization (cfr. the sixth chapter of
\cite{Glimm-Jaffe-87} and \cite{Haag-96}).

Indeed a quantum field theory satisfying the Osterwalder-Schrader
axioms obeys the Haag-Kastler axiom's too, but the conversely
doesn't hold.

This implies that the formal path-integral measures comparing in
euclidean field theories cannot in principle be made always
rigorous since they , mathematically rigorously, cannot always
exist

\smallskip

One the conceptually more fascinating examples of irriducibility
of Quantum Probability Theory to Classical Probability Theory to
a low order is given by the EPR-stuff we already introduced in
the example\ref{ex:Bayesian statistical inference for an
EPR-pair}.

Given the noncommutative probability space $ ( A , \omega )$,
with:
\begin{equation}
  A \; := \; M_{2} ({\mathbb{C}}) \bigotimes M_{2} ({\mathbb{C}})  \;
  = \;  M_{4}( {\mathbb{C}})
\end{equation}
\begin{equation}
  \omega ( \cdot ) \; : = \; Tr ( \rho_{| \psi > < \psi|} \cdot )
\end{equation}
\begin{equation}
 | \psi > \; := \; \begin{pmatrix}
   0 \\
   \frac{1}{\sqrt{2}} \\
    - \frac{1}{\sqrt{2}} \\
   0 \
 \end{pmatrix}
\end{equation}
\begin{equation}
  \rho_{| \psi > < \psi|} \; = \; \begin{pmatrix}
    0 & 0 & 0 & 0 \\
    0 & \frac{1}{2} & - \frac{1}{2} & 0 \\
    0 & - \frac{1}{2} & \frac{1}{2} & 0 \\
     0 & 0 & 0 & 0 \
  \end{pmatrix}
\end{equation}
let us consider the following noncommutative random variables:
\begin{equation}
  q^{A}_{1} \; := \; \sigma_{1} \bigotimes I \; = \; \begin{pmatrix}
    0 & 0 & 1 & 0 \\
    0 & 0 & 0 & 1 \\
    1 & 0 & 0 & 0 \\
    0 & 1 & 0 & 0 \
  \end{pmatrix}
\end{equation}
\begin{equation}
   q^{A}_{2} \; := \; \sigma_{2} \bigotimes I \; = \; \begin{pmatrix}
     0 & 0 & - i & 0 \\
     0 & 0 & 0 & - i \\
     i & 0 & 0 & 0 \\
     0 & i & 0 & 0 \
   \end{pmatrix}
\end{equation}
\begin{equation}
  q^{A}_{3} \; := \; \sigma_{3} \bigotimes I \; = \; \begin{pmatrix}
    1 & 0 & 0 & 0 \\
    0 & 1 & 0 & 0 \\
    0 & 0 & -1 & 0 \\
    0 & 0 & 0 & -1 \
  \end{pmatrix}
\end{equation}
\begin{equation}
  q^{B}_{1} \; := \; I \bigotimes \sigma_{1} \; = \; \begin{pmatrix}
    0 & 1 & 0 & 0 \\
    1 & 0 & 0 & 0 \\
    0 & 0 & 0 & 1 \\
    0 & 0 & 1 & 0 \
  \end{pmatrix}
\end{equation}
\begin{equation}
   q^{B}_{2} \; := \; I \bigotimes \sigma_{2} \; = \; \begin{pmatrix}
     0 & -i & 0 & 0 \\
     i & 0 & 0 & 0 \\
     0 & 0 & 0 & -i \\
     0 & 0 & i & 0 \
   \end{pmatrix}
\end{equation}
\begin{equation}
  q^{B}_{3} \; := \; I \bigotimes \sigma_{3} \; = \; \begin{pmatrix}
    1 & 0 & 0 & 0 \\
    0 & -1 & 0 & 0 \\
    0 & 0 & 1 & 0 \\
    0 & 0 & 0 & -1 \
  \end{pmatrix}
\end{equation}
having moments:
\begin{equation}
  M_{n}(  q^{A}_{i} ) \; = \; M_{n}(  q^{B}_{i} ) \; = \;
  \begin{cases}
    0 & \text{n even}, \\
    1 & \text{n odd}.
  \end{cases}  \; \; i= 1 , \cdots , 3
\end{equation}

\medskip

The first joint moments of the six noncommutative random
variables $ q^{A}_{1} \, , \,  q^{A}_{2} \, , \,  q^{A}_{3} \, ,
\, q^{B}_{1} \, , \, q^{B}_{2} \, , \, q^{B}_{3} $ are given by:
\begin{equation}
  E( q^{A}_{i} q^{A}_{j} ) \; = \; E( q^{B}_{i} q^{B}_{j} ) \; =
  \;\delta_{i,j} \; \; i,j = 1 , \cdots , 3
\end{equation}
\begin{equation}
  E ( q^{A}_{i}  q^{B}_{j} ) \; = \;  E ( q^{B}_{i}  q^{A}_{j} ) \; = \;
  \begin{cases}
    -1 & \text{$ i = j$}, \\
    0  & \text{$ i \neq j$}.
  \end{cases} \; \; i,j = 1 , \cdots , 3
\end{equation}

The contribution by John Bell was to show that \cite{Accardi-88}
\cite{Streater-95}, \cite{Streater-00a}:
\begin{theorem} \label{th:Bell's theorem}
\end{theorem}
BELL'S THEOREM:
\begin{equation*}
  Q \; = \; \{ q^{A}_{1} \, , \,  q^{A}_{2} \, , \,  q^{A}_{3} \, , \,
q^{B}_{1} \, , \, q^{B}_{2} \, , \, q^{B}_{3} \} \text{ is
irreducible to classical probability up to the } 2^{th} order
\end{equation*}
\begin{proof}
The simple numerical property:
\begin{equation}
  a b \, - \, b c \, + \, a c \; \leq \; 1 \; \; \forall a,b,c  \in \, [ - 1 \, , \, 1 ]
\end{equation}
implies  that:
\begin{align}
  |  & a b \, - \, b c  |  \; \leq \; 1 \, - \, a c \; \; \forall a ,b ,c \in [ 0 ,1 ] \\
  |  & a b \, + \, b c  | \, + \, |  a d \, - \, d c  |  \; \leq \; 1 \, + \, a c \; \; \forall a ,b ,c,d  \in [ 0 ,1 ] \\
\end{align}
from which it follows that it don't exist four random variables a , b , c , d  defined on a classical probability
space $ ( \, \Omega \, , \, P ) $ such that:
\begin{align}
   |  & E(  a b ) \, - \, E( b c )  |  \; \leq \; 1 \, - \, E( a c )  \\
   |  & E(  a b ) \, - \, E( b c )  |  \; \leq \; 1 \, + \, E( a c )  \\
    |  & E(  a b ) \, - \, E( b c )  | \, + \, |  E(  a d ) \, + \, E( d c )  | \; \leq \; 2
\end{align}
where E denotes expectation w.r.t. the P-measure:, i.e.:
\begin{equation*}
  E( F ) \; := \; \int_{\Omega} \, F \, dP
\end{equation*}
The thesis easily follows
\end{proof}

\begin{remark} \label{rem:Bell's theorem doesn't speak of locality}
\end{remark}
BELL'S THEOREM DOESN'T SPEAK OF LOCALITY:

Our way of presenting Bell's result is someway provocative, in
that it is completely different both from the form and from the
spirit of Bell's papers \cite{Bell-93}:

Bell's theorem was intended to be and is almost always looked as
\cite{Shimony-00} the proof that all local hidden
variables' theories imply an inequality which is incompatible with
some of the predictions of Quantum Mechanics.

Such inequality, anyway, is nothing but a consequence of the fact
that there does not exist a set of six classical random variables
$ \{ c^{A}_{1} \, , \, c^{A}_{2} \, , \, c^{A}_{3} \, , \,
c^{B}_{1} \, , \, c^{B}_{2} \, , \, c^{B}_{3} \} $ on a suitable
classical probability space, such that:
\begin{equation}
  M_{n}(  c^{A}_{i} ) \; = \; M_{n}(  c^{B}_{i} ) \; = \;
  \begin{cases}
    0 & \text{ if  $ n = 0 $ or $ n = 2 $}, \\
    1 & \text{if  $ n = 0 $}.
  \end{cases}  \; \; i= 1 , \cdots , 3
\end{equation}
\begin{equation}
  E( c^{A}_{i} c^{A}_{j} ) \; = \; E( c^{B}_{i} c^{B}_{j} ) \; =
  \;\delta_{i,j} \; \; i,j = 1 , \cdots , 3
\end{equation}
\begin{equation}
  E ( c^{A}_{i}  c^{B}_{j} ) \; = \;  E ( c^{B}_{i}  c^{A}_{j} ) \; = \;
  \begin{cases}
    -1 & \text{$ i = j$}, \\
    0  & \text{$ i \neq j$}.
  \end{cases} \; \; i,j = 1 , \cdots , 3
\end{equation}
The concept of locality appears nowhere and has nothing to do
with the physical meaning of theorem\ref{th:Bell's theorem}
concerning the irreducibility of entanglement to Classical
Probability Theory.

\begin{remark} \label{rem:Bell's theorem and functional integrals on superspaces}
\end{remark}
BELL'S THEOREM AND FUNCTIONAL INTEGRALS ON SUPERSPACES:

There exist a natural reaction to theorem\ref{th:Bell's theorem}, that could lead to think that there must be certainly a mistake in its proof:
considered a system of two uncoupled fermionic oscillators:
\begin{equation}
  \hat{H} \; := \; \frac{1}{2} (  \hat{a}_{1}^{\dagger} \hat{a}_{1} \, + \,  \hat{a}_{2}^{\dagger} \hat{a}_{2} )
\end{equation}
where:
\begin{align}
  \hat{a}_{i}^{2} & \; = \; ( \hat{a}_{i}^{\dagger} )^{2} \; = \; 0 \; \; i = 1,2  \\
  \hat{a}_{i}^{\dagger} & \hat{a}_{j} \, + \, \hat{a}_{j} \hat{a}_{i}^{\dagger} \; = \; \delta_{i,j} \; \; i = 1,2
\end{align}
every theoretical-physicists' textbook (cfr. e.g. the section 3.5 of \cite{Zinn-Justin-93}) tell us that we can compute all its correlation functions by functional derivatives
of the partition function:
\begin{multline} \label{eq:informal 2-fermionic oscillators' partition function}
  Z [ \bar{\eta}_{1} \, , \,  \eta_{1} \, , \, \bar{\eta}_{2} \, , \,  \eta_{2} ] \; := \; \int [ d c_{1}(t) d c_{2}(t) d \bar{c}_{1}(t) d \bar{c}_{2}(t)  ]  \exp [ - S( c_{1}, c_{2} , \bar{c}_{1} , \bar{c}_{2}) \, + \\
    \int d s \sum_{i=1}^{2}  \bar{\eta}_{i} (s) c_{i}(s) +  \bar{c}_{i} (s) \eta_{i}(s)  ]
\end{multline}
with euclidean action:
\begin{equation}
  S( c_{1}, c_{2} , \bar{c}_{1} , \bar{c}_{2}) \; := \; \int dt \sum_{i=1}^{2} \bar{c}_{i} (t) \dot{c}_{i} (t) \,  - \, \bar{c}_{i} (t) c_{i}(t)
\end{equation}
Isn't this fact an explicit confutation of theorem\ref{th:Bell's theorem}, implying the existence of the six classical random variables
$ \{ c^{A}_{1} \, , \, c^{A}_{2} \, , \, c^{A}_{3} \, , \, c^{B}_{1} \, , \, c^{B}_{2} \, , \, c^{B}_{3} \} $ we spoke about in the remark\ref{rem:Bell's theorem doesn't speak of locality}?

The reason why this is not the case is that the euristic measure of equation\ref{eq:informal 2-fermionic oscillators' partition function} cannot be defined in  a mathematically rigorous way.

Indeed, though being at the basis of many exciting mathematical results such as the proof of the Atiyah-Singer Index Theorem
by the computation of the index of the Dirac operator D on a spin-manifold $ ( M \, , \, g ) $ as the path-integral:
\begin{equation} \label{eq:path-integral's computation of the Dirac operator's index }
  Index(D) \; := \; \int_{p.b.c.} [ dx ] [ d \psi ]  \exp [ - \int_{0}^{\beta} dt L ]
\end{equation}
where:
\begin{equation}
  L \; := \; \frac{1}{2} g_{\mu , \nu} \dot{x}^{\mu} \dot{x}^{\nu} \, + \,  \frac{1}{2}  g_{\mu \nu} \psi^{\mu} \frac{D \psi^{\mu} }{D t}
\end{equation}
is the supersymmetric lagrangian of a spin-$\frac{1}{2}$ fermion living on $ ( M \, , \, g ) $ \cite{Alvarez-95}, a rigorous
mathematical formalization of functional integration on superspaces (going beyond informal time-splitting procedures such as that
of the fifth chapter of \cite{De-Witt-92}) doesn't exist yet.

It may be worth mentioning the possibility that it could  require an extension of the Kolmogorov's Axiomatization of Probability
rather than simply an application of it, and could in this way converge to Quantum Probability Theory, as the section5.3 of \cite{Khrennikov-99}
and the intellectual path of its author could suggest
\newpage
\section{Irreducibility of Quantum Computational Complexity Theory to Classical Computational Complexity
Theory} \label{sec:Irreducibility of Quantum Computational Complexity Theory to Classical Computational Complexity Theory}
\chapter{Quantum algorithmic randomness: where are we?} \label{chap:Quantum algorithmic randomness: where are we?}
\section{The unpublished ideas of Sidney Coleman and Andrew
Lesniewski} \label{sec:The unpublished ideas of Sidney Coleman
and Andrew Lesniewski}

The first people who began to investigate in a sistematic way the
interrelations between Quantum Theory and  the notion of
algorithmic randomness was certainly Paul Benioff in a serie of
1970's papers \cite{Benioff-73}, \cite{Benioff-74},
\cite{Benioff-77}, \cite{Benioff-78} in which he extensively
analyzed the algorithmic randomness status of the sequence of
outcomes of quantum measurements.

Benioff's intention was not, anyway, that of characterizng a
notion of quantum-algorithmic-randomness, but that of extracting
from Quantum Physics a new definition of classical-algorithmic
randomness.

Indeed, in those years, the great scientific revolution
concerning the incommensurability of quantum information and
classical information (we underlined in the example \ref{ex:the
Hilbert space of Quantum Information Theory} and in the
remark\ref{rem:the noncommutative combinatory information and the
definition of the qubit}) was not happened yet.

A very similar kind of investigation was then pursued by Sidney
Coleman and Andrew Lesniewski who tried to extend previous
considerations by Hartle, as well as by Sam Guttman
\cite{Guttmann-95}, \cite{Mittelstaedt-01}

Unfortunately Coleman and Lesniweski never published their thought
that is accessible only from the exposition of it made by John
Preskill in the section3.6 of his wonderful lecture notes
\cite{Preskill-98} as well as from the electronic correspondence
of Christopher Fuchs he gently gave to collectivity's disposition
(cfr. pagg.24-30 as well as pagg.106-110 of \cite{Fuchs-01}).

The starting point is the following analysis by Hartle
\cite{Hartle-68}:

the only point of the standard Copenhagen's axiomatization of
Quantum Mechanics in which the term $ << probability >> $ appears
is the \textbf{Postulate of Reduction}, stating that a
measurement of an observable $ \hat{A} \; := \; \sum_{a} a | a >
< a | $ on a quantum system prepared in the state $ | \psi > \;
:= \; \sum_{a} | a >  < a | \psi > $ has the following effects:
\begin{enumerate}
  \item
\begin{equation}
 Probability [measurement's \: outcome \; = \; a ] \; = \;  | < a | \psi
  >| ^{2}
\end{equation}
  \item if the measuremnt's outcome a occurs, then the state's
  system collapses istantaneously to the state $ | a > $
\end{enumerate}
where we have considered, for simplicity, the case when there is
no degeneration.

Hartle observed that the \textbf{Issue of the Interpretation of
Probability}   may be made to disappear  from the axiomatization
of Quantum Mechanics in the following way:
\begin{enumerate}
  \item one replaces the Postulate of Reduction with the weaker \textbf{Postulate of
Eigenstates}:
\begin{center}
  If we prepare a quantum state $ | a > $ such that $ \hat{A} | a
  > \, = \, a | a > $, and then immediately measure $ \hat{A} $,
  the outcome of the measurement is a with certainty
\end{center}
  \item the case of measurements performed in a state that is not
  an eigenstate of the measured observable is reconducted to
  the \textbf{Postulate of Eigenstates} by the assumption of a
  frequentistic interpretation of probability:

  suppose we want to make a statement about the probability of
  obtaining the result $ | \uparrow_{z} > $ when we measure $ \sigma_{z} $ in the state :
\begin{equation}
  | \psi > \; = \; a | \uparrow_{z} > \, + \, b | \downarrow_{z} >
\end{equation}
 Hartle imagines that one prepares an infinite number of copies, so that the state is:
\begin{equation}
  | \psi^{( \bigotimes \infty)} > \; :=  \; \bigotimes_{n=1}^{\infty} | \psi
  >
\end{equation}
and imagines that one measures $ \sigma_{z} $ for each of the
copies.

Introduced  the \textbf{average spin operator}:
\begin{equation}\label{eq:average spin operator}
  \bar{\sigma}_{z} \; := \; \lim_{n \rightarrow + \infty}
  \frac{1}{n} \sum_{i=1}^{n}  \bar{\sigma}_{z}^{(i)}
\end{equation}
Hartle claims that $  | \psi^{(\bigotimes n)} >  $  becomes an
eigenstate of  $ \bar{\sigma}_{z} $ with eigenvalue $ |a|^{2} -
|b|^{2} $ for $ n \rightarrow \infty $.

Then he appeals to the \textbf{Postulate of Eigenstates} to infer
that a measurement of $ \bar{\sigma}_{z} $ will yeld the result $
|a|^{2} - |b|^{2} $ with certainty, and that the fraction of all
the spins that point up is, therefore, $ | a |^{2} $.

In this sense $ | a |^{2} $ is the probability that the
measurement of $ \sigma_{z} $ yelds the outcome $ | \uparrow_{z}
> $.
\end{enumerate}

As an application of Hartle's strategy,  let us suppose , for
example, that:
\begin{equation}
  | \uparrow_{x}^{( \bigotimes n)} > \; := \; \bigotimes_{i=1}^{n}
  \frac{1}{\sqrt{2}} ( | \uparrow_{z} > + | \downarrow_{z} > )
\end{equation}
One has that:
\begin{equation}
  < \uparrow_{x}^{(\bigotimes n)} | \bar{\sigma}_{z} | \uparrow_{x}^{(\bigotimes n)} >  \; =
  \; 0
\end{equation}
\begin{equation}
  < \uparrow_{x}^{(\bigotimes n)} | \bar{\sigma}_{z}^{2} | \uparrow_{x}^{(\bigotimes n)} >  \; =
  \; \frac{1}{n}
\end{equation}
Thus, taking the limit $ n \rightarrow + \infty $, one concludes
that  $ \bar{\sigma}_{z} $ has vanishing dispersion about its mean
value so that, at least in this sense,  $ |
\uparrow_{x}^{(\bigotimes \infty)} > $ is an "eigenstate" of $
\bar{\sigma}_{z} $ with eigenvalue zero.

Coleman and Lesniewski has generalized Hartle's ideas observing
that indeed one can require that the sequence $ \cdot_{i}
\lambda_{i} $, where $ \lambda_{i} $ is the result of the
measurement of  the operator $ \sigma_{z}^{i} $, satisfies not
only the Law of Randomness of 1-Borel normality, but all the Laws
of randomess, i.e. that it is Martin L\"{o}f - Solovay - Chaitin
random.

So they introduce an orthogonal projection  operator $
\hat{\Pi}_{random} $ that acting on a state $ | \psi > $ that is
an eigenstate of each $ \sigma_{z}^{(i)} $ satisfies:
\begin{equation} \label{eq:the Coleman-Lesniewski operator acts as the identity on eigenstates of random sequences}
  \hat{\Pi}_{random} | \psi > \; =  \;  | \psi >
\end{equation}
if the sequence of eigenvalues of $ \sigma_{z}^{(i)} $ is
algorithmically-random, and:
\begin{equation} \label{eq:the Coleman-Lesniewski operator vanishes on eigenstates of not  random sequences}
  \hat{\Pi}_{random} | \psi > \; =  \; 0
\end{equation}
if the sequence of eigenvalues of $ \sigma_{z}^{(i)} $ is not
algorithmically-random.

Preskill reports that Coleman and Lesniewski discovered that
eq.\ref{eq:the Coleman-Lesniewski operator acts as the identity on
eigenstates of random sequences} and eq.\ref{eq:the
Coleman-Lesniewski operator vanishes on eigenstates of not random
sequences} properties, together with the condition that
 $ \hat{\Pi}_{random} $ is an orthogonal projection,  are not sufficient to determine how $
\hat{\Pi}_{random} $ acts on all $ {\mathcal{H}}_{2}^{\bigotimes
\infty} $, but that, with additional technical constrains, it
exists, it is unique, and has the property that:
\begin{equation}
  \hat{\Pi}_{random} | \uparrow_{x}^{ \bigotimes \infty } > \; =
  \; 1
\end{equation}
These considerations  seems to us rather strange, since,
according to us, the operator $ \hat{\Pi}_{random} $ may be
simply defined as:
\begin{definition} \label{Coleman-Lesniewski operator}
\end{definition}
COLEMAN-LESNIEWSKI OPERATOR:
\begin{equation}
  \hat{\Pi}_{random} \; := \; \int_{CHAITIN(\Sigma^{\infty})} dP_{unbaised} |
  \bar{x} > < \bar{x} |
\end{equation}
but than one has that:
\begin{multline}
  \hat{\Pi}_{random} | \uparrow_{x} ^{\bigotimes \infty} > \; \\
  = \; \hat{\Pi}_{random}  \bigotimes_{i=1}^{\infty} [
  \frac{1}{\sqrt{2}}( | 0 > + | 1 > ) \\
   = \; ( \lim_{n \rightarrow \infty } \frac{1}{ 2 \frac{n}{2} } ) \: \hat{\Pi}_{random}(  | 0^{\infty} > + | 1^{\infty} > ) \; = \;  0
\end{multline}

\smallskip
Introduced the following notion:
\begin{definition} \label{def:Coleman random sequences of qubits}
\end{definition}
COLEMAN RANDOM SEQUENCES OF QUBITS:
\begin{equation}
  COLEMAN-RANDOM( {\mathcal{H}}_{2}^{\bigotimes  \infty} ) \; := \;
   \{ | \psi > \in {\mathcal{H}}_{2}^{\bigotimes  \infty}
   ) \, :  \, \hat{\Pi}_{random}  | \psi > \; = \; | \psi > \}
\end{equation}

it would be clear why, according to our point of view, such a notion is completelly misleading as to the characterization of quantum algorithmic randomness:

as we extensively discussed in section\ref{sec:Why to treat sequences
of qubits one has to give up the Hilbert-Space Axiomatization of
Quantum Mechanics}, since the right space of qubits' sequences is
the noncommutative space  $ \Sigma_{NC}^{\infty} $ and not the
Hilbert space $ {\mathcal{H}}_{2}^{\bigotimes  \infty}
$, the space of algorithmically-random sequences of qubits is  a set of
objects of the form:
\begin{equation}
  RANDOM( \Sigma_{NC}^{\infty} ) \; \subset \;  \Sigma_{NC}^{\infty}
\end{equation}
and not a set of the form:
\begin{equation}
  RANDOM( {\mathcal{H}}_{2}^{\bigotimes \infty} ) \; \subset
  \; {\mathcal{B}}({\mathcal{H}}_{2}^{\bigotimes  \infty} )
\end{equation}
as $ COLEMAN-RANDOM( {\mathcal{H}}_{2}^{\bigotimes \infty} ) $.

Demanding to remark\ref{rem:difference between the raising of
commutative cardinality ad the raising of noncommutative
cardinality} and remark\ref{rem:the phenomenon of continuous
dimension from a logical point of view} for a complete analysis,
let us briefly recall that the passage from $ \Sigma_{NC}^{\star}
$ to $ \Sigma_{NC}^{\infty} $ corresponds to a genuine increasing
of \textbf{noncommutative cardinality} by one step, with the
resulting effect of \textbf{continuous dimension} and, hence,
\textbf{the lost of atomicity} of the underlying quantum logic,
while the passage from $ {\mathcal{B}} ( {
\mathcal{H}}_{2}^{\star} ) $ to $ {\mathcal{B}} ( {
\mathcal{H}}_{2}^{\infty} ) $ corresponds to an increasing  of
\textbf{commutative cardinality} by one step, that is different
from the correct required increasing of \textbf{noncommutative cardinality} by one step.

From a logico-mathematical point of view, this can be seen introducing the following:
\begin{definition}  \label{def:Coleman quantum propositions}
\end{definition}
COLEMAN PROPOSITIONS:
\begin{equation}
  CQP \; := \; \{ | \psi > <  \psi | \; : \;   | \psi > \in COLEMAN-RANDOM( {\mathcal{H}}_{2}^{\bigotimes  \infty} ) \}
\end{equation}
Clearly any Coleman quantum proposition is an \textbf{atomic quantum propositions} of the weak quantum logic $
{\mathcal{L}} ( {\mathcal{H}}_{2}^{\bigotimes \infty} ) $.

The  effects of erroneously supposing that the quantum logic of qubits' sequences has atomic propositions may be appreciated by the following:
\begin{remark} \label{rem:Coleman and the Entscheidungsproblem}
\end{remark}
THE HALTING-PROBABILITY'S  COLEMAN ATOMIC PROPOSITION WOULD SOLVE
THE COMMUTATIVE ENTSCHEIDUNGPROBLEM

Let us consider the following Coleman quantum proposition:
\begin{equation}
  p_{\Omega_{U}}  \; := \; | \Omega_{U} > < \Omega_{U} |
\end{equation}
where, according to definition\ref{def:halting probability}, $
\Omega_{U}$ denotes the Halting Probability w.r.t. the Chaitin
universal computer U.

Let us, then, introduce the \textbf{qubits' sequence operator}:
\begin{equation} \label{eq:qubits' sequence operator}
  \hat{q}^{\bigotimes \infty} \; := \; \bigotimes_{n \in
  {\mathbb{N}}} \hat{q}
\end{equation}
where $\hat{q}$ is the  qubit operator defined in
eq.\ref{eq:right definition of the qubit operator}. The
measurement of $ \hat{q}^{ \bigotimes \infty } $ in the state $  |
\Omega_{U} > $ results in the solution of the ($ C_{\Phi}$ -
classical, i.e. commutative)  Enstcheidungproblem (as David
Hilbert indicated the problem of determining whether or not a
given formula of the (Classical) Predicate Calculus is valid
\cite{Davis-65}, \cite{Odifreddi-89}).

We see, then, that the predicate  $ p_{\Omega_{U}} $ encodes the
solution of the Commutative Entsheidungsproblem.

So we would have that the Quantum Propositional Calculus admits an\textbf{ atomic} proposition (from which other
not-atomic i.e. not-elementary, propositions may be constructed
logically-connecting it with other propositions through the
connectives $ \bigvee , \bigwedge , \perp $), that, just alone, implies  a violation of the Church-Turing Thesis.

Assuming  the Church-Turing Thesis,  we have then to reject such a situation.

\smallskip

Let us explain, by the way,  more precisely the meaning of the expression Commutative Enschteidungsproblem  we used:

the impossibility of developing all Mathematical-Logics simply by the distributive orthocomplemended lattice of  Classical Predicate Calculus appears only when one wants to take into  account quantifications.
In this case, even restricting the analysis to First Order Theories in which one predicate cannot have other predicates  or functions as arguments and quantification on predicates or functions is forbidden,
 one has to pass to  Classical Formal Systems, their models and so on.

Also the Classical Predicate Calculus may, of course, be then embedded in such a more sophisticate language:

there exist many ways of axiomatize it as a classical formal system, and the general theory of formal systems may be applied to it
to conclude that, as a formal system (in any way we axiomatized it), Classical Predicate Calculus  is consistent though, as we have seen, undecidable.

As far as Quantum Logic is concerned, many people, and we among
them, tried to go beyond Quantum Predicate Calculus (i.e. the
theory of orthocomplemented orthomodular lattices) to deveop a
general theory of Quantum Formal Systems, the first attempts being of Von Neumann himself:
\begin{center}
\textit{"Dear Doctor Silsbee, It is with great regret that I am writing these lines to you, but I simply cannot help myself.
In spite of very serious attempts to write the article " Logics of Quantum Mechanics" I find it it completely impossible to do it at this time.
As you may know, I wrote a paper on this subject with Garrett Birkhoff in 1936 ("Annals of Mathematics", vol. 37 , pp. 823-843) and I have thought
a good deal on the subject since. My  work on continuous geometry , on which I gave the Amer. Math. Soc. Colloqiuim lectures of 1937, comes to a
considerable extent from this source. Also a good deal concerning the relationship between strict- and probability logics
(upon which I touched briefly in the Henry Joseph Lecture) and on the extension of this "Propositional calculus" work to "logic  with quantifiers"
(which I never discussed in public)"}
 (letter to Doctor Solbee ; July 2, 1945 ; cfr. \cite{von-Neumann-01})
\end{center}

Personally we tried to develop:
\begin{enumerate}
  \item a quantum corrispective of John Mc Carthy's  LISP
  \cite{Mc-Carthy-60}, i.e. more precisely, of Chaitin's version
  of it in which the evaluation operator \emph{"eval"} of
  syntax:
  \begin{center}
      eval  S-expression
  \end{center}
is replaced by a time-constrained \cite{Chaitin-98} version of it, whose syntax:
\begin{center}
         try time-limit S-expression
\end{center}
specifies the time-interval after which the computation halts
furnishing as output the partial computation performed
\footnote{Such a time-constraining is necessary since, otherwise, the request of evaluating a formal axiomatic system would never halt since, for all the not-trivial formal systems, the inferential chain of theorem-proving is infinite}

 We studied a new language, that we called the quantum-LISP,
 defined by the replacement of the instruction \emph{"try"} with a
 new instruction \emph{"quantum try"} with sintax:
 \begin{center}
          quantum-try time-limit quantum-S-expression
 \end{center}
where a \textbf{quantum-S-expression} is a list of the form:
\begin{equation}\label{eq:syntax of the quantum-try}
     (( S-expression_{G} \;  S-expression_{H} ) \; ( a \; b ))
\end{equation}
with $ a , b \, \in \, {\mathbb{C}}  \; | a |^{2} +  | b |^{2} \, = \, 1 $.

Under the command of eq.\ref{eq:syntax of the quantum-try} the computer chooses at random
one value of a binary random variable h such that:
\begin{align}
  Prob( & h = \frac{1}{2} ) \; = \; | a |^{2}\\
  Prob( & h = - \frac{1}{2} ) \; = \; | b |^{2}
\end{align}
and then operates as follows:
\begin{itemize}
  \item if occurs $ h = \frac{1}{2} $ then it sets the
  \textbf{halt-qubit-list} to $ (1 0)$ and operates as Chaitin-LISP would do under the instruction:
  \begin{center}
     try time-limit $ S-expression_{G} $
  \end{center}
  \item if occurs $ h = - \frac{1}{2} $ then it sets the \textbf{halt-qubit-list} to $ (0 1)$ and operates as Chaitin-LISP would do under the instruction:
\begin{center}
   try time-limit $ S-expression_{H} $
\end{center}
\end{itemize}

The idea underlying such a definition of  the quantum-try
instruction is to make it equivalent to a Deutsch's quantum
Turing machine in which the periodic monitoring of the
halting-qubit occurs at temporal-steps of time-interval \cite{Deutsch-85}.

By the impossibility of having a fair random generator extensively discussed in section\ref{sec:Irreducibility of Quantum Computational Complexity Theory to Classical Computational Complexity Theory}
Quantum-LISP is not implementable on a classical computer and, for practical purposes, must be replaced with a Virtual-quantum-LISP, i.e. a
language completelly identical to Quantum-LISP, but for the fact that the fair random generator is replaced with a PRG.

  \item proceeding euristically, we tried to characterize the notion of a  \textbf{quantum Post systems}
  associated to a classical Post system  \cite{Odifreddi-89} $ {\mathcal{G}} \; := \, ( \Sigma , A , Q )
  $ with the \textbf{axioms' set } $ A \subset \Sigma^{\star} $
  and \textbf{productions' set } Q as a triple $
  \hat{{\mathcal{G}}} \; := \; ( {\mathcal{H}}_{\Sigma}  \, , \,  {\mathcal{H}}_{A}
  \, , \, \hat{Q} )$, where $  {\mathcal{H}}_{\Sigma} \;
  = \;  {\mathcal{H}}_{2}^{ \bigotimes \star} $,  $  {\mathcal{H}}_{A} $ is an
  Hilbert sub-space of  $ {\mathcal{H}}_{\Sigma} $, while $
  \hat{Q} $ is the set of the \textbf{quantum productions}, i.e.
  operators  on  $  {\mathcal{H}}_{\Sigma} $ acting  as the
  productions Q of  $ {\mathcal{G}} $ on the computational basis.

  A theorem of a quantum Post system is then defined as an element
  of   $ {\mathcal{H}}_{2}^{\star} $ reachable by a vector
  belonging to $  {\mathcal{H}}_{A} $ by a finite number of
  application of suitable quantum productions giving rise to a
  plethora of logical-mathematical notions specularizing the
  classical ones.
\end{enumerate}

We then discovered that formalizations of the theory of Quantum Formal Systems already existed in the literature:
 in 1996  Philiph Maymin introduced a quantum analogue of Alonzo
 Church's Lambda Calculus \cite{Mayimin-96}, an idea already
independentely (and in a completely different way) developed by
David Finkelstein in the section 14.3.7 of his monograph
\cite{Finkelstein-97}. In 1997 Christopher Moore and James P.
Crutchfield  introduced quantum analogues of the whole Chomsky
hierarchy \cite{Moore-Crutchfield-97}.

A similar idea was concretelly implemented in his Masters Thesis by  Bernard \"{O}emer who  developed  QCL:
 an high-level, architecture independent programming language for quantum
 computing whose interpreter is downloadable from the author's
 homepage \cite{Oemer-98}; the syntactic structure of a QCL
 program is described by a context-free grammar, in a way
 concisely explained in the section4.19.4 of
 \cite{Calude-Paun-01}.

 A step forward in the formalization of what a General Theory of Quantum Formal Systems  has been done, according
 to us, by  Paul Benioff \cite{Benioff-98} who, introducing  (with the usual generalization on the computational basis) a quantum analogue of a toy-formal-system   by Raymond M.
 Smullyian (cfr. the first chapter of \cite{Smullyan-92})
 discusses not only its \textbf{syntax} but also its
 \textbf{semantic}.

 This is something new since, up to date, interpretations and models
 has been studied by the Quantum Logic Community only at the Quantum Propositional Calculus'
 level \label{Dalla-Chiara-Giuntini-01}.

 Now, exactly as the consideration of quantifications requires, in
 the classical case, to give up the simple lattice-theoretic
 Classical Propositional Calculus  passing to the more
 sophisticated language of Classical Formal Systems and arriving,
 on this way, to axiomatize Classical Propositional Calculus
 itself formalizing its Entscheidungsproblem (that we call the
 Commutative  Entscheidungsproblem) and discovering its unsolvability, we think the same must happen
 as to Quantum Propositional Calculus, whose Enstcheidungsproblem
 will be called the \textbf{Noncommutative Enstcheidungsproblem}
 from here and beyond.

  These preliminary, euristic considerations concerning \textbf{quantum formal systems} will be discussed, anyway, more
  explicitly in section\ref{sec:Karl Svozil's invention of Quantum Algorithmic
Information Theory} where we will extensively discuss the quantum
extension of the duality:
\begin{center}
   \textbf{languages versus automata}
\end{center}
and the consequential characterization of the notion of
\textbf{quantum formal system} obtained using such a duality at
the correct level of Moore's generalization of Chomsky's
hierarchy.

\newpage
\section{Karl Svozil's invention of Quantum Algorithmic Information
Theory} \label{sec:Karl Svozil's invention of Quantum Algorithmic
Information Theory}

In 1995 Karl Svozil first introduced the idea that the
irreducibility of Quantum Information Theory to the classical one
implies the necessity of developing a quantum analogue of
Classical Algorithmic Information Theory, namely Quantum
Algorithmic Information Theory, irreducible to the classical
theory \cite{Svozil-96}.

Given a quantum computer Q, i.e. a quantum-mechanical physical
system with Hilbert space $ {\mathcal{H}}_{2}^{\star} $  Svozil
affords the first issue:
\begin{center}
     \textbf{have the programs of Q to be coded in cbit or qubits?}
\end{center}
To obtain the quantum analogue of \textbf{prefix algorithmic
entropy}, Svozil claims, their lengths must satisfy the Kraft's
Inequality; but if we allowed Q's programs to be qubits' strings
instead of cbits' strings, than the Kraft sum would diverge.

As we will see this a key point, discussed also by Paul Vitanyi
\cite{Vitanyi-99}, \cite{Vitanyi-01} in his rediscovering of
Svozil's results and lying at the basis of the objections
Andr\'{e} Berthiaume, Wim van Dam and Sophie Laplante moved to
Vitanyi \cite{Berthiaume-van-Dam-Laplante-00} in their rediscovering of what Svozil had already
discussed years before.

The condition that the programs of Q are classical may be easily
formalized observing that any map:
\begin{equation}
  Q \in \stackrel{\circ}{MAP}( \Sigma^{\star} \, , \, {\mathcal{H}}_{2}^{ \bigotimes
  \star})
\end{equation}
may be equivalentely seen as a map:
\begin{equation}
  Q \; \in \;  \stackrel{\circ}{MAP}( {\mathbb{E}}_{\star} \, , \, {\mathcal{H}}_{2}^{ \bigotimes
  \star})
\end{equation}
identifying $ \Sigma^{\star} $ with the computational basis $
{\mathbb{E}}_{\star} $.

Assuming that the quantum computer is a \textbf{closed system} Q
will be clearly nothing but  the restriction to $
{\mathbb{E}}_{\star} $ of an \textbf{inner automorphism} of $
 {\mathcal{B}} ({\mathcal{H}}_{2}^{\star} )$.

Assumed the prefix-free condition:
\begin{equation}
  HALTING( Q ) \text{ is prefix-free}
\end{equation}
Svozil introduces the following:
\begin{definition} \label{def:Svozil's quantum algorithmic information}
\end{definition}
QUANTUM ALGORITHMIC INFORMATION OF $ | \psi > $ W.R.T. Q:
\begin{equation}
  I_{Q}( | \psi > ) \; := \;
  \begin{cases}
    \min \{ \vec{x} \in HALTING(Q) \: : \: Q ( \vec{x} ) =  | \psi > \} & \text{if $ \exists \vec{x} \in HALTING(Q) \: : \: Q ( \vec{x} ) =  | \psi > $ }, \\
    + \infty & \text{otherwise}.
  \end{cases}
\end{equation}

Then Svozil considers the definition of a quantum analogue of
Chaitin's Halting Probability.

To see how Svozil implements such a notion it is necessary, first
of all, to discuss his analysis of the Halting Problem for
Quantum Computers, i.e. his analysis of Quantum Diagonalization.

Diagonalization is a proof's technique introduced by Cantor to
prove that $ cardinality( 2^{{\mathbb{N}}} ) \, > \, \aleph_{0} $.

It may be formalized in the following way:
\begin{theorem} \label{th:diagonalization's theorem}
\end{theorem}
DIAGONALIZATION'S THEOREM:

\begin{hypothesis}
\end{hypothesis}
\begin{equation*}
  A \; \; set
\end{equation*}
\begin{equation*}
  R \; \subseteq \; A \times A \; \; \text{ binary relation on A}
\end{equation*}
\begin{equation} \label{eq:diagonal set of a binary relation}
  D \; := \; \{ a \in A \, : \, ( a , a) \notin R \} \; \; \text{ diagonal set for R}
\end{equation}
\begin{equation*}
  R_{a} \; := \; \{ b \in A : ( a , b ) \in R \} \; \; a \in A
\end{equation*}
\begin{thesis}
\end{thesis}
\begin{equation*}
  D \, \neq \, R_{a} \; \; \forall a \in A
\end{equation*}
\begin{proof}
Suppose ad-absurdum that:
\begin{equation}
  \exists \bar{a} \in A \; : \; D \, = \, R_{\bar{a}}
\end{equation}
i.e.:
\begin{equation} \label{eq:the diagonal is a suitable row}
   \exists \bar{a} \in A \; : \; D \, = \, \{ b \in A \, : \, (
   \bar{a}, b ) \in R \}
\end{equation}
Let us now consider the following question:
\begin{equation*} \label{eq:basic question in diagonalization}
  \bar{a} \; \in \;D \; ?
\end{equation*}
\begin{itemize}
  \item if the answer to the question in eq.\ref{eq:basic question in diagonalization}
  is \textbf{yes} it follows by eq.\ref{eq:diagonal set of a binary relation}
  that $ ( \bar{a} \, , \, \bar{a} ) \; \notin \; R $ that, by eq.\ref{eq:the diagonal is a suitable
  row}, implies that $ \bar{a} \; \notin \; D $ that is asburdum
  \item if the answer to the question in eq.\ref{eq:basic question in diagonalization}
  is \textbf{no} it follows by eq.\ref{eq:diagonal set of a binary
  relation} that $ ( \bar{a} \, , \, \bar{a} ) \; \in \; R $ that, by eq.\ref{eq:the diagonal is a suitable
  row}, implies that $ \bar{a} \; \in \; D $ that is again asburdum
\end{itemize}
\end{proof}

Cantor's argument runs than as follows:
\begin{theorem} \label{th:Cantor's theorem}
\end{theorem}
CANTOR'S THEOREM:
\begin{equation}
  cardinality( 2^{{\mathbb{N}}} ) \, > \, \aleph_{0}
\end{equation}
\begin{proof}
Let us suppose ad absurdum that $ 2^{{\mathbb{N}}} $ is
countable. Then there exists a a way of enumerating all members
of $ 2^{{\mathbb{N}}} $ as:
\begin{equation}
  2^{{\mathbb{N}}} \; = \; \{ R_{0} , R_{1} , R_{2} , \cdots  \}
\end{equation}
Introduced the relation on $ {\mathbb{N}} $ as:
\begin{equation}
  R \; := \; \{ ( i , j ) \, \in \, {\mathbb{N}} \times
  {\mathbb{N}} \: : \: j \in R_{i} \}
\end{equation}
the thesis immediately follows applying to R the
theorem\ref{th:diagonalization's theorem}
\end{proof}

\smallskip

Let us now pass to partial recursive functions and let us
introduce the following two sets:
\begin{definition} \label{def:first self-referential set}
\end{definition}
FIRST SELF-REFERENTIAL SET:
\begin{equation}
  SR_{1} \; := \; \{ i \in {\mathbb{N}} \, : \, i \in
  {\mathcal{W}}_{i} \}
\end{equation}
\begin{definition} \label{def:second self-referential set}
\end{definition}
SECOND SELF-REFERENTIAL SET:
\begin{equation}
  SR_{2} \; := \; \{ ( i , j ) \,  \in \, {\mathbb{N}} \times  {\mathbb{N}} \, : \, i \in
  {\mathcal{W}}_{j} \}
\end{equation}
Cantor's diagonalization argument immediately leads to the
following importan theorems:
\begin{theorem} \label{th:combinatorial core of the undecidability results}
\end{theorem}
COMBINATORIAL CORE OF THE UNDECIDABILITY RESULTS:
\begin{equation*}
  SR_{1} \; \text{ is r.e. but not recursive}
\end{equation*}
\begin{proof}
We have that:
\begin{equation}
  x \in SR_{1} \; \Leftrightarrow \; \varphi_{x}(x) \downarrow
\end{equation}
But theorem\ref{th:Goedel's numbering of partial recursive
functions} tells us that there exist a partial recursive $
\varphi $ such that:
\begin{equation}
  \varphi (x) \; = \; \varphi_{x} (x)
\end{equation}
and hence:
\begin{equation}
  SR_{1} \; = \; HALTING( \varphi )
\end{equation}
So $ SR_{1} $, being the halting set of a partial recursive
function, is a r.e. set.

The fact that $ SR_{1} $ is not recursive follows immediately by
applying theorem\ref{th:Goedel's numbering of partial recursive
functions} to the relation $ SR_{2} $.
\end{proof}

\begin{theorem} \label{th:unsolvability of the Halting Problem}
\end{theorem}
UNSOLVABILITY OF THE HALTING PROBLEM:
\begin{equation*}
  SR_{2} \; \text{ is r.e. but not recursive}
\end{equation*}
\begin{proof}
We have that:
\begin{equation}
  (i,j) \in SR_{2} \; \Leftrightarrow \; \varphi_{j}(i) \downarrow
\end{equation}
But theorem\ref{th:Goedel's numbering of partial recursive
functions} tells us that there exist a partial recursive $
\varphi $ such that:
\begin{equation}
  \varphi (i) \; = \; \varphi_{j} (i)
\end{equation}
and hence:
\begin{equation}
  SR_{2} \; = \; HALTING( \varphi )
\end{equation}
So $ SR_{2 } $, being the halting set of a partial recursive
function, is a r.e. set.

Let us then suppose by absurdum that $ SR_{2 } $ is recursive.
Since:
\begin{equation}
  x \in SR_{1} \; \Leftrightarrow \; ( x \, , \, x ) \in x \in SR_{2}
\end{equation}
this implies that $ SR_{1} $ is not recursive too, contradicting
theorem\ref{th:combinatorial core of the undecidability results}.
\end{proof}

\begin{remark} \label{rem:the first self-referential set and Russell's paradox}
\end{remark}
THE FIRST SELF-REFERENTIAL SET AND RUSSELL'S PARADOX

Bertrand Russell's Paradox is certainly the most famous example
of the many subtlities that  appear in the formalization of
\textbf{classes}, i.e. of \textbf{sets} whose elements are
\textbf{sets} themselves.

It runs as follows: considered the set:
\begin{equation}
  A \; := \; \{ x \, : \, x \notin x \}
\end{equation}
one has that:
\begin{equation}
  x \in A \; \Leftrightarrow \; x \notin x
\end{equation}
and thus:
\begin{equation}
  A \in A \; \Leftrightarrow \; A \notin A
\end{equation}
that is nonsense.

Let us now observe that the set $ {\mathbb{N}} \, - \, SR_{1} $
resembles Russell's set A: it is the set of numbers not belonging
to the r.e. set they code.

But the is no paradox here because Russel's argument simply shows
that such a set is not r.e. itself.

\smallskip

\begin{remark} \label{ref:programmation and meta-programmation}
\end{remark}
PROGRAMMATION AND META-PROGRAMMATION

The meaning of theorem\ref{th:unsolvability of the Halting
Problem} may be appreciated taking into account the concrete
programmation on the (classical, deterministic) computers we use
every day, observing that, by Church-Turing Thesis, the specific
hardware nature of the consided computer is irrilevant.

We can  divide the set of all programming languages for a generic
computer  in two classes, according to if they admit
\textbf{meta-programmation} or not.

By meta-programmation we mean the ability of programs to deal
with that particular kind of \textbf{objects}  made by program
themselves.

Indeed the more logico-mathematically featured programming
languages deal whith only one structure of objects (e.g.
\textbf{lists} in Mac Carthy's LISP \cite{Mc-Carthy-60} or
\textbf{expressions} in Wolfram's Mathematica).

So they automatically admit \textbf{meta-programmation} since
programs and the other objects on which they operate are of the
same (unique) structure.

What is important to observe is that the
\textbf{meta-programmation ability} realizes exactly that link
between \textbf{language} and \textbf{meta-language} that we
indicated in the remark\ref{Godel numbering and self-reference}
as the door leading (or better allowing) self-reference.

\smallskip

We can now explain  the  diagonalization argument lying behind
theorem\ref{th:unsolvability of the Halting Problem} in the
following more concrete way \cite{Svozil-93},
\cite{Davis-Sigal-Weyuker-94}, \cite{Lewis-Papadimitriou-98}:

Let us suppose, for example to enter a Mathematica session.

We could thus think that it is possible, using the
meta-programming in a clever way, to define, through a suitable
Mathematica expression:
\begin{equation} \label{eq:halt expression}
  In[1] \; := \; HALT[p_{-} , x_{-}] \: := \: \cdots
\end{equation}
a function HALT[p,x] that, when called, returns a cbit having the
value True or False according if, respectively, Mathematica halts
or doesn't halt under the input p[x]

If such a Mathematica expression HALT[p,x]  existed it could be
used to construct the following:
\begin{definition} \label{def:diagonal expression}
\end{definition}
DIAGONAL EXPRESSION:
\begin{equation}
  In[2] \; := \; DIAGONAL[x_{-}] \: := \: (Label[start] \, ; \, If[Halt[x,x]==True \, , \, Goto[start] \, , \,True )
\end{equation}

Notice what DIAGONAL[x] does: if the HALT program decides that the
program x would halt if presented with itself as input, then
DIAGONAL(x) loops forever; otherwise it gives as output True and
then halts.

From the function DIAGONAL[x] we could, then, construct the
following Mathematica expression:
\begin{definition} \label{def:paradox expression}
\end{definition}
\begin{equation}
  In[3] \; := \; PARADOX \: := \: DIAGONAL[DIAGONAL]
\end{equation}
Let us now give to Mathematica the following input:
\begin{equation}
  In[4] \; := \; PARADOX
\end{equation}

\smallskip

Will Mathematica halt giving the output Out[4] or not?

\smallskip

It will do it iff the input  HALT[DIAGONAL,DIAGONAL] gives as
output False; in other words Mathematica halts if and only if it
doesn't halt. That is a contradiction.

So we must conclude that the only hypothesis that started us on
this path is false, i.e. that there it doesn't exist any
Mathematical expression that put at the place of the dots in
eq.\ref{eq:halt expression} make the expression HALT[p,x] to be
defined so that it outputs 1 if the Mathematica expression p[x]
halts and zero otherwise.

\begin{remark} \label{rem:time-constrained halting function}
\end{remark}
TIME-CONSTRAINED HALTING FUNCTION:

Let us observe that a time-constrained version of the halting
function, i.e. a Mathematica expression $ HALT[p , x, T] $ that
outputs True if Mathematica  halts under the input p[x] in less or
equal than T seconds and outputs False otherwise, can be
implemented as follows:
\begin{equation} \label{eq:time-constrained halting function}
 In[1] \; := \;  HALT[p_{-} \, , \, x_{-} \, , \, T_{-}] \; := \;
  If[TimeConstrained[p[x],T] \, \neq \, \$Aborted \, , \, True \, ,
  \, False]
\end{equation}
Let us observe, anyway, that while a dialog  of the form:
\begin{align}
   In[2] \; := \; & HALT[p,x,T] \\
  Out[2] \; := \; & True
\end{align}
assures us that also the impossible HALT[p,x], if existed, would
give us True as output, a dialog of the form:
\begin{align}
   In[2] \; := \; & HALT[p,x,T] \\
  Out[2] \; := \; & False
\end{align}
tells us nothing because it is possible that there exist a time $
t > T $ such that:
\begin{align}
   In[3] \; := \; & HALT[p,x,t] \\
  Out[3] \; := \; & True
\end{align}
So the expression HALT[p,x,t] can't be used to solve the Halting
Problem.

\smallskip

Svozil has analyzed what happens when one supposes that the
halting-degree of freedom is codified by a qubit:
\begin{align}
  | Halt & > \; :=  \; c_{True} | True > \, + \, c_{False} |
 False > \; \in \; {\mathcal{H}}_{2}  \\
  | False > &  \; := \; | 0 > \\
  | True > &  \; := \; | 1 > \\
\end{align}
instead of by the cbit:
\begin{align}
   c_{halting} &  \; \in \Sigma \\
  False & \; := \; 0 \\
  True & \; := \; 1
\end{align}

we can rephrase his analysis replacing the halting Mathematica
expression HALT[p,x] such that:
\begin{equation}
 ( \,  In[n] \; := \; Halt[p,x] \, ) \; \Rightarrow \; ( \,  Out[n] \; := \;
  \begin{cases}
    True & \text{if $ p[x] \neq \uparrow  $ }, \\
    False & \text{otherwise}.
  \end{cases}
  \, )
\end{equation}
with an analogue expression QHALT[p,x] such that:
\begin{equation} \label{eq:behaviour of the quantum halting predicate}
 ( \,  In[n] \; := \; QHalt[p,x] \, ) \; \Rightarrow \; ( \,  Out[n] \; := \;
  \begin{cases}
    | True > & \text{if $ p[x] \neq \uparrow  $ }, \\
    | False > & \text{otherwise}.
  \end{cases}
  \, )
\end{equation}
Let us then suppose  that it is possible to implement such an
object by a suitable input of the form:
\begin{equation} \label{eq:Mathematica implementation of the quantum halting predicate}
 In[1] \; := \;  QHALT[p_{-} \, , \, x_{-} ] \; := \; \cdots
\end{equation}
If such a Mathematica expression HALT[p,x]  exists it can be used
to construct  quantum analogues of the \textbf{diagonal
expression} of definition\ref{def:diagonal expression}:
\begin{definition} \label{def:quantum diagonal expression of first kind}
\end{definition}
QUANTUM DIAGONAL EXPRESSION OF FIRST KIND:
\begin{equation}
  In[2] \; := \; QDIAGONAL1[x_{-}] \: := \: (Label[start] \, ; \, If[QHalt[x,x]== | True > \, , \,Goto[start] \, , \, True )
\end{equation}
\begin{definition} \label{def:quantum diagonal expression of second kind}
\end{definition}
QUANTUM DIAGONAL EXPRESSION OF SECOND KIND:
\begin{equation}
  In[2] \; := \; QDIAGONAL2[x_{-}] \: := \: (Label[start] \, ; \, If[QHalt[x,x]== | True > \, ,  Goto[start]  \, , \,  | True >
  )
\end{equation}
\begin{definition} \label{def:quantum diagonal expression of third kind}
\end{definition}
QUANTUM DIAGONAL EXPRESSION OF THIRD KIND:
\begin{equation}
  In[2] \; := \; QDIAGONAL3[x_{-}] \: := \: If[QHalt[x,x]== | True > \, , \,  | False > \, , |
  True >]
\end{equation}
as well as quantum analogues of the paradox function of
definition\ref{def:paradox expression}:
\begin{definition} \label{def:quantum paradox expression of first kind}
\end{definition}
QUANTUM PARADOX EXPRESSION OF FIRST KIND:
\begin{equation}
  In[3] \; := \; QPARADOX1 \: := \: QDIAGONAL1[QDIAGONAL1]
\end{equation}
\begin{definition} \label{def:quantum paradox expression of second kind}
\end{definition}
QUANTUM PARADOX EXPRESSION OF SECOND KIND:
\begin{equation}
  In[3] \; := \; QPARADOX2 \: := \: QDIAGONAL2[QDIAGONAL2]
\end{equation}
\begin{definition} \label{def:quantum paradox expression of third kind}
\end{definition}
QUANTUM PARADOX EXPRESSION OF THIRD KIND:
\begin{equation}
  In[3] \; := \; QPARADOX1 \: := \: QDIAGONAL1[QDIAGONAL1]
\end{equation}
Does the diagonalization proof of the not-existence of HALT[p,x]
hold also as to $ QHALT1[p,x] \, , \,  QHALT2[p,x] \, , \,
QHALT3[p,x] $?

Let us start analyzing QDIAGONAL1[x]: if the QHALT program
decides that the program x would halt if presented with itself as
input, then DIAGONAL(x) loops forever;otherwise it gives as
output True and then halts.

Up to the identification of $ \Sigma $ with the computational
basis $ {\mathbf{E}}_{2} $ of $ {\mathcal{H}}_{2} $, the
Mathematica QDIAGONAL1[x] is equivalent to DIAGONAL, giving rise
to the same diagonalization argument as to QPARADOX1.

Let us then pass to  QDIAGONAL2[x] whose action is the
following:  if the QHALT program decides that the program x would
halt if presented with itself as input, then DIAGONAL(x) loops
forever; otherwise it gives as output $ | True > $ and then halts.

Let us then look at what we expect Mathematica should output
under the input QPARADOX2: it halts outputting $ | True > $ iff
QHALT[QDIAGONAL2,QDIAGONAL2] outputs $ | False > $ so that once
again QDIAGONAL2 halts on itself iff it doesn't halt on itself
reproducing once again the paradox.

As a conclusion, the consideration of definition\ref{def:quantum
paradox expression of first kind} and definition\ref{def:quantum
diagonal expression of second kind} prove the impossibility of
substituting the dots in eq.\ref{eq:Mathematica implementation of
the quantum halting predicate} so that the implemented Mathematica
function behaves as eq.\ref{eq:Mathematica implementation of the
quantum halting predicate} proving that  QHALT  cannot exist too.

Svozil, instead, doesn't arrive to this conclusion, since he
considers only QDIAGONAL3[x] whose action is the following:  if
the QHALT program decides that the program x would halt if
presented with itself as input, then QDIAGONAL3[x] outputs $  |
False > $; otherwise it gives as output $ | True > $ and then
halts.

Since QDIAGONAL3 halts on every input, PARADOX3 simply outputs $
| False > $ so that, indeed, it can't be used to infer by
diagonalization the impossibility of QHALT.

Let us observe, by the way, that:
\begin{align}
  ( \, In[n] & \: := \: QDIAGONAL3[ | True > ] \, ) \; \Rightarrow \; ( \, Out[n] \: = \: | False >  \, ) \\
  ( \, In[n] & \: := \: QDIAGONAL3[ | False > ] \, ) \; \Rightarrow \; ( \,  Out[n] \: = \: | True
  > \, )
\end{align}
If QDIAGONAL3 was such that:
\begin{multline*}
  QDIAGONAL3 [ c_{True} | True > \, + \, c_{False} |
 False > ] \; = \; c_{True} QDIAGONAL3[  | True > ] \, + \\
  QDIAGONAL3[  | False > ] \; \;  \forall c_{True} , c_{False} \in {\mathbb{C}} \, :
  | c_{True} |^{2} \, + \, |  c_{False} |^{2} \, = \, 1
\end{multline*}
its restriction to inputs of the form $  c_{True} | True > \, +
\, c_{False} |  False > $ could indeed be represented by the NOT
gate $ \hat{\sigma}_{x} \; := \; \begin{pmatrix}
  0 & 1 \\
  1 & 0
\end{pmatrix} $.

Anyway this is not the case, since:
\begin{multline}
 ( \, c_{False} \neq 0 \; \Rightarrow \; QDIAGONAL3 [ c_{True} | True > \, + \, c_{False} | False > ] \;
  = \; | True > \, )  \\
   \forall c_{True} , c_{False}
 \in {\mathbb{C}} \, : \, | c_{True} |^{2} \, + \, |  c_{False}
 |^{2} \, = \, 1
\end{multline}

\smallskip

We have seen in remark\ref{rem:time-constrained halting function}
that a time-constrained version HALT[p,x,t] of the halting
program may indeed be easily defined.

This fact involves some subtility owing to the following
\cite{Arslanov-Calude-Chaitin-95}:
\begin{theorem} \label{th:on halting now or never again}
\end{theorem}
ON HALTING NOW OR NEVER AGAIN:
\begin{multline}
  \exists c \in {\mathbb{R}}_{+} \; : \\
    U(\vec{x}) \text{ has not yet halted at time } \Sigma( | \vec{x} | + c  )  \: \Rightarrow \\
     U(\vec{x}) \, = \, \uparrow
\end{multline}

where $ \Sigma(n) $ is the busy-beaver function of
definition\ref{def:busy beaver function}, whose proof lies on the
following \cite{Arslanov-Calude-Chaitin-95}:
\begin{lemma} \label{lem:bound on the halting time}
\end{lemma}

BOUND ON THE HALTING TIME:

\begin{hypothesis}
\end{hypothesis}
\begin{equation*}
  \vec{x} \in \Sigma^{\star} \; : \; U( \vec{x} ) \downarrow
\end{equation*}
\begin{equation*}
  t_{HALTING} \; := \; \min \{ t \in {\mathbb{R}}_{+} \, : \,
  \text{U has already halted on $ \vec{x} $ at time t} \}
\end{equation*}
\begin{thesis}
\end{thesis}
\begin{equation*}
  \exists c \in {\mathbb{R}}_{+} \; : \; I( t_{HALTING} ) \, \leq
  \, | \vec{x} | \, + \, c
\end{equation*}

Theorem\ref{th:on halting now or never again} would indeed seem to
contradict what we said in the remark\ref{rem:time-constrained
halting function} because it would seem to tell us that there
exist indeed a \textbf{dead-line-time function} $ ( p , x) \,
\stackrel{f}{\rightarrow} \, d[p,x] $ such that by a dialog of the
form:
\begin{align}
   In[n] \; := \; & HALT[p,x,d[(p,x)]] \\
  Out[n] \; := \; & False
\end{align}
(where HALT[p,x,T] is the time-constrained halting function of
eq.\ref{eq:time-constrained halting function}) one could infer
that p doesn't halt on x.

The function $ HALT[p,x,f[(p,x)]] $ would then seem to solve the
Halting Problem, since defining:
\begin{equation} \label{eq:first attempt of halting predicate}
  HALT[p_{-} \, , \,x_{-}] \; := \; HALT[p,x,d[(p,x)]]
\end{equation}
HALT[p,x] behaves exactly as the required halting predicate.

The bug in such a reasoning is owed to the following:
\begin{theorem} \label{th:not-recursivity of the busy-beaver function}
\end{theorem}
NOT-RECURSIVITY OF THE BUSY-BEAVER FUNCTION:
\begin{equation}
  \Sigma(n) \; \notin \; REC-MAP({\mathbb{N}}, {\mathbb{N}})
\end{equation}

Theorem\ref{th:not-recursivity of the busy-beaver function}
implies that the \textbf{dead-line-time function} function $ ( p ,
x) \, \stackrel{f}{\rightarrow} \, f[p,x] $ is itself
not-recursive, so that it cannot be implemented by an input of
the form:
\begin{equation} \label{eq:dead-line-time to halt}
  d[ p_{-} \, , \,  x_{-} ] \; := \; \cdots
\end{equation}
for a suitable substitution of the dots at the r.h.s. of
eq.\ref{eq:dead-line-time to halt} as it would be required for the
Mathematica implementation of the halting predicate in
eq.\ref{eq:first attempt of halting predicate} to be complete.

Anyway, at this point, one could object that it would be
sufficient to succeed in implementing an
\textbf{overestimated-dead-line-time-function}:
\begin{equation} \label{eq:overestimated-dead-line-time to halt}
  overestd[ p_{-} \, , \,  x_{-} ] \; := \; \cdots
\end{equation}
such that:
\begin{equation} \label{eq:overestimation of the dead-line-time}
  overestd[p \, , \, x ] \; \geq \; d[ p \, , \, x ] \; \; \forall \text{ Mathematica expression p , x}
\end{equation}
in order that:
\begin{equation} \label{eq:second attempt of halting predicate}
  HALT[p_{-} \, , \, x_{-}] \; := \; HALT[p,x,overestd[(p,x)]]
\end{equation}
behaves in the required way, implementing algorithmically the
halting predicate.

The bug in this reasoning is owed to the following (cfr. the
section8.3 of \cite{Svozil-93}):
\begin{theorem} \label{th:the busy beaver runs quickly than recursive functions}
\end{theorem}
THE BUSY BEAVER RUNS QUICKER THAN RECURSIVE FUNCTIONS:
\begin{equation}
  \forall f \in REC( {\mathbb{N}} \, , \,  {\mathbb{N}} ) \;
  \exists N \in {\mathbb{N}} \; : \; \Sigma(n) \, > \, f(n) \; \;
  \forall n > N
\end{equation}
that implies that there is no way of substituting the dots in
eq.\ref{eq:overestimated-dead-line-time to halt} so that
eq.\ref{eq:overestimation of the dead-line-time} is satisfied.

C. Calude, M.J. Dinneen and  K. Svozil consider, finally, the
situation in which on Mathematica has been implemented a
time-travel-algorithm: $ TimeTravel[t_{1} , t_{2} ] $ that (giving
suitable input to suitable hardware) allows to cause the
instantaneous time-travel:
\begin{equation}
  t = t_{1} \; \rightarrow \; t = t_{2}
\end{equation}
verified inside Mathematica by the dialog:
\begin{multline}
  In[n] \; := \; before \, = AbsoluteTime[ \, ]; \, TimeTravel[t_{1} , t_{2}
  ]; \\
   after \, = \, AbsoluteTime[ \, ]; \, \{ before \, , \, after \}
\end{multline}
\begin{equation}
  Out[n]  \; := \; \{ t_{1} \, , \,  t_{2} \}
\end{equation}

Though we are not interested here in the details of such an
hardware, it may be worth to observe that the existence of $
TimeTravel[t_{1} , t_{2} ] $ is not  incompatible with General
Relativity, if one assume that it is founded only on the
Cauchy-Problem for Einstein's equation, as it is proved by the
well-known G\"{o}del's solution (cfr. the section5.7 of
\cite{Hawking-Ellis-73}):
\begin{multline} \label{eq:Godel space-time}
  spacetime_{G\ddot{o}del} \; := \; ( {\mathbb{R}}^{4} \, , g _{G\ddot{o}del} \, := \, - dt
  \bigotimes dt \, + \, dx \bigotimes dx \, + \\
  \, dz \bigotimes dz \, - \frac{1}{2} \exp ( 2 \sqrt{2} \omega x ) dy \bigotimes
  dy - \, 2 \exp ( \sqrt{2} \omega x ) dt \bigotimes d y )
\end{multline}
for a  pressure-free perfect-fluid's matter-distribution with
energy momentum tensor $ T_{a b} \; = \; \rho u_{a} u_{b} $, where
the matter density $ \rho $ and the normalized  four-velocity
vector $ u_{a} $ are such that:
\begin{align}
  u^{a}_{0} & \; := \; \delta^{a}_{0} \\
  4  & \; \pi \, \rho \; = \; \omega^{2} \; = \; - \Lambda
\end{align}
where $ \Lambda $ is the cosmological constant, has closed
time-like curves. Indeed the censorship of causality violations
is usually added in the Foundations of General Relativity in the
following way (cfr. the cap.8 and the cap.12 of \cite{Wald-84}
and the cap.8 of \cite{Wald-94}):

given a space-time $ ( M \, , \, g_{a b} ) $ time-orientable, i.e.
such that the light-cone in any point may be divided in its
future and past halves in a way varying smoothly with the point:
\begin{definition} \label{def:chronological future of a space-time's point}
\end{definition}
CHRONOLOGICAL FUTURE OF $ p \in M $:
\begin{multline}
  I^{+} (p) \; := \; \{ q \in M \, : \, \exists \text{ a future-directed time-like curve
  } \lambda \, : \\
   [ 0 , 1 ] \rightarrow M : \lambda(0) = p \, ,
  \, \lambda(1)= q \}
\end{multline}
\begin{definition} \label{def:chronological past of a space-time's point}
\end{definition}
CHRONOLOGICAL PAST OF $ p \in M $:
\begin{multline}
  I^{-} (p) \; := \; \{ q \in M \, : \, \exists \text{ a past-directed time-like curve
  } \lambda \, : \\
   [ 0 , 1 ] \rightarrow M : \lambda(0) = p \, ,
  \, \lambda(1)= q \}
\end{multline}
\begin{definition} \label{def:causal future of a space-time's point}
\end{definition}
CAUSAL FUTURE OF $ p \in M $:
\begin{multline}
  J^{+} (p) \; := \; \{ q \in M \, : \, \exists \text{ a future-directed causal curve
  } \lambda \, : \\
   [ 0 , 1 ] \rightarrow M : \lambda(0) = p \, ,
  \, \lambda(1)= q \}
\end{multline}
\begin{definition} \label{def:causal past of a space-time's point}
\end{definition}
CAUSAL PAST OF $ p \in M $:
\begin{multline}
  J^{-} (p) \; := \; \{ q \in M \, : \, \exists \text{ a past-directed causal curve
  } \lambda \, : \\
   [ 0 , 1 ] \rightarrow M : \lambda(0) = p \, ,
  \, \lambda(1)= q \}
\end{multline}
\begin{definition} \label{def:chronological past of a space-time's region}
\end{definition}
CHRONOLOGICAL PAST OF $ S \subset M $:
\begin{equation}
  I^{-} (S) \; := \; \bigcup_{p \in S}  I^{-} (p)
\end{equation}
\begin{definition} \label{def:chronological future of a space-time's region}
\end{definition}
CHRONOLOGICAL FUTURE OF $ S \subset M $:
\begin{equation}
  I^{+} (S) \; := \; \bigcup_{p \in S}  I^{+} (p)
\end{equation}
\begin{definition} \label{def:causal past of a space-time's region}
\end{definition}
CAUSAL PAST OF $ S \subset M $:
\begin{equation}
  J^{-} (S) \; := \; \bigcup_{p \in S}  J^{-} (p)
\end{equation}
\begin{definition} \label{def:causal future of a space-time's region}
\end{definition}
CAUSAL FUTURE OF $ S \subset M $:
\begin{equation}
  J^{+} (S) \; := \; \bigcup_{p \in S}  J^{+} (p)
\end{equation}
\begin{definition} \label{def:achronal space-time's region}
\end{definition}
 $ S \subset M $ IS ACHRONAL:
\begin{equation}
   I^{+} (S) \, \bigcap \, S \; = \; \emptyset
\end{equation}
Given an achronal, closed set $ S \, \subset \, M $:
\begin{definition} \label{def:future domain of dependence of a space-time's region}
\end{definition}
\begin{multline}
  D^{+} ( S) \; := \; \{ p \in M \, : \, \gamma \text{ inextendible future causal curve through p
  } \; \Rightarrow \; Im(\gamma ) \bigcap S \neq \emptyset \}
\end{multline}
\begin{definition} \label{def:past domain of dependence of a space-time's region}
\end{definition}
\begin{multline}
  D^{-} ( S) \; := \; \{ p \in M \, : \, \gamma \text{ inextendible past causal curve through p
  } \; \Rightarrow \; Im(\gamma ) \bigcap S \neq \emptyset \}
\end{multline}
\begin{definition} \label{def:domain of dependence of a space-time's region}
\end{definition}
DOMAIN OF DEPENDENCE OF $ S \subset M $:
\begin{equation}
  D(S) \; := \; D^{+} ( S) \, \bigcap \,  D^{-} ( S)
\end{equation}
\begin{definition} \label{def:Cauchy surface}
\end{definition}
$ S \subset M $ IS A CAUCHY SURFACE:
\begin{equation}
  D(S) \; =  \; M
\end{equation}
We will say that:
\begin{definition} \label{def:globally hyperbolic space-time}
\end{definition}
$ ( \, M \, , \, g_{a b} \, ) $ IS GLOBALLY-HYPERBOLIC:
\begin{equation}
  \exists S \, \subset \, M \;  Cauchy-surface
\end{equation}
The impossibility of $ TimeTravel[ t_{1} \, , \, t_{2} ] $ may be
derived only assuming the following suppletive Roger Penrose's:
\begin{axiom} \label{ax:axiom of strong cosmic censorship}
\end{axiom}
AXIOM OF  STRONG COSMIC CENSORSHIP:
\begin{equation}
  ( M , g_{a b } ) \text{ physical space-time } \; \Rightarrow \; ( M , g_{a b }
  ) \text{ globally hyperbolic }
\end{equation}

If axiom\ref{ax:axiom of strong cosmic censorship} is required to
guarantee the correctness of the Cauchy's problem for Einstein's
equation is not yet clear \cite{Christodoulou-00}.

\smallskip

Closing this little parenthesis on the possibility of $
TimeTravel[ t_{1} \, , \, t_{2} ] $ the key observation by Calude,
Dinnen and Svozil is that one could use it, substantially, to
slow the increasing with time of the busy beaver function $
\Sigma(t) $ in order to surpass it recursively.

Let us suppose, for example, that we and our computer on which it
is running our Mathematica session, are enclosed into a big box
free-falling in the G\"{o}del's space-time of eq.\ref{eq:Godel
space-time}.

Threating the whole box as a massive particle of mass m, General
Relativity Theory tells us that its motion is described by the
action:
\begin{equation} \label{eq:action of a free-falling massive particle}
  S [ q( \tau ) ] \; := \; - m \int_{\tau_{1}}^{\tau_{2}} d \tau \sqrt{ g^{G\ddot{o}del}_{a b} \frac{ d q^{a}( \tau ) }{ d \tau } \frac{ d q^{b}( \tau ) }{ d \tau } }
\end{equation}
whose invariance under reparametrization of the paths leads to the
existence of the primary first-class constraint (the mass-shell
condition) (cfr. \cite{Henneaux-Teitelboim-92} for a general
presentation of the Theory of Dirac Constraints and
\cite{Arnold-Givental-01},\cite{McDuff-Salamon-98} for its
mathematical recasting in the framework of Symplectic Geometry):
\begin{equation}
  H \; = \; p_{a} g _{G\ddot{o}del}^{ a b } p_{b} \, + \, m^{2} \;
  \approx_{Dirac} \; 0
\end{equation}
identifying a coisotropic submanifold $ S \, \subset \,
{\mathbb{R}}^{8} $, called the \textbf{constraint surface}, of
the phase space $ {\mathbb{R}}^{8} $.

The space of all the \textbf{global-observables} of the box is $
C^{\infty} ( S_{red} ) $, where  $  S_{red} $ is the reduced
phase-space defined as the quotient of S w.r.t. the gauge orbits;
no element of it corresponds to time, since, from the own
reparametrization invariance of the action\ref{eq:action of a
free-falling massive particle}, it follows that, as to the
\textbf{exophysical} point of view (cfr. the $ 6^{th} $-chapter
of \cite{Svozil-93}) the box is a system with no time
\cite{Rovelli-88}.

So a gauge-fixing is required; we will choose the gauge $ \tau \,
= \, t $.

Anyway, as to the \textbf{endophysics} of we and the computer
enclosed in the box, the things are different: the internal-clock
of the computer by which Mathematica computes the function
AbsoluteTime[ ] selects the particular coordinate system $ ( \, t
\, , \, x \, , \,  y  \, , \,  z \, ) $ measuring the coordinate
t.

On  receiving the input $ TimeTravel[ t_{1} \, , \, t_{2} ] $
Mathematica tells to the hardware device to act on the whole box
in order of giving to it a suitable initial condition $ q^{a} (
\tau = 0 ) \, , \frac{ d q^{a} }{ d \tau} ( \tau = 0 ) $ function
of $ t_{1} \, , \, t_{2} $  so that the successive free-fall
motion of the box along a closed time-like curve  realizes the
required transition $ t = t_{1} \; \rightarrow \;  t = t_{2} $.

It should be noted, anyway, that the possibility exists that such
a process, for time-delaying such to allow to surpass recursively
the busy-beaver-function, may be not effectively implementable,
though we must confess we don't understand how this could happen.

\begin{remark} \label{rem:Godel against Godel}
\end{remark}
G\"{O}DEL AGAINTS G\"{O}DEL:

The funny think of such a use of the G\"{o}del's spacetime to
overpass the busy beaver function is that is a sort of fight
G\"{o}del against G\"{o}del, i.e.  G\"{o}del's solution of
Einstein's Equation against G\"{o}del's First Undecidability
Theorem.

Demanding to \cite{Odifreddi-89} and \cite{Davis-65} for
details,  G\"{o}del's First Undecidability Theorem may be
informally stated in the following way:
\begin{center}
  \textit{Every formal system which is sufficiently rich (i.e. contains Peano's Arithmetics), consistent(i.e. no false result
  may be proved by it) and recursively axiomatizable is  not only \textbf{incomplete}, i.e. there exist well-formed formulas in it that are nor
  provable neither refutable, but also \textbf{undecidable}, i.e. the set of its theorems is not recursive}
\end{center}
Such a theorem may be inferred by the Unsolvability of the Halting
Problem, namely by  theorem\ref{th:unsolvability of the Halting
Problem}, by simply observing that the decidability of the
involved classical formal system would allow to prove or disprove
any formula of the form $  << i \in {\mathcal{W}}_{j} >> $.

\smallskip

Calude, Dinnen and Svozil (cfr. \cite{Calude-Dinneen-Svozil-00}
as well as the final section of \cite{Calude-Paun-01})  have
recently analyzed the quantum analogue of this issue.

Their arguments is centered around the following:

\begin{conjecture} \label{con:on the dynamical evolution of the halting qubit}
\end{conjecture}
ON THE DYNAMICAL EVOLUTION OF THE HALTING QUBIT:

\begin{hypothesis}
\end{hypothesis}

 Q quantum computer with
unitary dynamics, whose \textbf{halting state} is specified by a
halting qubit:
\begin{multline} \label{eq:halting qubit at time t}
   | Halt > (t) \; := \; a_{True} (t) | True > \, + \,  a_{False}
  (t) | False >  \: \in \: {\mathcal{H}}_{2}  \\
  a_{True} (t) \, , \, a_{False} (t) \in
  {\mathbb{C}} :  | a_{True} (t)  |^{2} \, + \, | a_{False} (t)
  |^{2}= 1
\end{multline}
such that:
\begin{align*}
  Prob & [ \text{Q has halted at time t} ] \; = \; | a_{True} (t) |^{2} \\
  Prob & [ \text{Q has not halted yet at time t} ] \; = \; | a_{False} (t) |^{2}
\end{align*}
with the initial condition at $ t \, = \, 0 $:
\begin{equation*} \label{eq:initial condition for the halting qubit}
  | a_{True} (0) |^{2} \; = \; | a_{False} (0) |^{2} \; = \; \frac{1}{2}
\end{equation*}
\begin{thesis}
\end{thesis}
In the worst case $ | a_{True} (t) |^{2} $ decades temporally as:
\begin{equation*}
    \exists c \in {\mathbb{R}}_{+}  \; : \;  | a_{True} (t) |^{2} \, \propto \, \frac{1}{ \Sigma ( I(t) + c )}
\end{equation*}

that they claim to follow from lemma\ref{lem:bound on the halting
time} and the unitarity of Q's dynamics.

Given for granted the conjecture\ref{con:on the dynamical
evolution of the halting qubit} let us observe that:
\begin{enumerate}
  \item  Classical Computation may be seen as the particular case of Quantum
Computation in which no superposition of vectors of the
computational basis $ {\mathbf{E}}_{\star} $ occur.
  \item as to the \textbf{computability-issue} , \textbf{deterministic computation}
  and \textbf{classically-probabilistic  computation} are
  equivalent (cfr. e.g. the $ 6^{th} $ chapter of
  \cite{Balcazar-Diaz-Gabarro-95} ) \footnote{It is important, with this respect, not to make confusion between \textbf{classically-nondeterministic computability} and \textbf{classically probabilistic computability}: in terms of acceptance of a given input \textbf{classically-nondeterministic computability} requires that
  there exist at least one computational path leading to an accepting state, while \textbf{classically probabilistic computability} averages on all the computational paths requiring
  that the input is accepted with  probability greater than one half.}
  \item the monitoring of the halting-qubit gives a
  probabilistic-algorithm for solving the Halting Problem
\end{enumerate}

Also assuming that the informal arguments supporting
conjecture\ref{con:on the dynamical evolution of the halting
qubit} may be rigorously formalized, what constitues, according
to our modest opinion, the weaker point in the Calude - Dinnen -
Svozil's attach to  what Calude calls the \emph{Turing's barrier}
lies in a not enough sharp specification of their \textbf{halting
protocol}.

Considered the Hilbert space $ {\mathcal{H}}_{2}^{ \bigotimes
\star} \, \bigotimes \, {\mathcal{H}}_{2}^{Halting} $, to be
affected by the halting status of Q, the halting-qubit system H
must be someway coupled to the computer  so that, assuming the
compound system (computer + halting flag) to be closed, its
unitary dynamics starting from a disentangled state of the form $
\frac{1}{\sqrt{2}} ( | True > \, + \, | False > ) \, \bigotimes
\, | \vec{x} > $ will entangle them, so that:
\begin{enumerate}
  \item the computer will evolve as an open system (i.e. trough a suitable CPU-map of $ {\mathcal{B}} ( {\mathcal{H}}_{2}^{ \bigotimes \star})
  $) contradicting the assumed unitarity of Q
  \item the halting flag will evolve as an open system (i.e. through a suitable CPU-map of $ {\mathcal{B}} (
  {\mathcal{H}}_{2}^{Halting} )$ and hence, in particular, the
  halting state won't continue to be pure contradicting eq.\ref{eq:halting qubit at time t}
\end{enumerate}
The problem is similar to that concerning the consistence of
Deutsch's halting protocol for Quantum Turing Machines that what
questioned by Myers \cite{Myers-97} with the argument that, ought
to the entanglement between the \textbf{halting qubit} and the
\textbf{tape's and internal state's sector}, would cause the
periodic monitoring of the halting qubit to spoil the computation.

Unfortunately, part of the literature  generated by Myers'
objection \cite{Ozawa-98c}, \cite{Linden-Popescu-98},
\cite{Kieu-Danos-98}, \cite{Ozawa-98a} called the issue with the
misleading name of \emph{\textbf{"the Quantum Halting Problem of
Quantum Turing Machines"}}, an erroneous denomination, since the
the true \textbf{Quantum Halting Problem} is the problem of
finding a quantum algorithm that receiving as input an other
quantum algorithm, tell us if it halts or not.

The correct name of the discussed problem is \textbf{\emph{"the
consistence of the Halting Protocol of Quantum Turing Machines}}.

In the same way, the problem concerning our objection to the
Calude-Dinnen-Svozil's argument for solving the \textbf{Classical
Halting Problem} through a quantum computer may be called:
\textbf{\emph{"the consistence of the Calude-Dinnen-Svozil's
Halting Protocol"}}.

The difference between such a problem and the problem of the
consistence of the Halting Protocol of Quantum Turing Machines,
is that in the latter case a periodic monitoring of the quantum
qubit is involved, while in the former case, in any repetition of
the computation, it occurs only one measurement of the halting
qubit.

For this reason the problem of the consistence of the
Calude-Dinnen-Svozil's Halting Protocol cannot be be simply
resolved in terms of quantum nondemolition (QND) measurements
(for which cfr. \cite{Braginsky-Vorontsov-Thorne-83} and the $
12^{th} $ chapter of \cite{Peres-95}) as was made by Ozawa in
\cite{Ozawa-98c}.

A possible solution  could consist, mimicking the alternative
halting protocol for Quantum Turing Machines proposed by Ozawa in
\cite{Ozawa-98a} inglobing the halting qubit in the
quantum-state-register's Hilbert space, to identify the
halting-qubit's Hilbert space $ {\mathcal{H}}_{2}^{Halting} $
with the one qubit-sector of $ {\mathcal{H}}_{2}^{\star} $ by
posing:
\begin{align}
  | & True > \; := \; | 1 > \\
  | & False > \; := \; | 0 >
\end{align}
Such a strategy requires, first of all, that we replace the
initial condition of eq.\ref{eq:initial condition for the halting
qubit} for the halting-qubit with:
\begin{align}
  | &  a_{True} (0) | ^{2} \; = \; 0 \\
  | &  a_{False} (0) | ^{2} \; = \; 1
\end{align}
Whether the double rule that one qubit sector of $
{\mathcal{H}}_{2}^{\star} $ would assume can be in some way
managed consistentely is not, anyway, clear.

\smallskip

\begin{remark}
\end{remark}
DIRAC QUANTIZATION OF THE BOX IN G\"{O}DEL'S SPACETIME

As a matter of curiosity it may be finally interesting to
consider the  quantum analogue of the  gedanken experiment in
G\"{o}del spacetime previously introduced.

We can think to describe quantistically the massive box containing
we and the computer, exploiting the Dirac quantization of the
action of  eq.\ref{eq:action of a free-falling massive particle}
(cfr. the $ 13^{th} $ chapter of \cite{Henneaux-Teitelboim-92}
and \cite{Kirillov-01} for its mathematical recasting in the
framework of Symplectic Geometry, i.e. in terms of Geometric
Quantization):

introduced the canonical operators:
\begin{align}
  ( & \hat{x}^{\mu} \psi ) (x)   \; := \; x^{\mu} \psi (x)   \\
  ( & \hat{p}_{\mu} \psi ) (x) \; := \; - \, i \frac{\partial}{ \partial x^{\mu}
  } \psi (x)
\end{align}
on the Hilbert space $ {\mathcal{H}} \; := \; L^{2} (
{\mathbb{R}}^{8} , d \vec{x} ) $, the subspace of physical states
$ {\mathcal{H}}_{physical} \, \subset \, {\mathcal{H}} $ is given
by:
\begin{equation}
   {\mathcal{H}}_{physical} \; := \; Ker ( -
   \triangle_{g_{G\ddot{o}del}} \, + \, m^{2} )
\end{equation}

where $ \Delta_{g} $ denotes the Laplace-Beltrami operator of the
(pseudo)riemannian metric g.

The scalar product in $ {\mathcal{H}}_{physical} $ depends on the
choice of a gauge and is defined in terms of a suitable
self-adjoint operator $ \hat{O} $:
\begin{equation}
  < \psi_{1} | \psi_{2} >_{physical} \; := \; < \psi_{1} | \hat{O} |
  \psi_{2}> \; \;  | \psi_{1} > \, , \,  | \psi_{2} > \in {\mathcal{H}}
\end{equation}
having the effect of restricting the integral to the physical
degrees of freedom.

With the gauge-fixing  $ \tau \, = \, t $, what the classical
analysis of overpassing the busy beaver function becomes under
quantization of the box ?

Clearly such a treatment should be considered as the one-particle
approximation of the quantum field theory of a Klein-Gordon field
on G\"{o}del's space time whose consistence requires the involved
energies are less than the mass-gap of the theory
\cite{Deligne-Etingof-Freed-Jeffrey-Kazhdan-Morgan-Morrison-Witten-99a}.

It should be mentioned that while our previous formalization of
Calude-Dinnen-Svozil's classical considerations  is accademical
since the assumpution of axiom\ref{ax:axiom of strong cosmic
censorship} or weaker forms of it is rather compelling from a
physical point of view, the possibility of time-machines when
Quantum Mechanics is taken into account is a strongly debated
issue about which we have no competence and about which we demand
to \cite{Nahin-01}.

\smallskip

Considered again  the quantum computer Q of
definition\ref{def:Svozil's quantum algorithmic information}
Svozil have introduced the following quantities:
\begin{definition} \label{def:universal quantum algorithmic probability}
\end{definition}
UNIVERSAL QUANTUM ALGORITHMIC PROBABILITY OF $ | \psi
> \in {\mathcal{H}}_{2}^{\bigotimes \star} $:
\begin{equation}
  P_{Q} (   | \psi
> ) \; := \;  \sum_{\vec{x} \in \Sigma^{\star}} 2^{ - \frac{ | \vec{x} |
  }{2}}  | < True | Q( \vec{x} ) | ^{2}
\end{equation}
\begin{definition} \label{def:universal quantum halting amplitude}
\end{definition}
UNIVERSAL QUANTUM HALTING AMPLITUDE:
\begin{equation}
  \Omega_{Q} \; := \;  \sum_{\vec{x} \in \Sigma^{\star}} 2^{ - \frac{ | \vec{x} |
  }{2}}  | < True | Q( \vec{x} ) | ^{2}
\end{equation}
The next step of Svozil's contribution has been the conjecture
(cfr. the $ 17^{th} $ open problem of the list \cite{Calude-96})
that in Quantum Algorithmic Information Theory it should be
possible to formulate undecidability theorems analogues to the
two classical ones by Chaitin.

Let us then formalize the previously informally introduced notions
of \textbf{classical formal systems} and \textbf{quantum formal
systems} in the following way:
\begin{definition} \label{def:classical formal system}
\end{definition}
CLASSICAL FORMAL SYSTEM:

an r.e.  set:
\begin{equation}
  CFS \; := \; \{ ( \vec{a}_{i} \, , \, \vec{T}_{i}  ) \}_{i \in I}
\end{equation}
where:
\begin{align}
  card & ( I ) \in {\mathbb{N}} \\
   \vec{a}_{i} & \, , \, \vec{T}_{i} \in \Sigma^{\star} \; \; \forall i \in I
\end{align}

\smallskip

Given a classical formal system $ CFS \; := \; \{ ( \vec{a}_{i}
\, , \, \vec{T}_{i} ) \}_{i \in I} $ the meaning of
definition\ref{def:classical formal system} is the following:
every couple $ ( \vec{a}_{i} \, , \, \vec{T}_{i}  ) $ is a
\textbf{rule of inference} of CFS stating that the theorem
(indexed by the string) $ \vec{T}_{i} $  may be deduced from the
\textbf{axiom} (indexed by the string) $ \vec{a}_{i} $.

According to this interpretation we will express the fact that
the couple of strings $ ( \vec{a}_{i} \, , \, \vec{T}_{i}  ) $
belongs to CFS by the notation $ \vec{a} \, \vdash_{CFS} \vec{T}
$.

We are now ready for the following:
\begin{theorem} \label{th:first Chaitin's undecidability theorem}
\end{theorem}
FIRST CHAITIN'S UNDECIDABILITY THEOREM:

\begin{hypothesis}
\end{hypothesis}

CFS classical formal system with a unique axiom $ \vec{a} \in
\Sigma^{\star} $

\begin{thesis}
\end{thesis}

\begin{multline}
  \exists \, c_{CFS}^{(1)} \in {\mathbb{R}}_{+} \: : \\
  [ \, ( a \, \vdash _{CFS} \, I( \vec{x} ) > n ) \: \Rightarrow \: (I( \vec{x} ) > n
  ) \, ] \; \Rightarrow \\
  [ \, ( a \, \vdash _{CFS} \, I( \vec{x} ) > n ) \: \Rightarrow
  \: n \, < \, I( \vec{a} ) + c_{CFS}^{(2)} ]
\end{multline}
\begin{proof}
Consider the following Chaitin computer C:

for $ \vec{u} , \vec{v} \, \in \, \Sigma^{\star} $ such that:
\begin{equation}
  U( \vec{u} \, , \, \lambda ) \: = \: string(k) \; and \; U( \vec{v} \, , \, \lambda
  ) \: = \: \vec{a}
\end{equation}
put:

$ C ( \vec{u}  \cdot \vec{v} , \lambda ) \; := \; $  the first
string $ \vec{s} $ that can be shown    in CFS to have algorithmic
information greater than $ k + | \vec{v} | $.

Among the possible inputs for C we may find the minimal
self-delimiting descriptions for $ string(k) $ and $ \vec{a} $,
i.e.:
\begin{equation}
  u \, = \, (string(k))^{\star} \; ,  \; \vec{v} \, = \, \vec{a}^{\star}
\end{equation}
having algorithmic information $ I(string(k)) \, , \, I(\vec{a})
$ respectively.

If $ C( \vec{u}  \cdot \vec{v} \, , \, \lambda ) \; = \; \vec{s}
$, then:
\begin{equation}
  I_{C} ( \vec{s} ) \; \leq \; |  \vec{u}  \cdot \vec{v} | \; \leq
  \; |   (string(k))^{\star}  | \, |  \vec{a}^{\star} |
\end{equation}
On the other hand, for some constant d:
\begin{equation}
  k + | \vec{a} ^{\star} | \; < \; I( \vec{s} ) \; \leq \; |
  (string(k))^{\star} \,  \vec{a}^{\star} | \, + \, d
\end{equation}
We therefore get the following crucial inequalities:
\begin{equation}
   k + I( \vec{a} ) \; < \; I( \vec{s} ) \; \leq \; I( string(k))
   + I( \vec{a} ) + d
\end{equation}
This implies:
\begin{equation}
  k \; <  \; I( string(k)) + d \; = \; O( \log k )
\end{equation}
which can be true only for finitely many values of the natural k.

Pick now $ c_{F} \, = \, k $, where k is the value that violates
the above inequality. We have proven that $ \vec{s} $ cannot exist
for $ k \, = \, c_{F} $
\end{proof}

\begin{remark}
\end{remark}
FIRST CHAITIN'S UNDECIDABILITY THEOREM AND BERRY'S PARADOX:

Theorem\ref{th:first Chaitin's undecidability theorem} has a deep
conceptual meaning, suggesting that the mathematical phenomenon of
undecidability may have an information-theoretic nature:

in facts it tells us that a classical formal system CFS have an
explicative power ruled by its classical algorithmic information
I(CFS), in that it cannot be used to prove that some object has
classical algorithmic information substantially greater than
itself.

Exactly as the proof of G\"{o}del's First Undecidability Theorem
starts from a self-reference's paradox, i.e the Liar Paradox:
\begin{center}
 \textit{ $ << $ This statement is false $ >> $}
\end{center}
by:
\begin{enumerate}
  \item its modification in a form:
\begin{center}
 \textit{ $ << $ This statement is unprovable $ >> $}
\end{center}
that it is no more paradoxical (since assuming that statement is
indeed true and, hence, unprovable, no contradiction  arises)
  \item its formalization as as statement of Arithmetics by
  G\"{o}del numbering
\end{enumerate}
the proof of theorem\ref{th:first Chaitin's undecidability
theorem} lies on another self-reference's paradox, i.e the
Berry's paradox:
\begin{center}
 \textit{ $ << $ Let x be the least number that cannot be defined in less than 16 words $ >> $}
\end{center}
(whose paradoxical nature lies on the fact that such a statement
determines x by 15 words).

\smallskip

Let us now consider the sequence $ \bar{\Omega} \, := \{
\Omega_{n} \}_{n \in {\mathbb{N}}} $ of the binary digits of the
nonterminating dyadic expansion of the Halting Probability $
\Omega := \Omega_{U} $ w.r.t. to the fixed universal Chaitin
computer, i.e.:
\begin{equation}
   \bar{\Omega} \; := \; ({\mathcal{N}} | _{\Sigma^{\infty} \, -  \,
   \Sigma^{\star}})^{- 1} (\Omega)
\end{equation}
One has that:
\begin{theorem} \label{th:second Chaitin's undecidability theorem}
\end{theorem}
SECOND CHAITIN'S UNDECIDABILITY THEOREM:

\begin{hypothesis}
\end{hypothesis}

$ CFS \; := \; \{ ( \vec{a}_{i} \, , \, \vec{T}_{i} ) \}_{i \in
I} $ classical formal system  whose axiom's set $ A \; := \; \{
\vec{a}_{i} \}_{i \in I} $  is such that:
\begin{equation}
( \, A \, \, \vdash _{CFS} \, << \Omega_{n} \, = \, i >> \;
 \Rightarrow \; \Omega_{n} \, = \, i \, ) \; \; \forall i \in
 \Sigma , \forall n \in {\mathbb{N}}
\end{equation}
and whose theorems' set $ T \; := \; \{ \vec{T}_{i} \}_{i \in I}
$ is r.e.

\begin{thesis}
\end{thesis}
\begin{multline}
   \exists \, c_{CFS}^{(2)} \in {\mathbb{N}}_{+}  : \\
 [ \, ( \, A \, \, \vdash _{CFS} \,  << \cdot_{k=1}^{n} \Omega _{i_{k}} \, = \, \vec{x}
  >> \,)  \; \Rightarrow \; ( \, n \: < \: I(CFS) \, + \, c_{CFS}^{(2)}
  >> \, ) \, ]  \\
   \forall i_{1} , \cdots , i_{n} \in
  {\mathbb{N}}\, , \, \forall \vec{x} \in \Sigma^{n}
\end{multline}
\begin{proof}
If T provides k different cbits of $ \Omega $, then it gives us a
covering $ Cover_{k} $ of measure $ 2^{- k} $ which includes $
\Omega $.

Let us enumerate T until k cbits of $ \Omega $, $ \Omega_{i_{1}}
\, , \,  \cdots \, , \,  \Omega_{i_{k}} \; i_{1} \, < \, i_{2} \,
< \, \cdots \, < \, i_{k}   $ are determined.

Put:
\begin{multline}
   Cover_{k}  \; := \; \{ x_{1} \Omega_{i_{1}} \, , \, x_{2}
   \Omega_{i_{2}} \, , \, \cdots \, , \, x_{k} \Omega_{i_{k}} \, :  x_{1} \, , \, x_{2} \, ,  \, \cdots  \, ,  \, x_{k}  \in
   \Sigma^{\star} \\
   | x_{1} | \,  :=  \, i_{1} - 1 \, , \,   | x_{2} |  \, := \, i_{2} -
   i_{1} \, , \, | x_{k} |  \, := \, i_{k} -
   i_{k-1} \} \; \subset \; \Sigma^{\star}
\end{multline}
By construction $ Cover_{k} $ is a covering; furthermore:
\begin{equation}
  P_{unbiased} ( Cover_{k} \, \Sigma^{\infty} ) \; = \; \frac{2^{i_{k}} -
  k}{2^{i_{k}}} \; = \; 2^{ - k }
\end{equation}
So T yelds infinitely many cbits of $ \Omega $, contradicting
theorem\ref{th:Chaitin randomness of the halting probability}
\end{proof}
\begin{remark}
\end{remark}
USING THE DIGITS OF $ \Omega $ TO DECIDE ALL THE FINITELY
REFUTABLE CONJECTURES:

To appreciate the meaning of theorem\ref{th:second Chaitin's
undecidability theorem} let us observe that the knowledge of $
\vec{\Omega}(n) \, , \, n \in {\mathbb{N}} $ allows to solve the
halting problem for all the programs of length less or equal to
n, as can be proved in the following way:

since:
\begin{equation}
  \Omega \; \in \; ( \, {\mathcal{N}} ( \vec{\Omega}(n)
  0^{\infty} ) \, , \,  {\mathcal{N}} ( \vec{\Omega}(n)
  0^{\infty} ) + 2^{- n}
\end{equation}
one can simply make a sistematic search through all programs the
eventually halts until enough halting programs $ p_{i_{1}} \, , \,
p_{i_{2}} \, \cdots \,  p_{i_{k}} $, of length, respectively, $
l_{i_{1}} \, , \, l_{i_{2}} \, \cdots \,  l_{i_{k}} $, such that:
\begin{equation} \label{eq:programs balancing the first n cbits of the halting probability}
  \sum_{j=1}^{k} 2^{ - \, l_{i_{j}}} \; > \; {\mathcal{N}} ( \vec{\Omega}(n)
  0^{\infty})
\end{equation}
But then observe  eq.\ref{eq:programs balancing the first n cbits
of the halting probability} assure us that the collection of
programs $ p_{i_{1}} \, , \, p_{i_{2}} \, \cdots \,  p_{i_{k}} $
contains all the halting programs of length less or equal to n.

We saw in remark\ref{rem:Godel against Godel} that the Recursive
Unsolvability of the Halting Problem implies G\"{o}del's First
Undecidability Theorem.

With the same argument involved in such a proof one may
immediately infer that the solution of the  n-length Halting
Problem furnished by the knowledge of $ \vec{\Omega} (n) $
implies the decision of all finitely refutable conjectures that
can be expressed by at most n cbits.

\begin{remark}
\end{remark}
ON THE CONSTANTS IN CHAITIN'S UNDECIDABILITY THEOREM:

The recent LISP implementation by Chaitin of his undecidability
theorems \cite{Chaitin-98} shows constructively that two numbers
$ c_{CFS}^{(1)} $ and  $ c_{CFS}^{(2)} $ are concretelly
computable ( he computes them !).

It must be observed, anyway, that $ c_{CFS}^{((1)} $ and  $
c_{CFS}^{(2)} $ depend also on the fixed Chaitin Universal
computer U. Allowing U to vary we will have to denote them by  $
c_{CFS,U}^{(1)} $ and  $ c_{CFS,U}^{(2)} $.

Let us consider, in particular, the classical formal system ZFC
giving foundations to Mathematics:

Robert Solovay has recentely proved that \cite{Solovay-00}:
\begin{theorem} \label{th:second Solovay's theorem}
\end{theorem}
SECOND SOLOVAY'S THEOREM:

\begin{hypothesis}
\end{hypothesis}
ZFC is consistent

\begin{thesis}
\end{thesis}
\begin{equation*}
  \exists \; U \text{ universal Chaitin computer } \; : \;
  c_{CFS,U}^{(2)} \, = \, 0
\end{equation*}

\smallskip

Most of the conceptual deepness of theorem\ref{th:second Chaitin's
undecidability theorem} lies on its link with the tenth of the
celebrated list of 23 unsolved problem David Hilbert presented in
1900 at the Second International Congress of Mathematics in Paris
he considered would have been the germs of the incoming new
century.

Given a polynomial $ P( x \, ,  \, y_{1} , \cdots , y_{m} ) $
with integer coefficients (an  $ {\mathbb{N}} $-polynomial from
here and beyond):
\begin{definition}
\end{definition}
SOLUTIONS' SET OF P:
\begin{equation}
  SOL(P) \; := \; \{ x \in {\mathbb{N}} \, : \, P( x \, ,  \, y_{1} , \cdots , y_{m}
  ) \, = \, 0 \, \text{ for some }  y_{1} , \cdots ,  y_{m} \:
  \in \: {\mathbb{N}} \}
\end{equation}
\begin{definition}
\end{definition}
HILBERT'S TENTH PROBLEM:

to construct an algorithm $ HILBERT10[P] $ that  receiving as
input an $ {\mathbb{N}} $-polynomial P outputs:
\begin{equation}
  HILBERT10[P] \; := \;
  \begin{cases}
    True & \text{if  $ SOL(P) \, \neq \, \emptyset $}, \\
    False & \text{otherwise}.
  \end{cases}
\end{equation}

Considered now a set $ S \; \subseteq \;  {\mathbb{N}}  $:
\begin{definition}
\end{definition}
S IS DIOPHANTINE:
\begin{equation}
  \exists \; P \, {\mathbb{N}}-polynomial \; : \; S \, = \, SOL(P)
\end{equation}

The intellectual path of thirty years by Martin David, Hilary
Putnam and Julia Robinson was completed by Yuri Matiyasevich in
1970 through the proof of the following \cite{Matiyasevich-99}:
\begin{theorem} \label{th:Matijasevic's theorem}
\end{theorem}
MATIJASEVIC'S THEOREM:
\begin{equation}
  S \: Diophantine \; \Leftrightarrow \; S \: r.e.
\end{equation}
from which immediately follows that:
\begin{corollary}
\end{corollary}

the Hilbert's Tenth Problem is recursively-unsolvable.

\smallskip

Let us now  consider exponential ${\mathbb{N}}$-polynomials, i.e.
polynomials built not only by addition and multiplications, but
also by exponentiations.
\begin{definition}
\end{definition}
S IS EXPONENTIAL-DIOPHANTINE:
\begin{equation}
  \exists \; P \, exponential-{\mathbb{N}}-polynomial \; : \; S \, = \, SOL(P)
\end{equation}

Theorem\ref{th:Matijasevic's theorem} immmediately implies that:
\begin{corollary}
\end{corollary}
\begin{equation}
  S \: exponential-Diophantine \; \Leftrightarrow \; S \: r.e.
\end{equation}

It is possible, anyway, to prove a stronger result; introduced
the following:
\begin{definition}
\end{definition}
S IS SINGLE-FOLD EXPONENTIAL DIOPHANTINE:

$ S \; = \; SOL(P) $, where  the
exponential-Diophantine-polynomial $ P( x \, ,  \, y_{1} , \cdots
, y_{m} ) $ is such that:
\begin{equation}
  \forall x \in S \; \exists \, ! ( y_{1} , \cdots
, y_{m} ) \in {\mathbb{N}}^{m} \: : \: P( x \, ,  \, y_{1} ,
\cdots , y_{m} ) \, = \, 0
\end{equation}
one has that:
\begin{theorem} \label{th:Jones-Matijasevic's theorem}
\end{theorem}
JONES-MATIJASEVIC'S THEOREM:
\begin{equation}
  S \: single-fold-exponential-Diophantine \; \Leftrightarrow \; S \: r.e.
\end{equation}
Theorem\ref{th:Jones-Matijasevic's theorem} allows to prove the
following:
\begin{theorem}
\end{theorem}
LINK BETWEEN THE HALTING PROBABILITY AND HILBERT'S TENTH PROBLEM:

$ \exists \;  P( n \, , \, x \, , \,   \, y_{1} , \cdots , y_{m}
) $ exponential-Diophantine-polynomial such that for every $ k
\in {\mathbb{N}} $ the equation:
\begin{equation}
   P( n \, , \, x \, , \,   \, y_{1} , \cdots , y_{m} )
\end{equation}
has an infinite number of solutions iff the $ k^{th} $ bit of $
\Omega $ is 1

\begin{proof}
One has obviously that:
\begin{equation}
   {\mathcal{N}} |_{ \Sigma^{\infty}  - \Sigma^{\star} } \; \in \;
   {\mathbb{Q}} \; \; \forall n \in {\mathbb{N}}
\end{equation}
Furthermore:
\begin{equation}
  << \, ( \vec{\Omega}(k))_{m}  \: = \:  1  \, >>  \; \in REC(
  {\mathbb{N}} ) \; \; \forall n , k \in {\mathbb{N}} \, :  \,  m < n
\end{equation}
where we have followed the convention introduced in
section\ref{sec:The distinction between mathematical-classicality
and physical-classicality} of identifying \textbf{unary
predicates} with the sets of the elements satisfying them.

But then, using theorem\ref{th:Jones-Matijasevic's theorem}, one
gets an equation:
\begin{equation} \label{eq:Chaitin's exponential diophantine equation}
   P( n \, , \, x \, , \,   \, y_{1} , \cdots , y_{m} ) \; = \; 0
\end{equation}

having:
\begin{itemize}
  \item one solution $ y_{1} , \cdots , y_{m}  $ if the $ n^{th} $
cbit of $ \vec{\Omega}(k) $ is 1
  \item no solution $ y_{1} , \cdots , y_{m}  $ if the $ n^{th} $
cbit of $ \vec{\Omega}(k) $ is 0
\end{itemize}
The number of different  m-ples of natural numbers which are
solution of eq.\ref{eq:Chaitin's exponential diophantine
equation} is therefore infinite iff the $ n^{th} $ cbit of the
base-2 expansion of $ \Omega $ is 1.
\end{proof}

\smallskip

The first step forward the derivation of quantum amalogues of
Chaitin's Undecidability Theorems is to develop the theory of
\textbf{Noncommutative Formal Systems} expanding the short
preliminary considerations presented in section\ref{sec:The
unpublished ideas of Sidney Coleman and Andrew Lesniewski}.

Gereralizing the classical case one is led to the following:
\begin{definition} \label{def:noncommutative formal system}
\end{definition}
NONCOMMUTATIVE FORMAL SYSTEM:

a set:
\begin{equation}
  NCFS \; := \; \{ ( \vec{a}_{i} \, , \, \vec{\alpha}_{i}  ) \}_{i \in I}
\end{equation}
where:
\begin{align}
  card & ( I ) \in {\mathbb{N}} \\
   \vec{a}_{i} & \, , \, \vec{T}_{i} \in \Sigma_{NC}^{\star} \; \; \forall i \in I
\end{align}

\smallskip

Given a noncommutative formal system $ NCFS \; := \; \{ (
\vec{a}_{i} \, , \, \vec{T}_{i} ) \}_{i \in I} $ the meaning of
definition\ref{def:noncommutative formal system} is specular to
the commutative case: every couple $ ( \vec{a}_{i} \, , \,
\vec{T}_{i} ) $ is a \textbf{rule of inference} of NCFS stating
that the theorem (indexed by) $ \vec{T}_{i} $ may be deduced from
the \textbf{axiom} (indexed by) $ \vec{a}_{i} $.

Again we will express the fact that the couple of strings $ (
\vec{a}_{i} \, , \, \vec{T}_{i}  ) $ belongs to CFS by the
notation $ \vec{a} \, \vdash_{NCFS} \vec{T} $.

Before trying to derive  undecidability results quantifying the
explicative power of a quantum formal system by its quantum
algorithmic information, let us analyze a bit
definition\ref{def:noncommutative formal system} from the point
of view of Moore's generalization of \textbf{Chomsky's Hierarchy
}and of the \textbf{duality languages versus automata}.

Following the first chapter of \cite{Calude-Paun-01} let us start
from the following:
\begin{definition} \label{def:rewriting system}
\end{definition}
CLASSICAL REWRITING SYSTEM:

a couple of the form $ ( \,  \Sigma \,  , \, P \, ) $ where P is
a finite subset of $ \Sigma^{\star} \, \times \,
  \Sigma^{\star} $ whose elements are called the \textbf{productions}
  of $ ( \,  \Sigma \,  , \, P \, ) $.

\smallskip

Given a classical rewriting system $ \gamma \,  := \, ( \Sigma \,
, \, P ) $ let us introduce the following intuitive notation for
productions:
\begin{equation}
  ( \, \vec{x} \: \rightarrow \: \vec{y} \, ) \; := \;  ( \, \vec{x} \, , \,  \vec{y} \,
  ) \; \in \; P
\end{equation}
Given two words $  \vec{x} \, , \, \vec{y} \; \in \;
\Sigma^{\star} $:
\begin{definition} \label{def:immediate implication in a rewriting system}
\end{definition}
$ \vec{x} $ IMMEDIATELY IMPLIES $ \vec{y} $ IN $ \gamma $ ( $
\vec{x} \, \Rightarrow_{\gamma} \, \vec{y} $ ):

$  \exists \, \vec{x}_{1} \, , \, \vec{x}_{2} \, , \, \vec{u} \, ,
  \, \vec{v} \: \in \: \Sigma^{\star} \, $:
\begin{align}
  \vec{u} & \; \rightarrow \; \vec{v} \\
  \vec{x} & \; = \; \vec{x}_{1} \, \vec{u} \, \vec{x}_{2} \\
  \vec{y} & \; = \; \vec{x}_{1} \, \vec{v} \, \vec{x}_{2} \\
\end{align}
\begin{definition} \label{def:implication's relation of a rewriting system}
\end{definition}
IMPLICATION'S RELATION OF $ \gamma $ ( $ \Rightarrow_{\gamma} $ )

the symmetric and transitive closure of $ \gamma $

\smallskip

\begin{definition} \label{def:pure grammar}
\end{definition}
PURE GRAMMAR:

a therne $ G \, := \; ( \Sigma \, , \, P \, \, , \, \vec{a} ) $
such that:
\begin{itemize}
  \item $ ( \,  \Sigma \,  , \, P \, ) $ is a rewriting system
  \item  $ \vec{a} \in \Sigma^{\star} $ is called the \textbf{axiom} of
  G
\end{itemize}

Given a pure grammar $ G \, := \; ( \Sigma \, , \, P \, \, , \,
\vec{a} ) $:
\begin{definition} \label{def:language generated by a pure grammar}
\end{definition}
LANGUAGE GENERATED BY G:
\begin{equation}
  L(G) \; := \; \{ \vec{x} \in \Sigma^{\star} \; : \; \vec{a} \:
  \Rightarrow_{\gamma}^{\star} \: \vec{x} \}
\end{equation}
\begin{definition} \label{def:Chomsky's grammar}
\end{definition}
CHOMSKY'S GRAMMAR:

a quintuple $  G \, := \, ( N \, , \, T \, , \, S \, , \, P ) $
such that:
\begin{itemize}
  \item N and T are disjoint finite alphabet called, respectively,
  the \textbf{nonterminating alphabet} and the  \textbf{terminating alphabet}
  \item $ S \in N $ is called the \textbf{axiom} of G
  \item P is a finite subset of  $ ( N \bigcup T ) ^{\star}  \cdot
  N \cdot  ( N \bigcup T ) \cdot ( N \bigcup T ) ^{\star} $ called
  the \textbf{productions' set} of G
\end{itemize}
As for pure grammars $ \vec{u} \, \rightarrow \, \vec{v}$ will
stand for $ ( \vec{u} \, , \, \vec{v}) \in P $.

Given a Chomsky grammar  $ G \, := \, ( N \, , \, T \, , \, S \, ,
\, P ) $ and two words $ \vec{x} \, , \, \vec{y} \, \in \, ( N
\bigcup T ) ^{\star}  $:
\begin{definition} \label{def:immediate implication in a Chomsky's grammar}
\end{definition}
$ \vec{x} $ IMMEDIATELY IMPLIES $ \vec{y} $ IN G ( $ \vec{x} \,
\Rightarrow_{G} \, \vec{y} $ ):

$  \exists \, \vec{x}_{1} \, , \, \vec{x}_{2} \in \, ( N \bigcup
T ) ^{\star} \, , \, \vec{u} \, \rightarrow \, \vec{v} \; \in \;
P $ such that:
\begin{align}
  \vec{u} & \; \Rightarrow \; \vec{v} \\
  \vec{x} & \; = \; \vec{x}_{1} \, \vec{u} \, \vec{x}_{2} \\
  \vec{y} & \; = \; \vec{x}_{1} \, \vec{v} \, \vec{x}_{2} \\
\end{align}

The implication $ \Rightarrow_{G}^{\star} $ is then defined as
the symmetric and transitive closure of $ \Rightarrow_{G} $
exactly as for pure grammars.
\begin{definition} \label{def:sentential form in a Chomsky grammar}
\end{definition}
$ \vec{x} \, \in \, ( N \bigcup T^{\star}) $ IS A SENTENTIAL FORM
IN G:
\begin{equation}
  S \;  \Rightarrow_{G}^{\star} \; \vec{x}
\end{equation}
\begin{definition} \label{def:language generated by a Chomsky grammar}
\end{definition}
LANGUAGE GENERATED BY G:
\begin{equation}
  L(G) \; := \; \{ \vec{x} \in T^{\star} \; : \; S \:
  \Rightarrow_{G}^{\star} \: \vec{x} \}
\end{equation}

\smallskip

\begin{remark}
\end{remark}
LANGUAGES GENERATED BY PURE GRAMMARS AND CHOMSKY GRAMMARS

Comparing definition\ref{def:language generated by a Chomsky
grammar} with definition\ref{def:language generated by a pure
grammar} it is essential to observe that the language generated
by a Chomsky grammars contains only sentential forms that are
strings over the terminating alphabet.

\smallskip

\begin{definition}
\end{definition}
G   IS CONTEXT-SENSITIVE:
\begin{equation}
  | \vec{u} | \; \leq \;  | \vec{v} | \; \; \forall \, \vec{u}
  \rightarrow \vec{v} \, \in \, P
\end{equation}
\begin{definition}
\end{definition}
G   IS CONTEXT-FREE:
\begin{equation}
  \vec{u} \; \in \; N \; \; \forall  \vec{u}
  \rightarrow \vec{v} \, \in \, P
\end{equation}
\begin{definition}
\end{definition}
G  IS  LINEAR:
\begin{equation}
  \vec{u} \: \in \: N \; and \; \vec{v} \: \in \: T^{\star} \bigcup T^{\star}  \cdot N \cdot T^{\star} \; \; \forall  \vec{u}
  \rightarrow \vec{v} \, \in \, P
\end{equation}
\begin{definition}
\end{definition}
G  IS  REGULAR:
\begin{equation}
  \vec{u} \: \in \: N \; and \; \vec{v} \: \in \: T \bigcup T \cdot N \cdot \{ \lambda \} \; \; \forall  \vec{u}
  \rightarrow \vec{v} \, \in \, P
\end{equation}
\begin{definition}
\end{definition}
FINITE LANGUAGES:
\begin{equation}
   FIN \; := \; \{ L \subset A^{\star} \, : \, \max ( card(A) ,
   card(L) ) < \infty \}
\end{equation}
\begin{definition}
\end{definition}
REGULAR LANGUAGES:
\begin{equation}
  REG \; := \; \{ L(G) \, : \, G \text{ regular Chomsky grammar} \}
\end{equation}
\begin{definition}
\end{definition}
LINEAR LANGUAGES:
\begin{equation}
  LIN  \; := \; \{ L(G) \, : \, G \text{ linear Chomsky grammar} \}
\end{equation}
\begin{definition}
\end{definition}
CONTEXT-FREE LANGUAGES:
\begin{equation}
  CF  \; := \; \{ L(G) \, : \, G \text{ context-free Chomsky grammar} \}
\end{equation}
\begin{definition}
\end{definition}
RECURSIVELY ENUMERABLE LANGUAGES:
\begin{equation}
  RE  \; := \; \{ L(G) \, : \, G \text{  Chomsky grammar} \}
\end{equation}
One has that:
\begin{theorem} \label{th:Chomsky's hierarchy}
\end{theorem}
CHOMSKY'S HIERARCHY:
\begin{equation}
  FIN \; \subset \; REG \; \subset \; LIN \; \subset \; CF \; \subset
  \; CS \; \subset \; RE
\end{equation}

There exist a natural duality among languages and computing
devices; Given a language L and a computing device D:

\begin{definition} \label{def:device accepting a language}
\end{definition}
D ACCEPTS L:

the final internal state of D under the input x belongs or not a
subset of accepting states of D's set of internal states
according to whether the word x belong or not to the language L

\smallskip

Such a duality between languages and automata induces a hierarchy
of different kinds of computing devices able to accept the
various kinds of languages in Chomsky's hierarchy:

\smallskip

\begin{tabular}{|c|c|}
  LANGUAGES & AUTOMATA \\  \hline
  regular & finite \\
  linear & one-turn pushdown \\
  context-free & pushdown \\
  context-sensitive & linear-bounded \\
  recursively enumerable & Turing \\  \hline
\end{tabular}

\smallskip

\begin{remark}
\end{remark}
CHURCH'S THESIS AND CHOMSKY'S HIERARCHY

Reasoning in an opposite way from the usual one, i.e. looking at
the notion of \textbf{recursivity} as a derived  notion induced
from the primary notion of \textbf{recursive enumerability}
defining a recursive function as a function having an r.e. graph,
one can look at Church's thesis as the statement that Chomsky's
hierarchy stops with recusivively enumerable languages, i.e. that
there doesn't exist a grammar more powerful than Chosmky's
grammar as to effective generation of languages is concerned.

The duality between languages and automata identifies Turing
Machines,as the more powerful computational device for accepting
languages.

\smallskip

While we will define Turing machines in section\ref{sec:The
problem of characterizing mathematically the notion of a quantum
algorithm}, we won't introduce the other less powerful
computational devices for which we demand to
\cite{Davis-Sigal-Weyuker-94}, \cite{Lewis-Papadimitriou-98} (and
to the Mathematica packages developed by Jaime Rangel
Mondrag\'{o}n \cite{Mondragon-99} for their concrete
implementation on computer).

What it is relevant to our purposes is to observe that under
noncommutative generalization both the duality languages-automata
and the existence of Chomsky's hierarchy in the degree in
generating power of grammar preserve, as it has been shown by
Christopher Moore \cite{Moore-Crutchfield-97}.

Before of trying to use theorem\ref{def:noncommutative formal
system} to derive quantum analogues of theorem\ref{th:first
Chaitin's undecidability theorem} and theorem\ref{th:second
Chaitin's undecidability theorem}, one should check whether the
naife definition\ref{def:noncommutative formal system} is correct
from the point of view of the quantum Chomsky' hierarchy, i.e. if
it raelly characterizes those qwuantum languages that are
generated by quantum Chomsky grammars.

This is a \textmd{conditio sine qua non}, since only in this case
the definition\ref{def:noncommutative formal system} is the
mathematical-logic counterpart, via the
languages-automata-duality, of the computational device
formalizing Quantum Computation, i.e. the Quantum Turing Machine
we will introduce in section\ref{sec:The problem of
characterizing mathematically the notion of a quantum algorithm}

\newpage
\section{Yuri Manin's suggestion at the June 1999 Bourbaki seminar}
\section{Paul Vitanyi's Quantum Kolmogorov complexity}
\section{The objection by Berthiaume van Dam and Laplante to Vitanyi}
\section{Peter Gacs' quantum algorithmic entropy}
\section{The algorithmic approach to Quantum Chaos Theory: quantum algorithmic information versus quantum dynamical
entropies} \label{sec:The algorithmic approach to Quantum Chaos Theory: quantum algorithmic information versus quantum dynamical entropies}
\chapter{Typical properties in Quantum Probability Theory}
\section{Conformism in Quantum Probability Theory} \label{sec:Conformism in Quantum Probability Theory}
The approaches to Quantum Algorithmic Information Theory by Karl
Svozil, by Paul Vitanyi, by  Berthiaume van Dam and Laplante and
by Paul Gacs have a thing in common: they deal with
\textbf{strings of qubits}.

Starting from strings of qubit is seen, indeed, as simpler, the
case of sequences of qubits being seen as a derived, more
complicate issue to be derived in a second time by the case of
strings.

As we stressed in the remark\ref{rem:impossibility of a sharp
distinction between regularity and randomness for strings},
anyway, a sharp distinction between regularity and randomness in
the classical case is impossible for strings but only for
sequences.

This is essentially owed to the fact that
theorem\ref{th:foundation of the applicability of probability
theory to reality}, whose importance we underlined in the remark
\ref{rem:on why Classical Probability Theory applies to reality},
holds only for sequences.

So, despite of appearances, as to algorithmic randomness the
analysis of \textbf{sequences} is greatly conceptually clearer and
simpler than that of \textbf{strings}.

This implies that the  same happens as to quantum algorithmic
randomness where a quantum  analogue of theorem\ref{strong law of
excluded gambling strategies} can hold only in the case of
sequences.

As a conclusion, as to the characterization of \textbf{quantum
algorithmic randomness}, one has to start from \textbf{sequences
of qubits} and not from \textbf{strings of qubits}.

This was indeed the attitude of the approach by Coleman and
Lesniewski, who, anyway, failed in individuating the correct
space of qubits' sequences, with the consequences we reported in
the remark\ref{rem:Coleman and the Entscheidungsproblem}.

Our objective here consists in trying of characterizing the
correct notion of an \textbf{algorithmic random sequence of
qubits} by formulating a suitable quantum analogue of the
approach pursued in chapter\ref{chap:Classical algorithmic
randomness as satisfaction of all the classical algorithmic
typical properties}.

\smallskip

Given a noncommutative collectivity  $ S_{NC} $ made of $ N :=
n^{2} \in {\mathbb{N}} $ non-commutative people:
\begin{equation}
  cardinality_{NC} ( S_{NC} ) \; = \; n^{2}
\end{equation}
(i.e.  $  S_{NC} \; = \; M_{n} ({\mathbb{C}}) $) our objective is
to define a \textbf{typical property} of $ S_{NC} $.

We will have to face a certain number of issues conceptually
based on the double nature of a Von Neumann algebra:
\begin{itemize}
  \item as a commutative set
  \item as a noncommutative set
\end{itemize}
as we stressed in the remark\ref{rem:the metaphore by which we
can speak about noncommutative sets from within ZFC}.

The first issue we have to deal with is the following:
\begin{center}
  \textbf{FIRST ISSUE: Have we to consider ordinary, commutative, properties or noncommutative properties?}
\end{center}

\smallskip

We shall try to answer the following question following both the
alternatives and comparing them:

\begin{enumerate}
  \item Properties have to be considered as classical, i.e.
  ordinary, predicates over $ S_{NC} $.

Thus an other subproblem arises:
\begin{center}
   \textbf{SECOND ISSUE: How have one to count the number of the elements of $ S_{NC} $ having a given property ?}
\end{center}

Mimicking the classical case one would be tempted to consider the
quantity:
\begin{equation*}
  cardinality( \{ x \in S_{NC} \, : \, p(x) \; holds \}
\end{equation*}
But this doesn't seem to be such a good idea, as can be seen
considering, for example, the predicate:
\begin{equation} \label{eq:unitarity predicate}
  p_{unitarity} (x) \; := \; << x^{\star}  x \, = \, x x^{\star} \, = \, I  >>
\end{equation}
and observing that obviously:
\begin{multline}
   cardinality( \{ x \in S_{NC} \, : \, p_{unitarity}(x) \; holds \} )=
   \aleph_{1}^{ \frac{n^{2}}{2}} \\
   = \;  \aleph_{1} \; = \; \aleph_{1}^{n^{2}} \; = \; cardinality(S_{NC})
\end{multline}
So one infers that to count the number of elements having a given
property one has to use noncommutative cardinality.

This, anyway, requires the restriction to properties such that the
subset of all the elements of $ S_{NC} $ having such a property
is a sub-factor of $ S_{NC} $.

\begin{example} \label{ex:matrices with constrained first entry}
\end{example}
MATRICES WITH CONSTRAINED FIRST ENTRY:

Let us consider the following family of predicates:
\begin{equation}
  p_{\text{first entry not y}} (x) \; := \; << x_{1,1} \: \neq
  \: y >> \; \; y \in {\mathbb{C}}
\end{equation}
We would be tempted to say that the majority of people in $
S_{NC} $ have this property so that it must be considered a
\textbf{majoritary property} of $ S_{NC} $.

But if votes are counted by noncommutative cardinality, this may
be true only if $ y = 0 $,  since the set $ \{ x \in S_{NC} \, :
\, p_{\text{first entry not y}}(x) \; holds \} $ is not a
subfactor of $ S_{NC} $ for every $ y \neq 0 $ and
conseguentially:
\begin{equation}
  cardinality_{NC} ( \{ x \in S_{NC} \, : \,
p_{\text{first entry not y}}(x) \; holds \}) \; = \; \uparrow \;
\; \forall y \neq 0
\end{equation}

\smallskip

Let us consider, for example, the unitarity predicate of
eq.\ref{eq:unitarity predicate} for which we have that:
\begin{equation}
  cardinality_{NC}( \{ x \in S_{NC} \, : \, p_{unitarity}(x) \; holds \}
  \; = \; \frac{N}{2}
\end{equation}
Thus we conclude that $ p_{unitarity} $ is neither a
\textbf{majoritary property} nor a \textbf{minoritary property}.

Let us consider, instead, the following predicate:
\begin{equation}
  p_{\text{traceless skew-adjoint}} (x) \; := \; << x^{\star} \, =
   \, - x \: and \: \tau_{unbaised} (x) = 0 >>
\end{equation}
Since:
\begin{equation}
  cardinality_{NC}( \{ x \in S_{NC} \, : \, p_{\text{traceless skew-adjoint}}(x) \; holds
  \} ) \; = \; \frac{ N - 1 }{2}
\end{equation}
we conclude that $  p_{\text{traceless skew-adjoint}} $ is a
\textbf{minoritary property} of $ S_{NC} $, so that its negation
$ p_{\text{traceless skew-adjoint}}^{ \bot } $ is a
\textbf{majoritary property}.

As in the commutative case a  typical properties is then  defined
as a property $ p ( \cdot ) $ such that:
\begin{equation}
  cardinality_{NC}( \{ x \in S_{NC} \, : \, p_(x) \; holds \})
  \; \gg \; \frac{N}{2}
\end{equation}
This is the case, for example, of the property $p_{\text{first
entry not 0}} $ we introduced in example\ref{ex:matrices with
constrained first entry} provided $ N \gg 1 $, where, as in the
commutative case, the informality of such a notion derives from
the informal nature of the \textbf{very greater than} ordering
relation.

\smallskip

Let us now consider an infinite countable noncommutative
community $ cardinality_{NC} ( S_{NC} ) \; = \; \aleph_{0} $. As
in the commutative case the same notion of a \textbf{majoritary
property} loses its meaning.

Exactly as in the  classical case, anyway, one has that for an
infinite uncountable noncommutative community the notion of a
typical property may be defined provided  $ S_{NC} $ admits an
unbiased noncommutative probability distribution $
\tau_{unbaised} $ , i.e. provided $ Type( S_{NC} ) \; = \; II_{1}
$: in this case a \textbf{typical property} is defined as a
property holding $ \tau_{unbaised}$-almost everywhere

\item  Properties have to be considered as noncommutative
propositions over $ S_{NC} $, i.e. as elements of the quantum
  logic  $ QL(S_{NC}) $ of $ S_{NC} $.

 The set of typical properties of a noncommutative probability space of the form $ ( S_{NC} \, , \, \omega ) $ will then be subset
 of $ {\mathcal{P}}( S_{NC} ) $ and, clearly, won't satisfy the distributive law as to conjunction and disjunction of its elements.
\end{enumerate}

Let us now formalize these considerations, demanding to the
remark\ref{rem:the metaphore by which we can speak about
noncommutative sets from within ZFC} for their conceptual
foundations.

Given a noncommutative probability space $ ( A \, , \, \omega ) $:

\begin{definition} \label{def:commutative predicates of a nocommutative space}
\end{definition}
COMMUTATIVE PREDICATES OVER A:
\begin{equation}
  {\mathcal{P}}_{C}(A) \; := \; \{ p(x) \; \text{ statement concerning } x \in A
  \} \; = \; MAP( X , \{ 0 ,1 \} )
\end{equation}
\begin{definition} \label{def:noncommutative predicates of a nocommutative space}
\end{definition}
NONCOMMUTATIVE PREDICATES OVER A:
\begin{equation}
  {\mathcal{P}}_{NC}(A) \; := \; {\mathcal{P}}(A)
\end{equation}

\begin{example} \label{ex:commutative and noncommutative properties over the one qubit alphabet}
\end{example}
COMMUTATIVE AND NONCOMMUTATIVE PREDICATES OVER THE ONE QUBIT
ALPHABET

Given the one qubit alphabet $ \Sigma_{NC} \, = \, M_{2}
({\mathbb{C}}) $, let us consider the following commutative
predicates:
\begin{align*}
  p_{normality} & (x) \; := \; << x x^{\star} \, = \, x^{\star} x >>  \\
  p_{hermitianicity} & (x) \; := \; << x \, = \, x^{\star} >> \\
  p_{positivity} & (x) \; := \; << \exists y \, : \, x \, = \, y y^{\star} >> \\
  p_{projectivity} & (x) \; := \; <<  x \, = \, x^{\star} \, = \, x^{2}
  >> \\
  p_{unitarity} & (x) \; := \; <<  x x^{\star} \, = \, x^{\star} x
  \, = \, I >>
\end{align*}

belonging to the classical logic $ ( \Sigma_{NC} \, , \, \leq_{C}
\, , \, \bot_{C} ) $.

Let us observe that:
\begin{align}
   p_{projectivity} & \, \leq_{C} \,  p_{positivity}  \, \leq_{C} \, p_{hermitianicity} \, \leq_{C} \,  p_{normality} \\
   p_{projectivity} & \, \leq_{C} \,  p_{unitarity} \, \leq_{C} \, p_{normality}
\end{align}
so that the neither \textbf{orthomodularity} nor the
\textbf{modularity} of $ ( \Sigma_{NC} \, , \, \leq_{C} \, , \,
\bot_{C} ) $  would be by itself sufficient to guarantee, for
example, that:
\begin{multline}
  p_{hermitianicity} (x)   \, \bigvee \, ( p_{positivity}(x) \bigwedge  p_{projectivity}(x) ) \; =  \\
   ( p_{hermitianicity}(x) \, \bigvee \, p_{positivity}(x) ) \,  \bigwedge \, ( p_{hermitianicity}(x) \, \bigvee \, p_{projectivity}(x)
   ) \; \; \forall x \in M_{2} ( {\mathbb{C}} )
\end{multline}
that may inferred only by \textbf{distributivity}.

One has clearly that:
\begin{multline}
  {\mathcal{P}}_{NC}(\Sigma_{NC} ) \; = \; {\mathcal{P}}(\Sigma_{NC} ) \; =
  \; \{ x \in \Sigma_{NC} \, : \,  p_{projectivity}(x) \; holds \}
\end{multline}
A not-trivial projection has the form:
\begin{equation}
  p_{\vec{n}} \; := \; \frac{1}{2} ( I + \vec{\sigma} \cdot
  \vec{n} ) \; \;  \vec{n}^{2} \, = \, 1
\end{equation}
and, in the physical realization of $ \Sigma_{NC} $ as a spin $
\frac{1}{2} $ system, corresponds to the statement:
\begin{center}
 \emph{$<<$ a measurement of $\vec{\sigma}$ in the direction $ \vec{n} $ gives outcome $+1$ with certainty $>>$}
\end{center}
or, more concisely:
\begin{center}
 \emph{$<<$ the spin point in the direction $ \vec{n} >>$}
\end{center}
assuming with Einstein Podolski and Rosen
\cite{Einstein-Podolski-Rosen-83} that:
\begin{center}
  \emph{"If, without in any way disturbing a system, we can predict with certainty (i.e. with probability equal to unity) the value
  of a physical quantity, then there exist an element of physical reality corresponding to this physical quantity" }
\end{center}
A simple calculation shows that the rules of the game are:
\begin{enumerate}
  \item
\begin{equation}
   p_{\vec{n}}^{ \bot }  \; = \; p_{- \vec{n}} \; \; \forall \vec{n}
\end{equation}
  \item
\begin{equation}
  p_{ \vec{n}_{1} }  \, \bigvee \, p_{ \vec{n}_{2} } \, = \, 0 \; \; \forall \vec{n}_{1} \, \neq \,  \vec{n}_{2}
\end{equation}
  \item
\begin{equation}
  p_{ \vec{n}_{1} } \; \bigwedge \; p_{ \vec{n}_{2} } \, = \, 1 \;
  \; \forall \vec{n}_{1} \, \neq \,  \vec{n}_{2}
\end{equation}
\end{enumerate}
whose meaning is \cite{Thirring-01}:
\begin{enumerate}
  \item We are  sure that $ \vec{\sigma} $  does not point in
  the direction $ \vec{n} $ only if it points to $ - \vec{n} $
  \item The sharpest proposition which is implied by both  $ << \,
    \vec{\sigma}$ points to $ \vec{n}_{1}  \, >> $ and $ << \,
    \vec{\sigma}$ points to $ \vec{n}_{2}  \, >> $ is the
    tautology $ << \, $ the spin points somewhere $ \, >>  $
  \item The proposition $ << \,
    \vec{\sigma} $ points to $ \vec{n}_{1} \, and \, \vec{n}_{2} >>
    $ is always false
\end{enumerate}
They immediately imply the violation of the distributive law:
\begin{multline}
  p_{\vec{n}_{1}} \, \bigvee \, ( \vec{n}_{2} \, \bigwedge \,
  \vec{n}_{3} ) \; = \;  p_{\vec{n}_{1}} \, \bigvee \, 0 \; = \;
  p_{\vec{n}_{1}} \\
  \neq \; ( p_{\vec{n}_{1}} \, \bigvee \, p_{\vec{n}_{2}} ) \,
  \bigwedge \, ( p_{\vec{n}_{1}} \, \bigvee \, p_{\vec{n}_{3}} )
  \, = \, 1 \, \bigwedge \, 1 \; = \; 1 \; \; \forall \vec{n}_{1} \, \neq \,
  \vec{n}_{2} \, \neq \, \vec{n}_{3}
\end{multline}

\smallskip

Given a classical probability space $ CPS \, := \,( X \, , \, \mu
) $ and a commutative predicate over X $ p(x) \, \in \,
{\mathcal{P}}_{C}(X) $ let us introduce its fibre in zero:
\begin{equation}
  N_{p} \; := \; \{ x \in X \: : \: p(x) \, = \, 0 \}
\end{equation}
Looking at definition\ref{def:classical null set} it is clear
that:
\begin{equation}
  p \in {\mathcal{P}}_{TYPICAL}(CPS) \; \Leftrightarrow \; \mu(  N_{p} ) \, = \, 0
\end{equation}
Let us now look at CPS as the commutative probability space $ ( A
\, := \, L^{\infty} ( X \, , \mu ) \: , \: \omega_{\mu} ( \cdot )
\, := \, \int_{X} d \mu \cdot ) $.

We have clearly that:
\begin{equation}
   p \in {\mathcal{P}}_{TYPICAL}(CPS) \; \Leftrightarrow \;
   \chi_{N_{p}} \in {\mathcal{P}}_{NC}(A)
\end{equation}
so that we can express $ {\mathcal{P}}_{TYPICAL}(CPS) $ as:
\begin{equation}
 {\mathcal{P}}_{TYPICAL}(CPS) \; = \; \{ p \in {\mathcal{P}}_{NC}
 ( A) \: : \: \omega( p ) \, = \, 0 \}
\end{equation}

Given an algebraic probability space $ APS \, := \, ( A \, , \,
\omega ) $, each of the two possible answers given to the FIRST
ISSUE suggests a natural noncommutative generalization of the
definition\ref{def:typical properties of a classical probability
space}:
\begin{enumerate}
  \item looking at commutative properties of A one is lead to
  introduce the following:
\begin{definition}
\end{definition}
$ S \; \subset A $  IS A NULL SET OF APS:
\begin{equation}
  E( a )  \; = \; 0 \; \; \forall a \in S
\end{equation}
\begin{definition} \label{def:typical commutative properties of an algebraic probability space}
\end{definition}
TYPICAL COMMUTATIVE PROPERTIES OF APS:
\begin{equation}
 {\mathcal{P}}^{TYPICAL}_{C}(APS) \; \equiv \; \{ \, p \in  {\mathcal{P}} ( A ) \, :
   \{ a \in A \, : \, p ( a ) \text{ doesn't hold } \}
   \; \text{is a null set of APS} \}
\end{equation}
  \item
  looking at noncommutative properties of A one is led to
  introduce the following:
\begin{definition} \label{def:typical noncommutative properties of an algebraic probability space}
\end{definition}
TYPICAL NONCOMMUTATIVE PROPERTIES OF APS:
\begin{equation}
 {\mathcal{P}}_{NC}^{TYPICAL}(APS) \; = \; \{ p \in {\mathcal{P}}_{NC}
 ( A) \: : \: \omega( p ) \, = \, 0 \}
\end{equation}
\end{enumerate}

\smallskip

Let us now try to formalize the idea of randomness as satisfaction
of all typical properties, following again both the streets:
\begin{enumerate}
  \item
\begin{definition} \label{def:Kolmogorov commutatively-random elements of an algebraic probability space}
\end{definition}
SET OF THE  KOLMOGOROV  COMMUTATIVELY-RANDOM ELEMENTS OF APS:
\begin{multline}
  KOLMOGOROV-RANDOM_{C} ( APS ) \; := \\
   \{ \, a \, \in \, A \, : p ( a ) \; holds \; \; \forall p \in  {\mathcal{P}}_{C}^{TYPICAL}(APS) \}
\end{multline}
  \item
\begin{definition} \label{def:Kolmogorov noncommutatively-random elements of an algebraic probability space}
\end{definition}
SET OF THE  KOLMOGOROV  NONCOMMUTATIVELY-RANDOM ELEMENTS OF APS:
\begin{equation}
  KOLMOGOROV-RANDOM_{NC} ( APS ) \; := \bigvee_{p \in
  {\mathcal{P}}_{C}^{TYPICAL}(APS)} \, p
\end{equation}
\end{enumerate}
The efficacy of such a formalization will be investigated in
section\ref{sec:On absolute conformism in Quantum Probability
Theory}.
\newpage
\section{Constraint on infinite noncommutatively-independent tosses of a quantum coin}

We know by theorem\ref{th:foundation of the applicability of
probability theory to reality} that the correct notion of $
NC_{M}-C_{\Phi}$-randomness, namely Martin-L\"{o}f Solovay-Chaitin
randomness, satisfies the following intuitive condition:
 \begin{constraint} \label{con:independent sequence of tosses of a classical coin}
 \end{constraint}
\textsf{ON THE NOTION OF A RANDOM SEQUENCE ON THE COMMUTATIVE
ALPHABET $ \Sigma $ :}

\emph{Making infinite \textbf{independent} trials of the
experiment consisting on tossing a \textbf{classical coin} we
must obtain an algorithmically-random sequence with probability
one}

\smallskip

 So a reasonable strategy to obtain informations about the correct definition of a random sequence of qubits would consist in:
\begin{itemize}
  \item formulating an analogous constraint in terms of an
  infinite sequence of experiments consisting in tossing a quantum
  coin
  \item identifying the information that such a constraint gives on the correct way of
  of defining an algorithmically-random sequence of qubits
\end{itemize}

\smallskip

In the constraint\ref{con:independent sequence of tosses of a
classical coin} the commutative random variables $ c_{i} $ and $
c_{j}$ on the commutative probability space $ ( L^{\infty} (
\Sigma^{\infty} , P_{unbiased} ) , \tau_{unbiased} ) $
representing the results of the classical-coin tossings at times,
respectively, $ i , j \, \in \, {\mathbb{N}}$, are assumed to be
\textbf{independent}:
\begin{multline}
  \tau_{unbiased}( c_{i}^{n} c_{j}^{m} ) \; = \\
   \tau_{unbiased}(
  c_{i}^{n} ) \, \tau_{unbiased}(
  c_{j}^{m} ) \; \; \forall n,m \, \in \, {\mathbb{N}}
\end{multline}
By theorem\ref{th:dependence of noncommutating random variables}
this implies that:
\begin{equation}
  [ \, c_{i} \, , \, c_{j} \, ] \; = \; 0
\end{equation}

Let us now consider the quantum situation and let us adopt the
terminology introduced in  example\ref{ex:the hyperfinite finite
continuous factor} and example\ref{ex:Powers factors}.

The quantum coin tossing at time $ i \in {\mathbb{N}} $ will be
described by the noncommutative random variable $ c_{i} $
associated, in the passage from  $ A_{{\mathbb{Z}}} $ to  $ R \,
= \bigotimes_{n=0}^{\infty} ( M_{2} ( {\mathbb{C}})  \, , \,
\tau_{2} ) $, to a noncommutative random variable on the
noncommutative probability space $ ( A_{\{i\}} \, , \, \tau_{\{ i
\}} ) $.

By  theorem\ref{th:automatic independence on tensorial products}
we have that:
\begin{equation}
  c_{i} \; and \; c_{j} \text{ are independent } \; \forall i , j
  \in {\mathbb{N}}
\end{equation}
\begin{remark} \label{rem:commutative independence versus noncommutative independence}
\end{remark}
COMMUTATIVE INDEPENDENCE VERSUS NONCOMMUTATIVE INDEPENDENCE

The adoption of the classical, commutative notion of independence
in a quantum context is of great physical relevance, lying at the
heart of one of the most important features of Quantum Physics,
both foundationally and for applications: \textbf{entanglement}.

Anyway, from a conceptual viewpoint, it is reasonable to expect
that in Noncommutative Probability Theory there must exist a
notion of non-commutative independence according to  which there
may exist noncommutatively-independent random variables that don't
commute among themselves.

Among the many proposals the more radical is Dan Voiculescu's
Free Probability Theory \cite{Hiai-Petz-00} according to which
the own fact that two random variables commute and thus has some
kind of algebraic interrelation among them is seen as a lack of
genuine noncommutative-independence among them.

To introduce Free Probability Theory it is necessary to introduce
some notions concerning free groups \cite{Collins-Zieschang-98}.

Given a group G and a subsets of its $ \chi \subset G $:
\begin{definition} \label{def:system of generators of a group}
\end{definition}
$ \chi  $ IS A SYSTEM OF GENERATORS OF G:

the smallest subgroup of G containing $ \chi $ in G

\smallskip

\begin{definition} \label{def:rank of a group}
\end{definition}
RANK OF G:
\begin{equation}
  d(G) \; := \; \min \{ cardinality( \chi ) \, \chi \text{ system of generators of
  G} \}
\end{equation}
Given a system of generators $ \chi  $ of G let us consider an
alphabet $ \hat{\chi} $ such that there exist a bijection $ b \, :
\, \hat{\chi} \: \rightarrow \: \chi $.

Let us adopt capital letters $ X,Y,Z,A,B,C, \cdots $ to denote
elements of $ \hat{\chi} $ and the corresponding small letters $
x,y,z,a,b,c, \cdots $ for the corresponding elements of $ \chi$.

\begin{definition} \label{def:word}
\end{definition}
WORD OVER $ \hat{X} $ OF LENGTH $ k \in {\mathbb{N}}_{+} \; ( | W
| = k ) ) $:

a formal expression:
\begin{equation}
  W \, := \, W(\hat{\chi}) \, = \,  \prod_{i=1}^{k} X_{i} ^{\epsilon_{i}}  \; \; X_{i}
  \in \hat{\chi} , \epsilon_{i} \in \{ - 1 , 1 \} \forall i \in \{ 1
  , 2, \cdots, k  \}
\end{equation}

\smallskip

\begin{definition} \label{def:word representing a group's element}
\end{definition}
THE WORD $ W(\hat{\chi}) $ REPRESENTS THE ELEMENT $ g \in G $:
\begin{equation}
  g \; = \; W(\hat{\chi}) \; = \; \prod_{i=1}^{k} X_{i}^{\epsilon_{i}}
\end{equation}
Given two words $ W(\hat{\chi}) = g $ and $ V(\hat{\chi}) = h $
representing respectively the elements g and h of G:
\begin{definition} \label{def:product of words}
\end{definition}
PRODUCT OF $ W(\hat{\chi}) $ and $ V(\hat{\chi}) $:
\begin{equation}
  W(\hat{\chi}) V(\hat{\chi}) \; = \; g h \in G
\end{equation}

\smallskip

\begin{definition} \label{def:inverse word}
\end{definition}
INVERSE WORD OF $ W(\hat{\chi}) \, = \, \prod_{i=1}^{k}
X_{i}^{\epsilon_{i}} $:
\begin{equation}
   W(\hat{\chi})^{- 1} \, := \, \prod_{i=1}^{k}
X_{k-i}^{- \epsilon_{k - i}}
\end{equation}

Obviously if $  W(\hat{\chi}) $ represents $ g \in G $ then $
W(\hat{\chi})^{- 1} $ represents $ g^{- 1} \in G $.

We will denote the \textbf{trivial word}, namely the thy word
consisting of no letters and representing the identity $ e \in G
$ as I.

\begin{definition} \label{def:peak}
\end{definition}
PEAK:

a word of the form:
\begin{equation}
  X^{\epsilon} X^{- \epsilon} \; \; X \in \hat{\chi} \, , \,
  \epsilon \in \{ - 1 , 1 \}
\end{equation}

Given two words V and W:
\begin{definition} \label{def:free equivalent words}
\end{definition}
V AND W ARE FREELY EQUIVALENT:
\begin{equation}
  V \: \equiv \: W \; \; \text{ V may be obtained by W inserting or deleting peaks}
\end{equation}
\begin{definition} \label{def:relator}
\end{definition}
THE WORD $ R( \hat{\chi} ) \; := \; X_{1}^{\epsilon_{1}} \cdots
X_{k}^{\epsilon_{k}}  $ IS A RELATOR W.R.T. $ \chi $ AND G:
\begin{equation}
   x_{1}^{\epsilon_{1}} \cdots x_{k}^{\epsilon_{k}} \; = \; 1 \; \; \text{ in G}
\end{equation}

Given a system of relators $ {\mathcal{R}} $:
\begin{definition}
\end{definition}
$ {\mathcal{R}} $ IS A SYSTEM OF DEFINING RELATORS W.R.T. $ \chi
$ AND G:

every relator is a consequence of those in $ {\mathcal{R}} $,
i.e. is freely-equivalent to a word:
\begin{equation}
  L_{1}( \hat{\chi}) R_{1}( \hat{\chi})^{\eta_{1}}L_{1}( \hat{\chi})^{-
  1} \cdots  L_{k}( \hat{\chi}) R_{k}( \hat{\chi})^{\eta_{1}}L_{k}( \hat{\chi})^{-
  1} \; \; R_{j}( \hat{\chi}) \in {\mathcal{R}} \, , \, \eta_{j}
  \in \{ - 1 , 1 \} \, L_{j}( \hat{\chi}) \text{ word}
\end{equation}
\begin{definition} \label{def:presentation of a group}
\end{definition}
PRESENTATION OF G:

a couple ( $ \chi $ , $ {\mathcal{R}} $ ) such that:
\begin{itemize}
  \item $ \chi $ is a generating system of G
  \item $ {\mathcal{R}} $ is a system of defining relators w.r.t. G
  and $ \chi $
\end{itemize}
Since to assign a presentation of a group is equivalent to
assigning it, the fact that ( $ \chi $ , $ {\mathcal{R}} $ ) is a
presentation of the group G is usually indicated as:
\begin{equation}
  G \; = \; <   \chi  |  {\mathcal{R}}  >
\end{equation}
\begin{definition} \label{def:finitely generated group}
\end{definition}
G IS FINITELY GENERATED:
\begin{equation}
  dim(G) \; < \; \aleph_{0}
\end{equation}
\begin{definition} \label{def:finitely presented group}
\end{definition}
G IS FINITELY PRESENTED:
\begin{align}
  G \; &  = \; <   \chi  |  {\mathcal{R}}  > \\
  cardinality (\chi) & \; < \; \aleph_{0} \\
  cardinality( {\mathcal{R}} ) & \; < \; \aleph_{0}
\end{align}
Given a finitely presented group $ G \;   =  \; <   \chi  |
{\mathcal{R}}  > $:
\begin{definition} \label{def:word problem of a finitely presented group}
\end{definition}
WORD PROBLEM OF G:

the problem of determining if an arbitrary word in the elements of
$ \chi $ defines, as a consequence of $ {\mathcal{R}} $, the
identity element

\smallskip

\begin{definition} \label{def:free group}
\end{definition}
G IS A FREE GROUP:
\begin{equation}
  G \; = \; <   \chi  | -  > \; := \;  <   \chi  | \emptyset  >
\end{equation}
i.e. if it has a presentation free of defining relators.

In particular:
\begin{definition} \label{def:free group of rank n}
\end{definition}
\begin{equation}
  F_{n} \; := \; \text{ free group } \, : \, d( F_{n} ) \, = \, n
\end{equation}
\begin{definition} \label{def:free basis of the free group of rank n}
\end{definition}
FREE BASIS OF $ F_{n} $:
\begin{equation}
  \{ g_{1} \, , \, \cdots \, , \, g_{n} \} \; : \; F_{n} \; = \; < g_{1} , \cdots , g_{n} | - >
\end{equation}
\begin{remark} \label{rem:freeness and algebraic independence}
\end{remark}
FREENESS AS ALGEBRAIC INDEPENDENCE

The fact that there doesn't exist any algebraic relation among
the elements of a free-basis $ \{ g_{1} , \cdots , g_{n} \} $ of $
F_{n} \; = \; < g_{1} , \cdots , g_{n} | -
> $ up to their membership to $ F_{n} $  may be interpreted as a condition of algebraic independence
among them.

\begin{remark} \label{rem:random walks on free groups}
\end{remark}
RANDOM WALKS ON FREE GROUPS

Let us consider a drunk living  in a D-dimensional space  $
{\mathbb{R}}^{D} $, exiting  at time $ t \; = \; 0 $ from  the
tavern located in the origin and beginning to walk completelly at
random,  making at each temporal step $ n \in {\mathbb{N}} $ one
step of unit length.

Introduced the unit versors $ \vec{v}^{(i)} \; \in \;
{\mathbb{R}}^{D} \, \, i = 1, \cdots , D $:
\begin{equation}
  \vec{v}^{(i)}_{j} \; := \; \delta_{i , j} \; \; i ,j = 1, \cdots , D
\end{equation}
and the \textbf{near-neighbourhood matrix}:
\begin{equation}
  J_{\vec{x} , \vec{y}} \; := \; \sum_{i=1}^{D} \delta_{\vec{x} , \vec{y}
  + \vec{v}^{(i)}}
\end{equation}
we have clearly that the probability $ P_{t} ( \vec{x} ) $  that
at the temporal step  $ t \in {\mathbb{N}} $ he is located in the
lattice site $ \vec{x} \, \in \, {\mathbb{Z}}^{D} $ satisfies the
conditions:
\begin{align} \label{eq:probability recursion relation of the commutative random walk}
  P_{t} ( \vec{x} ) & \; = \; \frac{1}{2 D} \sum_{\vec{y} \in  {\mathbb{Z}}^{D}} J_{\vec{x} , \vec{y}} P_{t-1} ( \vec{y} )    \\
  P_{0} ( \vec{x} ) & \; = \; \delta_{\vec{x} , \vec{0}}
\end{align}
Introduced the Fourier-transform of the $ P_{t} ( \vec{x} ) $:
\begin{align}
   P_{t} ( \vec{x} ) &  \; = \; \int_{[ - \pi , \pi ]^{D}} \frac{d \vec{p}}{( 2 \pi )^{D}} e^{ i \vec{p} \cdot \vec{x}} \tilde{P}_{t} ( \vec{p} )  \\
  \tilde{P}_{t} ( \vec{p} )  &  \; = \; \sum_{ \vec{x} \in
  {\mathbb{Z}}^{D}} e^{- i \vec{p} \cdot \vec{x}} P_{t} ( \vec{x} )
\end{align}
eq.\ref{eq:probability recursion relation of the commutative
random walk} implies that:
\begin{equation}
  \tilde{P}_{t} ( \vec{p} ) \; = \; ( \frac{1}{D} \sum_{i=1}^{D}
  \cos( p_{i} ) )^{t}
\end{equation}
To analyze  the asympotic situation at large distances w.r.t. the
lattice spacing and for long times, it is convenient to rescale
and times using a lattice spacing a rather then 1 and a time
interval $ \tau $ rather than 1.

After the substitutions:
\begin{align}
  t &  \; \rightarrow \; \frac{t}{\tau} \\
  \vec{x} & \; \rightarrow \; \frac{ \vec{x} }{a} \\
  \vec{k} & \; \rightarrow \; a  \vec{k}
\end{align}
one obtains that:
\begin{equation}
   P_{t} ( \vec{x} )  \; = \; a^{D} \int_{[ - \frac{\pi}{a} ,  \frac{\pi}{a}
   ]^{D}} \frac{ d \vec{p} }{( 2 \pi )^{D}} e^{ i \vec{p} \cdot
   \vec{x}} ( \frac{1}{D} \sum_{i=1}^{D}
  \cos(a \, p_{i} ) )^{ \frac{t}{\tau} }
\end{equation}
Let us now take the limit as a and $ \tau $ go to zero keeping the
distance and time intervals fixed.

Following , for simplicity, the informal approach of
\cite{Itzykson-Drouffe-89a} let us consider  a volume $ \Delta
\vec{x} $ which is large w.r.t. the elementary lattice volume $
a^{D} $ but which is also sufficiently small to ensure that P
remain nearly constant within $ \Delta \vec{x} $; this last
requirement is fulfilled if $ \frac{t}{\tau} $ is also large.

This permits to introduce a probability density  $ p ( \vec{x} ,
t ) $ defined as:
\begin{equation}
  p( \vec{x} , t) \Delta \vec{x} \; := \; \sum_{\vec{x}' \in \vec{x} + \Delta \vec{x}
  } P( \vec{x}' , t ) \; \approx \; \frac{\Delta \vec{x} }{a ^{D}}
  P ( \vec{x} , t )
\end{equation}
so that:
\begin{equation}
  p( \vec{x} , t ) \; = \; \lim_{a , \tau \rightarrow 0} \int_{[ - \frac{\pi}{a} ,  \frac{\pi}{a}
   ]^{D}} \frac{ d \vec{p} }{( 2 \pi )^{D}} e^{ i \vec{p} \cdot
   \vec{x}} ( \frac{1}{D} \sum_{i=1}^{D}
  \cos( a \, p_{i} ) )^{ \frac{t}{\tau}}
\end{equation}
This limit is not trivial only when a and $ \tau $ vanish in such
a way that the ratio $ \frac{\tau}{a^{2}} $ is kept fixed, as
shown by the expansion of the cosine:
\begin{equation}
  ( \frac{1}{D} \sum_{i=1}^{D} \cos( a \, p_{i} ) )^{
  \frac{t}{\tau}} \; = \; ( 1 - \frac{ a^{2} }{ 2 D} \vec{p}^{2} +
  o ( \frac{1}{a} , \frac{1}{\tau} ))^{ \frac{t}{ \tau} }\; \rightarrow \; e ^{ - t \vec{p}^{2}}
\end{equation}
in which the time scale has been fixed using:
\begin{equation}
  \tau \; = \; \frac{a ^{2} }{ 2 D}
\end{equation}
Hence:
\begin{equation}
  p( \vec{x} , t ) \; = \; \int \frac{d \vec{p} }{ ( 2 \pi )^{D}}
  \exp [ - t \vec{p}^{2} + i \vec{x} \cdot \vec{p} ] \; = \;
  \frac{1}{ ( 4 \pi t )^{ \frac{D}{2}}} \exp ( - \frac{ \vec{x}^{2} }{4
  t})
\end{equation}
Fixed the notation for gaussian distributions in the following
way:
\begin{definition} \label{def:gaussian distribution}
\end{definition}
D-DIMENSIONAL GAUSSIAN  MEASURE OF MEAN $ \vec{m} $  AND
COVARIANCE MATRIX $ \hat{C} $:

the probability measure on $ {\mathbb{R}}^{d} $ of halting-set $
{\mathcal{F}}_{Borel}  $ with density:
\begin{equation}
  gauss(D , \vec{m} , \hat{C} ; \vec{x} ) \; := \; \frac{1}{ ( 2 \pi )^{\frac{D}{2}}
  } \exp [ - \frac{1}{2} ( \vec{x} - \vec{m} ) \, \cdot \, \hat{C}^{- 1} ( \vec{x} - \vec{m}
  ) ]
\end{equation}

\smallskip
we see that the asympotic probability distribution of the drunk's
position is  $ gauss(D , \vec{0} , \frac{\hat{I}}{2 \sqrt{t}} ;
\vec{x} ) $.

Up to translation and rescaling the essence of the asymptotic
behaviour of the random walk is thus encoded in the following:

\begin{definition}
\end{definition}
STANDARD GAUSSIAN  MEASURE:

the probability measure on $ {\mathbb{R}} $ of halting-set $
{\mathcal{F}}_{Borel}  $ with density $ gauss_{STANDARD} := g(1 ,
0 , 1 ; x ) $

i.e. by the sequence $ M_{n} [ gauss_{STANDARD} ] $ of its
moments:

\begin{theorem} \label{th:gaussian moments}
\end{theorem}
GAUSSIAN MOMENTS:
\begin{equation}
  M_{n} [ gauss_{STANDARD} ] \; := \; \int_{- \infty}^{+ \infty}
  dx x^{n} gauss_{STANDARD} (x) \; = \;
  \begin{cases}
    (2 m -1 )!!  & \text{if $ n  = 2m \, , \, m \in {\mathbb{N}} $}, \\
    0 & \text{otherwise}.
  \end{cases}
\end{equation}

$ M_{2m} [ gauss_{STANDARD} ] $ has an intuitive combinatorial
meaning: it is  the \textbf{number of pair partitions} of a set
of $ 2m $ elements.

\smallskip

Let us now consider another drunk making a random walk on the
free group  of rank D $ F_{D} \, = \, < g_{1} , \cdots , g_{D} | -
> $, starting at time $ t \; = \; 0 $ from the tavern located in
the identity element $ e $, going in one step from g to h g with
probability:
\begin{equation} \label{eq:transition probability of the free random walk}
  Prob[ g \; \rightarrow \; h g ] \; = \;
  \begin{cases}
    \frac{1}{2 D} & \text{if $ h \in \{ g_{1} \, , \, \cdots , g_{n} \, , \,  g_{1}^{-1} \, , \, \cdots \, , \, g_{n}^{-1} \} $}, \\
    0 & \text{otherwise}.
  \end{cases}
\end{equation}
Let us now recall that to any discrete group G one can associate
the following Hilbert space:
\begin{definition} \label{Hilbert space of a discrete group}
\end{definition}
HILBERT SPACE OF G:

the Hilbert space $ ( l^{2} (G) \, , \,  < \cdot | \cdot
> ) $  such that:
\begin{itemize}
  \item
\begin{equation}
  l^{2} (G) \; := \; \{ \xi \, : \, G \rightarrow {\mathbb{C}}  \,
  : \, \sum_{g \in G} | \xi(g) | ^{2} \; < \; + \infty \}
\end{equation}
  \item
\begin{equation}
  < \xi | \eta > \; := \; \sum_{g \in G}  \xi(g) \eta^{\star} (g)
  \; \; \xi , \eta \in l^{2} (G)
\end{equation}
\end{itemize}
Furthermore there exist natural action representations of G on
the Hilbert space $ l^{2} (G) $:
\begin{definition} \label{def:left regular representation}
\end{definition}
LEFT REGULAR REPRESENTATION OF G ON $ l^{2} (G) $:
\begin{equation}
  (L_{g} \xi ) ( \eta ) \; := \; \xi( g^{- 1} h ) \; \; \xi \in l^{2}
  (G) , g , h \in G
\end{equation}
\begin{definition} \label{def:right regular representation}
\end{definition}
RIGHT REGULAR REPRESENTATION OF G ON $ l^{2} (G) $:
\begin{equation}
  (R_{g} \xi ) ( \eta ) \; := \; \xi( h g ) \; \; \xi \in l^{2}
  (G) , g , h \in G
\end{equation}
from which one can define two Von Neumann algebras associated to
G:
\begin{definition} \label{def:left Von Neumann algebra of a discrete group}
\end{definition}
(LEFT) GROUP VON NEUMANN ALGEBRA OF G:
\begin{equation}
  {\mathcal{L}}(G) \; := \; \{ L_{g} \, , \, g \in G \}^{''}
\end{equation}
\begin{definition} \label{def:right Von Neumann algebra of a discrete group}
\end{definition}
(RIGHT) GROUP VON NEUMANN ALGEBRA OF G:
\begin{equation}
  {\mathcal{R}}(G) \; := \; \{ R_{g} \, , \, g \in G \}^{''}
\end{equation}
Introduced the following notation for the Kronecker's deltas on G
looking them as function on the second argument:
\begin{equation}
  \delta_{g} (h) \; := \; \delta_{g h} \; \; g , h \in G
\end{equation}
it may be proved \cite{Hiai-Petz-00} that the state $ \tau \in
{\mathcal{L}}(G) $ defined as:
\begin{equation}
  \tau ( a ) \; := \; < a  \delta_{e} \, , \,  \delta_{e} > \; \;
  a \in  {\mathcal{L}}(G)
\end{equation}
is a faithful, normal tracial state on $ {\mathcal{L}}(G) $.

 Returning at last to our drunk random walking on $ F_{D} $
let us introduce the following noncommutative random variables
over the noncommutative probability space $ (
{\mathcal{L}}(F_{D}) \, , \, \tau ) $:
\begin{equation}
  a_{i} \; := \; \frac{1}{\sqrt{2}} ( L_{g_{i}} + L_{g_{i}^{- 1}})
  \; \; i = 1 , \cdots , D
\end{equation}
and let us observe that, by eq.\ref{eq:transition probability of
the free random walk}, the probability that the drunk is again at
the tavern at the temporal step t may be expressed as the
following expectation value over the noncommutative probability
space:
\begin{equation}
  P_{t} ( e ) \; = \; E( ( \sum_{i=1}^{D} a_{i} ) ^{t} ) \forall t
  \in {\mathbb{N}}
\end{equation}
whose asympotic behaviour is given by:
\begin{equation}
  P_{ 2 t} ( e ) \; \approx \; \frac{1}{( 2 \pi )^{D}}
  \frac{1}{t+1} \begin{pmatrix}
     2 \, t \\
    t \
  \end{pmatrix}
\end{equation}
So, exactly as the asymptotic return-to-tavern probability for the
drunk living in the D-dimensional euclidean space is governed by
the gaussian distribution, the asymptotic return-to-tavern
probability for the drunk living living in the D-rank free group
is ruled by the probability measure on the real line having
vanishing odd moments and even moments given by the
\textbf{Catalan numbers}:
\begin{equation}
  C_{n} \; := \; \frac{1}{n+1} \begin{pmatrix}
    2 n \\
    n \
  \end{pmatrix}
\end{equation}
namely the \textbf{standard semicircle measure}, according to the
following:
\begin{definition}
\end{definition}
SEMICIRCLE MEASURE OF MEAN m AND VARIANCE $ \frac{r^{2}}{4} $:

the probability measure on $  {\mathbb{R}} $ with halting set $
{\mathcal{F}}_{Borel} $ and density:

$ sc(m ,r ; x ) \; := $
\begin{equation}
  \begin{cases}
    \frac{2}{\pi r^{2}}  \sqrt{ r^{2} - (x-m)^{2}} & \text{if $ m-r \leq x \leq m+r$}, \\
    0 & \text{otherwise}.
\end{cases}
\end{equation}
\begin{definition}
\end{definition}
STANDARD SEMICIRCLE MEASURE:

the probability measure on $  {\mathbb{R}} $ with halting set $
{\mathcal{F}}_{Borel} $ and density $ sc_{STANDARD} := sc(0 ,2; x
) $

\smallskip

One has indeed the following:
\begin{theorem} \label{th:semicircle moments}
\end{theorem}
SEMICIRLE MOMENTS:
\begin{equation}
  M_{n} [ sc_{STANDARD} ] \; := \; \int_{- \infty}^{+ \infty}
  dx x^{n} sc_{STANDARD} (x) \; = \;
  \begin{cases}
    C_{2 \, m}  & \text{if $ n  = 2m \, , \, m \in {\mathbb{N}} $}, \\
    0 & \text{otherwise}.
  \end{cases}
\end{equation}

The moments of the semicircle distribution, namely the Catalan
numbers, have a combinatorial meaning similar to that of the
gaussian moments: $ M_{2m} \, = \,  C_{2m} $ is the
n\textbf{umber of non-crossing pair partitions} of a linearly
ordered set of 2m objects (let's say the 2m-letters' alphabet $
\Sigma_{2m} $), according to the following:
\begin{definition}
\end{definition}
THE PARTITION $ {\mathcal{V}} \, := \, \{ V_{1} , \cdots , V_{s}
\} $ OF $ \Sigma_{n} $ IS NON-CROSSING:
\begin{multline}
  ( V_{i} \, = \, \{ v_{1} , \cdots , v_{p} \} \: and \: V_{j} \, = \, \{ w_{1} , \cdots , w_{q}
  \} ) \; \Rightarrow \\
  (   w_{m} < v_{1} < w_{m+1} \:
  \Leftrightarrow \: w_{m} < v_{p} < w_{m+1} \, \, m = 1 , \cdots
  , q-1 )
\end{multline}

\smallskip

Let us now observe that what rules the asympotic behaviour of the
drunk living in the D-dimensional euclidean space is the Central
Limit Theorem of Classical Probability Theory, stating that the
probability distribution:
\begin{equation*}
  \frac{ x_{1} \, + \, \cdots \, + \, x_{n}}{\sqrt{n}}
\end{equation*}
of a collection of identically distributed, independent random
variables tends to the gaussian distribution when $ n \rightarrow
\infty $.

In the same way what rules the asympotic behaviour of the drunk
living in the D-ranked free group is a noncommutative central
limit theorem stating that the average:
\begin{equation*}
  \frac{a_{1} + \cdots + a_{n} }{\sqrt{n}}
\end{equation*}
is a noncommutative random variable whose distribution, for $ n
\rightarrow \infty $, tends to the semicircle distribution.

\medskip

Given a free group $ F \; = \; < \chi | - > $ and a word W over $
\chi $:
\begin{definition} \label{def:reduced word of a free group}
\end{definition}
W IS REDUCED:

W doesn't contain peaks

\smallskip
An important problem of free groups is the solvability of their
word problem

\begin{theorem}  \label{th:solution of the world problem of free groups}
\end{theorem}
SOLUTION OF THE WORLD PROBLEM OF FREE GROUPS:

each element of $ F_{n} $ is represented by a unique reduced word

\medskip

Let us consider again the collection of noncommutative random
variables $ \{ a_{i} \} $ on the noncommutative probability space
$ ( {\mathcal{L}} ( F_{n} ) \, , \, \tau ) $  introduced in the
remark\ref{rem:random walks on free groups} on discussing the
random walk on $  F_{n} $ and the noncommutative central limit
theorem governing its asymptotic behaviour: its a collection of
mutually-noncommuting  random variables so that, by
theorem\ref{th:dependence of noncommutating random variables},
they are certainly not independent.

Formalizing the kind of algebraic independence of such a
collection, inherited from the free structure of $ F_{n} $, one
arrives to the following:

given a noncommutative probability space $ ( A \, , \, \omega ) $
and a family  $ \{ A_{i} \}_{i \in I} $ of subalgebras of A:
\begin{definition} \label{def:freeness}
\end{definition}
THE SUBALGEBRAS $ \{ A_{i} \}_{i \in I} $ ARE FREE:
\begin{multline}
  \forall n \in {\mathbb{N}} \, , \, \forall i(1) , \cdots , i(n)
  \in I \, : \, i(k) \neq i(k+1)  \:  ( 1 \leq k \leq n-1 ) \\
  a_{k} \in A_{i(k)} \, , \, \omega( a_{k} ) = 0 \, 1 \leq k \leq n  \; \Rightarrow   \omega( a_{1} a_{2} \cdots  a_{n} ) \: = \: 0
\end{multline}

\smallskip

Clearly definition\ref{def:freeness} is a generalization of the
simpler case in which $ I = {\mathbb{N}} $ and each $ \{ A_{i} \}
= span( \{ a_{i} \}) $, in which freeness implies the following
useful result:
\begin{theorem} \label{th:the vanishinh expectation value of a product of centered free random variables}
\end{theorem}
THE VANISHING EXPECTATION VALUE OF A PRODUCT OF CENTERED FREE
RANDOM VARIABLES

\begin{hypothesis}
\end{hypothesis}
\begin{equation*}
  ( A \; , \; \omega ) \;  \text{ noncommutative probability space}
\end{equation*}
\begin{equation*}
  \{ a_{1} \, , \, \cdots  \, , \, a_{n} \} \; \text{$n^{th}$-ple of free random variables}
\end{equation*}
\begin{thesis}
\end{thesis}
\begin{equation*}
  E( \sum_{i=1}^{n} a_{i} \, - \, E( a_{i} ) ) \; = \; 0
\end{equation*}
\begin{proof}
The family of subalgebras $ \{ A_{i} \}_{i=1}^{n} $ spanned by
every $ a_{i} \, i = 1 , \cdots n $ may be ordered in a way such
that $ a_{k}  \in A_{i(k)} $ with $ i(k) \neq i(k+1) \, k = 1 ,
\cdots , n-1 $.

Furthermore one has clearly that:
\begin{equation}
  E( a_{i} - E( a_{i} ) ) \; = \; 0 \; \; i = 1 , \cdots , n
\end{equation}
So the freness condition implies the thesis.
\end{proof}

We know that the independence of a collection  $ \{ a_{1} \, ,
\cdots \, , a_{n} \} $ of random variables on an algebraic
probability space $ ( A \, , \, \omega ) $ may be seen as a rule
for deriving the joint-moment $ E( \prod_{i=1}^{n} a_{i} ) $ for
the collection of expectation values $ \{ E ( a_{1} )  \, , \cdots
\, , E ( a_{n} ) \} $ according to:
\begin{equation}
  E( \prod_{i=1}^{n} a_{i} ) \; = \; \prod_{i=1}^{n} E (  a_{i} )
\end{equation}
Also freeness may be seen in this way, but with a different way of
computing the joint moments:
\begin{corollary} \label{cor:joint moment of free random variables}
\end{corollary}

\begin{hypothesis}
\end{hypothesis}
\begin{equation*}
  ( A \; , \; \omega ) \;  \text{ noncommutative probability space}
\end{equation*}
\begin{equation*}
  \{ a_{1} \, , \, \cdots  \, , \, a_{n} \} \; \text{$n^{th}$-ple of free random variables}
\end{equation*}
\begin{thesis}
\end{thesis}
\begin{equation*}
  E( \prod_{i=1}^{n} a_{i} ) \; = \; \sum_{i=1}^{n} \sum_{1 \leq k_{1} \leq \cdots \leq k_{r} \leq
  n} ( - 1 )^{r+1} E ( a_{k_{1}} ) \cdots E ( a_{k_{r}} ) E (
  a_{1} \cdots \hat{a}_{k_{1}} \cdots \hat{a}_{k_{r}} \cdots a_{n})
\end{equation*}
where $ \hat{\cdot} $ indicates terms that are omitted.

\smallskip

\begin{example} \label{ex:freeness versus independence for a couple of noncommutative random variables}
\end{example}
FREENESS VERSUS INDEPENDENCE FOR A COUPLE OF NONCOMMUTATIVE
RANDOM VARIABLES

Given two free algebraic random variables $ a  \, , \, b $ on a
noncommutative probability space $ ( A \, , \, \omega ) $,
Corollary\ref{cor:joint moment of free random variables} implies
that:
\begin{equation}
  E( a \, b ) \; =  \; E( a ) E( b )  \; = \, E ( b \, a )
\end{equation}
exactly as one would have if a and b were independent.

Anyway one has that:
\begin{equation}
  E ( a \, b \, a \, b ) \; = \; E ( a^{2} ) E ( b )^{2} \, + \, E(
  a)^{2} E ( b^{2} ) \, - \, (E(a))^{2} (E ( b ))^{2}
\end{equation}
and:
\begin{equation}
  E( a b^{2} a ) \; = \; E( a^{2} )   E( b^{2} )
\end{equation}
The latter identity shows in particular that free relation
precludes commutativity.

Indeed one has that:
\begin{corollary} \label{cor:on the vanishing variance of free commuting random variables}
\end{corollary}
ON THE VANISHING VARIANCE OF FREE COMMUTING RANDOM VARIABLES:
\begin{equation}
  ( a,b \text{ free } \; and \; [ a \, , \, b ] = 0 ) \;
  \Rightarrow \; ( Var(a) = 0 \; or \; Var(b) = 0 )
\end{equation}
\begin{proof}
By Corollary\ref{cor:joint moment of free random variables} and
commutativity one has that:
\begin{equation}
  E (  ( a - E(a)I )^{2} \, ( b - E(b) I )^{2} ) \; = \; 0
\end{equation}
from which the thesis immediately follows.
\end{proof}

\smallskip

\begin{example} \label{ex:the free flavour of the qubits' strings  Hilbert space}
\end{example}
THE FREE FLAVOUR OF THE QUBITS' STRINGS' HILBERT SPACE

Fock spaces have a deep connection with freeness we want here to
stress, starting from the following:
\begin{theorem} \label{th:Fock spaces as Hilbert spaces of free groups}
\end{theorem}
FOCK SPACES AS HILBERT SPACES OF FREE GROUPS

\begin{hypothesis}
\end{hypothesis}
\begin{equation*}
  {\mathcal{H}} \: : \: \text{ Hilbert space } \, : \, dim( {\mathcal{H}}
  ) = n
\end{equation*}
\begin{thesis}
\end{thesis}
\begin{equation*}
  {\mathcal{F}} ( {\mathcal{H}} ) \; = \; l^{2} ( F_{n} )
\end{equation*}

\smallskip

that implies, in particular, that the qubit-strings' Hilbert
space is nothing but the Hilbert space of the 2-ranked free group:
\begin{equation}
  {\mathcal{H}}_{2}^{ \bigotimes \star} \; = \; l^{2}( F_{2} )
\end{equation}

Let us formalize a bit the analysis of Fock spaces, considering
as prototype the qubit-strings' Hilbert space we are mainly
interested to.

\begin{definition} \label{def:vacuum vector of the qubit strings' Hilbert space}
\end{definition}
VACUUM VECTOR OF $ {\mathcal{H}}_{2}^{ \bigotimes \star} $:

the unit vector $ | \Phi > $ such that:
\begin{equation}
  {\mathcal{H}}_{2}^{\bigotimes 0} \; = \; {\mathbb{C}} | \Phi >
\end{equation}
\begin{definition} \label{def:number operator of the qubit strings' Hilbert space}
\end{definition}
(QUBITS') NUMBER OPERATOR:

the positive self-adjoint operator $ \hat{N} $ on $
{\mathcal{F}}_{S}( {\mathcal{H}}_{2}) $ such that $
{\mathcal{H}}_{2}^{\bigotimes n} $ is an eigensubspace
corresponding to the eigenvalue n ($ n \in {\mathbb{Z}}^{+} $).

\smallskip

Given $ | \psi > \, \in \, {\mathcal{H}}_{2} $:

\begin{definition} \label{def:creation operator w.r.t. to a qubit's state}
\end{definition}
CREATION OPERATOR W.R.T. $ | \psi >  $:
\begin{equation}
  c^{\star} ( | \psi >  )  | \eta > \; := \;
  \begin{cases}
    | \psi> & \text{if $ | \eta > = | \Phi > $ }, \\
    | \psi> \bigotimes | \eta > & \text{if $ < \eta | \Phi > \, = \, 0 $}.
  \end{cases}
\end{equation}
\begin{definition} \label{def:distruction operator w.r.t. to a qubit's state}
\end{definition}
DISTRUCTION OPERATOR W.R.T. $ | \psi >  $:

the operator $ c ( | \psi >  ) $

\smallskip

Let us now consider the noncommutative probability space $ (
{\mathcal{B}}( {\mathcal{H}}_{2}^{\bigotimes \star} ) \: , \:
\omega( \cdot ) \, := \, Tr[ \rho_{ | \Phi > < \Phi |} \cdot ] )$.

One has that:
\begin{theorem} \label{th:freeness of creation and distruction operators}
\end{theorem}
FREENESS OF CREATION AND DISTRUCTION OPERATORS OF A FAMILY OF
ORTHOGONAL STATES

\begin{hypothesis}
\end{hypothesis}
\begin{equation*}
  \{ | h_{i} > \in {\mathcal{H}}_{2} \}_{i \in I} \; : \;  < h_{i} | h_{j}
  > \; = \; 0 \, \, \forall i \neq j
\end{equation*}
\begin{thesis}
\end{thesis}
\begin{equation*}
  \{ c( | h_{i} > )^{\star}  \, , \, c( | h_{i} > ) \}_{i \in I} \text{ are in free-relation}
\end{equation*}

\smallskip

\begin{remark} \label{rem:unsovable word problems and freeness}
\end{remark}
UNSOLVABLE WORD PROBLEMS AND FREENESS:

Yet in the middle fifthies W.W. Boone and P.S. Novikov discovered
the existence of groups with recursively-unsolvable word-problem.

This happens, for example, for the group $ G \; =  <
{\mathcal{X}} |  {\mathcal{R}} > $:
\begin{equation}
   {\mathcal{X}}    \; := \; \{ a ,b ,d ,e , p ,q, r , t ,k \}
\end{equation}
\begin{multline}
 {\mathcal{R}} \; := \; \{ p^{10} a \; = \; a p \; , \;  p^{10} b  \; = \; b p \; , \;  p^{10} c  \; = \; c
  p \; , \;  p^{10} d  \; = \; d p \; , \;  p^{10} e  \; = \; e
  p \\
  q a \; = \; a q^{10} \; , \;  q b \; = \; b q^{10} \; , \;  q c \; = \; c
  q^{10} \; , \;  q d \; = \; d q^{10} \; , \;  q e \; = \; e
  q^{10} \\
  r a \; = \; a r \; = \;  r b \; = \; b r \; = \;  r c \; = \; c r \; =
  \;  r d \; = \; d r \; = \;  r e \; = \; e r \\
  p a c q r \; = \; r p c a q \; = \; p^{2} a d q^{2} r \; = \; r
  p^{2} d a q^{2} \; = \; p^{3} b c q^{3} r \; = \; r p^{3} c b
  q^{3}  \\
   p^{4} b d q^{4} r \; = \; r p^{4} d b q^{4} \; ,
  \; p^{5} c e q^{5} r \; = \; r p^{5} e c a q^{5} \\
  p^{6} d e q^{6} r \; = \; r p^{6} e d b q^{6} \; , \; p^{7} c d
  c q^{7} r \; = \; r p^{7} c d c e q^{7} \; \\
   p^{8} c a a a q^{8} r \; = \; r p^{8} a a a q^{8} \; , \; p^{9} d a a a q^{9}
  r \; = \; r p^{9} a a a q^{9} \\
  p t \; = \; t p \; , \; q t \; = \; t q  \; , \, p k \; = \; k p
  \; , \; q k \; = \; k q \; , \; k ( a a a )^{-1 } t ( a a a ) \;
  = \; ( a a a )^{- 1} t ( a a a ) k \}
\end{multline}
Considered the noncommutative probability space $ ( {\mathcal{L}}
(G) \, , \, \tau ) $, couldn't the unsolvability of the word
problem of G result in the recursive undecidability of freeness
for a suitable collection of noncommutative random variables?

\smallskip

Since we have seen that there exist (at least) two possibilities
of formulating a quantum analogue of the
constraint\ref{con:independent sequence of tosses of a classical
coin} we will pursue both:
\begin{enumerate}
  \item insisting in adopting the notion of \textbf{independence}
  also in the quantum case, we arrive at the following:

\begin{constraint} \label{con:independent sequence of tosses of a quantum coin}
\end{constraint}
\textsf{INDEPENDENCE-CONSTRAINT ON THE NOTION OF A RANDOM SEQUENCE
ON THE NONCOMMUTATIVE ALPHABET $ \Sigma_{NC} $ :}

\emph{Making infinite \textbf{independent} trials of the
experiment consisting on tossing a \textbf{quantum coin} we must
obtain an algorithmically-random sequence of qubits with
certainty}

Let us try to formalize this analysis; denoted by $ RANDOM(
\Sigma_{NC}^{\infty}) $ the space of algorithmically-random
sequences of qubits, exactly as constraint\ref{con:independent
sequence of tosses of a classical coin} results in the condition:
\begin{equation}
  p_{\text{Chaitin randomness}} \; \in \;
  {\mathcal{P}}_{C}^{TYPICAL}[ ( \Sigma^{\infty} \, , \, \tau_{unbaised} ) ]
\end{equation}
where:
\begin{equation}
  p_{\text{Chaitin randomness}} ( \bar{x} ) \; := \; <<  \bar{x}
  \: \in \: RANDOM(\Sigma^{\infty} ) >>
\end{equation}
the constraint\ref{con:independent sequence of tosses of a quantum
coin} results in the condition:
\begin{equation}
  p_{\text{noncommutative randomness}} \; \in \;
  {\mathcal{P}}_{C}^{TYPICAL}[ ( \Sigma_{NC}^{\infty} \, , \, \tau_{unbaised} ) ]
\end{equation}
where:
\begin{equation}
  p_{\text{noncommutative randomness}} ( \bar{x} ) \; := \; <<  \bar{x}
  \: \in \: RANDOM(\Sigma_{NC}^{\infty} ) >>
\end{equation}
  \item assuming that the sequence of quantum coin tosses must be  \textbf{free} instead of
  \textbf{independent}, one must give up the sequence $ \{ c_{i}
  \} $ replacing it with a sequence $  \{ \tilde{c}_{i}
  \} $ of free random variables.

  The more natural way of doing this is passing through the notion
  of \textbf{free product} of noncommutative probability spaces.

  Given two groups $ G_{1} \; := \; < \chi_{1} | {\mathcal{R}}_{1}
  > $ and $ G_{2} \; := \; < \chi_{2} | {\mathcal{R}}_{2}
  > $:
\begin{definition} \label{def:free product of groups}
\end{definition}
FREE PRODUCT OF $ G_{1} $ AND $ G_{2} $:
\begin{equation}
  G_{1} \, \ast \,  G_{2} \; := \; \frac{< \chi_{1} \bigcup \chi_{2}
  | {\mathcal{R}}_{1} \bigcup {\mathcal{R}}_{2} >}{{\mathcal{R}}_{1} \bigcap {\mathcal{R}}_{2}}
\end{equation}
\begin{example} \label{ex:free groups as free products}
\end{example}
FINITE FREE GROUPS AS FREE PRODUCTS

The set $ {\mathbb{Z}} $ may be seen as the 1-rank free group:
\begin{equation}
  {\mathbb{Z}} \; =  F_{1} \; = \;  < \Sigma_{1} | - >
\end{equation}
The higher rank free groups may be obtained simply by free
products:
\begin{equation}
  F_{n} \; = \; \ast_{i=1}^{n} {\mathbb{Z}}
\end{equation}

\smallskip

A different equivalent way of characterizing free products is
through the so called Universality Property; let us recall that
(cfr. the second chapter of \cite{Grillet-99}) given a  group G:

\begin{definition}
\end{definition}
COMMUTATOR SUBGROUP OF G:
\begin{equation}
  [ G \, , \, G ] \; := \; \{ [ x , y ] \, := \; x y x^{-1} y^{-1}
  \, , \, x , y \in G \}
\end{equation}

\begin{theorem} \label{th:universality property for directs sums of groups}
\end{theorem}
UNIVERSALITY PROPERTY FOR DIRECTS SUMS OF GROUPS
\begin{multline}
  G \; = \; \bigoplus_{i \in I} G_{i} \; \Leftrightarrow \;
  \forall \{ \varphi_{i} : G_{i} \rightarrow G \}_{i \in I} \text{ family of homomorphims
  } \, : \, [ \varphi_{j} ( x_{j} ) \, , \, \varphi_{k} ( x_{k} )
  ] = e \, \forall j \neq k \\
  \exists ! \varphi \bigoplus_{i \in I} G_{i} \, \rightarrow \, G
  \, : \, \varphi \circ i_{i} \, = \, \varphi_{i}
\end{multline}

Replacing in theorem\ref{th:universality property for directs sums
of groups} the range of the homomorphisms from G to its
commutator subgroup $ G' $ and removing the commutativity
condition, one results in the following:
\begin{theorem} \label{th:universality property for free products of groups}
\end{theorem}
UNIVERSALITY PROPERTY FOR FREE PRODUCTS OF GROUPS:
\begin{multline}
  G \; = \; \ast_{i \in I} G_{i} \; \Leftrightarrow  \; \exists \{ i_{i} :  G_{i} \rightarrow G\}_{i \in I} \text{ family of
  homomorphims}\, : \\
  ( \forall \{ \varphi_{i} : G_{i} \rightarrow G' \}_{i \in I} \text{
family of homomorphims
  } \, \\
  \exists ! \varphi \ast_{i \in I} G_{i} \, \rightarrow \, G
  \, : \, \varphi \circ i_{i} \, = \, \varphi_{i} )
\end{multline}
The free product of algebras may be introduces in  a similar way
(cfr. the third chapter of \cite{Cuculescu-Oprea-94}):

given a family of algebras $ \{ A_{i} \}_{i \in I} $ and an
algebra A
\begin{definition} \label{def:free product of algebras}
\end{definition}
A IS THE FREE PRODUCT OF  $ \{ A_{i} \}_{i \in I} \; (  A \; :=
\; \ast_{i \in I} \{ A_{i} \} ) $:
\begin{multline}
 \exists \{ i_{i} :  A_{i} \rightarrow A \}_{i \in I} \text{ family of
  homomorphims with $ i_{i}(I) = I $} : \\
  \forall B \, algebra \; , \; \forall  \{ \varphi_{i} : A_{i} \rightarrow B \}_{i \in I} \text{
family of homomorphims: $ f_{i}(I) = I $} \\
\exists ! h : A \mapsto B \; homomorphism \; : \; h \circ i_{i}
\, = \, f_{i}
\end{multline}

Definition\ref{def:free product of algebras} constructs the free
product in the category of algebras.

The definition of free product in the category of  Von Neumann
algebras requires the free product of algebra must be endowed of a
rule concerning involution and a norm. Furthermore a completion
procedure must be carried off.

While the rule concerning involution will be explicitly specified
in the through the theorem we are going to introduce , we won't
discuss the other sophisticated technicalities concerning the
norm and the completion, demanding the interested reader to the
section7.1 of \cite{Hiai-Petz-00}.

Let us  consider two algebraic probability spaces $ ( A_{1} \, ,
\, \varphi_{1} ) $ and $ ( A_{2} \, , \, \varphi_{2} ) $.
Decomposed $ A_{i} $ as $ A_{i} \; = \; {\mathbb{C}} I \bigoplus
A_{i}^{0} $, where:
\begin{equation}
  A_{i}^{0} \; := \; \{ a_{i} \in A_{i} : \varphi_{i} ( a ) \, =
  \, 0 \}
\end{equation}
it may be proved that:
\begin{theorem} \label{th:direct sum expansion of the free product algebra}
\end{theorem}
DIRECT SUM EXPANSION OF THE FREE PRODUCT ALGEBRA:
\begin{multline}
  A_{1} \, \ast \, A_{2} \; = \; {\mathbb{C}}I \, \oplus \,
  \bigoplus \{ A_{i(1)}^{0} \bigotimes A_{i(2)}^{0} \bigotimes
  \cdots \bigotimes A_{i(n)}^{0} \, : \\
   i(k) \in \{ 1,2 \} , 1 \leq k \leq n \, , \, i(k) \neq i(k+1) 1
  \leq k \leq n-1 \, , \, n \in {\mathbb{N}} \} \; = \\
  = \; {\mathbb{C}}I \, \bigoplus \, A_{1}^{0}  \, \bigoplus \,
  A_{2}^{0} \, \bigoplus \, (  A_{1}^{0} \bigotimes  A_{2}^{0} ) \\
 \bigoplus
  \, (  A_{2}^{0} \bigotimes  A_{1}^{0} ) \bigoplus (  A_{1}^{0} \bigotimes
  A_{2}^{0} \bigotimes A_{1}^{0} ) \bigotimes  (  A_{2}^{0} \bigotimes
  A_{1}^{0} \bigotimes A_{2}^{0} )  \bigotimes \\
   \bigotimes (  A_{1}^{0} \bigotimes
  A_{2}^{0} \bigotimes A_{1}^{0} \bigotimes A_{2}^{0} ) \\
   \bigoplus \,  (  A_{2}^{0} \bigotimes
  A_{1}^{0} \bigotimes A_{2}^{0} \bigotimes A_{1}^{0} ) \, \bigoplus
  \, \cdots
\end{multline}

Let us explicitly analyze the products' structure of $  A_{1} \,
\ast \, A_{2} $:

if $ a_{i} \in A_{i}^{0} \; i \in \{ 1 , 2 \} $, then their
product $ a_{1} \, \cdot \, a_{2} $ is $ a_{1} \, \bigotimes \,
a_{2} \; \in \; A_{1}^{0} \bigotimes  A_{2}^{0} $.

More generally:
\begin{multline}
  ( x \, = \, \bigotimes_{k=1}^{n} a_{k} \, , \, a_{k} \in A_{i(k)}^{0}  \, , \, i(k) \in \{ 1 , 2 \} ,
  1 \leq k \leq n ) \; and \;  ( y \, = \, \bigotimes_{l=1}^{m}
  b_{l} \,  \\
    b_{l} \in A_{j(l)}^{0} \, , \,  j(l) \in \{ 1 , 2 \} ,
  1 \leq l \leq m )   \; and \\
   (i(n) \neq j(1)) \; \Rightarrow x \cdot y \; = \; \bigotimes_{k=1}^{n} a_{k} \, \bigotimes  \, \bigotimes_{l=1}^{m}
  b_{l}
\end{multline}
If $ i(n) \, = \, j(1) $, anyway, $ x \cdot y $ can't be obtained
simply concatenating the two tensor products, since, in this case
it is not true, in general, that $ a_{n} b_{1} \in A_{i(n)}^{0}
$; precisely, decomposed $ a_{n} b_{1} $ as:
\begin{equation}
  a_{n} b_{1} \; = \; \lambda I + a \; \; a \in A_{i(n)}^{0}
\end{equation}
one has that:
\begin{equation}
  x \cdot y \; = \; \bigotimes_{k=1}^{n-1} a_{k} \bigotimes a \bigotimes_{l=2}^{m}
  b_{l}  \, + \, \lambda \bigotimes_{k=1}^{n-1} a_{k} \bigotimes_{l=2}^{m}
  b_{l}
\end{equation}
The promised definition of the involution  in the $
W^{\star}$-algebra $ A_{1} \ast A_{2} $ is straightforward:
\begin{equation}
  x^{\star} \; = \; \bigotimes_{k=1}^{n} a_{n - k}
\end{equation}

\begin{example} \label{ex:free product of right Von Neumann algebras}
\end{example}
FREE PRODUCT OF GROUPS VON NEUMANN ALGEBRAS

Given two discrete groups $ G_{1} $ and $ G_{2} $ the left (right)
group Von Neumann algebra w.r.t. their free product is equal to
the free product of their left (right) group Von Neumann algebras,
i.e.:
\begin{align}
    {\mathcal{L}} & ( G_{1} \, \ast \, G_{2}  ) \; = \; {\mathcal{L}}
  (  G_{1}  ) \, \ast \, {\mathcal{L}} (  G_{2}  ) \\
   {\mathcal{R}} & ( G_{1} \, \ast \, G_{2}  ) \; = \; {\mathcal{R}}
  (  G_{1}  ) \, \ast \, {\mathcal{R}} (  G_{2}  )
\end{align}

\smallskip

Let us now introduce the following:
\begin{definition} \label{def:free product of algebraic probability spaces}
\end{definition}
FREE PRODUCT OF THE ALGEBRAIC PROBABILITY SPACES $ ( A_{1} \, ,
 \, \varphi_{1} ) $ AND $ ( A_{2} \, ,
 \, \varphi_{2} ) $:
\begin{equation}
  ( A_{1} \, , \, \varphi_{1} ) \, \ast \,  ( A_{2} \, , \, \varphi_{2}
  ) \; := \; ( A_{1} \ast A_{2} \, , \, \varphi )
\end{equation}
with:
\begin{equation}
  \varphi ( \lambda I \, + \, \sum \bigotimes_{k=1}^{n} a_{k} ) \;
  := \; \lambda
\end{equation}
The free product $ \ast_{i \in I}  ( A_{i} \, , \, \varphi_{i} )
$ of an arbitrary collection $ \{ ( A_{i} \, , \, \varphi_{i} )
\} $ of algebraic probability spaces may then be easily obtained
by definition\ref{def:free product of algebraic probability
spaces} by iteration.

Each $ A_{i} $ is embedded in $ \ast_{i \in I} A_{i} $ by the
following:
\begin{definition}
\end{definition}
CANONICAL EMBEDDING of $ A_{i} $ IN $ \ast_{i \in I} A_{i} $:

the map $ i_{i} \, : \, A_{i} \; \mapsto \; \ast_{i \in I} A_{i}
$:
\begin{equation}
  i_{i} ( a_{i} ) \; := \; \varphi_{i} ( a_{i} ) I  \, \bigoplus
  \, (  a_{i} \, - \,  \varphi_{i} ( a_{i} ) I )
\end{equation}
Exactly as the notion of independence of classical random
variables may be based on the notion of product of classical
probability spaces, the notion of freeness of algebraic random
variable may be based on the notion of free product of
noncommutative probability spaces through the following:
\begin{theorem} \label{the:freeness through the free product of algebraic probability spaces}
\end{theorem}
FREENESS THROUGH FREE PRODUCT OF ALGEBRAIC PROBABILITY SPACES:

\begin{hypothesis}
\end{hypothesis}
\begin{equation*}
  ( A \, , \, \varphi ) \text{ noncommutative probability space}
\end{equation*}
\begin{equation*}
  \{ A_{i} \}_{i \in I} \text{ family of subalgebras of A }
\end{equation*}
\begin{equation*}
  ( B \, , \, \omega ) \; := \; \ast_{i \in I} (  A_{i} \, \varphi_{| \;
  A_{i}} )
\end{equation*}
\begin{thesis}
\end{thesis}
\begin{multline*}
  \{ A_{i} \}_{i \in I}  \text{ are free} \; \Leftrightarrow \;
  \varphi( \prod_{k=1}^{n} a_{k} ) \; = \\
   \omega( \prod_{k=1}^{n} i_{i(k)}( a_{k} )) \; \; a_{k} \in A_{i(k)} \, ,
  \, i(k) \in I \, , \, i(k) \neq i(k+1) \, , \, 1 \leq k \leq n-1
  \, , \, n \in {\mathbb{N}}
\end{multline*}
\end{enumerate}

\smallskip

Let us now repeat the construction of remark\ref{ex:the
hyperfinite finite continuous factor} using free products instead
of tensor products:

so, given the one dimensional lattice $ {\mathbb{Z}} $, let us
attach to the $ n^{th} $ lattice site the one-qubit $
W^{\star}$-algebra:
\begin{equation}
  A_{n} \; := \; M_{2} ({\mathbb{C}}) \, n \in {\mathbb{Z}}
\end{equation}
Given an arbitary set of sites $ \Lambda \subseteq {\mathbb{Z}}
$, let us define:
\begin{equation}
  ( A_{\Lambda}^{free} \, , \, \tau_{\Lambda}^{free} ) \; := \; \ast_{n \in
  \Lambda} ( A_{n} \, , \, \tau_{2} )
\end{equation}
Let us then introduce the following $ W^{\star}$-algebra:
\begin{equation}
   R_{free} \; := \;  \pi_{ \tau_{{\mathbb{Z}}}} ( A_{{\mathbb{Z}}}^{free} ) ''
\end{equation}

We can now try to implement mathematically the following:
\begin{constraint} \label{con:free sequence of tosses of a quantum coin}
\end{constraint}
\textsf{FREENESS-CONSTRAINT ON THE NOTION OF A RANDOM SEQUENCE ON
THE NONCOMMUTATIVE ALPHABET $ \Sigma_{NC} $ :}

\emph{Making infinite \textbf{free} trials of the experiment
consisting on tossing a \textbf{quantum coin} we must obtain an
algorithmically-random sequence of qubits with certainty}

\smallskip

The quantum coin tossing at time $ n \in {\mathbb{N}} $ will be
then represented by a noncommutative random variable $
\tilde{c}_{n} \in A_{n}^{free} $ such that $ \{ \tilde{c}_{n} \} $
is a free sequence or, more precisely, by the associated free
 sequence obtained in the passage to $ R_{free} $.

The mathematical implementation of constraint\ref{con:free
sequence of tosses of a quantum coin} is, anyway, affected by the
following two issues:
\begin{enumerate}
  \item it would be natural, at this, point, to think that constraint\ref{con:free
sequence of tosses of a quantum coin} is translated mathematically
by the statement:
\begin{equation}
  P_{unbaised} [ RANDOM (  R_{free}  ) ] \; = \; 1
\end{equation}
Anyway it must be observed that:
\begin{enumerate}
  \item since:
\begin{equation}
  Type [ R_{free} ] \; = \; III
\end{equation}
the unbiased noncommutative probability distribution doesn't exist
  \item also supposed to bypass such a problem, translating
  mathematically the constraint\ref{con:free
sequence of tosses of a quantum coin} by  a suitable statement:
\begin{equation}
  F [ RANDOM (  R_{free}  ) ] \; = \; 0
\end{equation}
it is not clear how to express it in terms of $
\Sigma_{NC}^{\infty} $
\end{enumerate}
A possible way to overcome the first problem may consist in
passing from the family of noncommutative random variables $ \{
\bar{c}_{n} \} $ to the family of noncommutative random variables
$ \{ av_{n}  \} $ defined as:
\begin{equation}\label{eq:average of the free quantum coin tossing}
   av_{n} \; := \; \frac{1}{\sqrt{n}} \sum_{i=1}^{n} \tilde{c}_{i}
\end{equation}
and investigating the typical properties of the sequence $ \{
av_{n} \} $ w.r.t. the central limit distribution.

This immediately leads to analyze the relationship between the
probabilistic-quantum-information and
alghorithmic-quantum-information, as we will discuss in the next
section.
\end{enumerate}

\smallskip

\begin{remark} \label{rem:about algorithmically-random matrices}
\end{remark}
ABOUT ALGORITHMICALLY-RANDOM MATRICES:

Most of the success of Voiculescu's Free Probability Theory and
the interest about it shown by theoretical physicists is owed to
the fact that random matrices of the most commom ensembles
(cfr.\cite{Pastur-00} or the section 10.3 of \cite{Itzykson-Drouffe-89b} for a review), are asymptotically free.

Given a classical probability space  $ CPS \, :=  \, ( X \, , \,
P ) $:
\begin{definition}
\end{definition}
ENSEMBLE W.R.T. CPS OF ORDER n:

the noncommutative probability space:
\begin{equation}
 ENSEMBLE[ CPS \, , \, n ] \; := \; (  RANDOM-MATRICES[ CPS \, , \, n ] \, , \, e\tau_{n} )
\end{equation}
where:
\begin{equation}
  RANDOM-MATRICES[ CPS \, , \, N ] \; := \; \{ a \in M_{n} (
  {\mathbb{C}}) \, : M_{k}( a_{i,j}) \in \bigcap_{1 \leq p \leq
  \infty} L^{p}(X , P ) \}
\end{equation}
\begin{equation}
  e\tau_{n} (a) \; := \; E( \tau_{n} (a) ) \; = \; \frac{1}{n} \sum_{i=1}^{n} E(a_{i,i})
\end{equation}
\begin{definition}
\end{definition}
EMPIRICAL EIGENVALUE DISTRIBUTION OF THE RANDOM MATRIX $ a \in
ENSEMBLE[ CPS \, , \, n ] $:

the random atomic measure:
\begin{equation}
  \mu_{emp} (a) \; := \; \frac{1}{n} \sum_{i=1}^{n} \delta(
  \lambda_{i}(a))
\end{equation}
where $ \lambda_{1}(a) \, , \, \cdots \, , \, \lambda_{n}(a) $
are the eigenvalues of a.

\smallskip

\begin{definition}
\end{definition}
MEAN EIGENVALUE DISTRIBUTION OF THE RANDOM MATRIX $ a \in
ENSEMBLE[ CPS \, , \, n ] $:
\begin{equation}
  \mu_{mean} (a)  \; := \; E ( \mu_{emp} (a) )
\end{equation}

\smallskip

\begin{example}
\end{example}
THE GAUSSIAN ORTHOGONAL ENSEMBLE (GOE) AND THE  GAUSSIAN
UNITARY ENSEMBLE (GUE)

Let us consider the random matrices' ensemble  describing a
random $ n \times $ real-entries symmetric matrix a such that $ \{
a_{i,j} \, : \, 1 \leq i \leq j \leq n \} $ is a family of
independent real Gaussian random variables such that:
\begin{align}
  E(a_{ij}) & \; = \; 0 \\
   E(a_{ij}^{2}) & \; = \; \frac{1 + \delta_{ij}}{n+1}
\end{align}

The variance has been chosen so that:
\begin{equation}
  e\tau_{n}( a^{2}) \; = \; 1
\end{equation}
Such an ensemble is usually called the \textbf{n-order gaussian
orthogonal ensemble (n-GOE)} , the name being owed to the
orthogonal invariance of the underlying classical probability
measure.

\smallskip

Another very important random matrices' ensemble is that
describing a random $ n \times $ complex self-adjoint matrix a
such that:
\begin{enumerate}
  \item $ \{ Re(a_{i,j}) \, : \, 1 \leq i \leq j \leq n \} \, \bigcup \, \{ Im(a_{i,j}) \, : \, 1 \leq i \leq j \leq n \}
  $ is an independent family of Gaussian random variables
  \item
\begin{align}
  E(a_{ij}) & \; = \; 0 \; \;  1 \leq i \leq j \leq n  \\
   E(a_{ii}^{2}) & \; = \; \frac{1}{n} \; \;  1 \leq i \leq n \\
   E( (\Re (a_{ij})^{2})) & \; = \; E( (\Im (a_{ij})^{2}))  \; =
   \; \frac{1}{2 n}  \; \;  1 \leq i \leq j \leq n
\end{align}
where the normalization has been chosen again in order that:
\begin{equation}
  e\tau_{n}( a^{2}) \; = \; 1
\end{equation}
\end{enumerate}
Such an ensemble is usually called the \textbf{n-order gaussian
unitary ensemble (n-GUE)}, the name being owed to the unitary
invariance of the underlying classical probability measure.

\smallskip

Let us now consider a sequence $ \{ ( A_{n} \, , \, \omega_{n} )
\}_{ n \in {\mathbb{N}}} $ of noncommutative probability spaces
and a sequence $ \{ a^{(n)}_{1} \, , \, \cdots  \, , \,
a^{(n)}_{k} \}_{ n \in {\mathbb{N}}} $ of $ k^{ples}$ of
noncommutative random variables over the $ \{ ( A_{n} \, , \,
\omega_{n} ) \}_{ n \in {\mathbb{N}}} $'s.

\begin{definition} \label{def:asymptotic freeness}
\end{definition}
THE NONCOMMUTATIVE RANDOM VARIABLES $ \{ a^{(n)}_{1} \, , \,
\cdots  \, , \,  a^{(n)}_{k} \}_{ n \in {\mathbb{N}}} $ ARE
ASYMPTOTICALLY FREE:
\begin{equation}
  E( \prod_{i=1}^{k} a^{(n)}_{i} ) \; = \; \sum_{i=1}^{k} \sum_{1 \leq k_{1} \leq \cdots \leq k_{r} \leq
  k} ( - 1 )^{r+1} E ( a^{(n)}_{k_{1}} ) \cdots E ( a^{(n)}_{k_{r}} ) E (
  a^{(n)}_{1} \cdots \hat{a}^{(n)}_{k_{1}} \cdots \hat{a}^{(n)}_{k_{r}} \cdots
  a^{(n)}_{k}) \, + \, O( \frac{1}{n})
\end{equation}
where $ \hat{\cdot} $ indicates again terms that are omitted.

It may be proved that \cite{Hiai-Petz-00}:
\begin{theorem} \label{th:asymptotic freeness of independent GOE's random matrices}
\end{theorem}
ASYMPTOTIC FREENESS OF INDEPENDENT GOE'S RANDOM MATRICES

\begin{hypothesis}
\end{hypothesis}
\begin{equation*}
   a^{(n)}_{1} \, , \, \cdots  \, , \, a^{(n)}_{n} \text{ independent n-GOE's random matrices}
\end{equation*}
\begin{thesis}
\end{thesis}
\begin{equation*}
  a^{(n)}_{1} \, , \, \cdots  \, , \, a^{(n)}_{n} \text{ are asymptotically free}
\end{equation*}
Theorem\ref{th:asymptotic freeness of independent GOE's random
matrices} gives a noncommutative-probabilistic foundation to the
following celebrated:
\begin{corollary} \label{cor:Wigner's theorem}
\end{corollary}
WIGNER'S THEOREM:

\begin{hypothesis}
\end{hypothesis}
\begin{equation*}
  a^{(n)} \, \in \, n-GOE
\end{equation*}
\begin{thesis}
\end{thesis}
\begin{equation*}
  d-\lim_{n \rightarrow\infty}  \mu_{mean} ( a^{(n)}) \; = \;  sc_{STANDARD}
\end{equation*}
\begin{proof}
Given $  b^{(n)}_{1} \, , \, \cdots  \, , \, b^{(n)}_{n} $
independent n-GOE's random matrices one has that the distribution
of the random matrix:
\begin{equation*}
  \frac{\sum_{k=1}^{n} b^{(n)}_{k}}{\sqrt{n}}
\end{equation*}
is the same as that of  $ a^{(n)} $.

The thesis immediately follows from theorem\ref{th:asymptotic
freeness of independent GOE's random matrices} and
theorem\ref{th:freeness' central limit theorem}
\end{proof}
\newpage
\section{From the communicational-compression of the Quantum Coding Theorems to the algorithmic-compression in Quantum Computation} \label{sec:From the communicational-compression of the Quantum Coding Theorems to the algorithmic-compression in Quantum Computation}
Among the many successes of Quantum Information Theory must be
certainly acknoledged the Quantum Coding Theorems, i.e. the
Schumacher's Theorem ruling the upper bound for the compression of
quantum information in a noiseless quantum-channel, and the
Holevo-Schumacher-Westmoreland Theorem ruling the upper bound for
the capacity of noisy quantum-channels used to transmit classical
information  \cite{Winter-99}, \cite{Nielsen-Chuang-00}, \cite{Werner-01}.

As we will show, from a structural point of view Coding Theorems, both classical and quantum, are
an application of \textbf{Algebraic Large Deviations' Theory}, a noncommutative generalization of \textbf{ Classical Large Devations's Theory} \cite{Varadhan-84}, \cite{Dembo-Zeitouni-98}.

Let us begin to introduce the necessary stuff starting from the
Central Limit Theorems we informally introduced in the
remark\ref{rem:random walks on free groups}.

\begin{theorem} \label{th:independence's central limit theorem}
\end{theorem}
INDEPENDENCE'S  CENTRAL LIMIT THEOREM:

\begin{hypothesis}
\end{hypothesis}
\begin{equation*}
  ( A \, , \, \omega ) \text{ algebraic probability space }
\end{equation*}
\begin{equation*}
  \{ a_{n} \}_{n \in {\mathbb{N}}_{+}}  \text{ sequence of \textbf{independent}
  algebraic random variables on $ ( A \, , \omega ) $}
\end{equation*}
\begin{equation*}
   E(a_{i}) \, = \, 0 \; , \;  Var(a_{i}) \, = \, 1 \; \; i \in {\mathbb{N}}_{+}
\end{equation*}
\begin{equation*}
  M_{k} ( a_{i} ) \; < \; + \infty \; \; \forall i,k \in {\mathbb{N}}_{+}
\end{equation*}
\begin{thesis}
\end{thesis}
\begin{equation*}
  d-\lim_{n \rightarrow \infty }  \frac{1}{\sqrt{n}}
  \sum_{i=1}^{n} a_{i} \; = \; gauss_{STANDARD}
\end{equation*}
where $ d-\lim $ denotes convergence in distribution.

\begin{theorem} \label{th:freeness' central limit theorem}
\end{theorem}
FRENESS' CENTRAL LIMIT THEOREM:

\begin{hypothesis}
\end{hypothesis}
\begin{equation*}
  ( A \, , \, \omega ) \text{ algebraic probability space }
\end{equation*}
\begin{equation*}
  \{ a_{n} \}_{n \in {\mathbb{N}}_{+}}  \text{ sequence of \textbf{free}
  algebraic random variables on $ ( A \, , \omega ) $}
\end{equation*}
\begin{equation*}
   E(a_{i}) \, = \, 0 \; , \;  Var(a_{i}) \, = \, 1 \; \; i \in {\mathbb{N}}_{+}
\end{equation*}
\begin{equation*}
  M_{k} ( a_{i} ) \; < \; + \infty \; \; \forall i,k \in {\mathbb{N}}_{+}
\end{equation*}
\begin{thesis}
\end{thesis}
\begin{equation*}
  d-\lim_{n \rightarrow \infty }  \frac{1}{\sqrt{n}}
  \sum_{i=1}^{n} a_{i} \; = \; sc_{STANDARD}
\end{equation*}
\begin{remark} \label{ref:the exponential decay of the deviation from the central limit}
\end{remark}
THE EXPONENTIAL DECAY OF THE DEVIATION FROM THE CENTRAL LIMIT

We have seen in chapter\ref{chap:Classical algorithmic randomness
as satisfaction of all the classical algorithmic typical
properties} that the Law of Randomness $ p_{\text{Borel normality
of order 1}} $ stating the Law of Large Numbers is not the whole
story.

Also the way in which such an asymptotic behaviour is reached, as
ruled by suitable Laws of Randomness, such as $ p_{\text{iterated
logarithm }} $ or $ p_{\text{infinite recurrence }} $, is
essential to the Foundation of Probability Theory in the sense of
remark\ref{rem:on why Classical Probability Theory applies to
reality}.

This situation has an immediate translation in terms of the
convergence in distribution  of the averages $ \frac{1}{\sqrt{n}}
\sum_{i=1}^{n} a_{i} $: not only the asymptotic behaviour, as
stated by the Central Limit Theorems, is important, but also the
way such asymptotic distribution is reached.

It is here where the Large Deviation Principle appears:

informally speaking, if it does exist a two-argument functional $
R(d_{A} ,d_{B} ) $ over the probability distributions, said a
\textbf{rate functional}, such that the probability that the
distribution $ d_{n} $ of the average $ \frac{1}{\sqrt{n}}
\sum_{i=1}^{n} a_{i} $ deviates from its central limit $
d_{\text{central limit}} $ decays exponentially as:
\begin{equation}
  Probability [ d_{n} \text{deviates from  } d_{\text{central
  limit}} ] \; \sim \; \exp [ - n \, R(d_{n} ,d_{\text{central
  limit}} ) ]
\end{equation}
then we will say that the convergence to the central limit obeys
the Large Deviation Principle w.r.t. the rate functional R.

In such situation a Probabilistic-information Coding Theorem
naturally occurs, since realizing a transmission coding neglecting
the exponentially improbable messages one obtains a trasmissional
compression of information with asymptotically vanishing
probability of error.

The maximum possible compression is ruled by the rate functional R

\smallskip

Let us now formalize these considerations starting from the
classical case \cite{Dembo-Zeitouni-98}.

Given a topological space X:

\begin{definition} \label{def:rate function}
\end{definition}
RATE FUNCTION OVER X:

a map $ I \in MAP[ X \, , \, [ 0 , \infty ) ] $:
\begin{equation}
  \{ x \in X \, : \, I(x)  \leq \alpha \} \text{ is closed } \;
  \forall \alpha \in [ 0 , \infty )
\end{equation}

Let us now consider a family of probability measures $ \{
\mu_{\epsilon} \} $ over X having all the same halting set:
\begin{equation}
  HALTING( \mu_{\epsilon} ) \; = \; {\mathcal{B}}
\end{equation}
where $ {\mathcal{B}} $ is a certain $ \sigma $-algebra over X,
not necessary equal to the Borel-$\sigma$-algebra of X.

We will say that:
\begin{definition} \label{def:large deviation principle}
\end{definition}
$ \{ \mu_{\epsilon} \} $ SATISFIES THE LARGE DEVIATION PRINCIPLE
WITH RATE FUNCTION I:
\begin{equation}
  - \inf_{x \in \Gamma^{0}} I(x) \; \leq \; \liminf_{\epsilon \rightarrow
  0} \epsilon \log \mu_{\epsilon} ( \Gamma ) \; \leq \; \limsup_{\epsilon \rightarrow
  0} \epsilon \log \mu_{\epsilon} ( \Gamma ) \; \leq \; - \inf_{x \in \bar{\Gamma}} I(x) \; \; \forall \Gamma \in   {\mathcal{B}}
\end{equation}
where $ \Gamma^{0} $ denotes the interior of $ \Gamma $ while $
\bar{\Gamma} $ denotes the closure of $ \Gamma $.

\smallskip

Denoted by  $ P _{\mu} $ the probability law $ \mu^{\infty} $ on $
\Sigma^{\infty} $ associated to a sequence of independent,
identically distributed random variables distributed following $
\mu \in {\mathcal{D}} ( \Sigma ) $, let us now introduce the
following notions:
\begin{definition} \label{def:type of a string}
\end{definition}
TYPE OF $ \vec{x} \in \Sigma^{n} $:

the probability measure $ L_{n}^{\vec{x}} \in {\mathcal{D}}
(\Sigma) $:
\begin{equation}
  L_{n}^{\vec{x}} ( i ) \; = \; \frac{ N_{i}(\vec{x}) }{n}
\end{equation}

We will denote the set of all possible types of strings of length
n by $ {\mathcal{L}}_{n} $, i.e.:
\begin{equation} \label{eq:set of types for strings of length n}
  {\mathcal{L}}_{n} \; := \; \{ L_{n}^{\vec{x}} \, , \, \vec{x}
  \in \Sigma^{n} \}
\end{equation}
Let us then denote by $ L_{n}^{\vec{X}} $ the random element of $
{\mathcal{L}}_{n} $ associated with the random string $ \vec{X} $.

Then:
\begin{theorem} \label{th:Sanov's theorem}
\end{theorem}
SANOV'S THEOREM:

The family of laws $ P_{\mu} ( L_{n}^{\vec{Y}} \in \cdot ) $
satisfies the Large Deviation Principle with rate function $
S_{Kullback-Leibler} ( \cdot | \mu ) $

\smallskip

where:
\begin{definition} \label{def:Kullback-Leibler relative entropy}
\end{definition}
KULLBACK-LEIBLER RELATIVE ENTROPY OF $ \mu $ WITH RESPECT TO $
\nu $:
\begin{equation}
  S_{Kullback-Leibler}( \mu , \nu ) \; := \;
  \begin{cases}
    \int_{X}  d \mu (x) &  \log \frac{ d \mu (x) }{ d \nu (x) }  \text{if $ \mu \prec \nu$}, \\
    + \infty & \text{otherwise}.
  \end{cases}
\end{equation}
$ \prec $ being the \textbf{absolute continuity} partial-ordering
relation among classical measures.

\smallskip

Theorem\ref{th:Sanov's theorem} lies at the heart of the proof of
theorem\ref{th:mathematically classical mathematically physical
noiseless coding theorem}.

To show it let us observe that:
\begin{enumerate}
  \item Shannon's entropy of definition\ref{def:Shannon entropy of a
distribution} may be considered as a derived quantity arising
from the Kullback-Leibler relative entropy of
definition\ref{def:Kullback-Leibler relative entropy}, as is
stated by the commutative case of a general procedure of
Algebraic Probability Theory by which probabilistic-information
may be derived from the notion of relative
probabilistic-information,i.e. relative entropy (cfr. the paper
"\emph{From relative entropy to entropy"} at pagg.149-158 of
\cite{Thirring-98} and the homonymous sixth chapter of
\cite{Ohya-Petz-93}):

given a $ W^{\star} $-algebra  A acting on an Hilbert space $
{\mathcal{H}} $:
\begin{definition} \label{def:Araki relative entropy}
\end{definition}
ARAKI'S RELATIVE ENTROPY OF $ \omega_{1} \in S(A) $ W.R.T.
$ \omega_{2} \in S(A) $:
\begin{equation}
  S_{Araki}( \omega_{1} \, , \,  \omega_{2} ) \; := \;
  \begin{cases}
    - ( \log \Delta ( \phi, & \omega'_{| \psi >}  ) | \psi >  , | \psi >  )   \text{if $ | \psi > \in supp(\phi)$}, \\
    + \infty & \text{otherwise}.
  \end{cases}
\end{equation}
where $ | \psi > \in {\mathcal{H}} $ is an  (arbitrary) auxiliary
vector and $ \omega'_{| \psi >} \in S( A' ) $ is the state on the
commutant $ A' $ of A induced by the state $ \omega_{| \psi
>} \in S( A ) $ associated to $ | \psi
> $.

One has that:

\begin{theorem}
\end{theorem}
ARAKI'S RELATIVE ENTROPY IS A NONCOMMUTATIVE GENERALIZATION OF
KULLBACK-LEIBLER RELATIVE ENTROPY:

\begin{hypothesis}
\end{hypothesis}
\begin{equation*}
  \mu , \nu \text{ probability measures on X: } HALTING(  \mu ) \,
  = \, HALTING(  \nu )
\end{equation*}
\begin{thesis}
\end{thesis}
\begin{equation*}
  S_{Araki}( \omega_{\mu} \, , \,  \omega_{\nu} ) \; = \;  S_{Kullback-Leibler}( \mu , \nu )
\end{equation*}

Definition\ref{def:Araki relative entropy} may sound rather
exotic; if $ \omega_{1} $  and $\omega_{2} $ are normal states,
anyway, it reduced to the more used Umegaki's expression:

\begin{theorem} \label{th:on Umegaki's relative entropy}
\end{theorem}
ON UMEGAKI'S RELATIVE ENTROPY:
\begin{multline*}
  \omega , \varphi \in S_{n} (A) \; \Rightarrow \\
   S_{Araki}( \omega_{1} , \omega_{2} )  \; = \; S_{Umegaki} ( \rho_{\omega} ,
\varphi_{\phi} ) \; = \; Tr [  \rho_{\omega} ( \log_{2}
\rho_{\omega} \, - \,   \log_{2} \rho_{\phi} ) ]
\end{multline*}
\begin{equation*}
  S_{Umegaki} ( \rho_{\omega} ,
\rho_{\varphi} ) \; := \;
  \begin{cases}
    Tr [  \rho_{\omega} ( \log_{2} \rho_{\omega} \, - \,   \log_{2} \rho_{\varphi} ) ] & \text{if $ supp( \rho_{\omega} ) \leq supp( \rho_{\varphi} ) $ }, \\
    + \infty & \text{otherwise}.
  \end{cases}
\end{equation*}

One of the most important tools in Quantum Information Theory is
the following:

\begin{theorem} \label{th:second Uhlmann's theorem}
\end{theorem}
SECOND UHLMANN'S THEOREM (MONOTONICITY'S THEOREM FOR RELATIVE
ENTROPY)

\begin{hypothesis}
\end{hypothesis}
\begin{equation*}
  A \, , \, B \; \; W^{\star}-algebras
\end{equation*}
\begin{equation*}
  \alpha \in CPU( A , B)
\end{equation*}
\begin{thesis}
\end{thesis}
\begin{equation*}
  S_{Araki}( \omega_{1} \circ \alpha  , \omega_{2}  \circ \alpha )
  \; \leq \; S_{Araki}( \omega_{1} , \omega_{2} )  \; \; \forall
  \omega_{1} ,  \omega_{2} \in S(A)
\end{equation*}

Another important property of Araki's relative entropy is the following:
\begin{theorem} \label{th:positivity of Araki's relative entropy}
\end{theorem}
POSITIVITY OF ARAKI'S RELATIVE ENTROPY:
\begin{enumerate}
  \item
\begin{equation}
  S_{Araki}( \omega_{1} , \omega_{2} ) \; \geq \; 0 \; \; \forall \omega_{1} , \omega_{2} \in S_{n}(A)
\end{equation}
  \item
  \begin{equation}
  ( \, S_{Araki}( \omega_{1} , \omega_{2} ) \, = \, 0 \; \Leftrightarrow \; \omega_{1} \, = \, \omega_{2} \, )  \; \; \forall \omega_{1} , \omega_{2} \in S_{n}(A)
\end{equation}
\end{enumerate}

\smallskip

\begin{remark} \label{rem:the link between quantum probabilistic information and quantum algorithmic information}
\end{remark}
THE LINK BETWEEN QUANTUM PROBABILISTIC INFORMATION AND QUANTUM
ALGORITHMIC INFORMATION

The \textbf{probabilistic approach} and  the \textbf{algorithmic
approach} to Information Theory, trying to focalize the same
object from different point of views, are strictly linked.

As to the cell $ C_{M}-C_{\Phi}$ of the diagram\ref{di:diagram of
computation} such a link is explicitly formalized by
theorem\ref{th:link between mathematically classical
mathematically physical probabilistic information and
mathematically classical mathematically physical algorithmic
information}.

It is very natural to suppose that an analogous situation occurs
as to the cell $ NC_{M}-NC_{\Phi}$,  in particular when the
underlying physical theory is Quantum Physics, so that   $
NC_{\Phi} $ may be read not only as \textbf{physically
not-classical}, but, equivalently, as \textbf{physically
not-commutative}.

It must be then possible to use such a link to sharp  the concept
and properties  of quantum algorithmic information.

\smallskip

\begin{definition} \label{def:decompositions of a state on a W-star algebra}
\end{definition}
DECOMPOSITIONS OF $ \omega \in S(A) $:
\begin{multline}
  DEC( \omega ) \; := \; \{ \{ ( \lambda_{i} ,  \omega_{i} ) \}_{i=1}^{n} \, :  \,  \lambda_{i} \in [ 0 , 1 ] \, , \, , \omega_{i} \in S(A) \, , \,  i = 1 , \cdots , n  \: : \\
   \sum_{i=1}^{n} \lambda_{i} = 1 \: , \: n \in {\mathbb{N}} \}
\end{multline}
\begin{definition} \label{def:extremal decompositions of a state on a W-star algebra}
\end{definition}
EXTREMAL DECOMPOSITIONS  $ \omega \in S(A) $:
\begin{equation}
   DEC_{EXT}( \omega ) \; := \; \{ \{ ( \lambda_{i} ,  \omega_{i} ) \} \in  DEC( \omega )  \, :  \,  \omega_{i} \in \Xi(A) \, \forall i \}
\end{equation}
\begin{definition} \label{def:othogonal decompositions of a state on a W-star algebra}
\end{definition}
ORTHOGONAL DECOMPOSITIONS OF $ \omega \in S(A) $:
\begin{equation}
  DEC_{\bot}( \omega ) \; := \; \{ \{ ( \lambda_{i} ,  \omega_{i} ) \} \in  DEC( \omega )  \, :  \,  supp( \omega_{i} )  \; \perp \;  supp( \omega_{j}
  ) \; \; \forall i \neq j \}
\end{equation}

\begin{example} \label{ex:Schatten decompositions of a normal state}
\end{example}
SCHATTEN DECOMPOSITIONS:

Given the density operator $ \rho_{\omega} \in {\mathcal{D}} ( {\mathcal{H}} )  $ of a normal state $ \omega \in S_{n}(A) $  on  a Von Neumann algebra $ A \, \subseteq \, {\mathcal{B}}( {\mathcal{H}}) $
acting on a separable Hilbert space $ {\mathcal{H}} $, let us consider the sequence $ 1 \, \leq \, \rho_{1} \, \leq \, \rho_{2} \, \leq \, \cdots \, \leq \, 0 $
of its eigenvalues, repeated according to their multiplicity , let us introduce the following:
\begin{definition} \label{def:Schatten decompositions of a normal state}
\end{definition}
SCHATTEN DECOMPOSITIONS OF $ \omega $:

\begin{multline}
  DEC_{Schatten} ( \omega ) \; := \; \{ {\mathcal{E}} = \{ \rho_{i} \, , \, | e_{i} > < e_{i} | \} \, \in \, DEC_{\bot}  ( \omega ) \bigcap DEC_{EXT}  ( \omega )  \, : \\
   \{ | e_{i} > \} \text{  is an orthogonal basis of eigenvectors of  $ \rho_{\omega} $} \}
\end{multline}

\smallskip

\begin{definition} \label{def:entropy of a state over a W-star algebra}
\end{definition}
ENTROPY OF $ \omega \in S(A) $:
\begin{equation}
  S( \omega) \; := \; \sup\{ \sum_{i} \lambda_{i} S ( \omega_{i} , \omega ) \, : \, \{ ( \lambda_{i} , \omega_{i} )  \} \, \in \, DEC(A) \}
\end{equation}

One has that:
\begin{theorem} \label{th:invariance of a state's entropy under inner automorphisms}
\end{theorem}
INVARIANCE OF A STATE'S ENTROPY UNDER INNER AUTOMORPHISMS:
\begin{equation}
  S( \alpha_{\star} ( \omega )) \; = \; S ( \omega ) \; \; \forall \alpha \in INN(A) \, , \,  \forall \omega \in S(A)
\end{equation}
\begin{theorem} \label{th:not-monotonicity of a state's entropy under channels}
\end{theorem}
NOT-MONOTONICITY OF A STATE'S ENTROPY UNDER CHANNELS:
\begin{enumerate}
  \item
\begin{equation}
   \alpha \in CPU(A) \, , \, \omega \in S(A) \; \nRightarrow \; S( \alpha_{\star} ( \omega )) \; = \; S ( \omega )
\end{equation}
  \item
\begin{equation}
   \alpha \in CPU(A) \, , \, \omega \in S(A) \; \nRightarrow \; S( \alpha_{\star} ( \omega )) \; > \; S ( \omega )
\end{equation}
  \item
\begin{equation}
   \alpha \in CPU(A) \, , \, \omega \in S(A) \; \nRightarrow \; S( \alpha_{\star} ( \omega )) \; < \; S ( \omega )
\end{equation}
\end{enumerate}

Theorem\ref{th:positivity of Araki's relative entropy} immediately implies the following:
\begin{theorem} \label{th:positivity of the entropy a state}
\end{theorem}
POSITIVITY OF A STATE'S ENTROPY:
\begin{enumerate}
  \item
\begin{equation}
  S( \omega ) \; \geq \; 0 \; \; \forall \omega \in S(A)
\end{equation}
  \item
\begin{equation}
   S( \omega ) \; = \; 0 \, \Leftrightarrow \, \omega \in \Xi(A)
\end{equation}
\end{enumerate}

Another important property of the entropy functional is the following:
\begin{theorem} \label{th:concavity of the entropy a state}
\end{theorem}
CONCAVITY OF THE ENTROPY FUNCTIONAL:
\begin{equation}
  S ( \sum_{i=1}^{n} \lambda_{i} \omega_{i} ) \; \geq \; \sum_{i=1}^{n}  \lambda_{i} S (  \omega_{i} ) \; \; \forall \{ ( \lambda_{i} \, , \, \omega_{i} ) \}_{i=1}^{n} \in DEC(\omega)
\end{equation}

One has furthermore that:

\begin{theorem}
\end{theorem}
THE ENTROPY OF A STATE IS A NONCOMMUTATIVE GENERALIZATION OF
SHANNON'S ENTROPY:

\begin{hypothesis}
\end{hypothesis}
\begin{equation*}
  \mu  \text{ probability measure on X}
\end{equation*}
\begin{thesis}
\end{thesis}
\begin{equation*}
  S( \omega_{\mu} ) \; = \;  H_{Shannon}( \mu )
\end{equation*}

\begin{theorem} \label{th:on Von Neumann's entropy}
\end{theorem}
ON VON NEUMANN'S ENTROPY:
\begin{multline*}
  \omega \in S_{n} (A) \; \Rightarrow \\
   S( \omega )  \; = \; S_{Von \; Neumann} ( \rho_{\omega}
   ) \; := \; - Tr ( \rho_{\omega} \log_{2} \rho_{\omega} )
\end{multline*}

In 1956 E.T. Jaynes proved what in our algebraic setting may be stated as the following: \cite{Petz-01}:
\begin{theorem} \label{th:Jaynes' theorem}
\end{theorem}
JAYNES' THEOREM:

\begin{hypothesis}
\end{hypothesis}
\begin{equation*}
   \omega \in S_{n} (A)
\end{equation*}
\begin{thesis}
\end{thesis}
\begin{equation*}
  {\mathcal{E}} \text{ is optimal } \; \; \forall {\mathcal{E}} \in DEC_{Schatten} ( \omega )
\end{equation*}
that immediately implies that:
\begin{corollary} \label{cor:von Neumann's entropy as the Shannon's entropy of the eigengvalues}
\end{corollary}
VON NEUMANN'S ENTROPY AS THE SHANNON'S ENTROPY OF THE EIGENVALUES

\begin{hypothesis}
\end{hypothesis}
\begin{equation*}
   \omega \in S_{n} (A)
\end{equation*}
\begin{thesis}
\end{thesis}
\begin{equation*}
  S( \omega ) \; = \; H ( \{ \rho_{i} \} )
\end{equation*}
\begin{remark} \label{rem:from the variational principle of minimum free energy to the variational principle of maximum entropy}
\end{remark}
FROM THE VARIATIONAL PRINCIPLE OF MINIMUM FREE ENERGY TO THE VARIATIONAL PRINCIPLE OF MAXIMUM ENTROPY

Having at last introduced Von Neumann's entropy we can discuss some conseguences of theorem\ref{th:positivity of Araki's relative entropy}.
Given a closed quantum dynamical system with observables' algebra the space $ {\mathcal{B}}( {\mathcal{H}} ) $ of all bounded
operators over an Hilbert space $ {\mathcal{H}} $  such that:
\begin{equation}
  cardinality_{NC} ( {\mathcal{B}}( {\mathcal{H}} ) ) \; < \; \aleph_{0}
\end{equation}
and dynamics described by the strongly-continuous one-parameter family
of $ {\mathcal{B}}( {\mathcal{H}} ) $'s inner automorphisms generated by the hamiltonian $ h \in   ({\mathcal{B}}( {\mathcal{H}} )) _{sa} $,
let us introduce the following notions:
\begin{definition}
\end{definition}
CANONICAL STATE W.R.T. h AND  $ \beta \in [ 0 \, , \, \infty ] $:
\begin{multline*}
  \omega^{(CAN)}_{ h , \beta} \, \in \, S ( {\mathcal{B}}( {\mathcal{H}} ) ) \\
   \omega^{(CAN)}_{ h , \beta} ( a ) \; := \; \frac{ e^{ - \beta \, h} \, a }{Tr e^{ - \beta \, h}}
\end{multline*}
\begin{definition} \label{def:free energy of  a state}
\end{definition}
FREE ENERGY OF $ \omega  \in S ( {\mathcal{B}}( {\mathcal{H}} ) ) $ W.R.T.  h AND  $ \beta \in [ 0 \, , \, \infty ] $:
\begin{equation}
  F_{ h , \beta} ( \omega ) \; := \; Tr ( \rho_{\omega} h ) \, - \, \frac{1}{ \beta} S( \omega )
\end{equation}
Then one has the following:
\begin{theorem} \label{th:variational principle of minimum free energy}
\end{theorem}
VARIATIONAL PRINCIPLE OF MINIMUM FREE ENERGY:
\begin{equation}
  F_{ h , \beta} (  \omega^{(CAN)}_{ h , \beta} ) \; \leq \;  F_{ h , \beta} ( \omega ) \; \; \forall \omega \in  S ( {\mathcal{B}}( {\mathcal{H}} ) )
\end{equation}
\begin{proof}
Let us observe that:
\begin{equation}
  S_{Araki} ( \omega^{(CAN)}_{ h , \beta}  \, , \, \omega  ) \; = \; F_{ h , \beta} ( \omega ) \, - \,  F_{ h , \beta} ( \omega ) ( \omega^{(CAN)}_{ h , \beta} ) \; \; \forall \omega \in  S ( {\mathcal{B}}( {\mathcal{H}} ) )
\end{equation}
Applying theorem\ref{th:positivity of Araki's relative entropy} the thesis follows.
\end{proof}

\begin{corollary}
\end{corollary}
VARIATIONAL PRINCIPLE OF MAXIMUM ENTROPY:
\begin{equation}
 ( Tr ( \rho_{\omega} h ) \, = \, Tr ( \rho_{\omega^{(CAN)}_{ h , \beta}} h ) ) \; \Rightarrow \;(  S( \omega ) \, \leq \, S ( \omega^{(CAN)}_{ h , \beta} )
\end{equation}
\begin{proof}
Theorem\ref{th:variational principle of minimum free energy} says that  the free entropy of  $ \omega^{(CAN)}_{ h , \beta} $ is less of equal to the free antropy of any
other state; so, in  particular, it is less or equal to the free entropy of any other state having the same energy.

By the definition\ref{def:free energy of  a state} the thesis immediately follows
\end{proof}

  \item the key point in the proof of theorem\ref{th:mathematically classical mathematically physical
noiseless coding theorem} is the Asymptotic Equipartition
Property.

Given a probability distribution $ P \in {\mathcal{D}} ( \Sigma )
$ let us introduce the following:
\begin{definition} \label{def: typical set of a distribution}
\end{definition}
TIPICAL SET OF P W.R.T. $ ( n \, , \, \epsilon ) $:
\begin{equation}
  A_{\epsilon}^{(n)} \; := \; \{ \vec{x} \in \Sigma^{n} \, : \,
  P^{n} ( \vec{x} ) \, \in \, ( 2^{- n ( H(P) + \epsilon ) } \, , \,
  2^{- n ( H(P) - \epsilon )} ) \}
\end{equation}
where we have denoted by $ P^{n} $ the product measure $
\times_{i=1}^{n} P $.

One has that:
\begin{theorem} \label{th:asymptotic equipartition property}
\end{theorem}
ASYMPTOTIC EQUIPARTITION PROPERTY:

\begin{enumerate}
  \item
\begin{equation*}
  H(P) - \epsilon \; \leq \; - \frac{1}{n} \log_{2} P^{n} ( \vec{x}
  ) \; \leq \; H(P) + \epsilon \; \; \forall \vec{x} \in
  \Sigma^{n} \, , \, \forall \epsilon \in {\mathbb{R}}_{+}
\end{equation*}
  \item
\begin{multline*}
  \forall \epsilon \in {\mathbb{R}}_{+} \, \exists N \in
  {\mathbb{N}}_{+} \; : \\
   P^{n} (  A_{\epsilon}^{(n)} ) \, > \, 1
  - \epsilon \; \; \forall n  > N
\end{multline*}
  \item
\begin{equation*}
  cardinality( A_{\epsilon}^{(n)} ) \; \leq \; 2^{n ( H(P) + \epsilon
  )} \; \; \forall \epsilon \in {\mathbb{R}}_{+} , \forall n \in {\mathbb{N}}_{+}
\end{equation*}
\begin{multline*}
   \forall \epsilon \in {\mathbb{R}}_{+} \, \exists N \in
  {\mathbb{N}}_{+} \; : \\
  cardinality( A_{\epsilon}^{(n)} ) \; \geq \; ( 1 - \epsilon )  2^{n ( H(P) - \epsilon
  )} \; \; \forall n  > N
\end{multline*}
\end{enumerate}

\begin{remark}
\end{remark}
THE RELEVANCE OF THE ASYMPTOTIC EQUIPARTITION PROPERTY FOR  DATA
COMPRESSION

The fact that theorem\ref{th:asymptotic equipartition property}
immediately implies theorem\ref{th:mathematically classical
mathematically physical noiseless coding theorem} may be
intuitivelly understood observing that is states substantially
that asymptotically:
\begin{itemize}
  \item the set $ \Sigma^{n} $, made of $ 2^{n} $ strings,  is the union
  of roughly $ 2^{n H(P)} $  \textbf{typical strings} and  $ 2^{n} - 2^{n H(P)}
  $ \textbf{nontypical strings}
  \item the \textbf{typical strings} are equiprobable, each
  having a probability roughly equal to $ 2^{- n H(P)} $
  \item the \textbf{nontypical strings} have a roughly vanishing
  probability of occurring
\end{itemize}
The lexicographic ordering on $ \Sigma^{n} $ induces, clearly, a
total ordering on $  A_{\epsilon}^{(n)} $.

Let us thus define a code D such that:
\begin{itemize}
  \item
\begin{equation}
  D( \vec{x} ) \; := \; \uparrow \; \; \forall  \vec{x} \in
  \Sigma^{n} -  A_{\epsilon}^{(n)}
\end{equation}
  \item D assign to each typical sequence its ordering-number
\end{itemize}
For $ n \ \rightarrow \infty $ the average code-word length $
L_{D,P} $ tends to $ H(P) $ that is conseguentially the required
minimal number of cbit per letter

  \item theorem\ref{th:asymptotic equipartition property}  may be easily shown to derive from Sanov's Theorem.

  Following the eight section of \cite{Varadhan-84} let us
  introduce the following notion:
\begin{definition}
\end{definition}
EMPIRICAL PROBABILITY MEASURE INDUCED BY $ \vec{x} \in
\Sigma^{\star} $:

the probability measure $ R_{\vec{x}} $ on $ \Sigma^{\infty} $:
\begin{equation}
  R_{\vec{x}} \; := \; \frac{1}{ | \vec{x} | } \sum_{i=1}^{n} \delta_{\sigma^{i} \vec{x}^{\infty}}
\end{equation}

Denoted by $ Q_{n} $ the distribution of $  R_{\vec{x}} $ in the
space of all the classical shifts over $ \Sigma $,
theorem\ref{th:Sanov's theorem} implies that the sequence of
distributions $ \{ Q_{n} \}_{n \in {\mathbb{N}}} $ satisfies the
Large Deviation Principle with rate function $ I(P) \; := \;  -
\, H(P) $, implying theorem\ref{th:mathematically classical
mathematically physical noiseless coding theorem}

\end{enumerate}

Let us at last observe that it must be possible to translate the
link among $ C_{M}- C_{\Phi} $-probabilistic information  and $
C_{M}- C_{\Phi} $-algorithmic information stated by
theorem\ref{th:link between mathematically classical
mathematically physical probabilistic information and
mathematically classical mathematically physical algorithmic
information} looking at the satisfaction of a \textbf{Large
Deviation Principle}  as a  \textbf{Law of Randomness}.

Let us consider again the case of a sequence of independent
tosses of a classical coin, the throw at time n  being described
by the Bernoulli(1/2) random variable $ x_{n} $.

Since the averages of the first tosses  $ av_{n} \; := \;
\frac{1}{n} \sum_{i=1}^{n} x_{i} $ take values in the whole
interval $ [ 0 ,1 ]  $ the sequence $ \bar{av} \; := \; \{ av_{n}
\}_{n \in {\mathbb{N}} } $ belong to $ [ 0 ,1 ]^{\infty} $.

On each copy of $ [ 0 ,1 ] $ it is defined the central limit
distribution $ P_{central \; limit} \; := \; gauss_{standard} $.

Taking the direct product of all them we obtain the probability
measure $ P_{central \; limit}^{\infty} $ by which we can define
the classical probability space $ (  [ 0 ,1 ]^{\infty} \, , \,
P_{central \; limit}^{\infty} ) $.

Sanov's Theorem may then be  seen as  a Law of Randomness $
p_{Sanov} \in {\mathcal{L}}_{RANDOMNESS} [  (  [ 0 ,1 ]^{\infty}
\, , \, P_{central \; limit}^{\infty} ) ] $.

\begin{remark}
\end{remark}
THE COMPUTABILITY ISSUE INVOLVED IN  THE CHARACTERIZATION OF THE
LAWS OF RANDOMNESS OF $ [  (  [ 0 ,1 ]^{\infty} \, , \,
P_{central \; limit}^{\infty} ) ] $:

While the definition of the typical properties of  $ [  (  [ 0 ,1
]^{\infty} \, , \, P_{central \; limit}^{\infty} ) ] $ is
standard, the definition of the Laws of Randomness of such  a
classical probability space is a subtle issue since it involves
the specification of $ NC_{M}-C_{\Phi}-\Delta_{0}^{0} $ or $
NC_{M}-NC_{\Phi}-\Delta_{0}^{0} $ objects with the inherent
problematics we discussed in section\ref{sec:The distinction
between mathematical-classicality and physical-classicality} and
will further analyze in chapter\ref{sec:On absolute conformism in
Quantum Probability Theory}.

\smallskip

The next step consists in generalizing noncommutatively the
definition\ref{def:large deviation principle}.

Let us consider again the sequence of independent tosses of a
classical coin of constraint\ref{con:independent sequence of tosses of a classical coin}.

Sanov's Theorem states, informally speaking,  that in a
statistical inferential process in  which we estimate the
probability distribution of the coin from the experimental result
of n tosses, the probability that we don't distinguish the true
unbiased probability distribution $ P_{unbiased} \, = \,
Bernoulli( \frac{1}{2} ) $ from a biased one P decades
exponentially for large n as:
\begin{equation}
  P[ not-distinguish( P , P_{unbiased} ) ] \; \sim \; \exp [ - n
  S_{Kullback-Leibler} ( P , P_{unbiased} ) ]
\end{equation}
It appears then natural to consider the same issue as to
constraint\ref{con:independent sequence of tosses of a quantum
coin} analyzing what V. Vedral, M. B. Plenio and P.L. Knight call the Quantum Sanov's Theorem (cfr. the section 6.4.6 "Statistical Basis of Entanglement Measure" of \cite{Vedral-Plenio-Knight-00}).

This  requires, anyway, to afford the subtle issue of
\textbf{distinguishability of quantum states}.

A proper way of making this is to introduce some generalization
of definition\ref{def:entropy of a state over a W-star algebra}
\cite{Ohya-Petz-93}, \cite{Ingarden-Kossakowski-Ohya-97},
\cite{Ohya-98}.

Given a sub-$W^{\star}$-algebra $ A_{accessible} \, \subset \, A
$ of a $W^{\star}$-algebra A:
\begin{definition} \label{def:entropy of  a subalgebra  of a W-star algebra w.r.t. a state}
\end{definition}
ENTROPY OF THE SUBALGEBRA  $ A_{accessible} $  W.R.T. THE STATE $ \omega \in S(A) $ :
\begin{equation}
  H_{\omega}( A_{accessible} ) \; := \; \sup\{ \sum_{i} \lambda_{i} S ( \omega_{i}|_{A_{accessible}} , \omega|_{A_{accessible}} ) \, : \, \{ ( \lambda_{i} , \omega_{i} )  \} \, \in \, DEC(A) \}
\end{equation}

We have clearly that:
\begin{theorem}
\end{theorem}
\begin{equation}
  H_{\omega}( A ) \; = \;  S( \omega )
\end{equation}

Furthermore Uhlmann's Monotonicity's Theorem for relative
entropy, namely  theorem\ref{th:second Uhlmann's theorem},
immediately implies the following:
\begin{theorem}
\end{theorem}
MONOTONOCITY'S THEOREM FOR THE ENTROPY OF A SUBALGEBRA W.R.T. TO A STATE:
\begin{equation}
  A_{1} \, \subset \,  A_{2} \, \subset \, A \; \Rightarrow \; H_{\omega}( A_{1} ) \, \leq
  \, H_{\omega}( A_{2} )
\end{equation}

Considered  another $W^{\star}$-algebra B and a channel $ \alpha
\in CPU( B , A ) $:
\begin{definition} \label{def:entropy of a channel w.r.t. a state}
\end{definition}
ENTROPY OF THE CHANNEL $ \alpha $ W.R.T. $ \omega \in S(A) $  :
\begin{equation}
  H_{\omega}( \alpha ) \; := \; \sup\{ \sum_{i} \lambda_{i} S ( \omega_{i} \circ \alpha , \omega \circ \alpha ) \, : \, \{ ( \lambda_{i} , \omega_{i} )  \} \, \in \, DEC(A)  \}
\end{equation}

Uhlmann's Monotonicity's Theorem for relative entropy, namely
theorem\ref{th:second Uhlmann's theorem}, can be used again to
derive another monotonicity's property:
\begin{theorem}
\end{theorem}
MONOTONOCITY'S THEOREM FOR THE ENTROPY OF A CHANNEL W.R.T. TO A STATE:
\begin{equation}
   H_{\omega} ( \beta \circ \alpha ) \; \leq \;  H_{\omega}(  \alpha ) \; \; \forall \omega \in S(A) \, , \, \forall \alpha ,
   \beta \in CPU(A)
\end{equation}

Let us now introduce the following quantitity:
\begin{definition} \label{def:mutual entropy of a state over a W-star algebra and a channel}
\end{definition}
MUTUAL ENTROPY OF THE STATE $ \omega \in S(A) $ AND THE CHANNEL $ \alpha \in CPU(B,A) $:
\begin{equation}
  I( \omega \, ; \, \alpha) \; := \; \sup\{ \sum_{i} \lambda_{i} S ( \omega_{i} \circ \alpha , \omega \circ \alpha ) \, : \, \{ ( \lambda_{i} , \omega_{i} )  \} \, \in \, DEC_{\bot}(\omega) \}
\end{equation}
\begin{remark}
\end{remark}
THE ENTROPY OF A CHANNEL W.R.T. A  STATE  VERSUS THE MUTUAL ENTROPY
OF A STATE AND A CHANNEL

Definition\ref{def:entropy of a channel w.r.t. a state} and definition\ref{def:mutual entropy of a state over a W-star algebra and a channel} differ only in the
\textbf{orthogonality constraint} for the involved decompositions.

Not surprisingly the two notions are then  intimately related, as
it is stated by the following:
\begin{theorem}
\end{theorem}
\begin{enumerate}
  \item
  \begin{multline*}
   ( A \; commutative \; and \; card_{NC}(A) < \aleph_{0} ) \;
  \Rightarrow \; ( H_{\omega }( \alpha ) \, = \; I( \omega , \alpha
  ) \\
   \forall \omega \in  S(A) \, , \, \forall \alpha \in CPU(
  B,A) \, \, \forall B \text{ algebraic space} )
\end{multline*}
  \item
\begin{equation*}
    H_{\omega }( \alpha ) \; = \;\sup \{ I ( \mu , \beta \circ
   \alpha ) \, : \, \omega \, = \, \mu \circ \beta \}
\end{equation*}
where $ \beta $ runs over all channels $ A \, \mapsto \,  C $
where C is finite-dimensional and commutative while $ \mu $ runs
over all the states on C such that $  \omega \, = \, \mu \circ
\beta $
\end{enumerate}

Given a state $ \omega \in  S(A) $ and a decomposition of its $ {\mathcal{E}} \, := \, \{ ( \lambda_{i} , \omega_{i} )  \}_{i=1}^{n} \: \in \: DEC( \omega ) $:
\begin{definition} \label{def:Holevo information of a decomposition}
\end{definition}
HOLEVO INFORMATION OF THE DECOMPOSITION $ {\mathcal{E}} $:
\begin{equation}
  I_{Holevo} ({\mathcal{E}}) \; := \; S ( \sum_{i=1}^{n} \lambda_{i} , \omega_{i} ) \, - \, \sum_{i=1}^{n} \lambda_{i} S ( \omega_{i} )
\end{equation}
Once more Uhlmann's Monotonicity's Theorem for relative entropy, namely
theorem\ref{th:second Uhlmann's theorem}, can be used  to
derive another monotonicity's property:
\begin{theorem} \label{th:monotonicity's theorem for the Holevo's information of a decomposition}
\end{theorem}
MONOTONOCITY'S THEOREM FOR THE HOLEVO'S INFORMATION OF A DECOMPOSITION:
\begin{equation}
  I_{Holevo} ( \alpha_{\star} {\mathcal{E}} ) \; \leq \; I_{Holevo} ( {\mathcal{E}} ) \; \; \forall \alpha \in CPU(A)
\end{equation}
where the action of  the dual channel $ \alpha_{\star} $ of $ \alpha $ on the decomposition $  {\mathcal{E}} $ is simply
defined as:
\begin{equation}
  \alpha_{\star} (  \{ ( \lambda_{i} , \omega_{i} )  \}_{i=1}^{n} ) \; := \;   \{ ( \lambda_{i} , \alpha_{\star} \omega_{i} )  \}_{i=1}^{n}
\end{equation}

By theorem\ref{th:positivity of the entropy a state} one has that:
\begin{theorem} \label{th:Holevo's information and entropy}
\end{theorem}
HOLEVO'S INFORMATION AND ENTROPY:
\begin{equation*}
  I_{Holevo} (  {\mathcal{E}} ) \; = \; S ( \omega ) \; \; \forall {\mathcal{E}} \in DEC_{EXT}( \omega)
\end{equation*}

Furthermore theorem\ref{th:concavity of the entropy a state} immediately implies the following:
\begin{theorem} \label{th:positivity of Holevo's information}
\end{theorem}
POSITIVITY OF HOLEVO'S INFORMATION:
\begin{enumerate}
  \item
\begin{equation}
  I_{Holevo} ( {\mathcal{E}} ) \; \geq \; 0 \; \; \forall {\mathcal{E}} \in DEC( \omega) \, , \, \forall \omega \in S(A)
\end{equation}
  \item
\begin{multline}
  ( I_{Holevo} ( {\mathcal{E}} ) \;  = \; 0 \; \Leftrightarrow \; ( \lambda_{i} , \lambda_{j} > 0 \; \Rightarrow \omega_{i} = \omega_{j} ) ) \\
   \forall  {\mathcal{E}} = \{ ( \lambda_{i} , \omega_{i} )  \} \in DEC(\omega) \, , \, \forall \omega \in S(A)
\end{multline}
\end{enumerate}

\smallskip

Let us now introduce the following notion:
\begin{definition} \label{def:Shannon's entropy of a decomposition}
\end{definition}
SHANNON'S ENTROPY OF THE DECOMPOSITION $ {\mathcal{E}} \, := \, \{ ( \lambda_{i} , \omega_{i} )  \}_{i=1}^{n} $:
\begin{equation}
  H_{Shannon} ( {\mathcal{E}} ) \; := \; H( \{ \lambda_{i}  \}_{i=1}^{n} )
\end{equation}

\smallskip

We have at last all the ingredients required to afford the \textbf{Distiguishability Issue} in a sistematic way, clarifying at last the
reason of the locution \textbf{channel} adopted to denote a completely-positive unital map (cfr. definition\ref{def:CPU-maps among two W-star algebras})

Let us start , at this purpose, from the the following simple situation:

Alice adopts a communicational-channel to  send to Bob a letter A of the n-letters' alphabet $ \Sigma_{n} $, choosing the letter A to send according to
to a certain probability distribution $ \vec{p}^{(A)} \in {\mathcal{D}} ( \Sigma_{n} ) $.

Let us suppose that the communicational-channel is noisy so that the letter B received by Bob is a classical random variable with probability distribution $ \vec{p}^{(B)} \in {\mathcal{D}} ( \Sigma_{n} ) $  different from A.

\begin{definition} \label{def:joint entropy of two classical random variables}
\end{definition}
JOINT ENTROPY OF THE CLASSICAL RANDOM VARIABLES A AND B:
\begin{equation}
  H( A \, , \, B ) \; := \; - \sum_{a,b \in \Sigma_{n}} p(a,b) \log_{2} p(a,b)
\end{equation}
Let us introduce a useful property we will use in the sequel:
\begin{lemma} \label{lem:useful inequality on distributions}
\end{lemma}
\begin{equation}
  \sum_{k} p_{k} \log_{2} q_{k} \; \leq \; \sum_{k} p_{k} \log_{2} p_{k} \; \; \forall \vec{p} , \vec{q} \in {\mathcal{D}} ( \Sigma_{n} )
\end{equation}
\begin{proof}
It follows by the inequality:
\begin{equation}
  \log_{2} x \; \leq \; x \, - \, 1 \; \; \forall x \in {\mathbb{R}}_{+}
\end{equation}
poning $ x \; := \; \frac{q_{k}}{p_{k}} $
\end{proof}

As we already preannounced in section\ref{sec:Why prefix entropy is better than
simple entropy} the joint entropy has the following important property:
\begin{theorem} \label{th:subadditivity property of the joint entropy}
\end{theorem}
SUBADDITIVITY PROPERTY OF THE JOINT ENTROPY:
\begin{equation}
H( A \, , \, B ) \; \leq \;  H(A) \, + \, H(B)
\end{equation}
\begin{proof}
The thesis immediately follows by lemma\ref{lem:useful inequality on distributions} and the the definition
of the involved objects
\end{proof}

Introduced the following more convenient notation for the Kullback-Leibler's relative entropy:
\begin{definition} \label{def:conditional entropy of two classical random variables}
\end{definition}
CONDITIONAL ENTROPY OF THE CLASSICAL RANDOM VARIABLES A W.R.T. THE CLASSICAL RANDOM VARIABLE B:
\begin{equation}
  H( A \, | \, B ) \; := \; S_{Kullback-Leibler} ( \vec{p}^{(A)} , \vec{p}^{(B)} )
\end{equation}
one has the following:
\begin{theorem} \label{th:chain rule fo the joint entropy}
\end{theorem}
CHAIN RULE FOR THE JOINT ENTROPY:
\begin{equation}
  H( A \, , \, B ) \; = \; H(A) \, + \, H( B \, | \, A )
\end{equation}
\begin{proof}
Introduced the marginal distributions of A and B:
\begin{align} \label{eq:marginal distributions}
  p(a)  & \; := \;  \sum_{ a \in \Sigma_{n}} p(a,b)   \\
  p(b)  & \; := \;  \sum_{ b \in \Sigma_{n}} p(a,b)
\end{align}
one has that:
\begin{multline}
  H( A \, , \, B ) \; = \; - \sum_{ a \in \Sigma_{n}} \sum_{ b \in \Sigma_{n}} p(a,b) \log_{2} p(a,b)  \; = \; - \sum_{ a \in \Sigma_{n}} \sum_{ b \in \Sigma_{n}} p(a,b) \log_{2} p(a) p( b | a)  \\
  \; = \; - \sum_{ a \in \Sigma_{n}} \sum_{ b \in \Sigma_{n}}  p(a,b) \log_{2} p(a) \, - \, \sum_{ a \in \Sigma_{n}} \sum_{ b \in \Sigma_{n}}  p(a,b) \log_{2} p(b | a) \\
  \; = \; - \sum_{ a \in \Sigma_{n}}  p(a) \log_{2} p(a) \, - \,  \sum_{ a \in \Sigma_{n}} \sum_{ b \in \Sigma_{n}} p(a,b) \log_{2} p(b | a) \; = \; H(A) \, + \, H( B \, | \, A )
\end{multline}
\end{proof}
\begin{definition} \label{def:mutual information of two classical random variables}
\end{definition}
MUTUAL INFORMATION OF THE CLASSICAL RANDOM VARIABLES A AND B:
\begin{equation}
  I( A \, ; \, B ) \; := \; H(A) \, + \, H(B) \, - \, H( A , B )
\end{equation}
\begin{theorem}
\end{theorem}
POSITIVITY OF THE MUTUAL INFORMATION:
\begin{enumerate}
  \item
\begin{equation}
  I( A \, ; \, B ) \; \geq \; 0
\end{equation}
  \item
\begin{equation}
  I( A \, ; \, B ) \, = \, 0 \; \Leftrightarrow \; \text{ A and B are independent}
\end{equation}
\end{enumerate}
\begin{proof}
Applying theorem\ref{th:subadditivity property of the joint entropy} and definition\ref{def:mutual information of two classical random variables}
the thesis follows
\end{proof}

The joint entropy $ H( A \, , \, B ) $ quantifies the total uncertainty we have about the pair $ ( A \, , \, B ) $.

The mutual information $  I( A \, ; \, B ) $, instead, quantifies how much information X and Y have in common; in fact, $ H(X) \, +  \,H(Y) $ is equal
twice the information common to X and Y plus the non-common information of both the variables: hence the information common to X and Y may be obtained
subtracting from  $ H(X) \, +  \,H(Y) $ their joint entropy.

Let us now observe that the information  about A that can be obtained testing B is precisely the information that is common to A and B, namely $  H( A \, ; \, B ) $ as
it is stated formally by the following:
\begin{theorem}
\end{theorem}
ON GETTING INFORMATION ON A RANDOM VARIABLE TESTING ANOTHER RANDOM VARIABLE
\begin{equation}
  I( A \, ; \, B ) \; = \; H(B) \, - H( A | B )
\end{equation}
\begin{proof}
It is sufficient to apply theorem\ref{th:chain rule fo the joint entropy} at the r.h.s. of definition\ref{def:mutual information of two classical random variables}
\end{proof}

Let us now suppose that Alice tries to maximize the classical information she can trasmit to Bob sending a letter:
she can do it choosing the distribution $ \vec{p}^{(A)} $ in a clever way, i.e. so that it maximizes the mutual information  $ I( A \, ; \, B ) $.

Also with that choice, anyway, she can't transmit more than:
\begin{equation}
  C_{classical} \; := \; \max_{\vec{p}^{(A)} \in \Sigma_{n}} I( A \, ; \, B )
\end{equation}
cbit per letter.

Let us now observe that the number C depends only on the joint-distribution $ p(a,b ) $ defining the trasmission-channel
adopted by Alice and Bob and is called the \textbf{classical-capacity} of the channel.

Clearly $  C_{classical} $ is maximal when the transmission-channel is noiseless, i.e.:
\begin{equation}
  p(b \, | \, a ) \; = \; \delta_{a,b}
\end{equation}
In this case one has that the general expression for the mutual entropy in terms of the marginal distribution of eq. \ref{eq:marginal distributions}
\begin{equation}
  I( A \, ; \, B ) \; = \; \sum_{ a \in \Sigma_{n}} \sum_{ b \in \Sigma_{n}} p(a,b) \log_{2} \frac{p(a,b)}{p(a)p(b)}
\end{equation}
(one immediately derives from the definition\ref{def:mutual information of two classical random variables}) reduces to the entropy of A:
\begin{multline}
   I( A \, ; \, B ) \; = \; \sum_{ a \in \Sigma_{n}} \sum_{ b \in \Sigma_{n}} p(a,b) \log_{2} \frac{p(a) \,  p(b \, | \, a )}{p(a)p(b)} \\
   \; = \;  \sum_{ a \in \Sigma_{n}} \sum_{ b \in \Sigma_{n}} p(a) \,  \delta_{a,b} \log_{2} \frac{\delta_{a,b}}{p(b)} \; = \; - \sum_{ a \in \Sigma_{n}} p(a) \log_{2} p(a) \; = \; H(A)
\end{multline}
since no misunderstanding-error by Bob can occur, a fact that we
can express saying that Bob can \textbf{distinguish} the letters
of Alice's adopted alphabet $ \Sigma_{n} $.

\smallskip

Let us now observe that to say that the transmission channel is specified by the joint probability distribution $ p(a,b) $ is equivalent
to say that it is specified by the linear map $ \alpha_{\star} : {\mathcal{D}} ( \Sigma_{n} ) \, \mapsto \,  {\mathcal{D}} ( \Sigma_{n} ) $ such that:
\begin{equation}
 \alpha_{\star} ( \vec{p}^{(A)} ) \; = \; \vec{p}^{(B)}
\end{equation}
By theorem\ref{th:category isomorphism at the basis of Noncommutative Probability} we can in an equivalent way look at $ \alpha_{\star} $ as a map:
\begin{equation*}
   \alpha_{\star} \, : \, S ( L^{\infty} ( \Sigma_{n} \, , \, \vec{p}_{unbiased} )) \, \mapsto \, S ( L^{\infty} ( \Sigma_{n} \, , \, \vec{p}_{unbiased} ))
\end{equation*}
and, by theorem\ref{th:on the relationship between positivity and complete-positivity}, to infer that it is indeed the dual of a channel $ \alpha \in CPU(L^{\infty} ( \Sigma_{n} \, , \, \vec{p}_{unbiased} )) $.

We have thus shown that a $ C_{\Phi}$-classical transmission-channel  may be  seen as a channel in the meaning of definition\ref{def:CPU-maps on a W-star algebra}.

Let us observe, furthermore, that instead of speaking about the mutual entropy $ I ( A \; ; \; B ) $  of the  two classical random variables A and  B one could speak, in an equivalent way, of the mutual information $ I( A \, , \, \alpha ) $ of the random variable
A and the channel $ \alpha $, i.e. of the mutual entropy  $ I ( \omega_{\vec{P}^{A} } \; ; \; \alpha ) $  of the state $ \omega_{\vec{P}^{A}} \in L^{\infty} ( \Sigma_{n} \, , \, \vec{p}_{unbiased} ) $
and the channel $ \alpha \in CPU(L^{\infty} ( \Sigma_{n} \, , \, \vec{p}_{unbiased} )) $ in  the meaning specified by the definition\ref{def:mutual entropy of a state over a W-star algebra and a channel} of which it is indeed a
particular case.

Let us now suppose that to send her classical information, i.e. a letter of the commutative alphabet $ \Sigma_{n} $ Alice uses a quantum-communicational channel in the following way:

she codifies the letter $ i \in \Sigma_{n} $ through a certain state $  \omega_{i} \in S({\mathcal{A}})  $, where $ {\mathcal{A}} $ is a certain noncommutative space.

Clearly Alice's state of affairs is described by the following decomposition:
\begin{equation}
  {\mathcal{E}}^{(A)} \; := \; \{ p^{(A)}_{i} \, , \,  \omega_{i} \}_{i=1}^{n} \; \in \; DEC( \sum_{i=1}^{n} p^{(A)}_{i}  \omega_{i} )
\end{equation}
Let us now suppose that Alice transmit the state corresponding to the chosen letter through a noiseless quantum-transmission channel, mathematically described by the dual identity-channel $ {\mathbb{I}}_{\star} $.

So the state arrives unchanged to Bob who would like, through a suitable experimental process to
distinguish it in order of recovering the classical information Alice sent him.

To do this he makes a measurement, described by a suitable
observational channel $ \alpha \in CPU(C , {\mathcal{A}} ) $  on
the noncommutative space $ {\mathcal{A}} $.

Supposing that $ cardinality_{NC} (C) = m $ so  that the observational channel is determined by an m-ary partition of unity, Bob gets as output
a classical random variable $ j \in \Sigma_{m} $ with classical probability distribution $ \vec{p}^{(B)} \in {\mathcal{D}} (\Sigma_{m}) $.

The classical information that Bob obtains in this way is clearly given by  $ I ( A \, ; \, B ) $.

Clearly Bob we will make his measurement in order of maximizing
such a quantity; the maximal classical information he can gain,
i.e. the classical capacity of the adopted
quantum-transmission-channel, is then given by:
\begin{equation}
  C_{classical} \; = \; \max_{\alpha} \,  I ( A \, ; \, B )
\end{equation}

Let us now  formalize this analysis; given  an algebraic probability space $ ( A \, , \, \omega ) $ and a decomposition $ {\mathcal{E}} \, = \, \{ \lambda_{i} \, , \, \omega_{i} \}_{i=1}^{n} \in \, DEC(\omega) $:
\begin{definition} \label{def:accessible information of a decomposition}
\end{definition}
ACCESSIBLE CLASSICAL INFORMATION OF $ {\mathcal{E}} $:
\begin{equation}
  I_{acc} (  {\mathcal{E}} ) \; := \;  \max_{{\mathcal{V}} \, \in \, OPU(A)} I ( \{ \lambda_{i}  \}_{i=1}^{n} \, ; \,  \{ \omega ( \alpha_{j} )( {\mathcal{V}} )  \}_{j=1}^{card({\mathcal{V}})}) \}
\end{equation}
The \textbf{distinguishibility issue} may be completely analyzed in terms of the Groenwold-Lindblad's inequality, nowadays reknown as the:
\begin{theorem} \label{th:Holevo's bound}
\end{theorem}
HOLEVO'S BOUND:
\begin{equation}
  I_{acc} (  {\mathcal{E}} ) \; \leq \;  I_{Holevo} (  {\mathcal{E}}
  ) \; \; \forall  {\mathcal{E}} \in DEC( \omega )
\end{equation}
\begin{proof}

The thesis follows applying theorem\ref{th:monotonicity's theorem for the Holevo's information of a decomposition} w.r.t. the reduction channel
of the involved operational partitions of unity
\end{proof}

\begin{remark} \label{rem:indistinguishibility of nonothogonal states}
\end{remark}
INDISTINGUISHIBILITY OF NONORTHOGONAL STATES:

Considering again the previous communicational situation among
Alice and Bob, let us suppose that Alice codifies the $ i^{th} $
letter of the n-letters alphabet through a normal, pure state $
\omega_{i} $:
\begin{equation}
  \rho_{\omega_{i}} \; = \; | \phi_{i} > <  \phi_{i} | \; \; i \in \Sigma_{n}
\end{equation}
where $ \{ | \phi_{i} > \}_{i=1}^{n} $ is a collection of \textbf{mutually-orthogonal} states over a suitable Hilbert space $ {\mathcal{H}} $:
\begin{equation}
 (  i \neq j \; \Rightarrow \; <  \phi_{i} | \phi_{j} > \, = \, 0 ) \; \; \forall  i , j \in \Sigma_{n}
\end{equation}
Clearly:
\begin{equation}
  \{ p_{i} \, , \, \omega_{i} \}_{i=1}^{n} \; \in \; DEC_{EXT} ( \sum_{i=1}^{n} p_{i}  \,  \omega_{i} ) \; \; \forall \vec{p} \in {\mathcal{D}}( \Sigma_{n})
\end{equation}
so that theorem\ref{th:Holevo's information and entropy} implies that:
\begin{equation}
  I_{Holevo} (  \{ p_{i} \, , \, \omega_{i} \}_{i=1}^{n} ) \; = \; S ( \sum_{i=1}^{n} p_{i} \,  \omega_{i} ) \; \; \forall \vec{p} \in {\mathcal{D}}( \Sigma_{n})
\end{equation}
By the orthogonality condition we have, furthermore, that:
\begin{equation}
  \{ p_{i} \, , \,  \omega_{i} \}_{i=1}^{n} \; \in \; DEC_{Schatten} ( \sum_{i=1}^{n} p_{i} \,  \omega_{i} )
\end{equation}
so that, by  corollary\ref{cor:von Neumann's entropy as the Shannon's entropy of the eigengvalues}, we have that:
\begin{equation}
  I_{Holevo} (  \{ p_{i} \, , \, \omega_{i} \}_{i=1}^{n} ) \; = \; H_{Shannon}  (  \{ p_{i} \, , \, \omega_{i} \}_{i=1}^{n} )
\end{equation}
Let us now consider the operational partition of unity $ {\mathcal{V}} \; := \; \{ | \phi_{i} > <  \phi_{i} | \}_{i=1}^{n} $; by the orthogonality-condition we have that:
\begin{equation}
  I ( \{ p_{i}  \}_{i=1}^{n} \, ; \,  \{ \omega ( \alpha_{j} )( {\mathcal{V}} )  \}_{j=1}^{n}) \; = \;  H_{Shannon}  (  \{ p_{i} \, , \, \omega_{i} \}_{i=1}^{n} )
\end{equation}
We have thus shown that all the classical information contained in the Schatten-decomposition $ \{ p_{i} \, , \, \omega_{i} \}_{i=1}^{n} $ is accessible:
\begin{equation}
   I_{acc} (  \{ p_{i} \, , \, \omega_{i} \}_{i=1}^{n} ) \; = \; I_{Holevo} (  \{ p_{i} \, , \, \omega_{i} \}_{i=1}^{n} ) \; = \; H_{Shannon}  (  \{ p_{i} \, , \, \omega_{i} \}_{i=1}^{n} )
\end{equation}
so that Bob can distinguish the state Alice sent him, i.e. the letter of the alphabet $ \Sigma_{n} $ she transmitted.

\smallskip

Let us now remove the hypothesis of orthogonality of the states $ \{ | \phi_{i} > \}_{i=1}^{n} $ used by Alice.

Theorem\ref{th:Jaynes' theorem} implies than that:
\begin{equation}
  S ( \sum_{i=1}^{n} p_{i} \,  \omega_{i} ) \; \leq \; H_{Shannon}  (  \{ p_{i} \, , \, \omega_{i} \}_{i=1}^{n} )
\end{equation}
so that:
\begin{equation}
  I_{acc} (  \{ p_{i} \, , \, \omega_{i} \}_{i=1}^{n} ) \; < \; I_{Holevo} (  \{ p_{i} \, , \, \omega_{i} \}_{i=1}^{n} ) \; < \;  H_{Shannon}  (  \{ p_{i} \, , \, \omega_{i} \}_{i=1}^{n} )
\end{equation}
In this situation Bob cannot distinguish the state Alice sent him and, conseguentially,  the letter of the alphabet $ \Sigma_{n} $ she trasmitted.

\smallskip

The situation of remark\ref{rem:indistinguishibility of nonothogonal states} is absolutely general:
\begin{theorem} \label{th:on the reachability of Holevo's bound}
\end{theorem}
ON THE REACHABILITY OF HOLEVO'S BOUND
\begin{equation}
  I_{acc} ( {\mathcal{E}} ) \, = \, I_{Holevo} ( {\mathcal{E}} ) \; \Leftrightarrow \; {\mathcal{E}} \, \in \, DEC_{\bot} (\omega)
\end{equation}
obviously implying, by theorem\ref{th:Holevo's bound}, that the elements of an arbitrary ensemble of states over a noncommutative space are distinguishable iff they are mutually orthogonal:
\begin{corollary} \label{cor:indistinguishibility of nonorthogonal states}
\end{corollary}
INDISTINGUISHIBILITY OF NONORTHOGONAL STATES:
\begin{equation}
 I_{acc} ( {\mathcal{E}} ) \, < \, I_{Holevo} ( {\mathcal{E}} )  \; \; \forall {\mathcal{E}}  \in ( DEC ( \omega ) -  DEC_{\bot} (
 \omega))
\end{equation}

\smallskip

\begin{remark}
\end{remark}
HOLEVO'S INFORMATION OF A DECOMPOSITION VERSUS THE ENTROPY OF A SUB-$W^{\star}$-ALGEBRA:

Let us consider the following two notions:
\begin{itemize}
  \item the Holevo's information $ I_{Holevo} ( {\mathcal{E}} )  $ of a decomposition $ {\mathcal{E}} \in DEC(\omega) $
  \item the entropy $ H_{\omega} ( B ) $ of a sub-$W^{\star}$-algebra B w.r.t. $ \omega $
\end{itemize}
where $ \omega \in S(A) $ is a certain state on the algebraic space A.

Both have the feature of not-depending alone by $ \omega $ but also by an other ingredient.

It would appear, then, rather natural to investigate their interrelation asking ourselves if they are not, trivially,
speaking about the same thing in different languages, i.e. if,given  an algebraic probability space $ ( A \, , \, \omega ) $, it is not trivially the case that there
exist a translator-bijection $ Translation \, : \, \{ \text{ sub-$W^{\star}$-algebras of A } \}  \, \rightarrow \,  DEC( \omega )  $ such that:
\begin{equation} \label{eq:link between entropy of a subalgebra ansd Holevo's information}
   H_{\omega} (B) \; = \;  I_{Holevo} ( Translation(B) ) \; \; \forall B \in \{ \text{ sub-$W^{\star}$-algebras of A } \}
\end{equation}
Related issues has been extensively analyzed by Fabio Benatti \cite{Benatti-93}, \cite{Benatti-96}, \cite{Benatti-Narnhofer-Uhlmann-99}.

Given a k-dimensional abelian sub-$W^{\star}$-algebra B of A let us denote by $ \{ \hat{n}_{i} \}_{i=1}^{k} $ its \textbf{minimal projections}.

It would then natural to pose:
\begin{equation}
  Translation(B) \; := \; \{ \omega ( \hat{n}_{i} ) \, , \, \frac{ \omega ( \hat{n}_{i}  \cdot ) }{\omega ( \hat{n}_{i} )} \}_{i=1}^{k}
\end{equation}
Benatti has proved that in this case one has that:
\begin{equation}
   H_{\omega} (B) \; = \;  I_{acc} (  Translation(B) )
\end{equation}
Since:
\begin{equation}
  Translation(B) \; \in \; DEC_{\perp} ( \omega)
\end{equation}
it follows by theorem\ref{th:Holevo's bound} and corollary\ref{cor:indistinguishibility of nonorthogonal states} that:
\begin{equation}
   H_{\omega} (B) \; = \;  I_{Holevo} (  Translation(B) )
\end{equation}
Unfortunately the argument doesn't generalize to noncommutative sub$W^{\star}$algebras.
\begin{remark} \label{rem:distinguishibility versus cloning}
\end{remark}
DISTINGUISHIBILITY VERSUS CLONING OF STATES

In 1982 two papers, one by D. Dieks and the other by W.K. Wooters and W.H. Zurek, introduced the following:
\begin{theorem} \label{th:no unitary-cloning theorem for nonorthogonal pure states on Hilbert spaces}
\end{theorem}
NO UNITARY-CLONING THEOREM FOR NONORTHOGONAL PURE STATES ON HILBERT SPACES:

\begin{hypothesis}
\end{hypothesis}
\begin{equation*}
  {\mathcal{H}} \text{ Hilbert space}
\end{equation*}
\begin{equation*}
  | \psi_{1} > \, | \psi_{2} > , | s >  \, \in  \ , {\mathcal{H}}
\end{equation*}
\begin{equation*}
   | \psi_{1} > \, \not\perp \, | \psi_{2} > \; and \; | \psi_{1} > \, \neq \, | \psi_{2} >
\end{equation*}
\begin{equation*}
  < s | s > \; = \; 1
\end{equation*}
\begin{thesis}
\end{thesis}
\begin{multline*}
  \nexists \hat{U} \, \in \, {\mathcal{U}} [  {\mathcal{B}} ({\mathcal{H}}) ] \; : \\
  \hat{U} | \psi_{i} > \bigotimes | s >  \: = \: | \psi_{i} > \bigotimes | \psi_{i} > \; \; i \, = \, 1,2
\end{multline*}
\begin{proof}
If, ad absurdum, the thesis holded, it would imply that the complex number:
\begin{multline*}
  < \psi_{1} | \bigotimes < s | {\hat{U}}^{\dagger} {\hat{U}}  | \psi_{2} > \bigotimes | s >  \; \\
   < \psi_{1} | \bigotimes < s |  \psi_{2} > \bigotimes | s > \; = \; < \psi_{1} |  \psi_{2} >
\end{multline*}
should be equal to:
\begin{multline*}
  < \psi_{1} | \bigotimes < \psi_{1} | {\hat{U}}^{\dagger} {\hat{U}}  | \psi_{2} > \bigotimes | \psi_{2} >  \; \\
   < \psi_{1} | \bigotimes < \psi_{1} |  \psi_{2} > \bigotimes |  \psi_{2} >  \; = \; | < \psi_{1} |   \psi_{2} > |^{2}
\end{multline*}
i.e.:
\begin{equation*}
  | < \psi_{1} | \psi_{2} > |^{2} \; = \; < \psi_{1} | \psi_{2} >
\end{equation*}
that implies that $   |  \psi_{1} > \,  \perp \, |  \psi_{2} > $ contradicting the hypothesis.
\end{proof}

Furthermore one has that (cfr. cap.12 of \cite{Nielsen-Chuang-00}):
\begin{theorem} \label{th:equivalence between the not-unitary-clonability of nonorthogonal pure states and their not-distinguishability}
\end{theorem}
EQUIVALENCE BETWEEN THE NOT-UNITARY-CLONABILITY OF NONORTHOGONAL PURE STATES AND THEIR NOT-DISTINGUISHABILITY:

Theorem\ref{th:no unitary-cloning theorem for nonorthogonal pure states on Hilbert spaces} is equivalent to the restriction
of corollary\ref{cor:indistinguishibility of nonorthogonal states} to extremal-decompositions over discrete noncommutative-probability-spaces

\begin{theorem} \label{th:no cloning theorem for nonorthogonal states on discrete noncommutative probability spaces}
\end{theorem}
NO CLONING THEOREM FOR NONORTHOGONAL  NORMAL STATES ON DISCRETE NONCOMMUTATIVE PROBABILITY SPACES:

\begin{hypothesis}
\end{hypothesis}
\begin{equation*}
  {\mathcal{H}} \text{ Hilbert space}
\end{equation*}
\begin{equation*}
   \omega_{1} \, , \, \omega_{2} \, , \, \omega_{s} \in S_{n}[{\mathcal{B}} ( {\mathcal{H}})]
\end{equation*}
\begin{equation*}
   \omega_{1} \,  \not\perp \, \omega_{2} \; and \; \omega_{1} \, \neq \, \omega_{2}
\end{equation*}
\begin{thesis}
\end{thesis}
\begin{equation*}
  \nexists \alpha \in CPU(A \bigotimes A)  \; : \; \alpha_{\star} ( \omega_{i} \cdot \omega_{s} ) \: = \;  \omega_{i} \cdot \omega_{i} \; \; i \, = \, 1 , 2
\end{equation*}

Up to date the only attempt to catch  the $ W^{\star}$-algebraic structure lying behind the No-cloning theorems
is a theorem by G\"{o}ran Lindblad \cite{Lindblad-99} that is, indeed, a generalization of theorem\ref{th:no cloning theorem for nonorthogonal states on discrete noncommutative probability spaces}
but is not yet a theorem poning a censorphip of cloning of suitable states on an arbitary noncommutative space.

Theorem\ref{th:equivalence between the not-unitary-clonability of nonorthogonal pure states and their not-distinguishability} can lead to suspect that such a (still lacking) theorem would be equivalent to
corollary\ref{cor:indistinguishibility of nonorthogonal states}.

\smallskip

\begin{remark} \label{rem:distinguishibility of states versus Maxwell's demonology}
\end{remark}
DISTINGUISHIBILITY OF STATES VERSUS MAXWELL'S DEMONOLOGY

In the $ 9^{th} $ chapter of \cite{Peres-95} Asher Peres claims that a violation of corollary\ref{cor:indistinguishibility of nonorthogonal states}
would imply a violation of the Second Law of Thermodynamics.

His argument, anyway, lies on the assumption that the \textbf{thermodynamical-entropy} of a quantum system is described by \textbf{Von Neumann's entropy}, assumption
that he deeply analyzes explicitly reporting the celebrated original calculus by which Von Neumann, in the section5.2 of \cite{Von-Neumann-83}, computed the
thermodynamical entropy of a mixture $ \{ p_{i} \, , \, | \phi_{i} > < \phi_{i} | \}_{i=1}^{n} \in DEC_{EXT} ( \sum_{i} p_{i} | \phi_{i} > < \phi_{i} | )  $  as if each  $| \phi_{i} > < \phi_{i} | $
was a specie of ideal gas enclosed in a large impenetrable box and inferring that the thermodynamical mixing entropy of the different
species is $ S_{Von Neumann} ( \sum_{i} p_{i} | \phi_{i} > < \phi_{i} | ) $.

The assumption $ S_{therm} ( \omega ) \; = \; S ( \omega ) $ would indeed seem to respect the Second Principle of Thermodynamics:
\begin{itemize}
  \item by theorem\ref{th:invariance of a state's entropy under inner automorphisms} and axiom\ref{ax:noncommutative axiom on closed dynamics}, the thermodynamical-entropy of a closed
  quantum dynamical system remains unchanged and thus, in particular, cannot decrease with time
  \item by theorem\ref{th:not-monotonicity of a state's entropy under channels}  the thermodynamical-entropy of an open quantum dynamical system can decrease with time
\end{itemize}
But here the problem of Maxwell's demon, inherited from the classical problem for the mixing of different species of different ideal gases,  appears.

Let us  introduce it with Maxwell's own words; in the section
"Limitation of The Second Law of Thermodynamics" of the $ 12^{th}
$ chapter of \cite{Maxwell-71} he writes:
\begin{center}
  \textit{"One of the best extablished facts in thermodynamics is that it is impossible in a system enclosed in an envelope
which permits neither change of volume nor passage of heat, and in which both the temperature and the pressure are everywhere the same, to produce any inequality
of temperature or of pressure without the expenditure of work. This is the second law of thermodynamics, and it is undoubtedly
true as long as we can deal with bodies only in mass, and have no power of perceiving or handling the separate molecules of which they are made up.
But if we conceive a being whose faculties are so sharpened that he can follow every molecule in its course, such a being, whose attributes
are still as essentially finite as our own, would be able to do what is at present impossible to us. For we have seen that the molecules in a vessel full of air at uniform temperature are moving with velocities by no means
uniform, though the mean velocity of any great number of them, arbitrary selected, is almost exactly uniform. Now let us suppose that such a vessel is divided in two portions, A and B, by a division in which there is a small hall, and that a being, who
can see the individual molecules, opens and closes this hole so as to allow only the lower ones to pass from B to A. He will see, thus, without expenditure of work, raise the temperature of B and lower that of A, in contradiction with the second law of thermodynamics" (cited in the first chapter of \cite{Leff-Rex-90})}
\end{center}
In our case, instead of leaving to pass or stopping molecules according to their velocity, it leaves to pass or stops Von Neuman's impenetrable boxes
according to the specie of the $ | \phi_{i} > < \phi_{i} | $ it pertains in the following way:
\begin{itemize}
  \item it leaves to pass from A to B only boxes pertaining to species with label $ i \geq \lfloor \frac{n}{2} \rfloor $
  \item it leaves to pass from B to a only boxes pertaining to species with label $ i < \lfloor \frac{n}{2} \rfloor $
\end{itemize}

Let us then leave aside for the moment our quantum situation and let us analyze the classical problem, as Maxwell presents it.
In the 220 years after the publication of Maxwell's book an enormous literature tried to exorcize it in different ways; an historical review may be found
in the first chapter "Overview" as well as in the "Chronological Bibliography with Annotations and Selected Quotations" of the
wonderful book edited by Harvey S. Leff and Andrew F. Rex \cite{Leff-Rex-90}.

All these exorcisms were based on the idea that, to accomplish his task, the Maxwell'demon necessarily causes a thermodynamical-entropy's raising causing
the Second Law to be preserved:

they anyway strongly differed in identifying the element of the demon's dynamical evolution which is \textbf{necessarily thermodinamically-irreversible}:

coming to recent times, most of the  Scientific Community
strongly believed in Leon Brillouin's exorcism
\cite{Brillouin-90}, identifying such an element in the
\textbf{demon's information-acquisition's process}.

When anyone thought that the "The-end" script had at last appeared to conclude "The Exorcist" movie, Charles H. Bennett showed in 1982 \cite{Bennett-90a}, \cite{Bennett-90b},  basing on the previous work by Rolf Landauer on the \textbf{Thermodynamics of Computation} \cite{Landauer-90a}, that:
\begin{enumerate}
  \item Maxwell's Demon was still alive  owing to the
nullity of Brillouin's exorcism: the demon's acquisition process may be done in a completelly thermodynamically-reversible way:
  \item the \textbf{necessarily thermodinamically-irreversible} element is instead \textbf{demon's information-erasure's process}
\end{enumerate}

Given two arbitrary sets A and B and a partial function f from A
to B $ f \in \stackrel{ \circ } {MAP}(A,B) $:

\begin{definition} \label{def:logical reversibility}
\end{definition}
f IS LOGICALLY REVERSIBLE:

it is injective, i.e.:
\begin{equation}
  cardinality ( f^{- 1} (b) ) \in \{ 0 , 1 \} \; \; \forall b \in B
\end{equation}

\begin{definition} \label{def:thermodinamical-reversible-computability}
\end{definition}
f IS THERMODINAMICALLY-REVERSIBLY-COMPUTABLE:
\begin{enumerate}
  \item there exist a physical device of any kind computing f $\cdots $:
\begin{equation*}
  f  \; \in \;  \Delta_{0}^{0}-\stackrel{ \circ } {MAP}(A,B)
\end{equation*}
  \item $\cdots $  in a thermodinamically reversible way (i.e. in such a way that the computational-process doesn't increase
  the thermodynamical entropy of the Universe)
\end{enumerate}

Both the theoretical analysis of specific computational models and the experiments strongly support the following:

\begin{axiom} \label{ax:Landauer's principle}
\end{axiom}
LANDAUER'S PRINCIPLE:

\begin{hypothesis}
\end{hypothesis}
\begin{equation*}
  f  \; \in \;  \Delta_{0}^{0}-\stackrel{ \circ } {MAP}(A,B)
\end{equation*}
\begin{thesis}
\end{thesis}
\begin{equation}
 \text{f is thermodynamically-reversibly-computable } \; \Leftrightarrow \; \text{ f is logically-reversible}
\end{equation}

Let us now  suppose that we want to compute a thermodinamically-irreversibly-computable function $ f \, : \, A \, \stackrel{\circ}{\rightarrow} \, B $.

We can exploit our computation  in a thermodinamically-reversible way at the prize of keeping memory of the input , e.g. computing
the thermodinamically-reversibly-computable function $ \tilde{f} \, : \, A \stackrel{\circ}{\rightarrow} \, B $:
\begin{equation}
  \tilde{f} (a) \; := \; ( a \, , \, f(a) )
\end{equation}
In Thermodynamics of Computation, the suppletive information on the input we have conserved is called \textbf{garbage}.

Let us now consider the process of information-erasing: it can be mathematically described by the following:
\begin{definition}
\end{definition}
ERASURE-FUNCTION OF A:
\begin{equation}
  er(A) \, : \, A \, \rightarrow \, \emptyset
\end{equation}
One has then that:
\begin{corollary} \label{cor:information-erasure is thermodynamically-irreversibly-computable}
\end{corollary}
INFORMATION-ERASURE IS THERMODINAMICALLY-IRREVERSIBLY-COMPUTABLE:

er(A)  is thermodynmamically-irreversibly-computable

\begin{proof}
Given a thermodinamically-irreversibly-computable function $ f \,
: \, A \, \stackrel{\circ}{\rightarrow} \, B $  let us suppose ad
absurdum that er(A) is thermodynmamically-reversibly-computable;
then $ \tilde{f} \circ er(A) $  would be
thermodinamically-reversibly-computable. Since:
\begin{equation}
  \tilde{f} \circ er(A) \; = \; f
\end{equation}
this would contradict the hypothesis
\end{proof}

Let us now return to the Maxwell's demon: conceptually it may be
formalized as a computer that:
\begin{enumerate}
  \item gets the input $ ( s , v ) $  from a device measuring both the side s from which the molecule arrives and its velocity
  \item computes a certain semaphore-function p such that $ ( s , v ) \stackrel{p}{\rightarrow}  p[(s,v)] $ giving as output a 0 if the molecule must be
  left to pass while gives as output a one if the molecule must be stopped
  \item gives the output p[(s,v)] to a suitable device that operates on the molecule in the specified way
\end{enumerate}
Both the first and the third phases  of this process, taking into account also the involved devices, may be made in a thermodinamically-reversible way.

As to the second step, anyway, let us observe that the semaphore-function p is logically-irreversible and hence, by axiom\ref{ax:Landauer's principle}, also
thermodinamically-irreversibly-computable.

As above specified, such a thermodinamically-irreversibility may be avoided conserving \textbf{garbage}; let us, precisely, suppose, that
the demon-computer computes the thermodynamically-reversibly-computable function $ \tilde{p} $.

Let us suppose to make operate the demon-computer n times on n different molecules.

When n grows then demon, with no expenditure of work, would  raise the temperature of B and lower that of A.

But let us now analyze more carefully Clausius's  formulation of the Second Principle:

\begin{axiom}   \label{ax:second principle of thermodynamics in Clausius' form}
\end{axiom}
SECOND PRINCIPLE OF THERMODYNAMICS IN CLAUSIUS' FORM:

No thermodynamical transformation is possible \textbf{that has as
its only result} the passage of heat from a body at lower
temperature to a body at higher temperature

\smallskip

In the above process the passage of heat from A to B is not the only result: another result is the
storage in the demon-computer's memory of the n-ple of inputs $ ( ( s_{1} , v_{1} ) \, , \, \cdots \, , \, ( s_{n} , v_{n} ) ) $.

To make the passage of heat from A to B to become the only result of the process we could think
that the demon, at the end, erases his memory; but, by corollary\ref{cor:information-erasure is thermodynamically-irreversibly-computable} this (and only this) cannot be done
in a thermodynamically-reversible way: such an erasure causes an increase of entropy that may be proved to be
greater than or equal to the entropy-decrease produced by the passage of heat from A to B (an explicit computation for the conceptually analogous situation of Szilard's engine \cite{Szilard-90}
may be found in the section8.5 of \cite{Li-Vitanyi-97}).

Bennett's exorcism of Maxwell's demon, has, anyway, a far
reaching conseguence; supposed that the gas is described by the
thermodynamical ensemble  $ ( X \, , \, P ) $, let us introduce
the following:
\begin{definition} \label{def:Bennett's entropy}
\end{definition}
BENNETT'S ENTROPY OF P:
\begin{equation}
  S_{Bennett}(P) \; := \; I_{probabilistic}(P) \, + \, I_{algorithmic}(P) \; = \; H(P) + I(P)
\end{equation}
where H(P) is Shannon's entropy of the distribution P (i.e. its
Gibbs' entropy in thermodynamical language), while I(P) is its
prefix-algorithmic-entropy, i.e. the length of the shortest
program computing it on the fixed Chaitin universal computer U.

One has that:
\begin{theorem} \label{th:Bennett's theorem}
\end{theorem}
BENNETT'S THEOREM:
\begin{equation*}
  S_{therm} ( P ) \; = \;  S_{Bennett}(P)  \; \neq H (P)
\end{equation*}
\begin{proof}
The thesis follows by  Bennett's exorcism of Maxwell's demon:
 it implies that when a physical system increases its prefix-algorithmic-entropy by n cbits, it has the capacity  to convert about $ n  T \ln 2 $ of wasten heat
into useful  work in its surrounding.

Conversely, the conversion of about $ n  T \ln 2 $ of work into heat in the surrounding is necessary to decrease a system's
prefix-algorithmic-entropy by n cbits
\end{proof}

It may be worth to observe that the additive constant by which
prefix-algorithmic entropy is defined, depending on the choice of
the fixed universal computer, doesn't matter as far as the first
two principles of thermodynamics are concerned.

Such a constant is anyway fixed  by the imposition of the Third
Principle of Thermodynamics:

so, the imposition that the thermodynamical-entropy vanishes at zero temperature, curiously results in a fixing
of a particular Chaitin's universal computer U and, conseguentially, of the associated Halting Probability $ \Omega_{U} $.

\smallskip

The generalization both of Bennett's exorcism of Maxwell's demon
and of the Bennett's Theorem to the quantum domain has been
extensively analyzed by Wojciech H. Zurek \cite{Zurek-89},
\cite{Zurek-90a}, \cite{Zurek-90b}, \cite{Zurek-99}.

Given a density operator $ \rho \in {\mathcal{D}} ({\mathcal{H}}) $ on an Hilbert space $ {\mathcal{H}} $:
\begin{definition} \label{def:Zurek's entropy of a density operator}
\end{definition}
ZUREK'S ENTROPY OF $ \rho $:
\begin{equation}
  S_{Zurek}( \rho ) \; := \;  S_{Von Neumann}( \rho ) \, + \, I ( \rho )
\end{equation}
with:
\begin{equation} \label{eq:quantum algorithmic information of first kind}
  I( \rho ) \; := \; \min _{\vec{x} \in \Sigma^{\star} \, : \, U(\vec{x} ) = \rho } | \vec{x} |
\end{equation}
Zurek claims that the assumption of the Church-Turing's Thesis
eliminates any dependence from the particular universal computer U
adopted by (or better constituting) the demon.

He, in particular, seems to claim that, by the Church-Turing's
Thesis, it doesn't matter if U is a \textbf{classical computer}
or a \textbf{quantum computer}, i.e. that the assumption of
Church-Turing's Thesis implies that Quantum Algorithmic
Information Theory collapses to Classical Algorithmic information
Theory, an assumption absolutely arbitary as we  will analyze more
completely in section\ref{sec:Quantum Algorithmic Information
Theory as a particular case of the abstract Uspensky's approach}.

We will therefore assume, from here and beyond, that the computer
U in definition\ref{def:Zurek's entropy of a density operator} is
a Universal Quantum Computer.

But let then observe that, in this way, one implicitely assumes
that the quantum algorithmic information of a quantum state must
be defined in terms of classical-descriptions of such a state, as
claimed by Svozil  \cite{Svozil-96} and  Vitanyi
\cite{Vitanyi-99}, \cite{Vitanyi-01}, and not in terms of quantum
descriptions as it is claimed by Berthiaume, Van Dam and
Laplante  \cite{Berthiaume-van-Dam-Laplante-00} as we discussed
in chapter\ref{chap:Quantum algorithmic randomness: where are
we?}.

The \textbf{garbage} accumulated by the demon is, therefore, made of cbits, i.e. of classical information.

Let us suppose, instead, that the inputs of the quantum computer
U are qubits and not cbits, i.e. let us suppose to modify
definition\ref{def:Zurek's entropy of a density operator}
replacing the defintion of $ I( \rho) $ given by
eq.\ref{eq:quantum algorithmic information of first kind} so that
U's inputs are qubits and not cbit: the \textbf{garbage}
accumulated by the demon would then  consist of \textbf{quantum
information} instead of \textbf{classical information}.

As far as the exorcism is concerned, anyway, such a replacement doesn't change anything, since the generality of corollary\ref{cor:information-erasure is thermodynamically-irreversibly-computable}
assures the thermodynamical-irreversible-computability  of the erasure of both classical and quantum information.

What is important here to stress is that Zurek's extension of
Bennett's exorcism to the quantum domain implies the quantum
analogue of theorem\ref{th:Bennett's theorem}:
\begin{theorem} \label{th:Zurek's theorem}
\end{theorem}
ZUREK'S THEOREM:
\begin{equation}
  S_{therm} ( \rho ) \; = \;  S_{Zurek}(\rho)  \; \neq \; S_{Von Neumann}(\rho)
\end{equation}

Let us now observe that, in the general framework of Quantum Probability Theory, Bennett's and Zurek's results may be unified in the following way:

given an algebraic probability space $ ( A \, , \, \omega ) \, , \, \omega \in  NC_{M}-NC_{\Phi}-\Delta_{0}^{0}-S(A) $ (according
to the the notion of computability we will extensively discuss in chapter\ref{chap:Quantum algorithmic randomness as satisfaction of all the quantum algorithmic typical properties}):
\begin{definition} \label{def:double approach entropy}
\end{definition}
DOUBLE-APPROACH ENTROPY OF $ \omega $:
\begin{equation}
  S_{\text{double approach}} ( \omega ) \; := \; S( \omega ) \, + \, I( \omega )
\end{equation}
where $ I( \omega ) $ is the \textbf{algebraic algorithmic information} of $ \omega $ we will introduce in  chapter\ref{chap:Quantum algorithmic randomness as satisfaction of all the quantum algorithmic typical properties})

The name in definition\ref{def:double approach entropy} has been
chosen to stress  that it involves, in the Kolmogorovian terms
introduced in section\ref{sec:Introduction}, \textbf{both the
probabilistic and the algorithmic approach to Information Theory}.

What it is important to stress is that Bennett's and Zurek's analyses generalize in  a straightforward way giving rise to the following:
\begin{theorem} \label{th:Bennett-Zurek's theorem}
\end{theorem}
BENNETT-ZUREK'S THEOREM:
\begin{equation}
  S_{therm} ( \omega ) \; = \; S_{\text{double approach}} ( \omega )  \; \neq \; S( \omega )
\end{equation}

The fact, stated by theorem\ref{th:Bennett-Zurek's theorem}, that
to the \textbf{thermodynamical entropy} of a system doesn't
contibute only its \textbf{probabilistic entropy} but also its
\textbf{algorithmic entropy} is according to us nothing but the
opening of a new chapter in the history of Thermodynamics, whose
still lacking precise mathematical formalization has prevented it
to be even taken in consideration by the Mathematical-Physicists'
Community.

Those of them has spent a life studying Rigorous Statistical
Mechanics, e.g. in \cite{Simon-93} \cite{Ruelle-99},
\cite{Bratteli-Robinson-97}, may be reassured that in any
situation involving no information-gathering-and-using-system
(IGUS) $  I( \omega ) $ vanishes so that all the theorems therein
contained apply.

Instead of reacting to theorem\ref{th:Bennett-Zurek's theorem}
with  that  typical reactionary attitude of the worst among
mathematicians and mathematical-physicists (but fortunately not
by the greatest minds of both Mathematics and
Mathematical-Physics), consisting in discarding any novelty since
is not presented specifying if functions are of class $ C^{(1)} $
or   $ C^{(2)} $ and have not all the bows around the $
\epsilon$'s and $ \delta$'s well posed, the lovers of
mathematical rigour should contribute to mathematically formalize
the proof of theorem\ref{th:Bennett-Zurek's theorem}, as well as
to mathematically formalize the notion of an IGUS.

We have now all the ingredients required to analyze Asher Peres' claim that a
violation of corollary\ref{cor:indistinguishibility of nonorthogonal states}
would imply a violation of the Second Law of Thermodynamics, returning to his analysis of Von Neumann's computation of the thermodynamical-entropy of the mixture $ \{ p_{i} \, , \, | \phi_{i} > < \phi_{i} | \}_{i=1}^{n} \in DEC_{EXT} ( \sum_{i} p_{i} | \phi_{i} > < \phi_{i} | )  $.
Peres reviews Von Neumann's procedure in the following way:
\begin{center}
  \textit{"It also assumes the existence of semipermeable membranes which can be used to perform quantum tests.
  \textbf{These membranes separate orthogonal states with perfect efficency}. The fundamental problem here is whether it is
  legitimate to treat \textbf{quantum states} in the same way as varieties of classical ideal gases.
  This issue was clarified by Einstein in the early days of the "old" quantum theory as follows: consider an ensemble of quantum
  systems, each one enclosed in a large impenetrable box, so as to prevent any interaction between them.
  These boxes are enclosed in an even larger container, where they behave \textbf{as an ideal gas}, because each box is so  massive
  that classical mechanics is valid for its motion ($\cdots $). The container itself has ideal walls and pistons which may be, according to our needs,
  perfectly conducting, or perfectly insulating, or with properties equivalent to those of semipermeable membranes.
  \textbf{The latter are endowed with automatic devices able to peak inside the boxes and to test the state} of the quantum system enclosed therein." (from the section9.3 of \cite{Peres-95})}
\end{center}
There is a point, anyway, of this review in which, deliberately, Peres moves away from Von Neumann's original treatment:

he doesn't assume that \textbf{the membranes separate nonorthogonal states with perfect efficiency} as, instead, Von Neumann does:
\begin{center}
  \textit{"Each system $ s_{1} , \cdots ,  s_{n} $ is confined in a box $ K_{1} , \cdots ,  K_{n} $ \textbf{whose walls are impenetrable to all transmission effects} -- which is possible for this system because
  of the lack of interaction" (from the section5.2 of \cite{Von-Neumann-83})}
\end{center}

The reason why Peres, contrary to Von Neumann, doesn't make such
an assumption is that, according to him, this would imply a
violation of the Second Law of Thermodynamics; his argument is
the following: if semi-permeable membranes which unambiguously
distinguish non-orthogonal states were possible, one could use
them to realize the following cyclic thermodynamical
transformation for a mixture of two species of  1-qubit's states,
the $ | 0 > < 0 |$-specie  and the $ \frac{1}{2} ( | 0 >  + | 1 >
) ( < 0 | + < 1 | ) \} $- specie, both with the same
concentration $ \frac{1}{2}$
\begin{itemize}
  \item in the initial state the two species occupy two chambers with equal volumes, with the  $ | 0 > < 0 | $- specie occupying the right-half of the left-half of the vessel and the  $ \frac{1}{2} ( | 0 >  + | 1 > ) ( < 0 | + < 1 | ) \}  $- specie occupying the
  left-half of the right-half of the vessel
  \item the first step of the process is an isothermal expansion by which the  $ | 0 > < 0 | $- specie occupies all the left-half of the vessel while the $ \frac{1}{2} ( | 0 >  + | 1 > ) ( < 0 | + < 1 | )  $-specie occupies all the right-half of the vessel; this expansion supplies an amount
  of work:
\begin{equation}
  \Delta L _{1} \; = \; +  n T \ln 2
\end{equation}
T being the temperature of the reservoir.
  \item at this stage the impenetrable partitions separating the two species are replaced by the "magic"-semi-permeable membranes having the ability of distinguish non-orthogonal states;
precisely one of them is transparent to the $ | 0 > < 0 | $-specie and reflect the $ \frac{1}{2} ( | 0 >  + | 1 > ) ( < 0 | + < 1 | ) \}  $-specie while the other membrane has the opposite properties; then, by a double frictionless piston,
it is possible to bring the engine, without expenditure of work or heat transfer, to a state in which all the two species occupy with the  same concentration only the left-hand of the vessel, the right-hand of the vessel  remaining  empty; we can represent mathematically the state of affairs of the system by the decomposition:
\begin{align}
   {\mathcal{E}}_{1} & \; := \;  \{ ( \frac{1}{2} , | 0 > < 0 |  ) \, , \, ( \frac{1}{2} , \frac{1}{2} ( | 0 >  + | 1 > ) ( < 0 | + < 1 | )  )  \} \, \in \, DEC_{EXT}( \rho ) \\
  \rho  & \; := \; \begin{pmatrix}
    \frac{3}{4} & \frac{1}{4} \\
    \frac{1}{4} & \frac{1}{4} \
  \end{pmatrix}
\end{align}
  \item since the state of the mixture-of-species is completelly determined  by $ \rho $, and not by a particular its decomposition, to represent the actual state of affairs by  $ {\mathcal{E}} $ or by the Schatten's decomposition of $ \rho $:
\begin{align}
  {\mathcal{E}}_{1} & \; \; := \; \{ ( \rho_{-}  , | e_{-} > < e_{-} | ) \, , \, ( \rho_{+}  , | e_{+} > < e_{+} | ) \} \in {DEC}_{Schatten}( \rho ) \\
  \rho_{\pm} & \;  := \;  \frac{1}{4} ( 2 \pm \sqrt{2} ) \\
   | e_{\pm} >  & \; := \; ( 1 \pm \sqrt{2} ) ( | 0 > \, + \, | 1 > )
\end{align}
is absolutely equivalent
  \item let us now replace the two "magic" membranes with ordinary membranes able to distinguish only orthogonal species;  since the $ | e_{-} > < e_{-} | $-specie and the $ | e_{+} > < e_{-} | $-specie are orthogonal, the reversible
diffusion of the two species separate them, with the  $ | e_{+} >
< e_{+} |$-specie occupying the left-half of the vessel and the $
| e_{-} > < e_{-} |$-specie occupying the right-half of the
vessel.
  \item finally an isothermal compression takes the system in a situation in which the volume and the pressure are the same of the initial state; such
  a compression requires an expenditure of work of:
\begin{equation}
  \Delta L _{2} \; = \; -  n T [ \rho_{1} \log  \rho_{1} \, +  \, \rho_{2} \log  \rho_{2} ]
\end{equation}
\item finally a suitable unitary evolution takes the system again  in the initial state.
\end{itemize}
The net work made by the engine during the cycle is:
\begin{equation}
   \Delta L \; = \; \Delta L _{1} + \Delta L _{2} \; > \; 0
\end{equation}
so that the whole thermodynamical cycle converts the heat
extracted by the reservoir in a positive amount of work of $
\Delta L $.

This, according to Peres, violates the Second Principle, proving
that \textbf{the the "magic"  membranes able to separate
nonorthogonal states with perfect efficiency} cannot exist.

But let us again look  more carefully at  the precise  formulation of the Second Principle, making use of the following well known theorem of Thermodynamics:
\begin{theorem}   \label{th:equivalence of Clausius' and Kelvin's formulation of the Second Principle}
\end{theorem}
EQUIVALENCE OF CLAUSIUS' AND KELVIN'S FORMULATIONS OF THE SECOND PRINCIPLE OF THERMODYNMAMICS:

Axiom\ref{ax:second principle of thermodynamics in Clausius' form} is equivalent to the following Kelvin's-formulation:

No thermodynamical transformation is possible \textbf{that has as its only result} the transformation of heat into work

Let us now analyze a cycle of Peres'-engine: is the conversion of heat into work \textbf{the only result} of the process ?

The answer is negative and lead immediately to the the conceptual
deepness of theorem\ref{th:Bennett-Zurek's theorem}, whose
complete comprehension requires  to explicitly analyze the bug in
Von Neumann's proof that $ S_{therm} ( \rho ) \, = \, S_{Von
Neumann} ( \rho ) $.

The key point lies in the own definition of the semi-permeables membranes of Einstein's method: as correctly observed by Peres \textbf{the semipermeable-membranes are endowed with automatic devices able to peak inside the boxes and to test the state}.

What Peres seems unfortunately not to catch is that  a semi-permeable membrane is  then an  IGUS operating in the following way:
\begin{enumerate}
  \item gets the input $ ( s , i ) $  from a device measuring both the side s from which the $ | \phi_{i} > < \phi_{i} | $-specie arrives and its kind, i.e. the classical information codified by its label i.
  \item computes a certain semaphore-function p such that $ ( s , i ) \stackrel{p}{\rightarrow}  p[(s,i)] $ giving as output a 0 if the $ | \phi_{i} > < \phi_{i} | $-specie must be
  left to pass while gives as output a one if the $ | \phi_{i} > < \phi_{i} | $-specie must be stopped
  \item gives the output p[(s,i)] to a suitable device that operates on the  $ | \phi_{i} > < \phi_{i} | $-specie in the specified way
\end{enumerate}
The argument of Bennett's exorcism concerning the necessity of taking into account the  prefix-algorithmic-information
of the sequences of successive recorded $ ( s , i ) $'s in the membrane's memory thus apply.

But this must be done, in particular, in the cases of Peres'-engine:

taking into account also the algorithmic-entropy of the semi-permeable's membranes, one sees that it is greater than or equal to the
universe's entropy decrease corresponding to the work made by the engine, so that, by theorem\ref{th:Bennett-Zurek's theorem}:
\begin{equation}
  \Delta S_{therm} \; \geq \; 0
\end{equation}
and Peres' arguments falls down.

\medskip

Having analyzed in details the many subtilities of the \textbf{Distinguishibility Issue}, we can
return to  Vedral, Plenio and Knight's  Quantum Sanov's Theorem with the objective of rephrasing it as a Quantum Law of Randomness
in the same way we sketched for the classical case

Informally speaking, the argument by Vedral, Plenio and Knight is that the probability of not distinguishing two quantum states $ \rho \, , \, \sigma \; \in \; {\mathcal{D}} ( {\mathcal{H}} ) $ after n measurements decades exponentially as:
\begin{equation}
  P[not-distinguish(\rho \, , \, \sigma) ] \; \sim \; \exp [ - n \, S_{Umegaki} ( \sigma \, , \rho ) ]
\end{equation}

As we preannounced, its mathematical formalization requires to generalize noncommutatively the
definition\ref{def:large deviation principle}.

Given an algebraic space A:
\begin{definition} \label{def:algebraic rate function}
\end{definition}
ALGEBRAIC RATE FUNCTION OVER A:

a map $ I \in MAP[ A \, , \, [ 0 , \infty ) ] $:
\begin{equation}
  \{ a \in A \, : \, I(a)  \leq \alpha \} \text{ is closed } \;
  \forall \alpha \in [ 0 , \infty )
\end{equation}

Given a family $ \{ \omega_{\epsilon} \} $ of states over A:

\begin{definition} \label{def:algebraic large deviation principle}
\end{definition}
$ \{ \omega_{\epsilon} \} $ SATISFIES THE ALGEBRAIC LARGE DEVIATION PRINCIPLE
WITH RATE FUNCTION I:
\begin{equation}
  - \inf_{a \in \Gamma^{0}} I(a) \; \leq \; \liminf_{\epsilon \rightarrow
  0} \epsilon \log \omega_{\epsilon} ( \Gamma ) \; \leq \; \limsup_{\epsilon \rightarrow
  0} \epsilon \log \omega_{\epsilon} ( \Gamma ) \; \leq \; - \inf_{a \in \bar{\Gamma}} I(a) \; \; \forall \Gamma  \subset A
\end{equation}
where $ \Gamma^{0} $ denotes the interior of $ \Gamma $ while $
\bar{\Gamma} $ denotes the closure of $ \Gamma $ w.r.t. a suitable topology on A (let's say the weak topology).

Denoted by $ P_{\omega}^{(NC)} \in S( \Sigma_{NC}^{\infty} ) $ the state over $ \Sigma_{NC}^{\infty} $ associated
to a sequence of independent, identically distributed noncommutative random variables distributed following
$ \omega \in S ( \Sigma_{NC} )$, the key not trivial point is how to generalize noncommutatively the definition\ref{def:type of a string}
of the \textbf{type} $ L_{n}^{\vec{a}} \in S ( \Sigma_{NC} ) $ of a qubit string $ \vec{a} \in S ( \Sigma_{NC} ) $ so that:
\begin{conjecture} \label{con:quantum Sanov's theorem}
\end{conjecture}
QUANTUM SANOV'S THEOREM:

The family of laws $ P_{\omega} ( L_{n}^{\vec{a}} \, \in \, \cdot ) $ satisfies the Algebraic Large Deviation Principle with
algebraic rate function $ S_{Araki} ( \cdot \, , \, \omega ) $.

\smallskip

The reformulation of \ref{con:quantum Sanov's theorem} as a Law of Randomness of $ ( \Sigma_{NC}^{\infty} \, , \, P_{\text{noncommutative central limit}}^{\infty} ) $ would be then straightforward.

\newpage
\section{On absolute conformism in Quantum Probability Theory} \label{sec:On absolute conformism in Quantum Probability Theory}
Given an algebraic probability space $ ( A \, , \, \omega ) $ we have seen in section\ref{sec:Conformism in Quantum Probability Theory} how, trying to generalize noncommutatively
Kolmogorov's approach of characterizing absolute conformism, one ends up in definition\ref{def:Kolmogorov commutatively-random elements of an algebraic probability space} and definition\ref{def:Kolmogorov noncommutatively-random elements of an algebraic probability space} according to one
takes into account, respectively, \textbf{commutative predicates} or \textbf{noncommutative predicates} over A.

We know, anyway, by  theorem\ref{th:not existence of Kolmogorov random sequences of
cbits} that absolute-conformism in Classical Probability Theory is impossible so that:
\begin{corollary}
\end{corollary}
\begin{equation*}
  A \, commutative \; \Rightarrow \; KOLMOGOROV_{C}(APS) \, = \, KOLMOGOROV_{NC}(APS) \, = \, \emptyset
\end{equation*}

At a classical level this fact has the conceptually-deep effect of not-allowing to define an individual random element
of a classical probability space with only measure-theoretic tools, requiring the introduction of ingredients
from Computations' Theory in order of selecting a  suitable subclass of the the typical properties, the Laws of Randomness, to
the  constraint on conformism is restricted.

Let us remind that the diagonalization-proof of theorem\ref{th:not existence of Kolmogorov random sequences of
cbits} lies on the fact that:
\begin{equation}
  p_{\text{difference from $ \bar{y} $}} \, \in \,  {\mathcal{P}}( \Sigma^{\infty} \, , \, P_{unbaised} )  \; \; \forall \bar{y} \in \Sigma^{\infty}
\end{equation}

Let us than analyze the state of affairs of the algebraic generalization of such difference predicates:

\begin{definition}
\end{definition}
COMMUTATIVE PREDICATE OF DIFFERENCE FROM $ b \in A $:
\begin{multline}
   p_{\text{difference from b}} \in {\mathcal{P}}_{C}(APS) \\
   p_{\text{difference from b}} (a) \; := \; << a \, \neq \, b >>
\end{multline}

Let us observe that, owing to the nonexistence of points (not
even hidden) on a noncommutative space we discussed in
section\ref{sec:The problem of hidden points of a noncommutative
space}, one has that:
\begin{equation}
   p_{\text{difference from b}} \; \notin \;
   {\mathcal{P}}_{C}^{typical}(APS) \; \; \forall b \in A
\end{equation}
so that the diagonalization proof of theorem\ref{th:not existence
of Kolmogorov random sequences of cbits} doesn't generalize
noncommutatively as to commutative predicates; obviously this is
true also as  as to noncommutative predicate where no natural
noncommutative predicate of difference from a single element
exists.

One could, conseguentially, think that  the notion of a random
sequence of qubits might be characterized through  $
KOLMOGOROV_{C}( \Sigma_{NC}^{\infty} ) $ or by $ KOLMOGOROV_{NC}(
\Sigma_{NC}^{\infty} ) $.

According to us, anyway, one has to impose again the
effectiveness's constraint on the considered typical properties,
as we will do in the next chapter

\chapter{Quantum algorithmic randomness as satisfaction of all the quantum algorithmic typical properties} \label{chap:Quantum algorithmic randomness as satisfaction of all the quantum algorithmic typical properties}
\section{The problem of characterizing mathematically the notion of a quantum algorithm} \label{sec:The problem of characterizing mathematically the notion of a quantum algorithm}
The greatest goal of Quantum Computation is the discovery of many \textbf{quantum algorithms} allowing to make \textbf{tractable} problems \textbf{intractable} by classical computers, in the sense specified in section\ref{sec:Irreducibility of Quantum Computational Complexity Theory to Classical Computational Complexity Theory}, in almost all cases reconducting it to particular istances of the \textbf{Abelian hidden subgroup problem} (cfr. the $ 5^{th} $ and $ 6^{th} $ chapters of \cite{Nielsen-Chuang-00})

Beside all that business the answer to the innocent question:
\begin{center}
  \textbf{what is a quantum algorithm ?}
\end{center}
is absolutely not known.

Let us, then, return to the discussion of the Computability Issue of  section\ref{sec:The distinction between mathematical-classicality and physical-classicality}
As far as the $ cell_{21} $ of the diagram\ref{di:diagram of computation} is concerned, let us return to the discussion whether
Quantum  Mechanics can lead to a violation Church-Turing's Thesis, i.e. it cannot be the case that:
\begin{equation*}
  Q_{\Phi} - C_{M} - \Delta_{0}^{0}-\stackrel{\circ}{MAP} ( \Sigma^{\star} , \Sigma^{\star} ) \; \supset \; REC-\stackrel{\circ}{MAP} ( \Sigma^{\star} , \Sigma^{\star} )
\end{equation*}
where by $ Q_{\Phi} \subset NC_{\Phi} $ we denote that the involved physically-nonclassical computational device obeys the Laws of \textbf{Nonrelativistic Quantum Mechanics} or those
of \textbf{Special-relativistic Quantum Mechanics}.

The problem has been recently analyzed by Michael Nielsen in the
following way:

given a not-recursive function $ f \in MAP( \Sigma^{\star} \, , \, \Sigma^{\star} ) $ let us consider the following operators on $ {\mathcal{H}}_{2}^{\star}$:
\begin{equation}
  \hat{f} \; := \; \sum_{\vec{x} \in \Sigma^{\star}}  f(\vec{x}) \, | \vec{x} > < \vec{x} |
\end{equation}
\begin{equation}
  \hat{U}_{f} \; := \;  \sum_{\vec{x} \in \Sigma^{\star}} | f(\vec{x}) > < \vec{x} |
\end{equation}

Let us then introduce the following:

\begin{definition} \label{def:first Nielsen's algorithm}
\end{definition}
FIRST NIELSEN'S ALGORITHM W.R.T. f ($ NIELSEN_{1}(f) $):
\begin{enumerate}
  \item take a closed quantum mechanical system with observables' algebra $ {\mathcal{B}} ( {\mathcal{H}}_{2}^{\star} ) $
  \item prepare it in the state $ | \vec{x} > $
  \item make a measurement of  $ \hat{f} $
  \item read the measurement's outcome
\end{enumerate}
\begin{definition} \label{def:second Nielsen's algorithm}
\end{definition}
SECOND NIELSEN'S ALGORITHM W.R.T. f ($ NIELSEN_{2}(f) $):
\begin{enumerate}
  \item take a closed quantum mechanical system with observables' algebra $ {\mathcal{B}} ( {\mathcal{H}}_{2}^{\star} ) $
  \item prepare it in the state $ | \vec{x} > $
  \item act on it the quantum gate $ \hat{U}_{f} $
  \item make a measurement of $ | \vec{x} > < \vec{x} | $
  \item read the measurement's outcome
\end{enumerate}

One has then the following:
\begin{theorem} \label{th:on the effective realizability of Nielsen's quantum algorithms}
\end{theorem}
ON THE EFFECTIVE-REALIZABILITY OF NIELSEN'S QUANTUM ALGORITHMS:

\begin{hypothesis}
\end{hypothesis}

$ NIELSEN_{1}(f) $ or $ NIELSEN_{2}(f) $ is effectively-realizable

\begin{thesis}
\end{thesis}
\begin{equation*}
  Q_{\Phi} - C_{M} - \Delta_{0}^{0}-\stackrel{\circ}{MAP} ( \Sigma^{\star} , \Sigma^{\star} ) \; \supset \; REC-\stackrel{\circ}{MAP} ( \Sigma^{\star} ,
  \Sigma^{\star})
\end{equation*}
\begin{proof}
$ NIELSEN_{1}(f) $ and  $ NIELSEN_{2}(f) $ compute the the not-recursive function f
\end{proof}

Theorem\ref{th:on the effective realizability of Nielsen's quantum algorithms} implies that assuming the preservation
of the Church's Thesis one can infer the existence of a new superselection-rule:
\begin{corollary} \label{cor:on Nielsen's superselection rule}
\end{corollary}
ON NIELSEN'S SUPERSELECTION RULE:

\begin{hypothesis}
\end{hypothesis}
\begin{equation*}
  Q_{\Phi} - C_{M} - \Delta_{0}^{0}-\stackrel{\circ}{MAP} ( \Sigma^{\star} , \Sigma^{\star} ) \; = \; REC-\stackrel{\circ}{MAP} ( \Sigma^{\star} ,
  \Sigma^{\star})
\end{equation*}
\begin{thesis}
\end{thesis}
\begin{equation*}
  \hat{f} \text{ is not an observable } \; \; \forall f \in  REC-\stackrel{\circ}{MAP} ( \Sigma^{\star} ,
  \Sigma^{\star})
\end{equation*}

\begin{remark} \label{rem:the mathematical peculiarity of Nielsen's superselection rule}
\end{remark}
THE MATHEMATICAL PECULIARITY OF NIELSEN'S SUPERSELECTION RULE

Though, according to the general definition  we gave in the remark\ref{rem:on superselection rules}, Nielsen's one is a superselection-rule, its mathematical structure is rather
peculiar

Let us analyze this issue starting from the Hilbert space axiomatizations of Quantum Mechanics and making w.r.t. it the same observations
that in remark\ref{rem:on superselection rules} we made concerning the Noncommutative Axiomatizations:

axiom\ref{ax:Hilbert space axiom on states} tells us that a pure  state of a quantum-mechanical system is represented by a ray
on an Hilbert space $ {\mathcal{H}} $ but doesn't say that any ray on  $ {\mathcal{H}} $ represents a pure physical state of the system.

In the same way axiom\ref{ax:Hilbert space axiom on observables}
tells us that an observable of a quantum-mechanical system is
represented by a self-adjoint operator on  $ {\mathcal{H}} $ but
it doesn't say that any self-adjoint operator on  $ {\mathcal{H}}
$  represents a physical observable.

Now the usual superselection structure of a quantum system QS
consists in the existence of a set $ \{ Q_{i} \}_{i \in I} $ of
mutually commutating observables of the system, called its
\textbf{superselection charges}, such that:
\begin{enumerate}
  \item a self-adjoint operator $ O $ is  an observable of the system iff:
\begin{equation}
  [ O \, , \, Q_{i} ] \, = \, 0 \; \; \forall i \in I
\end{equation}
  \item a ray $ | \psi > < \psi | $ on  $ {\mathcal{H}} $  is a physical pure state of the
  system iff:
 \begin{equation}
  [ | \psi > < \psi | \, , \, Q_{i} ] \, = \, 0 \; \; \forall i \in I
\end{equation}
\end{enumerate}
Introduced the following:
\begin{definition}
\end{definition}
SKEW INFORMATION OF $ \rho \in {\mathcal{D}} ({\mathcal{H}}) $
W.R.T. $ a \in {\mathcal{B}} ({\mathcal{H}}) $:
\begin{equation}
  I_{skew} ( \rho \, , \, a ) \; := \; \frac{1}{2} Tr( [ a \, , \,
  \rho^{\frac{1}{2}} ] \, [ \rho^{\frac{1}{2}} \, , \, a ] )
\end{equation}
one has then that \cite{Wightman-95}:
\begin{theorem} \label{th:on the rule of skew information w.r.t. superselection rules}
\end{theorem}
ON THE RULE OF SKEW INFORMATION W.R.T. SUPERSELECTION RULES:

\begin{hypothesis}
\end{hypothesis}
\begin{equation*}
  \rho \, \in \, {\mathcal{D}} ({\mathcal{H}})
\end{equation*}
\begin{thesis}
\end{thesis}
\begin{equation*}
  \rho \text{ is a physical state of QS} \; \Rightarrow \; I_{skew} ( \rho \, , \,
  Q_{i} ) \, = \, 0 \; \; \forall i \in I
\end{equation*}

The situation as to  Nielsen's Superselection Rule is, instead,
strongly different as it may be easily inferred observing that:
\begin{equation}
  [ \hat{I} \, , \, \hat{f} ] \; = \; 0 \; \; \forall f \in MAP (
  \Sigma^{\star} \, , \,  \Sigma^{\star} )
\end{equation}
so that, whichever putatitive candidate superselection charge:
\begin{equation}
  \hat{Q}_{Nielsen} \; = \; \sum_{\vec{x} \in \Sigma^{\star}} \sum_{\vec{y} \in \Sigma^{\star}} q_{Nielsen}
  (\vec{x} , \vec{y}) | \vec{x} > <  \vec{y} | \; \;
\end{equation}
cannot satisfy the condition:
\begin{equation}
 (  [  \hat{Q}_{Nielsen} \, , \, \hat{f} ] \; = \; 0 ) \;
 \Leftrightarrow \; f \in REC-\stackrel{\circ}{MAP} ( \Sigma^{\star} ,
  \Sigma^{\star})
\end{equation}

\smallskip

Remark\ref{rem:the mathematical peculiarity of Nielsen's
superselection rule} can lead to think that a suitable
formalization of Nielsen's Superselection Rule requires some kind
of effectivization of the compatibility-condition.

This requires the introduction of \textbf{relative recursivity},
i.e. of \textbf{recursivity w.r.t. oracles} \cite{Odifreddi-89}.

Given a partial function  $ f \, in \stackrel{\circ}{MAP} (
\Sigma^{\star} ,  \Sigma^{\star}) $:
\begin{definition} \label{def:relative recursivity of functions with a function as oracle}
\end{definition}
CLASS OF PARTIAL FUNCTIONS RECURSIVE IN f (ON STRINGS)
$(REC_{f}-\stackrel{\circ}{MAP} ( \Sigma^{\star} ,
\Sigma^{\star})) $:

the smallest class of partial functions:
\begin{enumerate}
  \item containing the initial functions:
\begin{align}
  {\mathcal{O}} & (x) \; := \; 0 \\
  {\mathcal{S}} & (x) \; := \; x+1 \\
  {\mathcal{I}}_{i}^{n} & (x_{1}, \cdots , x_{n}) \; := \; x_{i}
  \; \; i=1 , \cdots , n \, , \, n \in {\mathbb{N}} \\
  f
\end{align}
  \item closed under \textbf{composition}, i.e. the schema that given $
  \gamma_{1} , \cdots , \gamma_{m} , \psi $ produces:
\begin{equation}
  \varphi ( \vec{x} ) \; := \;  \psi (  \gamma_{1} ( \vec{x} ) ,
  \cdots ,  \gamma_{m} ( \vec{x} ) )
\end{equation}
  \item closed under \textbf{primitive recursion}, i.e. the schema
  that given $ \psi \, , \, \gamma $ produces:
\begin{align}
  \varphi( \vec{x} , 0 ) \; & := \; \psi ( \vec{x} )  \\
  \varphi( \vec{x} , y+1 ) \; & := \; \gamma ( \vec{x} , y , \phi(
  \vec{x}, y ))
\end{align}
  \item closed under \textbf{unrestricted $ \mu $-recursion}, i.e. the
  schema  that given $ \psi $ produces:
\begin{equation}
  \varphi( \vec{x} ) \; := \; \min \{ y \, : \, ( \psi( \vec{x} , z ) \downarrow \; \forall z \leq y  ) \: and
  \:  ( \psi( \vec{x} , y ) \, = \, 0  ) \}
\end{equation}
  where $ \varphi( \vec{x} ) \, := \, \uparrow $ if there is no
  such function
\end{enumerate}

Given a set $ S \, \subseteq \Sigma^{\star} $:
\begin{definition} \label{def:relative recursivity of functions with a set as oracle}
\end{definition}
CLASS OF PARTIAL FUNCTIONS RECURSIVE IN S (ON STRINGS)
\begin{equation}
  (REC_{S}-\stackrel{\circ}{MAP} ( \Sigma^{\star} ,
  \Sigma^{\star})) \; := \; (REC_{\chi_{S}}-\stackrel{\circ}{MAP} ( \Sigma^{\star} ,
  \Sigma^{\star}))
\end{equation}

Given an n-ary predicate $ R(x_{1} , \cdots , x_{n}) $ on $
\Sigma^{\star} $:
\begin{definition} \label{def:relative recursivity of predicates with a function as oracle}
\end{definition}
R IS RECURSIVE IN f:
\begin{equation*}
  \chi_{R} \; \in \; REC_{f}- MAP( \Sigma^{\star} ,
  \Sigma^{\star}))
\end{equation*}
\begin{definition} \label{def:relative recursivity of predicates with a set as oracle}
\end{definition}
R IS RECURSIVE IN S:
\begin{equation*}
  \chi_{R} \; \in \; REC_{S}- MAP( \Sigma^{\star} ,
  \Sigma^{\star}))
\end{equation*}

A standard pictorial  way of expressing the nature of the class of
functions $ REC_{f}-\stackrel{\circ}{MAP} ( \Sigma^{\star} ,
\Sigma^{\star}) $ introduced by Alan Turing is to say that they
are $ C_{\phi}-computable $ w.r.t. the \textbf{oracle} f.

Definition\ref{def:partial recursive functions on numbers} and
definition\ref{def:relative recursivity of functions with a
function as oracle} immediately imply the following:
\begin{theorem}
\end{theorem}
RECURSIVE ORACLES ARE USELESS:
\begin{equation}
   REC_{f}-\stackrel{\circ}{MAP} ( \Sigma^{\star} ,
   \Sigma^{\star}) \; = \;   REC-\stackrel{\circ}{MAP} ( \Sigma^{\star} ,
   \Sigma^{\star})  \; \; \forall f \in REC-\stackrel{\circ}{MAP} ( \Sigma^{\star} ,
   \Sigma^{\star})
\end{equation}

Relative-recursivity naturally induces a partial ordering over $
\stackrel{\circ}{MAP} ( \Sigma^{\star} ,
   \Sigma^{\star}) $; given $ f \, , \, g \, \in \, \stackrel{\circ}{MAP} ( \Sigma^{\star} ,
   \Sigma^{\star}) $:
\begin{definition} \label{def:Turing reducibility}
\end{definition}
f IS TURING REDUCIBLE TO g $ ( f \, \leq_{T} \, g ) $:
\begin{equation}
  f \, \in \, REC_{g}-\stackrel{\circ}{MAP} ( \Sigma^{\star} ,
\Sigma^{\star})
\end{equation}
\begin{definition} \label{def:Turing equivalence}
\end{definition}
f IS TURING EQUIVALENT TO g $ ( f \, \sim_{T} \, g ) $:
\begin{equation}
   f \, \leq_{T} \, g \; and \;  g \, \leq_{T} \, f
\end{equation}
Since $ \leq_{T} $ is a partial-ordering, $ \sim_{T} $ is an
equivalence relation so that we can introduce the following:
\begin{definition} \label{def:Turing degrees}
\end{definition}
TURING DEGREES $( {\mathcal{D}}_{T} \, , \, \leq_{T} )$:
\begin{align*}
  {\mathcal{D}}_{T} \; & := \; \frac{\stackrel{\circ}{MAP} ( \Sigma^{\star} ,
\Sigma^{\star})}{\sim_{T}} \\
  [f]_{T}  & \:  \leq_{T} \:[g]_{T}  \; := \; h  \: \leq_{T} \: l
  \; \; \forall h \in [f]_{T} , \forall l \in [g]_{T}
\end{align*}

Given then two functions $ f , g \, \in \, \stackrel{\circ}{MAP}
( \Sigma^{\star} , \Sigma^{\star}) $ we can introduce the
following:
\begin{definition} \label{def:effective commutator}
\end{definition}
EFFECTIVE COMMUTATOR OF $ \hat{f} $ AND $ \hat{g} $:
\begin{equation}
  [ \hat{f}  \, , \, \hat{g} ]_{EFF} \; := \;
  \begin{cases}
    [ \hat{f}  \, , \, \hat{g} ]   & \text{if $ f \sim_{T} g $}, \\
    \uparrow & \text{otherwise}.
  \end{cases}
\end{equation}
Let us introduce, furthermore, the following notion:
\begin{definition}
\end{definition}
EFFECTIVE SKEW INFORMATION OF $ \rho \in {\mathcal{D}}
({\mathcal{H}}) $ W.R.T. $ a \in {\mathcal{B}} ({\mathcal{H}}) $:
\begin{equation}
  I^{(EFF)}_{skew} ( \rho \, , \, a ) \; := \; \frac{1}{2} Tr( [ a \, , \,
  \rho^{\frac{1}{2}} ]_{EFF} \, [ \rho^{\frac{1}{2}} \, , \, a ]_{EFF} )
\end{equation}

\smallskip

\begin{conjecture}
\end{conjecture}
It is possible that the situation as to Nielsen's Superselection
Rule may then be recasted so that it resembles ordinary
superselection rules in the following sense:
\begin{enumerate}
  \item a  necessary condition for the self-adjoint operator  O  on $ {\mathcal{B}}( {\mathcal{H}}_{2}^{\star} ) $ to be an observable of the system is that:
\begin{equation}
  [ O \, , \, Q_{Nielsen} ]_{EFF} \, = \, 0
\end{equation}
  \item a a  necessary condition for the ray $ | \psi > < \psi | $ on  $ {\mathcal{H}}_{2}^{\star} $  to be a physical pure state of the
  system is that:
 \begin{equation}
  [ | \psi > < \psi | \, , \, Q_{Nielsen} ]_{EFF} \, = \, 0
\end{equation}
  \item a necessary condition for a density matrix  to be an
  observable is that:
\begin{equation}
  I^{(EFF)}_{skew} ( \rho \, , \, Q_{Nielsen}
  ) \, = \, 0
\end{equation}
\end{enumerate}

\smallskip

Let us now, anyway, move away the Hilbert space axiomatization
originally assumed by Nielsen and let us analyze his argument in
the general framework of the Noncommutative Axiomatization
introduced in section\ref{sec:Why to treat sequences of qubits one
has to give up the Hilbert-Space Axiomatization of Quantum
Mechanics}.

Given a quantum-mechanical system with observables' algebra A the
algebra generated by the superselection rule is nothing but $ A'
$.

The hypothesis that all the superselection-charges commute among
themselves may then be formalized as the following:
\begin{constraint}
\end{constraint}
CONSTRAINT OF COMMUTATIVE SUPERSELECTION RULES:
\begin{equation}
  {\mathcal{Z}}(A) \; = \; A'
\end{equation}

The previuosly discussed  peculiarity of Nielsen's Superselection
Rule is such that it doesn't fit in this picture as we will now
show, starting from the following:
\begin{definition} \label{def:Nielsen-computable part of a noncommutative space}
\end{definition}
NIELSEN-COMPUTABLE PART OF A:
\begin{equation}
  REC_{Nielsen}(A) \; := \; \{ a \in A \, : \, Sp(a) \subseteq REC({\mathbb{C}}) \}
\end{equation}
Given $ a , b \, \in \, A_{p.s.d.} $:
\begin{definition} \label{def:effective commutator of operators with discrete spectrum}
\end{definition}
EFFECTIVE COMMUTATOR OF a AND b:
\begin{equation}
  [ a \, , \, b ]_{EFF} \; := \;
  \begin{cases}
    [  a \, , \, b ] &  \text{if $ Sp(a) \, \sim_{T} \, Sp(b) $}, \\
    \uparrow & \text{otherwise}.
  \end{cases}
\end{equation}
Let us observe, anyway, that the generalization of
definition\ref{def:effective commutator of operators with discrete
spectrum} for arbitrary spectrum requires the introduction of
concepts lying outside the boundaries of $
C_{\Phi}-C_{M}$-Recursion Theory, i.e. the definition of the
Turing degrees in $ \Sigma^{\infty} $.

Demanding to the literature (e.g. the $ 5^{th} $-chapter of
\cite{Odifreddi-89} or \cite{Odifreddi-99b}, \cite{Slaman-99}) for
details it will be sufficient here to say that exactly as
definition\ref{def:relative recursivity of predicates with a set
as oracle} allows to introduce the partial ordering of relative
computability $ \leq_{T} $ over $ 2^{\Sigma^{\star}} $ and the
associated equivalence relation  $ \sim_{T} $ quotienting w.r.t.
which Turing degrees on $ 2^{\Sigma^{\star}} $ are defined, the
definition of a partial ordering relation of relative
computability $ \leq_{T} $ over $ 2^{\Sigma^{\infty}} $ and the
associated equivalence relation  $ \sim_{T} $ allows, by
quotienting, to define Turing degrees over $ 2^{\Sigma^{\infty}}
$, allowing to generalize the  definition\ref{def:effective
commutator of operators with discrete spectrum} of the effective
comutator from $ A_{p.s.d.} \; \times \; A_{p.s.d.} $ to the
whole  $ A \; \times \; A $.

Assuming that the Von Neumann algebra $ A \subseteq {\mathcal{B}}
({\mathcal{H}}) $  acts on the Hilbert space $ {\mathcal{H}} $
one would be tempted to suspect that a suitable algebraic
formalization of Nielsen's Superselection Rule requires the
intoduction of the following notion
\begin{definition} \label{def:effective commutant of a Von Neumann algebra}
\end{definition}
EFFECTIVE COMMUTANT OF A:
\begin{equation}
  (A')^{(EFF)} \; := \;  \{ a \in A \, : \, [ a , b ]_{EFF} \, = \, 0 \; \; \forall
  b \in B({\mathcal{H}}) \}
\end{equation}

\smallskip

Leaving aside, for the moment, Nielsen's Superselection Rule to
which we will return in the next section, we would now to analyze
an objection moved to Nielsen's analysis by Masanao Ozawa
\cite{Ozawa-98b} showing its nullity.

Let us start from the following well-known results of Mathematical
Logic \cite{Odifreddi-89}.
\begin{theorem} \label{th:on the representability in ZFC}
\end{theorem}
ON THE REPRESENTABILITY IN ZFC:

assuming the consistence of the formal system of Zermelo-Fraenkel
with the Axiom of Choice (ZFC) it follows that any function
representable in ZFC is recursive

\smallskip

\begin{theorem} \label{th:on the undecidability of the consistence ZFC from inside}
\end{theorem}
ON THE UNDECIDABILITY OF THE CONSISTENCE OF ZFC FROM INSIDE

the statement $ << $ \emph{ZFC is consistent} $ >> $ is
undecidable in ZFC

\smallskip

Ozawa observes that any book, article, review on the
recursive/not-recursive nature of Quantum Mechanics is written in
mathematical language and, then, its  statements  are formulas of
the formal system giving foundation to Mathematics, namely  ZFC.

This would be true, obviously, also for any statement of the form:
\begin{equation}
 s(QP,f) \; := \;  \emph{  $ << $ the quantum process QP computates the  function f $ >> $ }
\end{equation}
Ozawa's argument, then, runs as follow:
\begin{enumerate}
\item if    s(QP,f)  may be shown to be a physical truth, this implies  that, properly effectively-codified  as a numerical function, s(QP,f) must be representable in
ZFC
\item
\label{en:mathematica fides}
 inside ZFC, we cannot prove its consistence;  however for our mathematical and physical  work to be meaningful we must  assume the consistence of ZFC
\item
\label{en:false conclusion}
 but then, for the theorem\ref{th:on the representability in ZFC}, it follows that  f is a recursive function
\end{enumerate}
In other words, for Ozawa, the very fact that mathematically one
expresses the computability of a function by some physical device
implies its recursivity.

Both the argument  and the general conclusion, however, do not
appear to be correct: point \ref{en:false conclusion} does not
hold, in that it cannot be inferred from point\ref{en:mathematica
fides}, after which we may only infer that s(QP,f) must be a
recursive function.

Of course  recursivity of s(QP,f) does not  imply the recursivity
of f and the whole argument fails.

\smallskip

Indeed Masanao Ozawa himself seems to realize in \cite{Ozawa-98a}
that his reasoning that would lead to the automatic recursiveness
of a $ C_{M}$-map computed by a quantum computer must be wrong
somewhere;

In  section\ref{chap:Classical algorithmic randomness as classical
algorithmic incompressibility} we already mentioned how Ozawa
himself  acts in \cite{Ozawa-98a} in order of leave preserved
Church's Thesis; let us analyze the issue more precisely:
\begin{definition} \label{def:classical deterministic Turing machine}
\end{definition}
CLASSICAL DETERMINISTIC TURING MACHINE ( $ M \; := \; ( Q \, , \,
\Sigma \, , \, \delta )$):

a classical device:
\begin{itemize}
  \item whose hardware is composed by the following three elements:
\begin{enumerate}
  \item a \textbf{processor} consisting in a finite set Q of
  possible \textbf{internal states} containing two particular
  states: the \textbf{starting state} $ q_{START} $ AND THE \textbf{halting
  state} $ q_{HALT} $
  \item an infinite \textbf{tape} registering a binary-sequence
  $ \bar{t} \in ( \Sigma \bigcup \{ - \} ) ^{\infty} $ where $ - $
  is called the \textbf{empty symbol}
  \item a \textbf{head} whose position on the tape is parametrized
  by a variable h, having the chance of moving on it left and
  right
\end{enumerate}
  \item whose configuration space is:
\begin{equation}
  S \; := \; Q \, \times \, ( \Sigma \bigcup \{ - \} ) ^{\infty} \, \times
  \, {\mathbb{Z}}
\end{equation}
where the state:
\begin{equation}
  s(n) \; := \; ( q_{n} \, , \, T_{n} \, , \, h_{n} )
\end{equation}
is the configuration in which the \textbf{internal state} is q,
the tape contains the message:
\begin{equation}
  T_{n} \; = \; \{ T_{n}(i) \, , \, i \in {\mathbb{Z}} \}
\end{equation}
and the head is located on the $ h^{th} $ cell of the tape
 \item whose discrete-time dynamics is specified by a
 \textbf{local transition-function} $ \delta \, : \, Q \times \Sigma \times \{ -1 , 0 , 1
 \}$ through the following rule:
\begin{multline} \label{eq:evolution rule of a classical deterministic Turing machine}
  \delta ( q_{n} \, , \, T_{n} ( h_{n} ) ) \: = \: ( q \, , \, a
  \, , \, d ) \; \Rightarrow \\
  q_{n+1} \: = \: q_{n} \, , \, h_{n+1} \: = \: h_{n} + d \, , \,
  T_{n+1} (i) \, = \,
  \begin{cases}
    T_{n} (i)  & \text{if $ i \, \neq \, h_{n} $ }, \\
    a & \text{otherwise}.
  \end{cases}
\end{multline}
and by an \textbf{initial condition} of the form:
\begin{equation}
  s(0) \; = \; ( q_{START} \, , \, 0 \, , \, T_{INPUT} )
\end{equation}
where $ T_{INPUT} $ obeys the following conditions:
\begin{align}
  T_{INPUT} & (i)  \; = \; - \; \; \forall i < 0 \\
  \exists ! &  l(INPUT) \in {\mathbb{N}}_{+} \; : \; (  T_{INPUT}
  (i) = - \, \, \forall i > l(INPUT)
\end{align}
where $ l(INPUT) $ is called the \textbf{length} of the
\textbf{input}, this last notion been defined as:
\begin{equation}
  INPUT \; := \; \{ T_{INPUT}(i) \, , \, i \in \{ 0 , \cdots ,
  l(INPUT) \}
\end{equation}
\end{itemize}

At the first time the internal state gets the value $ q_{HALT} $
the output, constitued by a word of $ \Sigma^{\star} $ surrounded
by infinite $ - $'s from left and right, is read on the tape.

\smallskip

Let us now pass to Quantum Computation, introducing the following:
\begin{definition} \label{def:quantum Turing machine}
\end{definition}
QUANTUM TURING MACHINE $ ( \, \hat{M} \, := \, ( Q \, , \, \Sigma
\, , \, \delta \, ) )$

a quantum device:
\begin{itemize}
  \item whose hardware is the same of that of a classical
  deterministic Turing machine with alphabet $ \Sigma $ and set of
  internal states Q
  \item whose Hilbert space $ {\mathcal{H}} $ of quantum states is
  generated by the so called \textbf{computational basis} of $
  \hat{M} $:
\begin{equation}
  {\mathbb{E}} \; := \; \{ | q , h , T  > \, , \, q \in Q \, , \,
  h \in {Z} \, , \, T \in \Sigma^{\star} \}
\end{equation}
  \item whose discrete-time dynamics is, at the $ ( n + 1 )^{th} $
  step, made of two substeps:
\begin{enumerate}
  \item application to the current state $ | s(n) > $ of the
  unitary operator $ \hat{U} $ on $ {\mathcal{H}} $ identified by
  a suitable \textbf{local transition function} $ \delta \, : \, Q^{2} \, \times
  \, \Sigma^{2} \, \times \, \{ - 1 \, , \, 0 \, , \, 1 \} \: \mapsto \:
  {\mathbb{C}} $ through the relation:
\begin{equation}
  \hat{U} | q \, , \, h \, , T > \; := \; \sum_{a \in \Sigma \, , \, q_{2} \in Q \, , \, d \in \{ - 1 , 0 , 1
  \}} \delta ( q_{1} \, , \, T(h) \, , \, a \, , \, q_{2} \, , \,
  d ) | q_{2} \, , \, h+d \, , \, T_{h}^{a} >
\end{equation}
where:
\begin{equation}
  T_{h}^{a} (i) \; := \;
  \begin{cases}
    a & \text{if $ i = h $}, \\
    T(i) & \text{otherwise}.
  \end{cases} \; \; a \in \Sigma \, , \, h \in {\mathbb{Z}}
\end{equation}
  \item a measurement of the halting qubit:
\begin{equation}
  \hat{q}_{HALT} \; := \; | q_{HALT} > <  q_{HALT} |
\end{equation}
\end{enumerate}
\end{itemize}

\begin{remark} \label{rem:on local computation}
\end{remark}
ON LOCAL COMPUTATION:

Given a generic classical or quantum discrete-time dynamical
system M:
\begin{definition} \label{def:local dynamical system}
\end{definition}
M IS LOCAL:

in each temporal step it may be altered only a \textbf{finite
number} of \textbf{localized} cbits or qubits

\smallskip

Both classical and quantum Turing machines are local.

As to quantum Turing machines, their locality is essential to
guarantee that one never goes out from $
{\mathcal{H}}_{2}^{\star} $ with all the involved issues discussed
in  section\ref{sec:Why to treat sequences of qubits one has to
give up the Hilbert-Space Axiomatization of Quantum Mechanics}.

It may be finally worth to observe that , instead both the
Rasetti, Castagnoli, Vincenti's
\cite{Castagnoli-Rasetti-Vincenti-92}  and Nielsen's analysis
about eventual violations of Church's Thesis by Quantum Mechanics
involve \textbf{nonlocal quantum computations}

\smallskip

But here comes the following:
\begin{theorem}
\end{theorem}
OZAWA'S THEOREM:

\begin{hypothesis}
\end{hypothesis}
\begin{equation*}
  Q_{\Phi} - C_{M} - \Delta_{0}^{0}-\stackrel{\circ}{MAP} ( \Sigma^{\star} , \Sigma^{\star} ) \; = \; REC-\stackrel{\circ}{MAP} ( \Sigma^{\star} ,
  \Sigma^{\star})
\end{equation*}
\begin{equation*}
   \hat{M} \, := \, ( Q \, , \, \Sigma \, , \, \delta \, ) \text{ quantum Turing machine }
\end{equation*}
\begin{thesis}
\end{thesis}
\begin{equation*}
  \hat{M} \text{ physically possible } \; \Rightarrow \; Im(
  \delta ) \, \subseteq \, REC({\mathbb{C}})
\end{equation*}
\begin{proof}
Let us suppose, ad absurdum, that:
\begin{multline*}
  \exists \, ( q_{1} \, , \,  q_{2} \, , \, a_{1}  \, , \, a_{2}
  \, , \, d ) \, \in \, Q^{2} \times \Sigma^{2} \times \{ - 1 , 0
  , 1 \} \; : \\
  \delta ( q_{1} \, , \,  q_{2} \, , \, a_{1}  \, , \, a_{2}
  \, , \, d ) \, \in \, {\mathbb{C}} - REC({\mathbb{C}})
\end{multline*}
One could then compute $  |  \delta ( q_{1} \, , \,  q_{2} \, , \,
a_{1}  \, , \, a_{2}
  \, , \, d ) | $ as a relative frequency.

Since $  REC({\mathbb{C}}) $ is a field, one has that:
\begin{equation}
   \delta ( q_{1} \, , \,  q_{2} \, , \, a_{1}  \, , \, a_{2}
  \, , \, d ) \notin REC({\mathbb{C}}) \; \Rightarrow \; |  \delta ( q_{1} \, , \,  q_{2} \, , \, a_{1}  \, , \, a_{2}
  \, , \, d ) |^{2} \notin
  REC({\mathbb{R}}) \; \; \forall z \in {\mathbb{C}}
\end{equation}
Since the digits' sequence of $  |  \delta ( q_{1} \, , \,  q_{2}
\, , \, a_{1}  \, , \, a_{2}
  \, , \, d ) |^{2} $ may be seen as map on $ {\mathbb{N}} $ the
  thesis immediately follows
\end{proof}

\newpage
\section{From Church's Thesis to Pour El's Thesis}

Marian B. Pour-El and Jonathan Ian Richards have
\cite{Pour-El-Richards-89}, \cite{Pour-El-99} have developed a
very interesting extension of Computable Analysis consisting in a
Recursion Theory of Operators on Banach spaces.

Given a double sequence  $ \{ x_{n,k} \in \mathbb{R} \} $ and an
other sequence  $ \{ x_{n} \} $ of real numbers such that:
\begin{equation}
\lim_{ k \rightarrow \infty }  x_{n,k} = x_{n}  \forall n \in
\mathbb{N}
\end{equation}
\begin{definition} \label{def:recursive convergence}
\end{definition}
$ \{ x_{n,k} \} $ CONVERGES RECURSIVELY TO $ \{ x_{n} \} $ ($r-
\lim_{k \rightarrow \infty } x_{n,k} = x_{n} ) $:
\begin{multline}
\exists e \, \in REC-MAP ( {\mathbb{N}} \times {\mathbb{N}} \, ,
\, {\mathbb{N}}) :  \\
( k > e(n,N) \Rightarrow \mid r_{k} - x \mid
\leq  \frac{1}{2^{N}} )  \forall n \in { \mathbb{N} } , \forall N
\in { \mathbb{N} }
\end{multline}
Given a sequence of real numbers: $ \{ x_{n} \in {\mathbb{R}}
\}_{n \in { \mathbb{N} } } $:
\begin{definition} \label{def:recursive sequence of real numbers}
\end{definition}
$ \{ x_{n} \in {\mathbb{R}} \}_{n \in { \mathbb{N} } } $ IS
RECURSIVE:
\begin{equation}
\exists \{ r_{n,k} \in { \mathbb{Q} } \}_{n,k \in { \mathbb{N} } }
: \mid r_{n,k} - x_{n} \mid \leq \frac{1}{2^{k}}
\end{equation}

Given a sequence of complex numbers  $ \{ z_{n} \in {\mathbb{C}}
\}_{n \in { \mathbb{N} } } $:
\begin{definition}\label{def:recursive sequence of complex numbers}
\end{definition}
$ \{ z_{n} \in {\mathbb{R}} \}_{n \in { \mathbb{N} } } $ IS
RECURSIVE:
\begin{equation}
  \{ \Re (z_{n} ) \in {\mathbb{R}} \}_{n \in { \mathbb{N} } } \;
  and \{ \Im (z_{n}) \in {\mathbb{R}} \}_{n \in { \mathbb{N} } }
  \; \text{ are recursive }
\end{equation}

\begin{remark} \label{rem:the recursivity of a sequence of complex numbers is stronger than the recursivity of all its elements}
\end{remark}
THE RECURSIVITY OF A SEQUENCE OF COMPLEX NUMBER IS STRONGER THAN
THE RECURSIVITY OF ALL ITS ELEMENTS

Given a sequence  $ \{ x_{n} \} $ of complex numbers, the fact
that each element  of the sequence is recursive, and can,
conseguentially, be effectively approximated to any desired
degree of  precision  by a computer program $ P_{n} $ given in
advance, doesn't imply the recursivity of the whole sequence since
there might not exist a way of combining the sequence of programs
$ \{ P_{n} \} $  in an unique program P computing the whole
sequence $ \{ x_{n} \} $.

\smallskip

The starting point of the Pour El-Richard's Theory is the notion
of a \textbf{computability structure} on a Banach space B.

Owing to remark\ref{rem:the recursivity of a sequence of complex
numbers is stronger than the recursivity of all its elements} the
definition of a computability structure on B is made through a
proper specification of  the computable sequences in B and not,
simply, by  the specification of a suitable set of recursive
vectors.

\begin{definition} \label{def:computability structure on a Banach space}
\end{definition}
COMPUTABILITY STRUCTURE ON B:

 a specification of a subset $ {\mathcal{S}} $ of the set $ B^{\infty} $ of all
 the sequences  in B identified as the \textbf{set of the
 computable sequences on B} satisfying the following axioms:

\begin{axiom}  \label{ax:linear forms}
\end{axiom}
ON LINEAR FORMS:

\begin{hypothesis}
\end{hypothesis}
\begin{equation*}
   \{ | x_{n} > \}  \, , \,  \{ | y_{n} > \}  \; \in  \; {\mathcal{S}}
\end{equation*}
\begin{equation*}
   \{ \alpha_{n,k} \} \, , \,  \{ \beta_{n,k} \} \, \text{ recursive double
 sequences in } {\mathbb{C}}
\end{equation*}
\begin{equation*}
  d \; \in \; REC-MAP( {\mathbb{N}} \, , \, {\mathbb{N}} )
\end{equation*}
\begin{equation*}
  | s_{n} > \; := \; \sum_{k=0}^{d(n)} \alpha_{n,k} | x_{k} > + \beta_{n,k}
  | y_{k} >
\end{equation*}

\begin{thesis}
\end{thesis}
\begin{equation*}
  \{ | s_{n} > \} \in {\mathcal{S}}
\end{equation*}

\smallskip

\begin{axiom} \label{ax:limits}
\end{axiom}
ON LIMITS:

\begin{hypothesis}
\end{hypothesis}
\begin{equation*}
   \{ | x_{n,k} > \}   \text{ recursive double sequence in B } \; : \;
   r-\lim_{k \rightarrow \infty} | x_{n,k} > \, =  \,  |x_{n} >
\end{equation*}
\begin{thesis}
\end{thesis}
\begin{equation*}
  \{ | x_{n} > \} \in {\mathcal{S}}
\end{equation*}

\smallskip

\begin{axiom}  \label{ax:norms}
\end{axiom}
ON NORMS:

\begin{hypothesis}
\end{hypothesis}
\begin{equation*}
  \{ | x_{n} > \} \in {\mathcal{S}}
\end{equation*}

\begin{thesis}
\end{thesis}
\begin{equation*}
  \{ \| | x_{n} > \| \} \text{ is a  recursive sequence  in } {\mathbb{R}}
\end{equation*}

\smallskip

We will denote the set of all computability structures on a
Banach space B by COMP-ST(B).

\begin{remark}
\end{remark}
COMPUTABILITY STRUCTURE AS AN EFFECTIVIZATION OF BANACH'S
STRUCTURE:

The idea behind definition\ref{def:computability structure on a
Banach space}   lies in effectivizing the three ingredients a
Banach space is made of: a vector space V, a norm on V and the
condition of convergence of Cauchy's sequences.

Given a Banach space B endowed with a computability structure $
{\mathcal{S}} \; \in \;  COMP-ST(B) $:
\begin{definition} \label{Pour El-computable vectors of a Banach space}
\end{definition}
POUR EL - COMPUTABLE VECTORS OF B:
\begin{multline}
  REC_{Pour \; El} ( B \, , \, {\mathcal{S}} ) \; := \; \{ | \psi > \, \in \, B \; : \\
   (  | \psi > \, , \, | \psi
  > \, , \, \cdots ) \: \in \: {\mathcal{S}} \}
\end{multline}

We will  speak, more concisely, of  $ REC_{Pour \; El} ( B $ when
the assumed computability structure may be understood.

\smallskip

Unfortunately axiom\ref{ax:linear forms}
definition\ref{def:computability structure on a Banach space}
doesn't provide the axiomatic definition of a unique structure
for a Banach  space B  since in general, $ card[ COM-ST(B) \; >
\; 1 $.

This requires the existence of an additional condition by which
the univocity condition can be obtained.

Given $ \mathcal{S} \; \in \; COM-ST(B) $ on a Banach space B:
\begin{definition}
\end{definition}
EFFECTIVE GENERATING SET FOR B W.R.T. $ {\mathcal{S}} $:

 a computable sequence  $ \{ | e_{n} > \} \in {\mathcal{S}} $
whose linear span is dense in B

\begin{definition}
\end{definition}
B IS EFFECTIVELY-SEPARABLE:
\begin{equation}
  \exists \, \{ | e_{n} > \} \in {\mathcal{S}}  \text{ effective generating set of B w.r.t.
  } {\mathcal{S}}
\end{equation}

Pour-El and Richards  proved the following:
\begin{theorem} \label{th:theorem of stability}
\end{theorem}
THEOREM OF STABILITY:

\begin{hypothesis}
\end{hypothesis}
\begin{equation*}
   {\mathcal{S}}_{1} \, , \, {\mathcal{S}}_{2} \; \in \;
   COMP-ST(B)
\end{equation*}
\begin{equation*}
   \{ | e_{n} > \} \in {\mathcal{S}}_{1} \cap {\mathcal{S}}_{2} \text{ effective generating set }
\end{equation*}
\begin{thesis}
\end{thesis}
\begin{equation*}
   {\mathcal{S}}_{1} \; = \; {\mathcal{S}}_{2}
\end{equation*}

\smallskip

\begin{example} \label{ex:computability for continuous functions}
\end{example}
COMPUTABILITY FOR CONTINUOUS FUNCTIONS:

Given $ a \, , \, b \; \in \; REC-({\mathbb{R}}) \: : \: a \, <
\, b $ let us denote by  C[a , b] the set of all continuous
functions on the interval  $ ( a \, , \, b ) $.

C[a , b] is well known to be a Banach space w.r.t. the uniform
norm:
\begin{equation}
  \| f \| \; := \;\sup_{x \, \in \, [ a \, , \, b ]} | f(x) |
\end{equation}

It is natural to assume that the sequence $ \{ x^{n} \}_{n \in
{\mathbb{N}}} $ is computable.

Since it is then an effective generating set of C[a , b] for
theorem\ref{th:theorem of stability} it identifies a natural
computability structure on it

\smallskip

\begin{example}
\end{example}
$ L^{p} $-COMPUTABILITY

Given $ a \, , \, b \; \in \; REC-({\mathbb{R}}) \: : \: a \, <
\, b $ let us denote by $ L^{p} [a , b] $, with  $ p  \, \in \, [
1 \, , \, + \infty )$, the space of p-integrable functions, i.e.:
\begin{equation}
   L^{p} [a , b] \; := \; \{ f \, : \, \int_{a}^{b} dx | f(x)
   |^{p} \, < \, + \infty \}
\end{equation}
We will assume that $ p \in REC({\mathbb{R}}) \, \bigcap \, [ 1 \,
, \, + \infty ) $.

$ L^{p} [a , b] $ is well-known to be a Banach space w.r.t. the
norm:
\begin{equation}
  \| f \|_{p} \; := \; \int_{a}^{b} dx | f(x) |^{p}
\end{equation}
It is natural to assume that the sequence $ \{ x^{n} \}_{n \in
{\mathbb{N}}} $ is computable.

Since it is then an effective generating set of $ L^{p} [a , b] $
for theorem\ref{th:theorem of stability} it identifies a natural
computability structure on it

\smallskip

\begin{remark} \label{rem:the computability structures of quantum information-Hilbert space framework}
\end{remark}
THE COMPUTABILITY STRUCTURES OF QUANTUM INFORMATION: HILBERT
SPACE FRAMEWORK

Given a quantum-mechanical system with Hilbert space of states $
{\mathcal{H}} $ it is natural to assume that any complete basis
of eigenvectors of an effectively measurable  physical observable
is an effective generating set of $ {\mathcal{H}} $.

Since this is the case as to the qubits' string operators:
\begin{equation}
  \hat{q}^{n} \; := \; \bigotimes_{i=1}^{n} \hat{q} \; \; n \in {\mathbb{N}}
\end{equation}
(where  $ \hat{q} $ is the qubit operator defined in
eq.\ref{eq:right definition of the qubit operator}) it is then
natural to assume that the  computational basis $
{\mathbb{E}}_{\star} $ is an effective generating set of $
{\mathcal{H}}_{2}^{\bigotimes \star} $ that, by
theorem\ref{th:theorem of stability}, determines a computability
structure on $ {\mathcal{H}}_{2}^{\bigotimes \star} $ with which
we will assume it to be endowed form here and beyond.

The situation is, instead strongly subtler as to $
{\mathcal{H}}_{2}^{\bigotimes \infty} $ that, being \textbf{not
separable}, is clearly also\textbf{ non effectively separable}.

In our noncommutative  framework, anyway, the basic objects of
Quantum Information Theory are the noncommutative spaces $
\Sigma_{NC}^{\star} $ and $ \Sigma_{NC}^{\infty} $ of,
respectively, qubits' strings and qubits' sequences, so that what
would be relevant to us would be the identification of natural
computability structure on these spaces.

The analysis on how this may be performed requires the
introduction of some further notion of the Pour El Richards'
Theory

\smallskip

Given an Hilbert space $ {\mathcal{H}} $ endowed with a
computability structure $ {\mathcal{S}} \; \in \;
COMP-ST({\mathcal{H}}) $ and a closed linear operator T on $
{\mathcal{H}} $ \footnote{We recall that an operator T on a Banach
space B is called \textbf{closed} if its graph $ \Gamma ( T ) \;
:= \; \{ ( | \psi > , T | \psi > ) \, : \, | \psi > \in
HALTING(T) \, \} $ is a closed subset of $ {\mathcal{H}} \,
\times \, {\mathcal{H}} $.} :

\begin{definition}
\end{definition}
T IS EFFECTIVELY-DETERMINED:

it there exists  $ \, \{ | e_{n} > \in {\mathcal{S}} $ such that:
\begin{enumerate}
  \item
\begin{equation}
  \{ ( | e_{n} > , T |e_{n} > ) \} \; \in \; {\mathcal{S}} \times {\mathcal{S}}
\end{equation}
  \item
\begin{equation}
 span \{ ( | e_{n} > , T |e_{n} > ) \}  \; is \; dense \; in \;
 \Gamma (T)
\end{equation}
\end{enumerate}

We will denote the set of all effectively-determined linear
operators on $ {\mathcal{H}} $ w.r.t. the computability structure
$ {\mathcal{S}} $ as $ REC_{Pour \; El}-
 {\mathcal{O}} ({\mathcal{H}} \, , \,{\mathcal{S}} )$ or simply as
$  REC_{Pour \; El}- {\mathcal{O}} ({\mathcal{H}}) $ when the
assumed computability structure may be understood.

\smallskip

\begin{remark}
\end{remark}
EFFECTIVE DETERMINABILITY OF UNBOUNDED OPERATORS

for a bounded T operator on an Hilbert space $ {\mathcal{H}} $
one has that:
\begin{equation}
  HALTING(T) \; = \; {\mathcal{H}}
\end{equation}
and the notion of effective determinability simply requires that
the action of T on any Pour El-computable vector must be
effectively determinable; for an unbounded operator, anyway, one
has that:
\begin{equation}
  HALTING(T) \; \subset \; {\mathcal{H}}
\end{equation}
so that  we must be able to effectively determine if T halts on
the given Pour El-computable vector or not.

\smallskip

\begin{remark} \label{rem:the two strategies to assign computability on a noncommutative space}
\end{remark}
THE DOUBLE WAY INSIDE THE POUR EL-RICHARDS' THEORY OF SPECIFYING
COMPUTABILITY ON A NONCOMMUTATIVE SPACE

Given a Von Neumann algebra $ A \, \subseteq \, {\mathcal{B}} (
{\mathcal{H}}) $ acting on an Hilbert space $ {\mathcal{H}} $ let
us observe that in the Pour-El Richards' theory there exist two
ways of specifying which elements of A are computable:
\begin{enumerate}
  \item to assign a computability structure $ {\mathcal{S}}_{1} \in COMP-ST({\mathcal{H}})  $
  on $ {\mathcal{H}} $ and then to consider the effectively
  determined elements of it, resulting in $ A \, \bigcap \, REC_{Pour \;
  El}- {\mathcal{O}}( {\mathcal{H}} , {\mathcal{S}}_{1}) $
  \item to assign directly a computability structure $ {
  {\mathcal{S}}}_{2} \, \in \, COMP-ST(A) $
\end{enumerate}

\smallskip

\begin{remark} \label{rem:the computability structures of quantum information-noncommutative framework}
\end{remark}
THE COMPUTABILITY STRUCTURES OF QUANTUM INFORMATION:
NONCOMMUTATIVE FRAMEWORK

According to remark\ref{rem:the two strategies to assign
computability on a noncommutative space} there exist two ways of
specifying which elements of $ \Sigma_{NC}^{\star} $ and  $
\Sigma_{NC}^{\infty} $.

Let us start with $ \Sigma_{NC}^{\star} $; we can:
\begin{enumerate}
  \item consider the set $ REC_{Pour \; El}-
  {\mathcal{O}}({\mathcal{H}} \, , \, {\mathcal{S}}_{natural}) $,
  where $ S_{natural} \in COMP-ST( {\mathcal{H}}_{2}^{\bigotimes \star}) $ is the natural computability
  structure over $ {\mathcal{H}}_{2}^{\bigotimes \star} $ induced
  by the computational basis $ {\mathbb{E}}_{\star} $
  \item to identify directly a natural computability structure
  over $ \Sigma_{NC}^{\star} $
\end{enumerate}
In this case, anyway, these two strategies \textbf{partially
}collapse since a natural computability structure on $
\Sigma_{NC}^{\star} \; = \; {\mathcal{B}} (
{\mathcal{H}}_{2}^{\bigotimes \star}) $ may be immediately
derived from $ {\mathbb{E}}_{\star} $ considering the associated
system of projectors $ \{ \, | \vec{x}
> < \vec{x} | \, : \, \vec{x} \in \Sigma^{\star} \, \} $.

The situation is, anyway, more difficult as to $
\Sigma_{NC}^{\infty} $ since:
\begin{enumerate}
  \item the nonseparability of $ {\mathcal{H}}_{2}^{\bigotimes \infty}
  $ causes that, as we discussed in the remark\ref{rem:the computability structures of quantum information-Hilbert space
  framework}, the computational basis $ {\mathbb{E}}_{\infty} $
  doesn't induce a natural computability structure on it
  \item since $ \Sigma_{NC}^{\infty} \; \subset \;
{\mathcal{B}} ( {\mathcal{H}}_{2}^{\bigotimes \star}) $ the
specification of an effective generating basis of $
{\mathcal{H}}_{2}^{\bigotimes \star} $ doesn't induce, passing to
projectors, the specification of an effective generating set of $
\Sigma_{NC}^{\infty} $
\end{enumerate}

\smallskip

The main results of the Pour El-Richard's Theory are the
following theorems:

\begin{theorem} \label{th:on computability's preservation}
\end{theorem}
ON COMPUTABILITY'S PRESERVATION:

\begin{hypothesis}
\end{hypothesis}
\begin{equation*}
  X_{1} \; , \;  X_{2} \text{ Banach spaces}
\end{equation*}
\begin{equation*}
  {\mathcal{S}}_{i} \; \in \; COMP-ST( X_{i} ) \; \; i  \, = \, 1,2
\end{equation*}
\begin{equation*}
  \{ | e_{n} > \}  \text{ effective generating set of } X_{1}
\end{equation*}
\begin{multline*}
  T : X_{1}  \mapsto X_{2} \text{ closed linear operator  } \; :
  \\
  \{ | e_{n} > \} \, \subseteq \,  D(T) \: and \: \{ T | e_{n} >
  \} \in {\mathcal{S}}_{2}
\end{multline*}
\begin{thesis}
\end{thesis}
\begin{equation}
T ( REC_{Pour \; El} (X_{1})) \, \subseteq \,  REC_{Pour \; El}
(X_{2}) \; \Leftrightarrow \; T \; is \; bounded
\end{equation}

\smallskip

\begin{example} \label{ex:the wave-equation doesn't preserve computability}
\end{example}
UNCOMPUTABLE PROPAGATION OF WAVES:

Let us consider the Cauchy's problem for the wave-equation:
\begin{align} \label{eq:Cauchy's problem for the wave-equation}
  & ( \frac{  \partial^{2} }{ \partial x^{2} } \, + \,  \frac{\partial^{2} }{ \partial y^{2} } \, + \,   \frac{\partial^{2} }{ \partial z^{2} }  \, - \,  \frac{\partial^{2} }{ \partial t^{2} }) \, u \; = \; 0  \\
  u( x & ,y , z , 0) \;   \; \; = \; f( x , y, z ) \\
   & \frac{ \partial u}{ \partial t } ( x,y,z , 0 ) \; = \; 0
\end{align}
Introduced the cubes:
\begin{equation}
  D_{l} \; := \; \{ | x | \leq l \, , \, | y | \leq l \, , \, | z | \leq
  l \} \; \; l \in {\mathbb{R}}_{+}
\end{equation}
the immediate multidimensional generalization of
eq.\ref{ex:computability for continuous functions} allows to
infer that there exist a natural computability structure on any
Banach space $ C( D_{l} ) \, l \in {\mathbb{R}}_{+}  $ (endowed
with the uniform norm): that determined by the effective
generating set $ \{ x^{a} y^{b} z^{c} \, : \, a , b ,c \, \in \,
{\mathbb{N}} \} $.

Since eq.\ref{eq:Cauchy's problem for the wave-equation}
describes waves travelling with unitary velocity, one has that
for $ t \in ( 0 \, , \,  2 ) $ the solution of the wave equation
on $ D_{1} $ doesn't depend on the initial values $ u( x,y,z,0) $
outside $ D_{2} $.

Hence, we can look at the time-evolution operator $ T_{t} \, , \,
t \in ( 0 \, , \, 2 ) $:
\begin{equation} \label{eq:time evolution operator}
   u( \vec{x} , t ) \; := \; T_{t} \,  u(  \vec{x} , 0 ) \; \; t \in ( 0 \, , \, 2 )
\end{equation}
as a linear  operator from $ C(D_{2}) $ to $ C(D_{1}) $.

Explicitly:
\begin{equation}
   (T_{t} f )( \vec{x} ) \; = \; \int_{ \partial S^{(2)}} [ f(
  \vec{x} + t \vec{n} ) \, + \, t ( \vec{\nabla} f )(
  \vec{x} + t \vec{n} ) \cdot  \vec{n}  ] \, d \Omega ( \vec{n} )
\end{equation}

Owing to the gradient-term, $  T_{t} \, , \, t \in ( 0 \, , \, 2
) $ is unbounded; by theorem\ref{th:on computability's
preservation} it follows that there exist a Pour-El-computable
function $ f \; in \; REC_{Pour \; El}( C(D_{2})) $ such that $
T_{1} f \; \notin \; REC_{Pour \; El}(C(D_{2})) $.

So $ u( \vec{x} , 1 ) $ is not computable despite the
computability of the initial condition $ u( \vec{x} , 0 ) $.

\smallskip

\begin{example}
\end{example}
COMPUTABLE EUCLIDEAN EVOLUTION OF A QUANTUM FREE PARTICLE:

Let us consider the Cauchy's problem for the heat-equation:
\begin{align} \label{eq:Cauchy's problem for the heat equation}
  \frac{\partial u}{ \partial t } & \; = \; ( \frac{\partial^{2}}{\partial x^{2}} \, + \, \frac{\partial^{2}}{\partial y^{2}} \, + \,  \frac{\partial^{2}}{\partial z^{2}}) u \\
  u( x & ,y , z , 0) \;  = \; f( x , y, z )
\end{align}
Let us introduce the hyper-cubes:
\begin{equation}
  D_{\vec{a}} \; := \; \{ \vec{x} \in {\mathbb{R}}^{n} \, : \, |
  x_{i} | \leq a_{i} \; i = 1 , \cdots , n \} \; \; \vec{a} \in  {\mathbb{R}}_{+}^{n}
\end{equation}

It should be now clear that a natural computability structure on
any $ C (D_{ \vec{a} }) $ is that identified by the effective
generating set $ \{ \prod_{i=1}^{n} x_{i}^{n_{i}} \; \; n_{i} \in
{\mathbb{N}} \, , \, i = 1, \cdots , n \} $.

Let us  now consider the Banach spaces $ C_{0} ({\mathbb{R}}^{n})
$ of all the continuous compactly-supported functions over $
{{\mathbb{R}}^{n}} $ (endowed with the uniform norm): a
computability structure on it is specified by the condition that
a computable function $ f \in C_{0} ({\mathbb{R}}^{n}) $ must be
computable in any $ C (D_{ \vec{a} }) $ and:
\begin{equation}
  r-\lim_{ | \vec{x} | \rightarrow \infty} f( \vec{x} ) \; = \; 0
\end{equation}
The solution to eq.\ref{eq:Cauchy's problem for the heat
equation} is given by the time-evolution operator $ T_{t} : C_{0}
({\mathbb{R}}^{3}) \; \rightarrow \; C_{0} ({\mathbb{R}}^{4}) \;
t \geq 0 $:
\begin{align}
  (T_{t} & f) ( x,y,z ) \; := \; \int_{{\mathbb{R}}^{3}} dx' dy' dz' \, K_{t}( x -x' \, , \, y - y' \, , \,  z -z' ) \, f( x' , y' , z' )  \\
  K_{t} & ( x,y,z ) \; = \; ( \frac{1}{4 \pi t} )^{\frac{3}{2} }
  \exp( - \, \frac{ x^{2} + y^{2} + z^{2} }{4 \, t} )
\end{align}
Since  $ T_{t} $ is bounded it follows that, if the initial
condition is computable, $ u( x,y,z, t ) $ remains computable for
every time.

Let us observe, by the way, that eq.\ref{eq:Cauchy's problem for
the heat equation} may be seen as the euclidean Schr\"{o}dinger
equation for a free-particle of mass $ m \, = \, 2 $.

The Hilbert space $ {\mathcal{H} } \; = \; L^{2} (
{\mathbb{R}}^{3} \, , \, d \vec{x} ) $ may be endowed with a
natural computability structure in the same way such a business
was managed for $ C_{0}( {\mathbb{R}}^{3}) $;

introduced the natural computability structure on $ L^{2}(
{\mathbb{R}}^{3}) $ again defining previously the computability on
 any $  L^{2} ( D_{ \vec{a} } ) $ requiring that $ \{ \prod_{i=1}^{n} x_{i}^{n_{i}} \; \; n_{i} \in
{\mathbb{N}} \, , \, i = 1, \cdots , n \} $ is an effective
generating set and the defining a function $ f \; in \; REC_{Pour
\; El }( L^{2}( {\mathbb{R}}^{3})) $ if it belongs to any $
COMP-ST( L^{2} ( D_{ \vec{a} } )) $ and furthermore:
\begin{equation}
  r-\lim_{ | \vec{x} | \rightarrow \infty} f( \vec{x} ) \; = \; 0
\end{equation}
$ \{ T_{t} \}_{t \in {\mathbb{R}}_{+}} $ may then be seen as the
markovian strongly-continuous contraction semigroup
\cite{Fukushima-Oshima-Takeda-94} describing the euclidean
evolution of the $ m = 2 $ free particle preserving itself the
computability of the initial wave-function.

\smallskip

The next key  step of the Pour El - Richards' theory  is the
following:
\begin{theorem} \label{th:first theorem on the computability of the eigenvalues' sequence}
\end{theorem}
FIRST THEOREM ON THE COMPUTABILITY OF THE EIGENVALUES' SEQUENCE

\begin{hypothesis}
\end{hypothesis}
\begin{equation*}
  T \, : \, {\mathcal{H}} \, \rightarrow \,  {\mathcal{H}} \text{ effectively-determined self-adjoint operator }
\end{equation*}
\begin{thesis}
\end{thesis}

$ \exists \, \{ \lambda_{n} \}_{n \in {\mathbb{N}}} \; \in \;
COMP-ST({\mathcal{H}})  \; , \; A \subset {\mathbb{N}} \; r.e. $
such that:
\begin{enumerate}
  \item
\begin{equation}
   \lambda_{n} \in Sp(T) \; \forall n \in {\mathbb{N}}
\end{equation}
  \item
\begin{equation}
  Sp(T) \; = \; \overline{\{ \lambda_{n} \}_{n \in {\mathbb{N}}}}
\end{equation}
  \item
\begin{equation}
  Eigenvalues(T) \; = \; \{ \lambda_{n} \: , \: n \in  {\mathbb{N}} - A \}
\end{equation}
\end{enumerate}

\smallskip

\begin{theorem} \label{th:second theorem on the computability of the eigenvalues' sequence}
\end{theorem}
SECOND THEOREM ON THE COMPUTABILITY OF THE EIGENVALUES' SEQUENCE

\begin{hypothesis}
\end{hypothesis}
\begin{equation*}
  ( {\mathcal{H}} \, , \, {\mathcal{S}} ) \text{ Hilbert space endowed with the computability structure
  } \, {\mathcal{S}}
\end{equation*}
\begin{equation*}
  \{ \lambda_{n} \}_{n \in {\mathbb{N}}} \; \in \; COMP-ST({\mathbb{R}})  \; , \; A \subset {\mathbb{N}} \; r.e.
\end{equation*}
\begin{thesis}
\end{thesis}

$ \exists \, T \text{ effectively-determined operator on }
  {\mathcal{H}} \; : $
\begin{enumerate}
  \item
\begin{equation*}
  Sp(T) \, = \, Closure( \{ \lambda_{n} \}_{n \in {\mathbb{N}}} )
\end{equation*}
  \item
\begin{equation*}
  Eigenvalues(T) \; = \; \{ \lambda_{n} \: : \: n \in {\mathbb{N}} - A \}
\end{equation*}
  \item
\begin{equation*}
  \{ \lambda_{n} \}_{n \in {\mathbb{N}}} \, bounded \; \Rightarrow
  \; T \text{ can be chosen bounded }
\end{equation*}
\end{enumerate}

\smallskip

\begin{remark}
\end{remark}
WHEN EACH EIGENVALUE IS COMPUTABLE  BUT THE EIGENVALUES' SEQUENCE
IS NOT COMPUTABLE:

the meaning of theorem\ref{th:first theorem on the computability
of the eigenvalues' sequence} and  theorem\ref{th:second theorem
on the computability of the eigenvalues' sequence} is, roughly
speaking, that each single eigenvalue $ \lambda _{n} $ of an
effectively-determined operator is computable by a suitable
program $ P_{n} $, but these program cannot be fit together to
obtain a single program P computing the whole sequence $ \{
\lambda_{n} \} $.

\begin{corollary}
\end{corollary}
ON INCOMPUTABLE EIGENVALUES OF SHARP OBSERVABLES IN DISCRETE
NONCOMMUTATIVE SPACES:

\begin{hypothesis}
\end{hypothesis}
\begin{equation*}
  {\mathcal{S}}  \in COMP-ST({\mathcal{H}}) \; : \; {\mathcal{H}}
  \text{ Hilbert space }
\end{equation*}
\begin{thesis}
\end{thesis}
\begin{equation*}
  \exists T \in ({{\mathcal{B}}} ({{\mathcal{H}}}))_{sa} \; : \;
Eigenvalues(T) \, \notin \,  {\mathcal{S}}
\end{equation*}
\begin{corollary}
\end{corollary}
COMPUTABILITY OF THE EIGENVALUES OF NONCOMMUTATIVE INFINITESIMALS
IN DISCRETE NONCOMMUTATIVE SPACES:

\begin{hypothesis}
\end{hypothesis}
\begin{equation*}
  {\mathcal{S}}  \in COMP-ST({\mathcal{H}}) \; : \; {\mathcal{H}}
  \text{ Hilbert space }
\end{equation*}
\begin{equation*}
  T \in {\mathcal{C}} ({\mathcal{H}})
\end{equation*}
\begin{thesis}
\end{thesis}
\begin{equation*}
  Eigenvalues(T) \text{ (suitably ordered) is a recursive sequence of } {\mathbb{R}}
\end{equation*}

\smallskip

Finally, Pour El and Richard has proved the following:
\begin{theorem} \label{th:on the uncomputable eigenvector}
\end{theorem}
ON THE UNCOMPUTABLE EIGENVECTOR:

$ \exists \, T  $ compact, self-adjoint, effectively determined
operator on $  L^{2}( [0 ,1] ) $ (endowed with the natural
computability structure) such that:
\begin{equation*}
  Ker(T) \, \bigcap \, REC_{Pour \; El} ( L^{2}( [0 ,1] ) ) \; =
  \; \emptyset
\end{equation*}

\begin{example}
\end{example}
FREE PARTICLE ON A  INFINITE WELL AND ON THE CIRCLE:

Let us observe that, if not censored by a superselection-rule, one
would look at the operator T of theorem\ref{th:on the uncomputable
eigenvector}  as a physical observable for a free particle of
mass $ m = 2$ on the infinite well defined by the potential:
\begin{equation}
  V(x) \; := \;
  \begin{cases}
    0 & \text{if $ x \in (0 , 1 ) $}, \\
    + \infty & \text{otherwise}.
  \end{cases}
\end{equation}
whose hamiltonian is given by the self-adjoint operator on $
L^{2}( [0 ,1] )$:
\begin{align}
  (H & \, \psi) (x)  \; := \; - ( \frac{d^{2}}{d x^{2}} \psi) (x)  \\
  \psi & (0) \, = \, \psi (1) \, = \, 0
\end{align}
as well as for a particle moving on the circle $ S^{(1)} $ whose
Aharonov-Bohm coupling with the encircled magnetic flux tube $
\theta \in [ 0 \, , \, 2 \pi ) $  selects the self-adjoint
extension of the operator $  -  \frac{d^{2}}{d x^{2}} $  on $
C_{0}^{\infty} ( [0 ,1]  ) $ specified by the boundary condition
(cfr.  the $ 10^{th} $ chapter of \cite{Reed-Simon-75}, the $
23^{th} $ chapter of \cite{Schulman-81} and the $ 11^{th} $
chapter of \cite{Landi-97}):
\begin{equation}
  \psi(0) \; = \; e^{ i \theta} \psi(1)
\end{equation}

\smallskip

Up to this point we have discussed the Pour El-Richards' theory
from the pure mathematical side.

As to the \textbf{physical relevance} of the \textbf{mathematical
theory} of partial recursive functions, namely $ C_{M} $-Classical
Recursion Theory, it lies on:
\begin{enumerate}
  \item the \textbf{mathematical  observation} that all the
  different attempts of formalizing the notion of $
  C_{M}$-computability collapse to a unique notion
  \item the \textbf{physical observation} of the \textbf{experimental verification} of a \textbf{physical
  principle}: Church's Principle
\end{enumerate}
the eventual physical relevance of the \textbf{mathematical} Pour
El-Richards' theory should lie on:
\begin{enumerate}
  \item the \textbf{mathematical  observation} of a collapse to a
  unique mathematical structure of all the attempts to formalize
  the notion of Computability of \textbf{objects} belonging to
  Linear Algebra on Banach Spaces
  \item the \textbf{physical observation} of the \textbf{experimental verification} of a \textbf{physical
  principle} stating the nature of the objects \textbf{objects} of
  Linear Algebra on Banach Spaces that appear
  physically to be effectively computable w.r.t. the informal
  notion of effective computability
\end{enumerate}
The satisfaction of the first of these two requirements has been
explicitly  invoked  by Marian Pour El:

\begin{center}
  \textit{"Thus, we are able to achieve the intrinsic quality we associate with the notion
  of computability. The situation is reminiscent of the one in ordinary recursion theory, when the various definitions, proposed
  by Turing, Herbrand-G\"{o}del, Church, Post and others, all intuitively convincing, were proposed to be equivalent.
  The notion of a computability structure acts as a unifying concept, since seemingly different
  definitions of computability, are, in fact, equivalent because of this unicity. Thus we have a "Church's Thesis" for the given
  Banach space"; from  \cite{Pour-El-99} at pag.450}
\end{center}
The second point, strongly more relevant, leads us to analyze a
weakened form of such a putative physical principle, concerning a
unquestionably experimentally testable setting:
\begin{definition} \label{def:Pour El's thesis}
\end{definition}
POUR EL'S THESIS:

\begin{enumerate}
  \item the set of the effectively-computable elements of $ {\mathcal{H}}_{2}^{\bigotimes
  \star} $, w.r.t. to the informal notion of effective
  computability, is $ REC_{Pour \; El} ( {\mathcal{H}}_{2}^{\bigotimes
  \star} \, , \, {\mathcal{S}}_{natural} ) $, where $
  {\mathcal{S}}_{natural} $ is the natural computability structure
  introduced in the remark\ref{rem:the computability structures of quantum information-Hilbert space framework}
  \item the set of the effectively-computable elements of $
  \Sigma_{NC}^{\star} $, w.r.t. to the informal notion of effective
  computability, is $ REC_{Pour \; El} ( {\mathcal{H}}_{2}^{\bigotimes
  \star} \, , \, {\mathcal{S}}_{natural} ) $, where $
  {\mathcal{S}}_{natural} $ is the natural computability structure
  over $ \Sigma_{NC}^{\star} $ discussed in the remark\ref{rem:the computability structures of quantum information-noncommutative framework}
  \item the set of the effectively-computable elements of $ CPU(
  \Sigma_{NC}^{\star} ) $, w.r.t. to the informal notion of effective
  computability, is $ CPU(
  \Sigma_{NC}^{\star} ) \: \bigcap \: REC_{Pour \; El}- {\mathcal{O}} (
    {\mathcal{H}}_{2}^{\bigotimes
  \star} ) $, namely the set of the effectively-determined
    channels over $ \Sigma_{NC}^{\star} $.
\end{enumerate}

A first thing to do, in order of evaluating whether the two
mentioned conditions suggesting the assumption of Pour El's
Thesis are really satisfied, is to compare it with Nielsen's
Superselection Rule.

One has that:
\begin{theorem} \label{th:Nielsen's superselection rule versus Pour-El's thesis}
\end{theorem}
NIELSEN'S SUPERSELECTION RULE VERSUS POUR-EL'S THESIS:
\begin{equation}
  REC_{Pour \; El} (\Sigma_{NC}^{\star}) \; = \; REC_{Nielsen} (\Sigma_{NC}^{\star})
\end{equation}
\begin{proof}
Let us consider a basis $  {\mathbb{E}} \, := \, \{ | e_{n} >
\}_{n \in {\mathbb{N}}} $ of $ {\mathcal{H}}_{2}^{ \bigotimes
\star} $ such that:
\begin{equation}
  {\mathbb{E}} \; \notin \; COMP-ST( {\mathcal{H}}_{2}^{ \bigotimes
\star} )
\end{equation}
Given a recursive function $ f \, \in \, REC-MAP( {\mathbb{N}} \,
, \, {\mathbb{N}}) $ one has that:
\begin{equation}
  \hat{f} \; := \; \sum_{n=0}^{\infty} f(n) | e_{n} > < e_{n}|  \;
  \in \; REC_{Nielsen} (  \Sigma_{NC}^{\star} )
\end{equation}
though $ \hat{f} $    cannot be obtained as an effective linear
combination of the vectors of the computational basis
\end{proof}

Theorem\ref{th:Nielsen's superselection rule versus Pour-El's
thesis} suggest a negative feature of Nielsen's approach to
characterize the effectively computable elements of a
noncommutative space: to take into account only the spectrum is
not sufficient, since an operator can have recursive eigenvalues
w.r.t. an effectively-nondeterminable basis.

While the Nielsen's approach is more compelling from a physical
ground, arising from the anlaysis of the constraints required in
order Quantum Computation doesn't violate Church's Thesis, the
Pour El-Richards' Theory is, conseguentially, more refined from a
mathematical side.

As it happened in the thirties for $ C_{\Phi}-C_{M}
$-computability, we have to expect that the right answers will
arise comparing all the different pioneering attempts.

Among these it must be cited Robin Havea's work on Constructive
Operators' Theory \cite{Bridges-Havea-99}:

though, as it has been strongly stressed by Douglas S. Bridges,
the link between  \textbf{Constructive Mathematics} and
\textbf{Computability Theory} is rather subtle since if it is
true  that Constructive Mathematics  is based on the assumption:
\begin{equation} \label{eq:the assumption of Constructive Mathematics}
  EXISTENCE \; = \; COMPUTABILITY
\end{equation}
it is also true that the meaning of \textbf{computability} at the
r.h.s. of eq.\ref{eq:the assumption of Constructive Mathematics}
is something more radical than \textbf{recursive}, as it is
dramatically shown by Bridge's example concerning the recursive
function $ f : {\mathbb{N}} \, \rightarrow \,  {\mathbb{N}} $:
\begin{equation}
  f(n) \; := \;
  \begin{cases}
    0 & \text{if $ 2^{\aleph_{0}} \, = \, \aleph_{1}$}, \\
    1 & \text{otherwise}.
  \end{cases}
\end{equation}
that nobody would consider constructive owing to Cohen's
celebrated independence result \cite{Bridges-98}, yet such a link
does exist.

It is funny, with this respect, that the Constructive Theory of
Von Neumann's algebras is linked with a radicalization of the
meaning of the term \textbf{constructive} in the Constructive
Field Theory's program \cite{Glimm-Jaffe-87}, \cite{Jaffe-00}.

\smallskip

The  analysis on the concept of  quantum algorithm of the
previous sections should allow to charaterize the notion of
quantum algorithmic randomness as ownership of all the quantum
algorithmic typical properties in the following way:

given an algebraic probability space $ APS \, := \, ( A \, , \,
\omega ) $:
\begin{definition}
\end{definition}
COMMUTATIVE LAWS OF RANDOMNESS OF APS:
\begin{equation}
  {\mathcal{L}}_{RANDOMNESS}^{C} (APS) \; := \;  \{ p \in
  {\mathcal{P}}_{C}^{TYPICAL} \; Q_{\Phi}-\Delta_{0}^{0} \}
\end{equation}
\begin{definition}
\end{definition}
NONCOMMUTATIVE LAWS OF RANDOMNESS OF APS:
\begin{equation}
  {\mathcal{L}}_{RANDOMNESS}^{NC} (APS) \; := \;  \{ p \in
  {\mathcal{P}}_{NC}^{TYPICAL} \; Q_{\Phi}-\Delta_{0}^{0} \}
\end{equation}
It should then be finally possible to define $ RANDOM(
\Sigma_{NC}^{\infty} ) $ as the subset of all the elements of $
\Sigma_{NC}^{\infty} $ possessing all the laws of randomness of $
( \Sigma_{NC}^{\infty} \, , \, \tau_{unbaised} ) $.
\newpage
\section{Quantum Algorithmic Information Theory as a particular case of the abstract Uspensky's approach} \label{sec:Quantum Algorithmic Information Theory as a particular case of the abstract Uspensky's approach}
We saw in section\ref{sec:Karl Svozil's invention of Quantum
Algorithmic Information Theory} how Karl Svozil arrived to
introduce the idea of the quantum algorithmic information of the
qubits' strings' Hilbert space w.r.t. a quantum computer Q.

The consistence of such an approach lies on the existence of a
quantum analogue of theorem\ref{th:invariance theorem for prefix
algorithmic entropy} allowing to get rid of dependence from the
particular quantum computer Q.

This consideration suggests that the delineation of  Quantum
Algorithmic Information Theory should be made realizing it as a
particular instance of the Uspensky's abstract approach introduced
in lksection\ref{sec:Uspensky's abstract definition of algorithmic
information}.

The problem consists, clearly, in identifying:
\begin{itemize}
  \item a suitable aggregate $ A_{1} \, := \,  ( X_{1} \, , \, R_{1} ) $
  \item a suitable metric aggregate $ A_{2} \, := \, ( X_{2} \, , \, R_{2}  \, , \, \mu )  $
  \item a suitable universe of description of $ A_{2} $ through $ A_{1}
   \; : \;  {\mathcal{R}} \, \in \,  {\mathcal{D}}( A_{1} \, , \,
  A_{2}) $
\end{itemize}
The key point of such an approach is that it structurally founds
Quantum Algorithmic Information Theory, from the beginning, on:
\begin{enumerate}
  \item the mathematical characterization of the concept of \textbf{quantum algorithm}
  \item the  condition that there exist an \textbf{optimal quantum algorithm}
\end{enumerate}
Since, as we saw in section\ref{sec:Brudno algorithmic entropy
versus the Uspensky abstract approach}, even in the classical
case the applicability of the Uspensky's abstract approach to
sequences is strongly doubtful, also in the quantum case we will
apply it only for strings.

We will assume, conseguentially, that:
\begin{align}
  X_{1} & \; := \; \Sigma_{NC}^{\star} \\
  X_{2} & \; := \; \Sigma_{NC}^{\star}
\end{align}

\smallskip

The next key step  consists in assuming that the
\textbf{concordance relations} $ {\mathcal{R}}_{1} $ and $
{\mathcal{R}}_{2} $ are such that that the \textbf{modes of
descriptions} on $ \Sigma_{NC}^{\star} $ are the
\textbf{channels} on it, i.e.:
\begin{equation}
  {\mathcal{D}} ( A_{1} \, , \, A_{2} ) \; = \;
  CPU(\Sigma_{NC}^{\star})
\end{equation}

\begin{remark}
\end{remark}
QUANTUM ALGORITHMIC INFORMATION VERSUS QUANTUM REVERSIBLE
COMPUTATION

 Let us observe that, whichever choice we adopt for the
point measure $ \mu $ (e.g. the norm on $ \Sigma_{NC}^{\star} $
or some other candidate induced on it by duality from a distance
on S( $ \Sigma_{NC}^{\star}  $ ) such as a the noncommutative
geodesic distance w.r.t. some noncommutative riemannian metric)
one has that the resulting notion of  quantum algorithmic
information given by our particular instance of
definition\ref{def:algorithmic information}) is not-trivial iff
also logically-non reversible modes of descriptions, in the sense
of definition\ref{def:logical reversibility}, are taken into
account.

This immediately leads to the following:
\begin{constraint} \label{con:nontriviality's constraint on quantum algorithmic information theory}
\end{constraint}
NONTRIVIALITY'S CONSTRAINT ON QUANTUM ALGORITHMIC INFORMATION
THEORY:
\begin{equation}
  {\mathcal{R}} \; \nsubseteq \; AUT( \Sigma_{NC}^{\star} )
\end{equation}
The assumption of constraint\ref{con:nontriviality's constraint on
quantum algorithmic information theory} implies in particular
that:
\begin{corollary} \label{con:Quantum Algorithmic Information Theory is meaningful only for open systems}
\end{corollary}
QUANTUM ALGORITHMIC INFORMATION THEORY IS MEANINGFUL ONLY FOR
OPEN SYSTEMS:
\begin{equation}
  {\mathcal{R}} \; \nsubseteq \; INN( \Sigma_{NC}^{\star} )
\end{equation}
Corollary\ref{con:Quantum Algorithmic Information Theory is
meaningful only for open systems} tells us that Quantum
Algorithmic Information Theory is nontrivial only if  we allow
description methods corresponding, by axiom\ref{ax:noncommutative
axiom on closed dynamics}, to \textbf{open dynamics}.

\smallskip

As to the classical case, theorem\ref{th:simple algorithmic
information theory w.r.t. all the partial functions is not
meaningful}, was the first of a set of theorems showing that
certain quantities of Classical Algorithmic Information Theory
are meaningful only by effectivizing some notion.

As we saw in section\ref{sec:On absolute conformism in Quantum
Probability Theory} with regard to theorem\ref{th:not existence
of Kolmogorov random sequences of cbits}, the translation of
these results to the quantum domain is far from obvious.

In this case, anyway, it seems to us that the reasoning lying
behind theorem\ref{th:not existence of Kolmogorov random
sequences of cbits} may be suitably  translated concluding that
Algorithmic Information Theory w.r.t. $ CPU( \Sigma_{NC}^{\star}
) $ is not meaningful.

Requiring the effectiveness of the description modes, one is then
led to assume as universe of description the set of the
quantistically-computable channels over the noncommutative space
of qubits' strings:
\begin{equation}
  {\mathcal{R}} \; := \; Q_{\Phi} - NC_{M}-\Delta_{0}^{0}-CPU( \Sigma_{NC}^{\star} )
\end{equation}
In particular, assuming Pour-El's thesis, such an assumption means
that:
\begin{equation} \label{eq:universe of description of quantum algorithmic information theory assuming Pour-El's thesis}
  {\mathcal{R}} \; = \;  CPU( \Sigma_{NC}^{\star} ) \: \bigcap \: REC_{Pour \; El}-
{\mathcal{O}} ( {\mathcal{H}}_{2}^{\bigotimes \star} )
\end{equation}
The next step open problem consists in proving that Algorithmic
Information Theory is meaningful w.r.t. $  Q_{\Phi} -
NC_{M}-\Delta_{0}^{0}-CPU( \Sigma_{NC}^{\star} ) $.

\smallskip

\begin{remark}
\end{remark}
$ C_{\Phi}$-QUANTUM ALGORITHMIC INFORMATION THEORY VERSUS $
Q_{M}- C_{\Phi}$-CLASSICAL ALGORITHMIC INFORMATION THEORY

In section\ref{sec:The distinction between
mathematical-classicality and physical-classicality} we observed,
on analyzing the $ cell_{22} $ of the diagram\ref{di:diagram of
computation}, that Church's Thesis doesn't imply that the answers
to the question:
\begin{center}
   \textbf{what is computable?}
\end{center}
concerning the $ cell_{12} $  and the   $ cell_{22} $ of the
diagram\ref{di:diagram of computation} must be equal.

Denoting with $ Q_{M} $-computability the kind of $  NC_{M}
$-computability concerning noncommutative spaces and channels over
them, such an observation implies, in particular, that Church's
Thesis does not imply that $ C_{\Phi}-Q_{M}$-computability is
equal to $ Q_{\Phi}-Q_{M}$-computability.

This is particularly important to our purposes since if
Algorithmic Information Theory is defined in terms of Uspensky's
abstract approach, the condition:
\begin{equation}
  C_{\Phi}-Q_{M}-\Delta_{0}^{0} \; \neq \; Q_{\Phi}-Q_{M}-\Delta_{0}^{0}
\end{equation}
is necessary  and sufficient so that Quantum Algorithmic
Information Theory doesn't collapse to the Classical Algorithmic
Information Theory of $ Q_{M} $-quantities.

Such a collapse occurs, instead, assuming  Pour-El's thesis, from
which the violation of eq.\ref{eq:universe of description of
quantum algorithmic information theory assuming Pour-El's thesis}
may be immediately inferred.

\newpage
\chapter{Quantum algorithmic randomness and The Law of Excluded
Quantum Gambling Systems}
\section{On the frequentistic interpretation of Quantum
Probability} \label{sec:On the frequentistic interpretation of
Quantum Probability}
On introducing the concept of a
\textbf{quantum ensemble}, i.e. of a statistical ensemble
consisting of many individual quantum systems $ S_{1} \, , \,
\ldots \, , \, S_{n} $, in the section 4.1  \textit{"The
fundamental basis of the statistical theory"} of his basic 1932's
book, Von Neumann explicitely says that:
\begin{center}
\textit{  "Such ensembles, called collectives, are in general necessary
  for establishing probability theory as the theory of
  ferquencies. They were introduced by R. v. Mises, who discovered
  their meaning for probability theory, and who built up a
  complete theory on this foundation" note 156 of \cite{Von-Neumann-83}}
\end{center}
Indeed, in 1932, Von Mises' axiomatization of (classical)
probability was the only one available on the market, the now
standard  Kolmogorovian , measure-theoretic axiomatization
\cite{Kolmogorov-56} not been appeared yet.

As we saw in section\ref{sec:The problem of hidden points of a
noncommutative space}, on discussing the meaning of Bell's
Theorem, the irreducibility of Noncommutative Probability Theory
to the Commutative one may be seen, in an equivalent way, as the
impossibility, in general, of describing noncommutative random
variables on a \textbf{single classical probability space}, owing
to the lack of operational meaning of the joint moments.

With this respect Von Neumann thought to be able to mantain the
frequentistic interpretation of quantum probability inherited from
the frequentistic interpretation of classical probability getting
around such a problem,  in the following way:
\begin{center}
  \textit{"Even if two or more quantities R, S in a single system are not simultaneously measurable, their probability distribution
  in a given ensemble $ [ S_{1} \, , \, \cdots \, S_{N} ] $ can be obtained with arbitary accuracy
  if N is sufficiently large. Indeed, with an ensemble of N elements it suffices to carry out the statistical ispections, relative to the
  distribution of values of the quantity R, not on all N elements $ [ S_{1} \, , \, \cdots \, S_{N} ] $, but on any
  subset of M ( $ \leq N $ ) elements, say  $ [ S_{1} \, , \, \cdots \, S_{M} ] $, provided that M, N are both large,
  and that M is very small compared to N. Then only the M/N-th part of the ensemble is affected by the changes with result from the measurement.
  The effect is an arbitrary small one if M/N is chosen small enough - which is possible for sufficiently large N, even in the case
  of large M. $ \cdots $. In order to measure two (or several) quantities R,S simultaneously, we need two
  sub-ensembles, say $  [ S_{1} \, , \, \cdots \, S_{M} ] \; , \;   [ S_{M+1} \, , \, \cdots \, S_{2 \, M} ] \; ( 2 M \, \leq \, N ) $ of such a type that the first is employed obtaining the statistics of R, and the second
  is obtaining those of S. The two measurements therefore do not disturb each other, although they are performed in the same ensemble $ [ S_{1} \, , \, \cdots \, S_{N} ] $ and they can change this ensemble only by
  an arbitrarily small amount, if $ 2 M / N $ is sufficiently small, which is possible for sufficiently large N even in the
  case of large M  $ \cdots $". From the section4.1 of \cite{Von-Neumann-83}}
\end{center}
As it has been lucidly observed by Miklos Redei \cite{Redei-01},
implicit in this reasoning is the assumption that the subensembles
are representative of the larger ensemble in the sense that the
relative frequency  of every attribute is the same both in the
original and in the subensemble.

This not-trivial assumption is nothing but the requirement that
the a \textbf{quantum ensemble}, as a collective, statisfies the
\textbf{Law of stability of Statistic Relative Frequencies} and
the \textbf{Law of Excluded Quantum Gambling Strategies}, i.e.,
respectively, the Axiom of Covergence, namely axiom\ref{ax:axiom
of convergence}, and the Axiom of Randomness, namely
axiom\ref{ax:axiom of randomness}, of Von Mises's frequentistic
axiomatization of Classical Probability Theory, or some quantum
analogue of them.

As Miklos Redei observes:
\begin{center}
  \textit{"Von Neumann does not elaborate on the details and significance for his
  interpretation of quantum probability of the randomness requirement; apparentely
  hid did not see any problem with taking  advantage of this not trivial (and controversial) feature
  of Von Mises' interpretation." From \cite{Redei-01}}
\end{center}
One could then think of formalizing a frequentistic
axiomatization of Quantum Probability explicitely formulating
quantum analogues of the Axiom of Von Mises' Axiom of  Convergence
an the Axiom of Randomness, some way in the spirit of the
Hartle's and Lesniewski-Coleman's analysis discussed in
section\ref{sec:The unpublished ideas of Sidney Coleman and Andrew
Lesniewski}.

The corner-stone of such an axiomatization would be, clearly, the
Law of Excluded Quantum Gambling Systems, formalizing the
not-existence of winning gambling strategies in suitable quantum
casinos.

It must be observed, anyway, that the link of  such a
frequentistic formulation of Noncommutative Probability with the
noncommutative-measure theoretic one would be far from obvious,
owing to the fact that the space of states S(A) over a
noncommuative space A is not a Choquet's simplex, so that states
on A has multiple extremal decomposition.

\smallskip

We saw in chapter\ref{chap:Classical algorithmic randomness as
stability of the relative frequences under proper classical
algorithmic place selection rules} that, from inside the usual
measure-theoretic foundation of Classical Probability Theory, Von
Misese's theory results in the characterization of a notion of
classical algorithmic randomness, namely Church's randomness,
weaker than the Martin L\"{o}f-Solovay-Chaitin's one.

In an analogous way, our attitude in the next sections will
consist in attempting to extract the more possible information on
the set $ RANDOM( \Sigma_{NC}^{\infty} ) $ of random qubits'
sequences.
\newpage
\section{Different kind of quantum casinos}
Quantum Decision Theory was invented by P.A. Benioff
\cite{Benioff-72} and extensively developed by C.W. Helstrom
\cite{Helstrom-76} \cite{Holevo-99}. A renewed interest in such
field has recentely grown up in the framework of Quantum Game
Theory \cite{Eisert-Wilkens-Lewenstein-99}, \cite{Boukas-00}.

As in the classical case we can always interpret a quantum
decision problem as a quantum gambling situation, with the
utility function playing the rule of the payoff.

Let us consider a gambler going to a \textbf{\textit{Quantum
Casino}} in which the croupier, at each turn n , \textbf{throws a
quantum coin}.

Such a situation may be interpreted in different ways giving rise
to different types of \textit{Quantum Casinos}.
\begin{definition} \label{def:first kind quantum casino}
\end{definition}
FIRST KIND QUANTUM CASINO:

a quantum casino specified by the following rules:
\begin{enumerate}
  \item At each turn n the croupier extracts  with unbiased
probability a \textbf{pure state} $ | \psi > (n) \; \in \;
\mathcal{H}_{2} $ , where $  \mathcal{H}_{2} $ is the \textbf{one
qubit Hilbert space}.
  \item
Before each quantum coin toss the gambler can decide, according
to a \textbf{direct gambling strategy}, among the following
possibilities:

\begin{itemize}
  \item to bet one fiche on a vector $ | \alpha > \in
  \mathcal{H}_{2} $
  \item not to bet at the turn
\end{itemize}
  \item If he decides for the first option it will happens that:
\begin{itemize}
  \item he wins a fiche if the distance among $  | \psi > (n)  $ and $
  | \alpha > $ is less or equal to fixed quantity $
  \epsilon_{Casino} $.
  \item he loses the betted fiche if the  distance among $ | \psi > (n)   $ and $
  | \alpha > $ is greater than $ \epsilon_{Casino} $
\end{itemize}
\end{enumerate}

But it is also possible to see the result of a quantum coin toss
as a \textbf{mixed state}, resulting in the following:

\begin{definition} \label{def:second kind quantum casino}
\end{definition}
SECOND KIND QUANTUM CASINO:

a quantum casino specified by the following rules:
\begin{enumerate}
  \item At each turn n the croupier extracts  with unbiased (quantum) probability a density matrix $ \rho_{n} $ on the
\textbf{one qubit alphabet} $ \mathcal{H}_{2} $.

  \item Before each quantum coin toss the gambler can decide, according to a \textbf{direct gambling strategy}, among the
following possibilities:

\begin{itemize}
  \item to bet one fiche on a density matrix $ \sigma \in
  \mathcal{H}_{2} $
  \item not to bet at the turn
\end{itemize}
  \item If he decides for the first option it will happens that:
\begin{itemize}
  \item he wins a fiche if the distance among $ \sigma  $ and $
  \rho_{n}$ is less or equal to fixed quantity $
  \epsilon_{Casino} $
  \item he loses the betted fiche if the  distance among $ \sigma  $ and $
  \rho_{n}$ is greater than $ \epsilon_{Casino} $
\end{itemize}
\end{enumerate}

To complete the definition of first and second kind quantum
casinos  (definition\ref{def:first kind quantum casino} and
definition\ref{def:second kind quantum casino}) we have to
clarify:
\begin{enumerate}
  \item which is the adopted notion of distance among  states on
an Hilbert space $ \mathcal{H} $.
  \item what we mean by a \textbf{direct gambling strategy}
\end{enumerate}

In section\ref{sec:On the rule Noncommutative Measure Theory and
Noncommutative Geometry play in Quantum Physics} we saw that,
despite the noncommutative-information-geometric strategy of
considering the noncommutative geodesic distance w.r.t. a
suitable spectral triple, there exist two natural notions of
distance among two density operators on an Hilbert space, the
\textbf{ quantum trace distance} and the  \textbf{quantum angle
distance}, introduced respectively by definition\ref{def:naife
quantum trace distance} and definition\ref{def:naife quantum
angle distance}.

Fortunately they are qualitatively equivalent
\cite{Nielsen-Chuang-00}:
\begin{theorem}
\end{theorem}
QUALITATIVE EQUIVALENCE OF TRACE AND ANGLE DISTANCES ON STATES:
\begin{equation}
  1 -  F ( \rho_{1} \, , \, \rho_{2} ) \; \leq  \; D ( \rho_{1} \, , \, \rho_{2}
  ) \; \leq  \; \sqrt{1 -  F ( \rho_{1} \, , \, \rho_{2} )} \; \;
  \forall \, \rho_{1} \, , \, \rho_{2} \: \in \: {\mathcal{D}} (
  {\mathcal{H}}_{2})
\end{equation}
and, conseguentially, it doesn't matter which of them we use in
order to define a \textbf{second kind Quantum Casino}.

For the case we are interested to in which the underlying Hilbert
space is the \textbf{one qubit Hilbert space} $ {\mathcal{H}}_{2}
$, the adoption of the \textbf{trace distance} may be preferred
since it satisfies the following:
\begin{theorem}
\end{theorem}
QUANTUM TRACE DISTANCE IN TERMS OF THE BLOCH SPHERE:
\begin{equation}
   D (  Bloch ( \vec{r}_{1} )  , Bloch ( \vec{r}_{2} ) ) \; = \; \frac{ \| \vec{r}_{1} - \vec{r}_{2} \|
   }{2} \; \; \forall \vec{r}_{1} , \vec{r}_{2} \in Ball^{(2)}
\end{equation}
with:
\begin{definition} \label{def:Bloch sphere bijection}
\end{definition}
BLOCH SPHERE BIJECTION:

$ Bloch : Ball^{(2)} \; \rightarrow \; {\mathcal{D}}
({\mathcal{H}} _{2} )$ :
\begin{equation}
  Bloch ( \vec{r} ) \; := \; \frac{ {\mathbb{I}} \, + \, \vec{r} \cdot \vec{\sigma}}{2}
\end{equation}
where $ Ball^{(2)} \; := \; \{ \vec{r} \in {\mathbb{R}}^{3} \, :
\, \| \vec{r} \| \leq 1 \}  $ is the unit-radius 2-ball while $
\vec{\sigma} \; := \; \begin{pmatrix}
  \sigma_{x} \\
  \sigma_{y} \\
  \sigma_{z}
\end{pmatrix} $ is the vector of the Pauli matrices.

Let us observe that the extraction with unbiased probability of
an element of $ {\mathcal{D}} ({\mathcal{H}}_{2}) $  involved in
the definition\ref{def:second kind quantum casino} may be
reconducted, through the definition\ref{def:Bloch sphere
bijection}, to the extraction of a value of uniform-distributed
random point on the  unit radius 2-ball $  Ball^{(2)} $.

\bigskip

Let us, now, clarify what we mean by a \textbf{direct gambling
strategy}.

To make his decision at the $ n^{th} $ turn, the gambler can take
in consideration the result of all the previous n-1 quantum coin
tosses.

He can do this in two different ways:
\begin{itemize}
  \item he can think on the \textbf{direct products} of the
  previous outcomes; we will call such a strategy a \textbf{direct gambling strategy}
  \item he can think on the \textbf{tensor products} of the
  previous outcomes;  we will call such a strategy a \textbf{tensor gambling strategy}
\end{itemize}

In a \textbf{first kind} and \textbf{second kind Quantum Casino}
the gambler has to play according to a \textbf{direct gambling
strategy}.

We will introduce, later a \textbf{third kind of Quantum Casino},
in which the gambler has to play according to a \textbf{tensor
gambling strategy}.

The \textbf{direct gambling strategies} according to which the
gambler plays in a \textbf{first kind} and \textbf{second kind
Quantum Casino} will be called, respectively, \textbf{first kind}
and \textbf{second kind quantum gambling strategies} and defined
in the following way:
\begin{definition}
\end{definition}
FIRST KIND QUANTUM GAMBLING STRATEGY:

$ S : {\mathcal{H}}_{2}^{ \star} \stackrel {\circ}{\rightarrow}
{\mathcal{H}}_{2} $

\bigskip

\begin{definition}
\end{definition}
SECOND KIND QUANTUM GAMBLING STRATEGY:

$ S : \mathcal{D}({\mathcal{H}}_{2}^{ \star}) \stackrel
{\circ}{\rightarrow} \mathcal{D}({\mathcal{H}}_{2}) $

\bigskip

Let us now consider the sets $  {\mathcal{H}}_{2}^{\infty} $ and $
\mathcal{D}({\mathcal{H}}_{2})^{\infty} $ of sequences of,
respectively, \textbf{one qubit vectors} and \textbf{one qubit
density matrices}.

Our objective is to characterize two subsets $ {\mathcal{Q}}
{\mathcal{C}}ollectives^{1} \; \subset \;
{\mathcal{H}}_{2}^{\infty}  $ and  $ {\mathcal{Q}}
{\mathcal{C}}ollectives^{2} \; \subset \;
\mathcal{D}({\mathcal{H}}_{2})^{\infty} $, that we will call,
respectively, \textbf{first kind quantum collectives} and
\textbf{second kind quantum collectives}, defined by the condition
of satisfying Von Mises's axiom\ref{ax:axiom of randomness} when
the class of the \textbf{first kind quantum admissible gambling
strategies}  $ {\mathcal{Q}}{{\mathcal{S}}trategies}_{admissible}
( {\mathcal{Q}} {\mathcal{C}}ollectives^{1} ) $ and the class of
the \textbf{second kind quantum admissible gambling strategies}  $
{\mathcal{Q}}{{\mathcal{S}}trategies}_{admissible} (
{\mathcal{Q}} {\mathcal{C}}ollectives^{2} ) $ are chosen according
to a proper algorithmic-effectiveness characterization specular
to the classical one of eq.\ref{eq:Church admissible gambling
strategies}.

We arrive, conseguentially, to the following definitions:
\begin{definition}
\end{definition}
FIRST KIND QUANTUM COLLECTIVES:

$ {\mathcal{Q}} {\mathcal{C}}ollectives^{1} \; \subset \;
{\mathcal{H}}_{2}^{\infty}  $ induced by the axiom\ref{ax:axiom of
randomness} and the assumptions that the \textbf{first kind
quantum admissible gambling strategies}    are nothing but the
\textbf{quantum algorithms} on $   {\mathcal{H}}_{2}^{\infty} $:
\begin{equation}
  {\mathcal{Q}}{{\mathcal{S}}trategies}_{admissible}
( {\mathcal{Q}} {\mathcal{C}}ollectives^{1} ) \; := \;
Q_{\Phi}-\Delta_{0}^{0}-\stackrel{\circ}{MAP}(
{\mathcal{H}}_{2}^{\infty} )
\end{equation}

\begin{definition}
\end{definition}
SECOND KIND QUANTUM COLLECTIVES:

$ {\mathcal{Q}} {\mathcal{C}}ollectives^{2} \; \subset \;
\mathcal{D}({\mathcal{H}}_{2})^{\infty} $ induced by the
axiom\ref{ax:axiom of randomness} and the assumption that the
\textbf{second kind quantum admissible gambling strategies} are
nothing but the \textbf{quantum algorithms} on $
\mathcal{D}({\mathcal{H}}_{2})^{\infty}  $:
\begin{equation}
  {\mathcal{Q}}{{\mathcal{S}}trategies}_{admissible}
( {\mathcal{Q}} {\mathcal{C}}ollectives^{2} ) \; := \;
Q_{\Phi}-\Delta_{0}^{0}- \mathcal{D}({\mathbb{H}}_{2})^{\infty} )
\end{equation}

\smallskip

Let us finally introduce Third Kind Quantum Casinos.

Let us finally introduce the following notions:
\begin{definition} \label{def:algebraic quantum coin}
\end{definition}
 ALGEBRAIC QUANTUM COIN: a \textbf{quantum random variable} on the \textbf{quantum probability
 space} $ ( \, M_{2} (  {\mathbb{C}} )  \, , \, \tau_{2} ) $

 \medskip

\begin{definition} \label{def:third kind quantum casino}
\end{definition}
THIRD KIND QUANTUM CASINO:

a quantum casino specified by the following rules:
\begin{enumerate}
  \item At each turn n the croupier throws an \textbf{algebraic quantum coin} $ A_{n} $ obtaining a value $
  a_{n} \, \in \, \Sigma_{NC} $
  \item
Before each algebraic quantum coin toss the gambler can decide,
by adopting a \textbf{quantum gambling strategy}, among the
following possibilities:

\begin{itemize}
  \item to bet one fiche on an a letter  $ b  \,\in \,
  \Sigma_{NC} $
  \item not to bet at the turn
\end{itemize}
  \item If he decides for the first option it will happens that:
\begin{itemize}
  \item he wins a fiche if the distance among $  a_{n}  $ and b  $ d( a_{n} , b ) \; := \; \| a_{n} - b \| $ is less or equal to a fixed quantity $
  \epsilon_{Casino} $.
  \item he loses the betted fiche if the  distance among $  a_{n}  $  and
  a $ d( a_{n} , b ) \; := \; \| a_{n} - b \| $  is greater than $ \epsilon_{Casino} $
\end{itemize}
\end{enumerate}

\medskip

where the adoption of a \textbf{tensor gambling strategy} is
formalized in terms of the following notion:
\begin{definition} \label{def:third kind quantum gamling straregy}
\end{definition}
THIRD KIND QUANTUM GAMBLING STRATEGY:

$ S : \Sigma_{NC}^{ \star }  \stackrel {\circ}{\rightarrow} M_{2}
({\mathbb{C}}) $

\medskip

The concrete way in which the gambler applies, in every kind of
Quantum Casino, the chosen strategy S is always the same:
\begin{itemize}
  \item if S doesn't halt on the \textbf{previous game history} he doesn't bet
  at the next turn
  \item if S halts on on the past game history he bets  S(previous game
  history)

\end{itemize}

Lets us denote by  $ \bar{a}  \in \Sigma_{NC}^{ \infty} $ the
occured quantum sequence of qubits and with $ \vec{a}(n) \; :=
a_{1} \, \bigotimes  \, \cdots \, \bigotimes \, a_{n} \in
\Sigma_{NC}^{ n} $ its \textbf{quantum prefix of length n}, i.e.
the quantum string of the results of the first n quantum coin
tosses.

\begin{example}
\end{example}
BETTING ON PAULI MATRICES CHOOSING ACCORDING TO THE HEIGHT OF THE
UNBIASED QUANTUM PROBABILITY MEASURE

Let us consider the following \textbf{third kind quantum gambling
strategy}:
\begin{equation}
  S ( \vec{a}(n) ) \; := \;
  \begin{cases}
    \uparrow   & \text{if $ \vec{a}(n)  = \lambda $ }, \\
    \sigma_{x} & \text{if $ P_{unbiased} ( \vec{a}(n)^{\dagger} \vec{a}(n) ) \; = \; 0 $}, \\
    \sigma_{y} & \text{if $ P_{unbiased} ( \vec{a}(n)^{\dagger} \vec{a}(n) ) < 2^{n} $}, \\
    \sigma_{z} & \text{otherwise}.
  \end{cases}
\end{equation}
where $ \lambda $ denotes the empty quantum string.

Let us imagine that the results of the first three quantum coin
tosses are:
\begin{equation*}
  a(1) \; = \; \begin{pmatrix}
    5.21295-0.543424 I & -5.83373-1.51207 I \\
    -5.72507+5.64286 I & 0.264194-5.36408 I \
  \end{pmatrix}
\end{equation*}
\begin{equation*}
   a(2) \; = \; \begin{pmatrix}
     -2.21604-8.29818 I & 2.29687-9.22925 I \\
     -7.10612+4.25443 I & -8.19842+6.03258 I \
   \end{pmatrix}
\end{equation*}
\begin{equation*}
  a(3) \; = \; \begin{pmatrix}
    9.80519-7.0523 I & -7.72367-6.40421 I \\
    -0.227234+7.87254 I & 6.36604+6.81784 I \
  \end{pmatrix}
\end{equation*}
so that:
\begin{equation*}
  \vec{a}(1) \; = \; \begin{pmatrix}
    5.21295-0.543424 I & -5.83373-1.51207 I \\
    -5.72507+5.64286 I & 0.264194-5.36408 I \
  \end{pmatrix}
\end{equation*}
\begin{multline*}
   \vec{a}(2) \; = \; a(1) \, \bigotimes \, a(2) \; = \\ \begin{pmatrix}
     -16.0615-42.0538 I & 6.95806-49.3598 I & 0.380344+51.7601 I & -27.3546+50.3679 I \\
     -34.7319+26.0397 I & -39.4597+35.9027 I & 47.8881-14.0742 I & 56.949-22.7959 I \\
     59.5124+35.0029 I & 38.9296+65.799 I & -45.0976+9.69467 I & -48.8996-14.7589 I \\
     16.6759-64.4557 I & 12.8955-80.7994 I & 20.9437+39.2418 I & 30.1933+45.5707 I \
   \end{pmatrix}
\end{multline*}
\begin{multline*}
  \vec{a}(3) \; = \; a(1) \, \bigotimes \, a(2) \; \bigotimes \,
  a(3) \; = \\
  \begin{scriptsize}
    \begin{pmatrix}
    -454-299 I & -145+428 I & -280-533 I & -370+337 I & 369+505 I & 329-402 I & 87+687 I & 534-214 I \\
    335-117 I & 184-377 I & 387+66 I & 381-267 I & -408-8.8 I & -350+332 I & -390-227 I & -518+134 I \\
    -157+500 I & 435+21 I & -134+630 I & 535-25 I & 370-476 I & -460-198 I & 398-625 I & -586-189 I \\
    -197-279 I & -399-71 I & -274-319 I & -496-41 I & 100+380 I & 401+237 I & 167+454 I & 518+243 I \\
    830-76 I & -235-651 I & 846+371 I & 121-758 I & -374+413 I & 410+214 I & -584+200 I & 283+427 I \\
    -289+461 I & 140+629 I & -527+292 I & -201+684 I & -66-357 I & -353-246 I & 127-382 I & -211-427 I \\
    -291-750 I & -542+391 I & -443-883 I & -617+541 I & 482+237 I & 90-437 I & 617+234 I & 59-545 I \\
    504+146 I & 546-297 I & 633+120 I & 633-426 I & -314+156 I & -134+393 I & -366+227 I & -118+496 I \
    \end{pmatrix}
  \end{scriptsize}
\end{multline*}
where we have passed from four to zero decimal digits to save
space.

Gambler's evening to a third kind quantum casino may be told in
the following way:
\begin{itemize}
  \item at the beginning he has $ PAYOFF( 0 ) \; = \; 0 $ ; since
  at the first turn he doesn't bet we have obviously that $  PAYOFF( 1 ) \; = \; 0 $
  \item since:
\begin{multline*}
  P_{un} ( \begin{pmatrix}
    5.21295-0.543424 I & -5.83373-1.51207 I \\
    -5.72507+5.64286 I & 0.264194-5.36408 I \
  \end{pmatrix} ^{\dagger} \, \begin{pmatrix}
    5.21295-0.543424 I & -5.83373-1.51207 I \\
    -5.72507+5.64286 I & 0.264194-5.36408 I \
  \end{pmatrix} ) \\
   = \; P_{un} ( \begin{pmatrix}
    5.21295+0.543424 I & -5.72507-5.64286 I \\
    -5.83373+1.51207 I & 0.264194+5.36408 I \
  \end{pmatrix} \, \begin{pmatrix}
    5.21295-0.543424 I & -5.83373-1.51207 I \\
    -5.72507+5.64286 I & 0.264194-5.36408 I \
  \end{pmatrix} )\; = \\
   P_{unbiased} ( \begin{pmatrix}
    92.0884 & -61.3705+18.1664 I \\
    -61.3705-18.1664 I & 65.1619 \
  \end{pmatrix} ) \; = \; 157.25 \; > \; 2
\end{multline*}
he bets on $ \sigma_{z} $.

 \item since:
\begin{equation*}
  \| a(2) \, - \, \sigma_{z} \| \; = \; \| \begin{pmatrix}
    -3.21604-8.29818 I & 2.29687-9.22925 I \\
    -7.10612+4.25443 I & -7.19842+6.03258 I \
  \end{pmatrix}  \|  \; = \; 11.5984 \; > \; 10
\end{equation*}
he loses his fiche. Consequentially $ PAYOFF(1) \; = \; -1 $
  \item since:
\begin{multline*}
   P_{un} ( \begin{pmatrix}
     -16.0615-42.0538 I & 6.95806-49.3598 I & 0.380344+51.7601 I & -27.3546+50.3679 I \\
     -34.7319+26.0397 I & -39.4597+35.9027 I & 47.8881-14.0742 I & 56.949-22.7959 I \\
     59.5124+35.0029 I & 38.9296+65.799 I & -45.0976+9.69467 I & -48.8996-14.7589 I \\
     16.6759-64.4557 I & 12.8955-80.7994 I & 20.9437+39.2418 I & 30.1933+45.5707 I \
   \end{pmatrix} ^{\dagger} \\
    \begin{pmatrix}
     -16.0615-42.0538 I & 6.95806-49.3598 I & 0.380344+51.7601 I & -27.3546+50.3679 I \\
     -34.7319+26.0397 I & -39.4597+35.9027 I & 47.8881-14.0742 I & 56.949-22.7959 I \\
     59.5124+35.0029 I & 38.9296+65.799 I & -45.0976+9.69467 I & -48.8996-14.7589 I \\
     16.6759-64.4557 I & 12.8955-80.7994 I & 20.9437+39.2418 I & 30.1933+45.5707 I \
   \end{pmatrix} ) \\
   \; =  P_{un} ( \begin{pmatrix}
     -16.0615+42.0538 I & -34.7319-26.0397 I & 59.5124-35.0029 I & 16.6759+64.4557 I \\
     6.95806+49.3598 I & -39.4597-35.9027 I & 38.9296-65.799 I & 12.8955+80.7994 I \\
     0.380344-51.7601 I & 47.8881+14.0742 I & -45.0976-9.69467 I & 20.9437-39.2418 I \\
     -27.3546-50.3679 I & 56.949+22.7959 I & -48.8996+14.7589 I & 30.1933-45.5707 I \
   \end{pmatrix} \\
  \begin{pmatrix}
     -16.0615-42.0538 I & 6.95806-49.3598 I & 0.380344+51.7601 I & -27.3546+50.3679 I \\
     -34.7319+26.0397 I & -39.4597+35.9027 I & 47.8881-14.0742 I & 56.949-22.7959 I \\
     59.5124+35.0029 I & 38.9296+65.799 I & -45.0976+9.69467 I & -48.8996-14.7589 I \\
     16.6759-64.4557 I & 12.8955-80.7994 I & 20.9437+39.2418 I & 30.1933+45.5707 I \
   \end{pmatrix} ) \; = \\
P_{un} ( \begin{pmatrix}
     13110.4 & 14312.4+2902.95 I & -8737.18+2586.31 I & -10110.9+888.808 I \\
     14312.4-2902.95 I & 17870.7 & -8965.55+4758.04 I & -11909.6+3525.38 I \\
     -8737.18-2586.31 I & -8965.55-4758.04 I & 9276.95 & 10127.5+2054.13 I \\
     -10110.9-888.808 I & -11909.6-3525.38 I & 10127.5-2054.13 I & 12645.4 \\
   \end{pmatrix} ) \\
   = \; 26451.7 \; > \; 4
\end{multline*}
he bets on $ \sigma_{z} $.

 \item since:
\begin{equation*}
  \| a(3) \, - \, \sigma_{z} \| \; = \; \| \begin{pmatrix}
    8.80519-7.0523 I & -7.72367-6.40421 I \\
    -0.227234+7.87254 I & 7.36604+6.81784 I \
  \end{pmatrix} \| \; = \; 15.3175 \; > \; 10
\end{equation*}
he loses his fiche. Consequentially $ PAYOFF(2) \; = \; -1 $

\item since:
\begin{equation*}
  P_{un} ( \vec{a}(3)^{\dagger} \vec{a}(3) ) \; = \; 6.97591 \, 10^{6}
  \; > \; 8
\end{equation*}
he bets on $ \sigma_{z} $.

 \item since:
\begin{equation*}
  \| a(4) \, - \, \sigma_{z} \| \; = \; \| \begin{pmatrix}
    3.55982-1.58403 I & 2.19976-1.67009 I \\
    0.284886+2.77311 I & -7.06443-6.30601 I \
  \end{pmatrix} \| \; = \; 10.0665 \; > 10
\end{equation*}
he loses his fiche. Consequentially $ PAYOFF(3) \; = \; -2 $

\item since:
\begin{equation*}
  P_{un} ( \vec{a}(4)^{\dagger} \vec{a}(4) ) \; = \;  7.5079 \,
  10^{8} \; > \;  16
\end{equation*}
he bets on $ \sigma_{z} $.

\item since:
\begin{equation*}
   \| a(4) \, - \, \sigma_{z} \| \; = \; \| \begin{pmatrix}
     -8.49908+1.07129 I & -0.361299-7.07676 I \\
     9.60704+6.81686 I & -1.16288-3.10934 I \
   \end{pmatrix} \| \; = \; 14.1717 \; > \; 10
\end{equation*}
he loses his fiche. Consequentially $ PAYOFF(3) \; = \; -3 $
\end{itemize}

\bigskip

Exactly as it happened for the other kinds of Quantum Casinos,
the notion of a \textbf{third kind Quantum Casino} induces
naturally the notion of a \textbf{third kind collective}:

\begin{definition}
\end{definition}
THIRD KIND QUANTUM COLLECTIVES:

$ {\mathcal{Q}} {\mathcal{C}}ollectives^{3} \; \subset \;
\Sigma_{alg}^{\bigotimes \infty} $ induced by the
axiom\ref{ax:axiom of randomness} and the assumption that the
\textbf{third kind quantum admissible gambling strategies} are
nothing but the \textbf{quantum algorithms} on $
\Sigma_{alg}^{\bigotimes \infty}  $:
\begin{equation}
  {\mathcal{Q}}{{\mathcal{S}}trategies}_{admissible}
( {\mathcal{Q}} {\mathcal{C}}ollectives^{3} ) \; := Q_{\Phi}-
\Delta_{0}^{0}- \stackrel{\circ}{MAP}( \Sigma_{NC}^{\infty})
\end{equation}

\newpage

\section{The censorship of winning  quantum gambling strategies}
As we explained section\ref{sec:On the frequentistic
interpretation of Quantum Probability} our excursion in Quantum
Gambling Theory is motivated by the assumption of the following
constraing on $ RANDOM( \Sigma_{NC}^{\infty} ) $:
\begin{constraint}
\end{constraint}
QUANTUM ALGORITHMIC RANDOMNESS IS STRONGER THAN OWNERSHIP TO
THIRD KIND QUANTUM COLLECTIVES

\begin{equation}
   {\mathcal{Q}} {\mathcal{C}}ollectives^{3} \; \supseteq \; RANDOM_{NC}( \Sigma_{NC}^{\infty})
\end{equation}

Can we give an assurance to a third kind Quantum Casinos' owner
that in the long run he doesn't risk anything?

The positive answer is stated by the following:
\begin{conjecture}
\end{conjecture}
LAW OF EXCLUDED QUANTUM GAMBLING STRATEGIES FOR THIRD KIND QUANTUM
CASINOS

For $ n \rightarrow \infty $ the set of  the \textit{lucky-winning
strategies} tends to the null set $ \forall \epsilon_{Casino} \in
{ \mathbb{R}}_{+} $.
\part{Classical Algorithmic Randomness of the results of
quantum measurements} \label{part:Classical Algorithmic
Information Theory of the results of quantum measurements}
\newpage
\appendix
\chapter{Mathematica package for Quantum Algorithmic Information Theory}
Many of the concept discussed in this work may be concretelly
implemented by the alleged Mathematica notebook
\textit{Noncommutative Algorithmic Information Theory.nb} not
incorporated in the text owing to the bad behaviour of TexSave's
outputs on greek letters.

It is not a package (in the sense of section2.6.10 of
\cite{Wolfram-96}).

Furthermore many of instructions will still require some debugging
process.

\end{document}